\documentclass{tADP2e}
\usepackage{amsmath,amssymb,latexsym,subfigure,rotating,multirow,xfrac}
\numberwithin{equation}{section}
\numberwithin{figure}{section}
\numberwithin{table}{section}

\begin{document}

\markboth{P. Monceau}{Advances in Physics}

\title{Electronic crystals: an experimental overview}

\author{P. Monceau\thanks{mail: pierre.monceau@grenoble.cnrs.fr
\vspace{6pt}} \\ \vspace{6pt}  {\em{Institut N\'eel, Dept. MCBT, CNRS/University Joseph Fourier, BP 166,\\ 38042 Grenoble cedex 9, France}}\\ \vspace{6pt}
\received{\bf Preprint submitted in January 2012, to be published in Vol. 61, No 4, July-August 2012, p. 325-581 (2012)}
}

\maketitle

\begin{abstract}
This article reviews  the static and dynamic properties of spontaneous superstructures formed by electrons. Representations of such electronic crystals are charge density waves and spin density waves in inorganic as well as organic low dimensional materials. A special attention is paid to the collective effects in pinning and sliding of these superstructures, and the glassy properties at low temperature. Charge order and charge disproportionation which occur in organic materials resulting from correlation effects are analysed. Experiments under magnetic field, and more specifically field-induced charge density waves are discussed. Properties of meso- and nanostructures of charge density waves are also reviewed.

\begin{keywords}
Strongly correlated electronic systems; metal-insulator transition; one-dimensional conductors; charge density wave; spin density wave; quantum solids; collective modes; Fermi surface instabilities
\end{keywords}

\centerline{\bfseries Contents}\medskip

\hbox to \textwidth{\hsize\textwidth\vbox{\hsize18pc
\hspace*{-12pt} {1.}    Introduction\\
{2.}    Basic fundamental notions\\
\hspace*{10pt}{2.1.}   Peierls transition\\
\hspace*{10pt}{2.2.}   1D susceptibility\\
\hspace*{10pt}{2.3.}   Fr\"ohlich Hamiltonian\\
\hspace*{10pt}{2.4.}   Extended Hubbard model\\
\hspace*{10pt}{2.5.}   1D electron gas\\
\hspace*{10pt}{2.6.}   Tomonaga-Luttinger liquids\\
\hspace*{10pt}{2.7.}   Strong coupling limit\\
\hspace*{10pt}{2.8.}   Spin-Pieierls\\
\hspace*{10pt}{2.9.}   Intermediate Coulomb interactions:\\ \hspace*{27pt} charge ordering\\
\hspace*{10pt}{2.10.}  Coupling with the lattice\\
\hspace*{10pt}{2.11.}  Imperfect nesting\\
\hspace*{10pt}{2.12.}  Fluctuations\\
\hspace*{10pt}{2.13.}  Fr\"ohlich conductivity from moving\\  \hspace*{27pt} lattice waves\\
\hspace*{10pt}{2.14.}  Incommensurate phases in dielectrics\\
\hspace*{10pt}{2.15.}  Electronic ferroelectricity\\
{3.}    Materials\\
\hspace*{10pt}{3.1.}  MX$_3$ compounds\\
\hspace*{24pt} {3.1.1.}   NbSe$_3$\\
\hspace*{24pt} {3.1.2.}   TaS$_3$\\
\hspace*{24pt} {3.1.3.}   NbS$_3$\\
\hspace*{24pt} {3.1.4.}   ZrTe$_3$\\
\hspace*{10pt}{3.2.}  Transition metal tetrachalcogenides\\  \hspace*{24pt} (MX$_4$)$_n$Y\\
\hspace*{24pt} {3.2.1.}   (TaSe$_4$)$_2$I\\
\hspace*{24pt} {3.2.2.}   (NbSe$_4$)$_3$I\\
\hspace*{24pt} {3.2.3.}   (NbSe$_4$)$_{10}$I$_3$\\
\hspace*{24pt} {3.2.4.}   Tetratellurides\\}\hspace*{-24pt}\vbox{\noindent\hsize18pc
\hspace*{10pt} {3.3.}   Pressure effects\\
\hspace*{10pt} {3.4.}   Lattice dynamics\\
\hspace*{24pt} {3.4.1.}   (TaSe$_4$)$_2$I\\
\hspace*{24pt} {3.4.2.}   (NbSe$_4$)$_3$I\\
\hspace*{24pt} {3.4.3.}   NbSe$_3$\\
\hspace*{24pt} {3.4.4.}   Phonon Poiseuille flow\\
\hspace*{10pt}{3.5.}  A$_{0.3}$MoO$_3$ A: K, Rb, Tl\\
\hspace*{10pt}{3.6.}  Organic CDWs\\
\hspace*{24pt} {3.4.1.}   TTF-TCNQ\\
\hspace*{24pt} {3.4.2.}   (Per)$_2$M(mnt)$_2$ salts\\
\hspace*{24pt} {3.4.3.}   (Fluoranthene)$_2$X\\
\hspace*{10pt}{3.7.}  Bechgaard-Fabre salts\\
\hspace*{24pt} {3.7.1.}   Structure and low T-ground\\  \hspace*{47pt} state\\
\hspace*{24pt} {3.7.2.}   Charge order and ferroelectric\\ \hspace*{47pt} Mott-Hubbard ground state\\
\hspace*{24pt} {3.7.3.}   SDW amplitude\\
\hspace*{24pt} {3.7.4.}   Magnetic structure in\\ \hspace*{47pt} Bechgaard salts\\
\hspace*{24pt} {3.7.5.}   Superconductivity\\ \hspace*{47pt} under pressure\\
\hspace*{24pt} {3.7.6.}   Coexistence superconductivity\\  \hspace*{47pt} and C/S DWs\\
{4.}    Properties of the sliding density wave\\
\hspace*{10pt}{4.1.}   General properties\\
\hspace*{10pt}{4.2.}  Theoretical models for density wave\\  \hspace*{24pt} sliding\\
\hspace*{24pt} {4.2.1.}   Classical equation of motion\\
\hspace*{24pt} {4.2.2.}   Phase Hamiltonian\\
\hspace*{24pt} {4.2.3.}   Thermal fluctuations\\
\hspace*{24pt} {4.2.4.}   Numerical simulations\\
}}
\hbox to \textwidth{\hsize\textwidth\vbox{\hsize18pc
\hspace*{24pt} {4.2.5.}   Quantum models\\ 
\hspace*{24pt} {4.2.6.}   Quantum corrections for CDW\\  \hspace*{47pt} motion\\
\hspace*{24pt} {4.2.7.}   Local interference of local and\\  \hspace*{47pt} collective pinnings\\
\hspace*{24pt} {4.2.8.}   Pinning in SDW\\
\hspace*{10pt}{4.3.}   Experimental results on density wave\\  \hspace*{24pt} sliding\\
\hspace*{24pt} {4.3.1.}   Threshold fields\\
\noindent \hspace*{24pt} {4.3.2.}   Density wave drift velocity\\
\hspace*{10pt}{4.4.}   NMR spin echo spectroscopy\\
\hspace*{24pt} {4.4.1.}   Motional narrowing\\
\hspace*{24pt} {4.4.2.}   Phase displacement of the CDW\\  \hspace*{47pt} below $E_T$\\
\hspace*{10pt}{4.5.}   CDW long range order\\
\hspace*{24pt} {4.5.1.}   CDW domain size\\
\hspace*{24pt} {4.5.2.}   Friedel oscillations and CDW\\  \hspace*{47pt} pinning\\
\hspace*{10pt}{4.6.}   Electromechanical effects\\
{5.}    Phase slippage\\
\hspace*{10pt}{5.1.}   Static CDW dislocations\\
\hspace*{24pt} {5.1.1.}   Elastic deformations\\
\hspace*{24pt} {5.1.2.}   CDW dislocations\\
\hspace*{10pt}{5.2.}   Phase slips and dislocations in the\\  \hspace*{24pt} sliding CDW state\\
\hspace*{24pt} {5.2.1.}   Surface pinning\\
\hspace*{24pt} {5.2.2.}   Gor'kov model\\
\hspace*{24pt} {5.2.3.}   Phase vortices\\
\hspace*{24pt} {5.2.4.}   Frank-Read sources\\
\hspace*{10pt}{5.3.}   Direct observation of a single CDW\\  \hspace*{24pt} dislocation\\
\hspace*{24pt} {5.3.1.}   Coherent X-ray diffraction and\\  \hspace*{47pt} speckle\\
\hspace*{24pt} {5.3.2.}   Single CDW dislocation\\
\hspace*{10pt}{5.4.}   Phase slip voltage, $V_{\rm ps}$\\
\hspace*{24pt} {5.4.1.}   Shunting-non shunting\\  \hspace*{47pt} electrodes\\
\hspace*{24pt} {5.4.2.}   Length dependence of the\\  \hspace*{47pt} threshold field\\
\hspace*{10pt}{5.5.}   Breakable CDW\\
\hspace*{24pt} {5.5.1.}   Lateral current injection\\
\hspace*{24pt} {5.5.2.}   Long range CDW coherence\\
\hspace*{24pt} {5.5.3.}   Critical state model\\
\hspace*{24pt} {5.5.4.}   Thermal gradient\\
\hspace*{10pt}{5.6.}   Thermally activated phase slippage\\
\hspace*{24pt} {5.6.1.}   Phase slip rate\\
\hspace*{24pt} {5.6.2.}   Phase-slip and strain coupling\\
\hspace*{24pt} {5.6.3.}   CDW elastic constant\\
\hspace*{10pt}{5.7.}   X-ray spatially residual studies of\\  \hspace*{24pt} current conversion\\
\hspace*{24pt} {5.7.1.}   Stationary state\\
\hspace*{24pt} {5.7.2.}   Normal $\leftrightarrow$ CDW current\\  \hspace*{47pt} conversion model\\
\hspace*{24pt} {5.7.3.}   Bulk phase slippage\\
\hspace*{24pt} {5.7.4.}   Transient structure of sliding\\  \hspace*{47pt} CDW\\
\hspace*{24pt} {5.7.5.}   Phase slip in narrow\\  \hspace*{47pt} superconducting strips\\
\hspace*{24pt} {5.7.6.}   Controllable phase slip\\
\hspace*{24pt} {5.7.7.}   CDW deformations on\\  \hspace*{47pt}compounds with a semi-\\  \hspace*{47pt} conducting ground state\\
\hspace*{10pt}{5.8.}   $Q_1$-$Q_2$ coupling in NbSe$_3$\\{6.}    Screening effects \\
\hspace*{10pt}{6.1.}   Relaxation of polarisation\\
\hspace*{10pt}{6.2.}   Elastic hardening due to Coulomb\\  \hspace*{24pt} interaction\\
\hspace*{10pt}{6.3.}   Low frequency dielectric relaxation at\\  \hspace*{24pt}low temperature\\
\hspace*{24pt} {6.3.1.}   Dielectric relaxation of o-TaS$_3$\\
\hspace*{24pt} {6.3.2.}   Dielectric relaxation in the\\  \hspace*{47pt} SDW state of (TMTSF)$_2$PF$_6$\\
\hspace*{24pt} {6.3.3.}   Glassy state\\
\hspace*{24pt} {6.3.4.}   Competition between weak and\\  \hspace*{47pt} strong pinning\\}
\hspace*{-24pt}\vbox{\noindent\hsize18pc
\hspace*{10pt}{6.4.}   Second threshold at low temperature\\
\hspace*{24pt} {6.4.1.}   Rigid CDW motion\\
\hspace*{24pt} {6.4.2.}   Phase slip processes\\
\hspace*{24pt} {6.4.3.}   Low temperature transport \\  \hspace*{47pt} properties of (TMTSF)$_2$X salts\\
\hspace*{24pt} {6.4.4.}   Macroscopic quantum\\  \hspace*{47pt} tunnelling\\
\hspace*{24pt} {6.4.5.}   Metastable plastic deformations\
\hspace*{10pt}{6.5.}   Switching in NbSe$_3$\\
{7.}   Excitations\\
\hspace*{10pt}{7.1.}   Amplitudons and phasons\\
\hspace*{24pt} {7.1.1.}   Incommensurate dielectrics\\
\hspace*{24pt} {7.1.2.}   Screening of the phason mode\\
\hspace*{10pt}{7.2.}   Excess heat capacity\\
\hspace*{24pt} {7.2.1.}   Phason heat capacity\\
\hspace*{24pt} {7.2.2.}   Contribution from low\\ \hspace*{47pt} energy modes\\
\hspace*{24pt} {7.2.3.}   Low-energy ``intra''-molecular\\  \hspace*{47pt} phonon modes in Bechgaard-\\  \hspace*{47pt} Fabre salts\\
\hspace*{24pt} {7.2.4.}   Bending forces\\
\hspace*{24pt} {7.2.5.}   A possible phason contribution\\ \hspace*{47pt}in the specific heat of \\ \hspace*{47pt}(TMTSF)$_2$AsF$_6$\\
\hspace*{10pt}{7.3.}   Very low temperature energy\\  \hspace*{24pt}relaxation\\
\hspace*{24pt} {7.3.1.}   Low energy excitations\\
\hspace*{24pt} {7.3.2.}   Sub-SDW phase transitions\\
\hspace*{24pt} {7.3.3.}   Non-equilibrium dynamics\\
\hspace*{10pt}{7.4.}   Effet of a magnetic field\\
\hspace*{10pt}{7.5.}   Electronic excitations\\
\hspace*{24pt} {7.5.1.}   Optical conductivity in CDWs\\
\hspace*{24pt} {7.5.2.}   Optical conductivity in\\  \hspace*{47pt} Bechgaard-Fabre salts\\
\hspace*{24pt} {7.5.3.}   Photoemission\\
\hspace*{10pt}{7.6.}  Strong coupling model\\
{8.}   Field-induced density waves\\
\hspace*{10pt}{8.1.}   FISDW\\
\hspace*{10pt}{8.2.}   Pauli paramagnetic limit\\
\hspace*{10pt}{8.3.}   FICDW for perfectly nested Fermi\\  \hspace*{24pt} surfaces\\
\hspace*{10pt}{8.4.}   Fermi surface deformation by the\\  \hspace*{24pt} pinned CDW structure\\
\hspace*{10pt}{8.5.}   Interplay between Zeeman and orbital\\  \hspace*{24pt} effects\\
\hspace*{10pt} {8.6.}   CDW gap enhancement induced\\  \hspace*{24pt} by a magnetic field in NbSe$_3$\\
\hspace*{-7pt}
\hspace*{-7pt}{9.}   Mesoscopy\\
\hspace*{7pt}{9.1.}  Shaping\\
\hspace*{24pt} {9.1.1.}   Growth of thin films of\\  \hspace*{51pt} Rb$_{0.30}$MoO$_3$\\
\hspace*{24pt} {9.1.2.}   Patterning of whisker crystals\\
\hspace*{24pt} {9.1.3.}   Nanowires\\
\hspace*{24pt} {9.1.4.}   Focus-ion beam (FIB)\\  \hspace*{51pt} technique\\
\hspace*{24pt} {9.1.5.}   Topological crystals\\
\hspace*{7pt}{9.2.}  Aharonov effect\\
\hspace*{24pt} {9.2.1.}   Columnar defects\\
\hspace*{24pt} {9.2.2.}   CDW ring\\
\hspace*{24pt} {9.2.3.}   Circulating CDW current\\
\hspace*{7pt}{9.3.}  Mesoscopy CDW properties\\
\hspace*{24pt} {9.3.1.}   CDW transport\\
\hspace*{24pt} {9.3.2.}   Quantised CDW wave vector\\  \hspace*{51pt} variation\\
\hspace*{7pt}{9.4.}  Mesoscopic CDW junctions\\
\hspace*{24pt} {9.4.1.}   Carrier reflection at the\\  \hspace*{51pt} N/CDW interface\\
\hspace*{24pt} {9.4.2.}   CDW heterostructures\\
\hspace*{7pt}{9.5.}  Point contact spectroscopy\\
\hspace*{24pt} {9.5.1.}   Point contact with a semi-\\  \hspace*{51pt} conducting CDW\\
\hspace*{24pt} {9.5.2.}   Point contact with NbSe$_3$ \vspace*{1cm} \\
}}
\hbox to \textwidth{\hsize\textwidth\vbox{\hsize18pc
\hspace*{7pt}{9.6.}  Intrinsic interlayer tunnelling\\  \hspace*{24pt} spectroscopy\\
\hspace*{24pt} {9.6.1.}   Conductivity anisotropy along\\  \hspace*{51pt} the $a^\ast$ axis\\
\hspace*{24pt} {9.6.2.}   CDW gaps\\
\hspace*{24pt} {9.6.3.}   Zero bias conductance peak\\  \hspace*{51pt} (ZBCP)\\
\hspace*{24pt} {9.6.4.}   Intragap CDW states\\
\hspace*{24pt} {9.6.5.}   Phase decoupling\\
\hspace*{7pt}{9.7.}  Current-effect transistor and gate\\  \hspace*{24pt} effect\\
\hspace*{24pt} {9.7.1.}   Current effect transistor\\
\hspace*{24pt} {9.7.2.}   Gate effect\\
\hspace*{-7pt}{10.}   Conclusions \\
\hspace*{-7pt}{11.}   Epilogue\\
\hspace*{7pt}{11.1.}  Potassium\\}
\hspace*{-24pt}\vbox{\noindent\hsize18pc
\hspace*{7pt}{11.2.}  Luttinger liquid\\
\hspace*{7pt}{11.3.}  Quantum wires\\
\hspace*{7pt}{11.4.}  Unconventional density waves\\
\hspace*{7pt}{11.5.}  Electronic phase separation\\
\hspace*{24pt} {11.5.1.}   Charge-order stripe in 2D\\  \hspace*{51pt}(BEDT-TTF)$_2$X salts\\
\hspace*{24pt} {11.5.2.}   Oxides\\
\hspace*{7pt}{11.6.}  Search for sliding mode in\\ \hspace*{24pt} higher dimensionality\\
\hspace*{24pt} {11.6.1.}   Misfit layer Sr$_{14}$Cu$_{24}$O$_{18}$ \\
\hspace*{24pt} {11.6.2.}   2D (BEDT-TTF)$_2$X salts\\  
\hspace*{24pt} {11.6.3.}   Manganites\\
\hspace*{24pt} {11.6.4.}   2D electronic solids\\
\hspace*{24pt} {11.6.5.}   Sliding mode in RTe$_3$ \\ \hspace*{24pt}  compounds\\
}}
\end{abstract}

\section{Introduction}\label{sec1}

From the middle to the end of fifties, several theories were developed which demonstrated instabilities in the free electron model. R.E.~Peierls \cite{Peierls55} showed the instability of a one-dimensional (1D) metal interacting with the lattice towards a lattice distortion and the opening of a gap in the electronic spectrum, the so-called charge density wave (CDW). Related to this CDW state, H.~Fr\"ohlich \cite{Frohlich54}, just before the Bardeen-Cooper-Schrieffer (BCS) theory of superconductivity \cite{BCS57}, described a 1D model in which the CDW can slide if its energy is degenerate along the chain axis, yielding thus a collective current without dissipation and leading to a superconducting state. W. Kohn \cite{Kohn59} showed that the existence of sharp Fermi surfaces leads to anomalies in the phonon spectrum. A.W.~Overhauser and A.~Arrott \cite{Overhauser59} were the first to speculate that the localised spins observed in neutron scattering might be orientated by their interaction with a spin-density wave (SDW) in the conduction electron gas. The antiferromagnetism in chromium was identified by A.W.~Overhauser \cite{Overhauser62} as being a manifestation of a static SDW. W.M.~Lomer \cite{Lomer62} recognised that the large amplitude of the SDW is connected with specific geometric features of the Fermi surface of Cr which allows the nesting between electron and hole sheets having similar shape.

Some years later W.A.~Little \cite{Little64} proposed the design of a possible organic superconductor formed by a long macromolecule chain on which a series of lateral chains were attached. Superconductivity might result from an excitonic mechanism in which charge oscillations in the side chains can provoke an attractive interaction between electrons moving in the long chain. Although the realisation of such a design was unfruitful, this model has given some impetus for further researches on organic superconductivity.

Beyond concepts, the real breakthrough was realised when chemists were able to synthesise inorganic as well as organic materials formed with chains or weakly coupled planes in which charge and spin modulations were discovered. Many of these systems, essentially quasi-one-dimensional ones, exhibit quite universal general phenomena, although the underlying microscopic physical mechanisms are diverse and specific to each system. They belong to a larger class of materials entitled Electronic Crystals: those join together various cases of spontaneous structural aggregation of electrons in solids \cite{R20Brazovskii99,R21Brazovskii02,R22Brazovskii05,R23Brazovskii09,R24Brazovskii12}. Representatives of such electronic crystals are charge and spin density waves in low-dimensional materials, Wigner crystals of electrons formed in volume, at surfaces or in wires, stripe phases in conducting oxides including the family of high $T_c$ superconductors, various forms of charge order in organic quasi one- and two-dimensional materials, charged colloidal crystals. Related systems include vortex lattices in superconductors and domain walls in magnetic and ferroelectric materials. A large number of reviews and books has been already published \cite{R1Jerome77,R3Devresse79,R5Gruner94,R6Ishiguro98,R7Gorkov89,R9Schlenker89,R10Schlenker96,
R11Baeriswyl04,R12Boswell99,R14Lebed08,R16Monceau85,R17Rouxel86,R18Giamarchi03}.

Among all these electronic crystals, this review will be more specifically focused on systems with charge and/or spin modulations, as well as on charge order in some organic materials, with the aim at a global survey of physical properties of these inorganic and organic systems.

Basic fundamental notions are shortly described in section~\ref{sec2}. Structural properties, lattice dynamics, phase transitions and charge ordering are presented in section~\ref{sec3}. Section~\ref{sec4} is devoted to the sliding properties when the charge- or spin-density wave is free to move and then contributes to the current, following the mechanism for collective conduction proposed by H.~Fr\"ohlich \cite{Frohlich54}. Phase slippage discussed in section~\ref{sec5} occurs at any discontinuity in the macroscopic phase of the condensate, i.e at electrode injection, around strong pinning impurity, ... At low temperature, screening of the density wave deformations are less and less effective, increasing the rigidity of the density wave, as it is explained in section~\ref{sec6}. The order parameter of a density wave is complex with an amplitude and a phase; section~\ref{sec7} is devoted to phason and amplitudon dispersions using neutron and inelastic X-ray scattering and femtosecond spectroscopy. At very low temperature, properties of C/S DW are governed by low energy excitations which can be detected by specific heat measurements. Phase diagrams of CDWs under high magnetic field are presented in section~\ref{sec8}. The use of electronic lithography techniques opens the possibility of studying mesoscopic properties in CDWs as demonstrated in section~\ref{sec9}.

\section{Basic fundamental notions}\label{sec2}
\setcounter{figure}{0}
\setcounter{equation}{0}

Till the sixties, it was accepted that the ground state in the Hartree-Fock approximation of any electron sea in a jellium (homogeneous positive background) was the superposition of free electron states. However Overhauser \cite{Overhauser60} and Peierls \cite{Peierls55} showed that in one-dimensional (1D) systems the formation of a density wave --a charge density wave (CDW) or a spin density wave (SDW)-- by mixing two states of opposite wave vectors --one occupied, the other empty-- separated by $2k_{\rm F}$ lowers the energy and therefore is the new Hartree-Fock ground state, $k_{\rm F}$ being the Fermi momentum. Similarly Fr\"ohlich \cite{Frohlich54} in his attempt for explaining superconductivity has given the similar argument in 1D: the interaction between states $\bm{k}$ and $\bm{k}+2k_{\rm F}$ of an electron gas in a jellium can be attractive and lead to a condensate with an energy gap at $E_{\rm F}$, if the jellium is modulated with a wave vector $2k_{\rm F}$. This formalism is very similar to the Bardeen-Cooper-Schrieffer (BCS) \cite{BCS57}  theory for pairing of electron pairs.

\subsection{Peierls transition}\label{sec2-1}

\begin{figure}
\begin{center}
\includegraphics[width=8.5cm]{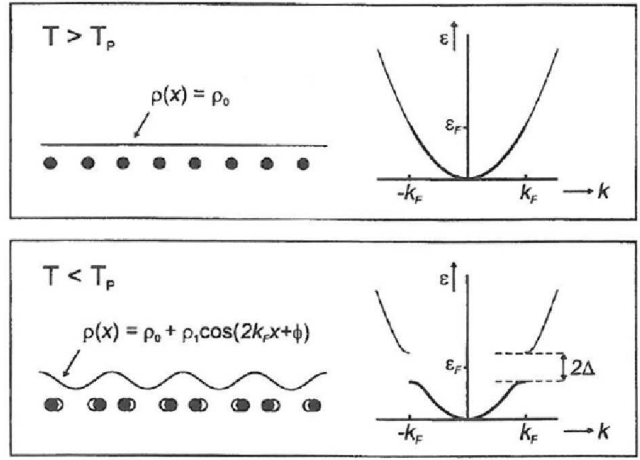}
\caption{(a) Upper: Conduction band for a linear chain of atoms at temperatures higher than the Peierls transition temperature ($T>T_{\rm P}$). (b)~Lower: $T\leq T_{\rm P}$, band structure for the insulator with opening of the gap $2\Delta$ at the Fermi level. The atomic position of the linear chain are modulated as $\rho$~= $\rho_0+\rho_1\cos(2k_{\rm F}x+\varphi)$ with the wave vector of the Peierls distortion: $Q$~= $2k_{\rm F}$.}
\label{fig2-1}
\end{center}
\end{figure}

In his attempt to explain the electronic properties of bismuth \cite{Peierls55}, Peierls pointed out that a one-dimensional metal is unstable at low temperatures and undergo a metal-insulator transition accompanied by the formation of a charge density wave \cite{Kagoshima88,R5Gruner94}. The mechanism of the Peierls transition can be simply explained as follows: it is possible to lower the electronic energy of a 1D system by opening a gap at the Fermi level from the coupling of a $2k_{\rm F}$ wave vector with the underlying lattice. The 1D-metallic conduction band is shown in figure~\ref{fig2-1}(a) filled up to the Fermi level $\varepsilon_{\rm F}$ at the Fermi wave vector $k_{\rm F}$. The introduction of a periodic potential due to atomic displacement having the periodicity of $2\pi/2k_{\rm F}$ introduces a new Brillouin zone at $\pm k_{\rm F}$; that consequently opens a band gap at $\varepsilon_{\rm F}$ as shown in figure~\ref{fig2-1}(b). For 1D systems the energy cost in distortion is always lower than the gain in electronic energy, making the transition favourable \cite{Kagoshima88,R5Gruner94}.

\subsection{1D susceptibility}\label{sec2-2}

\begin{figure}
\begin{center}
\includegraphics[width=7.5cm]{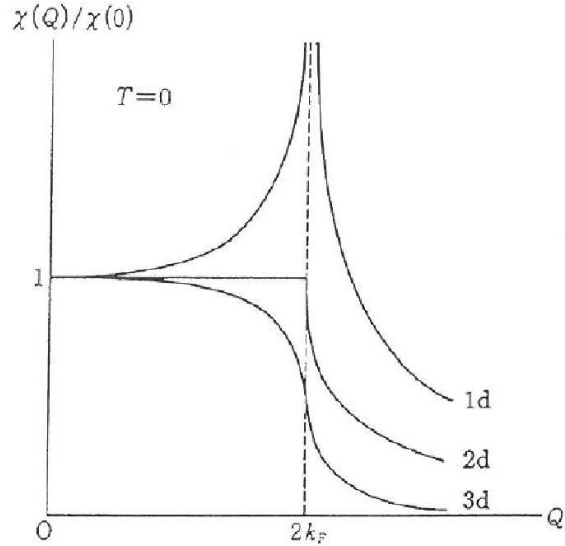}
\caption{Polarisation function $\chi(q)$ at absolute zero temperature at 1-2-3 dimensions.}
\label{fig2-2}
\end{center}
\end{figure}

As stated above, it is well-known that a system of conduction electrons in one-dimensional systems is  unstable with respect to spatially inhomogeneous perturbations. For non-interacting free electrons, a perturbation by a small periodic potential $V$ will cause a small fractional modulation of the electronic distribution such:
\begin{eqnarray}
&\Delta\rho(q)=\chi^0(q)V(q)\nonumber\\
\mbox{with }&\chi^0(q)\propto\displaystyle\sum_k\frac{f_{k+q}-f_k}{\varepsilon_{k+q}-\varepsilon_k}
\label{eq2-1}
\end{eqnarray}
where $\varepsilon_k$ is the energy of the state with wave vector $\bm{k}$, $f_k$ the Fermi occupation factor. $\chi^0(q)$ is the ``bare'' magnetic or electronic susceptibility of the conduction band and depends on the details of the band structure. The $q$-dependence of $\chi^0$ is shown in figure~\ref{fig2-2} 
at $T$~= 0 at three-two-one dimension. In one  dimension, $\chi^0(q)$ shows a logarithmic divergence for $Q$~= $2k_{\rm F}$. Divergence of $\chi^0(q)$ occurs when
\begin{equation}
\varepsilon_k-\mu=\varepsilon_{k+q}-\umu
\label{eq2-2}
\end{equation}
with $\mu$ the chemical potential. This condition indicates a perfect nesting of the hole and electron Fermi surface (FS), which in 1D are  planar parallel sections, by translation of the wave vector $Q$~= $2k_{\rm F}$ which spans the FS.

\begin{figure}
\begin{center}
\includegraphics[width=7cm]{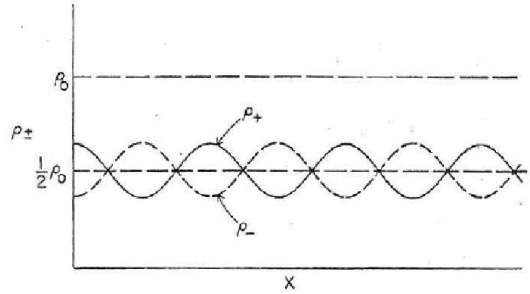}
\caption{Modulated electronic charge density with up- and down-spin electrons for a SDW ground state. The total charge density $\rho_0$ is constant.}
\label{fig2-3}
\end{center}
\end{figure}

If one include electron-phonon and electron-electron interactions, instabilities can occur when the generalised susceptibility $\chi(q)$ obtained \textit{via} a renormalisation using the random field approximation (RDA) diverges:
\begin{equation}
\chi(q)=\frac{\chi^0(q)}{1-X(q)\chi^0(q)}.
\label{eq2-3}
\end{equation}
$X(q)$ represents the interactions including exchange and correlations. Chan and Heine \cite{Chan73} have calculated the form of $X(q)$ depending of the interactions. They showed that the condition for a SDW is a sufficiently strong exchange interaction $U$ together with a large $\chi^0(q)$ which can be easily realised in 1D systems:
\begin{equation*}
\chi^0(Q)U=1
\end{equation*}
which is the generalised Stoner criterium. The Stoner criterium for ferromagnetism is recovered by taking $Q$~= 0. If $Q$ is commensurate with any reciprocal lattice vector the ground state is antiferromagnetic (AF). For incommensurate $Q$, the ground state is a SDW. Thus a SDW can be described as a kind of AF state with a spatial spin density modulation for which the difference between the density $\rho_\uparrow(x)$ of electron spins polarised upwards and the density $\rho_\downarrow(x)$ of electron spins polarised downwards is finite and modulated in space as a function of the position $x$ \cite{R5Gruner94}. As shown in figure~\ref{fig2-3}, the SDW has thus spatially inhomogeneous charge densities for both spin states, but out of phase by $\pi$ such:
\begin{equation}
\rho_\pm(x)=\frac{1}{2}\;\rho_0\left[1\pm\rho_1\cos Qx\right].
\label{eq2-4}
\end{equation}
The total charge $\rho_0$ is constant and independent of position. There is a net spin polarisation proportional $\rho_1\cos Qx$. Beyond this linearly polarised SDW, there is more complicated order found in the circularly polarised SDW with polarisation $Se^{iQx}$ \cite{Fazekas99}.

\begin{figure}
\begin{center}
\includegraphics[width=7cm]{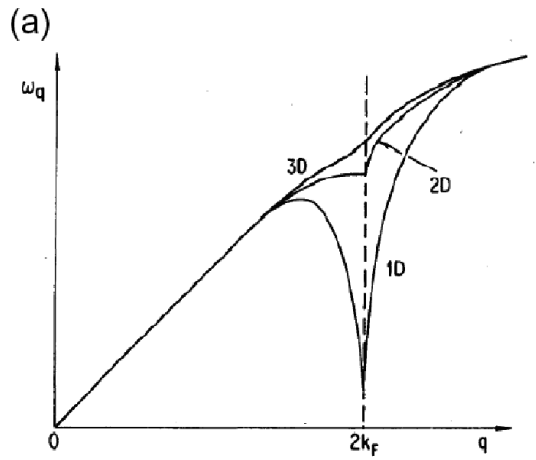}
\caption{Kohn anomaly in phonon spectra in each dimension.} 
\label{fig2-4}
\end{center}
\end{figure}

An alternative way of instability is by scattering against phonons through the el-ph interactions. When the el-ph interaction overwhelms the repulsive electrostatic term (Coulomb repulsion) the interaction becomes attractive and the instability is of a CDW type. The el-ph interaction renormalises the phonon frequencies such as:
\begin{equation}
\omega^2(q)=\Omega_0^2\left[1-X(q)\chi^0(q)\right]
\label{eq2-5}
\end{equation}
with $\Omega_0$ is the bare phonon spectrum. That leads to the softening of $\omega(q)$, known as the Kohn anomaly \cite{Kohn59}  in the appropriate phonon branch as shown in figure~\ref{fig2-4}. 
In ideal 1D, the Kohn anomaly is particularly strong and the phonon frequency is reduced to zero at $Q$ which induces the static lattice distortion. The phonons around $Q$~= $2k_{\rm F}$ are collective oscillations strongly coupled to the lattice and to the electronic density; they should be considered as macroscopically occupied with formation of a CDW of electrons and ion displacements.

For a CDW the charge density for the two spins are modulated in phase. Consequently the spin density is zero everywhere and the electronic charge density is:
\begin{equation}
\rho(x)=\rho_0\left[1+\rho_1\cos(2k_{\rm F}x+\varphi)\right]
\label{eq2-6}
\end{equation}
The lattice is periodically modulated in quadrature:
\begin{equation}
u(x)=u_0\sin 2k_{\rm F}x
\label{eq2-7}
\end{equation}
ensuring thus the neutrality condition. The phase $\varphi$ denotes the relative phase of the charge modulation with respect to the ion lattice. In this context with weak electron-phonon coupling, a CDW is intimately formed by a charge modulation concomitant with a periodic lattice distortion with identical wave length \cite{Friedel77}. While the mechanism of creation of electron-hole pairs --by mixing Bloch states whose $k$ vectors
\begin{eqnarray*}
k+Q\quad\uparrow,\qquad k\quad\downarrow\quad \mbox{for } SDW\\
k+Q\quad\uparrow,\qquad k\quad\uparrow\quad \mbox{for } CDW
\end{eqnarray*}
differ by $Q$-- due to properties of perfect nesting leading to the logarithmic divergence of the bare electronic susceptibility is similar, there are essential differences on the physical origin of the microscopic molecular field which induces the density wave (DW) instability. For SDW, the key ingredient is the exchange interaction, $U$, which is only relevant for electrons with antiparallel spins. On the other side, the CDW is essentially driven by the electron-phonon interaction which induces the Peierls transition.

There are basically two different approaches for the description of the physical properties of 1D conductors. The Fr\"ohlich model \cite{Frohlich54} taking into account the 1D coupled el-ph interaction is used for the study of the Peierls distortion, Kohn anomaly and phonon softening. On the other side, the effect of electronic correlations in molecular solids is better taken into account using the Hubbard model and its extensions.

\subsection{Fr\"ohlich Hamiltonian}\label{sec2-3}

This Hamiltonian for a quasi 1D linear chain of atoms separated by a distance $a$ includes the el-ph interaction but neglects Coulomb interactions between electrons:
\begin{equation*}
H=H_{\rm el}+H_{\rm ph}+H_{\rm el-ph}.
\end{equation*}
The electron part is given by:
\begin{equation}
H_{\rm el}=\sum_k\varepsilon(k)a^+_ka_k.
\label{eq2-8}
\end{equation}
In the tight binding approximation, the electron energy has the dispersion spectrum,
\begin{equation}
\varepsilon(k)=-2t_\parallel\cos ka\,,
\label{eq2-9}
\end{equation}
with $t_\parallel$ the longitudinal transfer integral along the 1D direction and $a$ the lattice constant. $4t_\parallel$ is equal to the bandwidth $W$. $a^+_k(a_k)$ creates (destroys) an electron in state $k$. The index $k$ includes both wave number and spin.

The phonon part is:
\begin{equation*}
H_{\rm ph}=\sum_k\hbar\omega_kb^+_kb_k
\end{equation*}
with $\hbar\omega_k$~= $ks$ ($s$: the sound velocity) and $b^+_k$ ($b_k$) are the phonon creation (destruction) operators.

\begin{figure}
\begin{center}
\includegraphics[width=7.5cm]{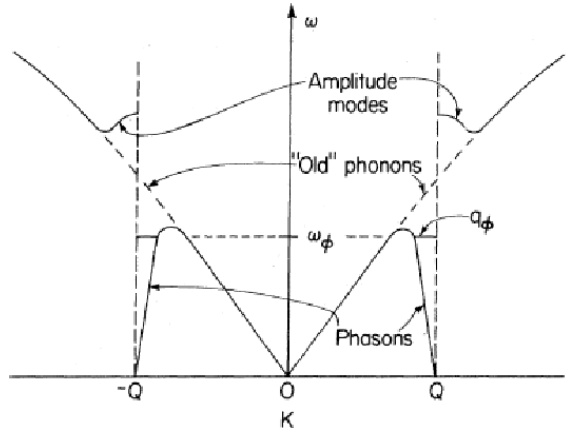}
\caption{Schematic illustration of the vibrational modes in a CDW compound. In the ideal case without pinning, the frequency of the phason branch goes to zero at $\pm Q$~= $2k_{\rm F}$(reprinted figure with permission from M.F. Bishop and A.W. Overhauser, Physical Review B 23, p. 3638, 1981 \cite{Bishop81}. Copyright (1981) by the American Physical Society).}
\label{fig7-1}
\end{center}
\end{figure}

The el-ph part takes the form:
\begin{equation*}
H_{\rm el-ph}=\sum_{p,k}g(k)a^+_{p+k}a_p\left(b_k+b^+_{-k}\right),
\end{equation*}
with $g(k)$ the el-ph coupling constant. An equivalent form of $H_{\rm el-ph}$ in terms of electron wave field operators is:
\begin{equation*}
H_{\rm el-ph}=\sum_k\int {\rm d}x\,g(k)\left(b_k+b^+_{-k}\right)e^{ikx}\psi_{(x)}\psi^+_{(x)}.
\end{equation*}
For $T<T_c$, the phonon modes $\omega_{2k_{\rm F}}$ of wave vector $\pm 2k_{\rm F}$ which connect the two parts of the FS become macroscopically occupied. The lattice undergoes a distortion with a $2\pi/2k_{\rm  F}$ modulation. The order parameter $\Delta$ which described this distortion is defined  by:
\begin{equation}
\Delta=g(2k_{\rm F})\langle b_{2k_{\rm F}}+b^+_{-2k_{\rm F}}\rangle e^{i2k_{\rm F}x}=|\Delta |e^{i2k_{\rm F}x},
\label{eq2-10}
\end{equation}
with $\langle\hdots\rangle$ indicating a thermal average. An energy gap is introduced in the electronic band structure so that the one-electron energies become:
\begin{eqnarray*}
E(k)&=&{\rm sgn}\,\varepsilon(k)\left[\varepsilon^2(k)+|\Delta |^2\right]^{1/2},\\
\mbox{with}\qquad{\rm sgn}\,\varepsilon(k)&= & + \mbox{ for }\varepsilon(k)> 0\\
 &&-\mbox{ for }\varepsilon(k)< 0.
\end{eqnarray*}
Rice et Str\"assler \cite{Rice73} found that in the tight binding approximation:
\begin{eqnarray}
k_{\rm B}T_c&=&2.28E_{\rm F}\,e^{-1/\lambda}\label{eq2-11}\\
\Delta(T=0)&=&\Delta_0=4E_{\rm F}\;e^{-1/\lambda},\nonumber\\
\mbox{with}\qquad\qquad\lambda&=&\frac{2N(0)|g(2k_{\rm F})|^2}{\hbar\omega_{2k_{\rm F}}}.\qquad\qquad\label{eq2-12}
\end{eqnarray}
From the complex nature of the CDW order parameter, amplitude and phase fluctuations have to be considered. They are collective modes associated with each kind of fluctuations, called amplitudon and phason. A phason can be considered \cite{Bishop81} as the superposition of two ``old" phonons, i.e. phonons of the undistorted lattice of wave vectors $q+Q$ and $q-Q$ ($Q=2k_{\rm F}$). The orthogonal linear combination of these same two ``old" phonons is the amplitude mode which occurs at high frequency. Away from $Q$ and $-Q$, the phase and the amplitude modes merge quickly in the phonon spectrum as shown schematically in figure~\ref{fig7-1}. 
At the frequency cut-off $\omega_\phi$ and at the corresponding wave-vector cut-off $q_\phi$, the phasons transform into phonons \cite{Bishop81}. At this cut-off one can define a characteristic phason temperature $\theta_\phi$ such:
\begin{equation}
\theta_\phi=\frac{\hbar\omega_\phi}{k_{\rm B}}.
\label{eq7-1}
\end{equation}
\begin{figure}
\begin{center}
\includegraphics[width=7.75cm]{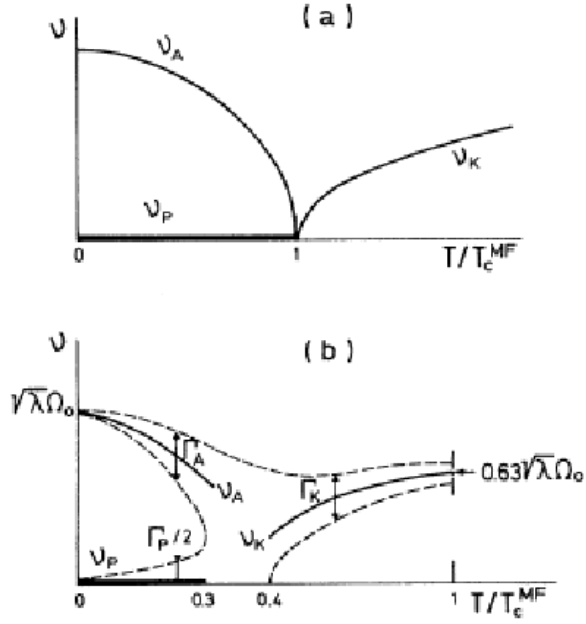}
\caption{Temperature dependence of the frequency of the Kohn anomaly $\nu_\kappa$, the amplitude $\nu_A$ and phase $\nu_P$ modes at the critical $Q$~= $2k_{\rm F}$ wave vector a)~in the mean field approximation, b)~from the numerical simulations \cite{Tutis91} for a strictly one-dimensional Peierls chain ($\Gamma_i$: damping of the various modes) (reprinted figure with permission from J.-P. Pouget \textit{et al.}, Physical Review B 43, p. 8421, 1991 \cite{Pouget91}. Copyright (1991) by the American Physical Society).}
\label{fig7-2}
\end{center}
\end{figure}
In the mean field approximation, the frequency $\nu_{\rm K}$ of the phonon mode which suffers the Kohn anomaly decreases to zero at $T_c$~= $T^{\rm MF}_c$. The separation between amplitude and phase modes holds immediately below $T_c^{\rm. MF}$ as shown in figure~\ref{fig7-2}(a), the amplitude mode decreasing also to zero at $T^{\rm MF}_c$ \cite{Rice73}. However, due to 1D fluctuations, the effective 3D phase transition occurs at $T_{\rm P}$ such as $T_{\rm P}/T_c^{\rm MF}\simeq$ 0.3--0.4 \cite{Lee73}. The response function of the damped harmonic oscillator $S(q,\nu)$ for a strictly one-dimensional Peierls chain has been obtained using numerical Monte-Carlo molecular dynamics simulations \cite{Tutis91} and schematically shown in figure~\ref{fig7-2}(b). When $T$ is reduced from $T_c^{\rm MF}$, the decrease of the phonon frequency $\nu_K$ of the Kohn anomaly and the increase of the damping $\Gamma_K$ due to anharmonicity leads \cite{Pouget91} to an overdamped response near $\sim 0.4T_c^{\rm MF}$. 

Lee, Rice and Anderson \cite{Lee74} gave the dispersion of amplitudons in the case of incommensurate CDW (I-CDW) such as:
\begin{equation}
\omega^2_+=\lambda\omega^2_Q+\frac{4}{3}\left(\frac{m}{M^\ast}\right) v_{\rm F}^2q^2,
\label{eq2-13}
\end{equation}
and that of phasons:
\begin{equation}
\omega^2_-=\frac{m}{M^\ast}v^2_{\rm F}q^2
\label{eq2-14}
\end{equation}
with $M^\ast$ the renormalised Fr\"ohlich mass of the CDW,
\begin{equation}
M^\ast=m_0\left(1+\frac{4\Delta^2}{\hbar\omega_Q\hbar^2}\right),
\label{eq2-15}
\end{equation}
$\lambda$ the electron-phonon coupling, $\omega_Q$ the frequency of the phonon mode at $Q$~= $2k_{\rm F}$.

\subsection{Extended Hubbard model}\label{sec2-4}

In this case, phonons are ignored. The 1D tight binding lattice Hamiltonian is:
\begin{equation*}
H=H_0+H_1+H_2,
\end{equation*}
\begin{eqnarray}
H_0 & = & t_\parallel\sum_{i,\sigma}c^+_{i+1,\sigma}c_{i,\sigma}+c^+_{i,\sigma}c_{i+1,\sigma}\label{eq2-16}\\
H_1 & = & U\sum_{i,\sigma}n_{i,\sigma}n_{i,-\sigma}\label{eq2-17}\\
H_2 & = & \sum_{i\neq j}V_{i-j} n_in_j,\label{eq2-18}
\end{eqnarray}
where $c^+_{i,\sigma}(c_{i,\sigma})$ is the creation (annihilation) operator of an electron of spin $\sigma$ at site $i$, $n_{i,\sigma}$~= $c^+_{i,\sigma}c_{i,\sigma}$ is the occupation number of this state. $n_i$~= $\sum_\sigma n_{i,\sigma}$ is the operator giving the total number of electrons on site i. $t$: the transfer integral between nearest-neighbour sites, $U$: the interaction of 2 electrons on the same site, $V_n$~= $V_{-n}$: the interaction of electrons on $n^{th}$ nearest neighbour sites. If one consider the near-neighbour interaction, then $H_2$~= $V\sum_in_{i+1}n_i$.

In the case of dimerisation along the chain, two transfer integrals should be distinguished and $H_0$ takes the form
\begin{equation}
H_0=t_1\sum_{i_{\rm even},\sigma}+\,t_2\sum_{i_{\rm add},\sigma}.
\label{eq2-19}
\end{equation}
Writing the operators $c^+(c)$ in Bloch state representation, i.e.
\begin{equation*}
a_i(k)=\frac{1}{\sqrt{N}}\sum^N_{i=1}\exp \left(-ika_i\right)c_{i,\sigma},
\end{equation*}
the interactionless part of the Hamiltonian becomes
\begin{equation}
H_0=\sum_{k,i}\varepsilon(k)a^+_i(k)a_i(k)
\label{eq2-20}
\end{equation}
with the same spectrum $\varepsilon(k)$~= $2t_\parallel\cos ka$ as in the Fr\"ohlich Hamiltonian, but with explicit spin in the index.

This Hubbard lattice Hamiltonian involving transfer integrals and potential terms which have large energy parameters corresponds to the strong coupling limit.

\subsection{1D electron gas}\label{sec2-5}

In the weak coupling limit, when $U$, $V\ll t$, correlation terms are considered as a perturbation of the one-electron formalism. Representing the interaction parts of the extended Hubbard Hamiltonian in terms of different coupling constants for different scattering processes yields the derivation of the usual 1D electron gas Hamiltonian in the continuum limit, also known as the $g$-ology model \cite{Solyom79,Firsov85,Voit95,Emery79}, $g_1$: backward scattering for large transfer $\sim 2k_{\rm F}$ across the FS, $g_2$: forward scattering with small transfer ($q\simeq 0$) and $g_3$: Umklapp scattering.

For a non-interacting electron gas, the $2k_{\rm F}$ charge density and the $2k_{\rm  F}$ spin density susceptibility have the same low-$T$ logarithmic divergence in $\log(E_{\rm F}/kT)$. When weak el-el correlations are added as a perturbation, a renormalisation group (RG) approach is needed to be rid of these logarithmic divergences. The strength of the single-chain susceptibility within the RG approach determines the ground state response. It is then obtained the well known $g_1-g_2$ phase diagram \cite{Solyom79}. It is found that superconductivity and DW orders are  separated by the line $g_1$~= $2g_2$. The SDW requires $g_2>0$ and the CDW susceptibility has a power law divergence if $2g_1-g_2$~$<0$. In the frame of the extended Hubbard model, one finds that:
\begin{eqnarray*}
g_1 & = & \frac{U+2V\cos\pi\rho}{\pi v_{\rm F}}\\
g_2 & = & \frac{U+2V}{\pi v_{\rm F}}
\end{eqnarray*}
with $\rho$ is the band filling. For a quarter filled band ($\rho$~= $\frac{1}{2}$) then $g_1$~= $U/\pi v_{\rm  F}$ and $g_2$~= $U+2V/\pi v_{\rm F}$.

\subsection{Tomonaga-Luttinger liquids}\label{sec2-6}

If the kinetic energy of electrons given in eq.~(\ref{eq2-9}) is linearised about the FS, which in strictly 1D systems consists of 2 planes at $k$~= $\pm k_{\rm F}$, one gets the dispersion relation (with $\hbar$, $k=1$) of the so-called L\"uttinger model:
\begin{equation}
\varepsilon(k)-\varepsilon_{\rm F}=v_{\rm F}\left(|k|-k_{\rm F}\right)
\label{eq2-21}
\end{equation}
with $v_{\rm F}$ the Fermi velocity:
\begin{equation*}
v_{\rm F}=2t_\parallel a\sin k_{\rm F}a.
\end{equation*}
Then the free Hamiltonian $H_0$ is written in the Tomonaga-Luttinger model (for a review see \cite{Voit95,Firsov85}) such as:
\begin{equation}
H_0=\sum_{k,\sigma}v_{\rm F}(k-k_{\rm F})a^+_{1,\sigma}(k)a_{1,\sigma}(k)-\sum_{k,\sigma}v_{\rm F}(k+k_{\rm F})a^+_{2,\sigma}(k)a_{2,\sigma}(k),
\label{eq2-22}
\end{equation}
where $a_{j,\sigma}(k)$ are electron operators related to the branch $k>0$ ($k\approx k_{\rm F}$) for $j=1$ and $k<0$ ($k\approx -k_{\rm F}$) for $j=2$.

Solutions of the Tomonaga-Luttinger model \cite{Voit95,Schulz91} has been made in restricting the interaction part of the Hamiltonian to small momentum transfers, i.e. only forward scattering is considered ($g_1,g_3$~= 0, there are no scattering of particles with antiparallel spins). It was found that the ground state is metallic and that there are no elementary excitations at low energies (Fermi-type of quasi-particles), but that all excitations are bosonic charge and spin fluctuations (without gap) with a linear dispersion law:
\begin{equation}
\omega_{\rho,\sigma}(k)=v_{\rho,\sigma}k
\label{eq2-23}
\end{equation}
with charge velocity $v_\rho$ different of spin velocity $v_\sigma$. Hence the usual quasi-particle picture breaks down. The momentum distribution function in the vicinity of $k_{\rm F}$ is:
\begin{equation}
n_k\approx n_{k_{\rm F}}-\beta \,{\rm sgn}(k-k_{\rm F})|k-k_{\rm F}|^\alpha
\label{eq2-24}
\end{equation}
and the single particle density of states follows the power law:
\begin{equation}
N(\omega)\approx |\omega|^\alpha
\label{eq2-25}
\end{equation}
with $\alpha=\frac{1}{4}\left[\kappa_\rho-1/\kappa_\rho-2\right]$. $\kappa_\rho$ is a coefficient which determines the power law decay of all correlation functions of the system. An incoming electron decays into charge and spin excitations which spatially separate with time. This is so-called spin charge separation can be interpreted in the frame of the Hubbard model (47); in the weakly interacting system it is found that $\kappa_\rho$~= $1-U/\pi v_{\rm F} + \ldots $

\subsection{Strong coupling limit}\label{sec2-7}

The $2k_{\rm F}$ Peierls scenario is derived without $el-el$ interaction. In the case of 1/4 filled band, 4 sites should share 2 electrons, which means that each site would be partially occupied.

Strong $el-el$ interactions impede such a double occupation; a given lattice site cannot be occupied by more than 1 electron (no orbital degeneracy). One are then faced to the problem of a spin-less 1D Fermi gas with the same energy dispersion than in eq.~(\ref{eq2-9}), but with the spin degrees of freedom lost. The same number of electrons fill the conduction band but from $-2k_{\rm F}$ to $+2k_{\rm F}$. This strongly correlated 1D electron gas is unstable to a charge modulation with a wave-vector $4k_{\rm F}$.

In the case of half-filling, $\rho$~= 1. Then $ 4k_{\rm F}$ is equal to a reciprocal lattice vector $G$ and Umklapp processes ($g_3$) enhances a $4k_{\rm F}$ CDW below a temperature $T_\rho$ below which a Mott-Hubbard gap is opened in the charge excitation spectrum. There is no gap for spin excitations.  This charge-spin separation which occurs in that case should not be confused with that resulting from the Luttinger liquid for which collective charge and spin excitations are without gap and are a strictly 1D property.

The $4k_{\rm F}$ instability can be viewed as a kind of Wigner crystallisation. The charge localisation forms like a Wigner crystal (described in 3D for small electron density). More generally \cite{Hubbard78} Hubbard has determined the ground state resulting from the distribution of a system of $\rho N$ electrons over the $2N$ orbitals of a chain of length $N$. The calculation was made in the case where interactions between electrons at near-neighbour sites have a dominant importance in determining the electronic structure (i.e. the bandwidth $t$ in eq.~(\ref{eq2-16}) is small and treated only as a perturbation). In the case of $\rho$~= $1/n$, it was shown that the lowest energy configuration is that in which all the electrons are equally spaced a distance $n$ neighbours apart. Configurations corresponding to any rational value of $\rho$~= $p/q$ where also found. In the case of $\rho$~= $\frac{1}{2}$ (1/4 filled band) two configurations are possible: if an occupied site is denoted by 1, and an empty one by 0, the 101010 configuration corresponding to a period 2 leads to the $4k_{\rm F}$ CDW; other possibility is the 1100 1100 arrangement with a period 4 corresponding to a tetramerisation. These various configurations can be considered as generalised-Wigner lattices. This regular arrangement of charges in a lattice is now called a charge ordered state (CO) with charge disproportionation.

\subsection{Spin-Peierls}\label{sec2-8}

In the case of a charge gap opened and no gap for spin excitations, the low energy excitations can be described by an antiferromagnetic spin chain 1/2 Heisenberg Hamiltonian \cite{Schulz77}:
\begin{equation}
H_{\rm spin}=\sum_i\,J\,{\bm S}_i\,{\bm S}_{i+1},
\label{eq2-26}
\end{equation}
where the ${\bm S}_i$ are spin 1/2 operators and $J$ an effective exchange constant such as $J$~= $2t^2/U$ for large $U$. If allowance is made for an elastic distortion of the lattice, the exchange integral $J$ depends on the lattice spacing. The spin system can lower its energy by dimerising (similarly to the canonic Peierls distortion). The phase transition leads to a non-magnetic (singlet) dimerised state $\chi(T=0)$~= 0, called spin-Peierls transition \cite{Pytte74,Schulz77,Bray83,Cross79}.

For 1/4 filled band, when dimerisation had already occurred ($4k_{\rm F}$), the dimer dimerises and forms a tetramerised lattice ($2k_{\rm F}$ transition). In this case, $e-e$ correlations have the effect to split the ``classical" Peierls transition into an electronic $4k_{\rm F}$ part and at low temperature a spin $2k_{\rm F}$  part. At $4k_{\rm F}$ the electronic degrees of freedom are lost and in the $2k_{\rm F}$ transition the spins are lost with the disappearance of the magnetic susceptibility.

\subsection{Intermediate Coulomb interactions: Charge ordering}\label{sec2-9}

For next-neighbour interaction $V$~= 0, it is known that the Hubbard model leads to a metallic state at any filling factor, except half-filling \cite{Lieb68}. The low energy excitations are of the L\"uttinger-like type \cite{Haldane81}. The presence of long range repulsion interaction with both $U$ and $V$ yields the insulating phase at large $U$ and $V$ while superconductivity becomes the most dominant fluctuations for large $V$ and small $U$.

If exact solutions of the Hubbard model can be derived, that is not the case for the extended Hubbard model. Numerical simulations are then necessary. Monte-Carlo simulations can show the interplay between $2k_{\rm F}$ and $4k_{\rm F}$ instabilities as a function of the Coulomb interactions and band filling. In the case of 1/4 filled band, it was concluded \cite{Hirsch83} that, in presence of a moderate nearest-neighbour interaction, the effect of Coulomb interaction suppresses the $2k_{\rm F}$ charge density Peierls transition and that the $4k_{\rm F}$ charge density peak is greatly enhanced. Schulz \cite{Schulz91} has given the exact description of the cross-over between weak and strong correlations and of the metal-insulating transition occurring when the average particle per site, $n$,approaches unity. The dependence of the correlation exponent $\kappa_\rho$ of the Hubbard model has been determined as a function of $U$ and $n$. For small $U$, the perturbation result $\kappa_\rho\sim 1-U/\pi v_{\rm F}$ is recovered. For $U\rightarrow\infty$, $\kappa_\rho\rightarrow\frac{1}{2}$ for all $n$ ($\kappa_\rho$~= 1 for non-interacting systems). The cross-over between $2k_{\rm F}$ and $4k_{\rm F}$ instabilities occur when $\kappa_\rho\lesssim\frac{1}{3}$.

Mila \cite{Mila93} has made the bridge between models at small energy scales (lower than the bandwidth) and extended Hubbard models with arbitrary amplitude of interactions. Exact solutions are only known for two limiting cases:
\begin{eqnarray*}
&V_c/t=2 \quad\mbox{at}\quad U\rightarrow\infty ,\\
\mbox{and} \qquad &\\
&U_c/t=4\quad\mbox{at}\quad V\rightarrow\infty.
\end{eqnarray*}

\begin{figure}
\begin{center}
\includegraphics[width=7.5cm]{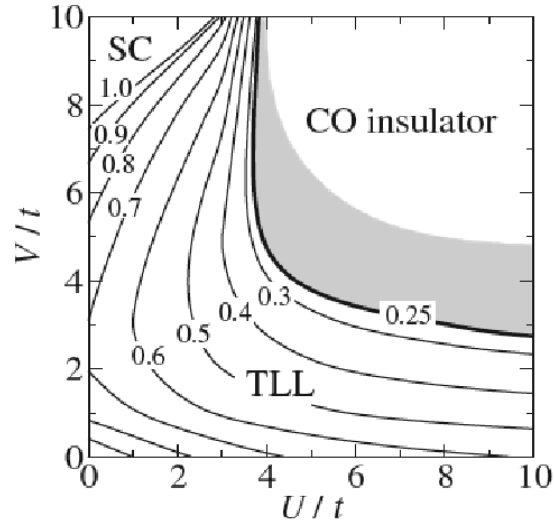}
\caption{U-V ground state phase diagram of the one-dimensional extended Hubbard model at quarter-filling. The bold line separates the charge order [CO]  insulator state to Tomonaga-Luttinger liquid metal states for different $\kappa_\rho$ curves. Superconductivity (SC) appears in the extreme upper part of the diagram (reprinted figure with permission from S. Ejima \textit{et al.}, Europhysics Letters 70, p. 492, 2005 \cite{Ejima05}. Copyright (2005) from EdpSciences).}
\label{fig2-6}
\end{center}
\end{figure}

In the $(U-V)$ phase diagram, as shown in figure~\ref{fig2-6}, there are a transition line from Luttinger liquid to a $4k_{\rm F}$ CDW insulator or to a charge order (CO) state \cite{Seo06,Ejima05}. This CO state is resulting from strong Coulomb interactions and is also called ``Wigner Crystal on Lattice" \cite{Seo04}.

Mila {\it et al.} have not taken into account a possible dimerisation along the 1D stacks. As shown in eq.~(\ref{eq2-19}) the transfer integral $t$ is splitted in two, $t_1$ and $t_2$. This dimerisation leads to the opening of a charge gap and to a Mott insulating state, where each carrier is located on a dimer pair \cite{Seo97} (this state is called dimer-Mott insulator). The dimerisation suppresses the metallic Luttinger state.

With many different techniques --mean field approximation \cite{Seo97,Seo06}, numericals, renormalisation group \cite{Nishimoto00}, bosonisation \cite{Yoshioka00}-- all show that an insulating state with CO is stabilised in the region of large $U$ and $V$ interactions. That is even the case with slightly dimerisation where the CO state competes with the dimer-Mott insulator \cite{Shibata01,Tsuchiizu01}. The spin susceptibility has been numerically calculated for finite $U/t$, $V/t$ values. This susceptibility is not affected by the opening of the CO gap \cite{Tanaka05}.

\subsection{Coupling with the lattice}\label{sec2-11}

In the Hubbard Hamiltonian, the effective coupling with the lattice is not apparent. In fact the interaction between electronic and structural degrees of freedom is established by the electron-phonon coupling. In the case of a molecular crystal intramolecular and intermolecular vibration modes may change respectively the intramolecular coordinates (and then affect the Coulomb potential $U$) or the intermolecular coordinates and then affect the $t_\parallel$ and $V$ energies.

One should then distinguish between the modulation of the intersite charge density --a CDW-- and the modulation of the intersite distance --a bond order wave (BOW)--. Each of these can have periodicities $2k_{\rm F}$ (period 4) and $4k_{\rm F}$ (period 2). The BOW can occur in two forms corresponding to different phase angles \cite{Ung94}. Depending on the band filling and the amplitude of the Coulomb interaction, the phase between CDW and BOW may not be the same \cite{Ung94,Mazumdar99,Clay07}. Bond alternation in half-filled band has been largely studied in the context of polyacetylene \cite{Su79,Dixit84}.

Depending on the strength of the lattice coupling coexistence between CDW, BOW and also SDW is found in some parts of the phase diagram \cite{Riera00}. From the extended Hubbard model [eqs~(\ref{eq2-16})-(\ref{eq2-18})] coupled to a classical phonon field, $H_{\rm ph}$~= $H_{\rm elas}$~= $\frac{1}{2} K_{\rm B}\sum_i(\delta_i)^2$, with $\delta_i$: the lattice displacement, and $K_{\rm B}$: the inverse of the strength of the lattice coupling, phase diagrams of 1/4 filled band were numerically derived. Two different types of structures are stable: in the weak coupling regime ($\sim U/t<3$), a strong $2k_{\rm F}$ BOW corresponding to a 11 00 sequence of bonds. This modulation coexists with a weaker $2k_{\rm F}$ site-centre CDW. At larger $U/t$ the phase corresponds to a spin-Peierls phase \cite{Riera00}. The influence of the anion potential on CO was also studied \cite{Riera01}.

\subsection{Imperfect nesting}\label{sec2-12}

\begin{figure}
\begin{center}
\includegraphics[width=13cm]{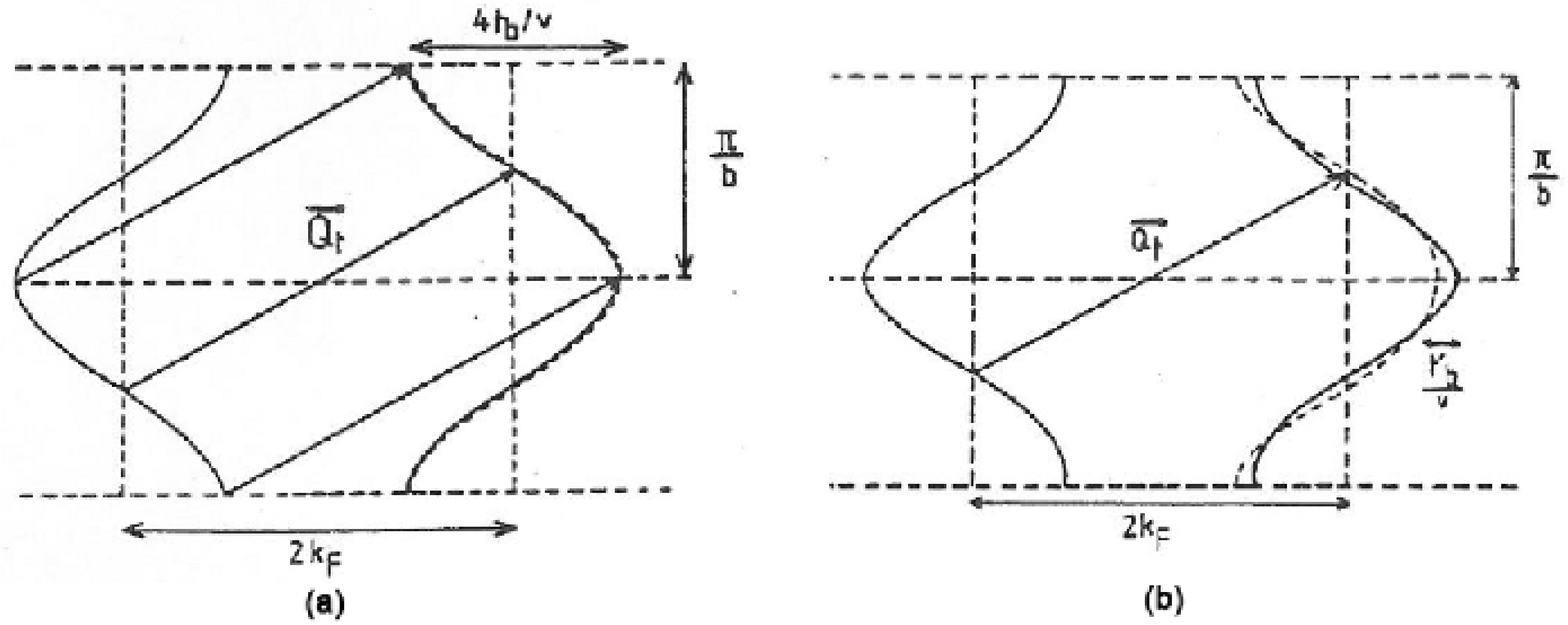}
\caption{One-dimensional Fermi surface with a finite transverse integral: $t_\perp$~= $-2t_b\cos(bk_b)-2t^\prime_b\cos 2bk_b$. With $t^\prime_b$~= 0 the Fermi surface has a sinusoidal warping with a perfect nesting $Q_0$ wave vector. For $t^\prime_b\neq 0$ the nesting is imperfect and some small electron-hole pockets remain at the FS.}
\label{fig2-8}
\end{center}
\end{figure}

All the considerations developed above were made for a strictly 1D material with an unique transfer integral $t_\parallel$ along the chain. However linear chains forming real materials are more or less strongly coupled together. One then should take into account the transverse dispersion relation \cite{Yamaji82} and eq.~(\ref{eq2-21}) becomes:
\begin{equation}
\varepsilon(k)-\varepsilon_{\rm F}=v_{\rm F}(|k_{\rm x}|-k_{\rm F})+t_\perp(k_{\rm y}),
\label{eq2-27}
\end{equation}
with $x$: the chain direction. In the tight binding approximation $t_\perp(k_{\rm y})$ takes the form \cite{Yamaji82,Montambaux88,Huang92,Mihaly97}:
\begin{equation}
t_\perp(k_{\rm y})=-2t_b\cos k_{\rm y}b,
\label{eq2-28}
\end{equation}
with $b$ the lattice constant in the transverse direction $y$ and $t_b$ the transverse transfer integral. With a finite $t_b$, the FS is no more formed of two parallel planes, but is wrapped as shown in figure~\ref{fig2-8}(a). 
\begin{figure}
\begin{center}
\subfigure[]{\label{fig2-10a}
\includegraphics[width=7cm]{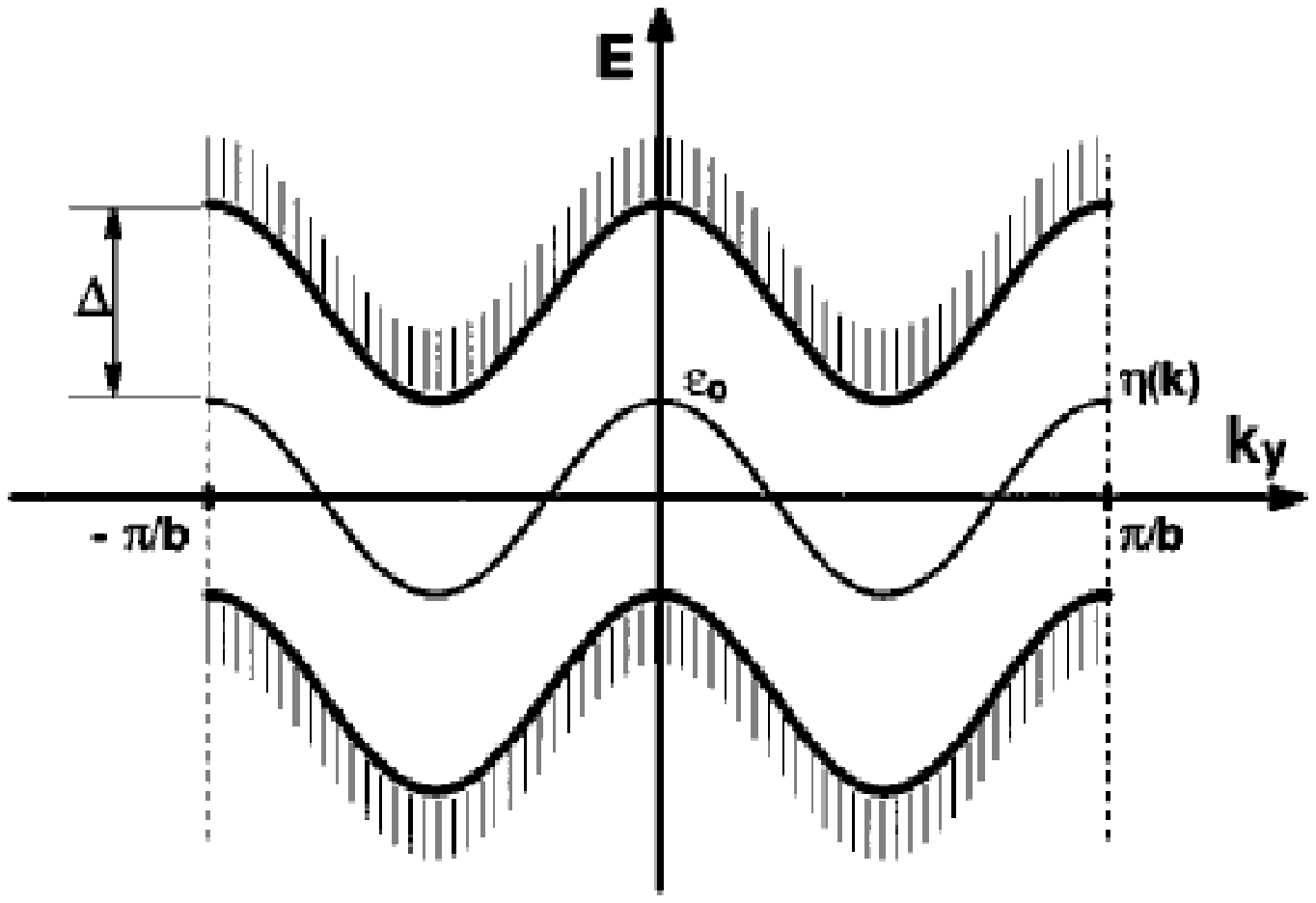}}
\subfigure[]{\label{fig2-10b}
\includegraphics[width=5cm]{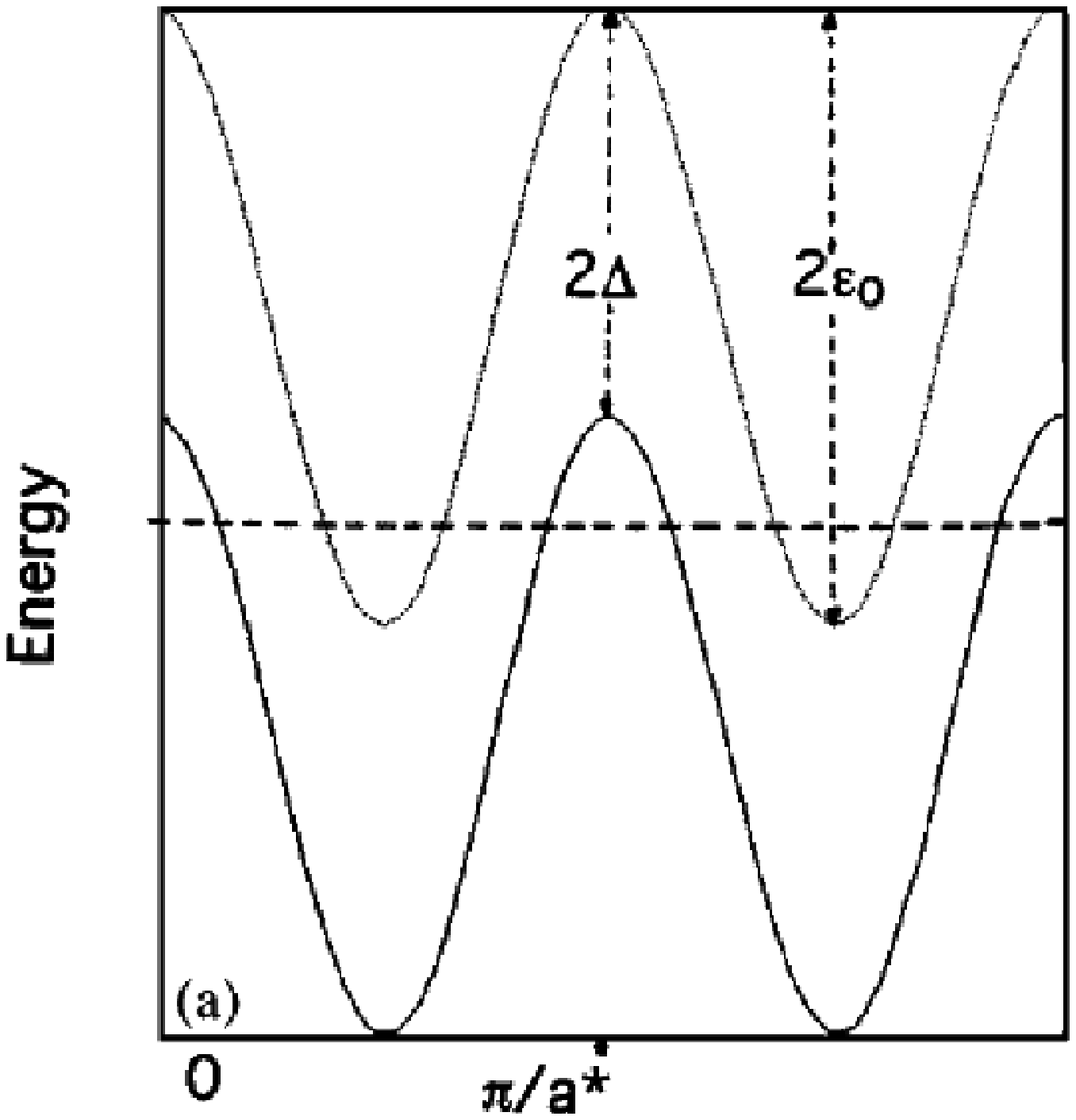}}
\caption{a)~Wrapping of the density wave gap as a function of the wave number $k_y$ perpendicular to the chain with $\eta(k)$~= $-2t'\cos(2bk_y)$~= $\epsilon_0\cos(2bk_y)$ (reprinted figure with permission from G. Mihaly \textit{et al.}, Physical Review B 55, R13456, 1997 \cite{Mihaly97}. Copyright (1997) by the American Physical Society). b)~Overlap of conduction and valence bands in the reciprocal space modulated perpendicularly to the nesting direction with large $\epsilon_0$, $\epsilon_0>\Delta$, yielding a semi-metallic state.}
\label{fig2-9}
\end{center}
\end{figure}
With eq.~(\ref{eq2-28}) this corrugation of amplitude $4t_b/v_{\rm F}$ is sinusoidal and a perfect nesting is still possible with the $Q$ vector ($2k_{\rm F}$, $\pi/b$).

More generally one also should take into account the other transverse transfer integral along the second transverse axis $c$ and then adds to the dispersion relation the term $2t_c\cos k_{\rm z}c$. Perfect nesting can be destroyed with $t_c$. Also if the dispersion (eq.~(\ref{eq2-28})) is not perfectly sinusoidal, for instance with an harmonic content \cite{Yamaji82} such:
\begin{equation*}
t_\perp(k_{\rm y})=-2t_b\cos k_{\rm y}b-\varepsilon_0\cos 2k_{\rm y}b,
\end{equation*}
with the imperfect nesting characterised by $\varepsilon_0$ given by:
\begin{equation}
\varepsilon_0=\frac{t^2_b\cos(ak_{\rm F})}{2[t_a\sin^2(ak_{\rm F})]},
\label{eq2-29}
\end{equation}
$a$: lattice constant along the a chain axis, $t_a=t_\parallel$ transfer integral along chains. The DW gap amplitude is modulated as a function of the wave number $k_y$ perpendicular to the chain as shown in figure~\ref{fig2-9}. For large value of $\epsilon_0$ conduction and valence bands can overlap that yields a semi-metallic character for conduction \cite{Yamaji82}.

For not too large value of $\varepsilon_0$, the ground state is still a DW with the nesting vector ($2k_{\rm F}$, $\pi/b$, $\pi/c$) (see figure~\ref{fig2-8}(b)).

\subsection{Fluctuations}\label{sec2-13}

The calculations expounded above were made in the mean field (or RPA) approximation. However it is well known that strictly 1D systems do not have a phase transition at finite temperature \cite{Sham79}. The structural phase transition obtained in the mean field approach is meared out by large fluctuations in 1D.

The effect of fluctuations is the best described using a Ginzburg-Landau type approach where the free energy per atom is expanded in powers of the order parameter $\Delta(x)$ and its derivatives. Near the mean-field transition temperature, the correlation function of the order parameter $\langle\Delta(x)\,\Delta(x')\rangle$ fall off exponentially with distance such as $\exp-(|x-x'|/\xi(T))$ where $\xi$ is the correlation-length. The temperature dependence of $\xi(T)$ was calculated exactly in 1D by Scalapino \textit{et al.} \cite{Scalapino72}. It was shown \cite{Lee73} that, for $T<\frac{1}{4}T^{\rm MF}$, $\xi(T)$ increases exponentially with temperature, indicating that a three-dimensional (3D) ordering occurs. The critical temperature for this ordering was estimated as $T_c\approx\frac{1}{4}T^{\rm MF}$.

Well below $T_c$, thermal fluctuations disappear. Assuming that quantum effects (zero-point oscillations of the phase) are negligible, the low temperature properties of the 3D ordered state are well described by the mean field approximation. Then $\Delta(T=0)$~$\sim\Delta^{\rm MF}(T=0)$. Consequently the BCS relation $2\Delta(T=0)$~= 3.52~$T_c$ does not hold but it is found that
\begin{equation*}
2\Delta(T=0)\gtrsim 3.52\,T_c.
\end{equation*}

In real materials one should also take into account the interchain coupling and consequently the non-zero correlation length, $\xi_\perp$, in the transverse directions \cite{Schulz77}. Sufficiently close to $T_c$, $\xi_\perp$ is large and fluctuations are 3D. With increasing temperature, $\xi_\perp$ decreases and at a ``cross-over" temperature, $T^\ast$, $\xi_\perp(T^\ast)\approx d$, the distance between chains. For $T>T^\ast$, the adjacent chains are essentially uncorrelated and behave as one-dimensional units exhibiting 1D fluctuations.

But, alternatively, it has been remarked \cite{McKenzie92} that, in Q1D compounds, the lattice zero-point motion $\delta u$~= $(\hbar/2M\omega_A)^{1/2}$ ($M$: the mass displaced and $\omega_A$: the amplitude mode frequency) is comparable to the lattice distortion. The lattice zero-point and thermal lattice motions are source of disorder. They have an effect on the electronic properties similar to that of a random potential with Gaussian correlations. The dimensionless disorder parameter $\eta$ was written as:
\begin{equation}
\eta=\lambda\frac{\pi\omega_{2k{\rm F}}}{2\Delta}\coth\left(\frac{\omega_{2k_{\rm F}}}{2T}\right).
\label{eq7-2}
\end{equation}
The problem to determine the gap parameter from this model was studied as mathematically equivalent to that of magnetic impurities in a superconductor. The disorder due the thermal lattice motion can destroy the Peierls state at a temperature well below the mean field value.

\begin{figure}
\begin{center}
\includegraphics[width=7.5cm]{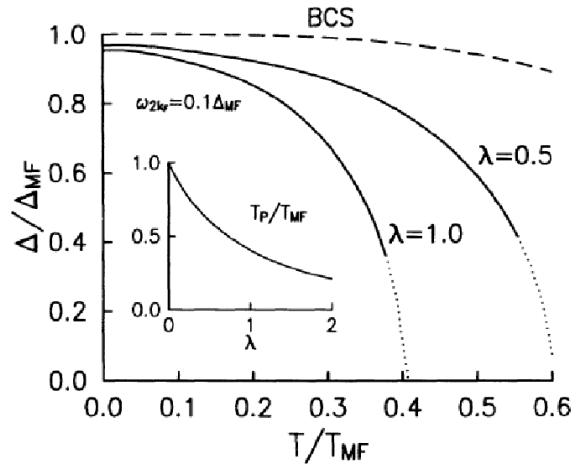}
\caption{Temperature dependence of the CDW gap normalised to the BCS mean field values ($T_{\rm MF}$: transition temperature, $\Delta_{\rm MF}$: zero temperature gap) for different value of $\lambda$: the dimensionless electron-phonon coupling. Increase of $\lambda$ manifests the strength of the disorder resulting from the lattice zero-point motion (see eq.~(\ref{eq7-2}). Inset: dependence of the Peierls transition temperature on $\lambda$ (reprinted figure with permission from R.H. McKenzie and J.W. Wilkins, Physical Review Letters 69, p. 1085, 1992 \cite{McKenzie92}. Copyright (1992) by the American Physical Society).}
\label{fig7-3}
\end{center}
\end{figure}

Figure \ref{fig7-3} shows that the reduction of the gap parameter below the mean field BCS value ($2\Delta$~= 3.52~$T^{\rm MF}$) increases with increasing the electron-phonon coupling $\lambda$ \cite{McKenzie92}. This reduction of $T_{\rm P}$ well below $T_{\rm MF}$ as seen in the inset of figure~\ref{fig7-3} contrasts the conventional view that $T_{\rm P}$ is determined by the competition between Q-1D thermodynamic fluctuations and interchain interactions.

\subsection{Fr\"ohlich conductivity from moving lattice waves}\label{sec2-14}

\begin{figure}
\begin{center}
\includegraphics[width=12cm]{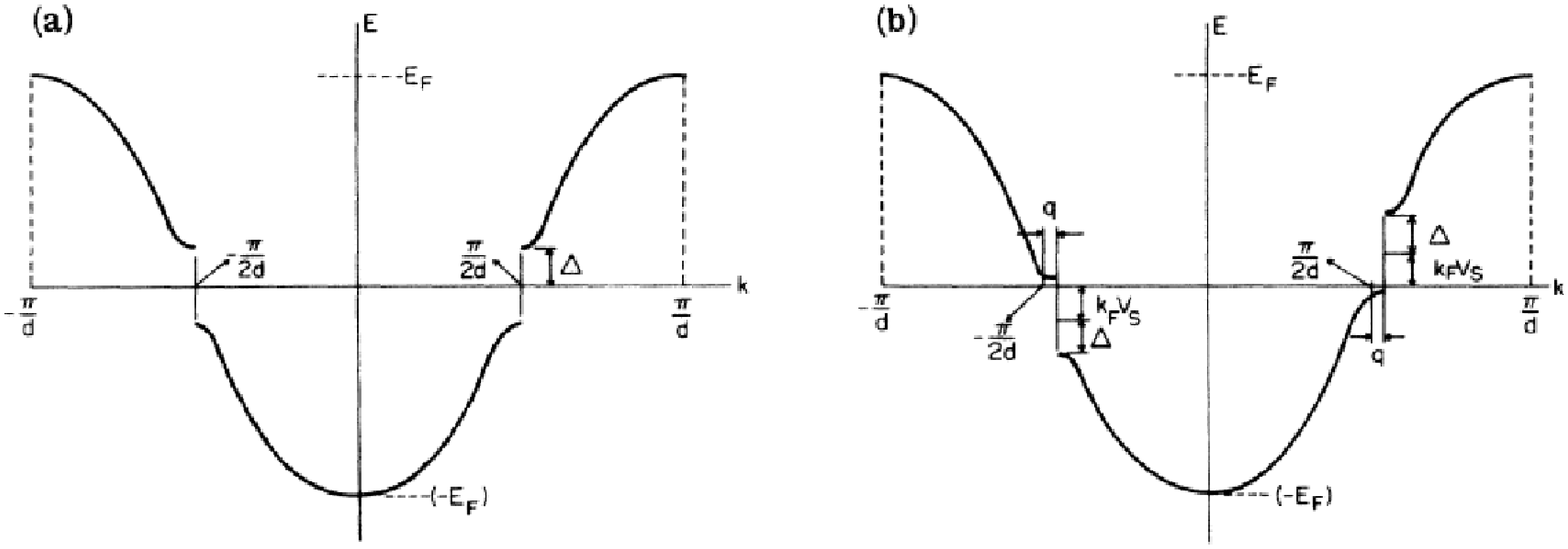}
\caption{(a) Tight binding band for a linear chain of atoms separated by $a$ with one electron per atom in the Peierls state with the gap $2\Delta$ opened at $k_{\rm F}$~= $\pm \pi/2a$. (b)~The same band as in (a) but displaced by $q$, which leads to the Fr\"ohlich current $J$~= $nev_s$ with $\hbar q$~= $m^\ast v_s$ ($m^\ast$: the Fr\"ohlich CDW mass) (reprinted figure with permission from D. Allender \textit{et al.}, Physical Review B 9, p. 119, 1974 \cite{Allender74}. Copyright (1974) by the American Physical Society).}
\label{fig2-10}
\end{center}
\end{figure}

Contrary to a semiconductor for which the energy gap at the Fermi level is due to the ionic potential, and therefore bound to the crystal frame, it was suggested by Fr\"ohlich \cite{Frohlich54} that there can be states with current flow if the energy gap is displaced with the electrons and remains attached to the FS. The Fr\"ohlich model has been studied again by Allender \textit{et al.} \cite{Allender74} in a tight binding model. If the lattice wave moves with the electrons with velocity $v_s$, the order parameter given in eq.~(\ref{eq2-10}) will vary as:
\begin{equation}
\Delta=|\Delta|e^{2ik_{\rm F}(x-v_st)}.
\label{eq2-30}
\end{equation}
Figure~\ref{fig2-10}(b) 
shows the Fermi distribution of the same chain as in figure~\ref{fig2-10}(a) when the modulation is displaced with a uniform velocity, $v_s$. The two planes of the FS are at $(-\frac{\pi}{k_{\rm F}})+q$ and $(+\frac{\pi}{k_{\rm F}})+q$ with
\begin{equation*}
m^\ast v_s=\hbar q,
\end{equation*}
where $m^\ast$ is the effective electronic mass. In the Galilean frame in motion with $v_s$, the system is again unstable with regard to a distortion which opens an energy gap at the new Fermi surface with the wave vector ${\bm Q}$~= $2k_{\rm F}$. All the electrons are below the gap and in this  frame the electronic current is zero. If the velocity, $v_s$, is small i.e. if the kinetic energy of the electrons in their translation, $\hbar k_{\rm F}v_s$, is small compared with the Peierls gap $\Delta$, in the static reference frame, the current in the sample will be:
\begin{equation}
J=-nev_s,
\label{eq2-31}
\end{equation}
where $n$ is the number of electrons per unit volume in the band affected by the CDW. Thus the energy gap reduces the elastic scattering of individual electrons because there is no state available for relaxing energy. The motion is therefore without dissipation and the compound in principle may become superconducting.

Note that the $\hbar k_{\rm F}v_s$ term makes the difference in energies between the left and right hand sides of the displaced distribution (figure~\ref{fig2-11}(b)). Also, when $\hbar k_{\rm F}v_s$ becomes greater than $|\Delta|$, electrons can be scattered back to the next higher band. Consequently the current will decrease rapidly to zero as in a pairing superconductor. This gives an effective upper limit for $v_s$.

\subsection{Incommensurate phases in dielectrics}\label{sec2-15}

A large number of dielectric crystals present perfect three-dimensional long range order but no translational periodicity at least in one direction being, like this, intermediate between classical ideal crystals and disordered or amorphous systems \cite{Blinc86}. Classes of materials include the A$_2$BX$_4$ family (typically K$_2$SeO$_4$), barium sodium nitrate, $\beta$ThBr$_4$, quartz molecular crystals as thiourea, biphenyl C$_{12}$H$_{10}$, etc. In these materials a local property such the electric polarisation, magnetisation, atomic position, \ldots is modulated with a periodicity $q$ incommensurate with the periodicity of the underlying lattice. Generally the translational lattice periodicity is restored at low temperature at a ``lock-in" incommensurate-commensurate (I-C) phase transition.

To explain the properties of these structurally incommensurate insulators, models have been developed in the context of the phenomenological Landau theory in which the frustration is the consequence of the fact that competition between short-range interactions favour different periodicities. The free-energy functional has been written in term of two competing order parameters of different symmetry and coupled through a Lifshitz-like gradient term which is non-vanishing only when the dimensionality of the order parameter is $\geq 2$.

\begin{figure}
\begin{center}
\includegraphics[width=6.5cm]{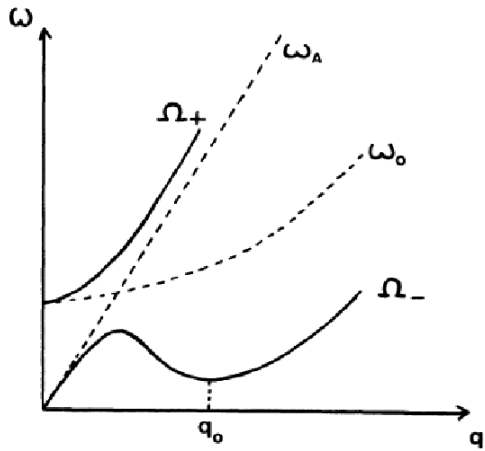}
\caption{Schematic representation of the dispersion curves resulting from the existence of a Lifshitz like invariant for incommensurate insulating materials such as quartz: dashed lines are uncoupled acoustic and soft optic modes and full lines correspond to the coupled modes (reprinted figure with permission from B. Berge \textit{et al.}, Journal de Physique (France) 45, p. 715, 1984 \cite{Berge84}. Copyright (1984) from EdpSciences).}
\label{fig2-11}
\end{center}
\end{figure}

In the case where the order parameter is unidimensional as for thiourea, NaNO$_2$, or quartz, it has been suggested that the incommensurate phase could result from the coupling of the order parameter with other degrees of freedom such as acoustic phonons. In the specific case of quartz the coupling occurs between a soft optic mode and the acoustic modes. This coupling leads to an ``anticrossing" of the phonon branches and the lower branch can present a minimum at the incommensurate $q_0$ wave-vector \cite{Berge84} as shown in figure~\ref{fig2-11}.

Near the I-C transition, the sinusoidal structure of the modulation is transformed in a periodic structure of domains of the commensurate phase separated by discommensurations (DC) or phase solitons. These solitons may be pinned by the basic crystal structure. A finite energy barrier is to be overcome to shift the pinned solitons \cite{MacMillan76}.

Very curiously, the physicists studying in the 70-80' years properties of incommensurate insulators often having ferroelectric ground states and those the phase transitions in quasi one-dimensional or two-dimensional materials were not really interacting. However the same experimental techniques --diffraction, EPR, NMR, neutron and light scattering-- were used. Although the microscopic phenomena are naturally different, physical concepts --soft mode, phason and amplitudon dispersion, pinning, hysteresis, devil'staircase, incommensurate-commensurate transition, chaos, etc.-- are very similar.

\subsection{Electronic ferroelectricity}\label{sec2-16}

It has been recently discovered that, in some types of materials, electron degrees of freedom with electronic interactions can give rise to a macroscopic electric polarisation and a ferroelectric transition. These systems are called electronic ferroelectric compounds \cite{Brink08,Ishihara10,Naka10}.

Ferroelectric transition caused by magnetic interaction and magnetic ordering such as in TbMnO$_3$ with all the multiferroic properties associated with the correlation between ferroelectricity and magnetism is out of the scope of this review (for a review see \cite{Arima11}).

\begin{figure}[h!]
\begin{center}
\includegraphics[width=7cm]{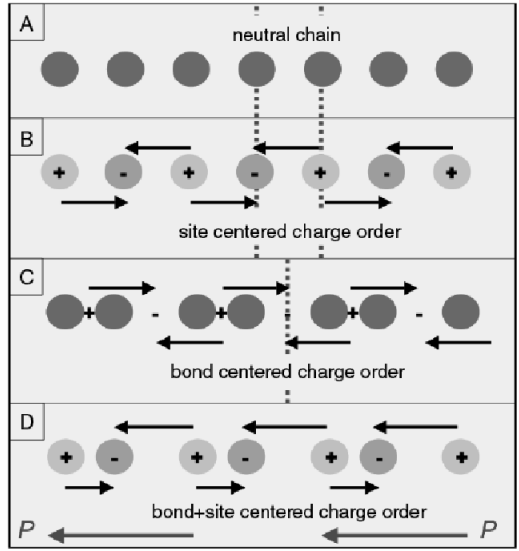}
\caption{Charge ordering and ferroelectricity. A: a neutral one-dimensional chain exhibiting, B: a site-centred charge ordering, C: band-centred charge ordering and D: a linear combination of these two that is ferroelectric. The arrows indicate the polarisation which is zero in B and in C but develops a macroscopic moment indicated by the red arrow in D. The red dashed lines in A, B and C indicate mirror planes of the system (reprinted figure with permission from J. van der Brink and D. Khomskii, Journal of Physics: Condensed Matter 20, p. 434217, 2008 \cite{Brink08}. Copyright (2008) by the Institute of Physics).}
\label{fig2-12}
\end{center}
\end{figure}

Hereafter will be consider the charge driven ferroelectricity which occurs from electron and electron-lattice interactions associated with charge ordering \cite{Brazovskii08}. Following van den Brink and Khomskii \cite{Brink08}, the mechanism by which charge ordering can lead to the appearance of ferroelectricity is as follows and as shown in figure~\ref{fig2-12}.

Figure~\ref{fig2-12}(a) represents an homogeneous one-dimensional crystal with equal charge on each site. Figure~\ref{fig2-12}(b) shows the same chain after charge ordering where the sites become unequivalent (site charge order). This process does not break the spatial inversion symmetry and the resulting state does not carry a dipole moment. Another type of charge ordering occurs when the system dimerises as shown in figure~\ref{fig2-12}(c) (bond charge ordered or bond order wave as explained above). In this case the sites remain equivalent but not the bonds which alternate as strong and weak bonds. Again the BOW is centrosymmetric and consequently not ferroelectric. The situation with simultaneous site-- and bond--charge order is shown in figure~\ref{fig2-12}(d). The inversion symmetry is broken in this case and each short bond develops a net dipole moment with the result that the whole system becomes ferroelectric. In previous section the possible coexistence of different charge order states was already discussed, but ferroelectricity is favourable if site and bond charge order occur simultaneously \cite{Brazovskii08}.

Ferroelectricity in perovskite manganites such Pr$_{1-x}$Ca$_x$M$_x$O$_3$, magnetite, can be explained by the occurrence of charge ordering. But with restricting oneselves to organic compounds, it has been shown that multi-component molecular systems can produce a displacive-type ferroelectric transition by the displacement of oppositely charged molecules. An example is given by charge-transfer (CT), formed from tetrathiafulvalene (TTF) with p-chloranil (tetrachloro-p-benzoquinone) TTF-CA organic compound which undergo a neutral-ionic (N-I) transition \cite{Torrance81,Torrance81b,Buron03,Lecointe95}. In these quasi-1D systems the non-polar alternation of electron donor ($D$) and acceptor ($A$) molecules along stacks $\ldots -D^0 A^0 D^0 A^0-\ldots$ can be symmetry-broken to form degenerate dimerised polar chains with DA dimers such $\ldots(D^+A^-)(D^+A^-)(D^+A^-)\ldots$ and $\ldots D^+(A^-D^+)(A^-D^+)(A^-D^+)(A^-\ldots)$. The loss of inversion symmetry leads to ferroelectric chains as detected by dielectric measurements \cite{Tokura89}. The electronic-structural transition is governed by the formation of CT exciton-strings which can either be several adjacent dimerised ionic molecular $(D^+A^-)$ pairs inserted in a $N$ chain or the opposite, several adjacent neutral molecular $D^0 A^0$ pairs inserted in the ferroelectric dimerised chain. These non-linear excitations are represented such as:
\begin{equation*}
D^0 A^0 D^0 A^0\underline{(D^+A^-)(D^+A^-)(D^+A^-)}D^0 A^0.
\end{equation*}
The relaxation of these CT exciton-string has been studied recently by photo-induced transformations \cite{Collet03}.

In contrast with the neutral-ionic transition which occurs in TTF-p-chloranil, the parent compound TTF-BA (tetrathiafulvalene-p-bromanil) is formed of TTF and BA molecules almost ionic. The $D^+A^-$ stack is then regarded as a 1D-Heisenberg chain with spin 1/2. TTF-BA undergoes a paramagnetic to a non-magnetic transition at $T_c\sim$ 53~K, consistent with the singlet formation in the 1D Heisenberg chain involving the spin-Peierls instability. A polarisation has been measured in this spin-Peierls state, which disappears when the singlet state is suppresses by a magnetic field. Thus TTF-BA would be the first material with a ferroelectric spin-Peierls ground state \cite{Kagawa10}.

\section{Materials}\label{sec3}
\setcounter{figure}{0}
\setcounter{equation}{0}

This section is devoted to a short presentation of the general structural features of families of quasi-one-dimensional electronic crystals the physical properties of which will be at length described in the following sections.

\subsection{MX$_3$ compounds}\label{sec3-1}

MX$_3$ crystals are synthesised with the transition metal M atom which belongs to group IV (Ti, Zr, Hf) or group V (Nb, Ta) and with chalcogenide atoms X such as S, Se, Te.

The basic constituent of the structure is a trigonal prism [MX$_6$] with a cross-section close to an isosceles triangle. The transition metal atom is located roughly at the centre of the prism. These trigonal prisms are stacked on top of each other by sharing their triangular faces forming metallic chains running along the $b$-axis. Chains are staggered with respect to each other by half the height of the unit prism. Therefore besides the six chalcogen atoms of the [MX$_6$] prism, each transition metal is bonded to two more X atoms from neighbouring chains and its coordination number is eight.

\begin{figure}[h!]
\begin{center}
\includegraphics[width=8.5cm]{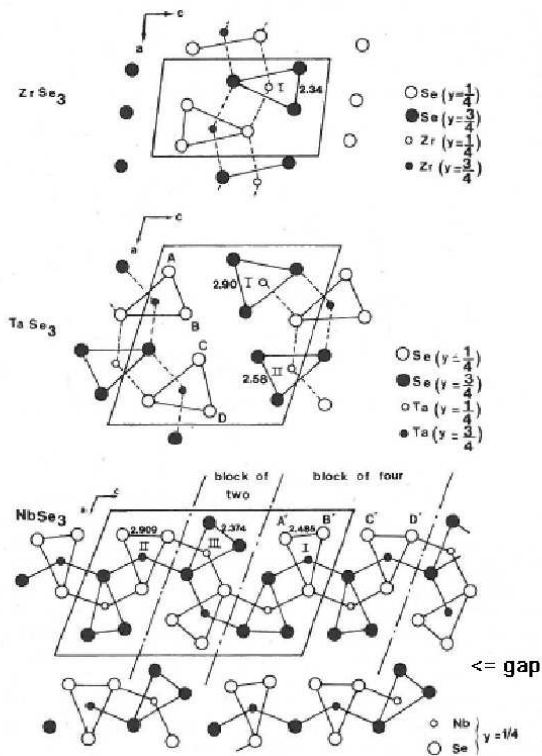}
\caption{Projection of the structure along the monoclinic $b$ axis of a)~ZrSe$_3$, b)~TaSe$_3$, c)~NbSe$_3$ and d)~TaS$_3$ with the monoclinic unit cell.}
\label{fig3-1}
\end{center}
\end{figure}

The simplest unit cell is of the ZrSe$_3$ type as shown in figure~\ref{fig3-1}(a) \cite{Rouxel89}. It consists of a unique type of MX$_3$ chain with covalent Se$_2^{2-}$ bonding (Se-Se distance of 2.34~$\AA$) between the nearest Se in the basis of the triangle of the prism. Cross-sections of the unit cell perpendicular to the chain axis for TaSe$_3$ and NbSe$_3$ are shown in figure~\ref{fig3-1}(b) and \ref{fig3-1}(c) \cite{Rouxel89}. MX$_3$ chains can be distinguished according to the strength of the chalcogen-chalcogen bond in the basis of the triangle of the chain. TaSe$_3$ is made up of two groups of two chains with an intermediate bond (Se-Se distance of 2.576~$\AA$) and a weak bond (Se-Se distance of 2.896~$\AA$).

\subsubsection{NbSe$_3$}\label{sec3-1-1}

NbSe$_3$ crystallises in a ribbon-like shape with the chain axis along $b$, the ribbon being parallel to the ($b,c$) plane. The typical size of the samples is a length of a several mm (even cm), a width along $c$ of 10--50~$\umu$m and  thickness of a few microns. The structure is monoclinic, with space group $P_{2_1}/m$, the unit cell parameters at room temperature being $a$~= 10.006~$\AA$, $b$~= 3.478~$\AA$, $c$~= 15.626~$\AA$, $\beta$~= 109.30$^\circ$ \cite{Meerschaut75,Hodeau78}. There are three types of chains in the unit cell (see figure~\ref{fig3-1}(c)): chains with strong Se-Se pairing (Se-Se distance of 2.37~$\AA$) are called chains III; those with intermediate pairing (Se-Se distance of 2.49~$\AA$) are chains I; and those with weaker bond (Se-Se distance of 2.91~$\AA$, the triangle of the chain being nearly equilateral) are chains II. The arrangement of NbSe$_3$ chains with respect to each other in the unit cell is derived from both the TaSe$_3$ and ZrSe$_3$ structures: blocks of four chains similar to those of TaSe$_3$ separated by groups of two chains of the ZrSe$_3$ type (see figure~\ref{fig3-1}(c)). The surface separating those blocks is parallel to the ($a,b$) plane. However, the distance between the Se atoms from chains I and II is only 2.73~$\AA$, indicating a relatively strong coupling between these chains.

\begin{figure}
\begin{center}
\subfigure[]{\label{fig3-2a}
\includegraphics[width=5cm]{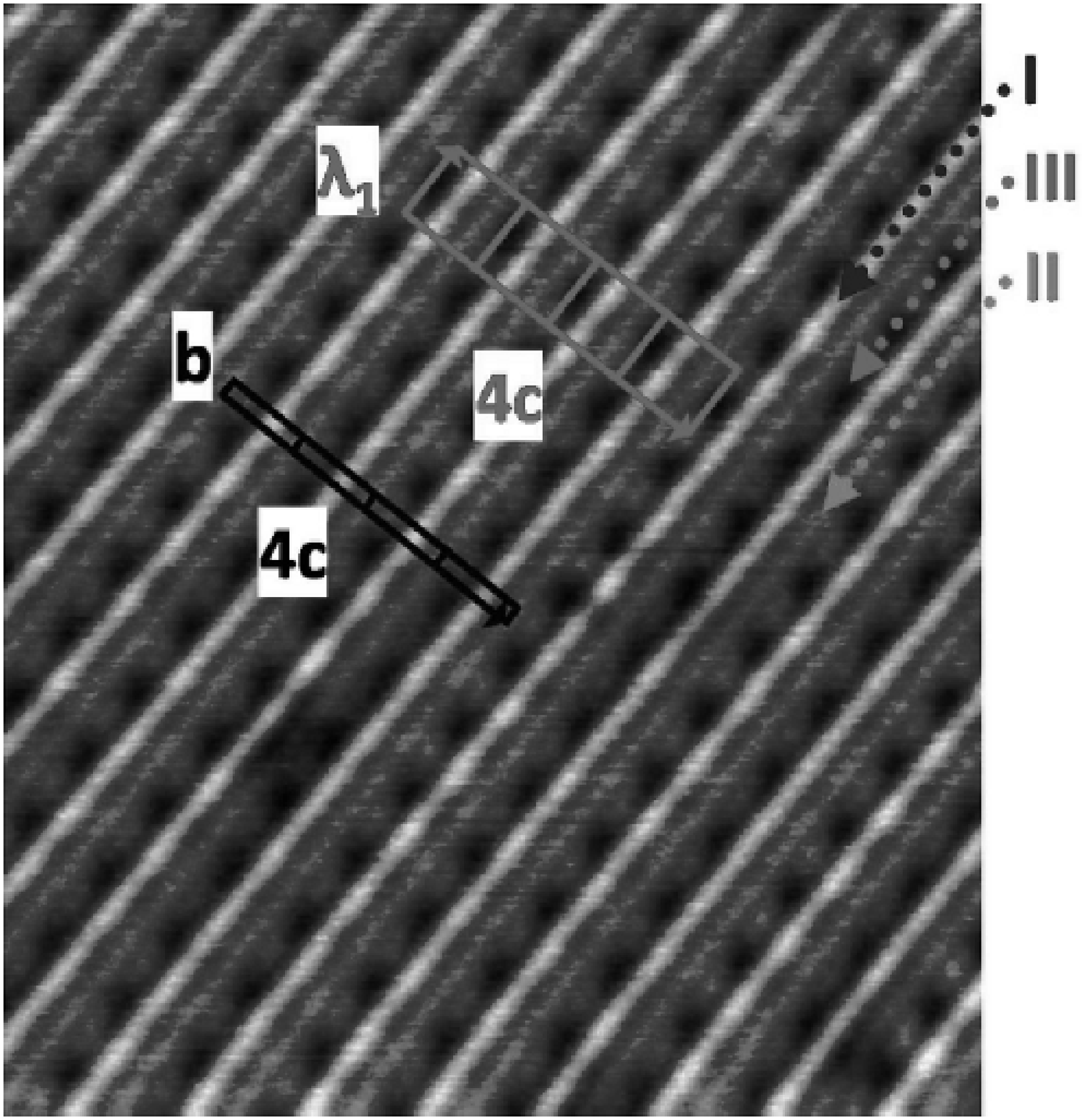}}
\subfigure[]{\label{fig3-2b}
\includegraphics[width=5cm]{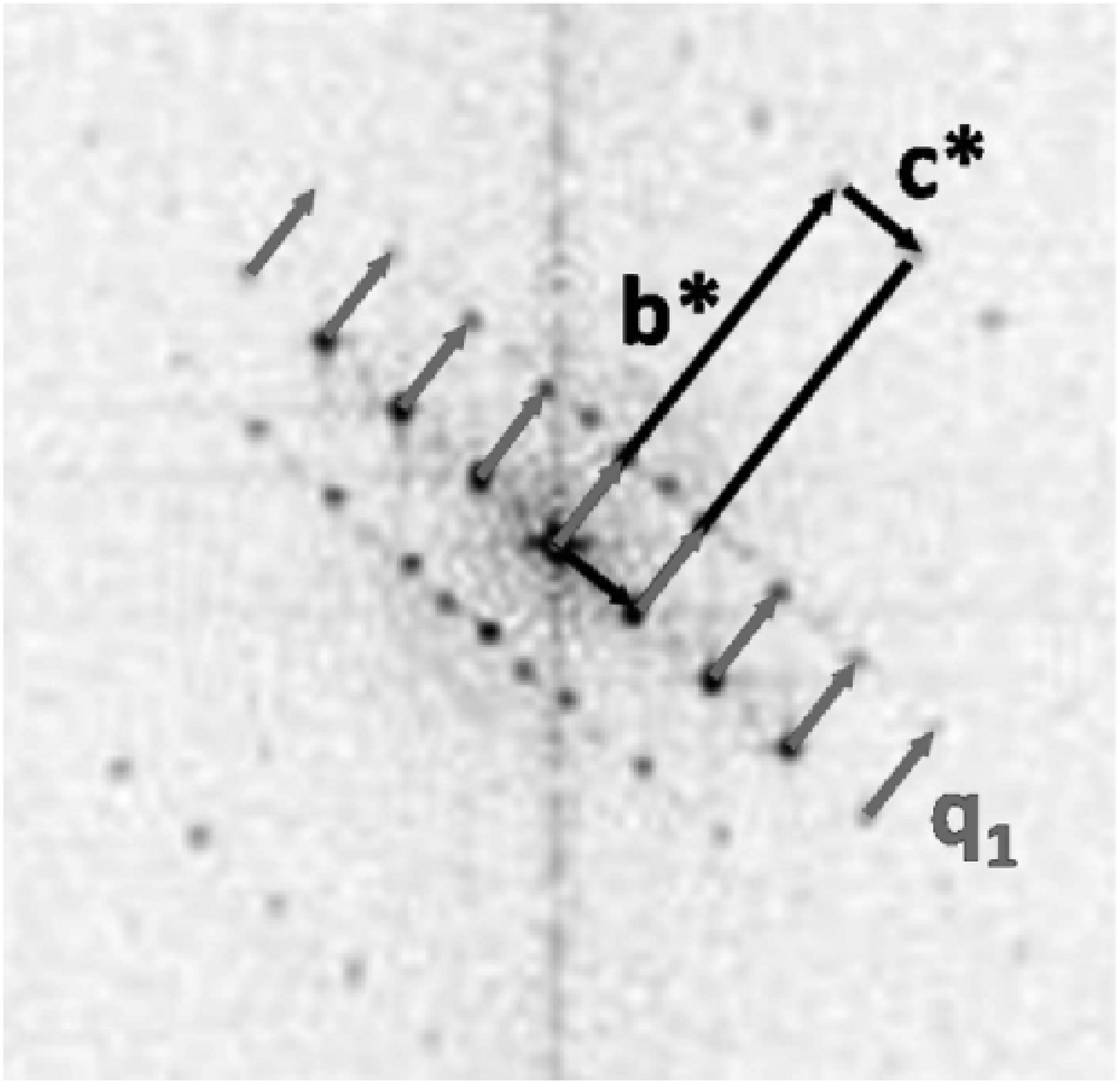}}
\subfigure[]{\label{fig3-2c}
\includegraphics[width=5.5cm]{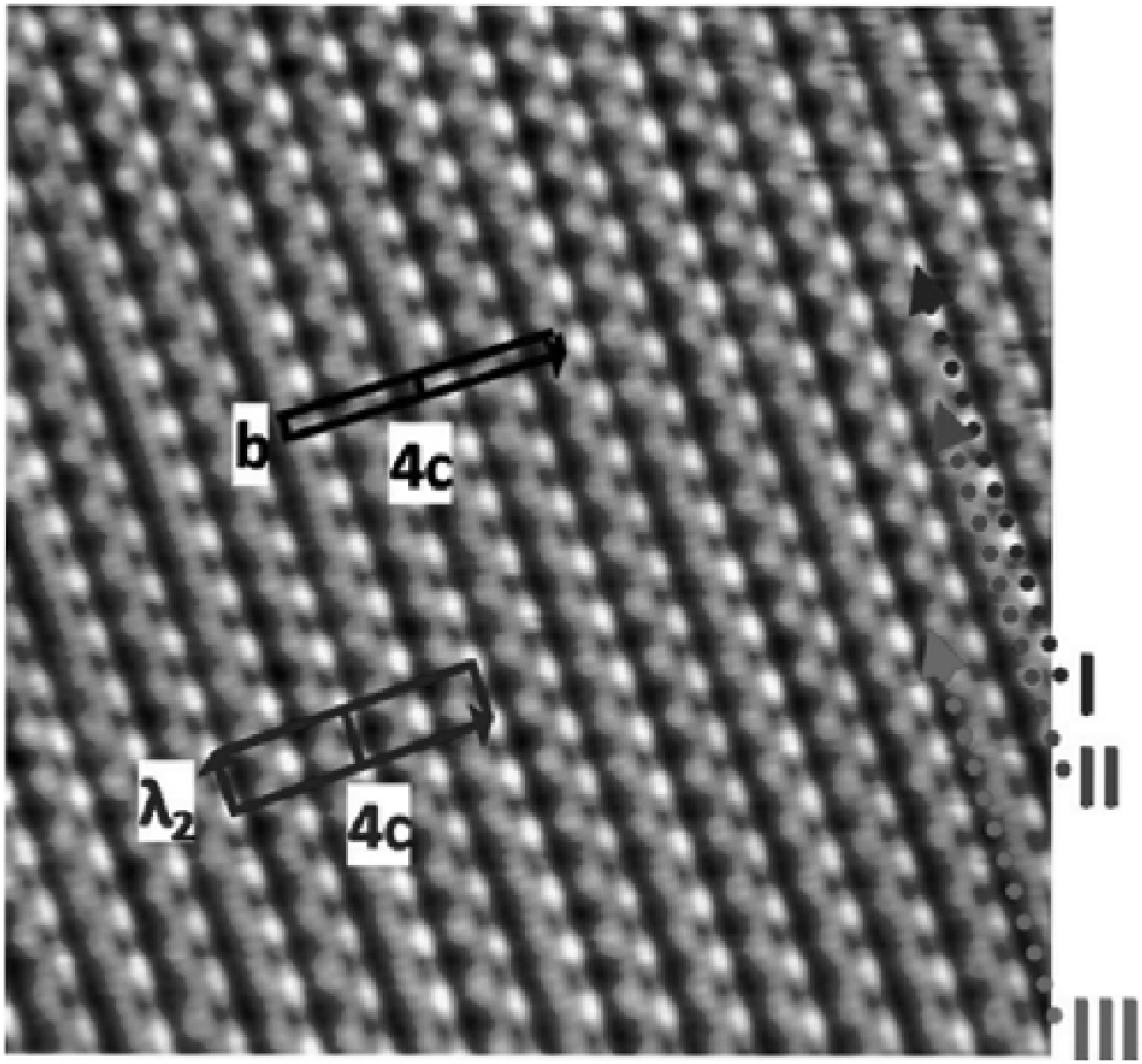}}
\subfigure[]{\label{fig3-2d}
\includegraphics[width=4.5cm]{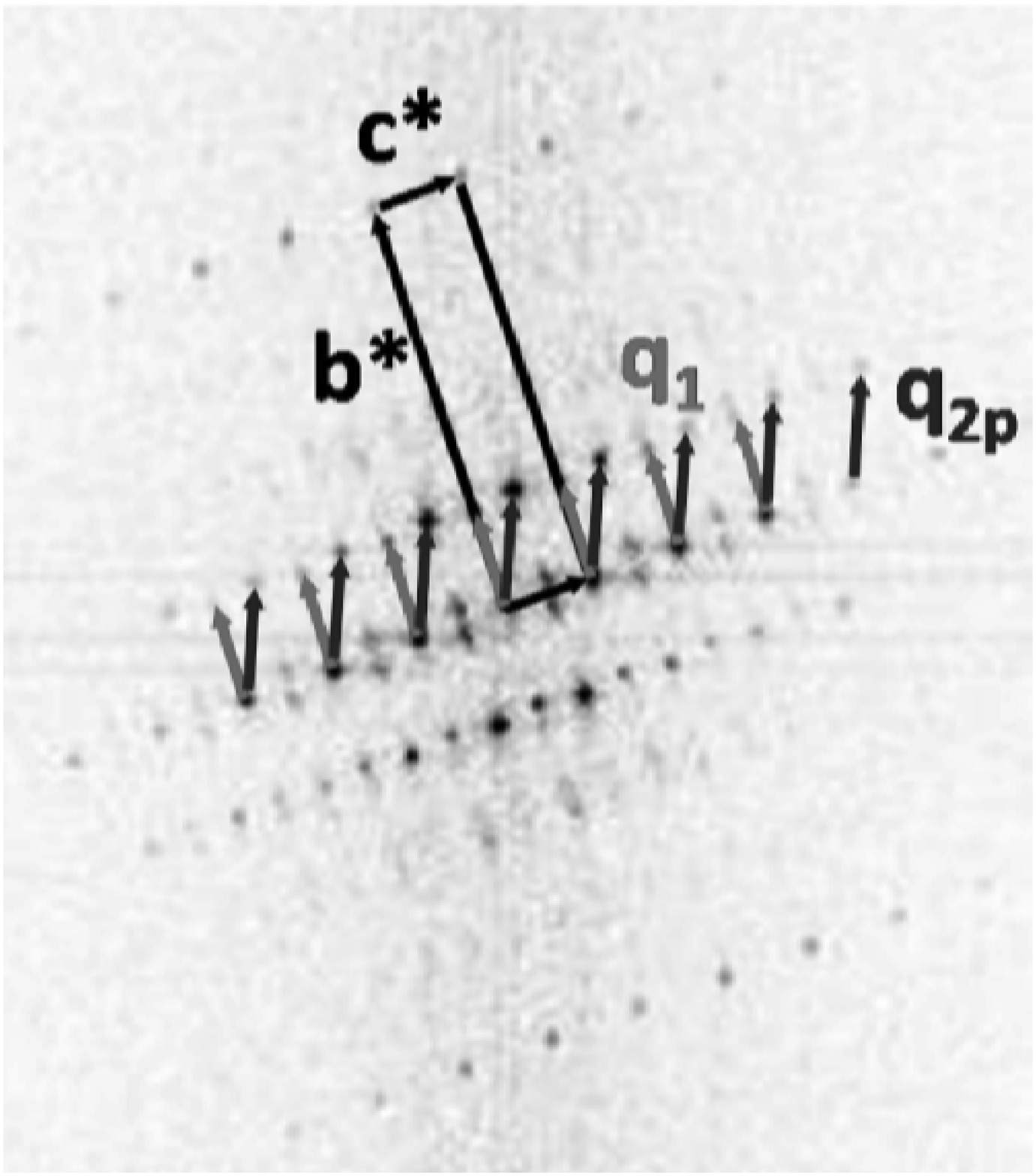}}
\caption{STM images of the ($b,c$) plane of \textit{in situ} cleaved NbSe$_3$. a)~$T$~= 77~K, scanned area 20$\times$20~nm$^2$. The unit cell is indicated together with the supercell of the $Q_1$ CDW. Three types of chains are identified, one of them carrying strong $Q_1$ CDW modulation is identified as chains III. CDW modulation forms bright maxima along the chains. $V_{\rm bias}$~= +100~mV, $I$~= 1~nA. b)~2D Fourier transform of the STM image in a). The lattice Bragg spots, corresponding to $b^\ast$ and $c^\ast$ reciprocal vectors as indicated in black arrows, $Q_1$ CDW superlattice spots by purple vectors. c)~$T$~= 5~K, $V_{\rm bias}$~= +200~mV, $I$~= 150~pA. The surface lattice unit cell and the $Q_2$ CDW surface superlattice are indicated by black and blue arrows ($\lambda_2$ is the $Q_2$ period along the chain axis. d)~2D Fourier transform of the STM image shown in c). $Q_1$ and $Q_2$ CDW superlattice spots are indicated respectively by purple and blue arrows (reprinted figure with permission from C. Brun \textit{et al.}, Physical Review B 80, p. 045423, 2009 \cite{Brun09}. Copyright (2009) by the American Physical Society).}
\label{fig3-2}
\end{center}
\end{figure}

NbSe$_3$ undergoes two successive Peierls transitions at $T_{\rm P_1}$~= 144~K and $T_{\rm P_2}$~= 59~K \cite{Chaussy76} with respective wave vectors $Q_1$~= (0, 0.241$b^\ast$, 0) and $Q_2$�= (0.5, 0.260$b^\ast$, 0.5) \cite{Fleming78}. Many experiments have shown that the $Q_1$ CDW mainly affects the Nb atoms of chains III while the $Q_2$ CDW mainly affects those of chains I. Identification of the three different chains was obtained from high resolution topographical images by means of low-temperature scanning tunnelling microscopy (STM) under ultra high vacuum on \textit{in situ} cleaved ($b,c$) surface \cite{Brun06}. Cleavage occurs inside the van~der~Waals gaps between the bilayer $b,c$ plane (see figure~\ref{fig3-1}(c) and below). After cleavage the top selenium are the highest away from the ($b,c$) plane, the nearest niobium atoms lying in the range of 1.8--2.4~$\AA$ below. Thus, the surface Se atoms are expected to determine the local electronic density of states and to contribute largely to the tunnelling current and hence to determine the contrast of the STM images \cite{Brun09}. Such an image measured at $T$~= 77~K is shown in figure~\ref{fig3-2}(a). The atomic lattice corrugation is resolved and the corresponding unit-cell vectors \textbf{b} and \textbf{c} are indicated, as well the CDW period $\lambda_1$~= $2\pi/Q_1$. The three types of chains are clearly visible and chains III carry strongly the $Q_1$ CDW modulation. It can be noted that chains II are also modulated by the $Q_1$ CDW but with a much weaker amplitude than that on chains III. That agrees with X-ray diffraction results which show that Se atoms from chains III are modulated by $Q_1$ \cite{vanSmaalen92}. Figure~\ref{fig3-2}(b) shows the 2D Fourier transform of the image presented in figure~\ref{fig3-2}(a). Both the surface lattice spots and the $Q_1$ CDW superlattice spots are observed, allowing to extract precisely the relation $Q_1$~= 0.24$b^\ast$.

It can also be noted the presence of satellites of smaller amplitude with vector \textit{$Q_2$}~= 0.26$b^\ast$+0.5$c^\ast$ clearly visible around the central peak. They correspond to the $Q_2$ CDW projected on the ($b,c$) plane. Observation of the $Q_2$ satellite spots at the surface of NbSe$_3$ almost 20~K above $T_{\rm P_2}$ indicates that the $Q_2$ CDW ordering occurs at higher temperature at the surface than in the bulk \cite{Brun10}. From measurements of in-plane correlation functions and extracting the inverse correlation lengths, a continuous evolution with $T$ was observed showing that the system is in a 2D regime between 88 and 62~K. This large regime of 2D fluctuations was analysed in the frame of a Berezinskii-Kosterlitz-Thouless type of surface transition \cite{Brun10}.

In some high-resolution images, locally a defect characterised by the occurrence of an extra CDW period on a particular chain, corresponding to a local dephasing of $\pi$ of the CDW, i.e. a local loss of the phase coherence was detected. On the STM profile measured along this chain, it is observed that the CDW amplitude is strongly reduced at this position of this defect. This defect can thus be identified as an amplitude soliton \cite{Brazovskii11}.

The STM results obtained at 5~K are shown in figure~\ref{fig3-2}(c) and the 2D Fourier transform of the image in figure~\ref{fig3-2}(d). Although the $Q_2$ CDW superlattice affects essentially chains I, surprisingly it also affects chains III, with an amplitude similar to the $Q_1$ contribution on these chains. This simultaneous double modulation on chains III leads \cite{Brun09} to a beating phenomenon between the $Q_1$ and $Q_2$ periodicities giving rise to a new domain superstructure developed along the chain axis characterised by the vector \textbf{u}~= $2\times(0.26-0.24)b^\ast$. This result is the microscopic manifestation of the coupling between the $Q_1$ and $Q_2$ CDWs in the pinned regime. This new periodic superstructure defined by the vector \textbf{u} has not been, up to now, reported in diffraction experiments. Is this coupling between $Q_1$ and $Q_2$ a surface effect or can it be detected in the bulk? Below $T_{\rm P_2}$ a phase locking of the coexisting CDWs has been anticipated \cite{Bruinsma80}. Indeed, the two NbSe$_3$ modulation wave vectors nearly satisfy the relation $2(Q_1+Q_2)\approx (111)\equiv 0$ suggesting a joint commensurability between the lattice and the two CDWs. However no anomaly in the $T$- dependence of $Q_1$ was detected \cite{Moudden90} in the vicinity of $T_{\rm P_2}$ and no lock-in transition to a true commensurate phase has been reported. A charge transfer between $Q_1$ and $Q_2$ in the sliding state \cite{Ayari04} will be discussed in section~\ref{sec5-8}.

A totally different model was suggested \cite{Prodan01,Prodan10}, based on the assumption that both $Q_1$ and $Q_2$ modulations are already together present below $T_{\rm P_1}$ indistinctly on chains III and I, being separated by anisotropic and unstable layered domains parallel to the ($b,c$) plane.
\begin{figure}
\begin{center}
\subfigure[]{\label{fig3-3a}
\includegraphics[width=5.5cm]{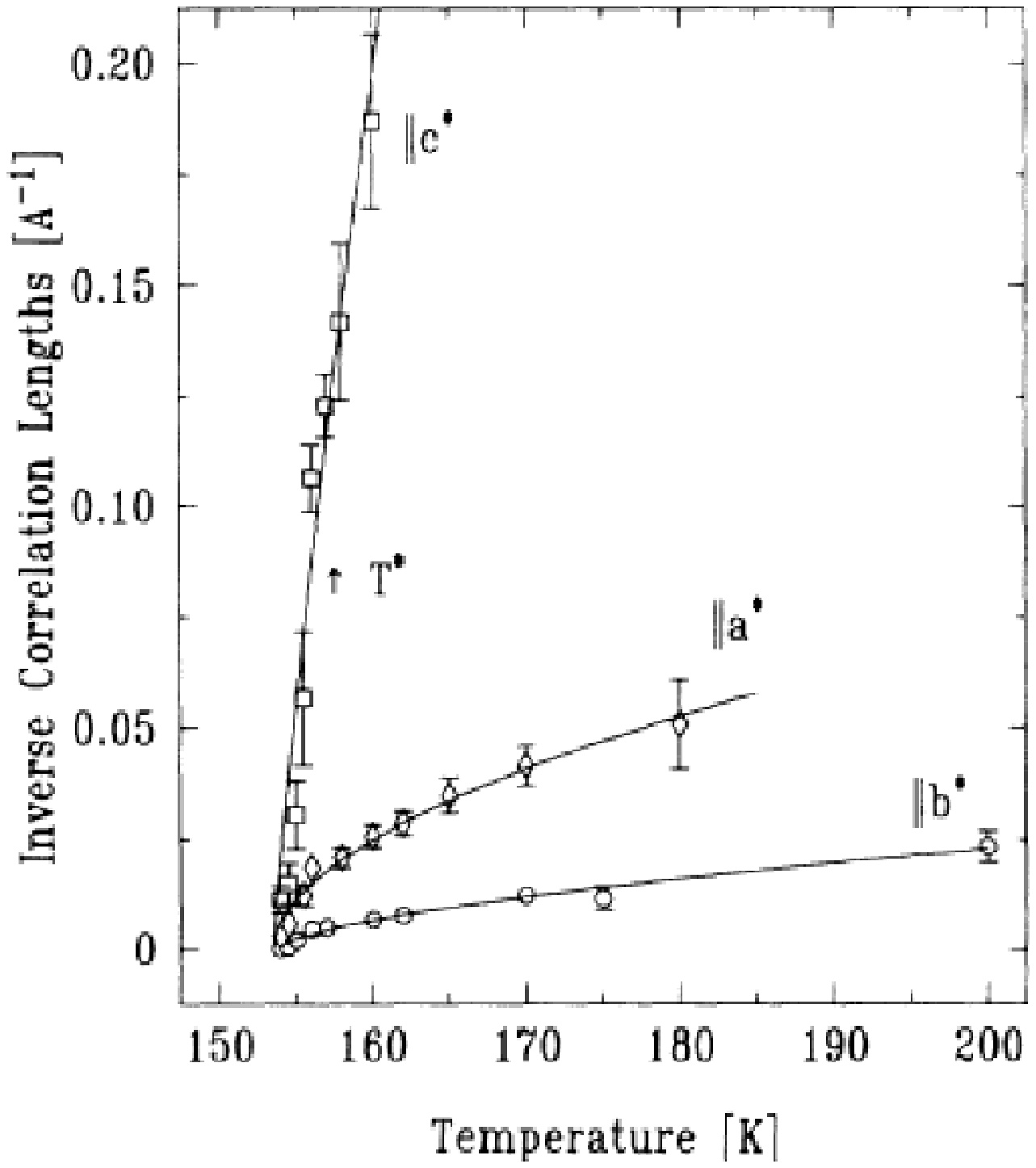}}
\subfigure[]{\label{fig3-3b}
\includegraphics[width=6.5cm]{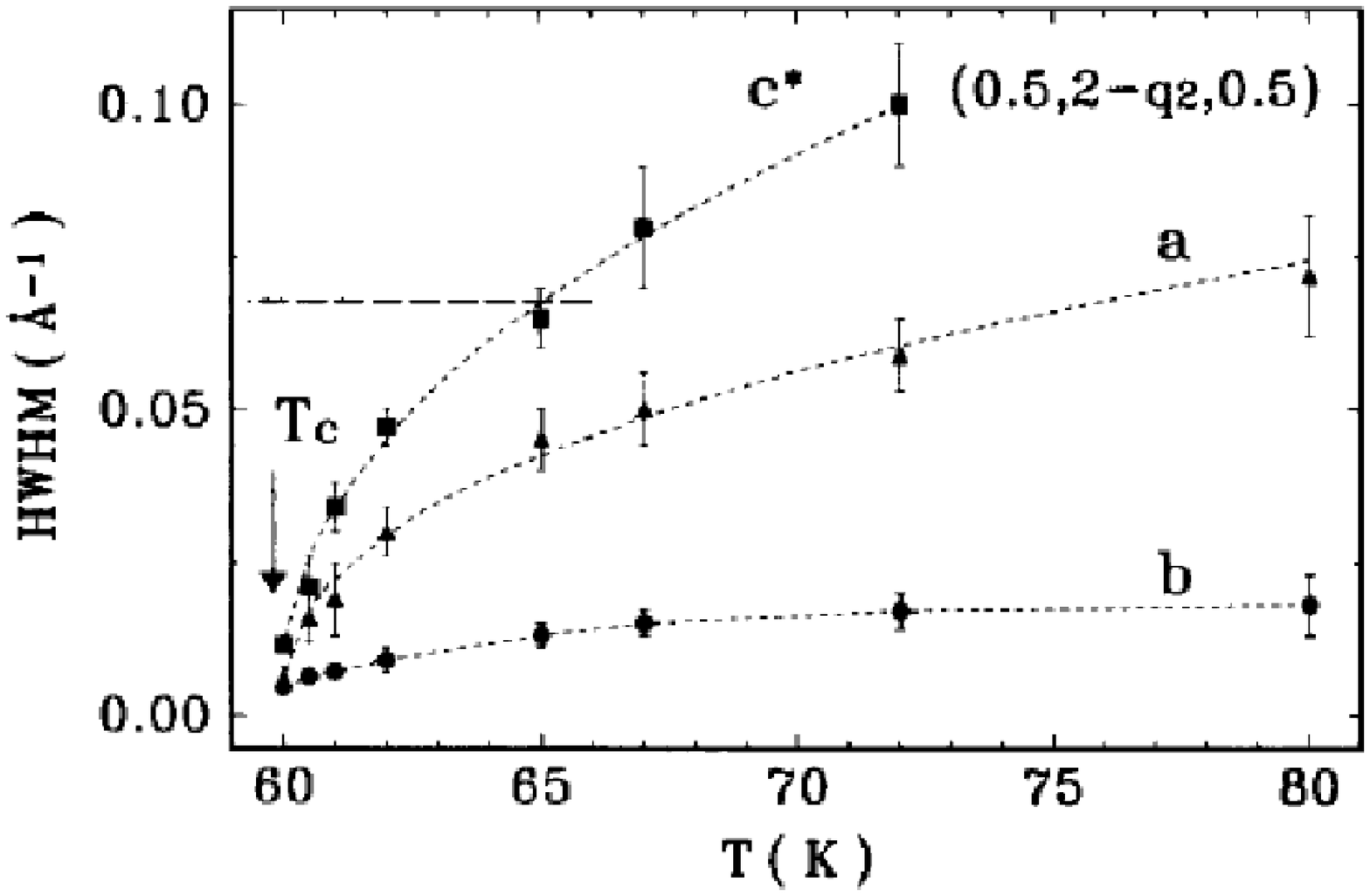}}
\caption{Temperature dependence of the inverse correlation lengths along three directions of NbSe$_3$: a)~through the $Q_1$ satellite reflection (reprinted figure with permission from A.H. Moudden \textit{et al.}, Physical Review Letters 65, p. 223, 1990 \cite{Moudden90}. Copyright (1990) by the American Physical Society); b)~through the $Q_2$ satellite reflection (reprinted figure with permission from S. Rouzi\`ere \textit{et al.}, Solid State Communications 97, p. 1073, 1996 \cite{Rouziere96}. Copyright (1996) with permission from Elsevier).}
\label{fig3-3}
\end{center}
\end{figure}

From X-ray studies, the scattering profiles of the pre-transitional CDW fluctuations have been measured in the ($b,a^\ast,c^\ast$) frame above $T_{\rm P_1}$ \cite{Moudden90} and extended above $T_{\rm P_2}$ \cite{Rouziere96} in the orthogonal ($b,a,c^\ast$) frame. The temperature dependence of the inverse correlation lengths (or Half-Width at Half Maximum (HWHM) of the satellite reflections) is shown respectively in figure~\ref{fig3-3}(a) 
and figure~\ref{fig3-3}(b). The anisotropy of the $Q_1$-pre-transitional fluctuations determined from the ratio of HWHM's was found to be (1 : 3.5 : 27) \cite{Moudden90} or (1 : 4 : 20) \cite{Rouziere96}, and that for the $Q_2$-fluctuations (1 : 3.5 : 6) \cite{Rouziere96}. It appears that the $Q_2$-fluctuations are more isotropic that the $Q_1$-ones. CDW pre-transitional fluctuations for both CDWs are essentially 2D with a cut-off just a few K above $T_{\rm P_1}$ and $T_{\rm P_2}$ when the HWHM along the $c^\ast$ direction is comparable to the inverse interchain distance (the dashed line in figure~\ref{fig3-3}(b), the arrow in figure~\ref{fig3-3}(a)). These data show that the planes of the 2D fluctuations of the $Q_1$ and $Q_2$ CDW are parallel to the ($a,b$) planes (along the ``block of two" in figure~\ref{fig3-1}). The intrachain correlation length $\xi_b$ and the interchain correlation lengths along $a$ behave similarly as a function of $T/T_c$ for both CDWs \cite{Rouziere96}; but $\xi_c$ ($T/T_c$) is a factor 4 smaller for the $Q_1$-fluctuations than for $Q_2$-ones. That was interpreted \cite{Rouziere96} by considering that the interaction between the $Q_1$ CDWs in the slabs parallel to the ($a,c$) plane is screened by the ``block of four" as shown in figure~\ref{fig3-1}(c) (i.e. the group of chains I and II) which, being still metallic, form a polarisable medium.

On the other hand, NbSe$_3$ (and other MX$_3$ compounds) can be described  as layers of chains two prisms thick weakly coupled through van~der~Waals bonds as shown in figure~\ref{fig3-1}(c) and \ref{fig3-7}. Elementary conducting layers can be defined in which the prisms are rotated and shifted with their edges towards each other. In these layers, the distances between the niobium perpendicularly to $b$ axis are relatively small, whereas the neighbouring conducting layers are separated by an insulating layer formed as a double barrier by the bases of the selenium prisms. Anisotropy of the chemical bonding has been demonstrated by the measurement of the anisotropy of compressibility. It was found \cite{Yamaya83} in kbar$^{-1}$ $K_a$~= 13.7$\pm 1.5\times 10^{-4}$, $K_b$~= 1.30$\pm 0.1\times 10^{-4}$ and $K_c$~= 5.85$\pm 0.3\times 10^{-4}$. Magnitudes of $K_b$ and $K_c$ are those of metals. The compressibility along $a$, nearly perpendicularly to the van~der~Waals gap, is large and comparable with the compressibility in layered compounds. In section~\ref{sec11}, it will be seen that intrinsic interlayer tunnelling occurs between these elementary layers.

\begin{figure}[h!]
\begin{center}
\includegraphics[width=9cm]{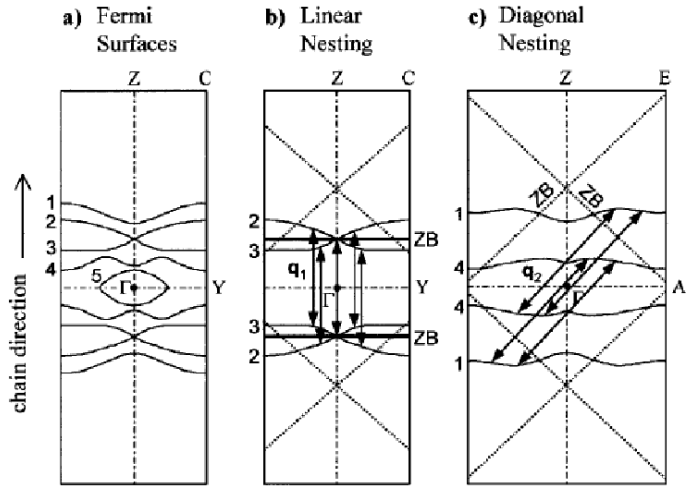}
\caption{Fermi surfaces (FS) of NbSe$_3$ calculated from density functional theory. a)~5 Fermi surfaces are seen in a cross-section along chain direction and $k_\perp$, b)~nesting of the upper CDW along the chains connecting FS 2 and 3; c)~nesting of the lower CDW in diagonal plane connecting FS 1 and 4. Dashed and solid lines: zone boundaries resulting from $Q_1$ and $Q_2$ (reprinted figure with permission from J. Sch\"afer \textit{et al.}, Physical Review Letters 87, p. 196403, 2001 \cite{Schafer01}. Copyright (2001) by the American Physical Society).}
\label{fig3-5}
\end{center}
\end{figure}

Several band structures have been proposed. It is considered that the band structure around $E_{\rm F}$ consists of six bands originating from the $dz_2$ states of the Nb atoms noticeably hybridised with the 4p states of the Se atoms. Four of these bands cross $E_{\rm F}$ corresponding to chains III and II. Different results have been presented for the position of the two bands corresponding to chains III, the bottom of one of these bands crossing \cite{Bullett79,Shima82,Schafer01} or not \cite{Canadell90} $E_{\rm F}$. Fermi surfaces (FS) of NbSe$_3$ were determined \cite{Schafer01} from density-functional calculations. Five FS were obtained as shown in figure~\ref{fig3-5}(a) 
in a cross-section along the chain direction in the ($a^\ast,b^\ast$) plane. Bands 2 and 3 cross $E_{\rm F}$ on the $\Gamma$-Z line at the same point (0.22~$\AA^{-1}$). Nesting of both CDWs is shown in figure~\ref{fig3-5}(b) and figure~\ref{fig3-5}(c), linear nesting connecting FS2 and FS3 for the $Q_1$ CDW and diagonal nesting connecting FS1 and FS4 for the $Q_2$ CDW.

\begin{figure}
\begin{center}
\includegraphics[width=7.5cm]{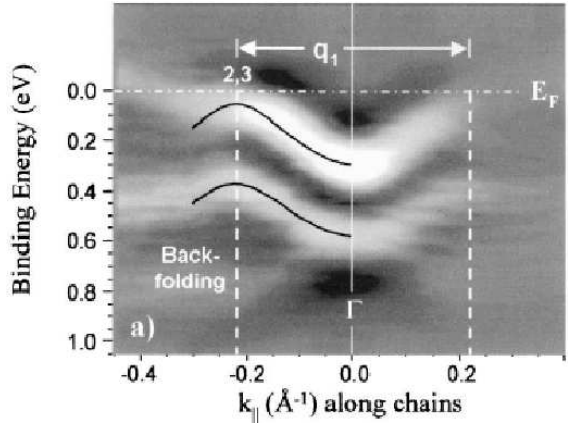}
\caption{Band map of NbSe$_3$ from photoemission spectra at room temperature along chains. Backfolding occurs at the $Q_1$ zone boundary with opening of a pseudo-gap (reprinted figure with permission from J. Sch\"afer \textit{et al.}, Physical Review Letters 87, p. 196403, 2001 \cite{Schafer01}. Copyright (2001) by the American Physical Society).}
\label{fig3-6}
\end{center}
\end{figure}

Photoemission spectra measured at room temperature do not exhibit a metallic crossing at $E_{\rm F}$. Instead backfolding and pseudogap is observed \cite{Schafer01} at $\pm 0.22~\AA$ along chains as shown in figure~\ref{fig3-6}. Thus CDW fluctuations are detectable at room temperature ($T>2T_{\rm P_1}$) with occurrence of a pseudo-gap. Faint diffuse intensity from X-ray diffraction corresponding at $Q_1$ and $Q_2$ CDW were also reported \cite{Pouget83}. That is conventionally ascribed to large fluctuations in one-dimensional systems \cite{Lee73}. However, as presented in section~\ref{sec4}, other data, especially tunnelling, are interpreted with a large corrugation of the FS perpendicularly to the chain direction resulting from imperfect nesting (as also shown in figure~\ref{fig2-9}).

NbSe$_6$ trigonal prismatic chains with the strongest Se-Se coupling (chains of type III in NbSe$_3$) form also a part of the structure of the FeNb$_3$Se$_{10}$ compound \cite{Cava81,Meerschaut81a}. The second group of chains is a double chain of edge-shared octahedra of selenium around both iron and niobium, randomly distributed within the chain. A metal-insulating transition of CDW type occurs at $T_c\approx 140$~K with a wave vector slightly $T$-dependent: $Q$~= (0, 0.270, 0) near 140~K and (0, 0.258, 0) near 6~K \cite{Hillenius81}. Thus the CDW in FeNb$_3$Se$_{10}$ is very similar to the upper CDW in NbSe$_3$, indicating again that the CDW is related to type III chains. The random potential created by the disordered adjacent Fe-Nb octahedral chain may explain the low temperature insulating state.

\subsubsection{TaS$_3$}\label{sec3-1-2}

The synthesis of TaS$_3$ yields two polytypes, one with a monoclinic, the other with an orthorhombic unit cell. Like NbSe$_3$, m-TaS$_3$ (monoclinic TaS$_3$) presents three types of chains which can be identified as chains I, II, III, two of them with a short S-S distances of 2.068~$\AA$ and 2.105~$\AA$ very close to the usual distance in $(S_2)^{2-}$ anions, the third one corresponding to a much larger distance of 2.83~$\AA$. The unit cell parameters are: $a$~= 9.515~$\AA$, $b$~= 3.3412~$\AA$, $c$~= 14.912~$\AA$, $\beta$~= 109.99$^\circ$. Two independent CDWs occurs \cite{Meerschaut79,Roucau80} in m-TaS$_3$, the upper one at $T_{\rm P_1}$~= 240~K with the wave vector $Q_1$~= (0, 0.253, 0) and the lower one at $T_{\rm P_2}$~= 160~K with $Q_2$~= (0, 0.247, 0). Contrary to NbSe$_3$, the low temperature ground state is semiconducting.
\begin{figure}[h!]
\begin{center}
\includegraphics[width=6.5cm]{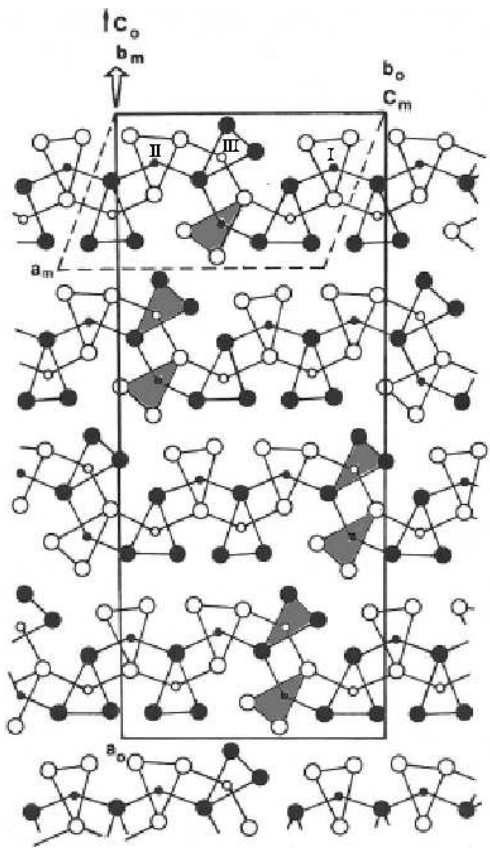}
\caption{Unit cell of orthorhombic TaS$_3$ with respect to monoclinic TaS$_3$. The chain direction $b_m$ and $C_0$ are perpendicular to the figure.}
\label{fig3-7}
\end{center}
\end{figure}

The structure of orthorhombic TaS$_3$ (o-TaS$_3$) is still unknown. The lattice parameters are very large in the directions perpendicular to the chain direction $c$-axis, namely $a$~= 36.804~$\AA$, $b$~= 15.173~$\AA$, $c$~= 3.34~$\AA$. The comparison of the unit cell parameters of both types of TaS$_3$ shows that $b_{\rm mono}\simeq c_{\rm ortho}$. The relative orientation of the two unit cells \cite{Roucau80} is shown in figure~\ref{fig3-7}. It was proposed \cite{Rouxel89} that the $C_{mcm}$ space group is more appropriate than $C_{222_1}$ as originally suggested and that the description of the unit cell should deal with four slabs built as shown in figure~\ref{fig3-7}.

Very mysteriously, although the m-TaS$_3$ polytype was really synthesised in the beginning of 80's, any further synthesis of TaS$_3$ has solely led to the orthorhombic phase. One of these two phases should be unstable and it appears that m-TaS$_3$ might be this phase (after many years kept at room temperature, some of single crystals of m-TaS$_3$ were found to be a mixture of the mono- and ortho-phases.

In o-TaS$_3$, a single CDW transition occurs \cite{Sambongi77} at $T_0$~= 215~K with a CDW wave vector temperature dependent; its components below $T_0$ are found \cite{Roucau83} to be [$0.5a^\ast,(0.125-\varepsilon)b^\ast,0.255c^\ast$] and to lock \cite{Roucau83} to commensurate  value $\varepsilon\rightarrow 0$ and $0.250c^\ast$.

The first synchrotron X-ray study of o-TaS$_3$ has shown \cite{Inagaki08} a splitting of the CDW wave vector between 130~K  and 50~K with the coexistence of an incommensurate CDW with $q_c$~= 0.252$c^\ast$ and a commensurate CDW with $q_c$~= 0.250$c^\ast$. The commensurate CDW begins to develop at around 130~K; both CDWs coexist until 30~K, at which the entire condensate becomes commensurate. These results were interpreted in terms of discommensurations, the incommensurate $c^\ast$ component being so close of being commensurate. It was also shown \cite{Inagaki08} that, by applying an electric field, the commensurate CDW is converted into the incommensurate one through the electric field induced generation of dislocations. These results have to be considered with references to previous ones where hysteresis in the temperature dependence of the resistivity between warming and cooling was observed \cite{Higgs83} (see section~\ref{sec11-3}), that being attributed to a non-equilibrium distribution of discommensurations interacting with defects. It should also be noted the difference in the temperature variation of the resistivity of both TaS$_3$ (shown in figure~\ref{fig3-7}). As a function of $1/T$, $\log\rho$ keeps a linear dependence for m-TaS$_3$, while a curvature at low temperatures (below 100~K) is observed in o-TaS$_3$ (this curvature do not exist in the transverse conductivity \cite{Nad85}). This extra source of carriers beyond normal carrier excitations through the CDW gap was ascribed \cite{Nad85} to non-linear excitations of soliton-type.

\subsubsection{NbS$_3$}\label{sec3-1-3}

NbS$_3$ exists also in the form of several polytypes. NbS$_3$ (type I) presents one type of chains in the unit cell (as ZrSe$_3$ in figure~\ref{fig3-1}(a)) with a true $(S)^{2-}$ pair. The symmetry is triclinic \cite{Rijnsdorp78} with lattice parameters: $a$~= 4.963~$\AA$, $b$~= 3.037+3.693~= 7.063~$\AA$, $c$~= 9.144~$\AA$, $\beta$~= 97.17$^\circ$, $\alpha=\gamma=90^\circ$. Dimerisation occurs along the chain $b$-axis with an alternation of short (3.037~$\AA$) and long (3.693~$\AA$) Nb-Nb distance. That results in a semiconducting state with an activation energy of 0.44~eV.

A second polytype of NbS$_3$ (called type II) was found \cite{Cornelissens78,Wang89} with a monoclinic cell with parameters $a^\prime$~= $2a,b^\prime$~= $b/2,c^\prime$~= $2c$, $\beta'=\beta$, eight chains forming the unit cell. Without dimerisation along the chain $b$-axis, the room temperature resistivity is $\sim 8\times 10^{-2}~\Omega$cm (to compare with 80~$\Omega$cm for NbS$_3$ type I). Two rows of superlattice spots are observed at room temperature corresponding to two distortion wave vectors defined \cite{Wang89} as:
\begin{eqnarray*}
Q_1=(0.5a^\ast,\;\; 0.298b^\ast,\;\; 0)\\
Q_2=(0.5a^\ast,\;\; 0.352b^\ast,\;\; 0)
\end{eqnarray*}

\begin{figure}
\begin{center}
\includegraphics[width=6cm]{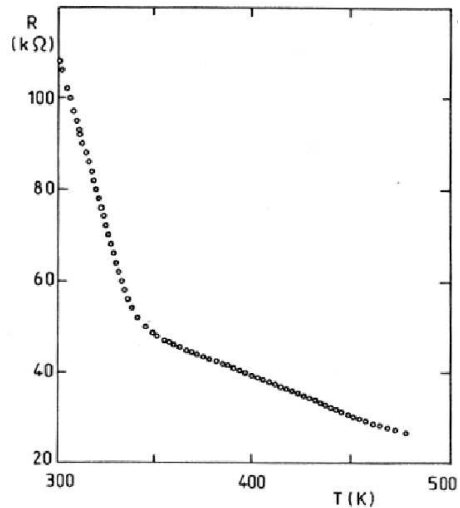}
\caption{Temperature dependence of the resistance of NbS$_3$ (type II). The CDW occurs at $T_{\rm P}$~= 340~K (reprinted figure with permission from Z.Z. Wang \textit{et al.}, Physical Review B 40, p. 11589, 1989 \cite{Wang89}. Copyright (1989) by the American Physical Society).}
\label{fig3-9}
\end{center}
\end{figure}

Electronic diffraction patterns as a function of temperature have shown that the intensity of $Q_1$ spots strongly decrease above 77$^\circ$C, diffuse streaks being still visible at 120$^\circ$C and disappearing totally at 175$^\circ$C. On the other hand the variation of the $Q_2$ spots has shown only a very weak temperature variation. It was concluded \cite{Wang89}  that the two sets of superlattice spots correspond at two independent CDWs on different chains occurring at different temperatures, one at 340~K, the second one at higher temperature (not measurable because the decomposition of NbS$_3$). Figure~\ref{fig3-9} shows the resistance variation as a function of temperature for a NbS$_3$ type II crystal in which the CDW transition corresponds to the abrupt change in the temperature-dependent resistance. Non-linear transport properties due to the CDW sliding were observed \cite{Wang89} below 340~K.

However more polytypes of NbS$_3$ may exist. A CDW (recognised by the non-linear transport properties) at $T_{\rm P}\simeq 150$~K was reported \cite{Zettl82}. Intergrowing of different phases may occur in a single whisker; thus two sets of non-linear curves were reported \cite{Zybtsev09} below 360~K and below 150~K. One may think that in the nanowhiskers considered (cross-section 730$\times$40~nm$^2$, length 33~$\umu$m), different chains with different properties were grown parallel to the $b$-axis.

\subsubsection{ZrTe$_3$}\label{sec3-1-4}

Only one type of chain is present in the unit cell of ZrTe$_3$ but in a variant (type B) with respect to the ZrSe$_3$ (type A) shown in figure~\ref{fig3-1}(a). The shape of the triangular basis is more distorted and the two interchain metal-chalcogen are unequivalent in type B. ZrTe$_3$ crystallises in a monoclinic structure with chains parallel to $b$. Resistivity measurements show an anomaly due to the transition at $T_{\rm P}$~= 63~K but only along the $a$ and $c$ directions \cite{Takahashi84}. The electrical resistivity is anisotropic with $\rho_a: \rho_b~: \rho_c$~= 1:1:10. Electronic diffraction have revealed \cite{Eaglesham84} superlattice CDW spots, with the wave vector $Q$~(0.07$a^\ast$,0,0.333$c^\ast$). This modulation with a small component  along $a^\ast$ and a tripling of the unit cell along $c^\ast$, without any component along the chain axis is very different of that in NbSe$_3$ or TaS$_3$. That results from the strong interchain bonding between Te atoms forming chains along the  $a$-axis. The Fermi surface of ZrTe$_3$ reported from ARPES consists \cite{Tokoya05} of quasi 1D and 3D surfaces, the quasi 1D electron-like Fermi surface sheets having its  origin in the Te-Te chains. The CDW modulation is stabilised \cite{Hoesch09} by the softening (Kohn anomaly) of an acoustic $a^\ast$ polarised phonon mode with a mostly transverse dispersion along $c^\ast$.

\subsection{Transition metal tetrachalcogenides (MX$_4$)$_n$Y}\label{sec3-2}

Halogened transition metal tetrachalcogenides of general formula (MX$_4$)$_n$Y with M: Nb, Ta; X: S, Se; Y: I, Br, Cl; $n$~= 2, 3, 10/3 provide a series of quasi-1D compounds. The iodine derivatives (MSe$_4$)$_n$I have been the most studied. These compounds crystallise with tetragonal symmetry and consist of MSe$_4$ chains parallel to $c$ axis and separated by iodine atoms. In an MSe$_4$ infinite chain, each metal is sandwiched by two rectangular selenium units. The dihedral angle between adjacent rectangles in 45$^\circ$, so that the stacking unit is an MSe$_8$ rectangular antiprism. The interaction between metal atoms is only through $d^2_z$ overlap along the chain (the shortest interchain metal-metal distance is about 6.7~$\AA$ to be compared to the intrachain average value $d$ of about 3.2~$\AA$. The shorter Se-Se side of rectangles is about 2.35--2.40~$\AA$ while the longer is about 3.50--3.60~$\AA$; the former value is typical of a Se$^{2-}_2$ pair. So that, if there were no iodine in the structure, the formal oxidation state would be M$^{4+}$(Se$^{2-}_2$)$_2$ (i.e a metal $d^1$ configuration). Iodine atoms being well separated from one another can be considered as I$^-$ ions. That leads to a decrease in the number of available $d$ electrons: for (MSe$_4$)$_n$I the average number of $d$ electrons on each metal ion is $(n-1)/n$ and the band filling of the $dz^2$ band is $f$~= $(n-1)/2n$. Thus (NbSe$_4$)$_3$I, (TaSe$_4$)$_2$I and (NbSe$_4$)$_{10}$I$_3$ would have 1/3, 1/4 and 7/20 filled bands. As $n$ increases, $f$ becomes closer to 1/2 which is the limit for $n\rightarrow\infty$ \cite{Gressier85b}.

(TaSe$_4$)$_2$I, (NbSe$_4$)$_2$I and (NbSe$_4$)$_{10}$I$_3$ undergo a Peierls transition respectively at $T_{\rm P}$~= 263~K \cite{Wang83a}, $T_{\rm P}$~= 210~K \cite{Fujishita84} and 285~K \cite{Wang83b} with, below $T_{\rm P}$, non linear transport properties. On the other hand (NbSe$_4$)$_3$I exhibits a ferrodistortive structural transition at $T_c$~= 274~K \cite{Gressier84a,Gressier84b,Izumi84a}.

\subsubsection{(TaSe$_4$)$_2$I}\label{sec3-2-1}

Band structure calculations \cite{Gressier84a,Gressier84b} suggest a single $d_z^2$ electronic band at the Fermi level. This band is 1/4 filled with one free electron per Ta$^{+4}$Ta$^{5+}$4Se$^{2-}$2I$^-$ formula unit. Consecutive Ta atoms occupy two alternating non equivalent sites, but the Ta-Ta distance is unique, $d_{\rm Ta-Ta}$~= 3.206~$\AA$. Due to the Se$_4$-unit rotation pattern, the crystallographic unit cell parameter is $c$~= $4d_{\rm Ta-Ta}$. On that basis, one may expect (TaSe$_4$)$_2$ to be an insulator, because the Fermi wave vector corresponds to a Brillouin zone boundary: $k_{\rm F}$~= $\frac{1}{4}(\pi/d_{\rm Ta-Ta})$~= $\pi/c$~= $c^\ast/2$. However it was argued \cite{Gressier84a,Gressier84b}, that due to the screw-symmetry of the undistorted chain, there is no energy gap associated with the $c^\ast/2$ zone boundary. In addition, interchain coupling, through the Se atoms, results in the splitting of this band leading to a Fermi vector $k_{\rm F}$ along the chain direction of 0.44$c^\ast$.

\begin{figure}
\begin{center}
\includegraphics[width=7cm]{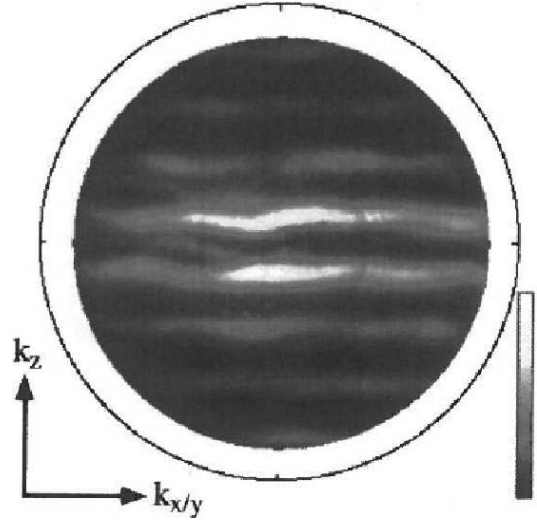}
\caption{Fermi surface mapping of (TaSe$_4$)$_2$I at $T$~= 300~K. The Fermi surface consists of nearly parallel planes perpendicular to the $k_z$ direction (reprinted figure with permission from S. H\"ufner \textit{et al.}, Journal of Electron Spectroscopy and Related Phenomena 100, p. 191, 1999 \cite{Hufner99}. Copyright (1999) with permission from Elsevier).}
\label{fig3-10}
\end{center}
\end{figure}

\begin{figure}
\begin{center}
\includegraphics[width=7.5cm]{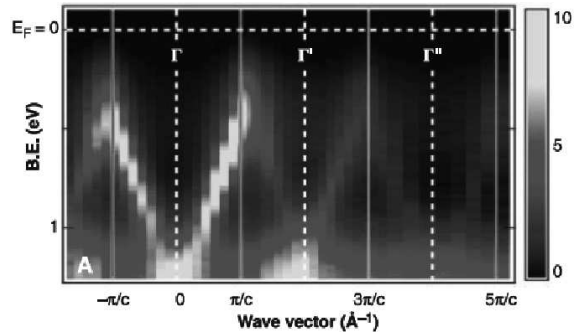}
\caption{ARPES intensity map of (TaSe$_4$)$_2$I [light $h\nu$~= 21~eV] at $T$~= 300~K along the 1D chain direction (B.E.: binding energy) (reprinted figure with permission from J. Voit \textit{et al.}, Science 290, p. 501, 2000 \cite{Voit00}. Copyright (2000) from the American Association for the Advancement of Science).}
\label{fig3-11}
\end{center}
\end{figure}

Fermi surface mapping based on measurements of the angular photoelectron intensity distribution (ARPES) has yielded \cite{Hufner99} the determination of the Fermi surface of (TaSe$_4$)$_2$I. It consists of parallel planes oriented perpendicularly to the chain direction ($c$-axis) (see figure~\ref{fig3-10}). 
Dispersion of the $d_z^2$ band has also been recorded \cite{Voit00} at $T$~= 300~K by ARPES data for a range of wave vector along the 1D chain direction as shown in figure~\ref{fig3-11}. 
The band shows a strong dispersion throughout the first Brillouin zone with minimum at $\Gamma$, the centre of the Brillouin zone, also detectable with a weaker intensity in the second and third Brillouin zones. The photoelectron intensity is peaked at $k_{\rm F}$ near the zone boundary at $\pm\pi/c$ (as explained above). The low intensity at $E_{\rm F}$ for wave vectors close to $k_{\rm F}$ is the indication of a pseudogap still present above $T_{\rm P}$ (manybody effects related to strong correlations will be discussed in section~\ref{sec7-4-1}.c).

Below $T_{\rm P}$, the CDW low temperature phase shows \cite{Fujishita84,Lee85} a set of eight satellite reflections at the positions $G+(\pm\delta H$, $\pm\delta K$, $\pm\delta L$) close to each main Bragg reflection $G=(H,K,L)$ with $\delta H$~= $\delta K$~= 0.045 and $\delta L$~= 0.085. The change of the CDW modulation of isoelectronically doped (Ta$_{1-x}$Nb$_x$Se$_4$)$_2$I ($0.1\%<x<1.2\%$) was studied in ref.~\cite{Requardt98b}. 

From band calculations, the periodic lattice distortion was considered \cite{Gressier84a,Gressier84b} to be a Ta-tetramerisation. However the satellite intensity selection rules are consistent \cite{Fujishita84,Lee85} with atomic displacements of the transverse acoustic type polarised in the basal plane in contrast to the expected $c$-polarised (-Ta-Ta) tetramerisation mode (this latter mode was later detected in ref.~\cite{Favre-Nicolin01}). 
\begin{figure}[h!]
\begin{center}
\subfigure[]{\label{fig3-12a}
\includegraphics[width=7cm]{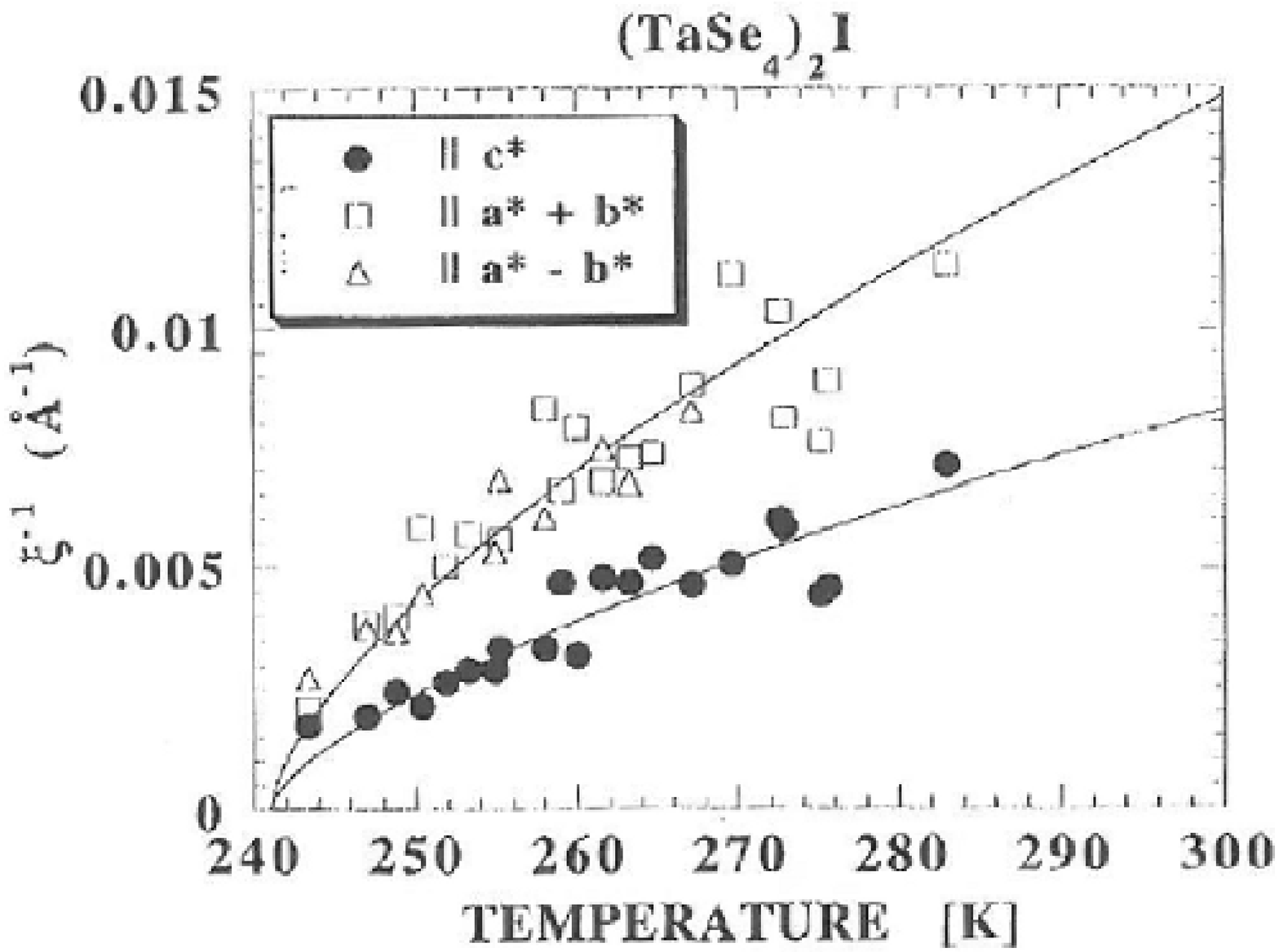}}
\subfigure[]{\label{fig3-12b}
\includegraphics[width=6cm]{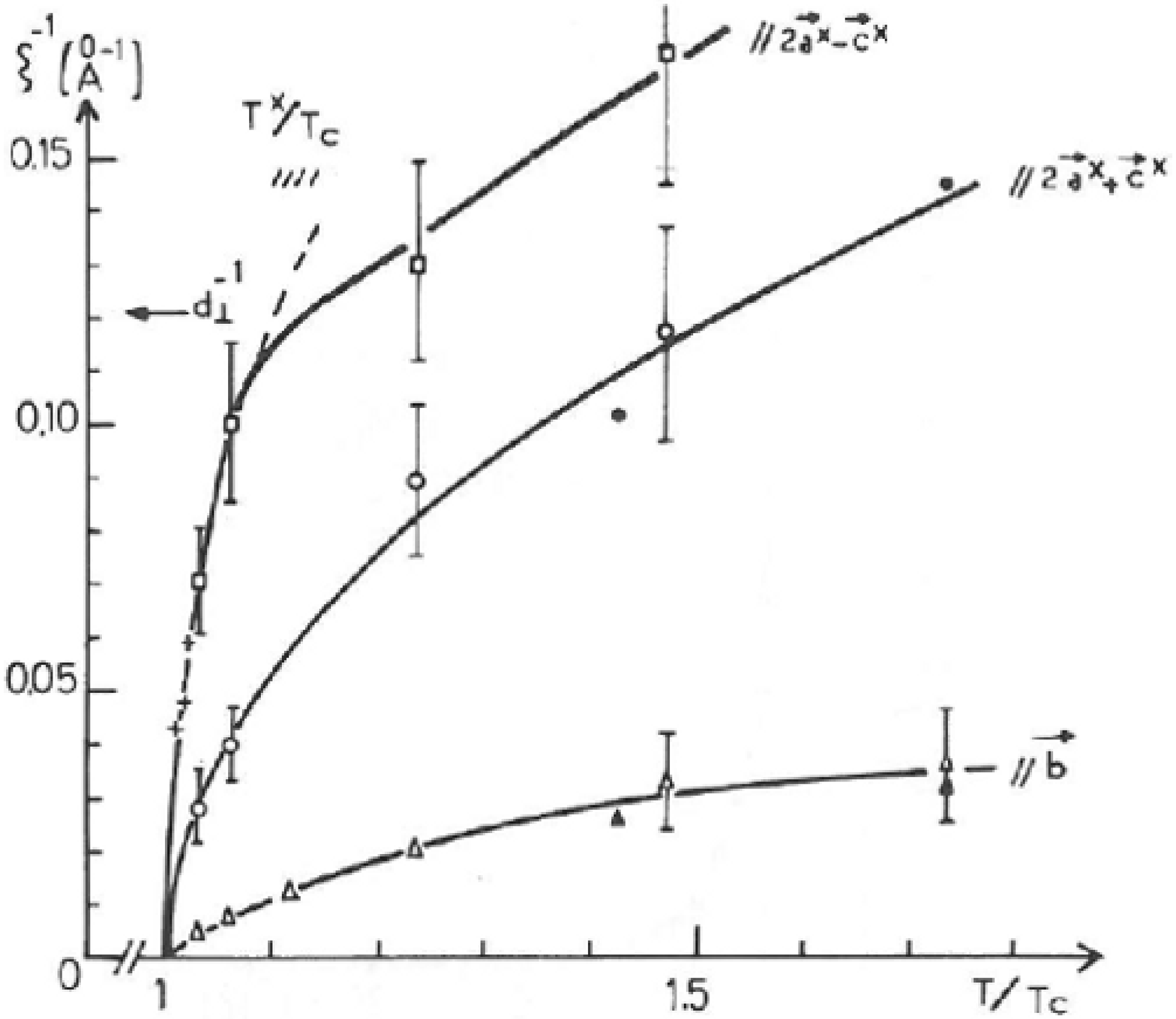}}
\caption{Temperature dependence of the inverse correlation lengths (or half-widths at half maximum (HWHM) of the satellite reflection) a)~of (TaSe$_4$)$_2$I along $a^\ast+b^\ast$, $a^\ast-b^\ast$, $c^\ast$ (reprinted figure with permission from J.E. Lorenzo \textit{et al.}, Journal of Physics: Condensed Matter 10, p. 5039, 1998 \cite{Requardt98b}. Copyright (1998) by the Institute of Physics), b)~of K$_{0.3}$MoO$_3$ along $b^\ast$, $2a^\ast+c^\ast$, $2a^\ast-c^\ast$. The 3D cross-over is indicated by the horizontal arrow (reprinted figure with permission from J.-P. Pouget \textit{et al.}, Journal de Physique (France) 46, p. 1731, 1985 \cite{Pouget85}. Copyright (1985) from EdpSciences).}
\label{fig3-12}
\end{center}
\end{figure}
Surprisingly, in spite of the 1D structure and anisotropy conductivity, a very small anisotropy of the correlation lengths parallel and normal to the chains was measured by neutron scattering \cite{Lorenzo88}. Inelastic neutron scattering data (see section~\ref{sec7-2-1}) show that at $T\rightarrow T_{\rm P}$, the main $T$-dependent precursor effect is the growth of a central component whose energy width is resolution limited. The corresponding inverse correlation lengths are shown in figure~\ref{fig3-12}(a) in comparison with those of K$_{0.3}$MoO$_3$ (see figures~\ref{fig3-3}(a) and \ref{fig3-3}(b) for NbSe$_3$). The in-chain ($\parallel c^\ast$) and in-plane $\parallel(a^\ast\pm b^\ast)$ correlation lengths are of comparable magnitude.

From the temperature dependence of critical scattering associated with the Peierls transition by means of high resolution X-ray scattering a finite anisotropy of the correlation lengths in the basal plane was found \cite{Requardt96}. An unusual exponent value, $\beta\approx 0.2$, was determined from the temperature dependence of the order parameter (integrated or peak intensity below $T_c$). This so small $\beta$ value is not predicted by a theoretical model for a continuous phase transition and may suggest a blurred discontinuous phase transition. All these unconventional results have pointed out to the need for a more sophisticated model for describing the Peierls mechanism in (TaSe$_4$)$_2$I (as it will be presented in section~\ref{sec7-2-1}.c). The temperature dependence of the resistivity of (TaSe$_4$)$_2$I and of its logarithmic derivative is shown in figure~\ref{fig3-13}(a). From the activation energy below $T_{\rm P}$~= 263~K, the ratio $2\Delta(0)/kT_{\rm P}$~= 11.4 was estimated.

\subsubsection{(NbSe$_4$)$_3$I}\label{sec3-2-2}

\begin{figure}[h!]
\begin{center}
\subfigure[]{\label{fig3-13a}
\includegraphics[width=4.5cm]{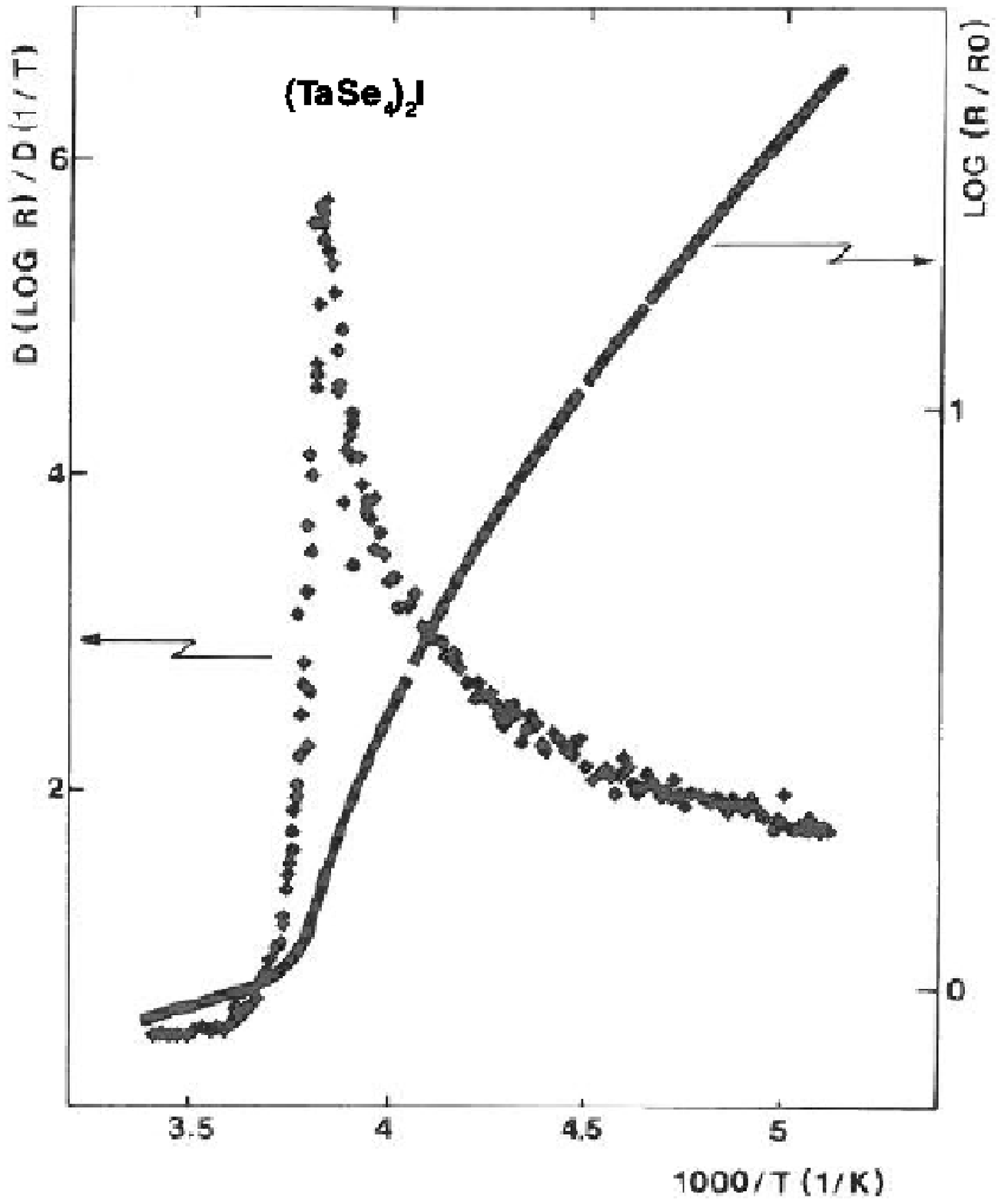}}
\subfigure[]{\label{fig3-13b}
\includegraphics[width=4cm]{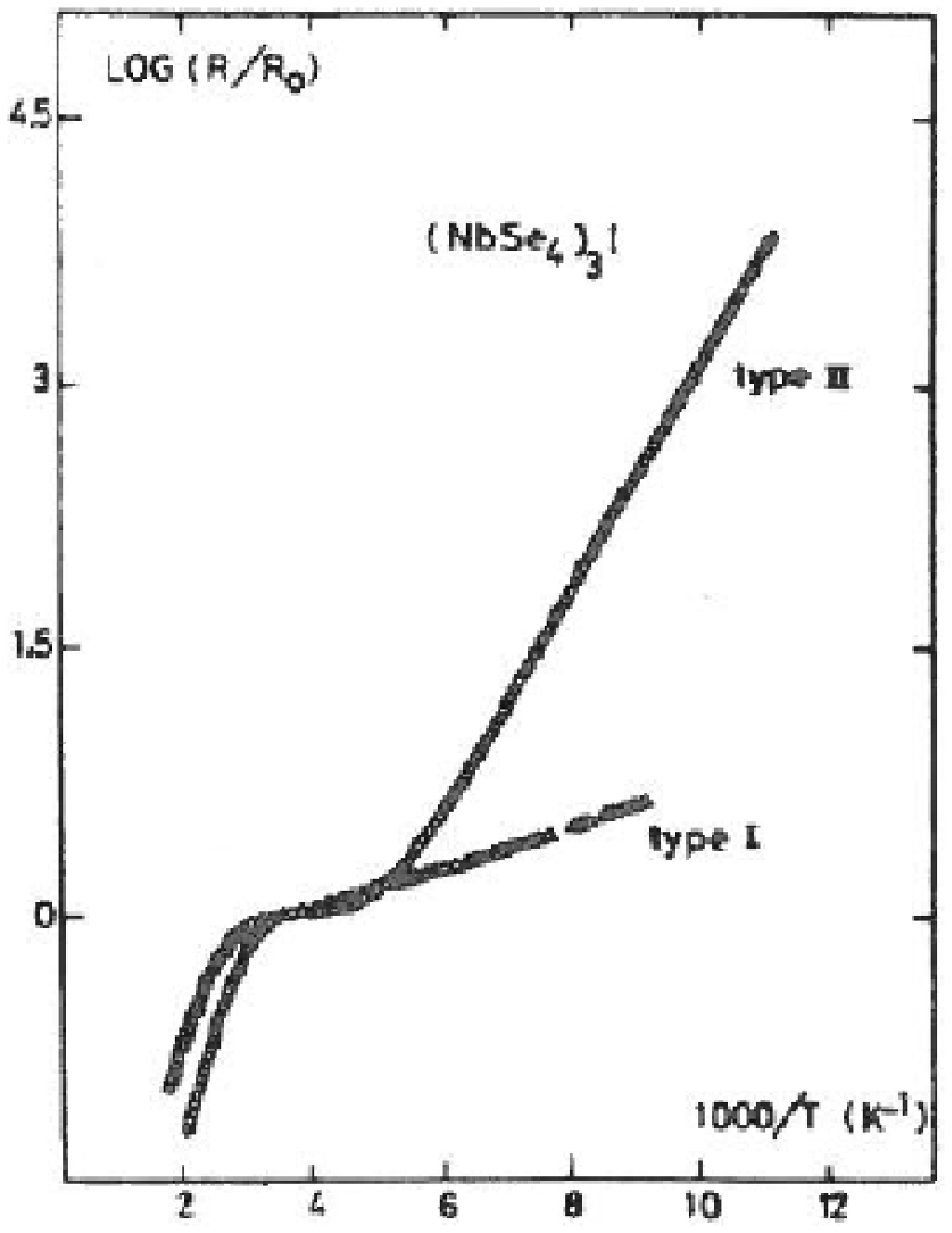}}
\subfigure[]{\label{fig3-13c}
\includegraphics[width=4.5cm]{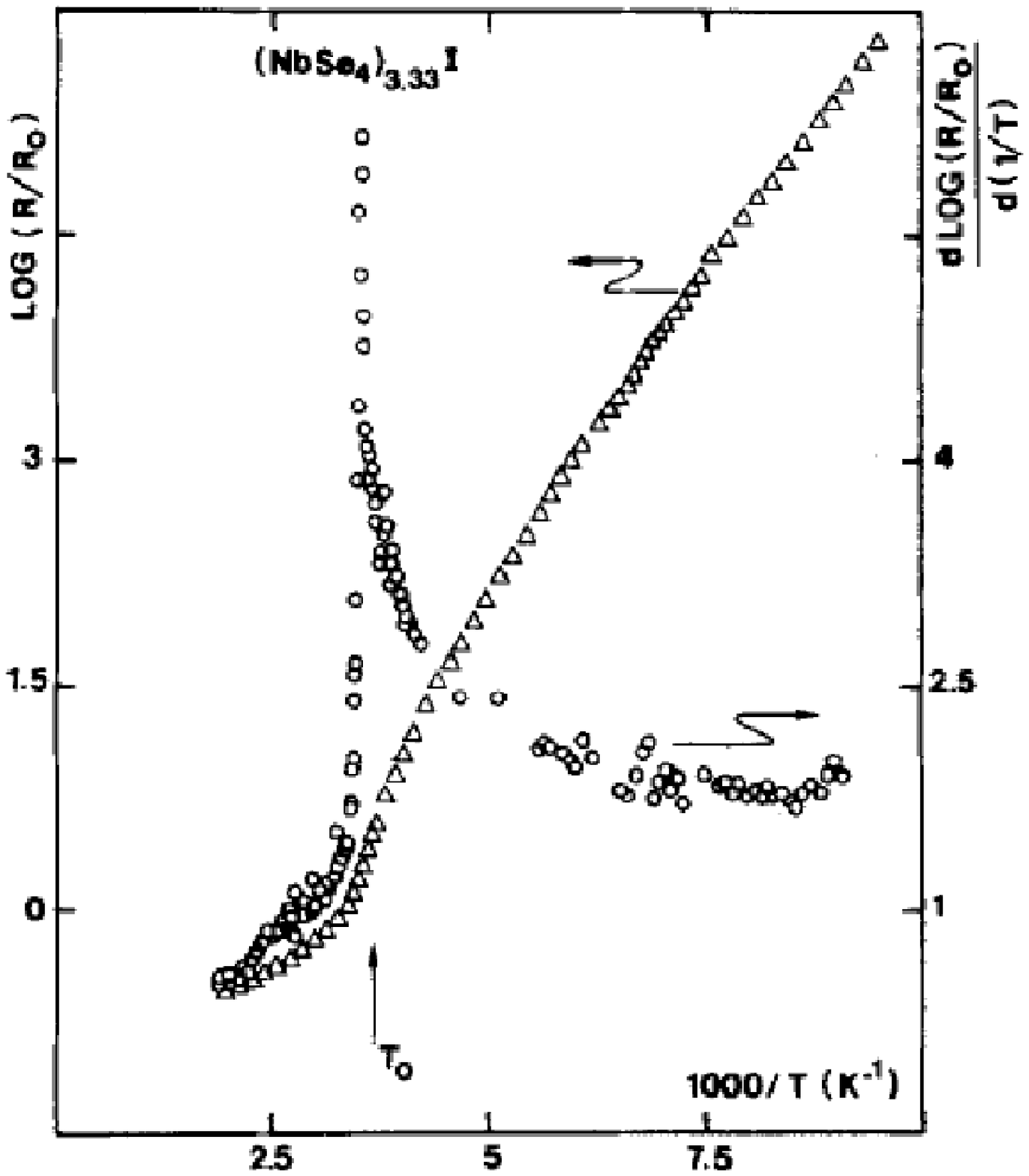}}
\caption{Temperature dependence of the resistivity and of the logarithmic derivative of (TaSe$_4$)$_2$I (a) and (NbSe$_4$)$_{10}$I$_3$ (c). Temperature dependence of the resistivity ($\log R/R_0$ versus $1/T$) of (NbSe$_4$)$_3$I. b)~type I and type II (reprinted figure with permission from P. Gressier \textit{et al.}, Journal of Solid State Chemistry 51, p. 141, 1984 \cite{Gressier84b}. Copyright (1984) with permission from Elsevier).}
\label{fig3-13}
\end{center}
\end{figure}

(NbSe$_4$)$_3$I crystallises at room temperature with tetragonal symmetry in the space group $P_4/mnc$. The unit cell parameters are $a$~= 0.949~$\AA$ and $c$~= 19.13~$\AA$. Along the two chains of a unit cell, six selenium rectangles and six niobium atoms are located within the $c$~= $6d_{\rm Ta-Ta}$ parameter. At room temperature a trimerisation of the metallic chain occurs with two different Nb-Nb distances (3.06~$\AA$ and 3.25~$\AA$), each short Nb-Nb bond being followed by two longer bands \cite{Gressier85}. The short Nb-Nb distance could be the consequence of a Nb$^{4+}$-Nb$^{4-}$ bond ($d^1-d^1$ configuration); these pairs are separated by a Nb$^{5+}$ cation ($d^0$ configuration) which produces two longer Nb-Nb distances. This agrees with the formula 2Nb$^{4+}$Nb$^{5+}$6(Se$_2$)$^{2-}$I$^-$. This trimerisation accounts for the semiconducting behaviour above $T_c$ with an energy gap of 0.21~eV. Below $T_c$ the activation energy is much smaller and reaches 0.08~eV \cite{Gressier84a,Gressier84b}. At lower temperatures ($T<120$~K) different activation energies have been reported on different types of crystals: crystals of type I keeping the very low value for the gap measured just below $T_c$ and crystals of type II for which the activation energy increases again to a value of 0.15~eV (see figure~\ref{fig3-13}(b) \cite{Gressier84a,Gressier84b}) (the physical parameters which distinguish between these two types are still unknown).

Structural analysis \cite{Izumi84b,Gressier85} of the two phases of (NbSe$_4$)$_3$I above and below $T_c$ show that the phase transition is a ferrodistortive second order transition. The two chains of the unit cell are found to be shifted in respect of each other. The metal bond in a chain is no more two long-one short bonds, but one long (3.31~$\AA$)-one intermediate (3.17~$\AA$) and one short (3.06~$\AA$). This reduced distortion accounts for the reduced energy gap below $T_c$. The space group change is $P_4/mnc\rightarrow P\bar{4}2_1c$. This low temperature space group belongs to the non-centrosymmetric crystal class $\bar{4}2m$, which shows optical activity with second harmonic generation. Although it is a non centrosymmetric space group, $P\bar{4}2_1c$ has no unique polar (symmetry) axis, so that neither pyroelectricity nor ferroelectricity or piezoelectricity under hydrostatic or uniaxial pressure are expected \cite{Gressier85}. However piezoelectricity under torsion which induces an electric moment along non polar directions (i.e. directions normal to x, y or z axis) was predicted \cite{Gressier85}. Nevertheless broadband dielectric spectroscopy measurements on (NbSe$_4$)$_3$I below $T_c$ resembles the behaviour of relaxor ferroelectrics \cite{Staresinic06}.

\subsubsection{(NbSe$_4$)$_{10}$I$_3$}\label{sec3-2-3}

(NbSe$_4$)$_{10}$I$_3$ crystallises in the same tetragonal symmetry and space group $P_4/mnc$. There are 10 selenium rectangles and 10 niobium atoms within one $c$~= $10d_{\rm Ta-Ta}$ parameter along each chain. While for (NbSe$_4$)$_3$I and (TaSe$_4$)$_2$I, iodine atoms are identically distributed in all their channels, for (NbSe$_4$)$_{10}$I$_3$ the channels $0,0,z$ and $1/2,1/2,z$ are differently occupied by four and two iodine atoms, respectively \cite{Meerschaut84}. In agreement with the band filling, the Nb-Nb bond alternation is represented by the sequence 3.17~$\AA$, 3.17~$\AA$, 3.23~$\AA$, 3.15~$\AA$ and 3.23~$\AA$. The chain distortion is much less pronounced than for (NbSe$_4$)$_3$I; that is reflected by a resistivity at room temperature two orders of magnitude lower: $10^{-2}~\Omega$cm for (NbSe$_4$)$_{10}$I$_3$ to be compared with 1~$\Omega$cm for (NbSe$_4$)$_3$I. A CDW transition occurs \cite{Wang83b} at $T_{\rm P}$~= 285~K ($13^\circ$C) with a semiconducting gap below $T_{\rm P}$ of 0.13~eV ($2\Delta(0)/kT_{\rm P}$~= 13.7 (see figure~\ref{fig3-13}(c)). CDW satellites were observed \cite{Roucau85} at 100~K, with the $Q$ vector ($0,0,0.487$). 

\begin{figure}
\begin{center}
\includegraphics[width=10cm]{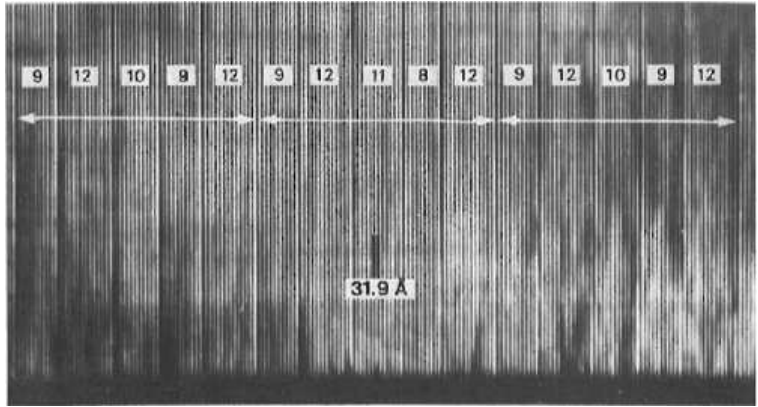}
\caption{Interference fringes and ``fault planes" in (NbSe$_4$)$_{10}$I$_3$ showing the complex sequence ($9c+12c+10c+9c+12c$) (reprinted figure with permission from Charge Density Waves in Solids, Lecture Notes in Physics, 217, C. Roucau and R. Ayrolles, p. 65, 1985 \cite{Roucau85}. Copyright (1985) from Springer Science + Business media).}
\label{fig3-14}
\end{center}
\end{figure}

Additional spots in diffraction patterns were studied by high resolution electron microscopy \cite{Roucau85}; interference fringes corresponding to atomic planes perpendicular to the $c$ axis with periodicity $c$~= 31.9~$\AA$ were observed. The periodicity of these fringes is disturbed by ``fault planes" which occasionally modify the distance between two consecutive lattice planes. This periodicity is about $11c$ and corresponds to the additional spots in diffraction. In fact the sequence is more complex and reappears periodically. In figure~\ref{fig3-14} the observed sequence is $(9c+12c+10c+9c+12c)$ namely $52c$~= 1660~$\AA$. However these plane defects do not vary with temperature.

X-ray study below $T_{\rm P}$ has shown that the lattice of (NbSe$_4$)$_{10}$I$_3$ exhibits \cite{Vucic96} a structural transformation from tetragonal to monoclinic at the transition temperature which coincides with the Peierls transition temperature $T_{\rm P}$. Below $T_{\rm P}$, the spontaneous monoclinic strain produces four domains corresponding to the four tetragonal $\langle 100\rangle$ directions. The monoclinic distortion was interpreted as a relative slip of (NbSe$_4$) chains along the chain, $c$-axis, the magnitude of which is measured by the monoclinic angle $\beta$. In the temperature interval between 285~K down to 130~K, the net change of $\beta$ was found \cite{Vucic96} to be small $\sim 2.1^\circ$. The monoclinic deformation of the tetragonal lattice, i.e. shear-strain was interpreted \cite{Vucic96} as an elastic response to the long wave length transverse modulation of the 3D-CDW order and qualitatively explained by taking into account the Coulomb interaction between CDWs on neighbouring chains and next-nearest neighbours.

The metal-metal sequences in (MX$_4$)$_n$Y compounds are listed in table~\ref{tab3-1}. 

\begin{sidewaystable}
\tbl{Various metal-metal sequences in the (MX$_4$)$_n$Y compound}
{\begin{tabular}{clccccccccccccc|c|cc}\toprule
\multicolumn{1}{c}{Compound} & \multicolumn{14}{c}{M-M sequence} & \multicolumn{1}{c}{$\rho_{\rm RT}$ ($\Omega$.cm)} & References \\ \hline
&&&&&&&&&&&&&&&&\\ 
(TaSe$_4$)$_2$I & & -----~ & Ta & 
$\begin{array}{c} \vspace{-4pt} 3.206 \\  \vspace{4pt} $--------$ \\  \end{array}$ & Ta &
$\begin{array}{c} \vspace{-4pt} 3.206 \\  \vspace{4pt} $--------$ \\  \end{array}$ & Ta &
$\begin{array}{c} \vspace{-4pt} 3.206 \\  \vspace{4pt} $--------$ \\  \end{array}$ & Ta & -----~& & & & & $1.5\times 10^{-3}$ & \cite{Gressier84a} \\
&&&&&&&&&&&&&&&&\\
(NbSe$_4$)$_3$I &  $\begin{array}{l}
 \\ \\T>T_c\\ \\ T_c=275~{\rm K} \end{array}$
& -----~ & Nb & 
$\begin{array}{c} \vspace{-4pt} 3.25 \\  \vspace{4pt} $--------$ \\  \end{array}$  & Nb &
$\begin{array}{c} \vspace{-4pt} 3.25 \\  \vspace{4pt} $--------$ \\  \end{array}$ & Nb &
$\begin{array}{c} \vspace{-4pt} 3.06 \\  \vspace{4pt} $--------$ \\  \end{array}$ & Nb & -----~& & & & & 
1 & \cite{Gressier85}\\
&&&&&&&&&&&&&&&&\\
& $T<T_c$& -----~ & Nb & 
$\begin{array}{c} \vspace{-4pt} 3.31 \\  \vspace{4pt} $--------$ \\  \end{array}$ & Nb &
$\begin{array}{c} \vspace{-4pt} 3.17 \\  \vspace{4pt} $--------$ \\  \end{array}$ & Nb &
$\begin{array}{c} \vspace{-4pt} 3.06 \\  \vspace{4pt} $--------$ \\  \end{array}$ & Nb & -----~& & & & & & \cite{Gressier84b} \\
&&&&&&&&&&&&&&&&\\
(NbSe$_{4}$)$_{10}$I$_3$ & & -----~ & Nb & 
$\begin{array}{c} \vspace{-4pt} 3.17 \\  \vspace{4pt} $--------$ \\  \end{array}$ & Nb &
$\begin{array}{c} \vspace{-4pt} 3.17 \\  \vspace{4pt} $--------$ \\  \end{array}$ & Nb &
$\begin{array}{c} \vspace{-4pt} 3.23 \\  \vspace{4pt} $--------$ \\  \end{array}$ & Nb &
$\begin{array}{c} \vspace{-4pt} 3.15 \\  \vspace{4pt} $--------$ \\  \end{array}$ & Nb &
$\begin{array}{c} \vspace{-4pt} 3.23 \\  \vspace{4pt} $--------$ \\  \end{array}$ & Nb &  -----~&  $ 10^{-2}$ & \cite{Meerschaut84}\\
\botrule
\end{tabular}}
\label{tab3-1}
\end{sidewaystable}

The temperature dependence of the resistivity of transition metal tri- and tetrachalcogenides was drawn in \cite{Monceau85}.

\subsubsection{Tetratellurides}\label{sec3-2-4}

Among the transition metal tetrachalcogenides NbTe$_4$ and TaTe$_4$ form a completely different class of compounds. As for halogenated tetrachalcogenides, the transition metal is sandwiched between two square layers of Te atoms forming a set of parallel chains. But Te atoms from neighbour chains form a strong covalent bond that can be considered as dimers. Due to this bonding between Te from adjacent chains, these two compounds are unlikely to be considered as Q1D, at least in their structural properties. The structure is modulated and the satellite pattern observed in diffraction experiments is characterised by three $Q$-vectors:
\begin{eqnarray*}
Q_1=(1/2,1,q_z)\\
Q_2=(1/2,0,q_z)\\
Q_3=(0,1/2,q_z)
\end{eqnarray*}
At room temperature, NbTe$_4$ is incommensurate with $q_z$~= 0.688 whereas TaTe$_4$ is commensurate with $q_z$~= 1/3. Upon cooling NbTe$_4$, the room temperature modulation gradually evolves through intermediate (LT1,LT2) phases until lock-in to a commensurate state occurs at $\sim 50$~K \cite{vanSmaalen86}. It was proposed that the temperature dependent driving mechanism towards commensurability involves the competition between nearest neighbour and next nearest neighbour column interactions. For a general review on NbTe$_4$, TaTe$_4$ and (Nb$_{1-x}$Ta$_x$)Te$_4$ see ref.~\cite{R12Boswell99}.

Measurements of the correlation lengths of the incommensurate structure fluctuations of NbTe$_4$ and of the determination of acoustic phonon dispersion were performed using neutron scattering \cite{Lorenzo92}. A CDW transition around 110~K was also reported on (Nb$_3$Te$_4$) \cite{Sekine87}.

\subsection{Pressure effects}\label{sec3-2-5}

The particular shape of the Fermi surface of Q-1D materials causes a very strong interaction between the carriers and certain phonons of their lattice. In a mean field approximation, the CDW transition temperature is given (see eqs.~(\ref{eq2-11}) and (\ref{eq2-12}) by:
\begin{equation}
T_{\rm CDW}\propto T_{\rm F}\exp(-1/\lambda)\propto T_{\rm F}\exp\left[-\,\frac{\omega_Q}{g^2N(E_{\rm F})}\right],
\label{eq3-1}
\end{equation}
where $T_{\rm F}$ is the Fermi temperature, $\omega_Q$ the bare phonon energy corresponding to the softened mode, $g$ the coupling constant between the carriers and phonons, $N(E_{\rm F})$ the electronic density of states at $E_{\rm F}$. The application of pressure has a strong effect on the CDW. The pressure hardening the bare phonon energies can render the distortion less favourable energetically, resulting in the lowering of $T_{\rm CDW}$ as can be deduced from eq.~(\ref{eq3-1}). Furthermore, the decrease in volume enhances the coupling between the 1D chains. Theoretically two behaviours can be expected. In the case of an extremely 1D material for which the transition temperature is drastically reduced by 1D fluctuations, an increase of the interchain coupling decreases the 1D character of the material, and favors the increase of $T_{\rm CDW}$ towards its mean field value. On the other hand, if in a given material the 1D fluctuations do not control $T_{\rm CDW}$, the reduction of the density of states $N(E_{\rm F})$ resulting of the increased interchain coupling causes a decrease of $T_{\rm CDW}$. At sufficient high pressure, the CDW can be suppressed. The strong electron-phonon interaction becomes then available for generating a superconducting state whose the transition temperature is now given by:
\begin{equation}
T_c=1.44\,\theta_{\rm D}\exp(-1/\lambda_s)=1.44\,\theta_{\rm D}\exp\left[-\,\frac{1}{N(E_{\rm F})V}\right],
\label{eq3-2}
\end{equation}
where $V$ is the electron pairing potential, $\theta_{\rm D}$ the Debye temperature.
\begin{figure}
\begin{center}
\includegraphics[width=7.5cm]{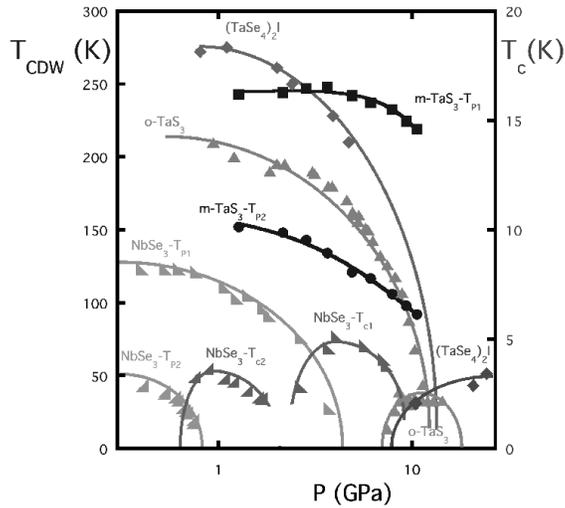}
\caption{Pressure dependence of the CDW and superconducting phase transitions of NbSe$_3$, o-TaS$_3$, m-TaS$_3$ and (TaSe$_4$)$_2$I (reprinted figure with permission from Synthetic Metals 55-57, M. Nunez-Regueiro \textit{et al.}, p. 2653, 1993 \cite{Nunez93}. Copyright (1993) with permission from Elsevier).}
\label{fig3-16}
\end{center}
\end{figure}

Pressure dependence up to 10~GPa \cite{Nunez92,Nunez93,Nunez11} of CDW and superconducting transition temperatures is shown in figure~\ref{fig3-16} for NbSe$_3$, both phases of TaS$_3$ and (TaSe$_4$)$_2$I. Superconductivity occurs in NbSe$_3$, o-TaS$_3$ and (TaSe$_4$)$_2$I when the CDW is suppressed. No zero resistivity was measured in o-TaS$_3$ indicating likely a filamentary type of superconductivity. $T_{\rm CDW}$ for the upper transition in m-TaS$_3$ and for (TaSe$_4$)$_2$I first increases with pressure, followed by a decrease with increasing pressure.

It was shown \cite{Ido90} that the lower CDW in NbSe$_3$ is suppressed at a critical pressure of about $P_2$~= 0.75~GPa while superconductivity appears, with $T_c$ rising rapidly to 3.3~K. It was noted that the logarithmic pressure slope of $T_{\rm P_2}$ is approximately the same as that of $T_c$, suggesting that both, the low temperature CDW and superconductivity above $P_2$, occur on the same crystallographic chains. The same behaviour is observed for the upper CDW suppressed \cite{Nunez92} around $P_1$~= 4~GPa. $T_c$ increases with increasing pressure above 2.5~GPa and exhibits the maximum value near $P_1$, while it decreases gradually up to 7.2~GPa. It was also suggested \cite{Nunez92} that the superconductivity above $P_1$ and the high $T$ CDW occur on the same chains (type III) and that SC and CDW coexist between 2.5 and 3.5~GPa.

\begin{figure}[h!]
\begin{center}
\includegraphics[width=7cm]{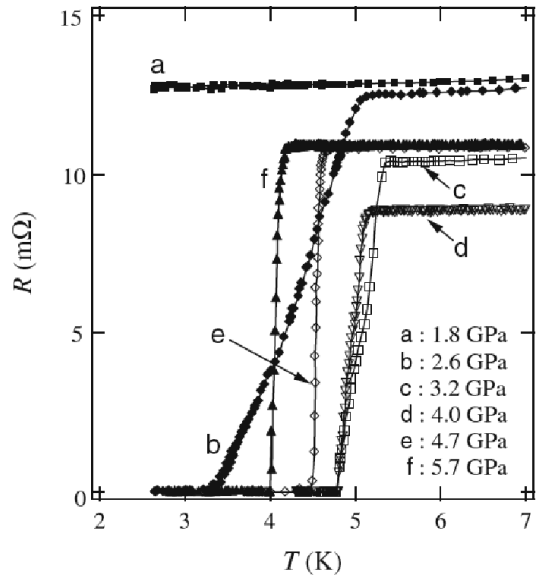}
\caption{Pressure dependence of the superconducting transition of NbSe$_3$ (reprinted figure with permission from Journal of the Physical Society of Japon 74, p. 1782, 2006, S. Yasuzuka \textit{et al.} \cite{Yasuzuka05}).}
\label{fig3-17}
\end{center}
\end{figure}

Data shown in figure~\ref{fig3-16} were obtained using an opposite-type anvil device with steatite as a solid pressure-transmitting medium \cite{Nunez92,Nunez93,Nunez11,Ido90}. That results in some pressure inhomogeneity leading to broad superconducting transitions. Homogeneous high pressure set-up has been developed \cite{Mori04} using a cubic anvil device with Daphne 7373 oil as a pressure-transmitting medium. The superconducting transition of NbSe$_3$ under pressure using the cubic anvils are shown in figure~\ref{fig3-17}. Extremely sharp superconducting transitions are observed \cite{Yasuzuka05} above 3.2~GPa. The broad transitions around 2.6~GPa cannot be attributed to the pressure inhomogeneity but correspond to the coexistence between CDW and superconductivity as shown in figure~\ref{fig3-30}(b).

\subsection{Lattice dynamics}\label{sec7-2}

Detailed information about CDW dynamics using inelastic neutron scattering technique are quite scarce. That is due, to a large extent, to the very small size of the available single crystals. However, results were obtained on the platinum chain compound K$_2$Pt(CN)$_4$Br$_{0.30}$.xH$_2$O (KCP), but intrinsic disorder has hampered the detailed analysis \cite{Comes79,Carneiro76}. The best documented results were obtained on K$_{0.3}$MoO$_3$ \cite{Pouget91,Hennion92}. Hereafter one presents lattice dynamics on-large crystals of the CDW (TaSe$_4$)$_2$I compound and of (NbSe$_4$)$_3$I. The technique of meV-energy resolution inelastic X-ray scattering is, by its small beam size and high photon flux, well adapted to small samples and has allowed to enlarge considerably the compounds accessible to phonon-dispersion studies, in particular, on NbSe$_3$ single crystals with a very small cross-section, typically 0.1~mm$\times$5~$\umu$m.

\subsubsection{(TaSe$_4$)$_2$I}\label{sec7-2-1}

In neutron scattering, the orientation of the sample was such to have an ($hh\ell$) horizontal scattering zone. Satellite reflections near the (224) fundamental Bragg reflection were monitored. The star of the modulation wave vector $\{q_s\}$ spans eight vectors \cite{Fujishita85,Lee85} ($n$~= 8). Among the eight satellite peaks at ($2\pm\eta$, $2\pm\eta$, $4\pm\delta$) with $\eta$~= 0.045 and $\delta$~= 0.085, the set of four which are located in the ($hh\ell$) scattering plane are essentially extinct, because the corresponding atomic displacements are along the normal to the scattering plane (for the structure of (TaSe$_4$)$_2$I, see section~\ref{sec3-2-1}).

\medskip
\noindent  \textit{3.4.1.a. Phonon dispersion}
\medskip 

\begin{figure}
\begin{center}
\includegraphics[width=7.25cm]{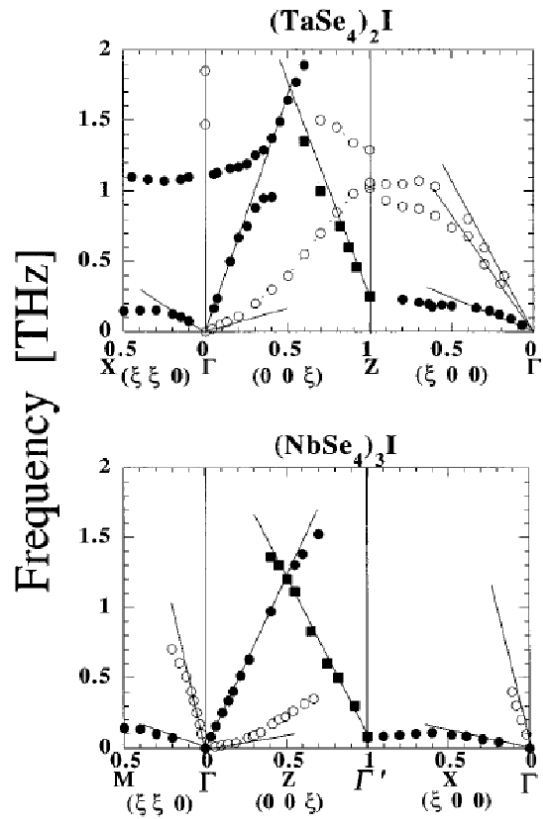}
\caption{Phonon dispersion curves along high-symmetry directions from single crystal data at room temperature. Modes polarised along (normal to) the chain axis as shown as closed (open) circles: upper: (TaSe$_4$)$_2$I (reprinted figure with permission from J.E. Lorenzo \textit{et al.}, Journal of Physics: Condensed Matter 10, p. 5039, 1998 \cite{Lorenzo98}. Copyright (1998) by the Institute of Physics) and lower: (NbSe$_4$)$_3$I (reprinted figure with permission from Physica B 156-157, P. Monceau \textit{et al.}, p. 20, 1989 \cite{Monceau89}. Copyright (1989) with permission from Elsevier). Full lines correspond to the sound velocities taken from refs~\protect\cite{Saint-Paul88a,Saint-Paul88b} for (TaSe$_4$)$_2$I and ref.~\cite{Saint-Paul88a} for (NbSe$_4$)$_3$I.}
\label{fig7-6}
\end{center}
\end{figure}
\begin{figure}
\begin{center}
\includegraphics[width=8cm]{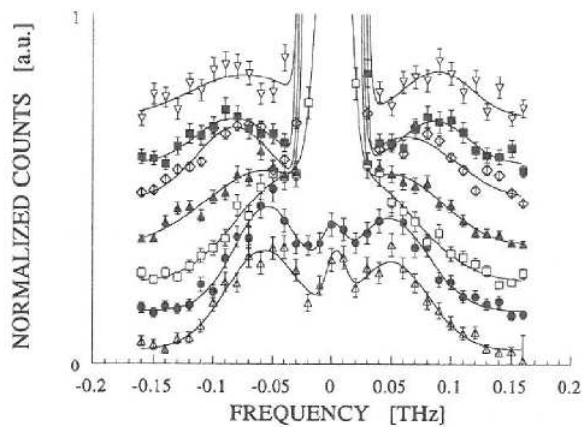}
\caption{Constant-$Q$ scans at the strong intensity satellite position above and below $T_{\rm P}$ in (TaSe$_4$)$_2$I ($T_{\rm P}$~= 253~K). The vertical scales are normalised using a factor proportional to $T$. Temperatures (in K): $\vartriangle$:  300; {\large$\bullet$}: 285; {\tiny$\square$}: 270; $\blacktriangle$: 230; $\lozenge$: 180; {\tiny$\blacksquare$}: 100; $\triangledown$: 50 (reprinted figure with permission from J.E. Lorenzo \textit{et al.}, Journal of Physics: Condensed Matter 10, p. 5039, 1998 \cite{Lorenzo98}. Copyright (1998) by the Institute of Physics.}
\label{fig7-7}
\end{center}
\end{figure}
The upper part of figure~\ref{fig7-6} presents the room temperature phonon dispersion curves measured along high-symmetry directions \cite{Lorenzo96,Lorenzo98}. There are several characteristic features of the acoustic dispersions in (TaSe$_4$)$_2$I:
\begin{itemize}
\item[-] The upward curvature of the TA-mode along (00$\xi$), characteristic of a large chain-bending stiffness. The sound velocity given by $v_{44}$~= $(C_{44}/\rho)^{1/2}$ follows a dispersion law of the type:
\begin{equation}
\omega^2(q)=v_{44}^2q^2+c^2q^4\,
\label{eq7-3}
\end{equation}
with $c$ denoting a chain-bending force. The same kind of behaviour is found for the related compound (NbSe$_4$)$_3$I at room temperature (lower part of figure~\ref{fig7-6}) and for many other Q-1D systems as discussed below in section~\ref{secbendingforces}. The Q-1D character of the (MSe$_4$)$_n$I structure is thus apparent in the elastic properties of all the compounds of the family, at least at high temperature.

\medskip
\item[-] The flat low-lying (0.2--0.3~THz) TA$_z$ branches polarised along the chains and propagating in the basal plane. The low-frequency plateau indicates that the (MSe$_4$)$_\infty$ column are relatively free to move along the $c$-direction, independently of each other, due to the weak Se-I-Se interchain bonds.

\medskip
\item[-] A detailed discussion of optic-phonon branches was given in ref.~\cite{Lorenzo93}. In figure~\ref{fig7-6} (upper part) one can see that the longitudinal acoustic branch along $c^\ast$ anticrosses at room temperature a flat mode in the region of 1.1~THz and $q_z$~= 0.25--0.45$c^\ast$. For $q$ along $a^\ast+b^\ast$, the 1.1~THz mode is observed up to the zone boundary. There is no detectable change in mode frequencies with temperature. In this frequency range there are no Raman-active modes \cite{Sekine85} whereas far-infrared experiments have detected a strong resonance at 1.1~THz  (36~cm$^{-1}$ \cite{Degiorgi91a}. A similar resonance has also been detected in K$_{0.3}$MoO$_3$ at 40~cm$^{-1}$ (1.2~THz) \cite{Degiorgi91b} and interpreted as a bound collective mode arising from the presence of polarisable impurities (see section~\ref{sec7-4-1}.a.).
\end{itemize}

\medskip
\noindent \textit{3.4.1.b. Search for Kohn anomaly}
\medskip

\begin{figure}[b]
\begin{center}
\includegraphics[width=7.5cm]{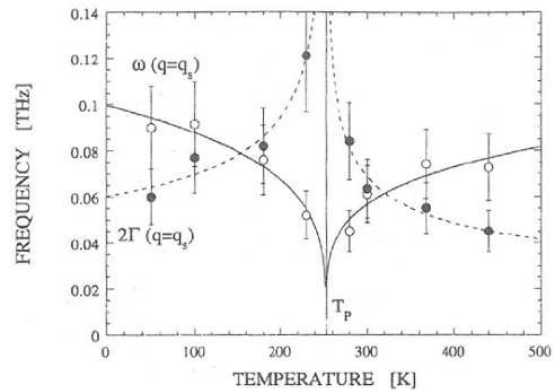}
\caption{Temperature dependence of the transverse acoustic phonon frequency and damping at the CDW satellite position in (TaSe$_4$)$_2$I ($T_{\rm P}$~= 253~K) (reprinted figure with permission from J.E. Lorenzo \textit{et al.}, Journal of Physics: Condensed Matter 10, p. 5039, 1998 \cite{Lorenzo98}. Copyright (1998) by the Institute of Physics).}
\label{fig7-8}
\end{center}
\end{figure}

The temperature dependence of the soft TA branch at the strong-satellite position is displayed in figure~\ref{fig7-7} \cite{Lorenzo98}. The elastic component corresponds to the satellite intensity at low temperatures, and to the central peak above $T_{\rm P}$. The inelastic response shows some softening as $T\rightarrow T_{\rm P}$, as well as an increase in phonon damping. The spectra in figure~\ref{fig7-7} have been analysed using a damped harmonic oscillator response function, convoluted with the instrument energy resolution. The results of the analysis are plotted in figure~\ref{fig7-8}. 
The phonon energy decreases close to $T_{\rm P}$ and at the same time the damping increases. Below $T_{\rm P}$ one recovers the same response as above, only with a slightly reduced structure factor. No indication for a change in line-shape associated with a splitting of the TA response into phase and amplitude modes was found as would be expected for a purely displacive transition. The softening of the acoustic phonon does not have a critical character; only a critical behaviour of the elastic central peak is observed: this leads to view the Peierls transition in (TaSe$_4$)$_2$I more as an ordering process than a displacive-type instability.

\bigskip
\noindent\textit{3.4.1.c. Acoustic-optic-coupled mode model}
\medskip

Band structure calculations \cite{Gressier84a,Gressier84b} indicate that (TaSe$_4$)$_2$I should be a zero-gap semiconductor with a Fermi wave vector close to $2k_{\rm F}$~= $c^\ast$, where $c$~= $4d$, $d$ being the in-chain Ta-Ta distance. It is then expected that the ionic displacements associated with the new modulated periodicity involve mostly z-polarised Ta displacements leading to an LLSS pattern of long (L) and short (S) Ta-Ta distances on each chain (Ta-tetramerisation). In lattice dynamical terms, the Peierls instability is expected to correspond to the condensation of a long-wavelength optic vibration.

However structural refinements \cite{Lee85} have led to the paradoxal result that the modulation has a strong acoustic character i.e., with all atoms moving along the same direction and with the same amplitude and that the displacements are mainly perpendicular to the chain axis. Since the electronic variables are unlikely to couple directly to such a long wavelength acoustic shear wave, a model was proposed \cite{Lorenzo98} in which the soft Ta-tetramerisation modes interact with the acoustic degrees of freedom and induce the condensation of a mixed acoustic/optic ionic modulation. The validity of this model was confirmed by the structural evidence for Ta-tetramerisation displacements from X-ray anomalous diffraction \cite{Favre-Nicolin01}. In this model, the values of the satellite wave vector components are not related to the topology of the conduction electron Fermi surface. They are determined by the strength of the gradient interaction terms between optical and acoustic degrees of freedom. Similarly to models developed in the context of incommensurate long-wavelength-modulated dielectrics such as quartz (see section~\ref{sec2-15}), the incommensurate structure arises from the presence of a pseudo-Lifshitz invariant involving an optical order parameter and the elastic deformations. The Ta-tetramerisation provides the key for understanding the mechanism of the phase transition at $T_{\rm P}$ which can be described as a Brillouin zone centre Peierls instability.

\subsubsection{(NbSe$_4$)$_3$I}\label{sec7-2-2}

\begin{figure}[h!]
\begin{center}
\includegraphics[width=7.5cm]{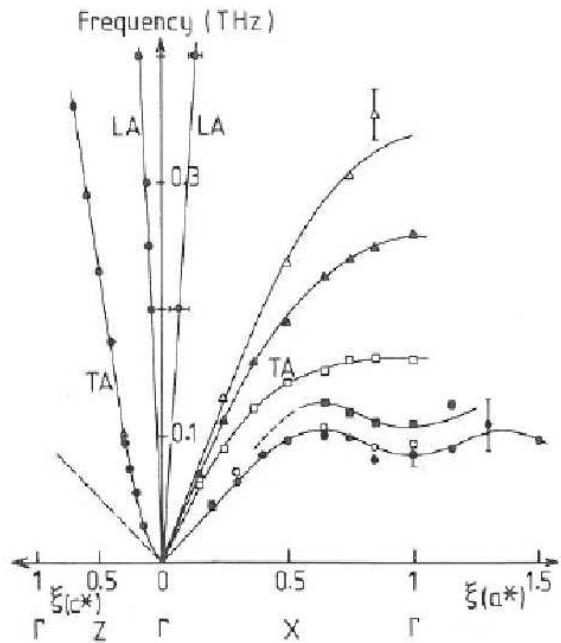}
\caption{Longitudinal (LA) and transverse (TA) acoustic dispersions in (NbSe$_4$)$_3$I. Left side: LA and TA modes along [001] direction at room temperature. Right side: LA and TA modes along [100] direction; LA dispersion at 300~K; TA dispersion at different temperatures (in K): {\large$\bullet$}: 300; {\large$\circ$}: 280; {\tiny$\blacksquare$}: 270, {\tiny$\square$}: 260; $\blacktriangle$: 200; $\vartriangle$: 100 (reprinted figure with permission from Physica B 156-157, P. Monceau \textit{et al.}, p. 20, 1989 \cite{Monceau89}. Copyright (1989) with permission from Elsevier).}
\label{fig7-9}
\end{center}
\end{figure}

Lattice dynamical studies of the second order ferrodistortive phase transition at $T_c$~= 274~K in (NbSe$_4$)$_3$I were performed by measuring the acoustic mode dispersions. The right-hand side (RHS) of figure~\ref{fig7-9} shows acoustic dispersions along the [100] direction for several temperatures between 300~K and 100~K. The whole branch between the two zone centres (006) and (106) is strongly temperature dependent. The initial TA slope varies from 230~m/s at 300~K up to 500~m/s at $T$~= 100~K in agreement with acoustic measurements \cite{Saint-Paul88a}. The dip at $\xi$~= 1 was identified as the soft mode of the ferrodistortive transition which hardens from 0.08~THz at 300~K to 0.37~THz at 100~K. The mode damping coefficient is found to be temperature independent ($\sim$ 0.2~THz). In addition, it should be noted that the frequency of the mode does not show any detectable variation between 300~K and the transition temperatures (274 K). The left-hand side (LHS) of figure~\ref{fig7-9} shows the LA and the doubly degenerate TA dispersions along the chain direction at room temperature. The upward curvature of the TA phonon frequency is the consequence of the strong intrachain bending forces (see section~\ref{secbendingforces}).

\subsubsection{NbSe$_3$}\label{sec7-2-3}

The growth of single crystals of NbSe$_3$ in the form of whiskers with very small cross-sections makes this compound essentially inaccessible for studies of the lattice- and CDW-related dispersions by neutron scattering techniques. A single experiment was reported \cite{Monceau87a} on the study of the longitudinal acoustic (LA) branch propagating along the chain direction on a bunch-like assembly of 10$^4$--10$^5$ NbSe$_3$ crystals aligned along the $b$ direction but randomly distributed perpendicularly to this direction.

\medskip
\noindent \textit{3.4.3.a. Phonon dispersion}
\medskip

\begin{figure}
\begin{center}
\includegraphics[width=6.5cm]{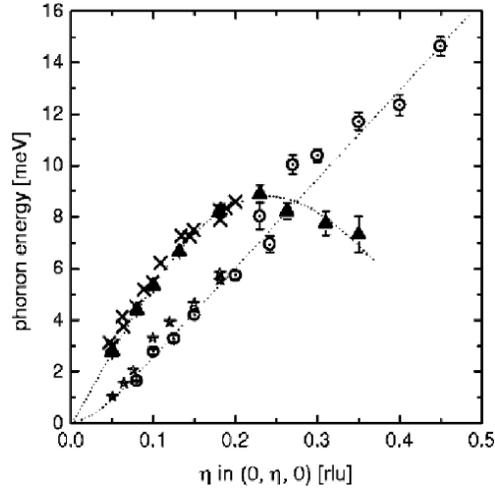}
\caption{Longitudinal (LA) and transverse (TA) phonon branches of NbSe$_3$ propagating along $b^\ast$ in the metallic phase ($T>T_{\rm P_1}$). Inelastic X-ray scattering data, $T$~= 295~K. $\odot$: TA$_1$; $\blacktriangle$: LA. The dotted lines are guides to the eye (from ref.~\cite{Requardt02a}). For comparison, triple-axis-neutron scattering data: {\large$\times$}: LA; {\large$\star$}: TA; $T$~= 200~K (reprinted figure with permission from Synthetic Metals 19, P. Monceau \textit{et al.}, p. 819, 1987 \cite{Monceau87a}. Copyright (1987) from Elsevier).}
\label{fig7-10}
\end{center}
\end{figure}

Ambient and low-temperature dynamics studies were possible on a single crystal of NbSe$_3$ using meV-resolution inelastic X-ray scattering (IXS) phonon dispersion \cite{Requardt02a}. Figure~\ref{fig7-10} 
shows the phonon dispersion curves obtained in the metallic state above the $T_{\rm P1}$ CDW transition. The solid triangles indicate the LA-dispersion data at $T$~= 295~K. The LA branch rises up to a maximum of about 9.3~meV at $\eta$~= 0.23 and continues in a smooth decrease with increasing $\eta$. It was not possible to follow the LA branch to the (0, 2, 0) Brillouin zone boundary since the phonon intensity decreases quickly and vanishes beyond $\eta\approx 0.35$. The crosses show the LA data from neutron measurements \cite{Monceau87a} at $T$~= 200~K following closely the IXS data. The longitudinal sound velocity calculated from the low-$\eta$ LA data yields 4.5~km/s for the IXS data ($T$~= 295~K) and 5.4~km/s for the neutron data ($T$~= 200~K), values consistent with that (4.8~km/s) derived from measurements of the room temperature Young's modulus \cite{Brill78}.

The transverse acoustic branch (TA)$_1$ was monitored with propagation direction along $b^\ast$ and the polarisation predominantly along $a^\ast$. The IXS data of the TA phonon are presented in figure~\ref{fig7-10} (open circles). The branch rises essentially linearly with $\eta$ with an upward curvature at low $\eta$ typical of a mode with chain-bending character. TA data from neutron scattering, observed via double-scattering events (Bragg scattering plus phonon) are shown as stars \cite{Monceau87a}.

\medskip
\noindent \textit{3.4.3.b. Search for Kohn anomaly}
\medskip

Among all the acoustic modes which can be possibly measured in NbSe$_3$, phonon softening was expected to occur in the TA$_1$ mode. In fact, from opposite extinction rules between the lattice- and CDW-satellite reflections, one can deduce that the atomic displacements due to CDW condensation have odd symmetry with respect to the system's $2_1$-screw axis. Then the LA mode propagating along the chain ($b^\ast$) is in the even representation and, thus, will not be concerned with the CDW. The TA modes along $b^\ast$ are in the odd representation and should therefore allow observation of the Kohn anomaly. In the case if the phonon softening does not occur directly in the TA($b^\ast$) modes, but in an odd-symmetry optic branch, it will necessary give rise to softening of the odd-symmetry modes at lower energies, including the TA branches, by the usual anticrossing mechanisms.
\begin{figure}
\begin{center}
\includegraphics[width=6.5cm]{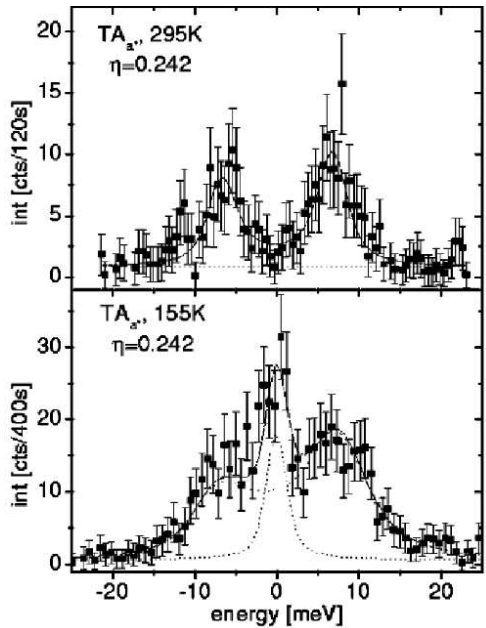}
\caption{Inelastic X-ray scattering energy scans of NbSe$_3$ taken at the upper-CDW satellite position ($\eta$~= 0.242); a)~at  $T$~= 295~K, b)~at $T$~= 155~K~= $T_{P_1}$+10~K. Dotted and dashed lines indicate the elastic and the phonon contribution, respectively (reprinted figure with permission from H. Requardt \textit{et al.}, Physical Review 66, p. 214303, 2002 \cite{Requardt02a}. Copyright (2002) by the American Physical Society).}
\label{fig7-11}
\end{center}
\end{figure}

Figure~\ref{fig7-11}(a) shows an energy scan of the TA$_1$ branch at room temperature at $\eta$~=  0.242 i.e. at the position of the $Q_1$ CDW satellite. At $T$~= 295~K, no elastic contribution, only phonon excitation, is observable. This profile has to be compared to figure~\ref{fig7-11}(b), showing an energy spectrum at the same $Q_1$-satellite position at $T$~= 155~K, i.e. ($T_{P_1}$+10~K). At 155~K, one observes an additional elastic contribution associated with the critical CDW fluctuations. The fit reveals one inelastic contribution which shows no decrease in phonon energy, but only a strong phonon line broadening.

\begin{figure}
\begin{center}
\includegraphics[width=7.5cm]{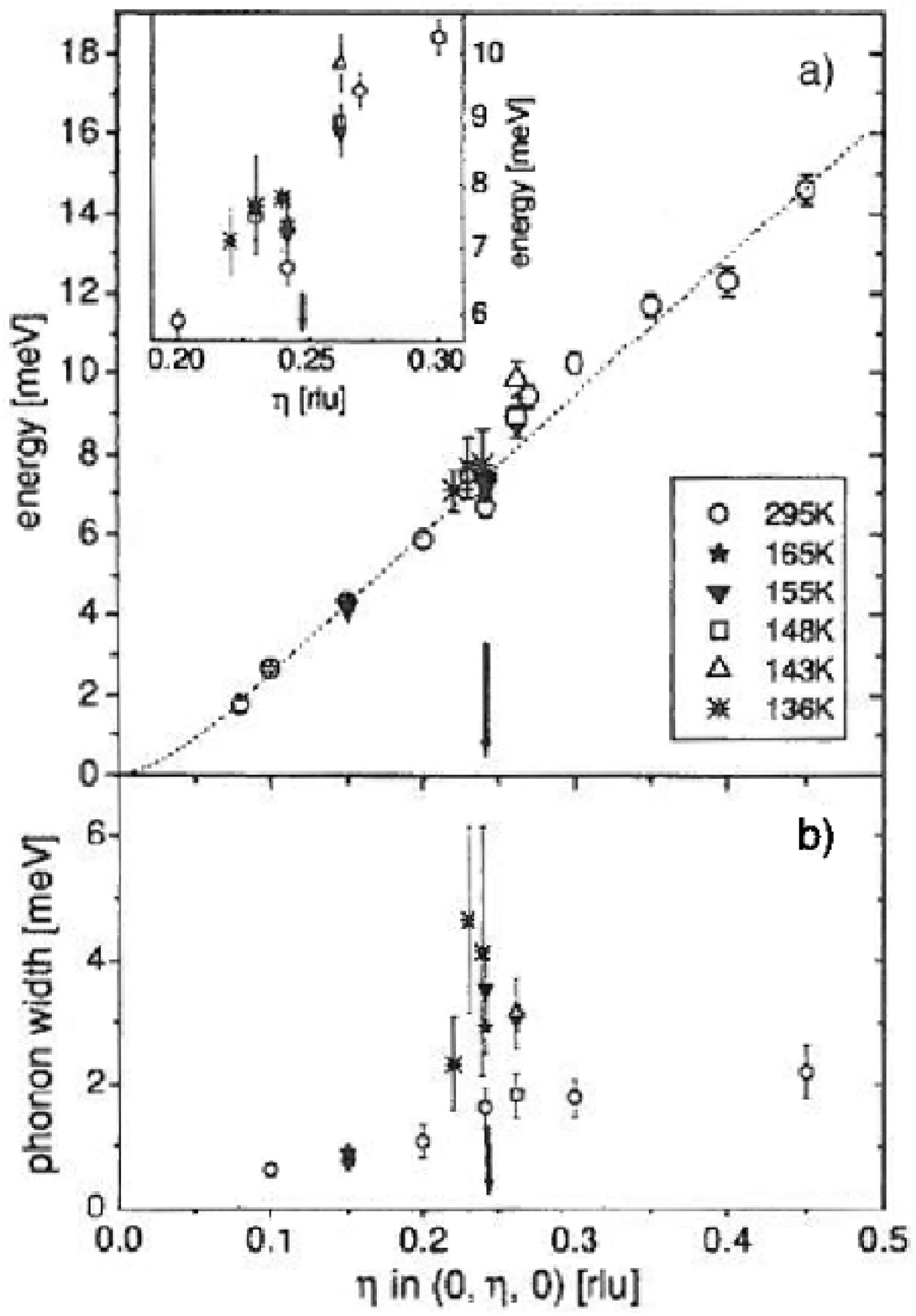}
\caption{a)~Transverse acoustic (TA) branch from $T$~= 295~K down to $T=136$~K (= ~$T_{P_1}$-9~K). The inset gives an enlarged view of the dispersion around the position of the CDW satellite. Note that the dispersion presents no sign of softening (Kohn anomaly) around the satellite position. b)~Linewidth of the TA phonon from $T$~= 295~K down to $T$~= 136~K. The black arrows indicate the position $\eta$ of the $Q_1$ CDW satellite (reprinted figure with permission from H. Requardt \textit{et al.}, Physical Review 66, p. 214303, 2002 \cite{Requardt02a}. Copyright (2002) by the American Physical Society).}
\label{fig7-12}
\end{center}
\end{figure}

Figure \ref{fig7-12}(a) 
shows the resulting TA$_1$-dispersion at room temperature and at low temperature down to $T$~= 136~K~= $T_{P_1}$-9~K. Indeed, no Kohn anomaly upon cooling towards the upper CDW transition is observable. As shown in figure~\ref{fig7-12}(b), the energy spectra reveal only a significant line broadening around the CDW-satellite position at $\eta$~= 0.241 (indicated by the black arrows. This line broadening was also observed in K$_{0.3}$MoO$_3$ but the absence of a Kohn anomaly shows that the dynamical behaviour of NbSe$_3$ differs markedly from that of K$_{0.3}$MoO$_3$ \cite{Pouget91} but resembles the behaviour observed in (TaSe)$_2$I (see figures~\ref{fig7-7} and \ref{fig7-8}). These differences in lattice-dynamical behaviour may result from possible differences in the strength of the electron-phonon interaction and a model of strong coupling may be appropriate for NbSe$_3$ (see section~\ref{sec7-5}).

\subsubsection{Phonon Poiseuille flow}\label{sec3-4-4}

\begin{figure}
\begin{center}
\includegraphics[width=7.5cm]{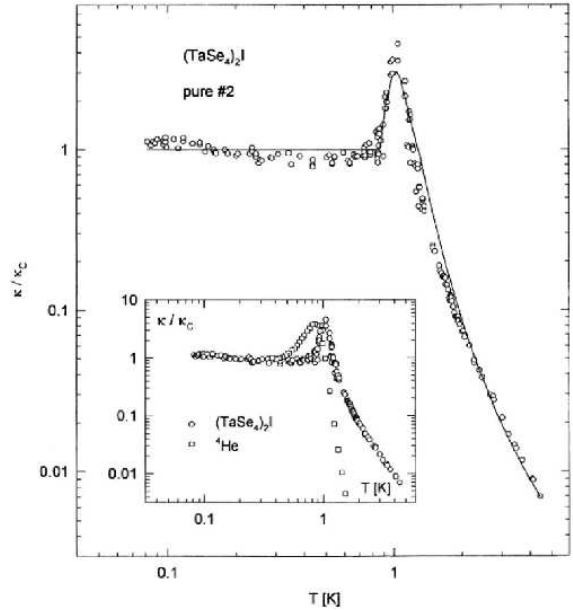}
\caption{Thermal conductivity $\kappa$ normalised to the Casimir-boundary scattering regime $\kappa_C=\frac{1}{3}c_v\Lambda\alpha T^3$ ($c_v$: specific heat, $\Lambda$: mean width of the sample) of (TaSe$_4$)$_2$I as a function of temperature in a log-log plot. $\kappa_c$~= $20T^3$ (Wm$^{-1}$K$^{-1}$). The continuous line is the theoretical calculation of $\kappa$ taking into account the phonon Poiseuille flow [Eq.~(11) in ref.~\cite{Smontara98}]. Inset shows data of the hexagonal $^4$He \cite{Hogan69} for comparison, with $\kappa/T^3$ normalised to the same Casimir value as for (TaSe$_4$)$_2$I (reprinted figure with permission from Journal of Low Temperature Physics 111, A. Smontara \textit{et al.}, p. 815, 1998 \cite{Smontara98}. Copyright (1998) from Springer Science and Business media).}
\label{fig7-9add}
\end{center}
\end{figure}

Thermal conductivity of (TaSe$_4$)$_2$I measured in the range 60 mK--6 K exhibits a strong deviation from the cubic variation expected at low $T$, as shown in figure~\ref{fig7-9add} representing the variation of $\kappa/T^3$ versus $T$ \cite{Smontara96,Smontara98}. The cubic regime is almost well obeyed below 0.8 K with $\kappa$~= 18--20/$T^3$~Wm$^{-1}$K$^{-1}$ in good agreement with the Casimir-boundary scattering regime estimated from  the mean width of the sample ($\Lambda$~= 0.7~mm) and the average sound velocities.

The main striking feature is the presence of a very sharp peak at $T$~= 1.1~K with $\kappa\simeq 100/T^3$~Wm$^{-1}$K$^{-1}$ which was explained by a phonon Poiseuille flow, originating from the strong anharmonicity amplified by the lattice anisotropy \cite{Smontara98}.

Indeed thermal resistivity is determined by phonon-phonon interactions with two types of scattering processes: the normal-process (N-process) in which the quasi-momentum of the phonon is conserved, and the Umklapp process (U-process) in which it is not. Since U-processes vanish exponentially at low $T$, a temperature region where N-processes dominate should theoretically exist \cite{Smontara98}. Consequently the thermal conductivity of phonon gas could be greater than predicted by the low-$T$ limit of Casimir boundary scattering: this is the temperature window for the observation of Poiseuille regime of phonon flow \cite{Berman76}. In fact the conditions for observation of this regime are so  stringent that it could be observed only in $^4$He up to now \cite{Hogan69,Mezhov66}.

The value of $\kappa$ at low $T$ in (TaSe$_4$)$_2$I agrees with the Casimir-boundary scattering regime (Casimir regime: $\kappa_{\rm C}$):
\begin{equation*}
\kappa_{\rm C}=\frac{1}{3}\,c_v\,\overline{v^2}\,\tau_c=\frac{1}{3}\,c_v\,\overline{v}\,\Lambda\,T^3
\end{equation*}
$\Lambda$: a mean width $\Lambda$, $\tau_c$: corresponding relaxation time.\\
The Poiseuille regime appears to be realised in (TaSe$_4$)$_2$I due to the very particular phonon properties in relation with the high anisotropy of the structure: low frequency flat TA branch for phonons propagating in the basal plane $[a^\ast,b^\ast]$ and polarised along $c^\ast$ with characteristic frequency $\nu_0\sim 0.15-0.20$~THz (see figure~\ref{fig7-6}(a)). That has a consequence on the very large Gr\"uneisen parameter $\gamma$ with increases as $\gamma\propto T^{-2}$ for these $C_{44}$ shear modes \cite{Saint-Paul96}; $\gamma$ is estimated to be of the order of $10^2$ at 1~K which means a large anharmonicity for phonon interaction N-processes. In (TaSe$_4$)$_2$I, the importance of the N-processes through the anharmonic interactions in this $T$-range is considerable, much more than for conventional solids. The second point for observing Poiseuille flow is the large anisotropy of phonon dispersion relations for the same $q$: the phonon frequency increases very rapidly as soon as there is a component along the $c^\ast$ axis enhancing N-processes.

Near 1~K the lifetime of the dominant thermal phonons which carry the heat flow along $c^\ast$ axis is essentially determined by anharmonic interaction with the modes of the flat TA branch with a characteristic frequency $\Delta_0/k_{\rm B}\equiv 10$~K.

3 relaxation times should be considered:\\
- $\tau_C$ related to the Casimir regime,\\
- $\tau_N$ related to N-processes: $\tau_N\sim Ae^{\Delta_0/T}$, $A$: strength of anharmonicity related to $\gamma^2$, the square of the Gr\"uneisen parameter.\\
- $\tau_r$ related to usual Umklapp or resistive processes: $\tau_r=\tau_0(e^{\Delta/T}-1)$.

Conditions for Poiseuille flow are such:
\begin{eqnarray*}\begin{array}{ll}
&\tau_N<\tau_C\\
\mbox{but with }&\tau_C<(\tau_N\tau_r)^{1/2}
\end{array}\end{eqnarray*}
The peak in $\kappa(T)$ results from the competition between N processes ($\tau_N$) and resistive one ($\tau_r$).

Sharp peaks in $\kappa/T^3$ were also observed \cite{Smontara98} in doped (TaSe$_4$)$_2$I: (Ta$_{1-x}$Nb$_x$Se$_4$)$_2$I with $x$~= 0.8\%, 1\%. For $x$~= 1\%, below 0.1~K down to 55~mK, the boundary regime, $\kappa_C$~= $10T^3$, is reached; but between 0.1~K and 1~K, a regime with a $T^{1.8}$ variation appears. That was attributed to ``linear or planar" defects oriented parallel to $c$-axis which do not destroy the Poiseuille flow known to be very sensitive to the presence of point defects. These defects parallel to $c$-axis play the role of a diffuse boundary scattering mechanism and they can be simulated \cite{Smontara98} by a $T$-dependent transverse geometry factor: $R(T)=BT^{-n}$ with $n$~= 1.2 which for $T<0.1$~K, $R(T\rightarrow 0)=\Lambda$.

\subsection{A$_{0.3}$MoO$_3$ A: K, Rb, Tl}\label{sec3-3}

\begin{figure}
\begin{center}
\includegraphics[width=8.5cm]{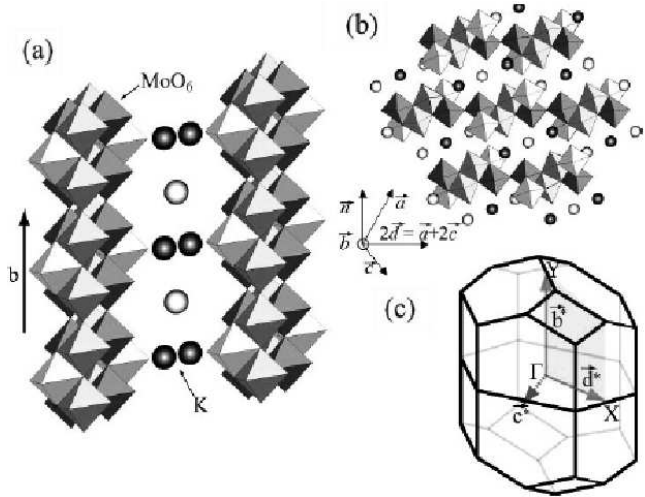}
\caption{Crystal structure of molybdenum blue bronze K$_{0.3}$MoO$_3$ showing the infinite chains of MoO$_6$ octahedra (a) and the cross-section of the infinite chains in the $a-c$ plane (b). Note that the chains are connected along the ($a+2c$) direction forming a layer parallel to the $b$, $a+2c$ plane with alkaline K between these layers. c) The brillouin zone of K$_{0.3}$MoO$_3$ for the simple monoclinic structure (reprinted figure with permission from H. Ando \textit{et al.}, Journal of Physics: Condensed Matter 17, p. 4937, 2005 \cite{Ando05}. Copyright (2005) by the Institute of Physics).}
\label{fig3-18}
\end{center}
\end{figure}

Among the large class of oxide bronzes of general formula A$_x$MO$_n$, mobybdenum bronzes have been intensively studied, specifically for the low dimensional conductivity properties associated with their structural anisotropy. Hereafter only the properties of transition metal oxides A$_{0.3}$MoO$_3$ (A: K, Rb), so-called blue bronzes, will be shortly described (for complete reviews, see ref.~\cite{R9Schlenker89,R10Schlenker96}) laying stress specifically on recent developments. They crystallise in the centred monoclinic $C2/m$ structure with lattice parameters for K$_{0.3}$MoO$_3$ $a$~= 18.25~$\AA$, $b$~= 7.560~$\AA$, $c$~= 9.855~$\AA$ and $\gamma$~= 117,53$^\circ$ (respectively, for Rb$_{0.3}$MoO$_3$, $a$~= 18.94~$\AA$, $b$~= 7.560~$\AA$, $c$~= 10.040~$\AA$ and $\gamma$~= 118.83$^\circ$) \cite{Graham66,Ghedira85,Ando05}. The crystal is built with MoO$_6$ octahedra bilayers, parallel to the ($\bar{2}01$) cleavage plane with alkali atoms located between the bilayers. A bilayer consists of clusters of ten distorted octahedra Mo$_{10}$O$_{30}$. A cluster contains three different types of molybdenum sites, namely Mo(1) related to type I MoO$_6$ octadedra (in white in figure~\ref{fig3-18}), 
Mo(2) related to type II MoO$_6$ (green in figure~\ref{fig3-18}) and Mo(3) related to type III MoO$_6$ (blue in figure~\ref{fig3-18}). MoO$_6$ octahedra of type II and III form chains along the $b$ direction (the direction of highest conductivity) through oxygen corner sharing. MoO$_6$ octahedra bilayers result from edge sharing clusters of neighbouring chains. However, the type I MoO$_6$ octahedra do not form continuous chains along the $b$ axis. Alkali cations are located between two bilayers at two crystallographically different sites, site 1 (white circles in figure~\ref{fig3-18}) at mid-distance between bilayer and site 2 (black circle in figure~\ref{fig3-18}) nearest a given bilayer.

The electrical resistivity at room temperature exhibits a large anisotropy with $\rho_b/\rho_{a+2c}\sim 100$ and $\rho_b/\rho_\perp(\bar{2}01)\sim 10^3$. It is also sample dependent, likely originating from localised states due to stoichiometry defects or low energy CDW excitations (see o-TaS$_3$). Nevertheless the ratio $2\Delta(0)/kT_{\rm P}$ is estimated to be $\sim 9$, again largely beyond the BCS mean field value.

Diffuse X-ray scattering measurements have revealed diffuse lines parallel to $b^\ast$ at $T>T_{\rm P}$ which condense as satellite reflections below $T_{\rm P}$ with a wave vector defined as:
\begin{equation*}
Q_{\rm CDW}=(0,\,1-q_b,\,1/2).
\end{equation*}
$1-q_b$ is temperature dependent with a variation from $\sim 0.70$ at 300~K to an incommensurate value very near to 0.75 at 100~K, remaining constant below 100~K, without any detectable incommensurate-commensurate transition \cite{Pouget85}.

The temperature dependence of the correlation lengths above $T_{\rm P}$ along three main directions are shown in figure~\ref{fig3-12}(b). Using low-temperature scanning tunnelling microscopy under ultra-high vacuum high-resolution topographical images were obtained on a cleaved ($\bar{2}01$) surface of Rb$_{0.3}$MoO$_3$ single crystal \cite{Brun05}. The three groups of MoO$_6$ octahedra were identified in one unit cell: two type I, four type II and four type III MoO$_6$ octahedra. The CDW modulation was measured at $T$~= 63~K and 78~K with the average projected components of the CDW wave vector on the ($\bar{2}\,0\,1$) surface in good agreement with X-ray and neutron results obtained on bulk samples. Molecular lattice and CDW superlattice were observed simultaneously in topographical constant current images.

However surprisingly, an unexpected  CDW-like modulation was detected by STM on type I MoO$_6$ octahedra which are only weakly involved in the CDW transition according to the superlattice structural study \cite{Schutte93}. Significant experimental inhomogeneities of the $b^\ast$ surface component on different locations separated by several $\umu$m were reported \cite{Brun05,Machado06} with $1-q_b$ ranged from 0.21 to 0.32 with respect to 0.25 for the bulk value. These observations have prompted to study STM images from first principles density functional theory calculations \cite{Machado06}. It was shown that the STM measurement probes the contribution of the uppermost O atoms of the surface, which is associated with the MoO$_6$ type I octahedra. Although the CDW modulation mostly affects the type II and type III MoO$_6$ octahedra, as a result of the strong hybridisation between the Mo and O orbitals, the orbital mixing associated with the CDW modulation affects the local density of states of these O atoms, leading to the observed profile of the image. It was noted that the distribution of type 1 alkali atoms at the surface plays a key role in the determination of the periodicity of the CDW modulation and is mainly responsible for the experimental deviations of $Q_{\rm CDW}$ at the surface. From calculations it was concluded \cite{Machado06} that different concentrations of alkali atoms at the surface generate a nearly rigid shift of the surface bands with respect to those of the bulk. Finally this inhomogeneity of the $b^\ast$ surface component detected by STM and not by other surface technique such as X-ray grazing incidence or angular resolved photoemission spectroscopy (ARPES) was understood \cite{Machado06} by noting  that STM probes $Q_{\rm CDW}$ only at the uppermost layer of the compound and very locally at the nanometer scale inside a single CDW domain, while X-ray and ARPES experiments probe the CDW wave vector $Q_{\rm CDW}$ over a macroscopical in-plane scale; that leads to an average value of $Q_{\rm CDW}$ in a macroscopical volume showing no surface inhomogeneities.

\begin{figure}
\begin{center}
\subfigure[]{\label{fig3-19a}
\includegraphics[width=5.5cm]{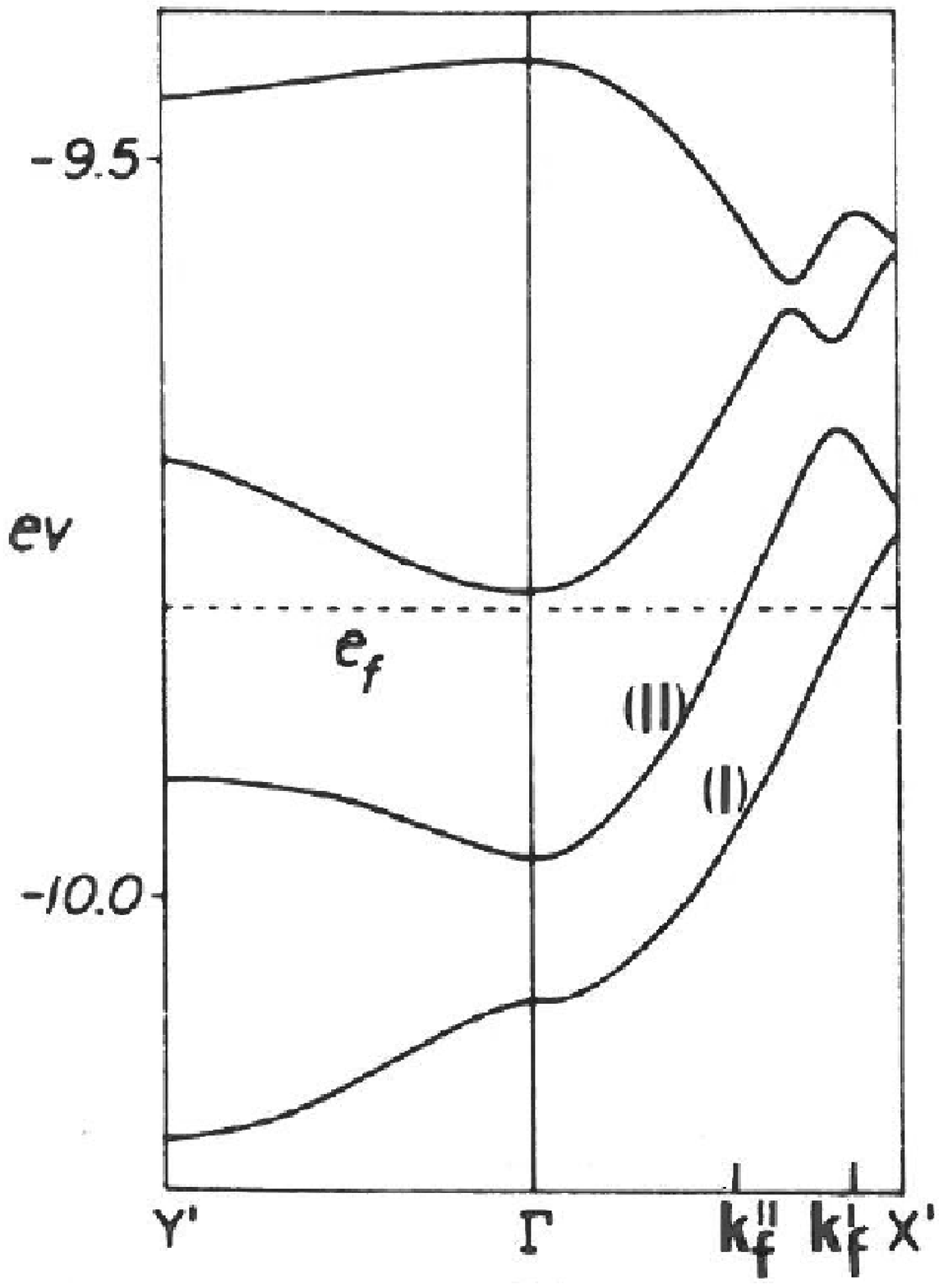}}
\subfigure[]{\label{fig3-19b}
\includegraphics[width=6.5cm]{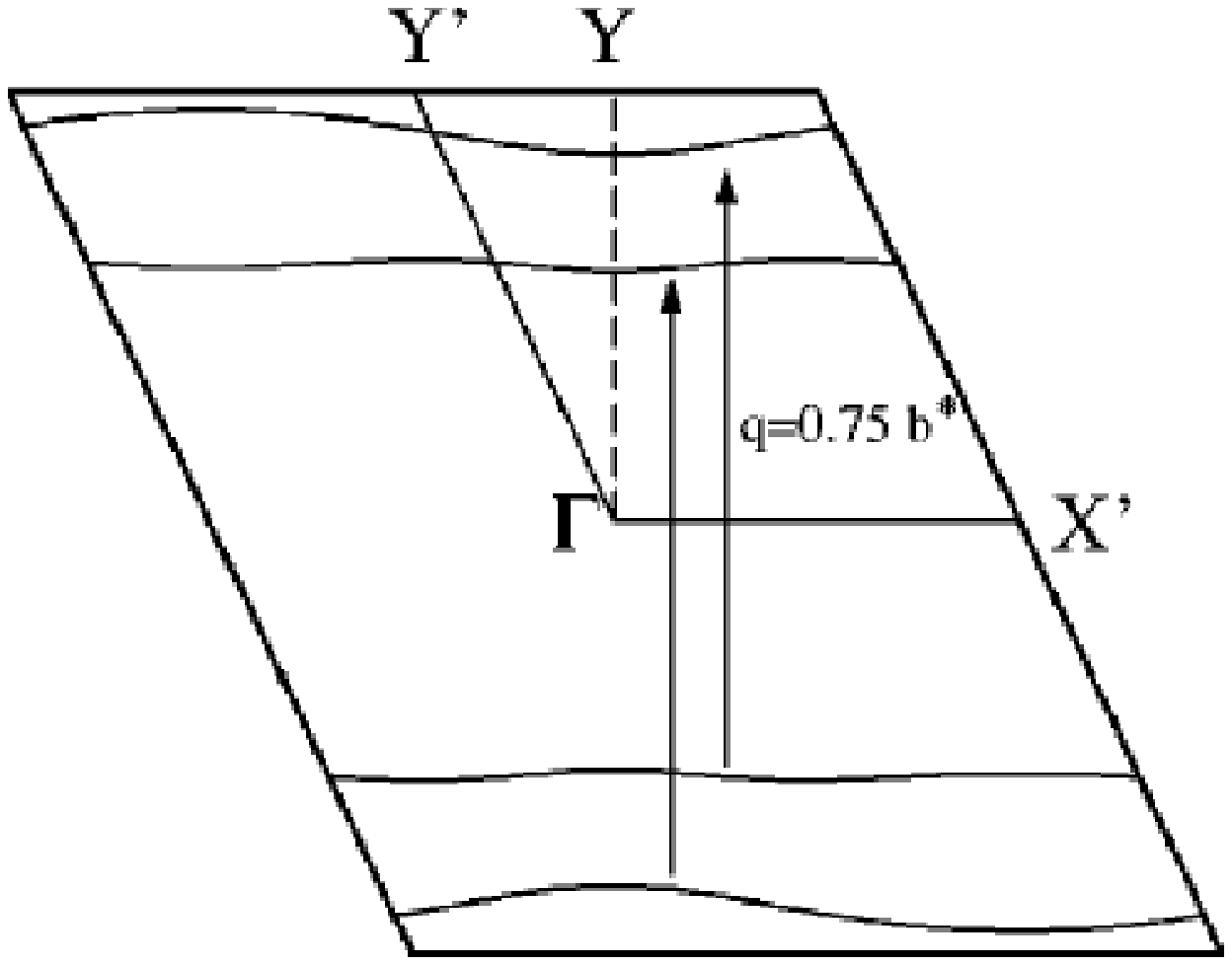}}
\caption{a) Band structure of K$_{0.3}$MoO$_3$ (reprinted figure with permission from M.-H. Whangbo and L.F. Schneemeyer, Inorganic Chemistry 25, p. 2424, 1986 \cite{Whangbo86}. Copyright (1986) by the American Chemistry Society), b)~calculated Fermi surface of K$_{0.3}$MoO$_3$ in the plane ($a'^\ast$, $b'^\ast$). $a'$~= $\frac{1}{2}(-a+b+c)$, $b'=b$. $\Gamma$~= (0, 0, 0), $X'$~= ($a'^\ast/2, 0, 0$), $Y'$~= ($0, b'^\ast /2, 0$) and $Y$~= ($0, b^\ast/2, 0$). The CDW nesting wave vector is indicated (reprinted figure with permission from J.-L. Mozos \textit{et al.}, Physical Review B 65, p. 233105, 2002 \cite{Mozos02}. Copyright (2002) by the American Physical Society).}
\label{fig3-19}
\end{center}
\end{figure}

Electronic band calculations \cite{Whangbo86} showed that two 1D bands cross the Fermi level respectively at $k_{\rm F_1}$~= 0.33$b^\ast$ and $k_{\rm F_2}$~= 0.42$b^\ast$, as shown in figure~\ref{fig3-19}(a). It was proposed that $Q_{\rm CDW}$ could be related to the nesting of one band with another by $q_{\rm CDW}$~= $k_{\rm F_1}+k_{\rm F_2}$~= 0.75$b^\ast$.

The calculated Fermi surface in the $a^\ast$, $b^{'\ast}$ plane of the Brillouin zone is shown in figure~\ref{fig3-19}(b) \cite{Mozos02} ($a'=\frac{1}{2}(-a+b+c)$; $b'=b$, $c'=c$). Angle-resolved photoemission spectroscopy (ARPES) is a technique than can directly investigate the topology of the Fermi surface. Recent considerable improvement in the momentum resolution of ARPES has brought it to the same level as that obtained in X-ray and neutron scattering experiments \cite{Hufner95}.
\begin{figure}
\begin{center}
\includegraphics[width=7.5cm]{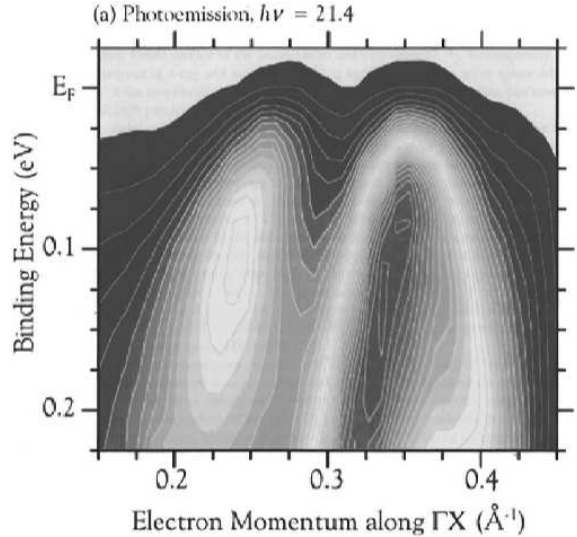}
\caption{A photoemission intensity map of K$_{0.3}$MoO$_3$ ($h\nu$~= 21.4) at room temperature; intensity increases from blue to red (reprinted figure with permission from A.V. Federov \textit{et al.}, Journal of Physics: Condensed Matter 12, p. L191, 2000 \cite{Federov00}. Copyright (2000) by the Institute of Physics).}
\label{fig3-20}
\end{center}
\end{figure}

Figure~\ref{fig3-20} displays a typical ARPES map of the electronic structure of K$_{0.3}$MoO$_3$ in the region close to the Fermi level crossings \cite{Federov00}; it represents a full, essentially continuous, three dimensional map of the photocurrent in this region of energy and momentum space. The experimental data reveal the existence of two bands that are parallel in the vicinity of $E_{\rm F}$. Measurements of the intensity of photoelectrons versus momentum at $E_{\rm F}$ from the ARPES map in figure~\ref{fig3-20} reveal peaks of highest intensity that are identified to the wave vectors $k_{\rm F_1}$ and $k_{\rm F_2}$. When $T$ is reduced, data showed \cite{Federov00} that the splitting between these peaks is reduced, implying an increase of $k_{\rm F_1}$+$k_{\rm F_2}$ in good agreement with X-ray and neutrons determination of $Q_{\rm CDW}$. More precisely $k_{\rm F_1}$ exhibits a strong $T$-dependence while no apparent shift with temperature of $k_{\rm F_2}$ is observed. Thus these data indicate a change in the electronic structure of K$_{0.3}$MoO$_3$ with a change in the coupling between chains belonging to different unit cells. A possibility suggested in ref.~\cite{Federov00} was a continuous change with temperature in the Mo-O distances in the basal plane. That excludes previous explanations of the $T$-dependence of the $Q$-vector attributed to charge transfer between the two bands crossing $E_{\rm F}$ and a third band \cite{Pouget85}, or a shift in the chemical potential.

ARPES studies in the whole Brillouin zone have led to a temperature dependent warping of the Fermi surface along the $(a+2c)^\ast$ direction \cite{Ando05}. Fermi surface curvatures become smaller at 70~K and nearly coincide indicating that the nesting condition of the Fermi surfaces varies from imperfect (200~K) to nearly perfect (70~K) when $T$ is lowered through $T_{\rm P}$.

\subsection{Organic CDWs}\label{sec3-4}

In the beginning of 70's, chemists succeeded to synthesise the first molecular crystal with a conductivity at room temperature approaching that of conventional metals, namely TTF-TCNQ (tetrathiafulvalene-tetracyanoquinodimethane). TTF-TCNQ consists of planar TTF and TCNQ molecules (see figure~\ref{fig3-26}) stacked to form segregated chains parallel to the crystallographic $b$ direction of the monoclinic structure. Charge transfer. 0.55~el/molecule from TTF to TCNQ at room temperature leads to a large metallic conductivity along the chains and to a large resistive anisotropy. A metal-semiconducting transition at low temperature was one of the first Peierls transition ever demonstrated \cite{Denoyer75,Kagoshima75} (for a review see \cite{Jerome82,Jerome04}).

\subsubsection{TTF-TCNQ}\label{sec3-4-1}

Several phase transitions occur in TTF-TCNQ. One at $T_H$~= 54~K with a CDW restricted to the TCNQ chains with a superstructure period $2a\times 3.3b\times c$. This modulation wave vector does not vary down to $T_M$, temperature at which a second CDW is established on TTF chains. Below $T_M$, the transverse modulation becomes incommensurate and temperature dependent, jumping to the commensurate $4a$ at $T_L$. The nature of the CDW in the temperature range between $T_M$ and $T_L$ where there is a coupling between both CDWs on TTF and TCNQ chains has been elucidated from STM under ultra high vacuum conditions  \cite{Wang03}. The measurement of the modulation wave vector along the $a$ direction has shown the existence of domains with a single plane wave modulated structure.

Non linear transport properties were measured \cite{Lacoe85} below $T_H$ with a threshold field around $E_T$~= 0.25~V/cm. A fast increase of $E_T$ is observed below $T_M$, due to the joint CDWs on TTF and TCNQ chains \cite{Lacoe87}.

Band mapping of the 1D bands corresponding to TTF and TCNQ chains has been performed from ARPES \cite{ClaessenPRL02,Claessen02,Zwick98}. Suppression of the spectral weight near $E_{\rm F}$ was observed, as for many other Q-1D materials (see section~\ref{sec7-4-3}).

\subsubsection{(Per)$_2$M(mnt)$_2$ salts}\label{sec3-4-2}

Based on perylene, many charge-transfer solids have been synthesised \cite{Almeida97}. Among them the $\alpha$-phases of the (Per)$_2$M(mnt)$_2$ compounds where Per~= perylene, mnt~= maleonitriledithiolate [S$_2$C$_2$(CN)$_2$] are two chain charge transfer salts. They consist in one-dimensional conducting chains of perylene molecules in the (Per)$_2^+$ oxidation state and  insulating chains of molecules in the M(mnt)$^-_2$ oxidation state with two formulas per unit cell giving rise to a $2k_{\rm F}$~= 3/4 filled band. The M(mnt)$_2$ chains are either diamagnetic for M~= Au, Cu or have some localised magnetic moments for M~= Ni, Pt, Pd, Fe.

\begin{figure}[h!]
\begin{center}
\includegraphics[width=8cm]{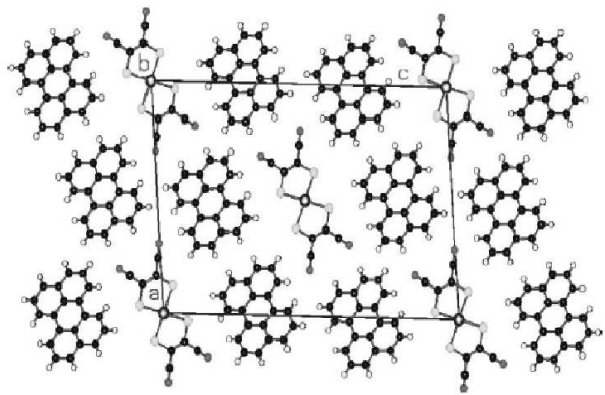}
\caption{Crystallographic projection of (Per)$_2$M(mnt)$_2$ along the perylene $b$-axis stacking direction (reprinted figure with permission from Solid State Communications 35, L. Alcacer \textit{et al.}, p. 945, 1980 \cite{Alcacer80}. Copyright (1980) with permission from Elsevier).}
\label{fig3-21}
\end{center}
\end{figure}

The crystallographic projection of (Per)$_2$M(mnt)$_2$ along the $b$-axis stacking direction is shown in figure~\ref{fig3-21}. [carbon atoms (black), M (blue), S (yellow), N (green)]  \cite{Alcacer80,Graf04a,Graf04b}. The conducting chains of perylene alternate in the ($a,c$) plane with chains of anions in such a way that each anion chain is surrounded by 6 stacks of perylene and each perylene stack has 3 stacks of perylene and 3 stacks of anions as nearest neighbours \cite{Alcacer80}.

\begin{figure}
\begin{center}
\includegraphics[width=6.25cm]{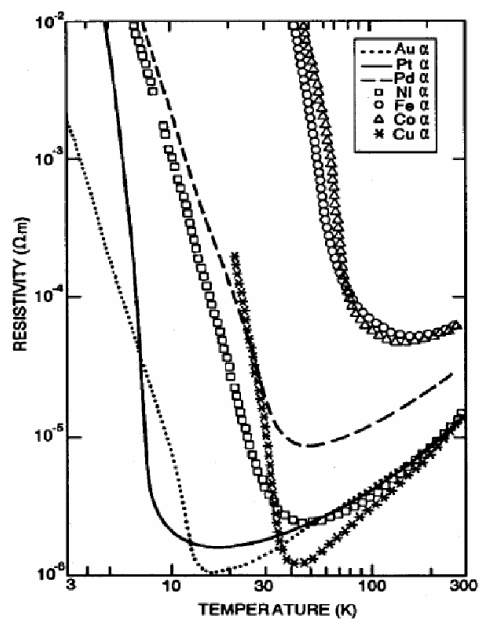}
\caption{Temperature dependence of the electrical resistivity of $\alpha$-(Per)$_2$M(mnt)$_2$ (reprinted figure with permission from Molecular Solids and Liquid Crystals 234, p. 171 (1993) \cite{Gama93b}. Copyright (1993) by Taylor and Francis).}
\label{fig3-22}
\end{center}
\end{figure}

The structure is monoclinic, with space group $P_{2_1}/c$. The lattice parameters of (Per)$_2$Pt(mnt)$_2$ were found \cite{Alcacer80} to be $a$~= 16.612~$\AA$, $b$~= 4.194~$\AA$, $c$~= 30.211~$\AA$ and $\beta$~= 118.70$^\circ$.
The temperature dependence of the electrical resistivity of several (Per)$_2$M(mnt)$_2$ compounds is shown in figure~\ref{fig3-22}.
The typical room temperature conductivity is of the order of 700~S/cm along the $b$-axis with the anisotropy in the ($a,b$) plane estimated to be $\sim 10^3$. All of them have a similar metallic behaviour with a metal-insulating transition at low temperature. This transition in the case of (Per)$_2$Au(Mnt)$_2$ was ascribed \cite{Lopes94} to a CDW from non-linear transport properties (see sections~\ref{sec4-1} and \ref{sec4-3}) similar to NbSe$_3$. The temperature of the Peierls transition ranges \cite{Gama93b} from 8~K for Pt to 73~K for Fe, with  12~K (Au), 25~K (Ni), 28~K (Pd), 32~K (Cu) and 53~K (Co).

For compounds with magnetic chains (M~= Ni, Pd, Pt), a spin-Peierls occurs simultaneously with the CDW transition associated with the dimerisation of the M(mnt)$_2$ chains. Although previous X-ray diffuse scattering measurements failed to detect the $2k_{\rm F}$ distortion in perylene chains on Pt and Au salts, for compounds with higher CDWs (Cu, Ni, Co, Fe) the $2k_{\rm F}$ distortion was clearly observed \cite{Gama93b}.

\begin{figure}[h!]
\begin{center}
\includegraphics[width=8cm]{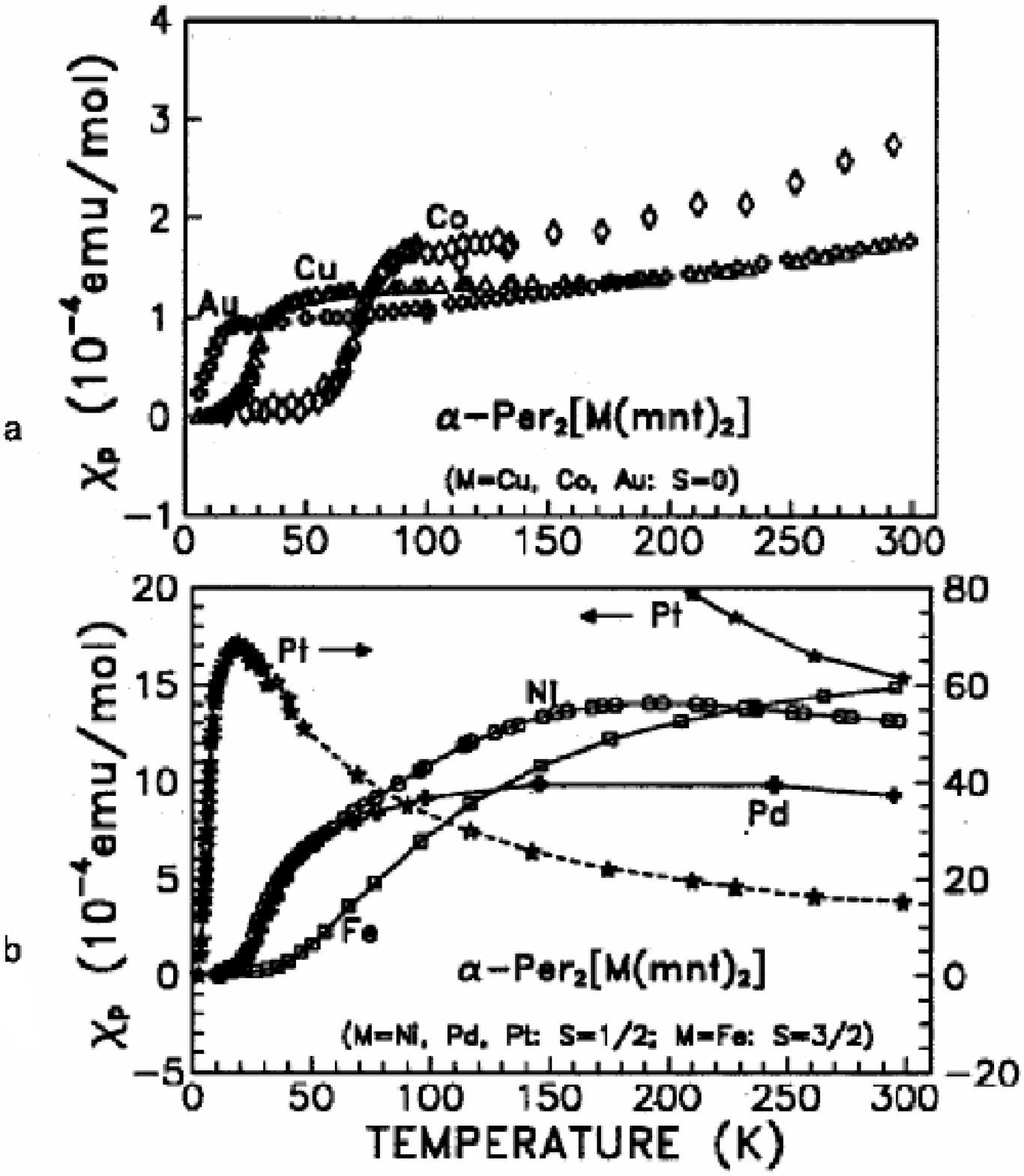}
\caption{Temperature dependence of the paramagnetic susceptibility of $\alpha$-(Per)$_2$M(mnt)$_2$ salts. a)~M~= Cu, Au, Co (diamagnetic anionic complexes). b)~M~= Ni, Pd, Pt ($S$~= 1/2) and Fe ($S$~= 3/2) (paramagnetic anionic complexes) (reprinted figure with permission from Molecular Solids and Liquid Crystals 234, p. 171, 1993 \cite{Gama93b}. Copyright (1993) by Taylor and Francis).}
\label{fig3-23}
\end{center}
\end{figure}

In addition to strong $b^\ast/2(4k_{\rm F})$ Bragg reflection in X-ray measurements \cite{Gama93a}, the spin-Peierls behaviour is clearly observed in susceptibility. While in the case of M~= Au, Co, Cu where the M(mnt)$^-_2$ unit is closed shell, the magnetic susceptibility is small and weakly temperature-dependent, in the case of M(mnt)$^-_2$ paramagnetic species (M~= Ni, Pd, Pt, Fe) the susceptibility is significantly larger due to the extra contribution of the chain of localised spins. This contribution vanishes \cite{Gama93b} at the same temperature than $T_{\rm MI}$ (see figure~\ref{fig3-23}) 
with the exception of Fe where Fe(mnt)$^-_2$ units are already chemically dimerised at room  temperature giving rise to a contribution typical of AF coupled pairs of spins 3/2.
So, the spin-Peierls transition inducing a dimerisation on the M(mnt)$_2$ chains and the Peierls transition inducing a tetramerisation on the perylene chains are strongly coupled and both occur simultaneously at the same temperature. It was suggested \cite{Henriques93,Gama93b} that a possible mechanism to trigger the dimerisation instability by a tetramerisation on perylene chains could be based on exchange interactions between the localised spins and conduction electrons with the coupling between them of RKKY type \cite{Bourbonnais91}. Magnetic and lattice fluctuations were also studied \cite{Bourbonnais91} by nuclear and electronic resonance.

Since there are four perylene stacks per unit cell there must have four bands crossing the Fermi level. The degeneracy  of these bands is lifted by taking into account the transverse interactions between perylene units in the unit cell. The calculated \cite{Canadell04} Fermi surface of Per M(mnt)$_2$ shows some warping of the sheets along the $a^\ast$ (estimated to be of the order of 2~meV) and practically no warping ($\sim~0.2$~mV) along $c^\ast$. It was suggested \cite{Canadell04} that the four Fermi surface sheets may interpenetrate each other, and by hybridisation between them, there could be regions with closed pockets. These effects are of importance essentially at low temperatures. It was found that the CDW state of Au and Pt compounds is suppressed under large magnetic field with, subsequently, a cascade of field induced transitions at higher field (see section~\ref{sec9-5}). These effects depend on the orientation of the magnetic field showing an orbital coupling with the electronic structure which may result from a finite interchain bandwidth.

\subsubsection{(Fluoranthene)$_2$X}\label{sec3-4-3}

Among the arene salts (Ar)$^+_2$X$^-$ where Ar is an aromatic molecule and X a monovalent anion such as PF$^-_6$, AsF$^-_6$, SbF$^-_6$, ... the fluoranthene radical cation salts (FA)$_2$X can be regarded as model systems for quasi-1D conductors \cite{Riess93}.
\begin{figure}
\begin{center}
\includegraphics[width=8cm]{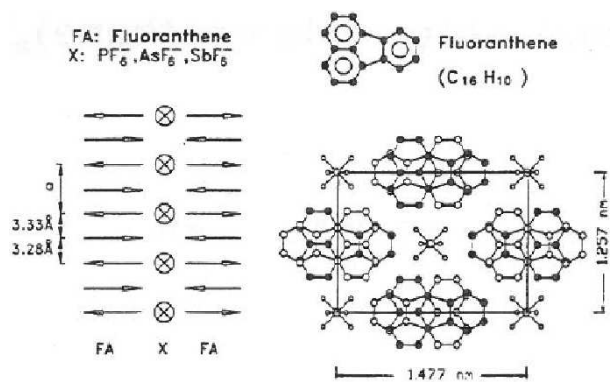}
\caption{Crystal structure of (FA)$_2$X at room temperature: projection on the $a$-$b$-$c^\ast$ and $a-c$ plane ; top: the aromatic hydrocarbon fluoranthene C$_{16}$H$_{10}$ (H-atoms are omitted) (reprinted figure with permission from W. Riess and W. Br\"utting, Physica Scripta T49, p. 721, 1993 \cite{Riess93}. Copyright (1993) by the Institute of Physics).}
\label{fig3-25}
\end{center}
\end{figure}

The basic features of the crystal structure of (FA)$_2$X salts is shown in figure~\ref{fig3-25}. (FA)$_2$X is composed of slightly dimerised stacks of fluoranthene molecules piled in a zigzag manner with strongly overlapping $\pi$-electron wave functions along the $a$-direction and weak overlapping in the transverse directions. The distances between FA molecules along $a$-axis are $d_1$~= 3.28~$\AA$ and $d_2$~= 3.32~$\AA$. Columns of anions X separate molecular FA stacks in the $b$ and $c$ directions. This leads to a Q1D electronic band structure along the chain direction. The room temperature conductivity is of the order of 1000~S/cm and the anisotropy $\sigma_a/\sigma_{b,c}\sim 10^3-10^4$ \cite{Brutting92}.

The stoichiometry 2:1 and the complete charge transfer of the donor to the acceptor stack lead to an average charge of 0.5 hole per FA and thus to $2k_{\rm F}$~= 0.75$a'^\ast$ or 0.5$a^\ast$ depending whether the average molecular periodicity $a'=(d_1+d_2)/2$ or the true crystallographic periodicity $a=2a'$ is considered. Then, either (FA)$_2$X can be described with a half-filled band by considering important the role of dimerisation or with a quarter filled band.

Below room temperature (FA)$_2$PF$_6$ undergoes two successive structural transitions \cite{Ilakovac93}. The upper one at $T_c$~= 194~K corresponds to a $A2/m\rightarrow P_{2_1}/c$ structural change. The lower one is a Peierls transition at $T_{\rm P}\sim 182$~K with the wave vector $Q$~= [1/2,0,0] in the half-filled band description. Below $T_{\rm P}$, collective transport phenomena were observed, ascribed to the sliding CDW \cite{Riess91}. The CDW energy gap was derived from the temperature dependence of the conductivity below $T_{\rm P}$. In the temperature range 120--50~K the thermally activated conductivity yields $\Delta_{\rm CDW}(0)$~= 60--90~meV for crystals from different batches. At lower temperature deviation from the thermally activated behaviour occur due to impurity level contribution within the gap. The ratio $2\Delta(0)/kT_{\rm P}$ is then found to be $7.7-11.5$ \cite{Brutting92}:

In section~\ref{sec4-3}, it will be shown that, although the CDW appears commensurate, the threshold field for CDW depinning is of the same order of magnitude than that for incommensurate CDW in K$_{0.3}$MoO$_3$ or TaS$_3$. That may mean that the quarter-filled band description of (FA)$_2$X is more appropriate, the commensurability pinning being then weaker, or more likely that defects such as deviations from the ideal 2:1 stoichiometry play an important role \cite{Ilakovac93}.

It is worth to note that the structure of (FA)$_2$X strongly resembles to that of Bechgaard (TMTSF)$_2$X where SDW transitions occur at low temperature at variance with the CDW transition at $T_{\rm P}\sim 180$~K in (FA)$_2$X. It was suggested that Coulomb interactions in the FA salts are efficiently screened by the strong polarisability of the Ar molecules composed of a large number of rings \cite{Ilakovac93}.

\subsection{Bechgaard-Fabre salts}\label{sec3-5}

\begin{figure}
\begin{center}
\includegraphics[width=7.5cm]{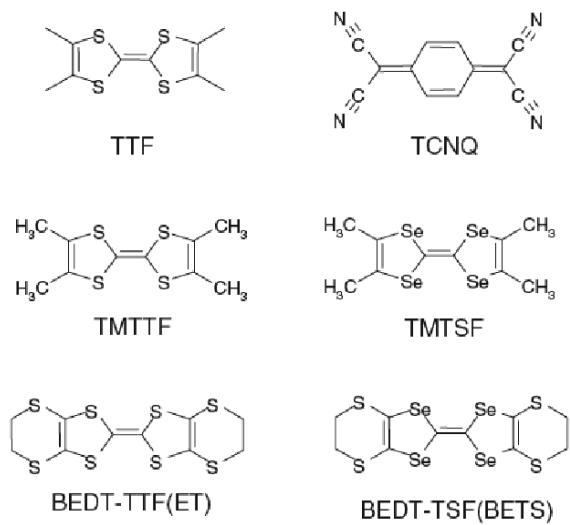}
\caption{Structure of organic molecules.}
\label{fig3-26}
\end{center}
\end{figure}

To avoid the constraint of the fractional stoichiometry on the TTF molecules, organic chemists \cite{Brun77,Galigne78,Galigne79} in Montpellier at the end of the 70's succeeded the synthesis of salts with single TMTTF (tetramethyltetrathiofulvalene) chains separated by anions such as: Br, PF$_6$, AsF$_6$, BF$_4$, SCN, ... Because the TMTTF chains are strongly dimerised, the room temperature conductivity of (TMTTF)$_2$X salts is low, $\sigma\sim 600~(\Omega$cm)$^{-1}$, and the temperature dependence of the conductivity was that of a semiconductor. Replacing S by Se \cite{Bechgaard81} has yielded a much less dimerisation along the TMTSF (tetramethyltetraselenafulvalene) chains, and consequently a metallic state at room temperature with a conductivity $\sigma\sim 20-100~(\Omega$cm)$^{-1}$. These compounds are used to be called Fabre salts for (TMTTF)$_2$X and Bechgaard salts for (TMTSF)$_2$X ones. The structure of TMTTF and TMTSF molecules are presented in figure~\ref{fig3-26}.

\subsubsection{Structure and low $T$-ground state}\label{sec3-5-1}

\begin{figure}
\begin{center}
\includegraphics[width=8cm]{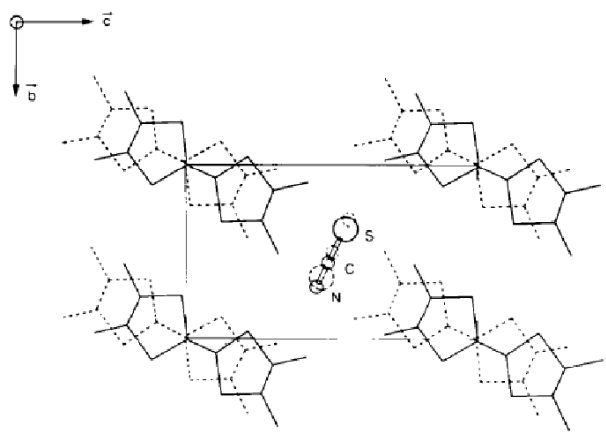}
\caption{Properties of the structure of (TMTTF)$_2$SCN parallel to the chain a-axis (from ref.~\cite{Galigne79}).}
\label{fig3-27}
\end{center}
\end{figure}

(TMTSF)$_2$X and (TMTTF)$_2$X salts are isostructural and belong to the same $P_{\bar{1}}$ space group. Nearly planar TMTSF or TMTTF molecules form zigzag stacks along the $a$ direction, being slightly tilted relative to the stacking direction: the space group is triclinic, with lattice constants (in $\AA$) for (TMTSF)$_2$PF$_6$ \cite{Thorup81}: $a$~= 7.297, $b$~= 7.711, $c$~= 13.522 and angles $\alpha$~= 83.39$^\circ$, $\beta$~= 86.27$^\circ$ and $\gamma$~= 71.01$^\circ$. It is then convenient to define a pseudo orthogonal lattice ($a$, $b'$, $c^\ast$) such $c^\ast\perp (a,b)$ and $b'\perp (a,c^\ast)$. The anions occupy loose cavities delimitated by the methyl groups of the organic molecules as shown for SCN in figure~\ref{fig3-27}.

These materials show almost all known ground states at low temperature: a metal, a paramagnetic insulator, spin/charge density wave, a spin-Peierls state and finally a superconducting state  \cite{Jerome82,Jerome04,R6Ishiguro98,R14Lebed08}. In parallel several different structural transitions occur due to anion ordering (AO) associated with slight arrangements in X chains \cite{Pouget96}. Their transition temperatures $T_a\sim 100-200$~K are much higher than those corresponding to magnetic transitions occurring in the range of $T_c\approx 1-20$~K. The electron transfer integral along the chain direction for both sulphur and selenium compounds is typically $t_a\approx 3000$~K, to be compared with the values in transverse directions: $t_b\approx 300$~K and $t_c\approx 10$~K. So, at temperatures higher than $t_b$, the system is in the confinement limit \cite{Giamarchi04}. A cross-over towards higher dimensionality is expected to occur at $T^\ast$ when $T$ becomes smaller than $t_b$.

The Bechgaard-Fabre conductors consist of molecular chains along the highest conductivity axis with two electrons per four molecules, which corresponds to a 1/4 filling in terms of holes. But these molecular chains are slightly dimerised; consequently the conduction band is splitted into a filled lower band separated from a half-filled upper band.

When globally considered, a generic pressure versus temperature phase diagram accounts for the general properties of (TMTSF)$_2$X and (TMTTF)$_2$X salts \cite{Jerome91}. The emerging picture from this phase diagram is the existence of a so-called 1D Mott-Hubbard charge localisation between room temperature and the magnetic ordering at low $T$ around 1-20~K. This insulating state exists for all the sulphur compounds with a charge gap having an activation energy of the order of $\Delta\rho$. Under pressure this correlation gap decreases, the bandwidth broadens, and the sulphur compounds behave similarly to the selenium ones. At low $T$, the increase of pressure induces a cascade of transitions with the ground state changing from Spin-Peierls to antiferromagnetic, then incommensurate SDW and finally superconductivity. Another transition in the Mott-Hubbard insulating state temperature range has been discovered, namely a transition into a charge ordered state generic for all the (TMTTF)$_2$X salts as described below. Spin-Peierls transition in (TMTTF)$_2$PF$_6$ has been studied by neutron scattering \cite{Foury04,Pouget06}.

The temperature dependence of the resistivity of several Fabre and Bechgaard salts is drawn in ref.~\cite{Dressel07}. All the (TMTTF)$_2$X are semiconducting below room temperature. (TMTSF)$_2$ undergoes a SDW transition at $T_{\rm SDW}$~$\sim 10$~K. Only (TMTSF)$_2$ClO$_4$ when slowly cooled remains metallic till approximately $T_c\simeq 1.2$~K where it becomes superconducting.

\subsubsection{Charge order and ferroelectric Mott-Hubbard ground state}

As discussed in section \ref{sec2-9}, in Q1D organic salt compounds, namely (TMTTF)$_2$X salts, electron-electron interactions play a leading role. As far as charge transfer is concerned, (TMTTF)$_2$X salts can be described as 1/4 filled band systems. However, structurally, they exhibit a significant dimerisation  of the intermolecular distance. In that case the dimerised 1D Hubbard model taking into account only the on-site Coulomb $U$ give rise to a dimer Mott insulating state. That is the consequence of the splitting of the conduction band into a filled lower band separated from a half-filled upper band by a dimerisation gap, $\Delta_\rho$, the resistivity showing an upturn from a metallic to a semiconducting behaviour at $T_\rho$. Quantitative estimates of the charge gap were made \cite{Mila95} for different values of $U$ and for respective values of the intermolecular transfer integrales along the stacks determined from quantum chemistry calculations \cite{Castet96}. However, considering the 1/4-filled band case, when a finite near-neighbour interaction $V$ is introduced, it was shown, using for instance the mean-field approximation \cite{Seo97,Seo00} that a $4k_{\rm F}$ superstructure occurs with charge disproportionation above a critical value $V_c$. The phase diagram $U/t$ versus $V/t$ with $t$: transfer integral between nearest-neighbour sites was shown in figure~\ref{fig2-6}. The charge order (CO) phase is stabilised for large $U/t$ and $V/t$ values, with a phase boundary between the dimer-Mott insulating state and the CO state. In the case of uniform molecule stacking, the dimer-Mott state is replaced by the metallic Tomonaga-Luttinger liquid phase \cite{Seo06}.

The first direct experimental evidence for the existence of CO, predicted in ref.~\cite{Seo97} was obtained by means of NMR studies in the Q1D (DI-DCNQI)$_2$Ag (in short DI-Ag) compound where DI-DCNQI is 2.5-diido-N,N'-dicyanoquinonediimine) \cite{Hiraki98}. It was shown that, with decreasing temperature below 220~K, the $^{13}$C-NMR spectra are splitted, pointing out the appearance of non-equivalent differently charged molecules along the chain axis.

Previous reports on Fabre salts, a change in slope of thermopower for (TMTTF)$_2$AsF$_6$ and (TMTTF)$_2$SbF$_6$ \cite{Coulon85}, a weak feature in the microwave dielectric permittivity in (TMTTF)$_2$SbF$_6$, (TMTTF)$_2$SCN and (TMTTF)$_2$ReO$_4$ \cite{Javadi88} were observed, pointing out a possible phase transition. However no corresponding change in crystal structure and no observation of superlattice reflections were detected \cite{Laversanne84}, making these transitions to be called ``structureless phase transitions". But none of these reports have identified the nature of these possible phase transitions.

Diffuse X-ray scattering experiments on (TMTTF)$_2$Br have revealed \cite{Pouget97} one-dimensional spin-Peierls fluctuations developing below 70~K and vanishing at the antiferromagnetic transition at $T_N$~= 13~K. In the case of (TMTTF)$_2$PF$_6$, $2k_{\rm F}$ displacive lattice instability of spin-Peierls type was found \cite{Pouget96,Pouget97} to develop below 100~K and to condense at the spin-Peierls phase transition $T_{\rm SP}$~= 19~K into satellite reflections. The initial experiences for studying the possible polarisation response of these fluctuating superstructures have led to the observation of a huge peak of the real part of the dielectric constant $\varepsilon'$ in (TMTTF)$_2$Br \cite{Nad98} and in (TMTTF)$_2$PF$_6$ \cite{Nad99} occurring at a temperature below $T_\rho$ for charge localisation. These results were then interpreted as a possible evidence for a charge-induced-correlated state \cite{Nad99}, the first step opening the field of charge ordering in (TMTTF)$_2$X salts, as presented hereafter.

\medskip
\noindent \textit{3.7.2.a. Low frequency ac conductivity}
\medskip

\begin{figure}
\begin{center}
\subfigure[]{\label{fig10-2a}   
\includegraphics[width=6.5cm]{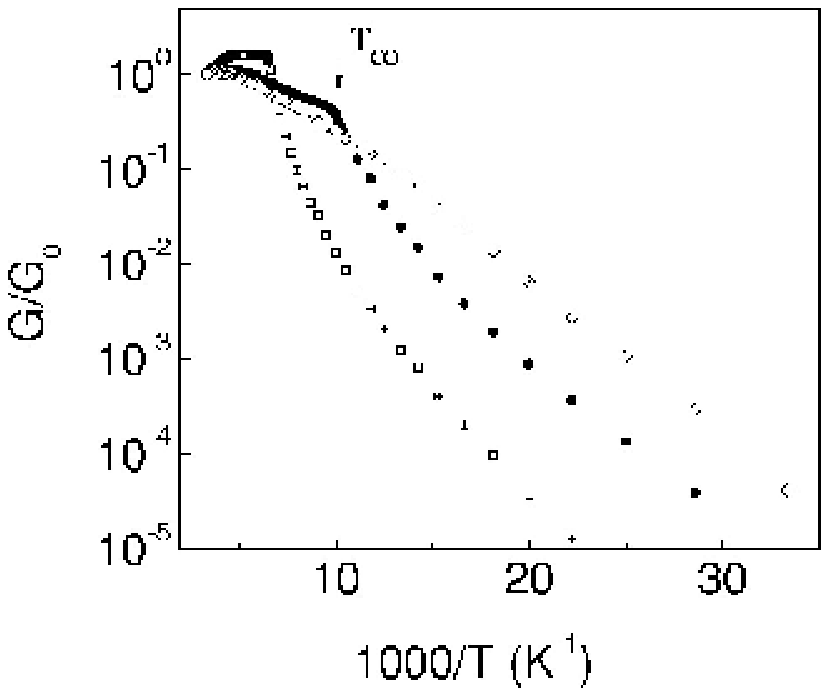}}
\subfigure[]{\label{fig10-2b}                    
\includegraphics[width=6.5cm]{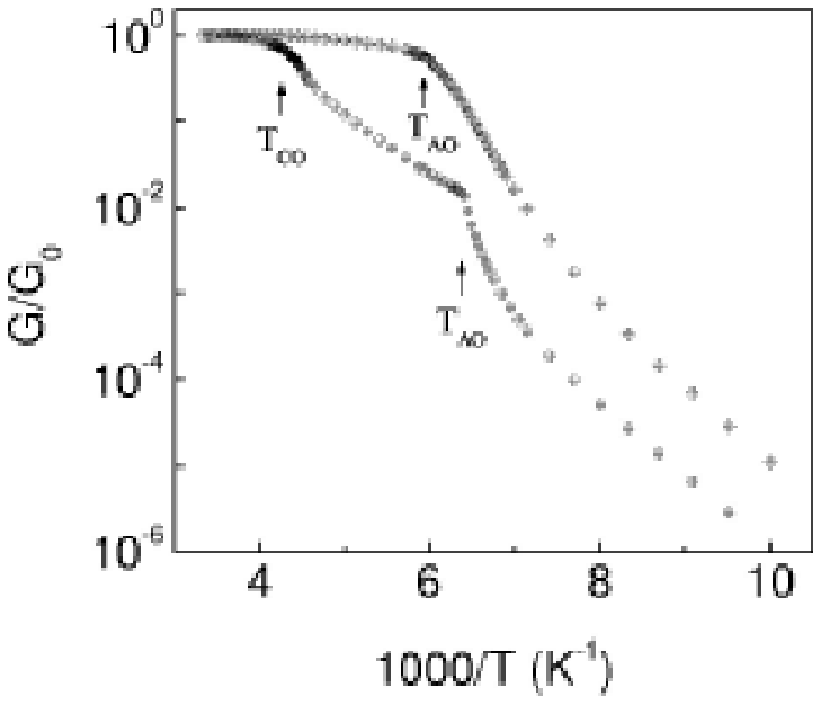}}
\caption{Real part of the conductance $G$ normalised to its room temperature value $G_0$ as a function of inverse temperature at frequency 1~kHz for (TMTTF)$_2$X salts with (a) centrosymmetric anions (CSA): X~= SbF$_6$ (squares), AsF$_6$ (circles), and PF$_6$ (diamonds) (reprinted figure with permission from Journal of the Physical Society of Japan 75, p. 051005, 2006, F. Nad \textit{et al.} \cite{Nad06a}), (b)~non-centrosymmetric anions (NCSA): X~= ReO$_4$ (circles), SCN (diamonds) (reprinted figure with permission from F. Nad \textit{et al.}, Journal of Physics: Condensed Matter 13, p. L717, 2001 \cite{Nad01a}. Copyright (2001) by the Institute of Physics).}
\label{fig10-2}
\end{center}
\end{figure}

Figure~\ref{fig10-2}(a)
shows the variation of the real part of conductance $G$, normalised by its room temperature value $G_0$, as a function of inverse temperature for (TMTTF)$_2$X samples with centro-symmetric anions (CSA) with X: PF$_6$, AsF$_6$ and SbF$_6$. The analogous dependence for samples with non centro-symmetric anions (NCSA) with X~= ReO$_4$ and SCN are shown in figure~\ref{fig10-2}(b). For the overwhelming majority of (TMTTF)$_2$X salts the temperature dependence of the conductance has common features: by cooling from room temperature the conductance is growing first, reaches a maximum at some temperature $T_\rho$ and below that it decreases approximately linearly in the $\log G/G_0(1/T)$ scale (figure~\ref{fig10-2}). The magnitude of $T_\rho$ and the slope of the $G/G_0(1/T)$ dependence, i.e., the ``activation energy" $\Delta_1$ in that Mott-Hubbard temperature range are listed in table~\ref{tab10-1}. One has to note that (TMTTF)$_2$SbF$_6$ shows a somewhat different behaviour: the initial usual growth of conductance is followed by a wide plateau without any strongly pronounced maximum (figure~\ref{fig10-2}).

In all these (TMTTF)$_2$X compounds a bend is observed below $T_\rho$ in the $G/G_0(1/T)$ dependences near some temperature $T_{\rm CO}$ indicating the opening of an additional gap. As can be seen from figure \ref{fig10-2}, after some transitional temperature interval at $T<T_{\rm CO}$, the $G/G_0(1/T)$ dependence again takes the form of a close linear variation with a new activation energy $\Delta_2$, the magnitude of which is also indicated in table~\ref{tab10-1}, as well as the appropriate value of $T_{\rm CO}$.

At lower temperatures, the behaviour of the CSA conductors differs from that of NCSA ones. In the case of CSA, the thermally activated decrease of conductance continues down to the transition temperature in the magnetic ordered state $T_{\rm MO}$. That corresponds to a transition into the spin-Peierls state (SP) for the PF$_6$ anion at $T_{\rm SP}$~= 19~K and into the antiferromagnetic state (AF) for the SbF$_6$ anion at $T_{\rm N}$~= 8~K \cite{Laversanne84}. In the case of (TMTTF)$_2$Br, features of the transition into the AF state at $T_{\rm N}$~= 15~K become apparent in the temperature dependences of the conductance and dielectric permittivity \cite{Nad98}. In Br salt the values of $T_{\rm CO}$~= 28~K and $T_{\rm N}$~= 15~K are relatively close to each other. For (TMTTF)$_2$X conductors with NCSA tetrahedral anions BF$_4$ and ReO$_4$, after some thermoactivated variation of $G(T)$ below $T_{\rm CO}$, an anion ordering occurs at $T_{\rm AO}$  with formation of a superstructure with wave vector ${\bf q}$~= (1/2, 1/2, 1/2) for both salts. As can be seen from figure~\ref{fig10-2}(b) the AO transitions in these salts have a more abrupt character than near $T_{\rm CO}$. In the case of (TMTTF)$_2$BF$_4$, it corresponds even to an increase of conductivity. In the case of the SCN linear anion (see the structure in figure~\ref{fig3-27}), no noticeable anomaly in the $G(T)$ dependence was observed below $T_\rho$ down to the AO transition temperature at $T_{\rm AO}$~= 169~K with ${\bf q}$~= (0, 1/2, 1/2) \cite{Coulon82a,Coulon82b}. As can be seen from figure~\ref{fig10-2}(a), ReO$_4$ and SCN salts show below $T_{\rm AO}$ a thermoactivated decrease of $G$ with a similar large activation energy $\Delta_3$~= 2000~K (table~\ref{tab10-1}). At lower temperature an AF transition occurs in the case of the SCN anion \cite{Coulon82a,Coulon82b}.

\begin{sidewaystable}
\tbl{Parameters of (TMTTF)$_2$X conductors}
{\begin{tabular}{@{}lcccccccl}\toprule
$\begin{array}{c}
\mbox{Anion}\\ X \end{array}$ &
$\begin{array}{c}
T_\rho \\ \mbox{(K)} \end{array}$  & 
$\begin{array}{c}
\Delta_1 \\ \mbox{(K)} \end{array}$ & 
$\begin{array}{c}
T_{\rm CO} \\ \mbox{(K)} \end{array}$ & 
$\begin{array}{c}
\Delta_2 \\ \mbox{(K)} \end{array}$  & 
$\begin{array}{c}
T_{\rm AO} \\ \mbox{(K)} \end{array}$ & 
$\begin{array}{c}
\Delta_3 \\ \mbox{(K)} \end{array}$ & 
$\begin{array}{c}
\mbox{Charge disproportionation} \\ \Delta_\rho=\rho_{\rm rich}-\rho_{\rm poor} \end{array}$ & 
$\begin{array}{c}
\mbox{Low}\\ \mbox{Temp.}\\ \mbox{state} \end{array}$ \\
\colrule
Br & $\approx 200$ & 75 & 28 & -- & -- & -- & -- & AFM \\
PF$_6$ & 250 & 300 & 70 &  370 & -- & -- & 0.28 \cite{Nakamura07} & SP  \\
AsF$_6$ & 230 & 175 & 100.6 & 360 & -- & -- &  0.34 \cite{Fujiyama06} & SP \\
SbF$_6$ & $\approx 170$ & -- & 154 & 500 & -- & -- & 0.50 \cite{Yu04}  & AFM \\
BF$_4$ & 220 & 580 & 83 & 750 & 39 & -- & -- & AFM  \\
ReO$_4$ & 290 & 800 & 227.5 & 1400 &  154 & 2000 & 0.34 \cite{Nakamura06} & SP \\
SCN & 265 & 500 & -- & -- & 169 & 2000 & 0.15 \cite{Nogami02,Nogami05} & AFM \\
\botrule
\end{tabular}}
\tabnote{$T_\rho$: temperature at which charge localisation occurs and the conductance is maximum, $\Delta_1$: the activation energy in conductivity below $T_\rho$, $T_{\rm CO}$: charge-ordering phase transition, $\Delta_2$: the energy in conductivity below $T_{\rm CO}$, $T_{\rm AO}$: anion-ordering phase transition, $\Delta_3$: activation energy in conductivity below $T_{\rm AO}$, $\Delta_\rho$: charge disproportionation in the charge order state.}
\label{tab10-1}
\end{sidewaystable}

\begin{figure}
\begin{center}
\subfigure[]{\label{fig10-4a}
\includegraphics[width=6.5cm]{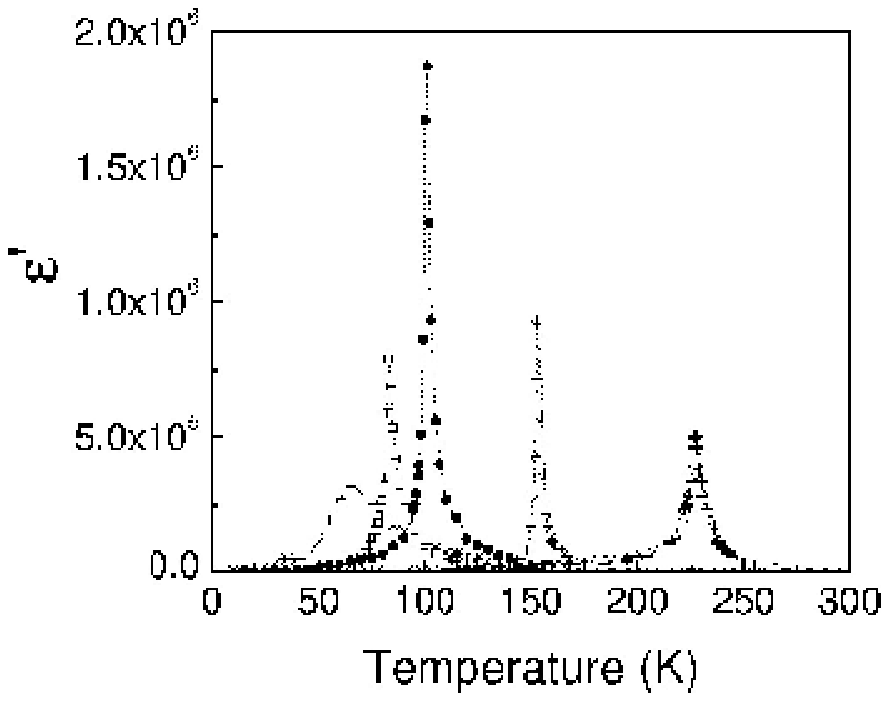}}
\subfigure[]{\label{fig10-4b}
\includegraphics[width=6.5cm]{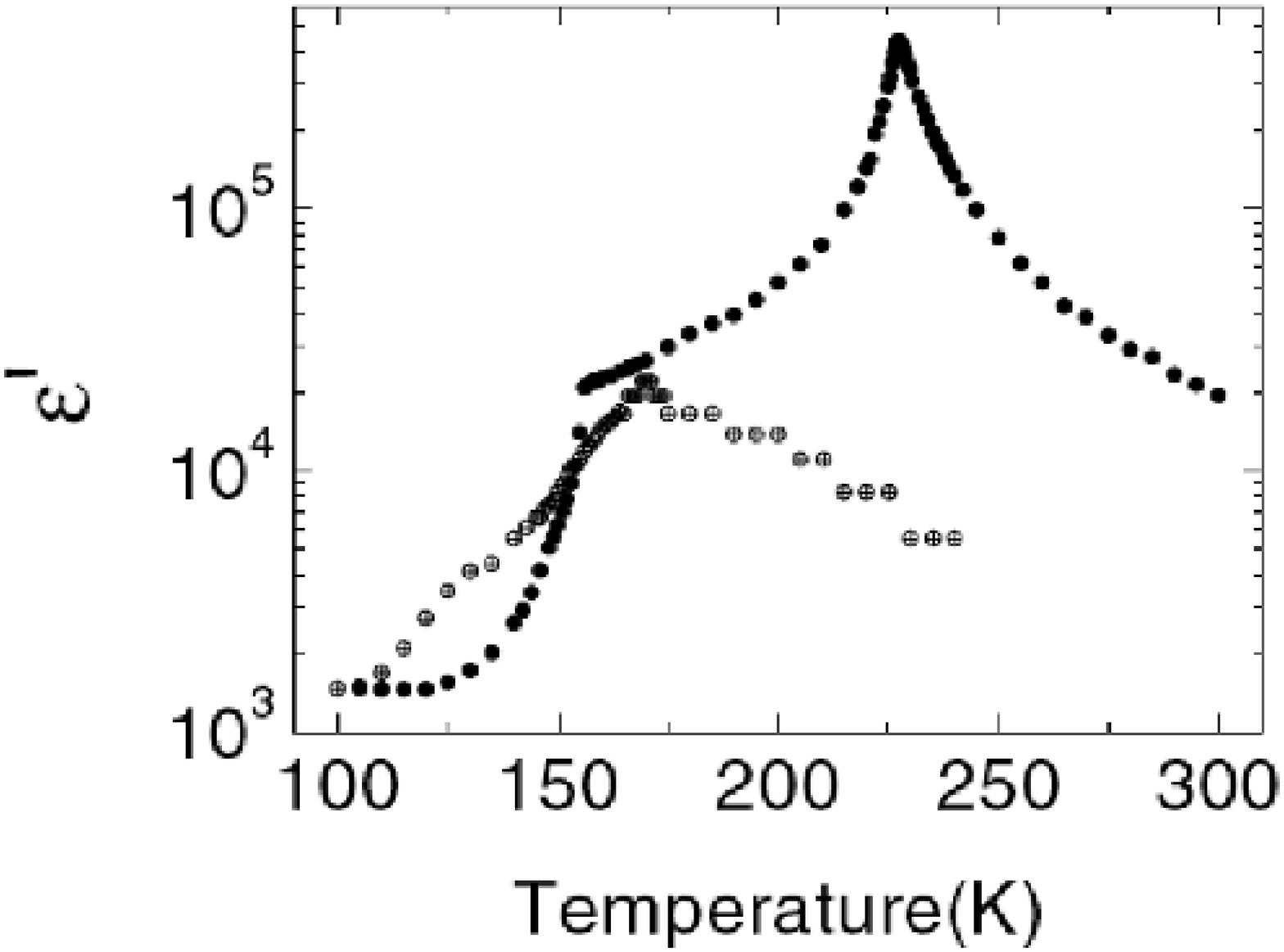}}
\caption{Temperature dependence of the real part of the dielectric permittivity $\varepsilon'$ at 100~kHz for (TMTTF)$_2$X salts a)~with X~= PF$_6$ (stars), BF$_4$ (squares), AsF$_6$ (circles), SbF$_6$ (diamonds) and ReO$_4$ (triangles) (reprinted figure with permission from Journal of the Physical Society of Japan 75, p. 051005, 2006, F. Nad \textit{et al.} \cite{Nad06a}), b)~on a semi-logarithmic scale at frequency 1~MHz: (TMTTF)ReO$_4$ (closed circles), (TMTTF)$_2$SCN (encircled) (reprinted figure with permission from F. Nad \textit{et al.}, Journal of Physics: Condensed Matter 13, p. L717, 2001 \cite{Nad01a}. Copyright (2001) by the Institute of Physics).}
\label{fig10-4}
\end{center}
\end{figure}

\begin{figure}
\begin{center}
\includegraphics[width=7.5cm]{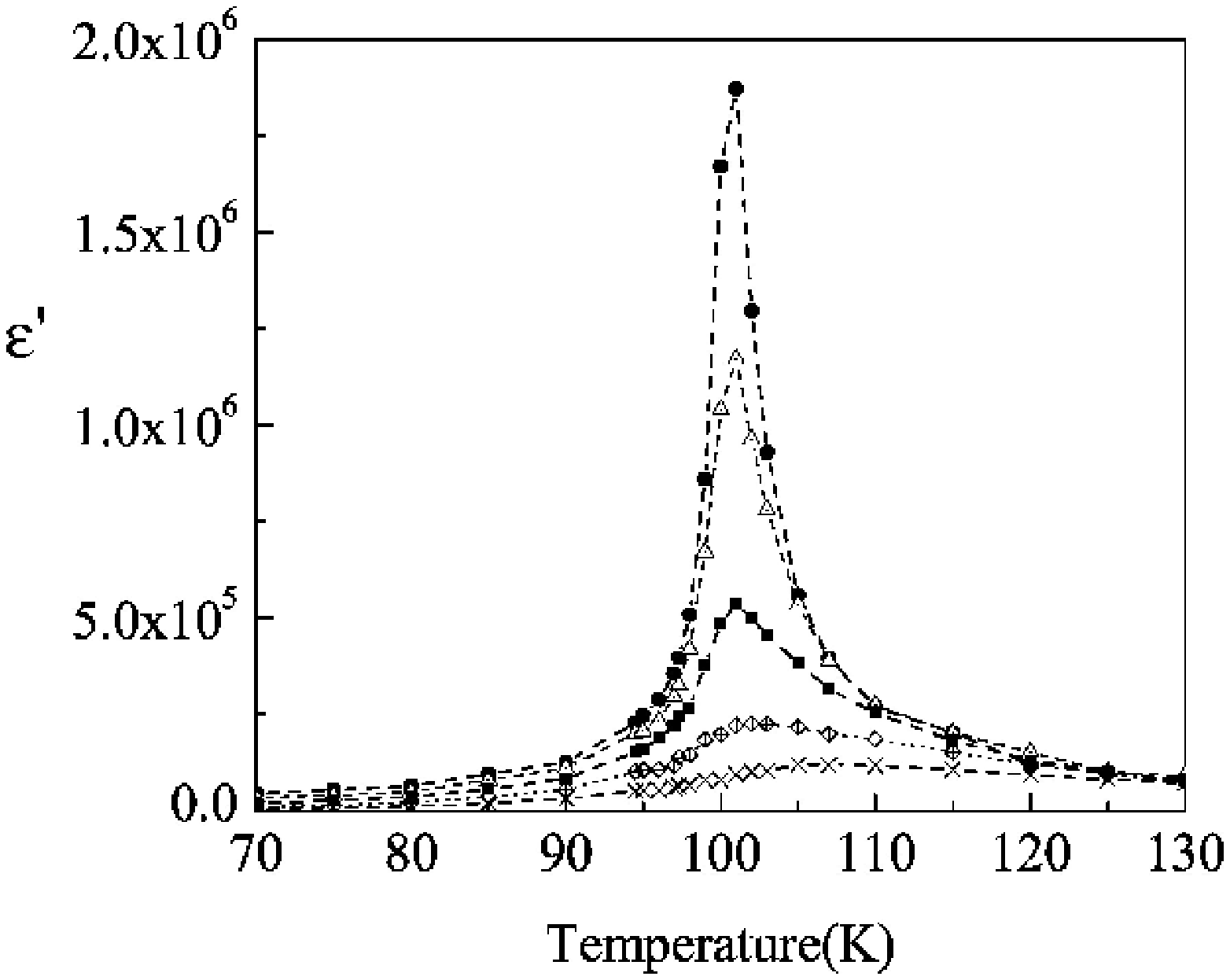}
\caption{Temperature dependence of the real part of the dielectric permittivity $\varepsilon'$ for (TMTTF)$_2$AsF$_6$ at frequencies: 100~kHz (black circles), 300~kHz (triangles), 1~MHz (black squares), 3~MHz (diamonds), 10~MHz (crosses) (reprinted figure with permission from F. Nad \textit{et al.}, Journal of Physics: Condensed Matter 12, p. 435, 2000 \cite{Nad00b}. Copyright (2000) by the Institute of Physics).}
\label{fig10-5}
\end{center}
\end{figure}

The magnitude of the dielectric permittivity was calculated by the standard equation $\varepsilon'$~= $\Im G/\omega$. Figure~\ref{fig10-4}(a) 
shows the temperature dependence of $\varepsilon'(T)$ at frequency 100~kHz for CSA: (TMTTF)$_2$PF$_6$ \cite{Nad99,Nad00}, (TMTTF)$_2$AsF$_6$ \cite{Nad00b} and (TMTTF)$_2$SbF$_6$ \cite{Nad01a,Nagasawa05} and for NCSA: (TMTTF)$_2$BF$_4$ \cite{Nad06a,Nad06b} and (TMTTF)$_2$ReO$_4$ and in figure~\ref{fig10-4}(b) in a logarithmic scale at frequency 1~MHz for NCSA: (TMTTF)$_2$ReO$_4$ (closed circles) and (TMTTF)$_2$SCN (encircled) (from ref.~\cite{Nad01a}). As can be seen in figure~\ref{fig10-4} the $\varepsilon'(T)$ dependences do not show any visible anomalies in the temperature range near the maximum of $G(T)$ at $T\approx T_\rho$. Narrow peaks in the $\varepsilon'(T)$ dependences are observed at $T_{\rm CO}$. Indeed, the maximum value of $\varepsilon'$ reaches the huge value of the order of 10$^6$. The data presented in figure~\ref{fig10-4} were obtained at some fixed frequency. The most pronounced features in the frequency dependence of $\varepsilon'(T)$ were obtained in high quality samples of (TMTTF)$_2$AsF$_6$ salt as shown in figure~\ref{fig10-5} \cite{Nad00b}. 
With increasing frequency the maximum value of $\varepsilon'$ decreases and the position of the $\varepsilon'(T)$ maximum shifts to higher temperatures. This shift occurs as a result of a so-called slowing down behaviour when the relaxation time $\tau$ becomes equal to the reciprocal cyclic frequency of the ac signal. However, below 1~MHz, the temperature at which $\varepsilon'(T)$ is maximum is frequency independent.

Relaxation processes are the best pronounced in the frequency dependence of dielectric losses, i.e. the imaginary part of the dielectric permittivity, $\epsilon"$. At $T>T_{\rm CO}$ the $\varepsilon"(f)$ curves for (TMTTF)$_2$AsF$_6$ have a nearly symmetric form \cite{Nad06b}. They are similar to each other in a double logarithmic scale, i.e. they superpose by an appropriate displacement along the two axis. With decreasing temperature, the maximum $\varepsilon_m$ of $\varepsilon"(f)$ at $f_m$ is shifted at low frequencies; the frequency dependence of $\varepsilon"(f)$ above $f_m$ approaches asymptotically a power law dependence $f^{-n}$ with $n\simeq 0.73$. All these features are associated with relaxation processes  of a Debye type \cite{Jonscher83}. On the contrary, for $T<T_{\rm CO}$~= 101~K, the $\varepsilon"(f)$ curve becomes wider with the development of an additional low-frequency shoulder. The low-frequency shoulder peak in $\varepsilon"(f)$ in a narrow temperature range below $T_{\rm CO}$ ($\sim 10$~K) has been ascribed to slower relaxation processes associated with the motion of domain walls within the domain ferroelectric structure developed at $T<T_{\rm CO}$ \cite{Nad06b}. At lower temperature the contribution of the domain wall motion in relaxation processes vanishes, that may be associated with freezing of the ferroelectric domain structure.

It is traditionally considered that the frequency $f_m$ corresponds to the mean relaxation time $\tau$~= $1/2\pi f_m$. For a very large part of the temperature range, the $\tau(1/T)$ dependence shows a growth close to a thermoactivated type $\tau\sim\exp[-\Delta/kT]$ with the activation energy $\Delta$ in good agreement with the value deduced from conductivity measurements (figure~\ref{fig10-2}). Near $T_{\rm CO}$, the $\tau(1/T)$ dependence shows an evident peak \cite{Nad06b}. The divergence of the relaxation time near $T_{\rm CO}$ corresponds to the theory of a classical ferroelectric transition for which it was shown \cite{Lines77} that $\tau\sim 1/|T-T_{\rm CO}|$. This divergence corresponds also to the softening of the oscillatory mode responsible for the observed ferroelectric transition.

In (TMTTF)$_2$X salts with CSA, there are no indication of anion ordering with the formation of a superstructure with a wave vector $q\neq 0$. That is due to the charge symmetry of these anions, when all their orientations in the cavity between TMTTF molecules are equivalent. In the case of NCSA such as ReO$_4$ or SCN, at $T>T_{\rm AO}$, their various orientations within the cavity are not equivalent, so that the center of symmetry is only preserved on average \cite{Pouget96}. In the frame of a model with two potential wells the anion can be in both wells with equal probability. With decreasing temperature the screening of the anion potential decreases and the interaction of anions with TMTTF molecules becomes stronger. The occurrence of short contacts between anions and $S$ atoms of TMTTF molecules favours their orientational ordering with the formation of a superstructure with $ q\neq 0$. The anion ordering transition in (TMTTF)$_2$ReO$_4$ at $T_{\rm AO}$~= 154~K leads to a doubling of the unit cell [${\bf q}=(1/2,1/2,1/2)$] and it corresponds to a first order transition as observed in X-ray studies. The wave vector of the superstructure developed in (TMTTF)$_2$SCN at $T_{\rm AO}$~= 169~K is ${\bf}=(0,1/2,1/2)$ \cite{Pouget96}.

As seen in figure~\ref{fig10-4}(b) in (TMTTF)$_2$ReO$_4$, $\varepsilon'$ exhibits a divergence at $T_{\rm CO}$ and a jump drop at $T_{\rm AO}$ \cite{Nad01a}. This $\varepsilon'(T)$ dependence resembles the behaviour of the dielectric permittivity in extrinsic ferroelectrics \cite{Lines77}. The $\varepsilon'(T)$ dependence of (TMTTF)$_2$SCN is qualitatively different from the jump-like dependence near $T_{\rm AO}$ for the other NCSA compounds. It has to be noted that the longitudinal component of the superstructure developed below $T_{\rm AO}$ is equal to zero. As a result the observed dependence of $\varepsilon'(T)$ has some features of the displacement ferroelectric transition of second-order analogous to the CO transition in CSA salts with $q$~=(0,0,0), allowing polarisation along a single stack. However, consequently to the 1/2, 1/2 transverse components of the anion ordering, the long range order of the electric dipoles corresponds to antiphase on adjacent stacks, yielding the antiferroelectric nature associated to the anion ordering \cite{Nad01a}.

(TMTTF)$_2$SbF$_6$ and (TMTTF)$_2$SCN are very specific among the Fabre salts, in the sense that metal-insulating transition and ferroelectric transition for the former, and anion ordering and antiferroelectric transition for the latter, occur at the same temperature.

\medskip
\noindent \textit{3.7.2.b. Charge disproportionation}
\medskip

The change of the activation energy at $T_{\rm CO}$ in electrical conductivity and the absence of any anomaly in magnetisation in the same temperature range \cite{Coulon82b} indicate that an additional charge gap opens below $T_{\rm CO}$. $^{13}$C NMR spectroscopy has demonstrated the inequivalence of the TMTTF molecules in the unit cell with unequal electron densities \cite{Chow98,Zamborsky02}. Above $T_{\rm CO}$, each molecule is equivalent but the two $^{13}$C nuclei in each molecule have inequivalent hyperfine coupling giving rise to two spectral lines. Below $T_{\rm CO}$, each of these two lines splits in two. This doubling results from two different molecular environments with unequal electron densities. The difference of the NMR frequencies inside each doubled line has been used for defining the order parameter of the CO phase transition. The CO transition appears to be of second order. For these NMR data $T_{\rm CO}$ was determined to be 101~K for (TMTTF)$_2$AsF$_6$ and 75~K for (TMTTF)$_2$PF$_6$, at the identical temperatures at which the dielectric constant $\varepsilon'$ exhibits a maximum.

\begin{figure}
\begin{center}
\includegraphics[width=7.5cm]{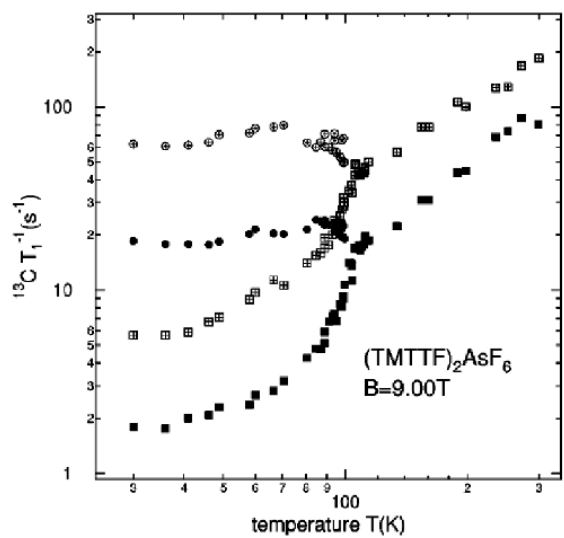}
\caption{Temperature dependence of $^{13}$C $T^{-1}_1$ relaxation rate in (TMTTF)$_2$AsF$_6$ (reprinted figure with permission from F. Zamborsky \textit{et al.}, Physical Review B 66, p. 081103(R), 2002 \cite{Zamborsky02}. Copyright (2002) by the American Physical Society).}
\label{fig10-9}
\end{center}
\end{figure}

The amplitude of the charge disproportionation (CD) can be estimated from the NMR spin-lattice relaxation rate. It is expected that $T^{-1}_1$ being dominated by hyperfine coupling is proportional to the square of the molecular charge: $T^{-1}_1\propto\rho^2$. From the data shown in figure~\ref{fig10-9} 
for (TMTTF)$_2$AsF$_6$ it was concluded \cite{Zamborsky02} that the ratio of CD in the CO state was 3:1, which corresponds to a degree of charge disproportionation of $\Delta\rho$~= 0.5. $\Delta\rho$ is defined as $\rho_{\rm rich}-\rho_{\rm poor}$, $\rho_{\rm rich,poor}$ being the charge on charge-rich (charge-poor) molecule site. A similar study on (TMTTF)$_2$AsF$_6$ yields the ratio of the charge densities to be about 2:1 \cite{Fujiyama06}. The determination of the CD from NMR spin-lattice relaxation experiments was recently reconsidered \cite{Hirose10}. It was noted that the NMR shift including both Knight and chemical shift, the $\Delta\rho$ values deduced from $T^{-1}_1$ are overestimated (because not taking into account the chemical shift) with respect to those deduced from the Knight shift. In that case $\Delta\rho$ was estimated to be 0.11 \cite{Hirose10}.

\begin{figure}
\begin{center}
\includegraphics[width=6.5cm]{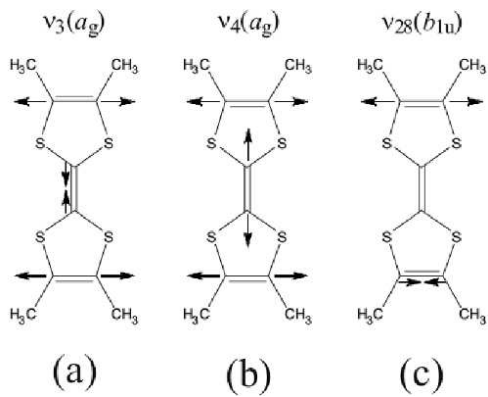}
\caption{Symmetric (a) and (b) and asymmetric (c) stretching modes of the TMTTF molecule.}
\label{fig10-10}
\end{center}
\end{figure}

\begin{figure}
\begin{center}
\includegraphics[width=7.5cm]{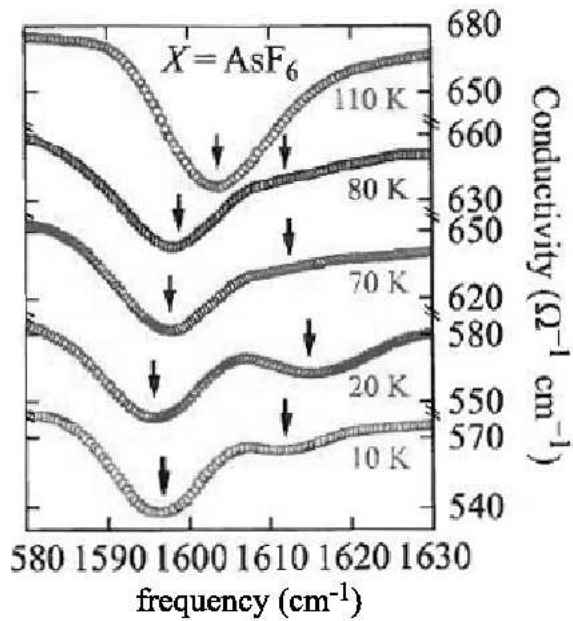}
\caption{Mid-infrared conductivity of (TMTTF)AsF$_6$ for light polarised parallel to the molecular stacks. The emv-coupled totally symmetric intramolecular $\nu_3$ ($a_g$) mode splits below $T_{\rm CO}$ (reprinted figure with permission from M. Dumm \textit{et al.}, Journal de Physique IV (France) 131, p. 55, 2005 \cite{Dumm05}. Copyright (2005) from EdpSciences).}
\label{fig10-11}
\end{center}
\end{figure}

Infrared spectroscopy is another powerful method for the determination of charge disproportionation. The stretching modes of the C=C bonds in the TMTTF molecule represent two symmetric modes $\nu_3(a_g)$ and $\nu_4(a_g)$ and one asymmetric mode $\nu_28(b_{1u}$ (see figure~\ref{fig10-10}). 
It is known that the frequency of these modes can yield a very sensitive measure of the charge per molecule \cite{Meneghetti84}. Although totally symmetric, $a_g$ modes are not infrared-active. Due to electron-molecular vibrational (emv) coupling, they can, however, be observed by infrared spectroscopy with the polarisation parallel to the 1D stacks. Figure~\ref{fig10-11} 
shows the temperature dependence of the mid-infrared conductivity around 1600~cm$^{-1}$ for (TMTTF)$_2$AsF$_6$. Above $T_{\rm CO}$, the antiresonance dip around 1600~cm$^{-1}$ was assigned to the Raman $a_g(\nu_3)$ mode \cite{Dumm05}. Below $T_{\rm CO}$, this mode is split and two features (indicated by arrows in figure~\ref{fig10-11}) are observed. From the neutral TMTTF C=C $a_g$ frequency ($\nu_0$~= 1639~cm$^{-1}$) and the monocationic frequency ($\nu_0$~= 1567~cm$^{-1}$) determined in ref.~\cite{Meneghetti84}, a difference of 0.37 between the charges 0.63e and 0.37e on the two unequal TMTTF molecules in the unit cell has been obtained \cite{Dumm06}.

Similarly infrared spectroscopy was performed with the focus on the asymmetric mode $\nu_{28}$ \cite{Hirose10} for avoiding complexities related to vibronic effects. Splitting of the $\nu_{28}$ mode was also observed indicating charge disproportionation. From the neutral and monocationic frequencies, it was deduced a charge distribution rich:poor as 0.58e:0.42e, i.e. with $\Delta\rho$~= 0.16.

Nevertheless, although there are some dispersions in the charge distribution (rich-poor) along the TMTTF stacks in (TMTTF)$_2$AsF$_6$ -- $\Delta\rho$~= 0.5 \cite{Zamborsky02}, 0.34 \cite{Fujiyama05}, 0.11 \cite{Hirose10} by NMR, and 0.26 \cite{Dumm05} and 0.16 \cite{Hirose10} by infrared spectroscopy, all these experiments clearly demonstrate charge disproportionation with the pattern rich-poor, i.e. a $4k_{\rm F}$ charge modulation.

Similar NMR experiments were performed on (TMTTF)$_2$PF$_6$ \cite{Nakamura07} and (TMTTF)$_2$ReO$_4$ \cite{Nakamura06}. The values of $\Delta\rho$ for (TMTTF)$_2$X salts are indicated in table~\ref{tab10-1}.

In the case of non-centrosymmetric anions : (TMTTF)$_2$ReO$_4$ and (TMTTF)$_2$SCN, refinements of the X-ray structures in the temperature range below the AO transition have allowed also to determine the charge distribution corresponding to the CO. The method used concerns the measurement of the intramolecular bond lengths \cite{Guionneau97} developed for BEDT-TTF molecules and expected to be also appropriate for TMTTF with regards to the similar TTF structure \cite{Nogami02}. Oxidation increases central $C=C$ and terminal $C=C$ double bonds of the TTF structure. The deduced CO pattern for (TMTTF)$_2$SCN below the AO transition corresponds to the $4k_{\rm F}$ pattern commensurate with the AO (0, 1/2, 1/2) superstructure with a charge modulation amplitude $\rho$ of 0.15e \cite{Nogami02,Nogami05}.

In the case of tetrahedral anion ReO$_4$, inequivalence of TMTTF molecules was clearly seen by NMR below $T_{\rm CO}$~= 225~K \cite{Nakamura06}. (TMTTF)$_2$ReO$_4$ undergoes a spin-singlet transition associated with the AO transition at $T_{\rm AO}$~= 158~K. A $2k_{\rm F}$ CO resulting from molecular tetramerisation \cite{Nogami05} with charge modulation amplitude $\rho$ of 0.25e has been observed together with the (1/2, 1/2, 1/2) AO with a possible redistribution of the charge pattern at $T_{\rm AO}$ \cite{Nakamura06}.

\medskip
\noindent \textit{3.7.2.c. Ferroelectric character of the charge order state}
\medskip

\medskip
\textit{- Curie law}
\medskip

\begin{figure}
\begin{center}
\includegraphics[width=7.5cm]{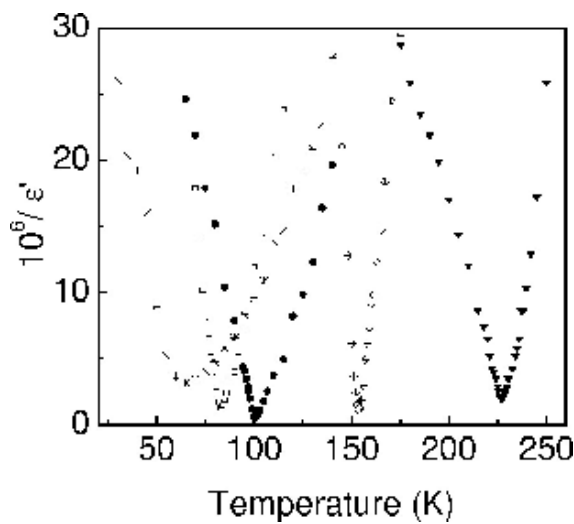}
\caption{Inverse of the real part of the dielectric permittivity $\varepsilon'$ as a function of temperature at frequency 100~kHz for (TMTTF)$_2$X: X~= PF$_6$ (stars), BF$_4$ (squares), AsF$_6$ (circles), SbF$_6$ (diamonds) and ReO$_4$ (triangles) (reprinted figure with permission from Journal of the Physical Society of Japan 75, p. 051005, 2006, F.Ya. Nad and P. Monceau \cite{Nad06a}).}
\label{fig10-12}
\end{center}
\end{figure}

The forms of the $\varepsilon'(T)$ dependence near $T_{\rm CO}$ indicate that the CO transition have a ferroelectric (FE) character and, in many aspects, they are similar to displacement type transitions in classical FE \cite{Lines77}. Indeed, the divergence of the dielectric permittivity $\varepsilon'$ near $T_{\rm CO}$ \cite{Monceau01} corresponds practically exactly to a FE second order transition usually described by a Curie law:
\begin{equation*}
\varepsilon'=\frac{A}{|T-T_{\rm CO}|}.
\end{equation*}
Figure \ref{fig10-12} 
shows the $1/\varepsilon'(T)$ dependence near $T_{\rm CO}$ for CSA (X~= PF$_6$, AsF$_6$ and SbF$_6$) and for NCSA (X~= BF$_4$ and ReO$_4$). As can be seen these dependences are really close to be linear in a relatively wide $T$ range near $T_{\rm CO}$. In the case of the most sharp transitions (AsF$_6$ and SbF$_6$) the slope of $1/\varepsilon'(T)$ branch (i.e. the magnitude of the Curie constant $A$) at temperatures below $T_c$ is twice that at $T>T_{\rm CO}$, in agreement with the theory of a second order FE transition. In the case of ReO$_4$ salt this ratio is equal to 1.5, probably because the influence of the orientational disorder of the semi-symmetric (tetrahedral) anions on the CO transition. For PF$_6$ anion, the CO transition is much broader, although the qualitative features of the $1/\varepsilon'(T)$ dependence are similar to those in other CSA.

\medskip
\textit{- Microscopic model}
\medskip

The charge disproportionation developed during the CO transition can influence the neighbour anion position in spite that the molecular interaction with anions may be not so large. Then an appropriate model for CO with ferroelectric characteristics should combine charge disproportionation and ionic displacements as schematically presented in section~\ref{sec2-16}.

There are two sources to the dimerisation and therefore to the charge gap $\Delta$ \cite{Brazovskii08,Brazovskii02}: the extrinsic one determined by the basic crystal structure and the intrinsic one spontaneously self-induced by the electronic subsystem. The gap $\Delta(U)$ appears as the consequence of both contributions to the Umklapp scattering $U$. At $T>T_{\rm CO}$ there is only the bond contribution $U_b$ which results from the bond alternation due to the tiny dimerisation. Below $T_{\rm CO}$, CO adds the on-site contribution $U_s$ from the non equivalence of sites. The charge gap is a function of the total amplitude $U$~= $\sqrt{U^2_b+U^2_s}$.

The energy change of the electronic system, $F_e$, due to both $U_b$ and $U_s$ depends only on the charge gap $\Delta$ and therefore on the total $U$. The energy of lattice distortions depends only on the spontaneous site component $U_s$: $F_e=\frac{1}{2}KU^2_s$, with $K$: an elastic constant. Thus, the total energy can be written in terms of the total $U$:
\begin{equation*}
F_{\rm tot}=F_e(U)+\frac{1}{2}KU^2-\frac{1}{2}KU^2_b.
\end{equation*}
The ground state is determined by its minimum over $U$, but with the constraint $U>U_b$.

From a more microscopic description, the Hamiltonian for the combined Mott-Hubbard state \cite{Monceau01,Brazovskii08} was derived:
\begin{equation*}
H_U=-U_s\cos 2\varphi-U_b\sin 2\varphi=-U\cos(2\varphi-2\alpha),
\end{equation*}
with $\tan 2\alpha$~= $U_b/U_s$. For a given $U_s$, the ground state is doubly degenerate between $\varphi=\alpha$ and $\varphi$~= $\alpha+\pi$. If the same phase is chosen for all the stacks, the ground state is ferroelectric while the state is antiferroelectric if the phase alternates on adjacent stacks. Then, the double action of the spontaneous charge disproportionation on the TMTTF molecules and of the X anionic potentials yields a $q=0$ ferroelectric phase transition.

\medskip
\textit{- Isotopic effect}
\medskip

While the global effect of an applied pressure will be presented in section~\ref{sec3-7-2-e. Coexistence versus competition between different ground states at low temperatur}, the local environment inside the TMTTF molecule, which may modify the long range Coulomb interactions, can be achieved by the exchange of hydrogen by deuterium in the methyl group of the TMTTF molecule.

\begin{figure}
\begin{center}
\includegraphics[width=7.5cm]{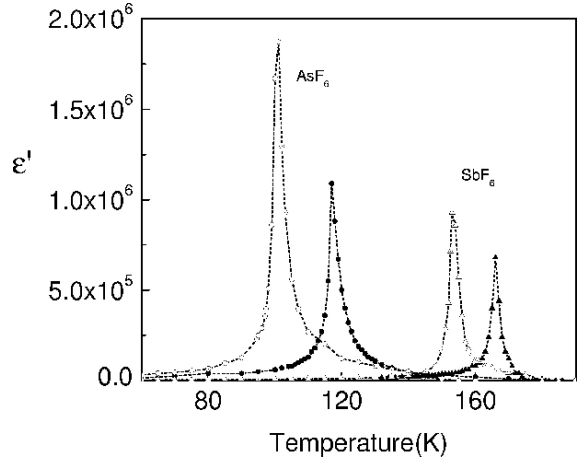}
\caption{Temperature dependence of the real part of the dielectric permittivity $\varepsilon'$ at frequency 100~kHz for deuterated (full circles) and hydrogenated circles with crosses) (TMTTF)$_2$AsF$_6$ and analogous dependencies for deuterated (full triangles) and hydrogenated (triangles with central dot) (TMTTF)$_2$SbF$_6$ (reprinted figure with permission from F. Nad \textit{et al.}, Journal of Physics: Condensed Matter 17, p. L399, 2005 \cite{Nad05}. Copyright (2005) by the Institute of Physics).}
\label{fig10-15}
\end{center}
\end{figure}

Figure \ref{fig10-15} 
shows the temperature dependences of $\varepsilon'(T)$ for deuterated and non-deuterated (TMTTF)$_2$AsF$_6$ and (TMTTF)$_2$SbF$_6$ samples \cite{Nad05}. The effect of deuteration leads to a considerable shift of the peaks in $\varepsilon'(T)$, signatures of the CO, to higher temperatures by 16~K for (TMTTF)$_2$AsF$_6$, 13~K for (TMTTF)$_2$SbF$_6$. The forms of $\varepsilon'(T)$ dependences near $T_{\rm CO}$ for deuterated samples are well described by the Curie-law, analogously to non-deuterated samples. In the case of deuterated (TMTTF)$_2$ReO$_4$, while the CO transition is increased by 7~K, the anion ordering transition $T_{\rm AO}$ is not affected by deuteration \cite{Nad05}. That favours the importance of long range Coulomb correlated interactions along molecular chains as the driving force for the CO transition. By X-ray measurements it was shown that the C-D bond length is shorter than that of C-H bond \cite{Furukawa05}. Deuteration thus plays the role of a ``positive" pressure effect, enhancing $T_{\rm CO}$. In addition the intermolecular distances $d_1$ and $d_2$ in the TMTTF molecule of (TMTTF)$_2$SbF$_6$ were measured \cite{Furukawa05,Nad05}. The dimerisation degree is evaluated from their ratio. This ratio increases with $T$ is reduced and was estimated to be 0.9855 at $T$~= 150~K. For deuterated (TMTTF)$_2$SbF$_6$, this ratio is large, near to 1 indicating that the dimerisation is strongly reduced by deuteration. The new synthesised (TMTTF)$_2$TaF$_6$ salt \cite{Iwase09} similarly to (TMTTF)$_2$SbF$_6$ exhibits CO at $T_{\rm CO}$~= 175~K and AFM transition at $T_{\rm N}\sim 9$~K.

\medskip
\textit{- Ferroelectricity in half-filled 1D charge-transfer compounds}
\medskip

The charge transfer (CT) complex are composed of mixed stacks of alternating donor ($D$) and acceptor ($A$) molecules \cite{Torrance81} (see section~\ref{sec2-16}). In the crystal the molecules are more or less electrically charged as $D^{+\rho}A^{-\rho}$ according to the degree of CT ($\rho$) within the stack. Fully ionic ($\rho$~= 1) compounds possess $S$~= 1/2 spin on each molecule, whereas the neutral ($\rho$~= 0) $DA$ stacks are diamagnetic \cite{Horiuchi08}.

The mixed-stack CT crystal, tetrathiafulvalene-p-chloranil (TTF-CA) [C$_6$H$_4$S$_4$-C$_6$Cl$_4$O$_2$] exhibits a ``neutral-ionic" (NI) phase transition at $T_c$~= 81~K. The original $D^{+\rho}A^{-\rho}D^{+\rho}A^{-\rho}$ ... sequence with a regular intermolecular separation is symmetry-broken to a polar chain by a lattice distortion forming dipolar $D^{+\rho'}A^{-\rho'}$ dimers. The CT degree $\rho$
 changes abruptly from $0.25\sim 0.3$ above $T_c$ at $\sim 0.65-0.7$ below $T_c$ indicating a first-order transition. Ferroelectricity accompanies this NI transition as proved by the loss of inversion centres that yields superlattice reflections \cite{Lecointe95} and by the  huge peak in the dielectric constant \cite{Okamoto91,Horiuchi08}. This ferroelectric behaviour has some similarities with that discussed on Fabre salts and even more on 2D (ET)$_2$X salts. The spatial distribution of the ferroelectric domains and their modification with application of an electric field was determined by microscopic electroreflectance spectroscopy \cite{Kishida09}.
 
\begin{figure}
\begin{center}
\includegraphics[width=8cm]{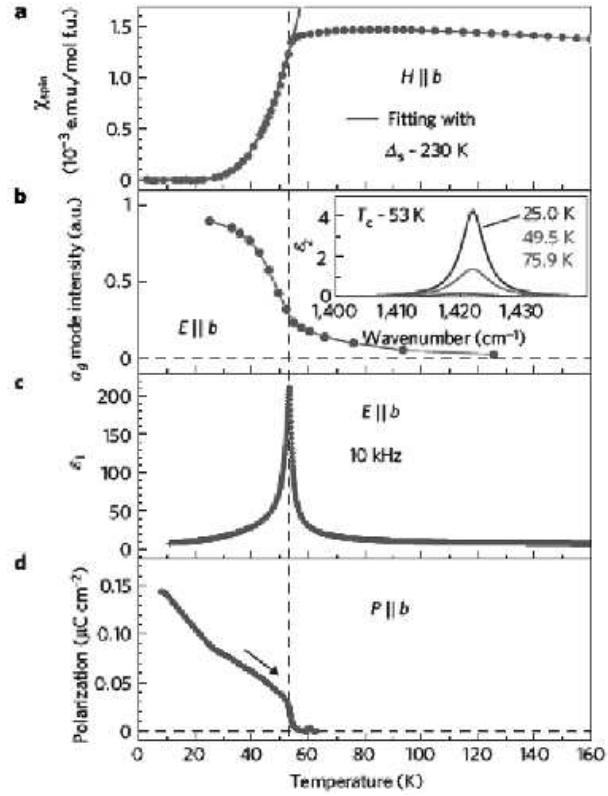}
\caption{Ferroelectric spin-Peierls in TTF-BA. Temperature dependence of a)~spin susceptibility with $\chi_{\rm spin}\sim\exp-\Delta_s/k_{\rm B}T$ below the spin-Peierls transition $T_{\rm SP}$~= 53~K, b)~normalised spatial weight of the $a_g$ mode as a measure of local $D^{+\rho}A^{-\rho}$ dimerisation, c)~dielectric constant at 10~kHz, d)~spontaneous polarisation (reprinted figure with permission from MacMillan Publishers Ltd: Nature Physics 6, p. 169, 2010 \cite{Kagawa10}. Copyright (2010)).}
\label{fig10-18}
\end{center}
\end{figure}

 Another type of transition has been observed in the 1:1 mixed-stack TTF-p-bromanil (TTF-BA) \cite{Tokura89}. The phase transition at $T$~= 53~K is characterised \cite{Garcia05} by the dimeric distortion of the initially equally distributed donor-acceptor stack similar to that observed in (TTF-CA). But the dimerisation does not change the high temperature ionic charge transfer $\rho\simeq 0.95$ \cite{Girlando85}. In this case the effective exchange interaction between spins on $D$ and $A$ molecules is antiferromagnetic. It is then well known that such an AF Heisenberg chain with $S$~= 1/2 is unstable towards a transition of spin-Peierls (SP) type involving a lattice dimerisation and going rise to a non-magnetic (spin-singlet) state of $D^{+\rho}A^{-\rho}$ pairs \cite{Bray83} (see section~\ref{sec2-8}). The SP state below $T_{\rm SP}$~= 53~K is accompanied by a ferroelectric behaviour \cite{Kagawa10}. The total polarisation was shown to be nearly along the $b$-axis, direction for which the dielectric constant shows a sharp peak. Electric field reversals of polarisation with the low frequency ac field applied along $b$ were also obtained as shown in figure~\ref{fig10-18}.
 
Furthermore by application of a huge magnetic field (pulsed fields up to 55~T), the dimerised ferroelectricity was shown to disappear \cite{Kagawa10} which occurs when the commensurate dimerisation corresponding to the singlet Peierls state is suppresses above a critical magnetic field. Because $T_{\rm SP}$ is large, the commensurate-incommensurate magnetic field given in section~\ref{sec9-3} as $H$~= 1.2~kT$_{\rm SP}$/g$\umu_{\rm B}$ is huge ($\sim 50$~T) in the case of (TTF-BA). Nevertheless the dependence of the polarisation on the singlet Peierls state may define (TTF-BA) as an organic multiferroic.

For (TTF-CA), the ferroelectric dimerised $D^{+\rho}A^{-\rho}$ stack exhibits degenerate polar ionic ground states such as:
 \begin{eqnarray*}
 I_+ \ldots \left(\underline{D^{+\rho}A^{-\rho}}\right)\quad\left(\underline{D^{+\rho}A^{-\rho}}\right) \quad \left(\underline{D^{+\rho}A^{-\rho}}\right) \ldots\\
 I_- \ldots \left(\underline{A^{-\rho}D^{+\rho}}\right) \quad \left(\underline{A^{-\rho}D^{+\rho}}\right)\quad\left(\underline{A^{-\rho}D^{+\rho}}\right) \ldots
 \end{eqnarray*}
 with opposite polarities of the domains. The domain walls at the boundary between segments should behave as mobile kink-type defects or solitons. The photo-induced paraelectric-to-ferroelectric phase transition was also directly observed \cite{Kagawa10a} by time-resolved X-ray diffraction. A 300 femtosecond laser pulse switches (TTF-CA) from the neutral to the ionic state on a 500 ps time scale. The X-ray data indicate a macroscopic ferroelectric reorganisation after the laser pulse \cite{Collet03}.

\medskip
\noindent \textit{3.7.2.d. Thermal expansion measurements}
\medskip

\begin{figure}
\begin{center}
\includegraphics[width=6.5cm]{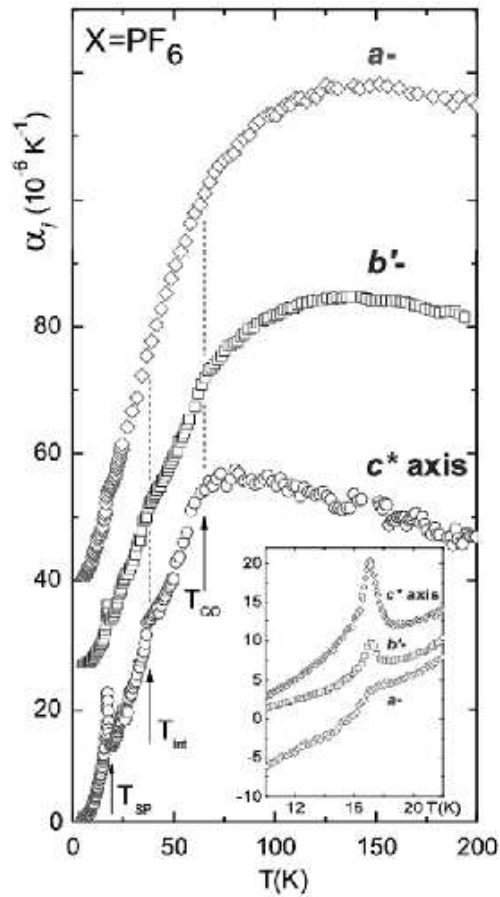}
\caption{Uniaxial expansion, $\alpha_i$, along three orthogonal axes of (TMTTF)$_2$PF$_6$. Arrows mark the spin-Peierls ($T_{\rm SP}$), the charge ordering ($T_{\rm CO}$) transition temperatures and another transition at $T_{\rm int}$. Inset shows details of the $\alpha_i$ anomalies at $T_{\rm SP}$ (reprinted figure with permission from M. de Souza \textit{et al.}, Physical Review Letters 101, p. 216403, 2008 \cite{Souza08}. Copyright (2008) by the American Physical Society).}
\label{fig10-13}
\end{center}
\end{figure}

The term ``structureless transition" was specifically applied in reference to (TMTTF)$_2$SbF$_6$ for which no structural effect was detected at the metal-insulating transition \cite{Coulon82b}, known, now, to be a charge order transition with a ferroelectric character. But ferroelectricity, breaking the inversion symmetry, should induce lattice effects. Those were obtained by the thermal expansion technique with a very high resolution ($\Delta\ell/\ell\sim 10^{-10}$) \cite{Souza08}. Uniaxial thermal expansion, $\alpha_i$, were measured along three orthogonal axis: the chain $a$ axis, $b'$ and $c^\ast$. At variance with ordinary lattice expansion, $\alpha_i(T)$ decrease when temperature increases above $T_{\rm CO}$, as shown in figure~\ref{fig10-13} 
for (TMTTF)$_2$PF$_6$, indicating the action of a negative contribution ascribed to rigid-unit modes of PF$_6$ anions coupled to the charge order \cite{Souza08}. The contribution of these  modes to the negative contribution to $\alpha_i$ was considered to be suppressed when those are frozen below $T_{\rm CO}$. Anomalies in $\alpha_i(T)$ are clearly observed at $T_{\rm CO}$, $T_{\rm SP}$ and also at an intermediate temperature $T_{\rm int}$ (it was noted \cite{Souza08} that the dielectric permittivity $\varepsilon'$ in a semilogarithmic scale shows some feature at the same temperature (see figure~3 in ref.~\cite{Souza08}); more work is needed to elucidate the nature of this anomaly). However the strongest effect occurs along the interstack $c^\ast$ with alternation of ($a,b$) planes of (TMTTF) molecules and of planes of anions (see structure of (TMTTF)$_2$X salts in figure~\ref{fig3-27}). This result reveals the intimate coupling between (TMTTF) molecules and anions and confirm the model of uniform shift of anions (transition at $q=0$) breaking the inversion symmetry with the consequence of ferroelectricity.

Neutron diffraction experiments on (TMTTF)$_2$PF$_6$ powder have determined an increase in the thermal variation of the intensity of several Bragg (210 and 201) reflections (typically $\pm 15$\%) at $T_{\rm CO}$ with respect to the regular extrapolated thermal evolution \cite{Foury10}. This structural effect has been analysed as resulting from the change of intermolecular bonds and the concomitant displacement of anions.

In the case of (TMTTF)$_2$SbF$_6$ the anomaly of $\alpha_{c^\ast}$ at $T_{\rm CO}$
 presents a $\lambda$-like negative shape \cite{Souza10} indicative of strong fluctuations and remarkably  different from the mean field like anomalies at $T_{\rm CO}$ in (TMTTF)$_2$PF$_6$ and (TMTTF)$_2$AsF$_6$ \cite{Souza08}. This difference was ascribed \cite{Souza10} to the enhanced screening in (TMTTF)$_2$SbF$_6$ for which $T_{\rm CO}$ coincides with a metal-insulator transition.
 
\medskip\vbox{
\noindent \textit{3.7.2.e. Coexistence versus competition between different ground states at low temperature}
\medskip

As discussed above, in addition to the $4k_{\rm F}$ transition into the CO state in the range of $\sim 100$~K, involving only charge degrees of freedom, there occurs at lower temperatures around $T\sim 10-20$~K another symmetry breaking transition involving spin degrees of freedom into either an antiferromagnetic AFM state or a spin-Peierls (SP) state. The study of the coexistence or the exclusion between CO and AFM/SP states has been the object of many works \cite{Kuwabara03,Clay07}.}

\medskip
\textit{- Charge redistribution}
\medskip

The amplitude of charge disproportionation measured by infrared spectroscopy was shown to be slightly reduced, but still finite in (TMTTF)$_2$AsF$_6$ below $T_{\rm SP}$ \cite{Dumm05}. $^{13}$C NMR investigations in the spin-Peierls state of (TMTTF)$_2$AsF$_6$ have brought contradictory results. First, it was stated that charge disproportionation between the two inequivalent TMTTF molecules is suppressed in the SP phase \cite{Fujiyama06} resulting from the observation that only two distinct NMR lines were visible in the NMR spectra and not four as above $T_{\rm SP}$. The same result was obtained \cite{Nakamura07}
for (TMTTF)$_2$PF$_6$. However, it was noted \cite{Zamborsky02} that the methods for determining the CO order parameter below $T_{\rm CO}$ (see figure~\ref{fig10-9}) do not work in the SP phase, because the paramagnetic shifts being nearly absent for the singlet ground state, all the $^{13}$C sites are equivalent. Only at higher magnetic field $B>B_c$ above the commensurate-incommensurate spin-Peierls transition (see section~\ref{sec9-3}), NMR absorption may reveal the contribution of different sites. It was then concluded \cite{Zamborsky02} that CD remains large in the SP ground state and consequently that the two orders --CO and SP-- coexist at low temperature.

\begin{figure}
\begin{center}
\includegraphics[width=7.5cm]{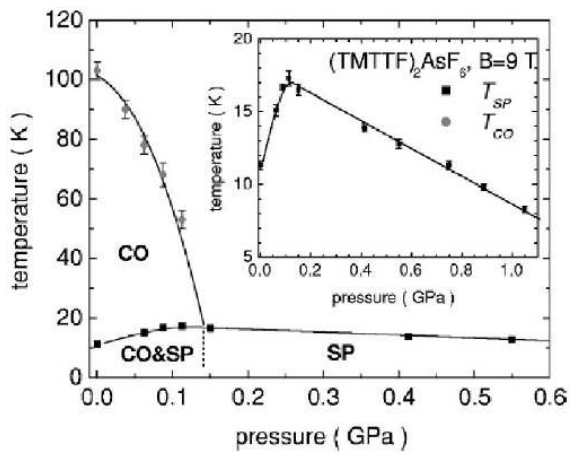}
\caption{Phase diagram of (TMTTF)$_2$AsF$_6$ under pressure showing coexistence between charge order and spin-Peierls ground states in the low pressure range (reprinted figure with permission from F. Zamborsky \textit{et al.}, Physical Review B 66, p. 081103(R), 2002 \cite{Zamborsky02}. Copyright (2002) by the American Physical Society).}
\label{fig10-16}
\end{center}
\end{figure}

CO is rapidly destabilised by application of a modest pressure. For (TMTTF)$_2$AsF$_6$, while $T_{\rm CO}$ is decreasing, $T_{\rm SP}$ increases significantly to about 150\% of its ambient pressure. Above $P$~= $P_C$~= 0.15~GPa, CO is suppressed. The maximum of $T_{\rm SP}$ at $P_C$ indicates that CO and SP states are competing (see figure~\ref{fig10-16}). 
It can thus be noted that higher $T_{\rm CO}$ is accompanied by lower $T_{\rm SP}$.

\medskip
\textit{- Case of (TMTTF)$_2$SbF$_6$}
\medskip

That was described above concerns, in fact only, among CSA samples, (TMTTF)$_2$PF$_6$ and (TMTTF)$_2$AsF$_6$. For (TMTTF)$_2$Br charge and magnetic orderings occur almost simultaneously and it is difficult to distinguish the specific effects on each order \cite{Coulon07}. Although SbF$_6$ is a CSA anion, the behaviour is quite different and (TMTTF)$_2$SbF$_6$ does not follow the general model. The resistivity is that of a metal down to 155~K without charge localisation, temperature at which appears the continuous ferroelectric-CO metal-insulator transition. (TMTTF)$_2$SbF$_6$ was shown to have the smallest degree of dimerisation in the transfer integral $t$. The large charge disproportionation in the CO state ($\Delta\rho\sim 0.5$) \cite{Yu04} results from the strong coupling to the anions. SF$_6$ is the largest monovalent anion and has the more compact packing in the ($b,c$) plane. The S-F distance (3.21~\AA) is smaller than the sum of the sulphur and fluorine van der Waals radii (3.27~\AA) \cite{Laversanne84}.

CO in (TMTTF)$_2$SbF$_6$ is specifically strong and the SP state is suppressed. The low temperature ground state is antiferromagnetic as identified by the peak in $^1$H spin relaxation rate at $T_{\rm N}$~= 7~K \cite{Yu04}. $T_{\rm CO}$ and $T_{\rm N}$ decrease under pressure. Above 0.5~GPa, the CO amplitude is no more detectable by NMR. At higher pressure, there is no evidence for the AF state ; instead from 2D NMR technique the spectra are characteristic of a singlet SP phase \cite{Yu04}. Then, not only CO tend to suppress the SP order and favours the ambient AF state when the CO state is strong, but as soon as CO is weakened by pressure, the SP state is re-established \cite{Yu04}.

Interestingly, it was also shown that the $^{19}$F NMR line broadens and shifts with the decrease of CO. This broadening resulting from slow anionic motions with respect to the time scale of the inverse of the NMR frequency was associated with dynamics of some methyl groups and $^1$H-$^{19}$F coupling. Then, the reduction or suppression of the anion motion leads to the destabilisation of the charge disproportionation on the donor molecules \cite{Yu04}.

\medskip
\textit{- Phase diagram}
\medskip

To account for the dependence of properties of Bechgaard-Fabre salts as a function of interstack interactions, a phase diagram Temperature-Pressure was proposed \cite{Jerome91}. The more 1D salts as (TMTTF)$_2$X were located on the left hand side exhibiting a Mott-Hubbard localisation at $T_\rho$ slightly below room temperature and a SP or AFM transition at low temperature. (TMTSF)$_2$X salts were on the right hand side with a metallic behaviour and a SDW phase transition at low $T$. Under pressure (TMTSF)$_2$PF$_6$ becomes a superconductor \cite{Jerome80}. However the essential role of anions was not included in this phase diagram as noted in the early times \cite{Brazovskii85}.

\begin{figure}
\begin{center}
\includegraphics[width=8cm]{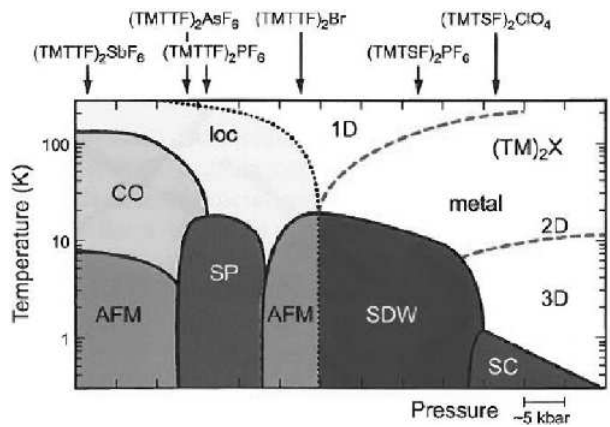}
\caption{Phase diagram of the Bechgaard-Fabre (TMTTF)$_2$X and (TMTSF)$_2$X salts first suggested in ref.~\cite{Jerome91} and further developed by including the charge order state (ref.~\cite{Yu04}). For the different compounds the ambient-pressure position in the phase diagram is indicated. Going from the left to the right, the materials are less one-dimensional (reprinted figure with permission from M. Dressel, Naturwissenschaften 94, p. 527, 2007 \cite{Dressel07}. Copyright (2007) with permission from Elsevier).}
\label{fig10-17}
\end{center}
\end{figure}

A revised phase diagram taking into account CO is shown in figure~\ref{fig10-17}. 
As said above, the large CO amplitude in (TMTTF)$_2$SbF$_6$ imposes the AFM ground state at low temperature, although the dimerisation in transfer integral along stacks is the smallest among Fabre salts. A question not yet solved concerns the nature of both AFM states located apart of the SP range. It has been proposed that the two AFM regions that straddle the SP phase have different charge occupancies: the same modulation pattern as the Wigner crystal that exists in the CO state in the case of (TMTTF)$_2$SbF$_6$ \cite{Clay07}, a redistribution of charge through a bond-charge-spin-density wave with the ...1100... charge pattern \cite{Clay03,Clay07} (1: charge rich, 0: charge poor) for the AFM region on the right hand side of the SP phase. 

By taking into account the lattice coupling in both TMTTF chains and anion chains, a very rich variety of phases was identified with mixed CDW-BOW states (BOW: bond order wave) \cite{Riera00} (see section~\ref{sec2-11}).

\medskip
\noindent \textit{3.7.2.f. Charge order in other systems}
\medskip

The importance to be attached to the slight dimerisation in the Fabre salts and consequently to the respective role of 1/4 and/or 1/2 Umklapp electron scattering processes in the electron localisation has continuously raised a long debate. It became however clear, as discussed at length above, that CO in Fabre salts results from large Coulomb interactions in a 1/4 filled band system. However, 1/4 filled-band organic compounds without dimerisation have been synthesised.

\;$\delta$-[EDT-TTF-CONMe$_2$]$_2$X\;\;\;(4,5-ethylenedithio-4'-(N,\;\,N-dimethylcarbamoyl)-tetrathiafulvalene) with $X$~= Br, AsF$_6$ are Q-1D systems (3/4-filled with electrons or 1/4-filled with holes) are Mott insulators at room temperature and exhibits AFM transition at $T_{\rm N}\simeq 8$~K \cite{Henze03}. In the Br sample a lattice distortion associated with an orthorhombic-monoclinic phase transition was identified below 190~K \cite{Zorina09}. The room temperature insulating state corresponds to a CO state as revealed by the doubling of the NMR $^{13}$C lines indicating two different sites which a charge ratio between them in the ratio 9:1 \cite{Auban-Senzier09,Zorina09}. The observation of superlattice reflections has made possible the refinement of the CO structure in $\delta$-(EDT-TTF-CONMe$_2$)$_2$Br using high resolution synchrotron radiation \cite{Zorina09}. It was concluded that, at room temperature, neutral and oxidised molecules alternate both along the stacking $a$ and transverse $b$ directions in orthorhombic non centro-symmetric space group $P_{2nn}$. The coupling between anion and cation sublattices was thus, again as for (TMTTF)$_2$X salts, demonstrated for the CO stabilisation, the bromide anion displacements exhibiting a static modulation along $b$, but with antiphase along $c$. It was also noted that in the CO state the stacks appear to be essentially uniform \cite{Zorina09}. Under pressure, both CO and AFM in $\delta$-(EDT-TTF-CONMe$_2$)$_2$AsF$_6$ are suppressed at $P_c\simeq 6$~GPa. Above $P_c$ the observed insulating state was ascribed to be a SDW before being destabilised at higher pressure, but without trace of superconductivity down to 70~mK  \cite{Auban-Senzier09}. In these 1/4 filled systems without dimerisation, the sequence of phases under pressure is then CO-density wave insulator-metal (Fermi liquid).

Another ideal 1/4 filled $\pi$-band without dimerisation was formed by the uniform stacking of DI-DCNQI (2,5-diido-dicyanoquinonediimine) molecules along the c-axis with monovalent Ag ions in between. (DI-DCNQI)$_2$Ag is a paramagnetic insulator below room temperature at ambient pressure and exhibits an AFM order at 5.5~K. $^{13}$C-NMR spectra have clearly demonstrated the inequivalence of sites below 220~K pointing out a $4k_{\rm F}$ Wigner crystal type of CO on DCNQI molecules with 3:1 charge disproportionation \cite{Hiraki98}. These results were confirmed by X-ray studies \cite{Nogami99} with the observation of $4k_{\rm F}$ satellite reflections at 30~K. The refinement of the structure in the CO state was performed using synchrotron radiation X-ray diffraction \cite{Kakiuchi07}. This structure exhibits three kinds of columns with different orders: in the first one, a $4k_{\rm F}$ CO but without lattice displacement, in the second, a $4k_{\rm F}$ bond order wave (BOW) with dimerisation, and in the third one a mixed state of CO and BOW. The origin of the 3 columns was estimated to be caused by the geometrical frustration of the (DCNQI)$_2$Ag structure resulting from the four fold screw symmetry \cite{Kakiuchi07}

\subsubsection{SDW amplitude}\label{sec3-5-2}

\begin{figure}
\begin{center}
\includegraphics[width=8cm]{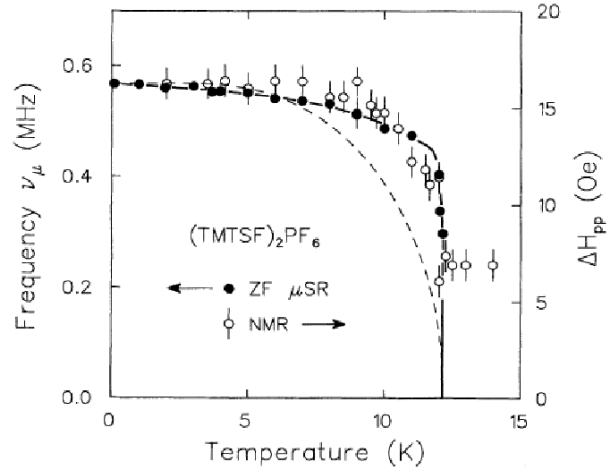}
\caption{Temperature dependence of the zero-field muon spin precession frequency $\nu_\mu$ (which is proportional to the SDW amplitude) observed in (TMTSF)$_2$PF$_6$. Proton-NMR results of the peak-to-peak width of absorption derivative $\Delta H_{\rm p.p.}(T)$ are also indicated. Dashed line: the energy gap in the mean field theory (reprinted figure with permission from L.P. Le \textit{et al.}, Physical Review B 48, p. 7284, 1993 \cite{Le93}. Copyright (1993) by the American Physical Society).}
\label{fig3-29}
\end{center}
\end{figure}

For (TMTSF)$_2$PF$_6$ and (TMTSF)$_2$AsF$_6$, the ground state is a SDW. From a precise analysis of $^1$H-NMR lineshapes the wave vector of the SDW of (TMTTF)$_2$PF$6$ was estimated \cite{Takahashi89,Takahashi86,Delrieu86} to be $Q$~= $0.5a^\ast,\,0.24\pm 0.03b^\ast,\,(-0.06\pm 0.20)c^\ast$. The amplitude of the SDW at low $T$ was also estimated to be 8\% $\umu_{\rm B}$/molecule. Muon-spin rotation ($\umu$SR) technique allows the determination of the field distribution without applied magnetic field. The temperature dependence of the SDW amplitude has been studied \cite{Le93} by deriving the frequency of the oscillation measured in the time spectra. Figure~\ref{fig3-29} 
shows the temperature dependence of the zero-field muon-spin precession frequency $\nu_\mu$, which is proportional to the SDW magnetisation, measured in (TMTSF)$_2$PF$_6$. The proton-NMR results from \cite{Takahashi86} are also plotted to comparison. The dashed line is the $T$-dependence of the SDW order parameter in the case of the mean field theory. Near $T_{\rm SDW}$, the transition does not follow the mean field behaviour, but the order parameter rapidly increases right below $T_{\rm SDW}$, indicating a nearly first-order transition.

In the case of (TMTTF)$_2$Br, the commensurate wave vector of the antiferromagnetic ground state was found \cite{Nakamura95} from $^1$H NMR studies to be:
\begin{equation*}
Q=(0.5a^\ast,\,0.25b^\ast,\,0c^\ast),
\end{equation*}
with the amplitude of the spin density 0.14~$\umu_{\rm B}$. The SDW and AF ground states of (TMTSF)$_2$PF$_6$, (TMTSF)$_2$AsF$_6$ and (TMTTF)$_2$Br were also studied by electron spin resonances \cite{Dumm00}.

\subsubsection{Magnetic structure in Bechgaard salts}

The SDW magnetic structure in Bechgaard salts has never been visualised by scattering measurements or by imaging. Additionally, CDW satellites were observed \cite{Pouget97,Kagoshima99} on (TMTSF$_2$)AsF$_6$ and (TMTSF)$_2$PF$_6$ by X-ray scattering below the SDW phase transition. The search for these satellite reflections was performed using elastic neutron scattering \cite{Danneau03}. The crystals were oriented in order to have the $2a^\ast+b^\ast,c^\ast$ axes of the majority twin in  the horizontal scattering plane of the spectrometer (IN14 at the La\"ue Langevin Institut). For d-(TMTSF)$_2$AsF$_6$ (1.5~mg) the reciprocal space volume covered was $0.46\leq q_H\leq 0.54$; $0.23\leq q_\kappa\leq 0.27$; $-0.50\leq q_L\leq +0.50$. Measurements were performed at 1.6~K with an incident neutron wave length of 4.8~$\AA$. The experiment allowed to rule out the presence of a low temperature CDW or SDW superlattice reflection with an intensity larger than $5\times 10^{-5}$ that of the (210) reflection. In the case of d-(TMTSF)$_2$PF$_6$ (3~mg) the twin volume ratio was more favourable and found to be around 8/1 

The best geometry to be considered for further experiments \cite{Currat03} would be to placed the sample in a horizontal field cryomagnet. Applying a magnetic field along the (210) direction will align the staggered SDW magnetic moments in a direction normal to (210). Since the (210) direction corresponds to the average direction of the scattering vector, this would optimise the SDW structure factor in the neutron measurements, a field of 1~T being sufficient to flip the spins from their (unknown) natural orientation to the orientation normal to the applied field. Nevertheless the sensitivity level ($5\times 10^{-5}$) in the best conditions in neutron experiments is short to the estimated value $\sim 10^{-6}$ with respect to the (210) intensity assuming that the magnetic moments are perpendicular to the wave vector $Q$. Thus, the experiment is not far to become successful, assuming an improvement of the signal/noise ratio, and/or the synthesis of a high crystallographic quality deuterated (TMTSF)$_2$PF$_6$ or (TMTSF)$_2$AsF$_6$ of at least 10~mg.

An alternative method would be to use polarised neutrons but with the overall loss in intensity of the order of a factor 5 or 10.

The other possibility would be to use spin-polarised scanning tunnelling microscopy (for a review see \cite{Wulfhekel07,Wiesendanger09}). This technique was shown to be very powerful for the determination of magnetic structures of antiferromagnetic thin films and especially chromium \cite{Hanke05} films with a direct imaging of spiral terraces \cite{Kawagoe05}. However, the tip characterisation is crucial for analysing spin-polarised measurements, and consequently the spin sensitivity of the tip. Spin polarisation of the tip was studied by coating the tungsten tip apex by a monocrystal chromium epitaxially grown in W \cite{Rodary11}. Atomic resolution has yet to be reached. In spite of difficulties for surface preparation (which have been resolved in the case of TTF-TCNQ \cite{Wang03}), the first magnetic structure in Bechgaard-Fabre salts would likely be the antiferromagnetic structure of (TMTTF)$_2$Br.

\subsubsection{Superconductivity under pressure}\label{sec3-5-3}

In Bechgaard salts, the nearly perfect nesting of the Fermi surface and the resulting divergence of the electronic susceptibility at $Q$~= $2k_{\rm F}$ are at the origin of the SDW phase transition. The modulated phase can be destabilised by application of pressure which modify the lattice constants and the transfer integrals. Thus the first organic superconductor was (TMTSF)$_2$PF$_6$ with $T_c$~= 0.9~K under a pressure 1.1~GPa \cite{Jerome80}. Anion ordering is also important for superconductivity. When the anion is non-centrosymmetric such as ClO$^-_4$, it can take the orientation of one of the two possible directions at higher temperature. When (TMTSF)$_2$ClO$_4$ is cooled rapidly ($\sim$~6~K/s) through the anion ordering temperature $T_{\rm AO}$~= 24~K, anion orientations are frozen randomly in the two specific orientations and in this quenched state, the low $T$ ground state is a SDW. But, when it is slowly cooled (0.1~K/mn) anion ordering takes place with the wave vector (0,~1/2,~0) and in this relaxed state, the low $T$ ground is superconducting with $T_c$~= 1.2~K \cite{Bechgaard81}.

In Fabre salts, much higher pressure is needed to observe superconductivity. Then (TMTTF)$_2$Br was found superconducting with a maximum $T_c$~= 0.8~K at $P$~= 2.6~GPa \cite{Balicas94}. Application of pressure on (TMTTF)$_2$PF$_6$ first suppresses the spin-Peierls state into an AF or SDW ground state stable up to around 4~GPa \cite{Jaccard01}. At higher pressure, (TMTTF)$_2$PF$_6$ is superconductor \cite{Adachi00,Adachi01,Jaccard01} with a maximum $T_c$~= 2.2~K at $P$~= 4.73~GPa. Similarly (TMTTF)$_2$AsF$_6$ \cite{Itoi07} has a maximum $T_c$~= 2.64~K at 5~GPa, (TMTTF)$_2$BF$_4$ \cite{Auban03} is superconducting with $T_c\sim 1.38$~K in the pressure range 3.35--3.75~GPa. In the case of (TMTTF)$_2$SbF$_6$, the superconducting phase with $T_c$~= 2.8~K exists \cite{Itoi08} in a wide pressure range with two local maxima at 6 and 9~GPa.

While $T_c$ in (TMTTF)$_2$X salts is $\sim 2$ times higher than $T_c$ for Bechgaard salts, it is still in the range of 2.2--2.6~K. The present highest $T_c$ among organic materials \cite{Taniguchi03} is 14.2~K under the pressure of 8~GPa in the two-dimensional compound $\beta^\prime$-(BEDT-TTF)$_2$ICl$_2$.

\subsubsection{Coexistence between superconductivity and C/S DWs}\label{sec3-5-4}

The most remarkable feature of phase diagrams derived from the temperature dependence of the C/S DW transitions and the occurrence of superconductivity is the reentrance of the superconducting phase in the modulated one. This coexistence is a general property in strongly correlated systems and has also been found in high $T_c$ superconductors and in heavy fermions.

\begin{figure}
\begin{center}
\subfigure[]{\label{fig3-30a}
\includegraphics[width=7cm]{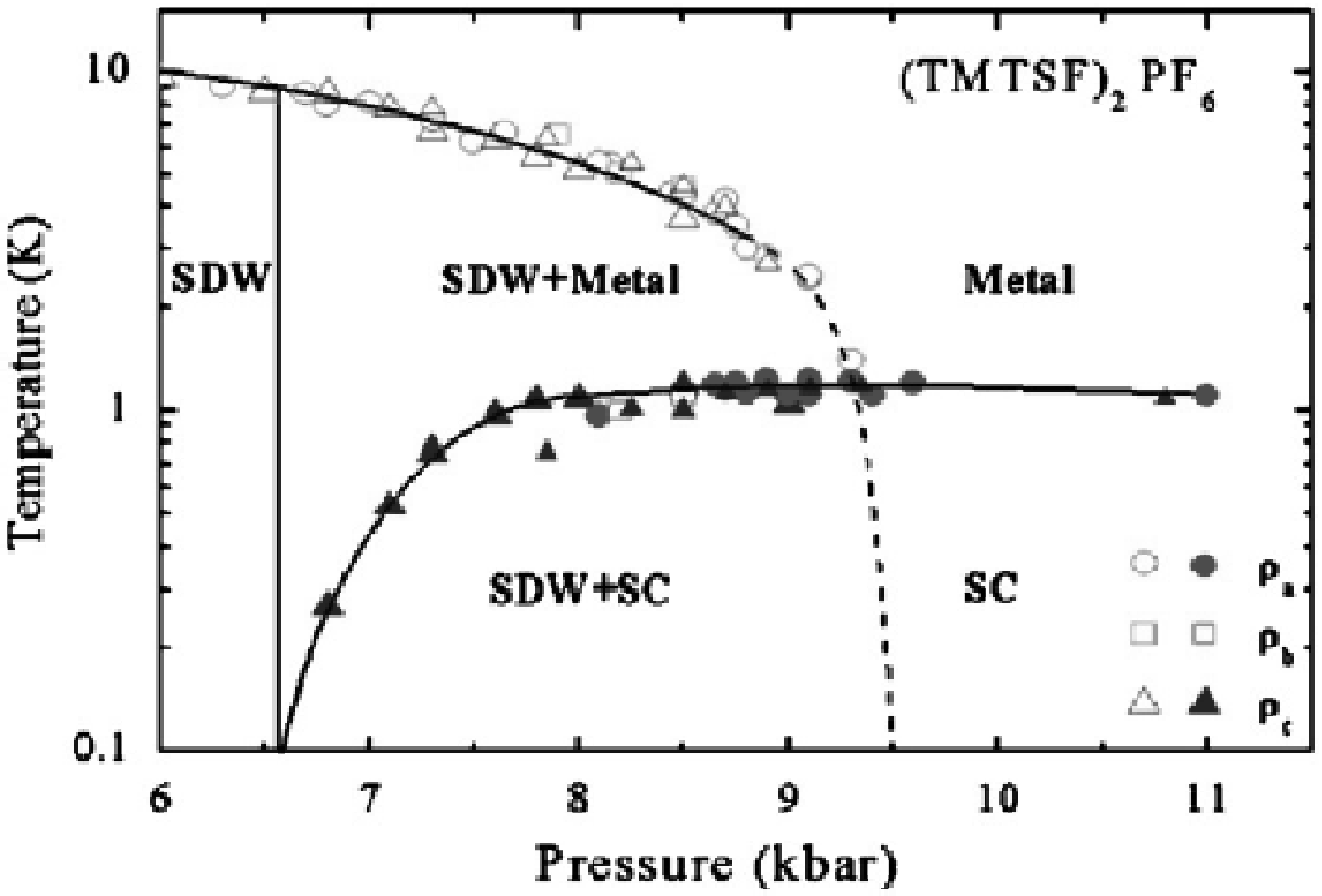}}
\subfigure[]{\label{fig3-30b}
\includegraphics[width=6.25cm]{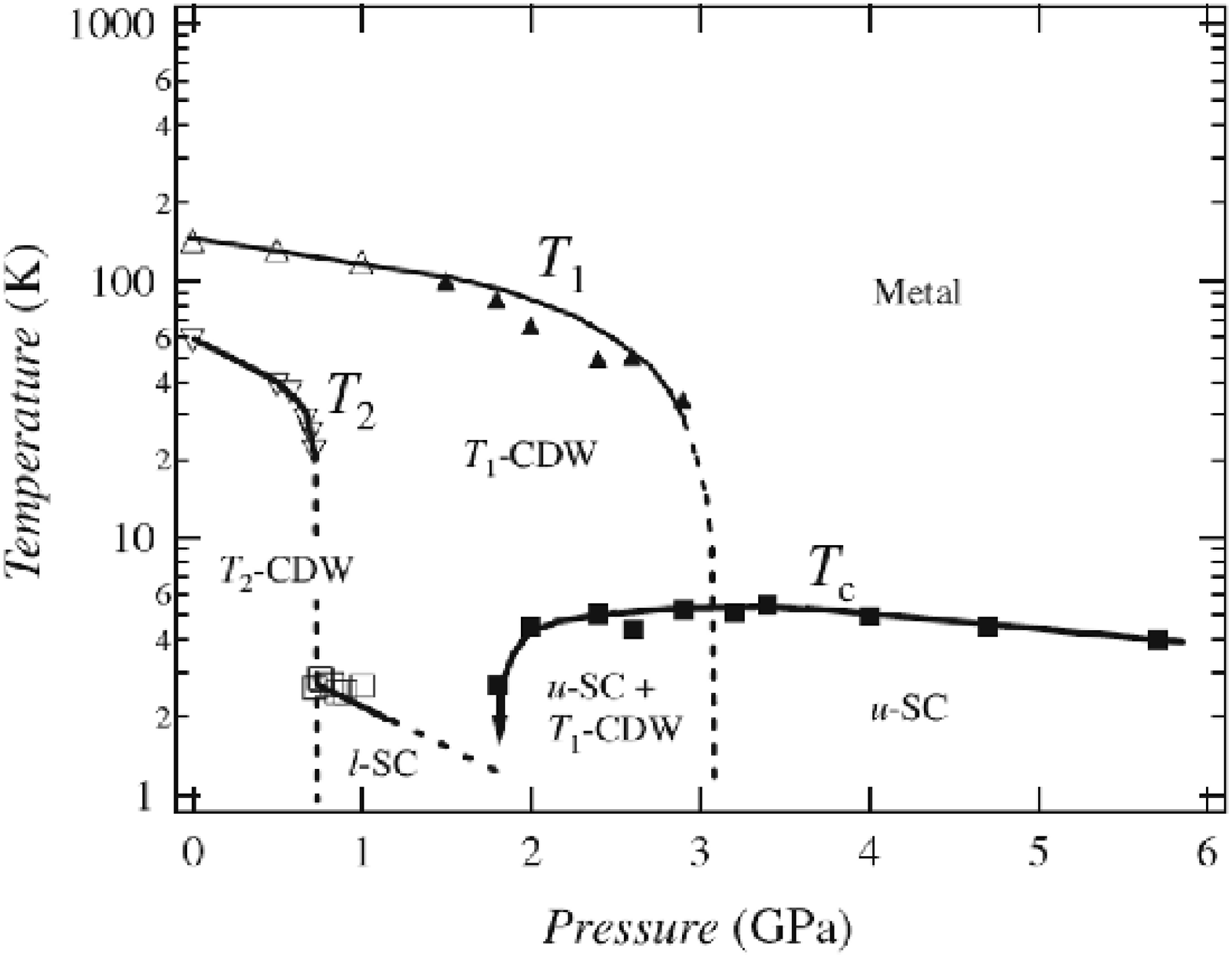}}
\caption{Phase diagram pressure-temperature of a)~(TMTSF)$_2$PF$_6$ (reprinted figure with permission from Physica B 404, B. Salameh \textit{et al.}, p. 476, 2009 \cite{Salameh09}. Copyright (2009) with permission from Elsevier) as determined from resistivity measurements along the three axes $a$, $b'$, and $c^\ast$, b)~NbSe$_3$. $T_1$-CDW: upper CDW, $T_2$-CDW: lower CDW, $\ell$-Sc: superconductivity when $T_2$-CDW is suppressed, $u$-Sc: superconductivity at higher pressure when $T_1$-CDW is suppressed (reprinted figure with permission from Journal of the Physical Society of Japon 74, p. 1782, 2006, S. Yasuzuka \textit{et al.} \cite{Yasuzuka05}).}
\label{fig3-30}
\end{center}
\end{figure}

Figure~\ref{fig3-30} shows the phase diagram of (TMTSF)$_2$PF$_6$ at low temperature \cite{Salameh09,Vuletic02,Kang10} between 0.6 and 1.1~GPa and that of NbSe$_3$ for both CDWs transitions \cite{Yasuzuka05} up to 6~GPa. With the use of the cubic-anvil device which minimises the pressure inhomogeneity, it has been observed that the $T_1$-CDW and the u-superconductivity [u for upper superconductivity with $T_c$~= 5~K, nearly two times higher than $T_c$~= 2.3~K for $\ell$-superconductivity ($\ell$ for lower) when only the $T_2$-CDW is suppressed] coexist between 2.0 and 2.9~GPa. The intrinsic nature of the u-SC phase transition coexisting with the $T_1$-CDW phase is detected by the very broad u-SC transition at 2.6~GPa (see figure~\ref{fig3-17}) while at higher pressure the transition width becomes much sharper. Phase separation between u-SC domains and CDW domains was suggested, the average size of u-SC domains increasing with pressure. Quantum tunnelling of Cooper pairs between the u-SC domains separated by the CDW domains was speculated \cite{Yasuzuka05}.

Inhomogeneity of the SC phase in (TMTSF)$_2$PF$_6$ was studied by measurements of resistivity $\rho_a$, $\rho_b'$, and $\rho_c^\ast$ along the three axis $a$, $b'$ and $c^\ast$. In the pressure range between 0.6 and 0.94~GPa, several phases were identified \cite{Kang10}. Just above 0.6~GPa the onset of superconductivity occurs first along $c^\ast$ while the resistivity along the perpendicular directions is still insulating. At slightly higher pressure, SDW and SC domains were found surprisingly in series along $b'$, i.e superconductivity slabs are perpendicular to the most conducting $a$-axis. Above 0.94~GPa the superconductivity is homogeneous.

Structural data of quasi-1D materials quoted in this review, including space group and unit cell parameters are collected in table~\ref{tab3-2}. 
\begin{sidewaystable}
\tbl{Structural data for quasi one-dimensional materials quoted in this review}
{\begin{tabular}{llllllllll}\toprule
\multicolumn{4}{l}{} & \multicolumn{4}{l}{Unit cell parameters} & \multicolumn{2}{l}{} \\ 
\multicolumn{4}{l}{} & \multicolumn{4}{l}{---------------------------------------------------------------------} & \multicolumn{2}{l}{} \\
& & Symmetry & Space group & $a(\AA)$ & $b(\AA)$ & $c(\AA)$ & $\beta(^\circ)$ & $\begin{array}{l} \mbox{Chains} \\ \mbox{per unit cell} \end{array}$ & References\\ \hline
&&&&&&&&&\\
\multirow{2}{0.5cm}{\vspace*{-0.35cm} NbS$_3$} & $\left\{ \begin{array}{l} \mbox{type I} \\  \\ \mbox{type II} \end{array}\right.$ &
$\begin{array}{l} \mbox{triclinic} \\ \\ \mbox{monoclinic} \end{array}$ &
$\begin{array}{l} P\bar{1} \\  \\ \\ \end{array}$ &
$\begin{array}{l} ~4.963 \\ \\ ~9.9\end{array}$ &
$\begin{array}{l} 2\times 3.365 \\ \\ ~3.4 \end{array}$ &
$\begin{array}{l} ~9.144 \\ \\ 18.3 \end{array}$ &
$\begin{array} {l}
\begin{array}{c}\vspace*{-8pt}\left\{ \begin{array}{l} 97.17 \\ \alpha=\gamma=90 \end{array}\right. \end{array} \\  \\ ~~~~97.2 \end{array}$ &
$\begin{array}{l} (2\times\underline{1}) \\  \\ 8 \end{array}$ & 
$\begin{array}{l} \mbox{\cite{Rijnsdorp78}} \\ \\ \mbox{\cite{Roucau83}} \end{array}$\\
&&&&&&&&&\\
TaSe$_3$ & & monoclinic & $P2_{1/m}$ & 10.402 & ~3.495 & ~9.829 & ~~~106.26 & $2\times\underline{2}$ & \cite{Bjerkelund66} \\
&&&&&&&&&\\
NbSe$_3$ & & monoclinic & $P2_{1/m}$ & 10.006  & ~3.478 & 15.626 & ~~~109.30 & $2\times\underline{3}$ & \cite{Meerschaut75,Hodeau78} \\
&&&&&&&&&\\
TaS$_3$ & & \hspace*{-0.35cm} $\left\{ \begin{array}{l} \mbox{monoclinic} \\ \mbox{orthorhombic} \end{array}\right.$ & $\begin{array}{l} P2_{1/m} \\ C222_1 \end{array}$ & $\begin{array}{l} ~9.515 \\ 36.804 \end{array}$ & $\begin{array}{l} ~3.3412 \\ ~3.34 \end{array}$ & $\begin{array}{l} 14.912 \\ 15.173 \end{array}$ & $\begin{array}{l} ~~~109.99 \\ \\ \end{array}$ & $\begin{array}{l} 2\times\underline{3} \\ 24 \end{array}$ & $\begin{array}{l} \mbox{\cite{Meerschaut81a}} \\ \mbox{\cite{Bjerkelund64}} \end{array}$ \\
&&&&&&&&&\\
\multicolumn{2}{l}{(Fe$_{1+x}$Nb$_{1-x}$)Nb$_2$Se$_{10}$} & monoclinic & $P2_{1/m}$ & ~9.213 & ~3.4382 & 10.292 & ~~~114.46 & & \cite{Cava81,Meerschaut81a}\\
&&&&&&&&&\\
ZrTe$_3$ & & monoclinic & $P2_{1/m}$ & ~5.89 & ~3.93 & 10.09 & ~~~~97.8 & 2 & \cite{Furuseth91} \\
&&&&&&&&&\\
(TaSe$_4$)$_2$I & & tetragonal & $I422$ & ~9.531 & & 12.824 & & $2\times\underline{1}$ & \cite{Gressier82} \\
&&&&&&&&&\\
(NbSe$_4$)$_{10}$I$_3$ & & tetragonal & $P4/mcc$ & ~9.461 & & 31.91 & & $2\times\underline{1}$ & \cite{Meerschaut84} \\
&&&&&&&&&\\
(NbSe$_4$)$_3$I & & tetragonal & $P4/mnc$ & ~9.489 & & 19.13 & & $2\times\underline{1}$ & \cite{Meerschaut77} \\
&&&&&&&&&\\
TaTe$_4$ & & tetragonal & $P4cc$ & $2\times 6.513$ & & $3\times 6.812$ & & & \cite{Selte64,Boswell83} \\
&&&&&&&&&\\
NbTe$_4$ & & tetragonal & $P4cc\mbox{ or }P4/mcc$ & ~6.496 & & ~6.823 & & & \cite{Selte64,Bohm99} \\
&&&&&&&&&\\
ZrTe$_3$ & & orthorhombic & $C_{mcm}$ & ~3.988 & 14.502 & 13.727 & & $2\times\underline{2}$ & \cite{Furuseth73,Furuseth91} \\
&&&&&&&&&\\
K$_{0.3}$MoO$_3$ & & monoclinic & $C2/m$ & 18.25 & ~7.56 & ~9.855 & ~~~117.53 & & \cite{Ghedira85,Graham66} \\
&&&&&&&&&\\
Rb$_{0.3}$MoO$_3$ & & monoclinic & $C2/m$ & 18.94 & ~7.56 & 10.040 & ~~~118.83 & & \cite{Ghedira85} \\
&&&&&&&&&\\
(Per)$_2$M(mnt)$_2$ & & monoclinic & $P2_{1/m}$ & 16.612 & ~4.194  & 30.211 & ~~~118.10 & 2 & \cite{Alcacer80} \\
&&&&&&&&&\\
(FA)$_2$PF$_6$  & &   \hspace*{-0.35cm} $\left\{ \begin{array}{l} \mbox{room temperature} \\  T<200~{\rm K} \end{array} \right. $ & $\begin{array}{l} A2/m \\ P2_{1/c} \end{array}$  &$\begin{array}{l} ~3.3 \\ \\ \end{array}$ & $\begin{array}{l} 12.57 \\  \\ \end{array}$ & $\begin{array}{l} 14.77  \\ \\  \end{array}$ & & $\begin{array}{l} 2 \\ \\  \end{array}$ & $\begin{array}{l} \mbox{\cite{Brutting92}} \\ \\  \end{array}$ \\
&&&&&&&&&\\ 
(TMTTF)$_2$Br & & triclinic & $P\bar{1}$ & ~7.034 & ~7.358 & 12.622 & $\begin{array}{l} \alpha=90.09 \\ \beta=93.12 \\ \gamma=109.07 \end{array}$ & 1 & \cite{Galigne78} \\
&&&&&&&&&\\ 
(TMTSF)$_2$PF$_6$ & & triclinic & $P\bar{1}$ & ~7.297 & ~7.711 & 13.522 & $\begin{array}{l} \alpha=83.39 \\ \beta=86.27 \\ \gamma=71.01 \end{array}$ & 1 & \cite{Thorup81} \\
\botrule
\end{tabular}}
\label{tab3-2}
\end{sidewaystable}
Data of room temperature resistivity, C/SDW transition temperature, $T_p$, gap in the low temperature modulated structure, ratio $2\Delta/kT_{\rm P}$ components of the superstructure in the  reciprocal space, superconducting transition temperature are collected in table~\ref{tab3-3}. 
\begin{sidewaystable}

\tbl{\hspace{0.25cm}Data concerning Peierls or superconducting transitions in quasi-1D materials quoted in this review. Listed are: the resistivity at room temperature, the Peierls or the superconducting temperature, the amplitude of the Peierls gap and its ratio with the Peierls temperature, the nature of the ground state at low temperatures and the components of the superlattice structure on the reciprocal axes.}
{\hspace{1cm}
{\begin{tabular}{lllllllllllllll}\toprule
& & 
\multicolumn{2}{l}{$\begin{array}{l} \rho(\Omega\times{\rm cm})\mbox{ at} \\ \mbox{room temperature} \end{array}$}
& \multicolumn{2}{l}{$\begin{array}{l} \mbox{Peierls} \\ \mbox{temperature} \\ {\rm (K)} \end{array}$}
& $\begin{array}{l} 2\Delta \\ {\rm (K)} \end{array}$ 
& $\displaystyle\frac{2\Delta}{kT_c}$ 
& \multicolumn{4}{l}{$\begin{array}{c} \mbox{Surstructure}\\ $---------------------------------------------------------$ \\ {\hspace*{-2.25cm}\rm a*}\quad\quad {\hspace*{0.35cm}\rm b*}\qquad\quad\hspace*{0.15cm}{\rm c*}\quad \end{array}$}
& \multicolumn{2}{l}{$\begin{array}{l} \mbox{Superconducting} \\ \mbox{temperature} \end{array}$} 
& $\begin{array}{l} \mbox{Ground state at} \\ \mbox{low temperature} \end{array}$ \\ \hline
NbS$_3$ $\left\{ \begin{array}{l} \mbox{type I} \\  \\ \mbox{type II} \end{array}\right.$ && $\begin{array}{l} 80  \\ \\ 8\times 10^{-2} \end{array}$ && $\begin{array}{l} \\ \\ 330 \end{array}$  && $\begin{array}{l} \\ \\ 4400 \end{array}$ & $\begin{array}{l} \\ \\ 13.3 \end{array}$ & $\begin{array}{l} \\ \\ \left\{\begin{array}{l} 0.5 \\ 0.5 \end{array}\right. \end{array}$ & $\begin{array}{l} \\ \\ \begin{array}{l} 0.298 \\ 0.352 \end{array} \end{array}$ & $\begin{array}{l} \\ \\ \begin{array}{l} 0 \\ 0 \end{array} \end{array}$ & $\begin{array}{l}\\ \\  \mbox{\cite{Roucau83}} \end{array}$ & & & $\begin{array}{l}  \mbox{Insulating} \\ \\  \mbox{Semiconducting} \end{array}$ \\
&&&&&&&&&&&&&&\\
TaSe$_3$ && $6\times 10^{-4}$ & & & & & & & & & & 2.2 & \cite{Sambongi77b} & Superconducting \\
&&&&&&&&&&&&&&\\
NbSe$_3$ && $2.5\times 10^{-4}$ & & $\hspace*{-0.25cm}\left\{\begin{array}{r} 145 \\ 59 \end{array}\right. $ & & $\begin{array}{l} \\  ~700 \end{array}$ & $\begin{array}{l} \\ 11.9 \end{array}$ & \hspace*{0.25cm}$\begin{array}{l} 0 \\ 0.5 \end{array}$ & $\begin{array}{l} 0.24117 \\ 0.26038 \end{array}$ & $\begin{array}{l} 0 \\ 0.5 \end{array}$ & \cite{Fleming78} & $\begin{array}{l} 3.5\mbox{ under} \\ \mbox{0.55~GPa} \end{array}$ & \cite{Nunez93} & $\begin{array}{l} \mbox{Metallic} \\ \end{array}$ \\
&&&&&&&&&&&&&&\\
TaS$_3$ $\left\{\begin{array}{l} \mbox{orthorhombic} \\ \\ \\\ \mbox{monoclinic} \end{array}\right. $ & & $\begin{array}{l} 3.2\times 10^{-4}  \\ \\  \\ 3\times 10^{-4} \end{array}$ & & $\begin{array}{l}  \raisebox{-2ex}{215}\\  \\ \\  \hspace*{-0.25cm} \left\{  \begin{array}{l} 240   \\160 \end{array}\right. \end{array}$ & & $\begin{array}{l} 1600  \\ \\  \\ 1900 \end{array}$ & $\begin{array}{l}~7.44 \\ \\  \\ 11.9 \end{array}$ & $\begin{array}{l} \left\{ \begin{array}{l} ? \\ 0.5 \end{array}\right. \\ \\ \left\{\begin{array}{l} 0 \\ 0.5 \end{array}\right. \end{array}$ & $\begin{array}{l} \begin{array}{l} 0.1 \\  0.125  \end{array} \\ \\ \begin{array}{l} 0.253 \\ 0.247 \end{array} \end{array}$ & $\begin{array}{l} \begin{array}{l} 0.255 \\ 0.250~(T< 130\mbox{~K})   \end{array} \\ \\ \begin{array}{l} 0 \\ 0.5 \end{array} \end{array}$  & $\begin{array}{l} \mbox{\cite{Wang83c}} \\  \\ \\ \mbox{\cite{Roucau80}} \end{array}$ & & $\begin{array}{l} \mbox{\cite{Nunez93}} \\ \\ \\ \\  \end{array}$ & $\begin{array}{l} \mbox{Semiconducting} \\   \\ \\ \mbox{Semiconducting} \end{array}$ \\
&&&&&&&&&&&&&&\\
\multicolumn{2}{l}{(Fe$_{1+x}$Nb$_{1-x}$Nb$_2$Se$_{10}$} & $10^{-3}$ & & \hspace*{-0.25cm}$\sim 140$ & & ~~360 & ~~2.55 & ~$\left\{\begin{array}{l} 0 \\ 0.5 \end{array}\right.$ & ~$\begin{array}{l} 0.27 \\ 0.33 \end{array}$ & ~$\begin{array}{l} 0 \\ \ell \end{array}$ & ~\cite{Hillenius81} & & & ~Semiconducting \\
&&&&&&&&&&&&&&\\
ZrTe$_3$ & & $2.5\times 10^{-4}$ & \cite{Takahashi84} & ~~63 & & & & ~~~~0.07 & ~~0 & ~0.333 & ~\cite{Eaglesham84} & $T<2$~K & & ~Superconducting \\
&&&&&&&&&&&&&&\\
(TaSe$_4$)$_2$I && $1.5\times 10^{-3}$ & & ~263 & & ~3000 & ~11.4 & ~~~~0.05 & ~~0.05 & ~0.084 & ~\cite{Roucau84} & & & ~Semiconducting \\
&&&&&&&&&&&&&&\\
(NbSe$_4$)$_{10}$I$_3$ & & $1.5\times 10^{-2}$ & & ~285 & & ~3900 & ~13.7 & ~~~~0 & ~~0 & ~0.487 & ~\cite{Roucau84} & & & ~Semiconducting \\
&&&&&&&&&&&&&&\\
NbTe$_4$& & $1.2\times 10^{-4}$ & & & & & & ~$\left\{\begin{array}{l} 0.5 \\ 0 \\ 0.5 \end{array}\right.$ & ~$\begin{array}{l} 0.5 \\ 0 \\ 0 \end{array}$ &  ~$\begin{array}{l} 0.344 \\ 0.311 \\ 0.333 \end{array}$ & ~$\begin{array}{l} \mbox{\cite{Mahy83}} \\ \mbox{\cite{R12Boswell99}} \end{array}$ & & & ~Metallic \\
&&&&&&&&&&&&&&\\
TaTe$_4$ & & $1.2\times 10^{-4}$ & & & & & & ~$\left\{\begin{array}{l} 0.5 \\ 0 \\ 0.5 \end{array}\right.$ & ~$\begin{array}{l} 0.5 \\ 0 \\ 0 \end{array}$ &  ~$\begin{array}{l} 0.333 \\ 0.333 \\ 0.333 \end{array}$ & & & &  ~Metallic \\
&&&&&&&&&&&&&&\\
K$_{0.3}$MoO$_3$ & & $5\times 10^{-3}$ & \cite{R10Schlenker96} & ~180 & \cite{Dumas93} & & ~~~~9 & ~~~~0 & $\begin{array}{l} \\ \hspace*{0.15cm}0.70 \\ \hspace*{-0.15cm}\sim 0.75 \end{array}$ &  $\begin{array}{l} \\ ~0.5 \\ \hspace*{-0.45cm}\mbox{($T<100$~K)} \end{array}$ & ~~\cite{Pouget85} & & & ~CDW Semiconducting\\
&&&&&&&&&&&&&&\\
(Per)$_2$Pt(mnt)$_2$  & & $1.3\times 10^{-3}$ &  & ~~~~8 & \cite{Gama93b} & 8.6~meV & & ~~~~0 & ~~0.375 & ~~0 & ~~\cite{Canadell04} &  & & ~Spin-Peierls \\
&&&&&&&&&&&&&&\\
(Per)$_2$Au(mnt)$_2$  & & &  & ~~~12 & \cite{Gama93b} & 3.5~meV & & & & & &  & & ~CDW \\
&&&&&&&&&&&&&&\\
(FA)$_2$PF$_6$  & & $10^{-3}$&  \cite{Riess93} & ~~179 & & &  &~~~~0.5 & ~~~0 & ~~~0 & ~~\cite{Ilakovac93} & &  &  ~CDW \\
&&&&&&&&&&&&&&\\
(TMTSF)$_2$PF$_6$  & & & & ~~~12 & & &  &~~~~0.5 & ~~~0.24 & ~~-0.06 & $\left\{\begin{array}{l} \mbox{\cite{Takahashi86}} \\ \mbox{\cite{Takahashi89}} \end{array}\right.$ & $\begin{array}{l} T_c=0.9~{\rm K} \\ P_c=0.9~{\rm GPa} \end{array}$ & ~~$\begin{array}{l} \mbox{\cite{Jerome80}} \\ \\ \end{array}$ &  ~SDW \\
&&&&&&&&&&&&&&\\
(TMTSF)$_2$Br$_6$  & & & & ~~~10 & & &  &~~~~0.5 & ~~~0.25 & ~~~0 & ~~~\cite{Nakamura95} & $\begin{array}{l} T_c=0.8~{\rm K} \\ P_c=2.6~{\rm GPa} \end{array}$ & ~~~\cite{Balicas94} &  ~SDW \\
&&&&&&&&&&&&&&\\
\botrule
\end{tabular}}
}
\label{tab3-3}
\end{sidewaystable}

\section{Properties of the sliding density wave}\label{sec4}
\setcounter{figure}{0}
\setcounter{equation}{0}

Non-linear collective transport properties below charge and spin density wave (C/S  DW) transition temperatures have been observed in several types of inorganic as well as organic compounds, however formed only with chain structures, namely for CDW: transition metal trichalcogenides as NbSe$_3$, TaS$_3$, NbS$_3$, halogened transition metal tetrachalcogenides as (TaSe$_4$)$_2$I, (NbSe$_4$)$_2$I, (NbSe$_4$)$_{10}$I$_3$, molybdenum oxides --blue bronze K$_{0.30}$MoO$_3$ and Rb$_{0.30}$MoO$_3$--, organic compounds TTF-TCNQ, perylene and (FA)$_2$X compounds, for SDW: essentially in Bechgaard (TMTSF)$_2$X salts with X~= ClO$_4$, AsF$_6$, PF$_6$, NO$_3$.

\subsection{General properties}\label{sec4-1}

The general properties of this current-carrying state can be summarised as follows:
\begin{itemize}
\item[-] The dc electrical conductivity increases above a threshold electric field $E_T$. As shown in figure~\ref{fig4-1}, 
\begin{figure}
\begin{center}
\includegraphics[width=5cm]{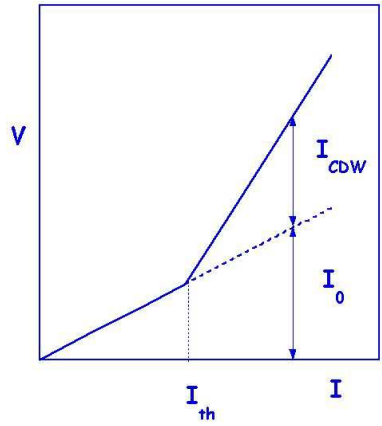}
\caption{Schematic I-V characteristic in the non-linear state of a C/S DW. Above the threshold voltage $V_T$, the total current $I$~= $I_{\rm n}+I_{\rm CDW}$, $I_{\rm n}$: the current carried by normal carriers; $I_{\rm CDW}$: the extra C/S DW current.}
\label{fig4-1}
\end{center}
\end{figure}
in the non-linear state the total current can be written as:
\begin{equation*}
I=I_n+I_{\rm C/S\,DW},
\end{equation*}
where $I_n$ corresponds to the current carried by carriers not condensed below the C/S DW gap and $I_{\rm C/S\,DW}$ the extra current delivered by the C/S DW motion \cite{Monceau76,Fleming79,Monceau80}.

\smallskip
\item[-] The conductivity is strongly frequency-dependent in the range of a few MHz till 10--100~GHz \cite{Ong77}. The electrodynamics of C/S DW has been studied in a very large frequency range \cite{Donovan94,Degiorgi91a,Degiorgi91b,Kim91}. Figure~\ref{fig4-2} 
\begin{figure}[h!]
\begin{center}
\includegraphics[width=7.5cm]{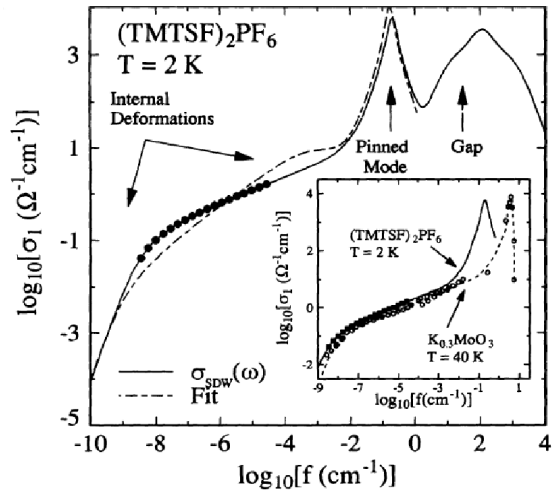}
\caption{Frequency dependence of the real part of the SDW conductivity, $\sigma$, in a log-log plot of (TMTSF)$_2$PF$_6$ at $T$~= 2~K exhibiting features due to the contribution of the internal deformations, the pinned mode and the gap. Inset shows the comparison of the SDW conduction of (TMTSF)$_2$PF$_6$ with the CDW conductivity of K$_{0.3}$MoO$_3$ (reprinted figure with permission from S. Donovan \textit{et al.}, Physical Review B 49, p. 3363, 1994 \cite{Donovan94}. Copyright (1994) by the American Physical Society).}
\label{fig4-2}
\end{center}
\end{figure}
shows the real part of the SDW conductivity for (TMTSF)$_2$PF$_6$ at 2~K  with in the inset that for K$_{0.30}$MoO$_3$ at $T$~= 40~K. In both cases there are a well-defined resonance: 100~GHz for K$_{0.30}$MoO$_3$ and 3~GHz for (TMTSF)$_2$PF$_6$ which is referred to as the pinned mode resonance due to $q$~= 0 oscillations of the DW condensate. At low frequencies there is a broad structure due to internal deformations of the collective mode; this one is strongly influenced by the screening from normal carrier and consequently this low frequency response is strongly temperature dependent (see section~\ref{sec6}).

\smallskip
\item[-] Above the threshold field $E_T$, a periodic time dependent voltage (called improperly narrow band noise (NBN) is generated in the crystal as well as a broad band noise (BBN) following a $1/f$ variation \cite{Fleming79,Richard82a,Bhatta87}. Figure~\ref{fig4-3} shows the Fourier transformed voltage spectrum in NbSe$_3$ as a function of frequency showing many harmonics \cite{Monceau90}.
\end{itemize}
\begin{figure}
\begin{center}
\includegraphics[width=8.5cm]{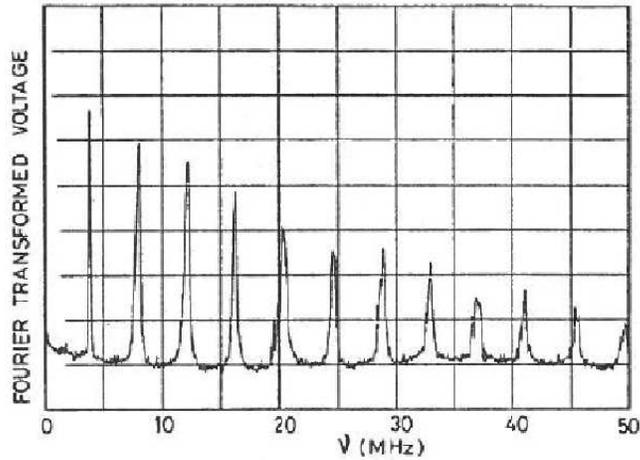}
\caption{Fourier transformed voltage spectrum (so-called narrow band noise-NBN) as a function of frequency in the non-linear state. NbSe$_3$, $T$~= 42~K (reprinted figure with permission from Applications of Statistical and Field Theory Methods to Condensed Matter, NATO ASI Series 218, P. Monceau, p. 357, 1990 \cite{Monceau90}. Copyright (1990) from Springer Science and Business media).}
\label{fig4-3}
\end{center}
\end{figure}

Following the Fr\"ohlich model for CDW motion (section~\ref{sec2-14}) the extra current density $J_{\rm C/S\,DW}\equiv$ $I_{\rm C/S\;DW}/A$, $A$: the cross-section of the sample, is $J_{\rm C/S\,DW}$~= $ne\,v_s$, $v_s$: the drift DW velocity, $ne$: the total electronic density condensed below the CDW gap. In reference~\cite{Monceau80} it was shown that, assuming the pinning forces to be periodic with the CDW phase, the CDW motion is the superposition of a continuous drift and a modulation due to the pinning potential at a recurrence frequency:
\begin{equation*}
\nu=(\frac{Q}{2\pi})v_{\rm drift},
\end{equation*}
with $Q$: the CDW wave vector. Then
\begin{equation}
J_{\rm CDW}=ne\,\frac{2\pi}{Q}\nu= ne\,\lambda_{\rm CDW}\nu.
\label{eq4-1}
\end{equation}
\begin{figure}[h!]
\begin{center}
\includegraphics[width=7.5cm]{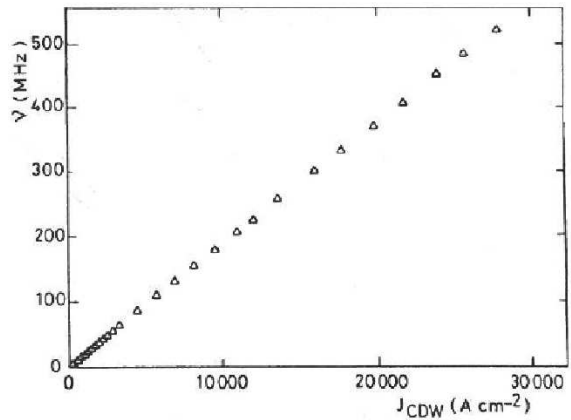}
\caption{Variation of the $J_{\rm CDW}$ carried by the CDW as a function of the fundamental frequency measured in the Fourier-transformed voltage. The slope $J_{\rm CDW}/\nu$~= $ne\lambda_{\rm CDW}$ leads to the number of carriers condensed below the CDW gap: o-TaS$_3$; $T$~= 127~K (reprinted figure with permission from Applications of Statistical and Field Theory Methods to Condensed Matter, NATO ASI Series 218, P. Monceau, p. 357, 1990 \cite{Monceau90}. Copyright (1990) from Springer Science and Business media).}
\label{fig4-4}
\end{center}
\end{figure}
The current carried by the CDW is linearly dependent of the periodic of the time-dependent voltage. Figure~\ref{fig4-4} shows the linear relationship between $J_{\rm CDW}$ and $\nu$ for an orthorhombic TaS$_3$ sample which is still valid with a CDW current density of 30.000~A/cm$^2$. According to eq.~(\ref{eq4-1}) the slope of $J_{\rm CDW}/\nu$ is a measurement of the number of electrons condensed below the CDW gap. $J_{\rm CDW}$ is measured from the non-linear V(I) characteristics as shown in figure~\ref{fig4-1} and the estimation of the cross-section of the sample. $\lambda_{\rm CDW}$ is obtained from the periodicity of the CDW modulation. The number $ne$ deduced from eq.~(\ref{eq4-1}) is of the order of the electron concentration (in the bands affected by the CDW condensation as it can be calculated from band structures or from chemical bonds). This result is thought to be the proof of the Fr\"ohlich conductivity, i.e. the collective motion of all the electrons trapped below the CDW gap.

\begin{itemize}
\item[-] When an ac field of frequency $\nu_{\rm ext}$ is superposed on a dc field $E>E_T$, interference occurs between the frequency of this ac field and the voltage oscillation frequency $\nu_{\rm d}$ induced by the CDW motion \cite{Monceau80,Richard82}. Interference seen as plateaux in the (I-V) curves or steps in the ${\rm d}V/{\rm d}I$ characteristics (Shapiro step \cite{Shapiro72}) are observed at $p\nu_{\rm ext}$~= $q\nu_{\rm d}$ ($p$ and $q$ integers). Figure~\ref{fig4-5} shows mode-locked Shapiro steps in NbSe$_3$ at 4.2~K with the rich sub-harmonic structure \cite{Hall84,Thorne88,Richard82}. 

\begin{figure}[h!]
\begin{center}
\includegraphics[width=7.5cm]{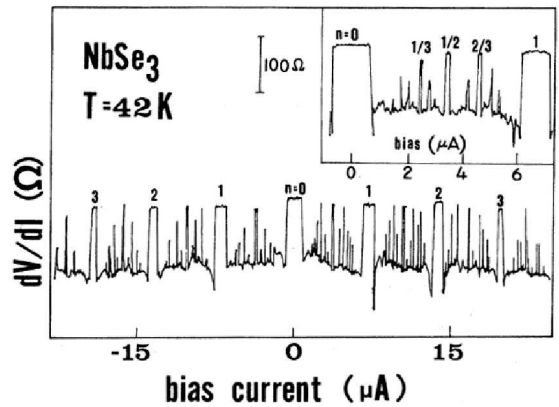}
\caption{Mode locked Shapiro steps in NbSe$_3$ at $T$~= 42~K showing a rich spectrum of harmonic and subharmonic (more detail in inset) steps. $rf$ frequency: 5~MHz (reprinted figure with permission from R.P. Hall and A. Zettl, Physical Review B 30, p. 2279, 1984 \cite{Hall84}. Copyright (1984) by the American Physical Societys).}
\label{fig4-5}
\end{center}
\end{figure}

\item[-] When the CDW is swept upwards and downwards through $E_T$ and when the polarity of the dc current is changed, metastability and hysteretic effects are often observed \cite{Littlewood87,Littlewood87b,Duggan85}, principally at low temperatures.
\end{itemize}

\subsection{Theoretical models for density wave sliding}\label{sec4-2}

As seen in figure~\ref{fig7-1}, two collective excitations occur in a CDW. For an incommensurate CDW, at the Peierls transition temperature, the soft mode at $2k_{\rm F}$ splits into an optical mode --amplitudon-- and an acoustic mode --phason--. For ideal pure materials, the energy of phase fluctuations tends to zero in the limit of large wavelengths: $\omega(q=0)$~= 0. This Golstone mode leads to the Fr\"ohlich superconductivity. However, in real systems, the translation invariance is broken by varied pinning phenomena --impurity pinning or commensurability pinning-- which introduce a gap in the phason spectrum and prohibits the Fr\"ohlich superconductivity. But, a force supplied by application of an electric field of a sufficient strength can overcome the pinning of the CDW phase (for a recent review on CDW pinning see \cite{Brazovskii04}.

Bardeen was the first \cite{Bardeen79,Bardeen79b} to interpret the non-linear conductivity in CDW systems \cite{Monceau76,Ong77} as the Fr\"ohlich conduction induced by the CDW motion. Moreover, the existence of a threshold field $E_T$ for depinning the CDW \cite{Fleming79} has to be considered. As for the vortex pinning in type II superconductors \cite{Larkin79}, a true threshold can only be accounted if one introduce some elasticity of the phase. A random distribution of impurities or dislocations is the more plausible cause of the pinning force. But, by its interaction with a completely rigid lattice, the force summation will be random leading to a $\Omega^{1/2}$ resultant ($\Omega$ being the volume, proportional to the number of impurities) and then to $E_T$ going to zero in the thermodynamic limit ($\Omega\rightarrow\infty$). On the contrary, some elasticity allows for a deformation of the phase and a finite second order effect due to the individual pinning forces. Thus Fukuyama, Lee and Rice (FLR) \cite{Lee79,Fukuyama78,Lee78} have shown that the phase coherence between distant points tends to zero with the distance if an arbitrary small elasticity is introduced in the random pinning problem. This is due to the accumulation of small phase disturbances over a great number of disturbing centres. The similar model for collective pinning was derived for superconducting flux lines \cite{Efetov77,Larkin79}. A domain size was defined, called now the Fukuyama-Lee-Rice length $L_{\rm FLR}$, such that the phase deviation to ideality is of the order of $\pi$, with an internal random summation of the pinning forces in the domain. Experimentally the value of $L_{\rm FLR}$ is only a few microns and so, any experiment is concerned with a many domain problem. If \textit{it can be assumed that the domains act independently}, the result will be the sum over a great number of domains and the total pinning force will be proportional to the total length, with $E_T$ independent of the length. In this description, the order parameter is zero at the borders of each domain so that coupling between adjacent domains requires only conservation of the electric current and continuity of potential. However, coupling of Josephson-type between phases in different domains was introduced for explaining switching effects \cite{Inui88}. The situation described above corresponds to the so-called weak pinning as introduced by FLR in the case where the impurity potential $V$ is weak. In the opposite strong pinning case, the CDW phase is locked at each impurity site and the CDW phase coherence is equal to the average distance between impurities.

\subsubsection{Classical equation of motion}\label{sec4-2-1}

If one consider a single domain, the CDW can be described with a unique dynamical variable, $\phi(t)$. A domain is associated with an equivalent mass, some dissipative mechanism, $\Gamma$ (thermalisation of the phase motion by the phonon bath), a net charge and a resultant pinning force periodic in $\phi$. This leads to the equation of motion \cite{Monceau80,Gruner81}, referred as the classical equation of motion:
\begin{equation}
\ddot{\phi}+\Gamma\dot{\phi}+\omega^2_{\rm p}\sin\phi=Q\,\frac{eE}{M^\ast},
\label{eq4-2}
\end{equation}
where $E$ is the applied electric field, $Q$~= $2\pi/\lambda_{\rm CDW}$, $\omega_{\rm p}$ the pinning frequency, $M^\ast$ the Fr\"ohlich mass and $\Gamma$ the CDW damping.

The equation~(\ref{eq4-2}) leads to at least qualitative explanations of many CDW phenomena and has been studied at length:
\begin{itemize}
\item[-] A threshold field defined by $E_T$~= $\displaystyle\frac{\lambda_{\rm CDW}}{2\pi}\,\frac{M^\ast\omega^2_{\rm p}}{e}$.\\

\item[-] If $E=E_0\cos\omega t$ and $E_0\ll E_t$ a linearisation of the $\sin\phi$ term gives a linear response. $\Re\sigma(\omega)$ and $\Im\sigma(\omega)$ are expressed such as:
\begin{eqnarray*}
\Re\sigma(\omega) & = & \frac{ne^2}{M^\ast\Gamma}\left(1+\frac{\omega^2_{\rm p}}{\Gamma\omega}\right)^{-2}\\
\Im\sigma(\omega) & = & \frac{ne^2}{M^\ast\Gamma}\,\frac{\omega_{\rm p}^2}{\Gamma}\left(1+\frac{\omega_{\rm p}^2}{\Gamma\omega}\right)^{-2}.
\end{eqnarray*}
From $\varepsilon(\omega)$~= $4\pi\Im(\sigma(\omega)/\omega)$, the dielectric constant in the low frequency range is given by:
\begin{equation*}
\varepsilon_{\omega\rightarrow 0}=\frac{4\pi ne^2}{M^\ast\omega^2_{\rm p}},
\end{equation*}
combined with the expression for $E_T$, the following relationship is derived:
\begin{equation*}
\varepsilon_{\omega\rightarrow 0}E_T=2ne\lambda_{\rm CDW}.
\end{equation*}
In this simplified overdamped model, the product of the dielectric constant (in the limit of $\omega\rightarrow 0$) by the threshold field is independent of temperature in good agreement with experimental results \cite{Wu84,Wu86}.\\

\item[-] For a dc field $E>E_T$, the ``$\sin\phi$'' force term gives rise to a velocity modulation at a fundamental frequency $\nu$ and its harmonics. A consequence of the classical equation of motion is that for $E>E_T$ the extra current $J_{\rm CDW}$$\sim(E-E_T)^{1/2}$.  Experimentally results show a power law $(E-E_T)^\gamma$ with $\gamma$ nearly 3/2. The 1/2 exponent results from the single domain approximation of the equation of motion. But with a distribution of threshold field from many domains which compose a given sample, the infinite derivative at $E_T$ of $J_{\rm CDW}(E)$ is removed. Some attempts have also been made to explain the regime near $E_T$ with some analogy with the critical behaviour of a second order phase transition which leads to the 3/2 exponent \cite{Fisher83}.\\

\item[-] If in eq.~(\ref{eq4-1}), $E$ is the superposition of a dc field $E>E_T$ and a small ac field $E_0\cos\omega t$, the non linear $\sin\phi$ term gives a frequency linking between the sliding CDW and the applied $\omega_{\rm ac}$, if the eigen frequency is near of $\omega_{\rm ac}$ (Shapiro step). During this synchronisation the CDW drift velocity $v_{\rm d}$ is independent of the continuous dc field and therefore the differential conductivity equals the linear ohmic value (complete mode locking) \cite{Richard82} as shown in figure~\ref{fig4-5}.
\end{itemize}

\subsubsection{Phase Hamiltonian}\label{sec4-2-2}

Starting from the Fr\"ohlich Hamiltonian or the anisotropic Hubbard model \cite{Yamaji82} for CDW and SDW respectively, the C/S DW phase Hamiltonian in d dimension can be written in a similar way \cite{Maki89a,Maki90b}:
\begin{eqnarray}
H_\phi & = &\int\, d^dx\left\{\frac{1}{4}N_0f\left[\frac{M^\ast}{m}\left(\frac{\upartial\phi}{\upartial t}\right)^2+v^2\left(\frac{\upartial\phi}{\upartial x}\right)^2+v^2_2\left(\frac{\upartial\phi}{\upartial y}\right)^2+v^2_3\left(\frac{\upartial\phi}{\upartial z}\right)^2\right]\right. \nonumber\\
&&\left. -e\rho fQ^{-1}\phi E\right\} + V_{\rm pin}(\phi),
\label{eq4-3}
\end{eqnarray}
where $N_0$ is the density of states at the Fermi level, $Q$~= $2k_F$, $\rho$ the electron density, $f$ the condensate density, $v$, $v_2$ and $v_3$ the anisotropic Fermi velocities. The phason mass $M^\ast$ has been defined in eq.~(\ref{eq2-15}) while $m^\ast/m$~= 1 for a SDW.

The slow spatial-temporal distortion of the phase $\phi$ generates an electric charge and a current given by:
\begin{eqnarray}
&\rho_c=e\rho fQ^{-1}\displaystyle\frac{\upartial\phi}{\upartial x} \hspace{5cm}\label{eq4-4}\\ 
\mbox{and}\hspace{5cm} & \nonumber \\
& j_c=-e\rho fQ^{-1}\displaystyle\frac{\upartial\phi}{\upartial t},\label{eq4-5}\hspace{5cm}
\end{eqnarray}
which satisfy the charge conservation (Poisson law)
\begin{equation}
\frac{\upartial\rho_c}{\upartial t}+\frac{\upartial j_c}{\upartial x}\equiv 0.
\label{eq4-6}
\end{equation}
The electric charge carried by the condensate is strictly conserved at all temperatures in the absence of topological defects (phase vortices, dislocation lines). As it will be seen below in section~\ref{sec5}, only topological defects can convert the electric charge carried by the condensate to the one carried by quasi particles.

Neglecting the time dependence in eq.~(\ref{eq4-3}) and using the scale transformation to transform eq.~(\ref{eq4-3}) in an isotropic form, one gets:
\begin{equation}
H(\phi)=f\int d^d_x\left\{\frac{1}{4}N_0\,\tilde{v}^2(\triangledown\phi)^2-e\rho Q^{-1}\phi E\right\}+V_{\rm pin}(\phi),
\label{eq4-7}
\end{equation}
where $\tilde{v}^2=\eta v^2$ with $\eta$~= $v_2v_3/v^2$. The impurity concentration $n_i$ is also rescaled as $\eta n_i$~= $\tilde{n}_i$.

For a CDW, neglecting quantum effects, $V_{\rm pin}(\phi)$ can be approximated by the form \cite{Lee79}:
\begin{equation}
V_{\rm pin}(\phi)=-\,\frac{2N_0V}{\lambda}\Delta(T)\sum_i\cos\left[Qx_i+\phi(x_i)-\phi_i\right],
\label{eq4-8}
\end{equation}
where $x_i$ is the impurity site, $V$ the Fourier component of the potential difference between the impurity and the host atom, $\lambda$ the electron-phonon coupling constant and $\phi_i$ the phase that the impurity at site $x_i$ tends to accommodate.

For a SDW \cite{Maki90a,Maki90b}:
\begin{equation}
V_{\rm pin}(\phi)=-\,\frac{\pi}{2}(N_0V)^2\Delta(T)\tanh\frac{\Delta(T)}{2T}\sum_i\cos\left\{2[Qx_i+\phi(x_i)-\phi_i]\right\}.
\label{eq4-9}
\end{equation}
Following FLR \cite{Fukuyama78,Lee79}, two types of pinning should be distinguished: strong pinning caused either by large amplitude pinning potential with the CDW phase being fixed at each impurity site or by dilute impurity concentration, or weak pinning in which a phase-coherent domain contains many impurities. Evaluation of the impurity strength parameter, $\varepsilon$, defined as the ratio between the characteristic pinning energy per impurity and the CDW elastic energy per impurity, introduced by Lee and Rice \cite{Lee79}, has been extended in $d$ dimensions \cite{Matsukawa88,Abe85} for the case of weak pinning.

For CDW in the strong pinning case, $\varepsilon\gg 1$, the threshold field is given by \cite{Lee78,Maki90b}:
\begin{equation}
E^S_{T-{\rm CDW}}(0)=\frac{2Q}{e\lambda}\,\frac{n_i}{n}(N_0V)\Delta(0).
\label{eq4-10}
\end{equation}
In the case where $\varepsilon\ll 1$, the elastic energy dominates. A random walk summation over impurities leads to a gain in impurity pinning energy but increases the elastic energy. By minimising the sum of the elastic and the impurity pinning energy, the phase-phase correlation length, or the Fukuyama-Lee-Rice length, $L_{\rm FLR}$, and the threshold field are related by:
\begin{eqnarray}
&E^W_T=&\frac{4-d}{d}\,\frac{Q}{e\rho}\,\frac{N_0}{L_{\rm FLR}^2},\label{eq4-11}\hspace{4.5cm}\\
\mbox{and }\hspace{3cm} &&\nonumber \\
& E^W_{T-{\rm CDW}}(0)\propto &n_i^{2/4-d}\Delta_0^{4/4-d}. \hspace{4cm}\label{eq4-12}
\end{eqnarray}
For $d=3$, the threshold field is proportional to the square of the impurity concentration but varies linearly with $n_i$ in the case $d=2$.

\subsubsection{Thermal fluctuations}\label{sec4-2-3}

The effect of thermal fluctuations of $\phi(x,t)$, reducing the impurity interaction potential, can modify drastically the temperature dependence of $V_{\rm pin}(\phi)$. Maki \cite{Maki86,Maki90b} incorporated this effect by multiplying $V_{\rm pin}(\phi)$ by a Debye-Waller factor such as $\exp-\frac{1}{2}\langle\phi\rangle^2$~= $\exp(-T/T_0)$ for a CDW (the bracket meaning a thermal average) taking into account short-range fluctuations $q\gg L^{-1}_{\rm FLR}$. Then the temperature dependence of $E^S_T(T)$ normalised to its value at $T=0$ becomes:
\begin{equation}
\frac{E_{T-{\rm CDW}}^S(T)}{E^S_{T-{\rm CDW}}(0)}=e^{-T/T_0}\left[\frac{\Delta(T)}{\Delta(0)}\right]f^{-1},
\label{eq4-13}
\end{equation}
where $f$ is the condensate density. $f$ is a complicated function of $\omega$ and $q$, the wave vector associated with the fluctuations of $\phi(x,t)$. It has been shown \cite{Maki89a,Maki89b,Rice79} that the limiting value of $f$ for $T\rightarrow T_c$ has different limits depending on the ratio between $\omega$ and $vq$ such as: $f$~= $f_1\propto\Delta^2(T)$ in the static limit when $\omega/vq\ll 1$ and $f$~= $f_0\propto\Delta(T)$ in the dynamic limit with $\omega/vq\gg 1$. In the FRL model in which the long range fluctuations of $\phi(x,t)$ are neglected, the latter limit is used which is appropriate for describing the microwave conductivity and the dc conductivity at high field, $E\gg E_T$. However the $f_1$ limit should be used for the analysis of the conductivity near the threshold, $E\approx E_T$, since the spatial fluctuations of $\phi(x,t)$ are of the extent of the FLR length. Plots of $f_0$ and $f_1$ as a function of $T/T_c$ are drawn in \cite{Maki90b}.

For SDW the threshold in the strong pinning case at $T=0$ is estimated \cite{Maki90a} as:
\begin{equation}
E^S_{T-{\rm SDW}}=\frac{Q}{e}\,\frac{n_i}{n}(\pi N_0V)^2\Delta(0),
\label{eq4-14}
\end{equation}
and the normalised temperature dependence as:
\begin{equation}
\frac{E^S_{T-{\rm SDW}}(T)}{E^S_{T-{\rm SDW}}(0)}=\left[\frac{\Delta(T)}{\Delta(0)}\right]\tanh\left[\frac{\Delta(T)}{2T}\right]\,f^{-1}_1.
\label{eq4-15}
\end{equation}
For SDW the effect of thermal fluctuations is nearly negligible since most of the SDW transition temperatures are around 10~K in contrast with many CDW transition temperatures which are around 100--200~K. For weak pinning, the threshold value for SDW is similar to that for CDW as in eq.~(\ref{eq4-11}).

It has been also shown that:
\begin{equation}
\frac{E^W_{T-{\rm SDW}}(T)}{E^W_{T-{\rm SDW}}(0)}=\left[\frac{E^S_{T-{\rm SDW}}(T)}{E^S_{T-{\rm SDW}}(0)}\right]^{4/4-D}.
\label{eq4-16}
\end{equation}
Finally commensurability of the CDW with the lattice yields a potential which can pin the CDW phase. The commensuration potential of order $N$~= 4 gives rise to a pinning energy roughly one order of magnitude smaller than that due to impurities.

\begin{figure}
\begin{center}
\subfigure[]{\label{fig4-6a}
\includegraphics[width=6cm]{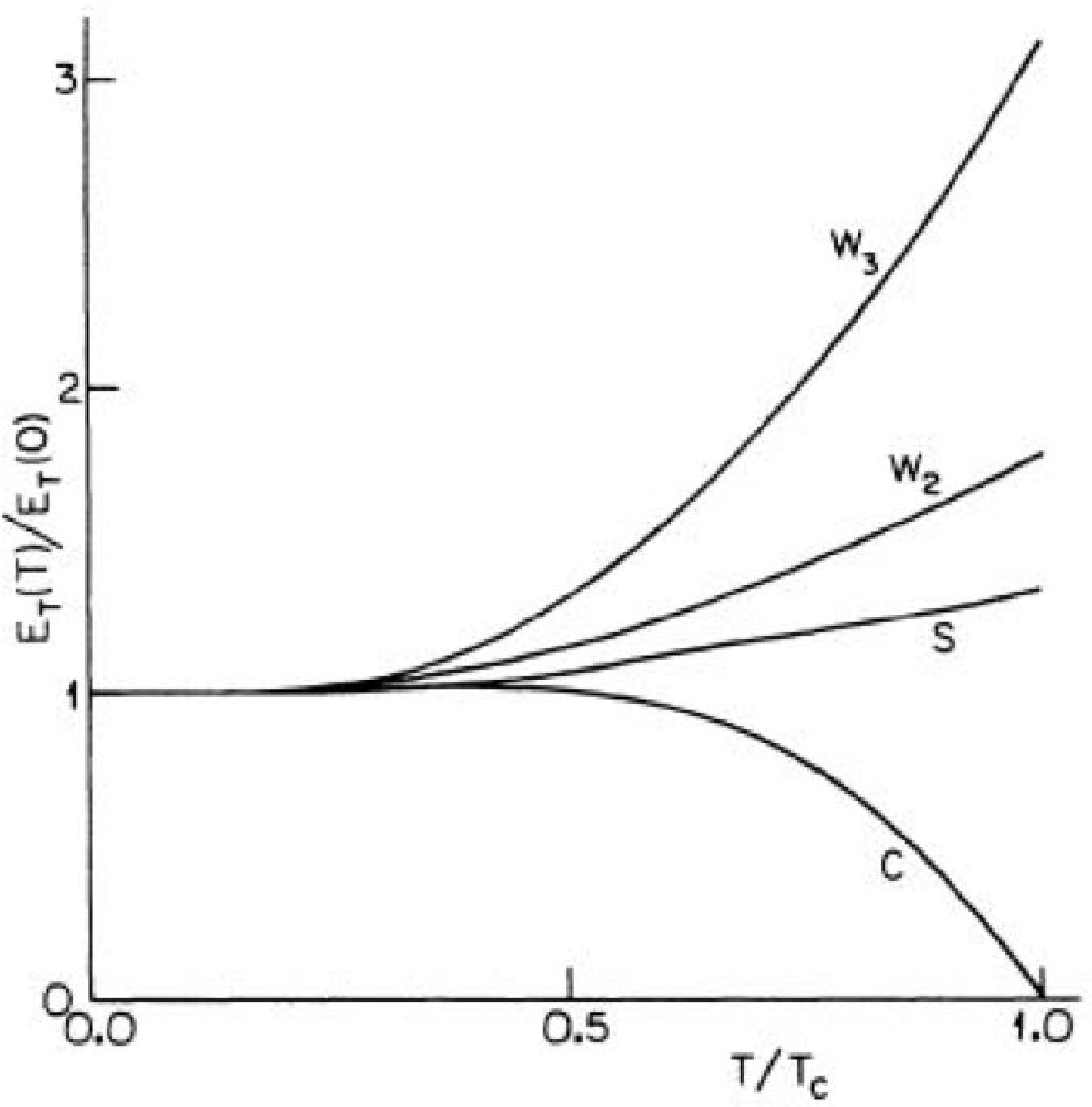}}
\subfigure[]{\label{fig4-6b}
\includegraphics[width=6.5cm]{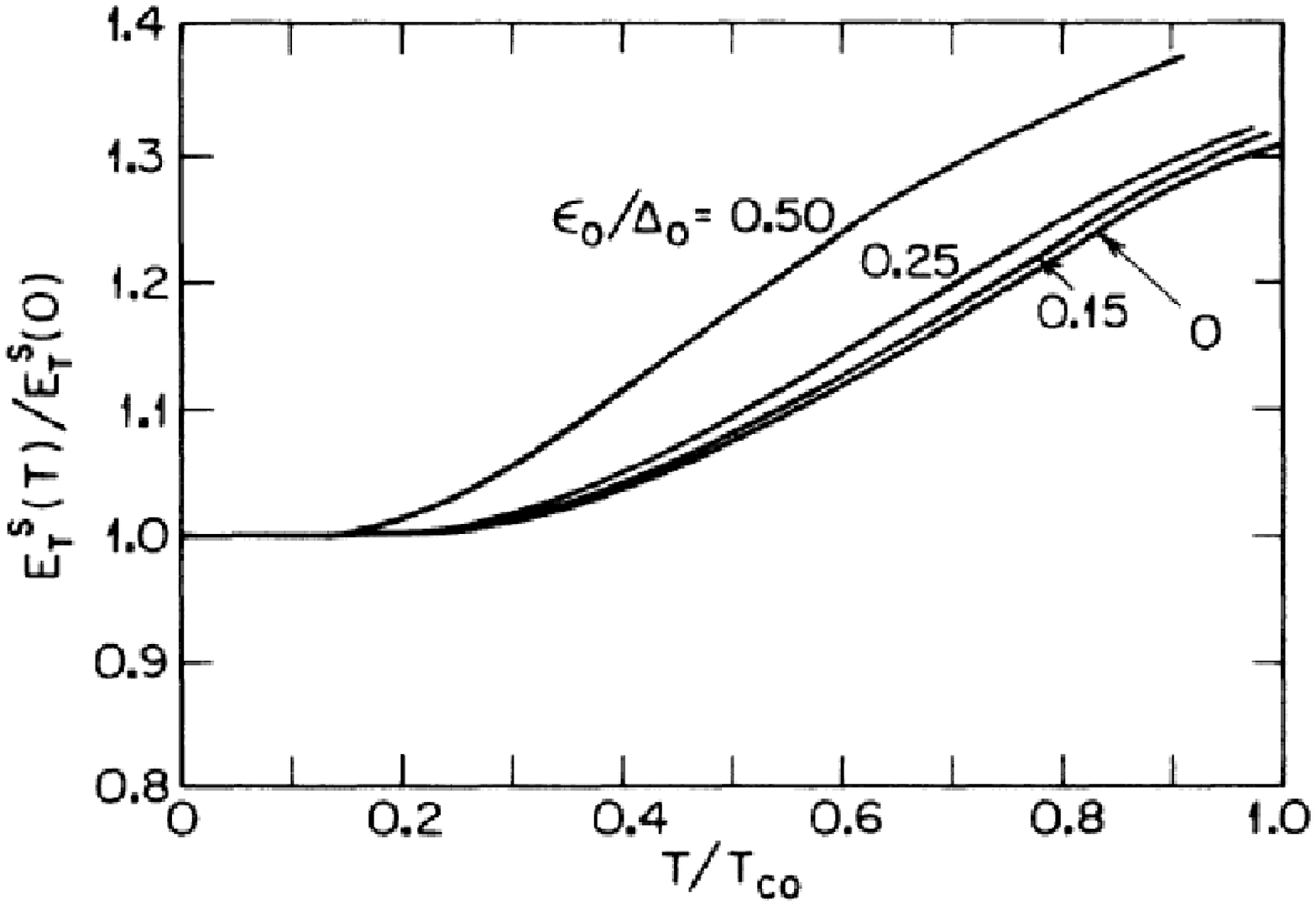}}
\caption{(a) Temperature dependence of the threshold field $E_T(T)$ for a SDW: in the strong-pinning limit ($s$), weak-pinning limit with $D$~= 3 ($W_3$) and $D$~= $2(W_2)$ and in the case of a commensurate pinning $c$ (reprinted figure with permission from K. Maki and A. Virosztek, Physical Review B 42, p. 655, 1990 \cite{Maki90b}. Copyright (1990) by the American Physical Society). (b)~temperature dependence of $E_T(T)$ for a SDW in the strong-pinning limit in the case of imperfect nesting evaluated by $\varepsilon_0/\Delta_0$ (reprinted figure with permission from X. Huand and K. Maki, Physical Review B 42, p. 6498, 1990 \cite{Huang90}. Copyright (1990) by the American Physical Society).}
\label{fig4-6}
\end{center}
\end{figure}

Following \cite{Maki90b} figure~\ref{fig4-6}(a) shows the temperature dependence of the threshold electric field for a SDW in the strong pinning limit ($s$), the weak pinning with $d$~=  3 (w3) and $d$~= 2 (w2) and in the commensurate case ($c$). It is seen that $E_{T-{\rm SDW}}$ is almost constant for $T\lesssim\frac{1}{2}T_c$. Imperfect nesting defined by the parameter $\varepsilon_0$ (eq.~(\ref{eq2-29})) destroys the two-dimensionality of the electronic bands and leads to a different temperature variation of $E_T(T)$; for instance between curve W2 and curve W3 for 2D and 3D weak pinning case (figure~\ref{fig4-6}(a)). Figure~\ref{fig4-6}(b) shows the $E_T(T)$ dependence for strong pinning for different $\varepsilon_0/\Delta_0$.

Thus, for SDW, $E_T$ decreases monotonously with lowering temperature whatever the regime of impurity pinning (strong or weak). $E_T$ does not diverge towards $T_c$ because of the second order coupling in contrast to CDW where the impurity pinning is of first order. $E_T$ is independent of $T$ below $T_c/2$.

\subsubsection{Numerical simulations}\label{sec4-2-4}

Non-linear dynamics of sliding CDW is representative of a class of collective transport phenomena in which an elastic medium is pulled through a random potential by an applied force. Thus the study of CDW dynamics involves many dynamical degrees of freedom and therefore is a very complicated task due essentially to the summation in the impurity potential in the FLR equation as written in eq.~(\ref{eq4-8}). In the models of rigid CDW motion, although the CDW interacts with impurities, the CDW is described with a unique dynamical variable $\phi(t)$ (section~\ref{sec4-2-1}). This approximation leads to the overdamped harmonic oscillator equation of motion which describes only some qualitative features of the non-linear state.

The sharp transition at $E_T$ between the pinned and the sliding CDW state has led to the interpretation of the depinning transition as a type of critical phenomena \cite{Fisher83,Fisher85,Pietronero91,Parisi91}. The order parameter for the depinning transition is the average steady-state CDW velocity $v$, which follows the power law:
\begin{equation*}
v\sim f^\xi,
\end{equation*}
with $f$ is the reduced driving force $f$~= $F-F_T/F_T$ where $F$ is the applied force, $F_T$ the critical force at the depinning transition. In mean field theory $\xi$ is equal to 3/2.

Extensive numerical simulations on a variety of models have been performed. Many of them consider a lattice version of the FLR Hamiltonian \cite{Myers93,Coppersmith90,Middleton93}
\begin{equation*}
H = \frac{1}{2} J\,\sum_{i,j}(\phi_j-\phi_i)^2-\frac{V}{2\pi}\,\sum^V_{i=1}\cos 2\pi(\phi_i-\beta_i)-F(t)\,\sum^V_{i=1}\phi_i,
\end{equation*}
where the phases $\phi_i(t)$ of the CDW at impurity sites $i$~= $1\ldots N$ are chosen to lie on a regular, square or cubic lattice of dimension $d$. The disorder due to impurity pinning is effective from the quenched random pinning phase $\beta_i$ at each lattice site. $J$ is the elastic coupling strength, $V$ the pinning potential, $F$ the spatially uniform driving force proportional to the electric field. The equations of motion are purely relaxational:
\begin{equation*}
\frac{d\phi_i}{dt}  =-\frac{\upartial H}{\upartial\phi_i}=J\Delta\phi_i-V\sin\left[2\pi(\phi_i-\beta_i)\right]+F
\end{equation*}
with
\begin{equation*}
\Delta\phi_i=\sum_\delta\left(\phi_{i+\delta}-\phi_i\right),
\end{equation*}
and $\delta$ the nearest-neighbour $\langle j\rangle$ to site $i$. Thus the model treats the CDW as an array of overdamped, harmonically coupled, randomly pinned oscillators under the driving force $F$. Simulations of this CDW FLR lattice have been performed by numerical integration of the equation of motion \cite{Pietronero83,Sibani90}, an automaton model \cite{Myers93} and a second-order predictor-corrector method \cite{Middleton93}. Another approach was using the renormalisation-group expansion \cite{Narayan92}.

Neglecting phase slippage, above $E_T$, the sliding motion of the CDW phase is a unique periodic steady state. Below $E_T$, the CDW is stuck to one of the many locally state configurations. Increasing the force results in a series of avalanches which get larger and larger until reaching $E_T$.

Critical exponent $\xi$ describing the scaling of the CDW velocity  was estimated in $d$~= 1, 2 and 3 dimensions and all the methods lead to approximately similar values. Also the velocity-velocity correlation length $\xi$ was found to be $\frac{1}{2}$ from renormalisation group theory and very near from $\frac{1}{2}$ and independent of the dimension from numerical simulations. However all these estimations neglect the possibility that strain on the CDW induces dislocations or phase slips. Phase slippage leads to a spatially non-uniform time-averaged velocity which destroys the criticality of the pinned-sliding transition \cite{Coppersmith90}.

\subsubsection{Quantum models}\label{sec4-2-5}

The first non-linear (I-V) characteristics \cite{Monceau76} on NbSe$_3$ showed a dependence on the applied dc electric field as:
\begin{equation*}
I\propto E\exp-E_0/E.
\end{equation*}
This form of conductivity suggested that tunnelling of Zener-type may be involved across a small pinning gap. Maki \cite{Maki77} using a Sine-Gordon-like equation for the CDW phase on a single chain found a similar expression for the probability to create a soliton-antisoliton pair by an electric field. However, the electrical conductivity resulting from quantum mechanical tunnelling was thought to be only observable at extremely low temperatures, where the thermal soliton density in the CDW is negligible.

However Bardeen \cite{Bardeen79,Bardeen85,Bardeen89,Bardeen89b} showed that tunnelling can take place when applied to the coherent motion of the CDW in a large-phase coherent volume (the L$_{\rm FLR}$ domain) containing many parallel chains. The pinning energy in such a volume can be much larger than $kT$, so that depinning cannot occur thermally. In this context, a CDW should be treated as a macroscopic quantum system as a superconductor or superfluid helium \cite{Bardeen90}.

Following the Fr\"ohlich model for superconductivity as described in section \ref{sec2-14}, in stationary conditions the Peierls gap opens at $\pm k_{\rm F}$. When the CDW is sliding at velocity $v$, the gaps appear at the boundaries of the displaced Fermi sea: $-k_{\rm F}+q$, $+k_{\rm F}+q$ with $\hbar q$~= $mv_d$, $m$ the band mass of electrons. The energy difference between opposite sides of the displaced Fermi sea is $\hbar\omega_d$~= $2\hbar k_{\rm F}v_d$, the energy of a macroscopically occupied phonon in the moving Fermi sea.

Bardeen has uncessantly insisted, without having too much been understood, that a quantum tunnelling step is required to increase the momentum of the CDW in an electric field because the momentum is quantified in units of $2\hbar k_{\rm F}$. He developed then a model with many analogies with the tunnelling of a supercurrent through a Josephson junction in which the tunnelling probability that enters is that of single electrons maintaining their phase coherence in the tunnelling process. Bardeen \cite{Bardeen89} also stated that it was not possible to arrive at an energy gap by treating impurities as a perturbation in any order of perturbation. The analogous problem in superconductivity is that one cannot arrive to the superconducting ground state by treating the electron-phonon interaction and impurity scattering in perturbation theory. In this quantum tunnelling model \cite{Bardeen90} the momentum to accelerate the CDW is given by the difference between the number of $2k_{\rm F}$ phonons moving to the right ($N_{\rm R}$) and the number moving on the left ($N_{\rm L}$) such as:
\begin{equation*}
(N_{\rm R}-N_{\rm L})2\hbar k_{\rm F}=N_eM^\ast v_d,
\end{equation*}
with $N_e$ the number of electrons in the volume concerned, $M^\ast$ the Fr\"ohlich mass and $v_d$ the CDW drift velocity. Although $N_{\rm R}-N_{\rm L}$ can be large, the fundamental point of the model is that, with impurity pinning, even to increase the momentum by one unit of $2\hbar k_{\rm F}$, a tunnel step is required. The step event consists in removing an electron with wave vector $-k_{\rm F}$ from one FLR domain and places it with wave vector $+k_{\rm F}$ in an adjacent domain, so that $k_{\rm F}$ is added to the wave vectors in each of the two domains.

The tunnelling probability is found as $I(E)=\exp(-E_0/E)$ as experimentally measured, with:
\begin{eqnarray*}\begin{array}{cc}
& E_0=\displaystyle\frac{\pi E_g^2}{4\hbar e^\ast v_{\rm F}}\hspace{7cm} \\
& \\
\mbox{and}\hspace{3.5cm} & E_g=\hbar\,\omega_p,\qquad \displaystyle\frac{e^\ast}{e}=\displaystyle\frac{m^\ast}{m^\ast+M^\ast}.\hspace{7cm}\end{array}
\end{eqnarray*}
The tunnel step allows the possibility of photon-assisted tunnelling \cite{Tucker79} as expressed in the ac conductivity and the effects of combined ac and dc fields \cite{Bardeen79}.

\subsubsection{Quantum corrections for CDW motion}\label{sec4-2-6}

The phase Hamiltonian (eqs~(\ref{eq4-3}) and (\ref{eq4-8})) picturing the interaction between the sliding CDW and the impurity in Ginzburg-Landau (G-L) theory describes the phase $\varphi(x)$ of the CDW as a slowly varying function of the space coordinate $x$ in which the large scale average CDW phase remains correlated over a long range phase coherence, the FLR distance, $\xi_{\rm FLR}$. The microscopic calculation of the local impurity-CDW interaction has been first derived by Barnes and Zawadowski \cite{Barnes83} and later on by T\"utto and Zawadowski \cite{Tutto85} (the picture of interactions is very different from that in G-L theory) which shows that Friedel oscillations compete with CDW charge modulations over an atomic distance of a characteristic size of the amplitude coherence length known as the BCS length: $\xi_0$~=  $v_{\rm F}/\Delta_0$.

If the impurity potential is strong enough, these Friedel oscillations dominate the charge modulation within this highly localised region and the phase of the oscillations become locked to its optimum value. The CDW on the other hand dominates the charge modulation at distances $x_0\gg\xi_0$. The force acting between the impurity and the CDW is determined by the mismatch in the region $\xi_0<x<x_0$ (for recent experimental results, see \cite{Brazovskii11}). Each impurity deforms the CDW well inside the range of the length $\xi_0$; then the pinning  periodic potential is strongly perturbed and does not keep its simple sinusoidal dependence. Quantum corrections to the distortion of the CDW around impurities have been calculated \cite{Barnes83,Tutto85}.

\begin{figure}[h!]
\begin{center}
\includegraphics[width=7.5cm]{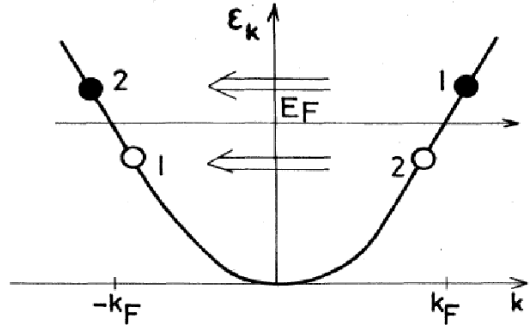}
\caption{1-D dispersion curve with the two macroscopic quantum states formed by electron-hole pairs (labelled by 1 and 2) forming the CDW. The arrows indicate the two backscatterings on the impurity which represent a transition between the two different types of pair (reprinted figure with permission from I. T\"utto and A. Zawadowski, Physical Review B 32, p. 2449, 1985 \cite{Tutto85}. Copyright (1985) by the American Physical Society).}
\label{fig4-7}
\end{center}
\end{figure}

It was shown that the CDW condensate can be thought of as the superposition of two macroscopic quantum states formed of electron-hole bound pairs with total momentum $\pm Q$. The motion of the CDW condensate splits  the two macroscopic quantum states. The first order perturbation theory leads to a (classical) energy density, periodic in space with the periodicity ($\sin 2k_{\rm F}x$). The second order term corresponds to two impurity scattering with large momentum transfer of an electron-hole pair with momentum $+2k_{\rm F}$ into an electron-hole pair with momentum $-2k_{\rm F}$ from the same side of the dispersion curve to the opposite one by backward scattering at the impurity (see figure~\ref{fig4-7}). 
This second-order perturbation theory shows a strong similarity with the Josephson effect. The result is an energy density $\sin 4k_{\rm F}x$ with a periodicity equal to half the CDW wavelength.

This quantum theory of local perturbation of the CDW by impurities is similar to the Bardeen theory in the sense of describing the CDW as two macroscopic quantum states. In the quantum approach of CDW depinning of Bardeen, tunnelling in the vicinity of impurities is responsible for the weak coupling between these two macroscopic quantum states. The approach of T\"utto and Zawadowski has a strong formal analogy with the theory of Josephson junctions. The strength of pinning depends on the forward scattering $T$ of the local electron-impurity interaction potential $V$ ($T$~= $V(q=\pm Q)$). In the case of weak pinning ($T/2\hbar v_{\rm F}\ll 1$) the CDW is only slightly disturbed by the impurity and the phase of the charge modulation at the impurity site advances continuously with the CDW phase displacement. For large values of scattering ($T/2\hbar v_{\rm F}\gg 1$) the charge modulation at the impurity site becomes fixed by Friedel oscillations.

In the equation of motion both potentials in $\sin 2k_{\rm F}x$ and $\sin 4k_{\rm F}x$ may contribute. The importance of the ``Josephson'' term can be estimated \cite{Tutto85} from the temperature dependence of the ratio of intensities of harmonics in the periodic voltage oscillations (NBN) generated in the non-linear state; the effect of non-sinusoidal potential can also be detected in Shapiro steps. With the aim to ascribe essentially the CDW pinning by strong pinning impurities, Tucker \textit{et al.} \cite{Tucker89,Tucker93} have developed a theory which combines the microscopic impurity CDW potential calculated by T\"utto and Zawadowski \cite{Tutto85} within the large scale Ginzburg-Landau framework by Lee and Rice \cite{Lee79}. Thus, the local CDW pinning will always be ``strong'' with the phase of the CDW at the impurity site close to the optimum value which matches the localised Friedel oscillations. But, the large scale average CDW phase far away from the impurity sites will be weakly pinned over correlated volumes of FLR type, as expected for ``weak" pinning.

\subsubsection{Pinning in SDW}\label{sec4-2-8}

\begin{figure}
\begin{center}
\includegraphics[width=7.5cm]{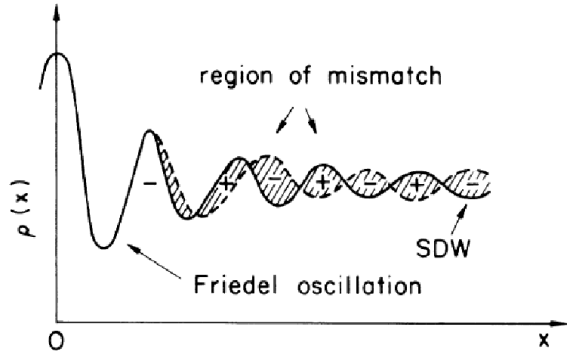}
\caption{Schematic plot of the electron densities of up (solid line) and down (dashed line) spin electrons near a non-magnetic impurity. Far from the impurity the SDW is not deformed. In the vicinity of the impurity Friedel oscillations are formed. In the cross-over region, the mismatch takes place, which is responsible for pinning (reprinted figure with permission from I. T\"utto and A. Zawadowski, Physical Review Letters 60, p. 1442, 1988 \cite{Tutto88}. Copyright (1988) by the American Physical Society).}
\label{fig4-9}
\end{center}
\end{figure}

In first order in the electron-impurity interactions, there is no pinning of the SDW by non magnetic impurities because the total electron density is constant \cite{Takeda84}. Introduction of a second-order term representing a Josephson-type mechanism as proposed by Barnes and Zawadowski \cite{Barnes83} for CDWs gives rise to a weakly pinned SDW with a periodicity half the SDW wavelength \cite{Tua85a,Tua85b}. However the interaction between impurities and conduction electrons might generate a small distortion of the charge density near the impurity sites in the form of Friedel oscillations. As in the CDW case, the mismatch between the charge Friedel oscillations in the extreme vicinity of the impurity and the undeformed SDW far from the impurity is responsible for the pinning \cite{Tutto88}. That can be viewed from the decomposition of a linearly polarised SDW into two CDW formed with spins up and spin down out of phase (as shown in figure~\ref{fig4-9}). The SDW pinning potential is written as:
\begin{equation*}
\Omega_{\rm SDW}(\phi)=\Omega_{\rm CDW}(\phi)+\Omega_{\rm CDW}(\phi+\pi).
\end{equation*}
As the pinning potential $\Omega_{\rm CDW}(\phi)$ is not sinusoidal, there is no cancellation of $\phi$ for $\Omega_{\rm SDW}$ and pinning occurs.

Magnetic impurities without magnetic field also pin the SDW in second order by exchange interactions, even for undistorted SDW condensate, with a periodicity half the SDW wavelength \cite{Tua84}. When a strong magnetic field is applied, all impurity spins are aligned in the same direction and then the energy density regains the periodicity of the SDW.

It should be noted that SDW sliding state  will not occur in a circularly polarised SDW because impurities will not be coupled to the SDW.

\subsection{Experimental results on density wave sliding}\label{sec4-3}

\subsubsection{Threshold fields}\label{sec4-3-1}

\noindent \textit{4.3.1.a. Determination of threshold fields}
\medskip

The threshold field is determined in the (I-V) curve when the deviation of the ohmic law  is observed. For the best crystals, the characteristic $dV/dI$ shows a sharp discontinuity. But for many compounds, the curve $\sigma(E)$ is rounded near $E_T$ and the threshold is then determined by extrapolation.

\begin{figure} [h!]
\begin{center}
\includegraphics[width=7.5cm]{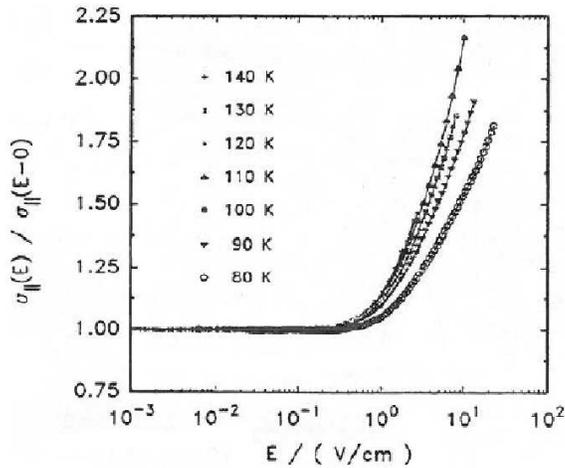}
\caption{Electric-field dependence of the conductivity normalised to its ohmic value for a (fluoranthene) PF$_6$ crystal at different temperatures from 140 to 80~K (reprinted figure with permission from W. Riess and W. Br\"utting, Physica Scripta T49, p. 721, 1993 \cite{Riess93}. Copyright (1993) by the Institute of Physics).}
\label{fig4-13}
\end{center}
\end{figure}
\begin{figure}[h!]
\begin{center}
\includegraphics[width=8cm]{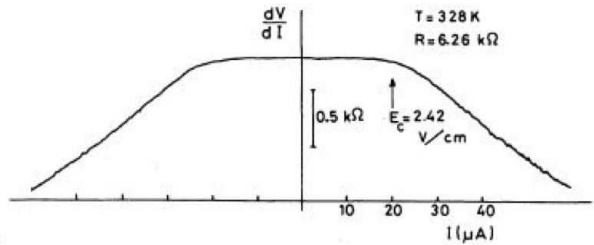}
\caption{Variation of the differential resistivity of a NbS$_3$ (phase II) crystal as a function of the applied current demonstrating CDW sliding at 328~K (reprinted figure with permission from Z.Z. Wang \textit{et al.}, Physical Review B 40, p. 11589, 1989 \cite{Wang89}. Copyright (1989) by the American Physical Society).}
\label{fig4-14}
\end{center}
\end{figure}

\begin{figure}[h!]
\begin{center}
\includegraphics[width=7.5cm]{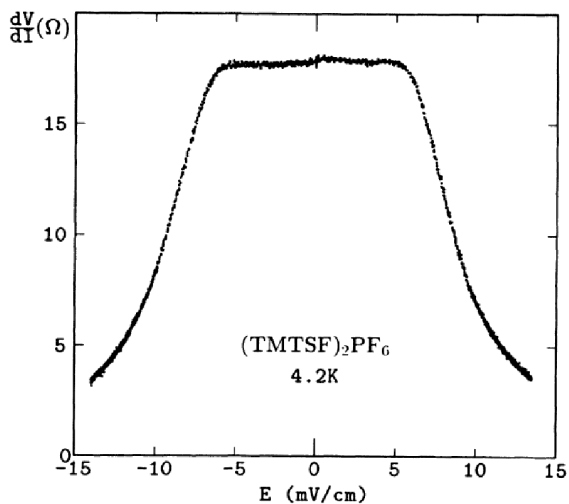}
\caption{Variation of the differential resistance of a (TMTSF)$_2$PF$_6$ crystal as a function of the applied electric field at $T$~= 4.2~K (reprinted figure with permission from W. Kang \textit{et al.}, Physical Review B 43, p. 1264, 1991 \cite{Kang91}. Copyright (1991) by the American Physical Society).}
\label{fig4-15}
\end{center}
\end{figure}

Figure~\ref{fig4-13} shows the non linear characteristic $\sigma(E)$ of the (fluoranthene)$_2$PF$_6$ --(FA)$_2$PF$6$-- at different temperatures \cite{Riess93}. As seen there are uncertainties to evaluate precisely $E_t$. Figure~\ref{fig4-14} shows \cite{Wang89} the differential resistivity of NbS$_3$ (phase II) at 328~K for which the Peierls transition is 340~K. In figure~\ref{fig4-15} 
 is drawn ${\rm d}V/{\rm d}I$ of the SDW compound (TMTSF)$_2$PF$_6$. This sharp characteristic \cite{Kang91} is obtained in very specific conditions for preparing the sample; many other results on (TMTSF)$_2$X show a $\sigma(E)$ characteristic similar to that in figure~\ref{fig4-13} \cite{Tomic89,Sambongi89}.
 
\begin{figure}
\begin{center}
\includegraphics[width=8cm]{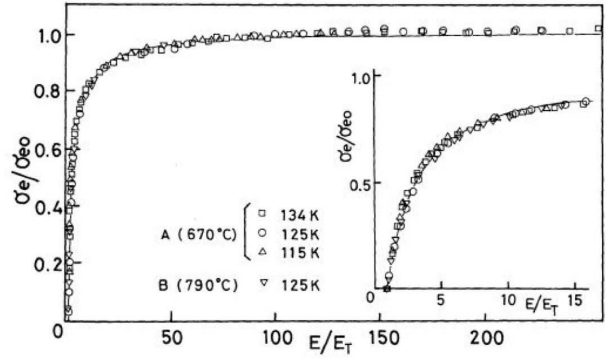}
\caption{Normalised conductivity-electric field curves at different temperatures for the upper CDW of NbSe$_3$: conductivity is normalised by its high field limiting value and $E$ by the threshold value. Inset shows detailed variation for lower $E/E_T$ (reprinted figure with permission from Journal of the Physical Society of Japan 56, p. 2503, 1987, M. Ido \textit{et al.} \cite{Ido87}).}
\label{fig4-16}
\end{center}
\end{figure}

Many results have been reported for NbSe$_3$ \cite{Monceau85} for which $\sigma(E)$ was fitted with an exponential variation: $\sigma(E)\propto\exp-E/E_0$. This variation modified in the low $E$ limit by the existence of the threshold $E_T$ was one of the first impulse for Bardeen to develop his tunnelling model: Figure~\ref{fig4-16} 
shows \cite{Ido87} that this exponential variation is valid up to 200~$E_T$. A similar plot at different temperatures is shown in figure~\ref{fig4-7} of ref.~\cite{Bardeen90}. But for many other compounds measurements were not performed to enough large $E/E_T$ for establishing firmly the universal validity of this functional dependence. In the case of K$_{0.3}$MoO$_3$, $\sigma(E)$ is still increasing for $E/E_T\sim 70$ and the non-linear conductivity is better described  \cite{Mihaly88d} with a power law: $\sigma_{\rm CDW}$~= $\sigma_nk\,E_T/E(E/E_T-1)^\alpha$ with $\alpha$ temperature dependent ($\alpha\sim 1.4$ at $T$~= 78~K).

\medskip
\noindent \textit{4.3.1.b. Impurity pinning}
\medskip 

As explained in section~\ref{sec4-2-2}, two types of pinning are distinguished: strong pinning for which the CDW phase is pinned at each impurity site and weak pinning the CDW phase being kept in the range $\pm\pi$ by fluctuations of the impurity potential on a length, $L_{\rm FRL}$, much larger than the average impurity spacing. The threshold field $E_T$ reflects the force needed to unpin the CDW phase and then to initiate the CDW sliding. For weak pinning, $E_T$ is predicted to vary as the square of the impurity concentration, $E_T\propto n_i^2$, for strong pinning: $E_T\propto n_i$.

Isoelectronic substitutional impurities as Ta for NbSe$_3$ are expected to be weak pinning centres, while non-isoelectronic impurities as Ti should strongly pin the CDW phase. For blue bronze two types of substitutional doping were performed; either the disorder is introduced on the alkali sub-lattice such as K$_{0.30-x}$Rb$_x$MoO$_3$, or the isoelectronic element $W$ is substituted for molybdenum K$_{0.30}$Mo$_{1-x}$W$_x$O$_3$. Another method for creating defects is irradiation with protons or electrons.

Results on the impurity dependence of $E_T$ are sometimes opposite. A $c^2$ dependence of $E_T$ was reported \cite{Brill81,McCarten92,Ido87} in Ta-doped NbSe$_3$, but a linear $c$ dependence was also found \cite{Underweiser87}. Rb doping in blue bronze was found to be a weak pinning centre ($c^2$ dependence in $E_T$) while a linear $c$ dependence is reported for $W$ doping \cite{Schneemeyer84}. The same $c$ dependence was found for Ti doping \cite{Ido87} in NbSe$_3$, electron irradiation in TaS$_3$ \cite{Mutka84b} and K$_{0.30}$MoO$_3$ \cite{Mutka84a}, as well with proton irradiation in NbSe$_3$ \cite{Fuller81}.

The great difficulty in doping is the more appropriate evaluation of the defect concentration. Chemical nominal concentration is often used, although there are no proofs that all substitutional impurities are included in the final sample. For NbSe$_3$ which remains metallic at low temperature, the residual resistance ratio RRR, assuming the validity of the Matthiesen law, although questionable, is often used for the evaluation of defects. However, the best method, when possible, is irradiation with fast electrons (2--3~MeV) which results in well controlled and varying defect concentration from the ppm level. Displacement of atoms through elastic collisions is the main process for defect production. The number of displacements per atom is $c$~= $\sigma_{\rm d}\phi$ where $\sigma_{\rm d}$ is the cross-section for displacement and $\phi$ the dose. Models have been developed for calculating the displacement cross-section in a polycrystal with different types of atoms \cite{Quelard76}.

\begin{figure}
\begin{center}
\includegraphics[width=6cm]{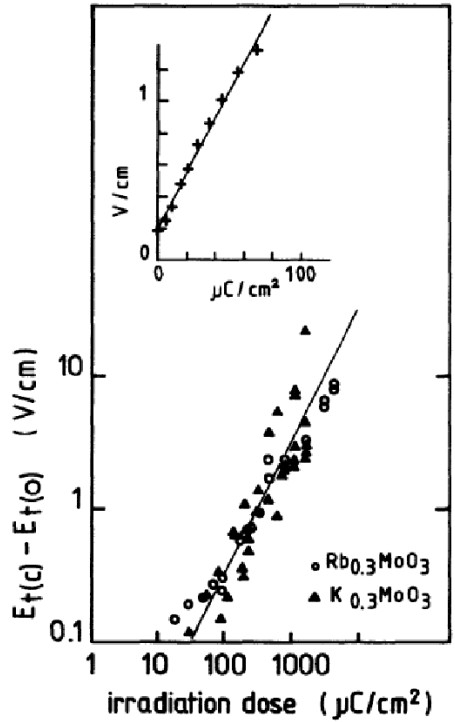}
\caption{Increase of the threshold field $E_T(c)-E_T(0)$ as a function of the irradiation dose with 2.5~MeV electrons for the same crystal of Rb$_{0.3}$MoO$_3$ at $T$~= 77~K (logarithmic scales). Inset shows the linear increase of $E_T$ as a function of the dose for another crystal of K$_{0.3}$MoO$_3$ (linear scales) (reprinted figure with permission from H. Mukta \textit{et al.}, Journal de Physique Lettres 45, p. 729, 1984 \cite{Mutka84a}. Copyright (1984) from EdpSciences).}
\label{fig4-10}
\end{center}
\end{figure}
\begin{figure}[h!]
\begin{center}
\includegraphics[width=7.5cm]{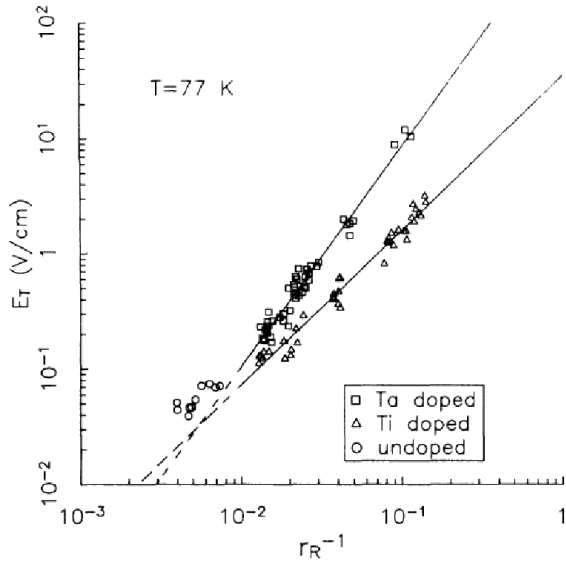}
\caption{Variation of the threshold field $E_T$ as a function of the inverse of the resistance ratio $r_{\rm R}^{-1}$ for thick Ta and Ti doped NbSe$_3$ crystals at $T$~= 77~K. The bold lines represent least-squares fits of the form $E_T$~= $A(r_{\rm R}^{-1})^b$ with $b$~= 1.9 for Ta-doped crystals and $b$~= 1.35 for Ti-doped crystals (log-log scales) (reprinted figure with permission from J. McCarten \textit{et al.}, Physical Review B 46, p. 4456, 1992 \cite{McCarten92}. Copyright (1992) by the American Physical Society).}
\label{fig4-11}
\end{center}
\end{figure}

In blue bronze an estimation of the irradiation defects with 2.5~MeV electrons has been made from the Curie law of the EPR lines related to deviation from stoichiometry and attributed to Mo$^{5+}$. The excess defect concentration was estimated \cite{Mutka84a} to be of the order of 10$^{-5}$ for an irradiation dose of 1~mC/cm$^2$. Figure~\ref{fig4-10} shows the increase of the threshold $E_T(c)-E_T(0)$ as a function of the irradiation dose for Rb$_{0.30}$MoO$_3$ and K$_{0.30}$MoO$_3$ crystals. The inset shows clearly the linear dependence of $V_T$~= $E_T/L$ for initial dose, indication of strong pinning of created defects.

For NbSe$_3$ a similar irradiation has been performed \cite{Monceau81}. The defect concentration was estimated to be $\sim 6\times10^{-3}$ displacement per atom for the maximum flux of $19\times10^{17}$ electrons cm$^{-2}$. Although the resistance of the sample was increased by a factor $\sim 25$, $E_T$ was weakly affected (increase of $\sim 2$). Irradiation from a Cu X-ray tube on a (TMTSF)$_2$PF$_6$ crystal has also shown that strong pinning defects are created \cite{Kang91}.

Figure~\ref{fig4-11} shows \cite{McCarten92} the comparison of $E_T$ versus (RRR) for Ta and Ti doped NbSe$_3$ samples at $T$~= 77~K. The best least-squares fit the data such as:
\begin{equation*}
E_T=A({\rm RRR}^{-1})^b,
\end{equation*}
with $b$~= 1.9$\pm$0.1 for Ta doped samples and $b$~= 1.35$\pm$0.1 for Ti doped samples. The exponent $b$ for Ta doped is very close to the value expected for weak pinning. The power-law exponent for Ti doped is intermediate between strong and weak pinning. A controversy took place on the consequence of these results. DiCarlo \textit{et al.} \cite{DiCarlo90} estimated that $({\rm RRR})^{-1}$ is given by $({\rm RRR})^{-1}$~= $({\rm RRR}_0)^{-1}+b_i^{-1}n_i$, $n_i$ the substitutional impurities (Ta or Ti) and $({\rm RRR}_0)^{-1}\propto n_0$ the residual concentration of defects and dislocations after that all introduced Ta or Ti impurities have been removed. These $n_0$ impurities were considered to be the 100~ppm of Ta included in the Nb powder used for the growth of NbSe$_3$ crystals. Adding these two components to the threshold value DiCarlo \textit{et al.} \cite{DiCarlo93a,DiCarlo93b,DiCarlo90,McCarten92} stated that for both Ta and Ti impurities the pinning is weak, even if the Ti pinning strength is $\sim 40$ times larger than the value of Ta.

On the contrary, in the model developed by Tucker \textit{et al.} \cite{Tucker88} each impurity pins strongly the CDW phase but the large scale average CDW phase away from the impurity sites remains correlated over large volumes as in nominally pure crystals. From crude arguments concerning the volume of average phase coherence, taking into account the interplay between ``weak" and ``strong" aspects of the impurity pinning, they made rough numerical estimations of the average depinning field for strong impurities:
\begin{equation*}
E_0(T)\propto\frac{\Delta}{L^{4/3}},
\end{equation*}
$L$ the phase coherence length along the chain direction. The 4/3 power law dependence is nearly identical to the 1.35 exponent obtained by Di~Carlo \textit{et al}. Then the general conclusion of Tucker \cite{Tucker93} is that the pinning of CDW is strong at any impurity site and that the dc CDW motion is made possible by phase slippage. 

These two totally opposite statements are by some way extreme and each system has to be studied for its own.

\medskip
\noindent \textit{4.3.1.c. Temperature dependence of $E_T$}
\medskip

\begin{figure}
\begin{center}
\includegraphics[width=7cm]{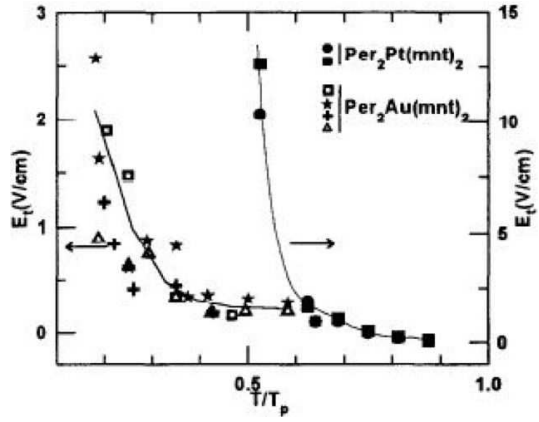}
\caption{Variation of the threshold field $E_T$ of (Per)$_2$(Mnt)$_2$ with M~= Au (left) and Pt (right) as a function of the reduced temperature $T/T_{\rm P}$ (reprinted figure with permission from Synthetic Metals 86, E.B. Lopes \textit{et al.}, p. 2163, 1997 \cite{Lopes97}. Copyright (1997) with permission from Elsevier).}
\label{fig4-18}
\end{center}
\end{figure}

For transition metal tri-- and tetrachalcogenures, $E_T$ presents a minimum below $T_c$ and increases at low temperature with an exponential dependence $e^{-T/T_0}$ (for a review, see ref.~\cite{Monceau85}). The same increase at low temperature is observed in TTF-TCNQ \cite{Forro87,Forro84} and in Per X (Mnt)$_2$ with X~= Au and Pt \cite{Lopes97} as seen in figure~\ref{fig4-18}.

\begin{figure}
\begin{center}
\includegraphics[width=12.5cm]{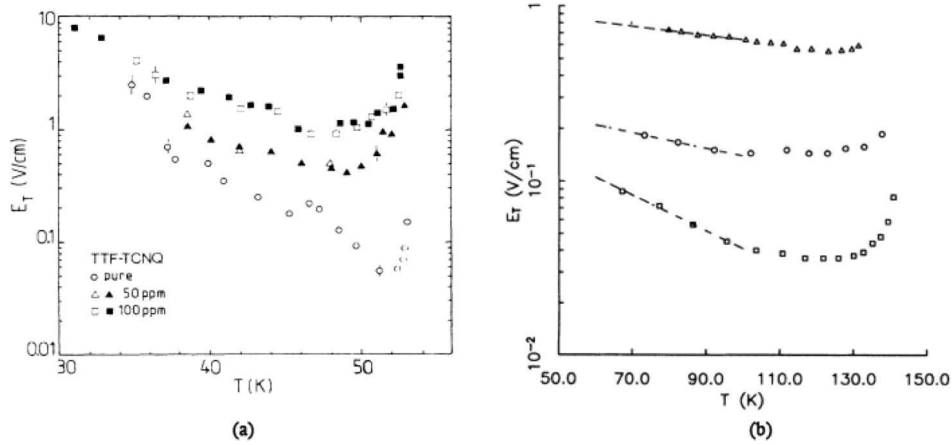}
\caption{Variation of the threshold field $E_T$ on log scale as a function of temperature. (a)~For pure and electron-irradiation doped TTF-TCNQ crystals; ({\large $\circ$}) - pure, ($\vartriangle$, $\blacktriangle$ - 50~ppm, ({\tiny$\square$}, {\tiny$\blacksquare$}) - 100~ppm (reprinted figure with permission from L. Forro \textit{et al.}, Physical Review B 35, p. 5884, 1987 \cite{Forro87}. Copyright (1987) by the American Physical Society). (b)~For NbSe$_3$ crystals containing different Ta concentrations. Dashed lines indicate the fit $E_T\propto\exp(-T/T_0)$ (reprinted figure with permission from J. McCarten \textit{et al.}, Physical Review B 46, p. 4456, 1992 \cite{McCarten92}. Copyright (1992) by the American Physical Society).}
\label{fig4-19}
\end{center}
\end{figure}

The effect of doping on the temperature variation of $E_T$ is presented in figure~\ref{fig4-19}(a) for TTF-TCNQ doped by irradiation and in figure~\ref{fig4-19}(b) for Ta doped NbSe$_3$ samples (respective curves from low to high $E_T$ corresponding to samples with RRR from 200, 71 and 29). The effect is very similar: strong increase of $E_T$ and decrease of the thermal fluctuation effect. As explained in section~\ref{sec4-2-3}, Maki \cite{Maki86} has incorporated the thermal fluctuations of the CDW phase $\phi(x)$ in the $T$ dependence of $E_T$: $E_T(T)$~= $E_Te^{-\langle\phi\rangle^2/2}$. The thermal average $\langle\phi^2\rangle$ was calculated by an integral with a cut-off at $k$~= $\xi^{-1}$ with $\xi$ the BCS amplitude coherence length. The evaluated value of $T_0$ is one or two orders of magnitude larger than the experimental one ($T_0\sim 15$~K for instance in pure NbSe$_3$). Then a higher cut-off momentum is then required \cite{Maki89a}, but without too much ground justification. Moreover $T_0$ in the Maki's calculation is independent of doping, which is not experimentally found as seen in figure~\ref{fig4-19}. The strong increase of $E_T$ measured in Per X(Mnt)$_2$ is intriguing, because with a Peierls transition, so low: 12~K for X-Pt and 8~K for X-Au, fluctuations of the CDW phase should be negligible as it has been stated for SDW in (TMTSF)$_2$X with transition temperatures in the same range ($\sim 10$~K).

The dimensionality of the pinning potential was also evaluated from fits to theoretical calculations \cite{Maki89a} leading in NbSe$_3$ to of a 2D weak pinning potential for the upper CDW while a strong-pinning potential is more appropriate for the lower CDW.

\begin{figure}
\begin{center}
\includegraphics[width=7.5cm]{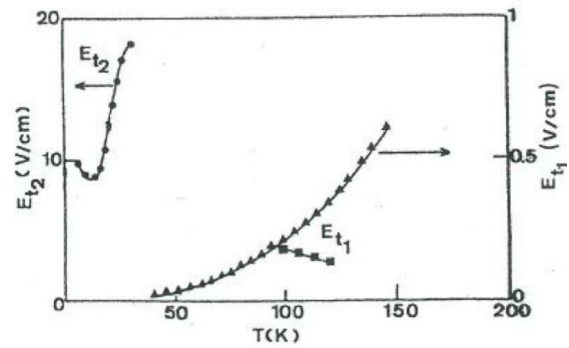}
\caption{Variation of the threshold field $E_T$ for K$_{0.3}$MoO$_3$ as a function of temperature (reprinted figure with permission from J. Dumas and C. Schlenker, International Journal of Modern Physics B 7, p. 4045, 1993 \cite{Dumas93}. Copyright (1993) by World Scientific).}
\label{fig4-21}
\end{center}
\end{figure}

The case of K$_{0.3}$MoO$_3$ is different: all the reported results \cite{Tsutsumi84,Shimizu91,Fleming86,Dumas93,Schlenker89b,Maeda85} indicate a decrease of $E_T$ between 100~K and 50~K; but in ref.~\cite{Schlenker89b} $E_T$ has a sharp peak at around 100~K and decreases when $T\rightarrow T_p$ while for \cite{Tsutsumi84,Shimizu91} the increase of $E_T$ continues up to 150~K. Figure~\ref{fig4-21} 
\begin{figure}
\begin{center}
\includegraphics[width=6.5cm]{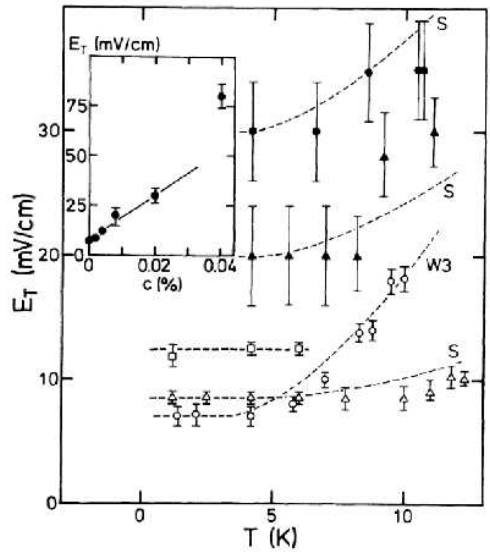}
\caption{Variation of the threshold field, $E_T$, as a function of temperature for pure and X-ray irradiated (TMTSF)$_2$PF$_6$ crystals with given molar concentration of defects: ({\large$\circ$}) - pure, ($\vartriangle$) - 0.002\%, ({\tiny$\square$}) - 0.004\%, ($\blacktriangle$) - 0.008\%, ({\large$\bullet$}) - 0.02\%. Dashed lines correspond to the theoretical calculations (see figure~\ref{fig4-6}(a)) in the strong-pinning limit (S) or the 3D weak-pinning limit (W3). Inset shows the variation of $E_T$ versus the defect concentration (reprinted figure with permission from W. Kang \textit{et al.}, Physical Review B 43, p. 1264, 1991 \cite{Kang91}. Copyright (1991) by the American Physical Society).}
\label{fig4-22}
\end{center}
\end{figure}
shows the last results of Dumas \textit{et al.} \cite{Dumas93}. A discontinuity in $E_T$ occurs at low temperature with a huge threshold below 40--50~K. It has been proposed that in this low $T$ state, the CDW motion occurs with almost no damping (see section~\ref{sec6-4}). Recent ultrasonic measurements \cite{Saint-Paul09} on K$_{0.3}$MoO$_3$ have revealed anomalies of the temperature dependence of the elastic constants in the vicinity of 50~K, such as a step-like increase of the sound velocity for the longitudinal mode propagating along the [$\bar{2}01$] direction.
 
\begin{figure}
\begin{center}
\includegraphics[width=6.5cm]{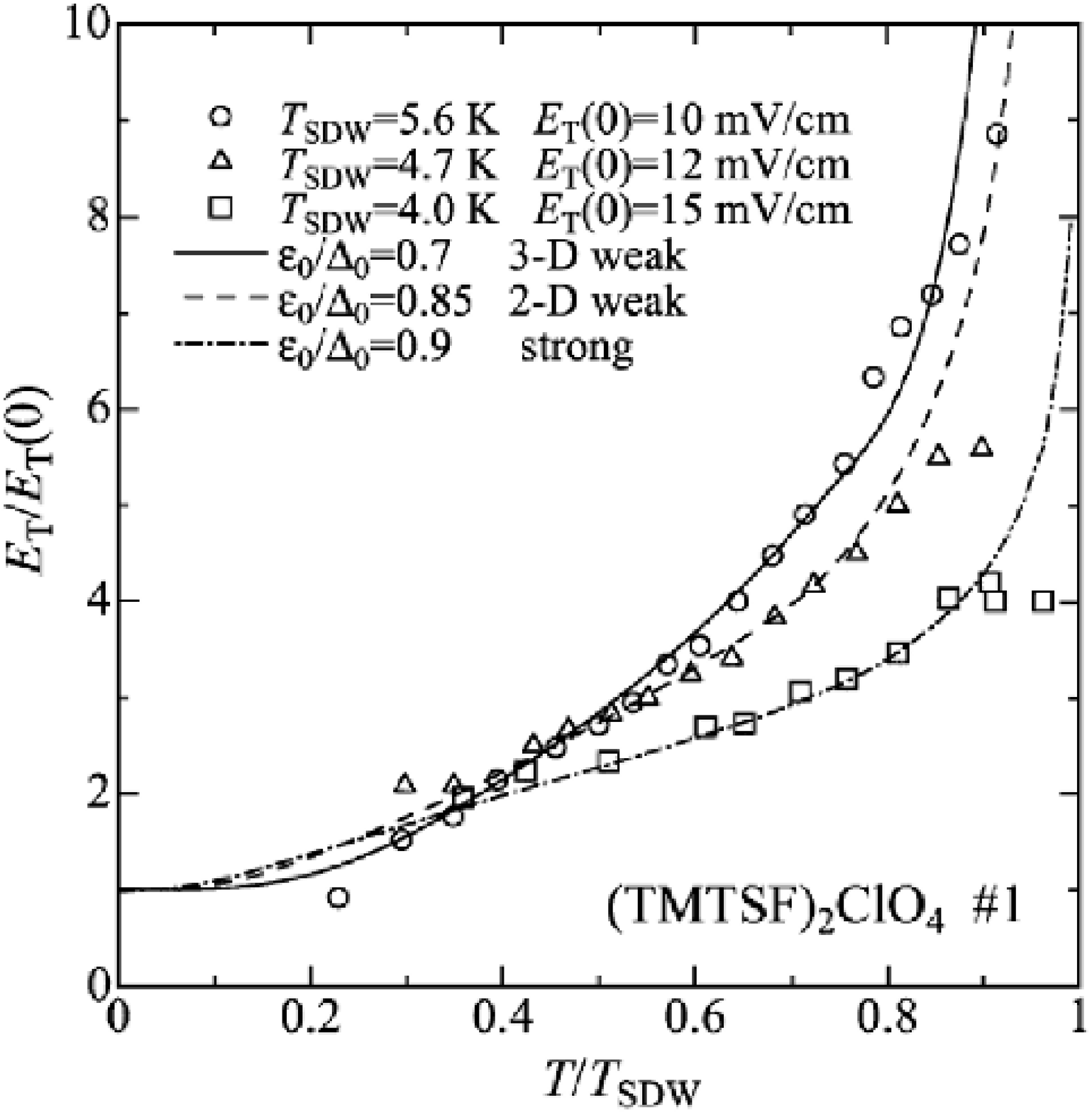}
\caption{Normalised temperature dependence of the threshold field of (TMTSF)$_2$ClO$_4$. $E_T$ is normalised to its low temperature value and the temperature by the SDW transition temperature which varies depending on the quenching process. Experimental data: ({\large$\circ$}) for $T_{\rm SDW}$~= 5.6~K, ($\vartriangle$) for $T_{\rm SDW}$~= 4.7~K and ({\tiny$\square$}) for $T_{\rm SDW}$~= 4.0~K. Theoretical calculations taking into account the imperfect nesting ($\varepsilon_0/\Delta_0$) solid line - 3D in the weak-pinning limit, dashed line - 2D in the weak-pinning limit, dashed and dotted line in the strong limit pinning (reprinted figure with permission from Journal of the Physical Society of Japan 69, p. 1457, 2000, A. Hoshikawa \textit{et al.} \cite{Hoshikawa00}).}
\label{fig4-23}
\end{center}
\end{figure}

For SDW all measurements show that $E_T(T)$ is independent of $T$ below $T_p/2$ and increases when $T\rightarrow T_c$ without any divergence at $T_c$. The effect of technique of contacting was revealed in ref.~\cite{Kang90} where with clamped contacts the $E_T(T)$ variation was that of a commensurate SDW while with painted contacts the variation indicates a weak-pinning dependence. Figure~\ref{fig4-22} shows the variation $E_T(T)$ for pure and doped by X-ray irradiation (TMTSF)$_2$PF$_6$ crystals. Pure crystal follows the 3D weak pinning variation (W3 curve) calculated in ref.~\cite{Maki90b} as shown in figure~\ref{fig4-6}(a). After defect creation the pinning becomes strong and follows the dependence indicated by dashed S lines. Another method to vary pinning was to change the SDW phase transition of (TMTSF)$_2$ClO$_4$ controlling the anion ordering by well defined quenching process \cite{Hoshikawa00}. The variation of $E_T(T)$ drawn with normalised temperature is plotted in figure~\ref{fig4-23} for three cooling rates implying three different SDW phase transitions.
It is seen  that the change of pinning from weak to strong is revealed by the increase of the imperfectness parameter $\epsilon_0$ (see section~\ref{sec2-12} and figure \ref{fig4-6}(b)) induced by the temperature quench which reduces $T_{\rm SDW}$ down to  $\sim 4.0$~K. The strong pinning in that case comes probably from the disorder in the anion lattice.

\medskip
\noindent \textit{4.3.1.d. Size effects on threshold}
\medskip

\begin{figure}
\begin{center}
\includegraphics[width=5.5cm]{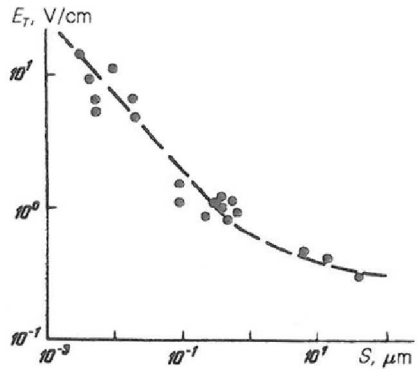}
\caption{Variation of the threshold field $E_T$ as a function of the cross-section area for o-TaS$_3$ at $T$~= 120~K (log-log scales) (reprinted figure with permission from JETP Letters 43, D.V. Borodin \textit{et al.}, p. 625, 1986 \cite{Borodin86a}. Copyright (1986) from Springer Science and Business media).}
\label{fig4-12}
\end{center}
\end{figure}

For high quality samples which have a whisker morphology, it was shown that the threshold field $E_T$ increases at $A^{-1/2}$ with $A$ the cross-section of the sample \cite{Borodin86a,Yetman87,Borodin86b} as indicated in figure~\ref{fig4-12} 
for o-TaS$_3$. On NbSe$_3$ RRR was shown to become dependent on the  thickness, $t$, of the sample when its width $w\gg t$; that occurs at $t<2\mu$ for RRR~$\sim$~200. $E_T$ has its bulk value for thick samples and depends linearly with $t^{-1}$ for thin samples. The characteristic thickness $t_c$ from the cross-over to size-independent behaviour occurs when $t_c$ is comparable with the transverse correlation lengths \cite{McCarten89,McCarten92}. The cross-over thickness $t_c$ decreases with the increase of impurity concentration. The explanation of this size effect by only taking into account the thickness of the sample was criticised in \cite{Tucker90} and \cite{Gill90a,Gill90b}. The dependence of $E_T$ on the length between electrodes will be addressed in section~\ref{sec5-4-2}.

\subsubsection{Density wave drift velocity}\label{fig4-3-2}

\noindent \textit{4.3.2.a. Ac voltage oscillations above $E_T$}
\medskip 

\begin{figure}
\begin{center}
\subfigure[]{\label{fig4-24a}
\includegraphics[width=5.5cm]{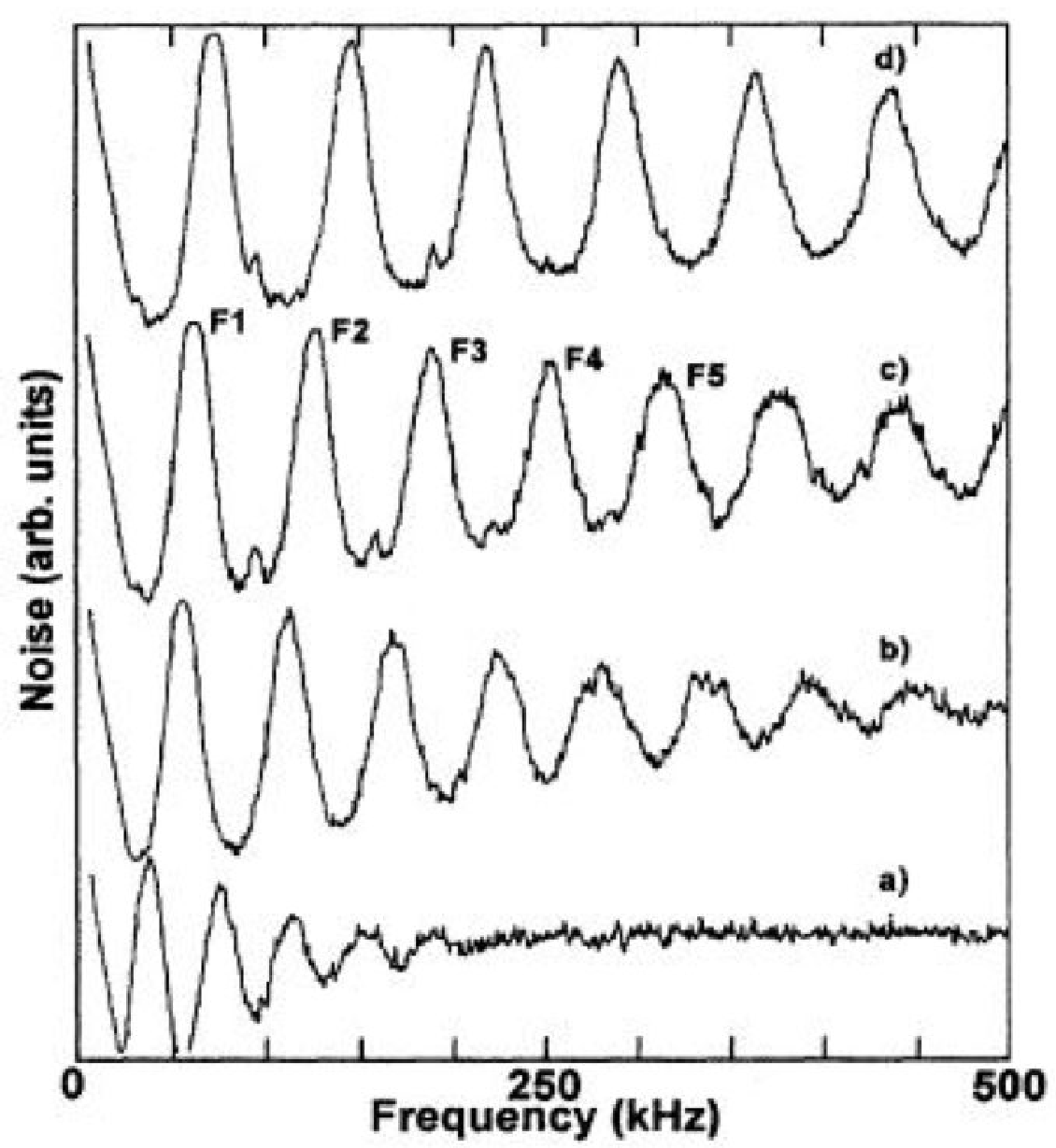}}
\subfigure[]{\label{fig4-24b}
\includegraphics[width=6cm]{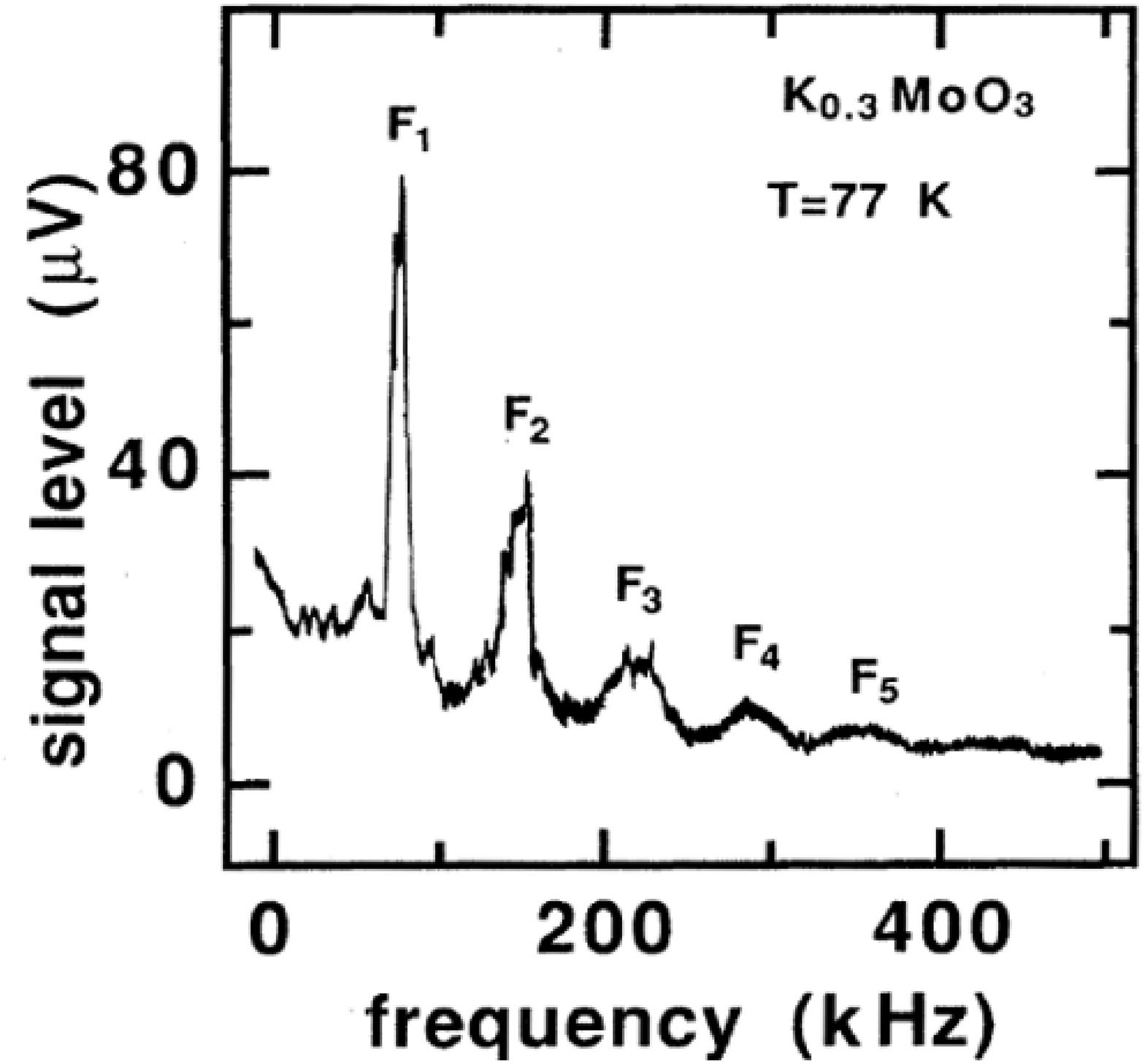}}
\subfigure[]{\label{fig4-24c}
\includegraphics[width=5.5cm]{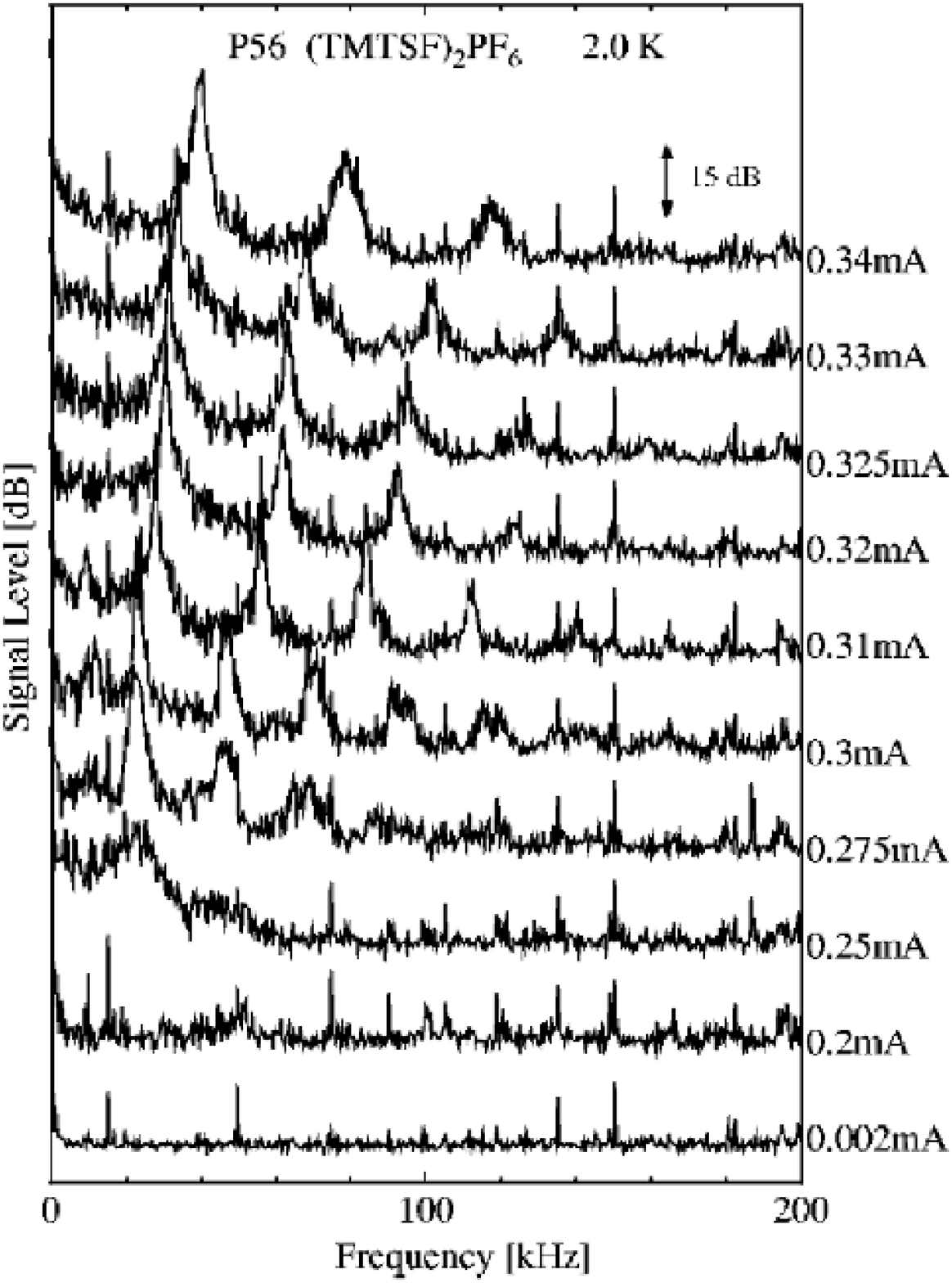}}
\subfigure[]{\label{fig4-24d}
\includegraphics[width=5.5cm]{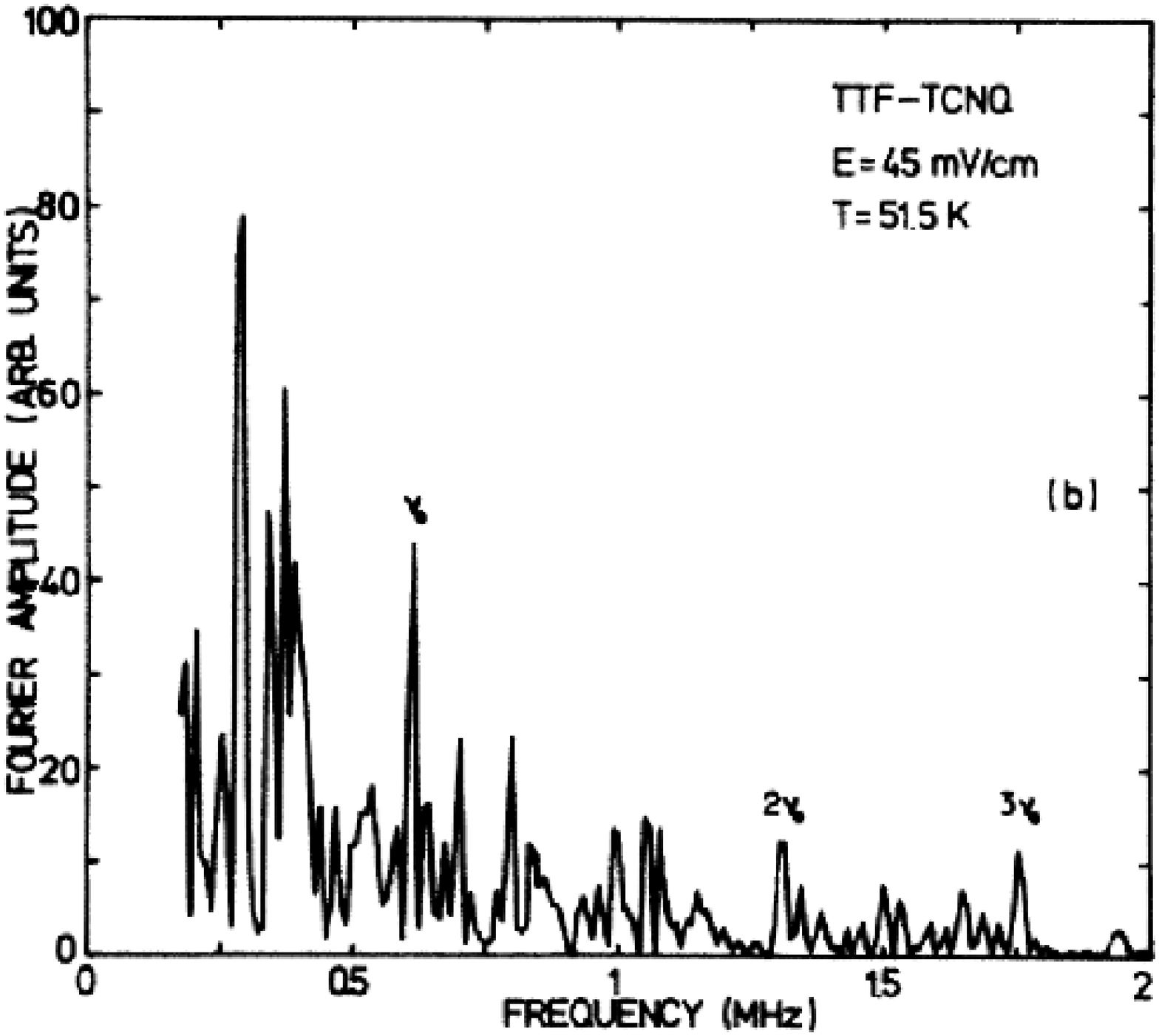}}
\vspace{-0.5cm}
\caption{Fourier spectra of the voltage oscillations (narrow band noise) as a function of frequency: (a)~for (perylene)$_2$Pt(Mnt)$_2$ at 4.2~K at different dc bias (reprinted figure with permission from E.B. Lopes \textit{et al.}, Physical Review B 52, p. R2237, 1995 \cite{Lopes95}. Copyright (1995) by the American Physical Society); (b)~for K$_{0.3}$MoO$_3$ at 77~K (reprinted figures with permission from M.F. Hundley and A. Zettl, Physical Review B 39, p. 3026, 1989 \cite{Hundley89}. Copyright (1989) by the America Physical Society); (c)~for (TMTSF)$_2$PF$_6$ at 2.0~K at different dc bias (reprinted figures with permission from T. Sekine \textit{et al.}, Physical Review B 70, p. 214201, 2004 \cite{Sekine04}. Copyright (2004) by the America Physical Society); (d)~for (TTF-TCNQ) at 51.5~K (reprinted figures with permission from S. Tomi\'c \textit{et al.}, Physical Review B 37, p. 8468, 1988 \cite{Tomic88}. Copyright (1988) by the America Physical Society).}
\label{fig4-24}
\end{center}
\end{figure}

\begin{figure}
\begin{center}
\includegraphics[width=6.5cm]{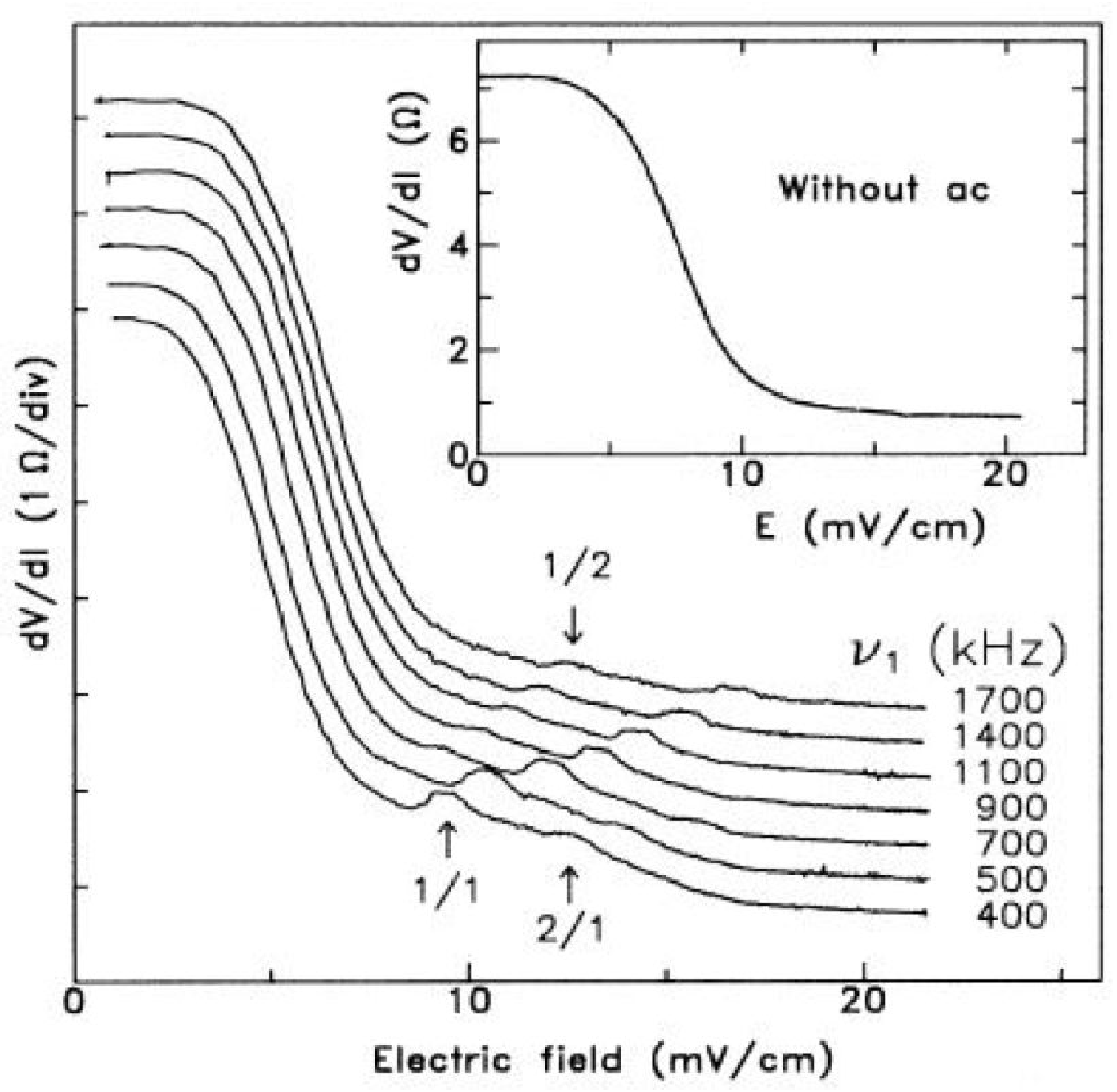}
\caption{Variation of the differential resistance of (TMTSF)$_2$AsF$_6$ as a function of the electric field with the superposition of an ac field of frequency $\nu_0$. Interference effects are identified (arrows) corresponding to ac-dc coupling (Shapiro step). Inset shows ${\rm d}V/{\rm d}I$ without ac field (reprinted figure with permission from G. Kriza \textit{et al.}, Physical Review Letters 66, p. 1922, 1991 \cite{Kriza91b}. Copyright (1991) by the American Physical Society).}
\label{fig4-25}
\end{center}
\end{figure}

One characteristic of the sliding DW state is the generation of a voltage oscillation the frequency of which increases linearly with $J_{\rm DW}$. In figure~\ref{fig4-24} 
are gathered the Fourier-transformed spectra of the ac voltage (NBN) in several DWs: thin samples of K$_{0.3}$MoO$_3$ \cite{Hundley89}, (perylene)$_2$ Pt(Mnt)$_2$ \cite{Lopes95}, TTF-TCNQ \cite{Tomic88} and on (TMTSF)$_2$PF$_6$ \cite{Sekine04} (for NbSe$_3$ see figure~\ref{fig4-3}). The other technique to detect this ac voltage is the ac-dc mode locking. Figure~\ref{fig4-25} shows the differential resistance versus frequency in (TMTSF)$_2$AsF$_6$ \cite{Kriza91b} (for NbSe$_3$ see figure~\ref{fig4-5}) . 

As explained above, the current carried by the DW in motion is given by $J_{\rm DW}$~= $nev$. The DW velocity, $v$, is written as the product of the fundamental frequency of the ac voltage generated in the sample and the pinning periodicity, $\lambda_p$, as follows:
\begin{equation*}
J_{\rm DW}/\nu=ne\lambda_p,
\end{equation*}
\begin{eqnarray*}\begin{array}{llll}
\mbox{with}\qquad\qquad\qquad&\lambda_p & =\lambda_{\rm CDW}&\quad\mbox{for a CDW}\\
&&&\\
&\lambda_p & =\lambda_{\rm SDW}/2&\quad\mbox{in the case of non-magnetic impurities}.\qquad
\end{array}
\end{eqnarray*}
The factor 2 comes from the second order in the impurity scattering (see section~\ref{sec4-2-8}). $\lambda_p$~= $\lambda_{\rm SDW}$ for a SDW with magnetic impurities.

The increase of $\nu$ with increase of $J_{\rm DW}$ is observed in all the sliding DW states with approximately a linear dependence, sometimes in a relatively narrow frequency range. However some curvature in the $\nu(J_{\rm DW})$ may occur if the DW current distribution is inhomogeneous and, then, when the fraction between the sliding volume and the pinned value may change when the injected current is increased.

The carrier concentration involved in the DW sliding has also to be calculated for each compound. $ne$ for a CDW is expressed as:
\begin{equation*}
n=p\,\frac{2b}{\lambda_{\rm CDW}}\,\frac{1}{v_{\rm unit\,cell}},
\end{equation*}
where $p$ is the number of bands affected by the CDW and $v_{\rm unit\,cell}$ the volume of the unit cell. This expression is derived assuming that, the Fermi level being at $Q$~= $2k_{\rm F}$~= $2\pi/\lambda_{\rm CDW}$, there are, in absence of the Peierls transition, two electrons in the band filled up to b$^\ast$~= $2\pi/b$.

From experiments values of $\nu/J_{\rm CDW}$ in kHz~cm$^2$/A have been obtained for several CDW compounds: 11$\pm$3 for K$_{0.3}$MoO$_3$ \cite{Butaud90}, 18 for (perylene)$_2$Pt(Mnt)$_2$ \cite{Lopes95}, 19.6 for o-TaS$_3$ (see figure~\ref{fig4-4} and ref.~\cite{Brown85}), 25 for the upper CDW in NbSe$_3$ and 29.4 for the lower one \cite{Richard93}. With estimation of $ne$, all these data converge to ascertain that for CDW $\lambda_{\rm pin}\equiv\lambda_{\rm CDW}$. Thus the previous results \cite{Monceau83} which estimated that the two impurity scattering process proposed by Barnes and Zawadowski was predominant with respect to the single impurity process are not confirmed. The variation of $J_{\rm CDW}/\nu$ with temperature near $T_c$ is similar to that of the CDW order parameter obtained for instance from X-rays studies.

All the models developed so far are considering a two fluid model for which in the non-linear state the normal (quasi particle) current and the DW current are independent. However this model may not be strictly valid. Thus, the non-linear Hall effect observed in o-TaS$_3$ \cite{Artemenko84} was explained by taking into account the effect of a moving CDW on the distribution of quasi particles. For NbSe$_3$ which remains metallic at low temperature, it was proposed that the CDW motion induces a back-flow normal current proportional to $J_{\rm CDW}$ \cite{Richard93}.

\begin{figure}
\begin{center}
\includegraphics[width=6cm]{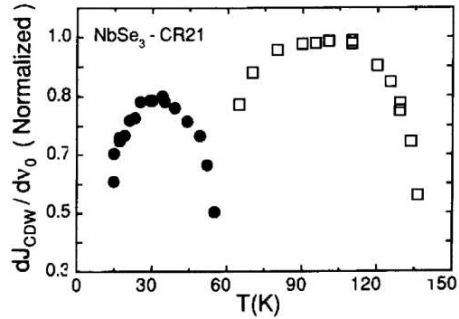}
\caption{Temperature variation of the slope of the CDW current-narrow band noise frequency characteristic (as shown in fig.~\ref{fig4-4}) for both CDW transitions of NbSe$_3$. ${\rm d}J_{\rm CDW}/{\rm d}\nu$ is normalised to the maximum value reached for the upper CDW (reprinted figure with permission from Solid State Communications 85, J. Richard \textit{et al.}, p. 605, 1993 \cite{Richard93}. Copyright (1993) with permission from Elsevier).}
\label{fig4-26}
\end{center}
\end{figure}

This back-low current can be the cause of the decrease of the effective CDW condensate at low temperature for both CDWs in NbSe$_3$ as shown in figure~\ref{fig4-26}. A great care was taken to ascertain a linear $J_{\rm CDW}(\nu)$ variation by pulse measurements with the same crystal immersed in different cryogenic liquids such as argon, nitrogen, neon or hydrogen in order to restrict at minimum any possible heating \cite{Richard93}.

It has also been considered \cite{Rice79} that the effective condensate density which determines the CDW current in the relation $J_{\rm CDW}$~= $\rho_{\rm eff}ev$ differs from the ``bare" condensate density, $\rho_0$, when CDW-normal carriers interaction is taken into account such as:
\begin{equation*}
\rho_{\rm eff}=\rho_0+\rho_n\left(\frac{1}{1+\tau/\tau_K}\right),
\end{equation*}
with $\rho_0$ and $\rho_n$ the condensed and the normal carrier fractions, $\tau_K$ the normal-carrier lattice relaxation time, $\tau$ the normal carrier CDW relaxation time. This normal-carrier-CDW interaction term acts on the total force suffered by the CDW. Temperature dependence of $\tau/\tau_K$ may lead to a temperature dependence of $\rho_{\rm eff}$. Figure~27 in ref.~\cite{McCarten92} shows a temperature dependence of $\rho_{\rm eff}$ for the upper CDW of NbSe$_3$ very similar to that in figure~\ref{fig4-26}. Moreover McCarten \textit{et al.} \cite{McCarten92} have taken argument of this $\rho_{\rm eff}$-temperature dependence to tentatively explain the increase of $E_T$ at low temperature to the detriment of the phase fluctuations approach \cite{Maki86}.

For SDW the determination of $\lambda_p$ has often been hampered by the badly defined fractional cross-sections of the sample involved in the sliding state \cite{Kriza91a,Kriza91b,Nomura89,Sekine04}. However the joint conduction noise and $^{13}$C NMR measurements in the non-linear state can yield the velocity of the SDW (see section~\ref{sec4-4}).

\medskip
\noindent \textit{4.3.2.b. Temporal coherence in the CDW sliding state}
\medskip

In the sliding state the finite width of the NBN frequency obtained by the Fourier analysis of the noise is believed to reflect the distribution of velocities within the sample. However a spectrum analyser generally involves time averaging whereas the amplitude, phase and frequency of the oscillations may fluctuate in time.

From a direct frequency-domain study it was shown that the NBN is dominated by slow temporal velocity fluctuations rather than a time-invariant velocity fluctuations \cite{Link88,Bhattacharya87}. Even in the complete mode-locked state when the NBN drastically becomes narrow, amplitude fluctuations remain \cite{Bhattacharya87}.
Another technique using wavelet analysis has been used \cite{Dumas95,Preobrazhenskii99}. The wavelet analysis allows to reduce strongly time averages to a minimum and to yield informations on short-lived events or intermittence.

The wavelet analysis has been performed on the transient voltage response of the CDW molecular conductor (Perylene)$_2$Pt(Mnt)$_2$ which exhibits quasi-periodic voltage oscillations with a strong amplitude in the response to rectangular current pulses applied above $E_T$ at 4.2~K \cite{Dumas95}.

This wavelet analysis is a powerful technique to reveal time-resolved effects in CDW dynamics, not observable by conventional Fourier analysis such as fast temporal variations of the CDW velocity, non-coherent oscillations in the CDW motion \cite{Preobrazhenskii09} in a short time interval corresponding to transient effects. These time-resolved analysis indicate also strongly that the CDW sliding state is not characterised by a unique periodic steady state, as often in numerical simulations but that local discontinuity in velocity and phase slippage at the domain borders have to be taken into account.

\subsection{NMR spin echo spectroscopy}\label{sec4-4}

\subsubsection{Motional narrowing}\label{sec4-4-1}

The motional narrowing and the appearance of sidebands of the NMR spectra under current are the microscopic evidences of the sliding charge density waves \cite{Ross86,Ross90,Segransan86,Butaud90,Janossy87,Nomura89}. In absence of electric field, the width and the shape of the NMR line are determined by the spatial modulation of the electric field gradient (EFG) tensor at the nuclei sites due to CDW. When the CDW slides with the velocity $v$, the CDW phase $\phi$ evaluate with time $\phi$~= $-\Omega t$ and $\Omega$~= $v/\lambda_{\rm CDW}$; the time average of the EFG at all nuclei becomes equal and the NMR spectrum motionally narrows. Well defined sidebands appear at $\omega$~= $\pm n\Omega$ for a uniform CDW motion as observed in blue bronze \cite{Segransan86,Janossy87}.

\begin{figure}
\begin{center}
\subfigure[]{\label{fig4-27a}
\includegraphics[width=6.75cm]{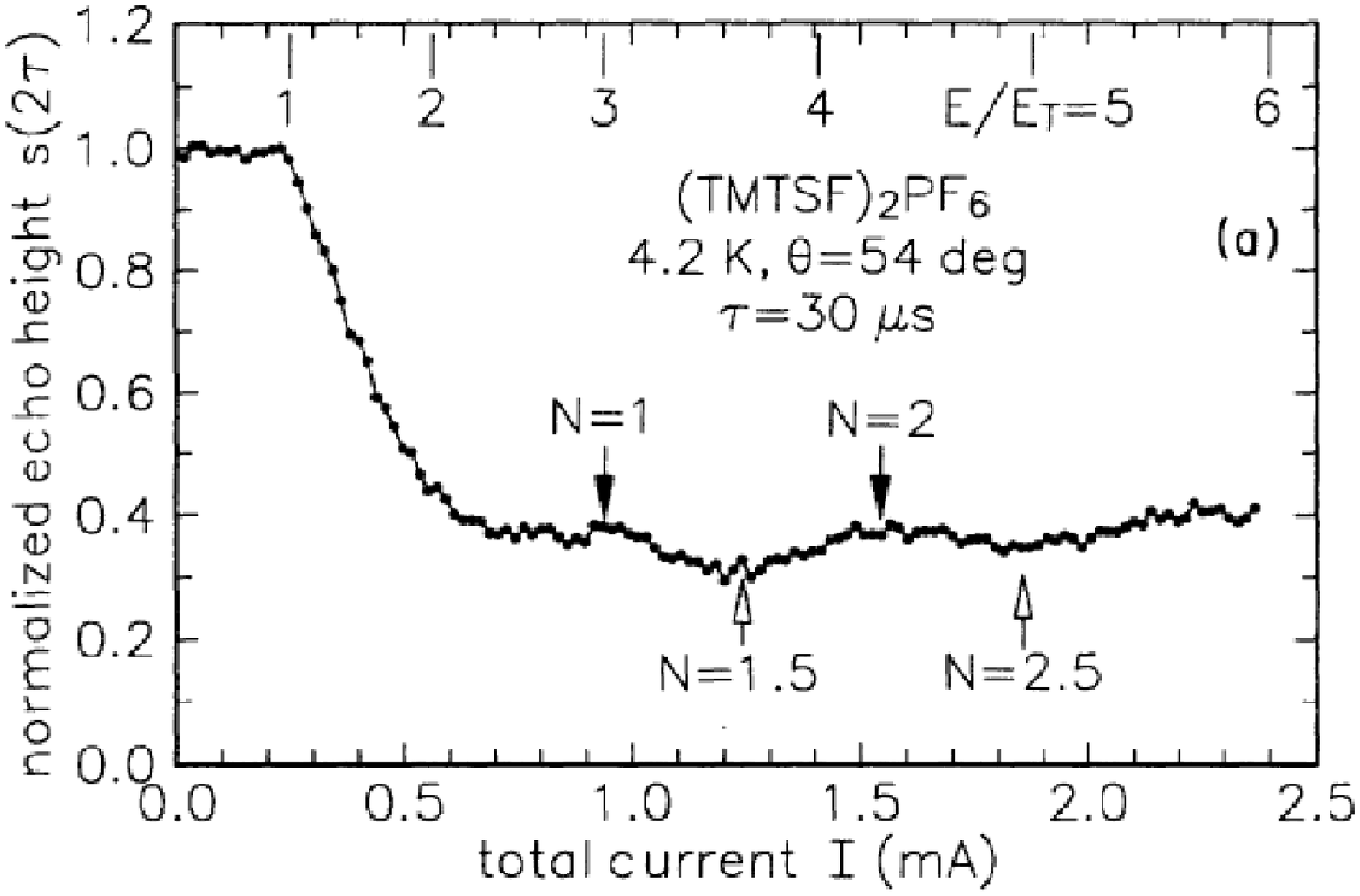}}
\subfigure[]{\label{fig4-27b}
\includegraphics[width=6.5cm]{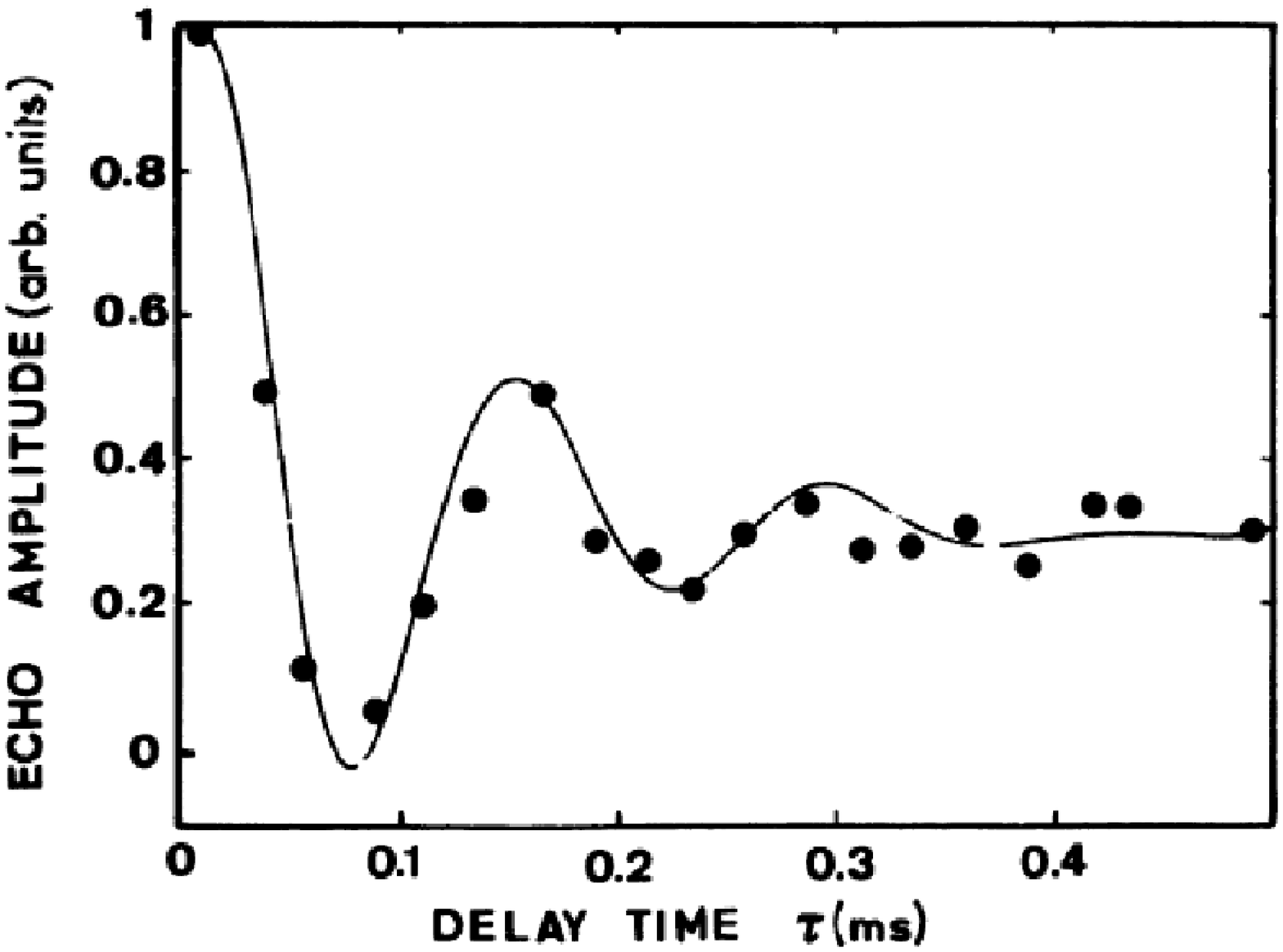}}
\caption{Variation of the amplitude of the spin echo (a)~in (TMTSF)$_2$PF$_6$ as a function of total current at 4.2~K (corresponding values of $E/E_T$ are indicated on the top axis) with a pulse separation of 30~$\umu$s (reprinted figure with permission from W.G. Clark \textit{et al.}, Physical Review B 49, p. 11895, 1994 \cite{Clark94}. Copyright (1994) by the American Physical Society), (b)~in R$_{0.3}$MoO$_3$ at $E/E_T$~= 20 and $T$~= 49~K as a function of the delay time $\tau$ (reprinted figure with permission from P. Butaud \textit{et al.}, Journal de Physique (France) 51, p. 59, 1990 \cite{Butaud90}. Copyright (1990) from EdpSciences).}
\label{fig4-27}
\end{center}
\end{figure}

Similar motional narrowing of the $^{13}$C NMR lines have been measured for SDW, bringing the unambiguous proof of the sliding SDW for current above threshold \cite{Wong93,Barthel93,Clark94}. A method to evaluate the SDW velocity and to compare it with NBN measurements is obtained from measurements of the modulation of the spin-echo by the sliding SDW as a function of the delay time. After application of a $\frac{\pi}{2}-\tau-\pi$ rf pulse sequence, the spin-echo signal with height $h(2\tau)$ is recorded at the time $2\tau$ after the first pulse as a function of the current. The echo amplitude oscillations of the spin echo from the protons in the (TMTSF) molecules of (TMTSF)$_2$PF$_6$ is shown in figure~\ref{fig4-27}(a) as a function of the total applied current at 4.2~K for a fixed delay time $\tau$ \cite{Clark94}. The oscillation index, when plotted as a function of $\tau I_{\rm SDW}$ shows a linear dependence with the slope $(neA\lambda_{\rm SDW})^{-1}$, $A$ the cross-section of the sample (for a uniform SDW velocity the maximum echo height should reach the value without SDW conduction). For comparison in figure~\ref{fig4-27}(b) is plotted the variation of the spin echo amplitude as a function of the delay time $\tau$ for a Rb$_{0.30}$MoO$_3$ sample at $E/E_T\sim20$ and $T$~= 49~K which emphasises  the time coherence of the CDW phase \cite{Butaud90}. The dip minimum observed at $\tau$~= 80~$\umu$s corresponds to the resonance condition between $\tau$ and the velocity of the CDW.

By observing the coherent temporal variation of the local magnetic (SDW) or electric field (CDW) at the nuclei sites, NMR is directly measuring the phase winding rate: $\nu_d$~= $1/2\pi\;{\rm d}\varphi/{\rm d}t$ and the C/S DW velocity is given by $v_s$~= $\lambda_{\rm C/S DW}\nu_d$. From transport properties, the C/S DW velocity is derived from the ac voltage frequency (NBN) times the pinning wavelength. The two experiments combining spin-echo and conduction noise found divergent results: $\lambda_p$~= $\lambda_{\rm SDW}/2$ in ref.~\cite{Clark94} and $\lambda_p$~= $\lambda_{\rm SDW}$ in ref.~\cite{Barthel93}. A possible origin of this discrepancy may be the difficulty to have an estimation of the cross-section area in the sliding state and also to take into account the very broad SDW velocity distribution in the samples. However the experimental determination of the $\lambda_{\rm SDW}/2$ wavelength for the pinning fits well with the so small pinning observed in SDW materials which is accounted for the second order in the impurity scattering.

\subsubsection{Phase displacement of the CDW below $E_T$}\label{sec4-4-2}

\begin{figure}
\begin{center}
\includegraphics[width=10cm]{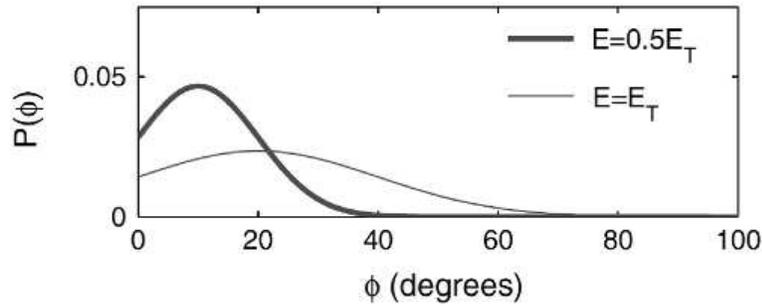}
\caption{Distribution of CDW phase displacements in NbSe$_3$ evaluated from the spin-echo decay curves in pressure of electric field pulses $E$~= $1/2\,E_T$ bold line) and $E$~= $E_T$ (thin lines). $T$~= 130~K (reprinted figure with permission from S. Suh \textit{et al.}, Physical Review Letters 101, p. 136407, 2008 \cite{Suh08}. Copyright (2008) by the American Physical Society).}
\label{fig4-28}
\end{center}
\end{figure}

Electric field-induced phase displacements of a NbSe$_3$ single crystal were measured using the same $^{93}$Nb NMR spin echo technique for applied electric field lower than $E_T$ \cite{Suh08}. If the CDW displaces in response to the electric field, the local environment of each $^{93}$Nb nucleus will change by an amount depending upon the coupling to the CDW and the magnitude of the local displacement. The displacement of the CDW will cause a destructive phasing in coherently precessing spins and induce a decay of the echo amplitude. Decay of the echo amplitude was measured as a function of the width $\tau_1$ of an electric field pulse of amplitude $E$~= $\alpha E_T$ applied between a $\frac{\pi}{2}-\pi$ rf pulse sequence. From the echo decay the phase displacements were evaluated exhibiting a broad distribution \cite{Suh08} as shown in figure~\ref{fig4-28}. 
The phase displacement exhibits a linear response with $E\leq E_T$; doubling $E$ from 0.5~$E_T$ to $E_T$ produces an echo that decays roughly twice as fast, corresponding to a phase displacement distribution with twice the mean value. The main and width of the phase distribution are 10 and 22~deg at $E$~= 0.5~$E_T$ and 20 and 44~deg at $E$~= $E_T$. Previous reports indicated a phase shift of 2~deg at $E$~= 0.75~$E_T$ \cite{Ross86}. The more likely cause of this difference may lead on the use of a single crystal with a uniform cross-section rather a multicrystalline sample formed with many crystals with probably different threshold.

These results show no evidence for diverging polarisation near threshold that might be expected if CDW depinning were a dynamical critical phenomenon. There is the possibility that the critical regime is confined in the externe vicinity of $E_T$ ($\ll 0.1~E_T$) around threshold and that the response outside is linear.

\subsection{CDW long range order}\label{sec4-5}

\subsubsection{CDW domain size}\label{sec4-5-1}

The influence of a random distribution of impurities on the CDW order can be studied by X-ray scattering experiments. Defects destroy the long range order and the satellite reflections at $\pm Q$ broaden. The intrinsic domain size of the CDW ordering is extracted by fitting the experimental scans by a convolution of a Gaussian experimental resolution and a Lorentzian-squared profile (sometimes a Lorentzian profile is also used). The half-width half maximum (HWHM) is inversely proportional to the CDW correlation length (see section~\ref{sec3}) ($\ell$
= $1/HWHM$ for a Lorentzian profile, $\ell$~= $0.6436/HWHM$ for a Lorentzian squared profile).

High resolutions X-ray scattering have been performed on NbSe$_3$ satellites of the high temperature CDW of samples doped with  isoelectronic Ta (0.25\% and 2\%) and non isoelectronic Ti (1\%) impurities \cite{Rouziere99}; in electron irradiated blue bronze \cite{DeLand91}, and on V-doped (0.28\%, 1.44\% and 2.8\% and W-doped (2\%) blue bronze \cite{Rouziere96}. Typically at 20~K the CDW coherence lengths 
$\ell_{b^\ast}$, $\ell_a$ and $\ell_{c^\ast}$ are respectively 27, 14 and 7~$\AA$ for the Ti-doped NbSe$_3$ and 170, 40 and 15~$\AA$ for the 2\% Ta-doped NbSe$_3$, much less than the values 4000, 1350 and 300~$\AA$ measured for a 0.25\%-doped NbSe$_3$ \cite{DiCarlo94}. It is thus found  that the domain of coherence is more isotropic for the Ti-doped samples than for the Ta-doped samples.

From the values of coherence lengths the number of impurities per domain can be evaluated. It is found that typically for Ti impurity, each CDW domain in NbSe$_3$ contains of the order of one impurity or less (that depends if the Ti atoms are substituted on all the chains or preferentially on a given type of chains). The Ti impurities act as strong pinning centres. For the 2\% Ta-doped NbSe$_3$, it is estimated that each CDW domain contains between 8 or 24 Ta atoms depending similarly on the number of chains involved in the Ta substitution. This results indicate that, even for 2\% of doping, Ta impurities are weak pinning centres. That is all the more true for less impurity concentration. These coherence distances obtained by X-rays are worth to be compared with the thickness of the samples where size effects on the amplitude of the threshold field were reported \cite{McCarten92}.

Similar results were obtained for doped blue bronze: more isotropic domains than for undoped sample \cite{Girault88,Rouziere97}, strong pinning for non-isoelectronic V-doping with a single impurity per domain, weak pinning for isoelectronic W-doping with around 20 impurities in the coherent domain.

\subsubsection{Friedel oscillations and CDW pinning}\label{sec4-5-2}

Moreover in V-doped blue bronze, detailed analysis of the superlattice profiles have given access to the local properties of the CDW and on the microscopic nature of the pinning. It has been shown that when a spatial coherence between the impurity positions and the CDW periodic lattice distortion exists, interferences occur between the wave scattered by the impurity (taken as a reference wave) and the wave scattered by the atomic displacements originated by the same impurity \cite{Rouziere00,Ravy06}. Such interferences produce an asymmetry of the intensity within the $\pm 2k_{\rm F}$ pair of satellite reflections from which the phase of the periodic lattice distortion at the impurity site can be extracted. Such strong intensity asymmetries (called ``white-line effect" \cite{Brazovskii97}) have been observed in organic conductors \cite{Rouziere97}. A second effect is also observed where phase deformations around impurities resulting from Friedel oscillation phase shift lead to a profile asymmetry of the CDW satellite.

\begin{figure}
\begin{center}
\subfigure[]{\label{fig4-29a}
\includegraphics[width=6.5cm]{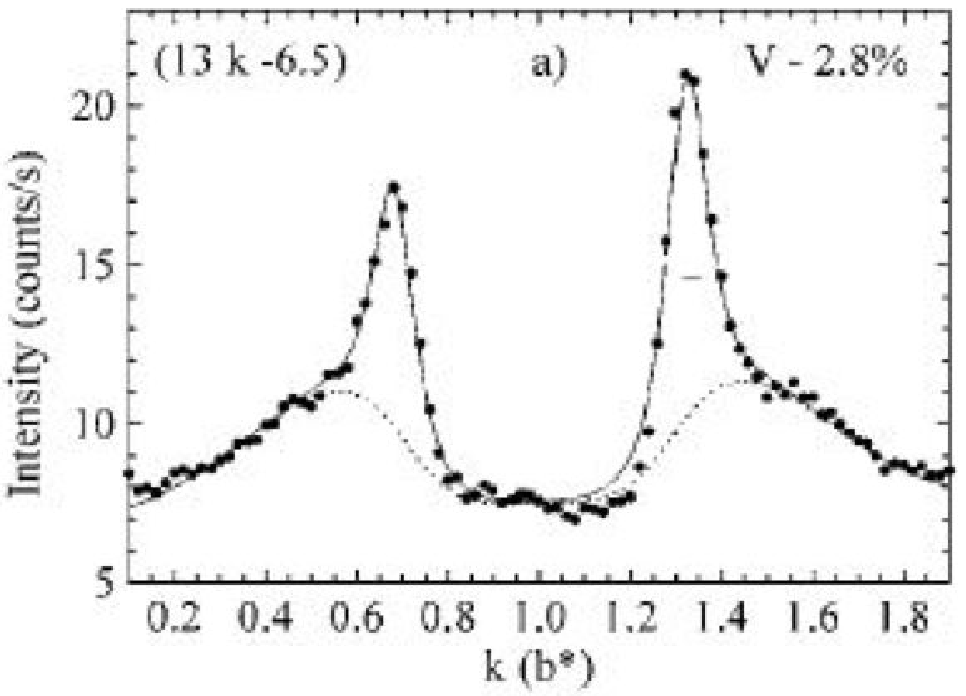}}
\subfigure[]{\label{fig4-29b}
\includegraphics[width=6.5cm]{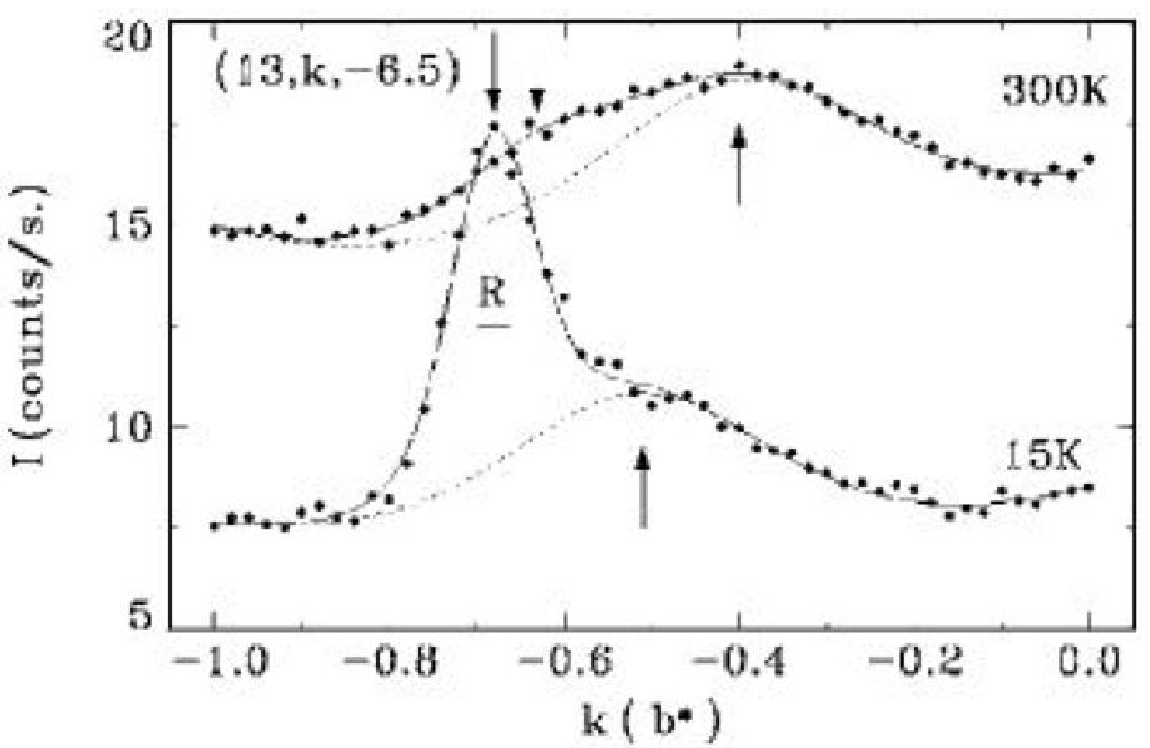}}
\caption{(a)~Transverse $k$ scan through the satellite positions (13, +2$k_{\rm F}$, 6.5) and (13, 2-2$k_{\rm F}$, 6.5) at $T$~= 15~K for V-2.8\% doped blue bronze crystal. (b)~Transverse $k$ scans through the satellite position (13, -2$k_{\rm F}$, $\overline{6.5}$) and the secondary scattering (13, $q_s$, $\overline{6.5}$) at $T$~= 15 and 300~K for a V-2.8\% doped blue bronze crystal: position of satellite reflection (downward arrow), position of secondary scattering (upward arrow) (reprinted figure with permission from S. Ravy \textit{et al.}, Physical Review B 74, p. 174102, 2006 \cite{Ravy06}. Copyright (2006) by the American Physical Society).}
\label{fig4-29}
\end{center}
\end{figure}

In figure~\ref{fig4-29}(a) is plotted a typical $k$ scan around the reciprocal position (13, 1, $\overline{6.5}$) at $T$~= 15~K in the V-2.8\% doped blue bronze sample. One can note the asymmetry of the satellite reflections with a stronger intensity in the $b^\ast$ direction towards the associated Bragg reflection position. The visible profile asymmetry is due to the presence of a second scattering at $q_s$ close to the satellite reflection as shown in more detail in figure~\ref{fig4-29}(b) which displays the right side of the scan in (a). The intensity of the secondary scattering remains constant even at room temperature, while the diffuse scattering located at $2k_{\rm F}$ disappears upon heating. A model has been developed in which the $q_s$ scattering corresponds to the Fourier transform of the Friedel oscillations around charged impurities \cite{Rouziere99,Ravy06}. The V$^{5+}$ atom provides a negative charge ($Z$~= -1) with respect to the molybdenum Mo$^{6+}$ background. This charge has to be screened by a hole requiring a total phase shift of the wave  function according to the Friedel sum rule.

In W-2\% doped blue bronze samples a slight profile asymmetry indicating a small phase deformation around impurities has been observed, although impurities are in limit of weak pinning. That may mean that W impurities, even isoelectronic, bring a small charge to be screened. That may also be the case for Ti doping in NbSe$_3$ where no asymmetry was found, that due to the too small impurity concentration to give rise to the effect.

\begin{figure}
\begin{center}
\includegraphics[width=8cm]{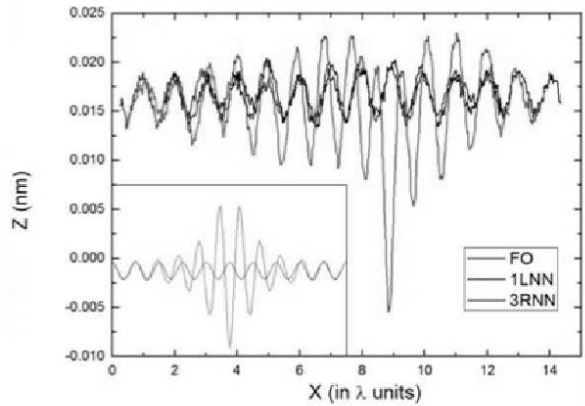}
\caption{From a STM image of the (b,c) plane of NbSe$_3$ measured at 110~K showing a defect giving rise to 1D Friedel oscillations, corresponding profiles along the chain hosting the defect and along the first left-nearest-neighbour (1 LNN) and third right-nearest-neighbour (3 RNN) of this defected chain. Inset: theoretical profile of a pure Friedel oscillation in presence of the CDW which modulation is recovered after a few CDW periods, as experimentally found (reprinted figure with permission from S. Brazovskii \textit{et al.}, Physical Review Letters 108, p. 096801, 2012 \cite{Brazovskii11}. Copyright (2012) by the American Physical Society).}
\label{fig4-sup}
\end{center}
\end{figure}

The experimental evidence of the scattering due to Friedel oscillations and of the pinning mechanism at the impurity site confirms nicely the theoretical work of T\"utto and Zawadowskii \cite{Tutto85}. The matching between Friedel oscillations and CDW at a single impurity site has been recently revealed on STM images \cite{Brazovskii11}, as shown in figure~\ref{fig4-sup}.

\subsection{Electromechanical effects}\label{sec4-6}

In addition to the non-linear transport properties induced by CDW sliding, it was shown that the depinning of the CDW induces elastic softening of the host crystal, demonstrating the intimate interaction between the CDW and the lattice. Experiments were performed on CDW whiskers with flexural and torsional resonant techniques in which the resonant frequency is proportional to the square root of an elastic stiffness modulus \cite{Brill84,Mozurkewich85,Xiang89}. The measurements with frequencies between 100~Hz and 1~kHz showed that at $E>E_T$ the relative decrease in shear modulus ($G$) was about an order of magnitude more than that in Young modulus ($Y$), implying that interchain coupling is greatly affected by the CDW motion. In o-TaS$_3$, the shear modulus ($G$) decreases \cite{Xiang87} by 20\% above $E_T$ while $Y$ decreases \cite{Brill86} by 1-2\%. The softening in $G$ and $Y$ is accompanied by an increase of the internal friction: $\Delta(1/Q)$ [$Q$: the quality factor]. In NbSe$_3$, the changes are much smaller and strongly sample dependent (\cite{Xiang89} and references therein).

Extension of the vibrating reed technique up to 100~kHz by studying flexural overtone showed that, above 1~kHz, the Young's modulus softening decreases such as: $|\Delta Y/Y|\propto\omega^{-3/4}$; this frequency dependence is reminiscent to that of the ac conductivity: $\omega^\alpha$ \cite{Wu86}. On the other side, quasi-static measurements \cite{Tritt91,Maclean92} of the Young modulus using a stress-strain device showed the absence of softening of $Y$ in the extreme low frequency limit.

Two approaches have been used in attempting to account for these elastic changes (for a review see \cite{Mozurkewich92,Brill01}). The first model \cite{Brill84} suggested that the measured softening was due to relaxation of some defects in the CDW. It was then proposed \cite{Mozurkewich90} that the relaxing quantity was the phase configuration (FLR-type domains) of the CDW, the relaxation strength being determined by the strain dependence of the CDW wave vector. Sliding motion facilitates phase reconfiguration: when the CDW is pinned, the average time relation $\tau\gg 1/\omega$, so the measured modulus is unrelaxed. When the CDW slides, $\tau$ decreases and the sample can relax; in the static limit, the modulus is always relaxed and no anomaly is expected (as experimentally found). Thus the frequency dependence of the Young modulus should present a maximum in the same arbitrary low frequency value. On the other hand, elastic anomalies were discussed \cite{Maki87} in terms of the ability of the CDW to screen phonons; while the pinned CDW could not contribute to screening, the depinned CDW could. The relative decrease in modulus with CDW depinning is then proportional to the relevant electron-phonon coupling constant \cite{Maki87}. In both models, either a large distribution of relaxation times or a distribution of threshold fields are necessary to account for the experimental data.

\begin{figure}
\begin{center}
\subfigure[]{\label{fig4-31a}
\includegraphics[width=4.5cm]{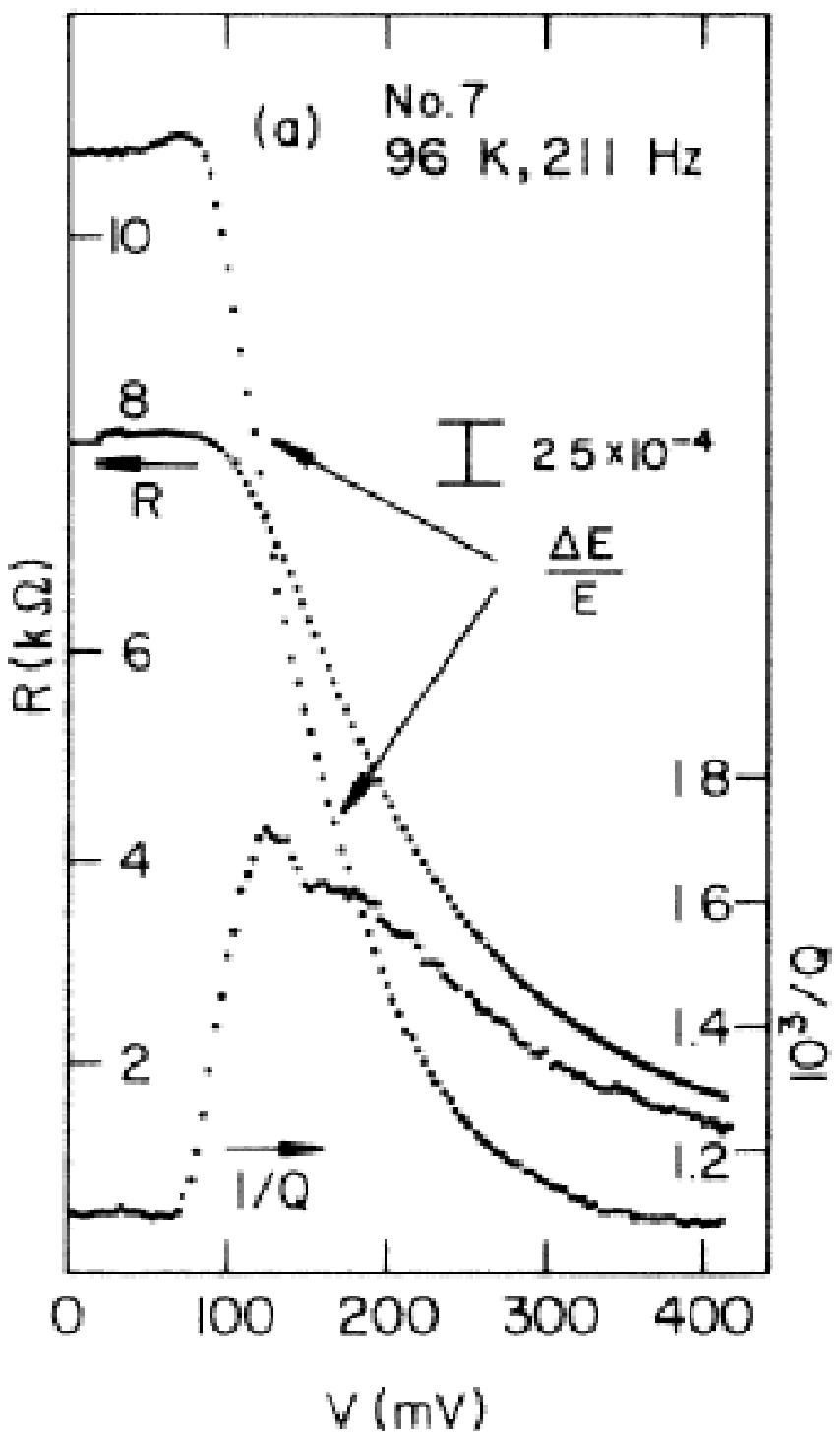}}
\subfigure[]{\label{fig4-31b}
\includegraphics[width=6.5cm]{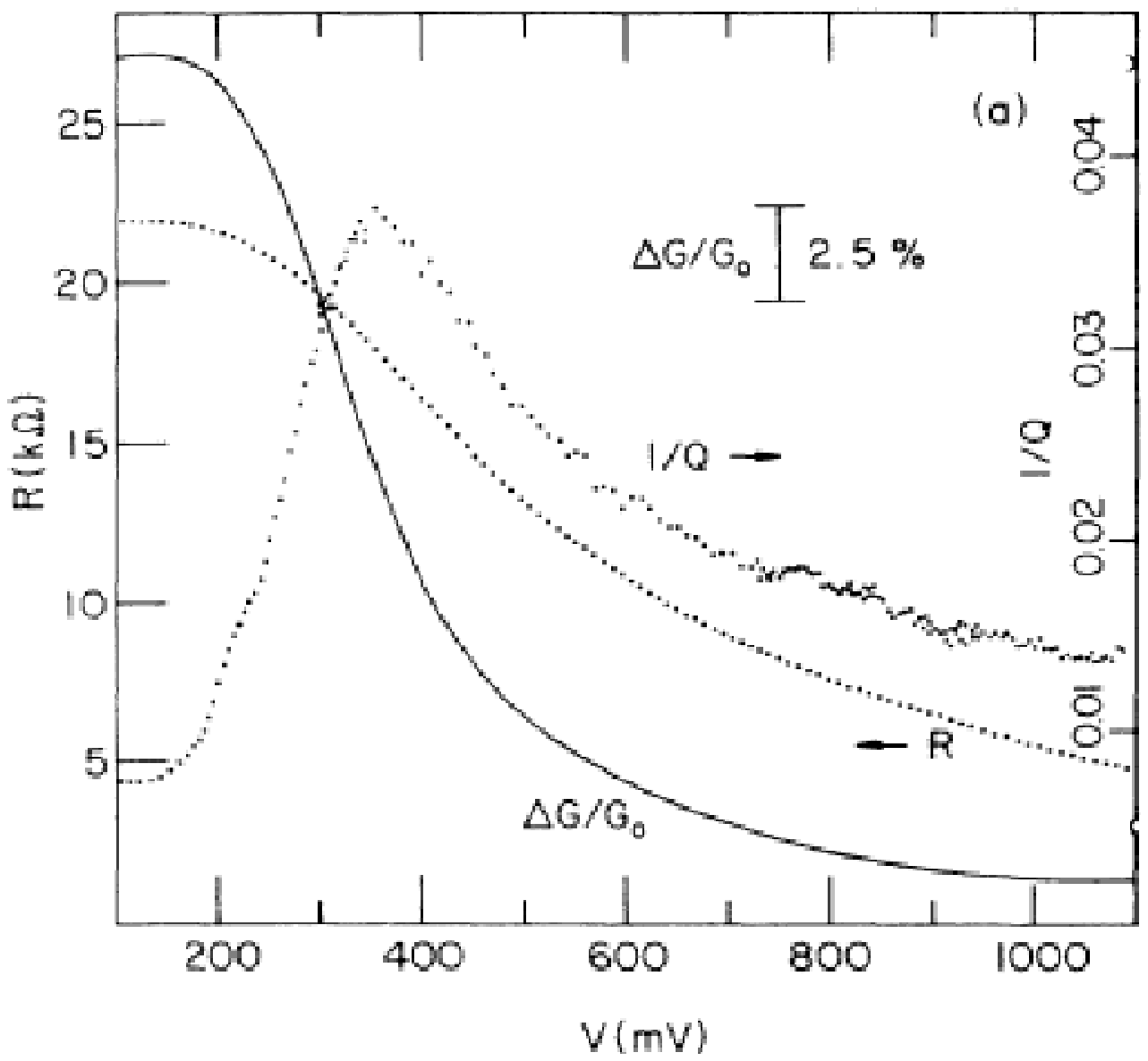}}
\subfigure[]{\label{fig4-31c}
\includegraphics[width=8.25cm]{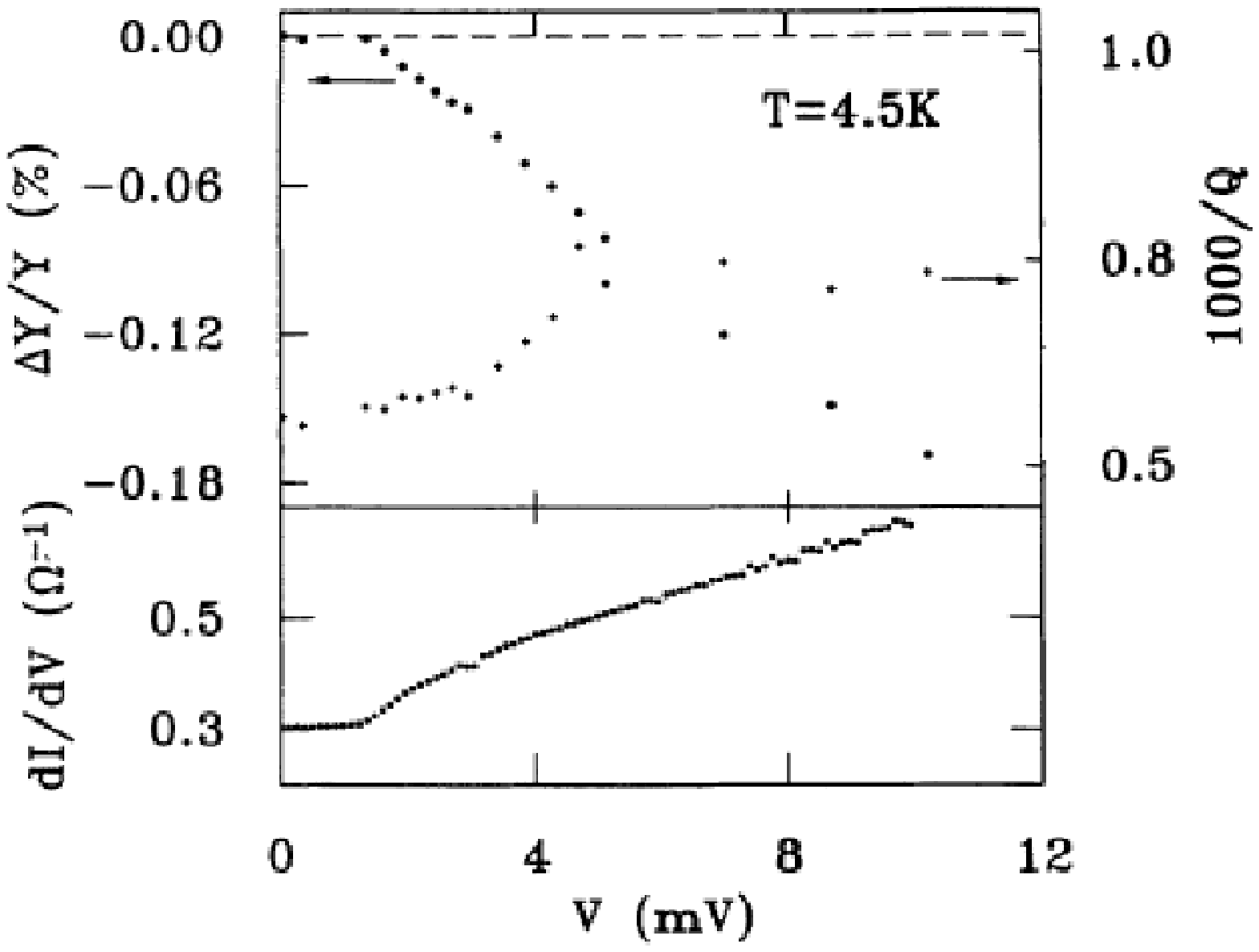}}
\caption{Variation of resistance, Young ($Y$) or shear ($G$) modulus and internal friction $10^3/Q$. a)~$Y$; o-TaS$_3$; $f$~= 211~Hz; $T$~= 96~K (reprinted figure with permission from J.W. Brill \textit{et al.}, Physical Review B 33, p. 6831, 1986 \cite{Brill86}. Copyright (1986) by the American Physical Society). b)~$G$; o-TaS$_3$; $f$~= 220~Hz; $T$~= 103~K (reprinted figure with permission from X.-D. Xiang and J.W. Brill, Physical Review B 36, p. 2969, 1987 \cite{Xiang87}. Copyright (1987) by the American Physical Society). c)~$Y$; (TMTSF)$_2$PF$_6$; $f$~= 300~Hz; $T$~= 4.5~K (reprinted figure with permission from S.E. Brown \textit{et al.}, Physical Review B 46, p. 1874, 1992 \cite{Brown92}. Copyright (1992) by the American Physical Society).}
\label{fig4-31}
\end{center}
\end{figure}

Similar softening of the Young's modulus and increase in the internal friction was observed \cite{Brown92} in (TMTSF)$_2$PF$_6$, demonstrating that the depinning of the SDW has a profound effect on the lattice. Variations of resistance and modulus $Y$ and $G$ and associated internal friction of o-TaS$_3$ and (TMTSF)$_2$PF$_6$ as a function of the applied voltage are drawn in figure~\ref{fig4-31}(a,b,c).

\begin{figure}
\begin{center}
\includegraphics[width=7.5cm]{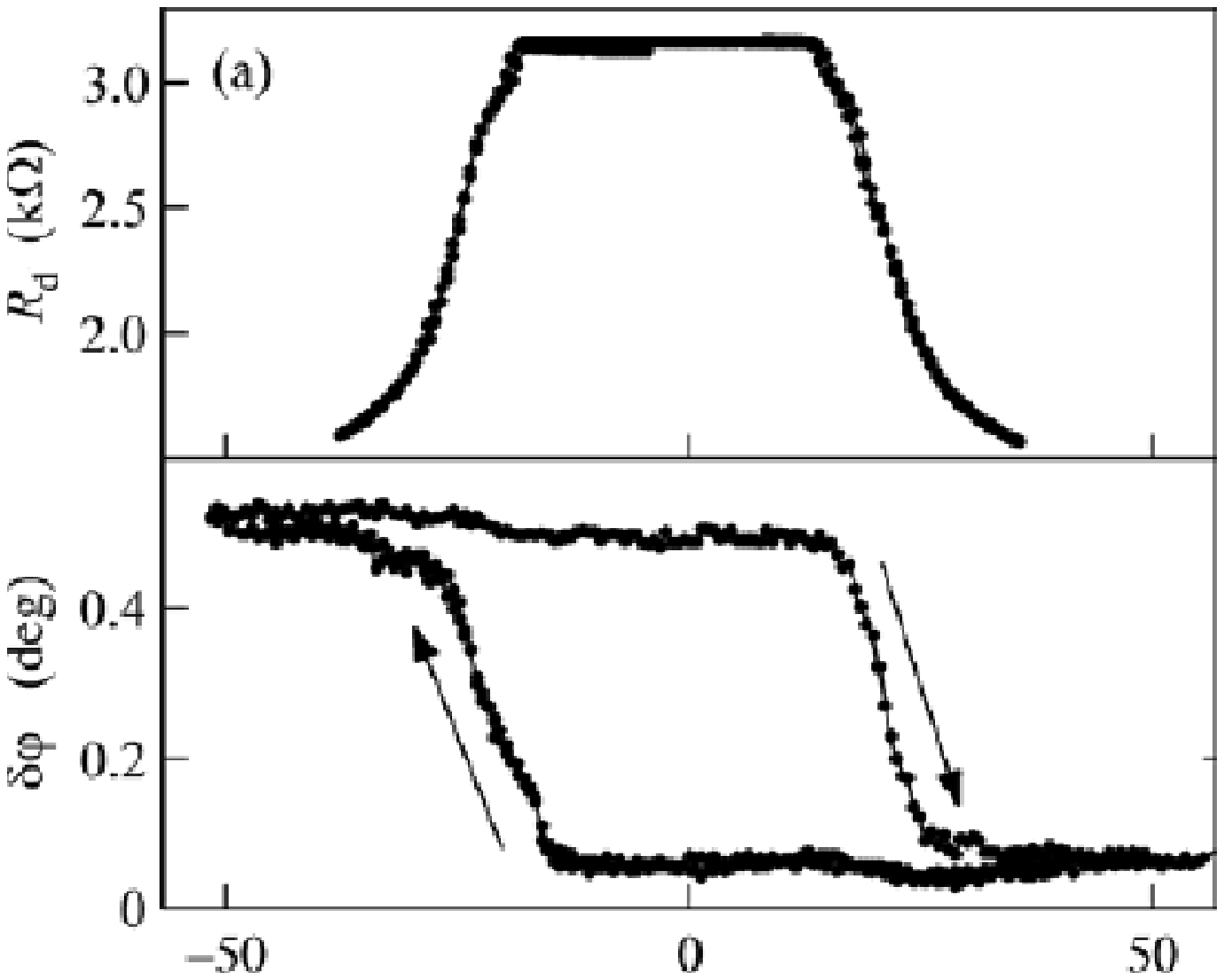}
\caption{Variation of the rotation angle $\delta\phi(I)$ and differential resistance $R_d(I)$ measured simultaneously of a o-TaS$_3$ sample with one contact freely suspended (reprinted figure with permission from JETP Letters 92, S.G. Zybtsev \textit{et al.}, p. 405, 2010 \cite{Zybtsev10}. Copyright (2010) from Springer Science and Business media) at $T$~= 121~K.}
\label{fig4-32}
\end{center}
\end{figure}

Deformations of the host crystal resulting from CDW deformations have been observed as a change of the length of the crystal by application of an electric field \cite{Hoen92} or by thermal cycling \cite{Golovnya02} or by a rotation \cite{Pokrovskii07}. The torsional strain under an electric field was studied optically, the deformation being traced by the deflection of a laser beam reflected from micromirrors stuck to the sample \cite{Pokrovskii07}. When a current near threshold is applied to a crystal of o-TaS$_3$ which has a contact mechanically clamped and the other contact freely suspended, it was shown that the free end rotates with respect to the fixed end by an angle $\delta\phi\sim 1^\circ$ \cite{Pokrovskii07,Day07} demonstrating the occurrence of a torsional deformation. The twist direction reverses for a current of the opposite polarity which yields hysteresis loops \cite{Pokrovskii07,Zybtsev10,Nichols09} in $\delta\phi(I)$ about $0.5^\circ$ wide as shown in figure~\ref{fig4-32}. 
The $\delta\phi(I)$ saturates rapidly above the threshold current and the torsion angle remains almost unchanged when the current is removed from the sample. This spectacular effect was attributed to surface pinning which can give rise to shear deformations of the CDW in the planes parallel to the $c$-axis. The giant torsional strain corresponds \cite{Pokrovskii07} to shear exceeding $10^{-4}$ at the surface and a piezoelectric coefficient of $10^{-4}$~cm/V, a value much larger than that of piezoelectric materials. Dynamics of this voltage-induced torsional strain appear to be very sample dependent \cite{Nichols09,Nichols10} which points out the role of extended lattice defects in this torsional strain.

\section{Phase slippage}\label{sec5}
\setcounter{figure}{0}
\setcounter{equation}{0}

In preceding  sections the essential assumption of models of Fukuyama-Lee-Rice type is that the CDW deformations are approximately elastic. While the energy of the long wavelength phase excitations can be arbitrary low, the energy of the amplitude fluctuations is much higher, being of the order of magnitude of the CDW condensation energy. They were therefore neglected and the CDW was considered as an elastic medium having only phase degrees of freedom. But FLR models fail in many cases resulting of the CDW being solely assumed to behave elastically, essentially when at strong and isolated crystal defects (electric electrodes, grain boundaries, \dots) the CDW velocity passes through sharp discontinuities. The conflict between different phase winding rates at the interface between regions with different velocities should be relieved by phase slippage, i.e. by the creation of singularities as vortices in superfluids \cite{Anderson66}. In a path surrounding these vortices, the phase changes by $2\pi$:
\begin{equation}
\oint\triangledown\varphi\,\bm{d\ell}=2\pi.
\label{eq5-1}
\end{equation}

\subsection{Static CDW dislocations}\label{sec5-1}
Feinberg and Friedel \cite{Feinberg88,Feinberg89} have developed a theory of elasticity and plasticity of CDWs in which the deformations of the pinned CDW under an electric field are compared with mechanical deformations of crystals under external stresses \cite{Dumas86}. However this description, in fact, assumes that all long range Coulomb degrees of freedom are screened by either normal carriers or thermally excited carriers above the gap for systems with a semiconducting ground state at low temperature. Thus, such a picture breaks down at very low temperature in the insulating state.

\subsubsection{Elastic deformations}\label{sec5-1-1}

They introduce a local displacement of the CDW
\begin{equation*}
u_x=\phi/Q,
\end{equation*}
parallel to the chains (x axis) equivalent to a phase shift $\phi$. They define the distorted CDW by its strain and stress fields such as:

\begin{itemize}
\item[-] Strain tensor [$e$] given by:
\begin{equation}
e_{\rm xx}=\frac{1}{Q}\,\frac{\upartial\phi}{\upartial {\rm x}},\quad e_{\rm xy}=e_{\rm yx}=\frac{1}{2Q}\,\frac{\upartial\phi}{\upartial {\rm y}},\quad e_{\rm xz}=e_{\rm zx}=\frac{1}{2Q}\,\frac{\upartial\phi}{\upartial {\rm z}}.
\label{eq5-2}
\end{equation}

\item[-] Stress tensor [$\Sigma$] given by:
\begin{equation}
\Sigma_{\rm xx}=Q^2K_xe_{\rm xx},~\Sigma_{\rm yx}=\Sigma_{\rm yx}=Q^2K_ye_{\rm xy},~\Sigma_{\rm xz}=\Sigma_{\rm zx}=Q^2K_{\rm z}e_{\rm xz}.
\label{eq5-3}
\end{equation}
\end{itemize}

The elastic energy stored by the CDW distortions is then:
\begin{equation}
U_{\rm ela}=\frac{1}{2}\left[\Sigma\right] \left[e\right].
\label{eq5-4}
\end{equation}
Two types of distortions can be defined in the long wavelength limit. The first type corresponds to a compression or a dilatation involving a longitudinal strain $e_{\rm xx}$ such $e_{\rm xx}$~= $\delta\lambda/\lambda$. The variation of the Fermi vector ($k_{\rm F}$~= $\sfrac{1}{2}\, Q$) creates an uncompensated charge density given by:
\begin{equation}
\delta\rho_s=\frac{n_se}{Q}\,\frac{\upartial\varphi}{\upartial {\rm x}}=\,-(n_se)e_{\rm xx}.
\label{eq5-5}
\end{equation}
The second type of distortion is a neutral shear involving $e_{\rm xy}$ and $e_{\rm xz}$.

\subsubsection{CDW dislocations}\label{CDWdislocations}

The topological defects of incommensurate CDWs, or dislocations, are very similar to vortices in superfluids and their properties are close to those of dislocation lines in crystals \cite{Feinberg88,Feinberg89}.

The Burgers vector of CDW dislocations parallel to the chain axis is given by:
\begin{equation*}
B=\oint\triangledown u\,dr=-\frac{x}{Q}\oint_c\triangledown\varphi\,dr.
\end{equation*}
There are simple cases of dislocations as shown in figure~\ref{fig5-1}:
\begin{figure}
\begin{center}
\subfigure[]{\label{fig5-1a}
\includegraphics[width=5.5cm]{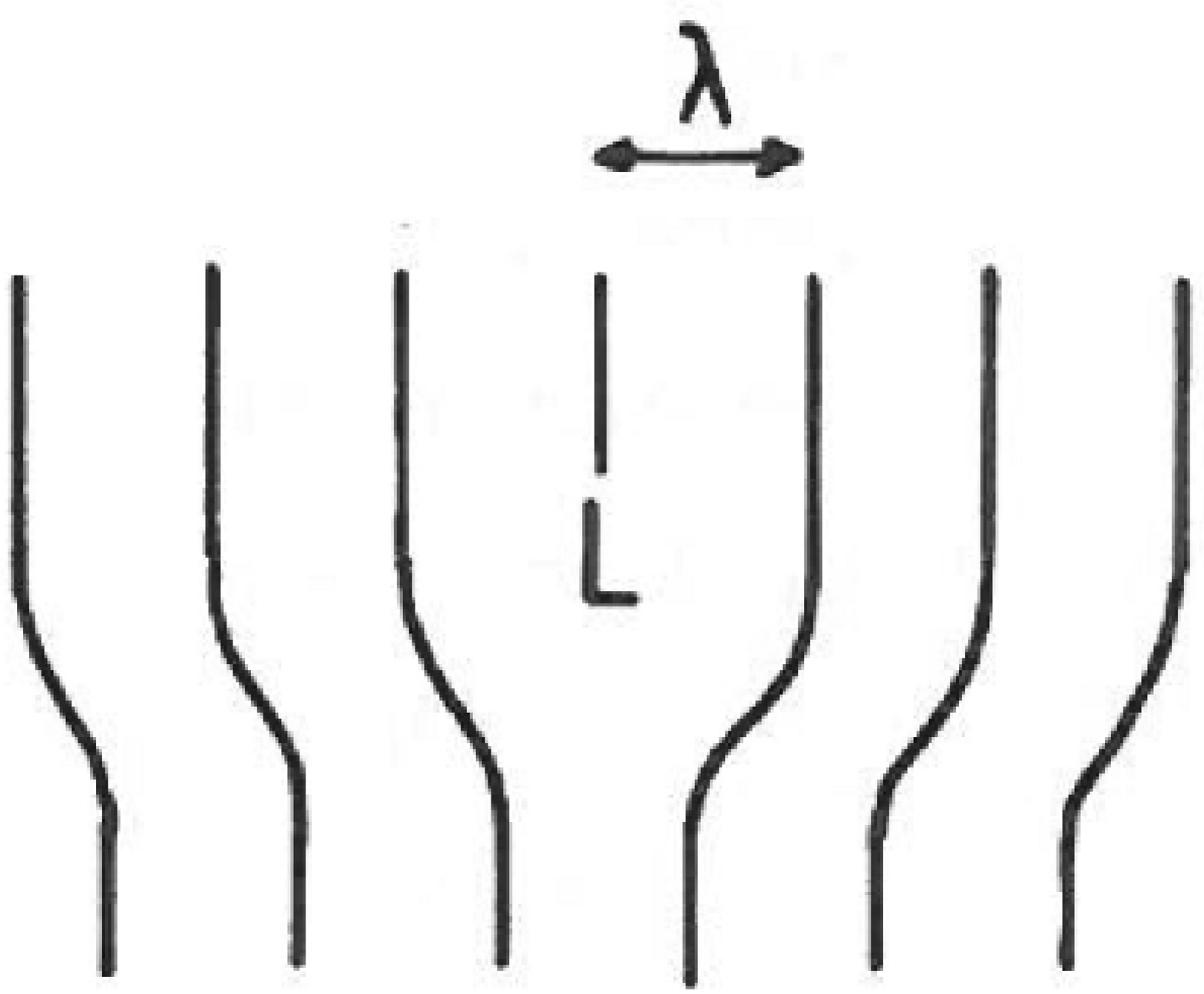}}
\subfigure[]{\label{fig5-1b}
\includegraphics[width=6.5cm]{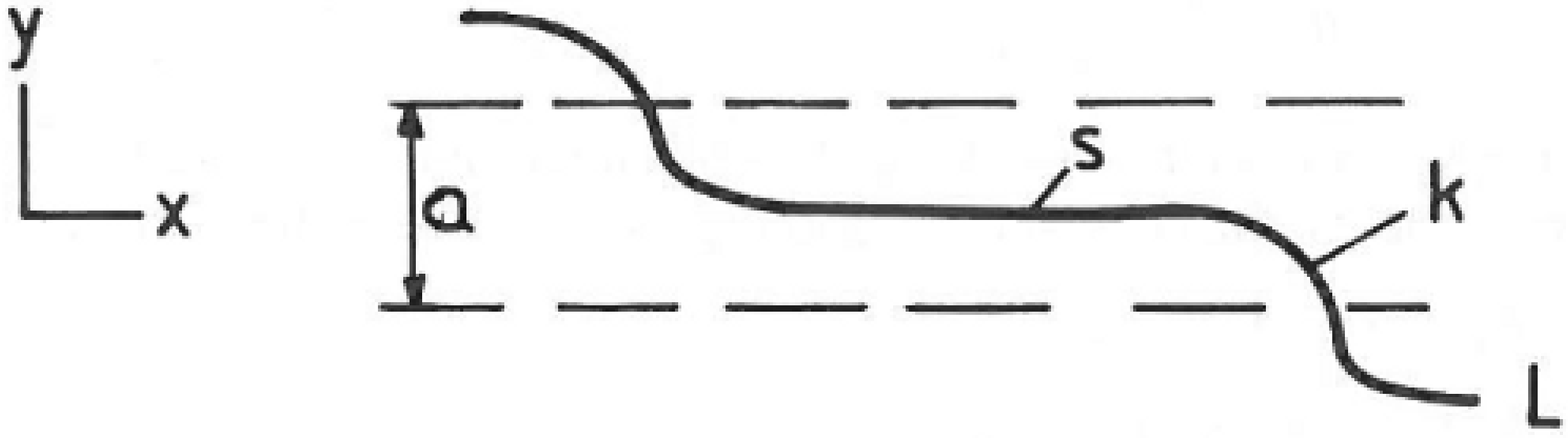}}
\caption{(a)~Edge dislocation line $L$ of a CDW, (b)~Kinked dislocation line $L$ on a glide plane parallel to the chains ($S$: screw part, $k$: kink).}
\label{fig5-1}
\end{center}
\end{figure}
screw dislocations (parallel to the chains) and edge dislocations (perpendicular to the chains). Screw dislocations involve only shear and have smaller energies than edge ones which involve compressions and dilatations. That results from the strong anisotropy of the dislocation core with dimensions $\xi_\perp\xi_\perp$ for a screw dislocation, and $\xi_\perp\xi_\parallel$ for a edge one, $\xi$ being the BCS amplitude coherence length. Thus the screw dislocation costs a $\xi_\perp/\xi_x$ less energy.

Edge dislocation loop introduces a dipolar charge redistribution, while the screw dislocation loop keeps the charge neutrality. It was also shown that, due to the discreetness of the lattice a screw dislocation can be transformed spontaneously into a zigzag form with screw parts connected by kinks (see figure~\ref{fig5-1}(b)). The question of the stability of these dislocations in the static regime i.e. without CDW collective transport has been addressed \cite{Bjelis89}. It was concluded that a dislocation line or a dislocation loop cannot survive in the strainless environment. The transverse motion of a screw dislocation can be stopped due to the chain discreetness, especially when the transverse amplitude coherence length $\xi_\perp$ is of the same order than the interchain distance. Another mechanism is the pinning of these static dislocations at some defects.

\subsection{Phase slips and dislocations in the sliding CDW state}\label{sec5-2}

\subsubsection{Surface pinning}\label{sec5-2-1}

\begin{figure}
\begin{center}
\includegraphics[width=7.5cm]{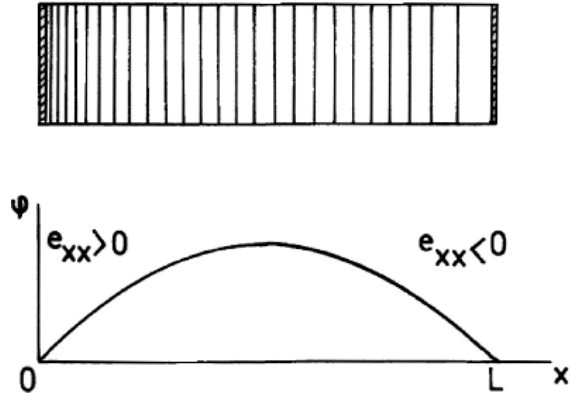}
\caption{Boundary (contact) pinning in a one-dimensional configuration (reprinted figure with permission from D. Feinberg and J. Friedel, Journal de Physique (France) 49, p. 485, 1988 \cite{Feinberg88}. Copyright (1988) from EdpSciences).}
\label{fig5-2}
\end{center}
\end{figure}

The most direct way for initiating nucleation of dislocations is to apply a strong enough external force, essentially an electric field. Keeping only the elastic part, with application of an electric field between electrodes located at $x$~= 0 and $x=L$, the Lee-Rice Hamiltonian for CDWs in a 1D configuration can be written as:
\begin{equation}
K_x\,\frac{d^2\varphi}{dx^2}+\frac{n_seE}{Q}=0,
\label{eq5-6}
\end{equation}
with $\varphi(0)$~= $\varphi(L)$~= 0. Then:
\begin{equation}
\varphi(x)=\frac{n_seE}{eQK_x}\,x(L-x).
\label{eq5-7}
\end{equation}
As shown in figure~\ref{fig5-2} 
the parabolic profile of $\varphi(x)$ resulting from the strong phase pinning at electrodes yields the maximum strain at $x=0$ and $x=L$, positive (compression) at one electrode and negative (dilatation) at the other such as:
\begin{equation}
e_{\rm max}=\frac{n_seEL}{2Q^2K_x}.
\label{eq5-8}
\end{equation}
The CDW will then start to move when the maximum stress $\Sigma_c$ at the electrodes will overcome the local pinning stress $e_c$ ($\sigma_c$~= $Q^2K_xe_{\rm max}$). That will occur when the applied voltage $E$:
\begin{equation}
V>V_c=E_cL=\frac{2\Sigma_c}{n_se}.
\label{eq5-9}
\end{equation}
$\Sigma_c$ depends on the surface condition, the contact geometry and the nature of CDW-metal interface of the electrode.

\subsubsection{Gor'kov model}\label{sec5-2-2}

An alternative way for boundary conditions was, in fact, previously proposed by Gor'kov in a model \cite{Gorkov84} in which the conversion between sliding CDW and strong pinning at the electrodes proceeds through a phase slip (PS). Analogy can be done with finite resistivity in thin superconducting wires \cite{Langer67,Ivlev78}. PS is a fast and localised annihilation of the order parameter amplitude during which the phase slips by $2\pi$. In the first microscopic analysis of the PS process in CDWs, Gor'kov used the time-dependent Ginzburg-Landau equations, including phase and amplitude, strictly valid for a gapless CDW conductor with the Peierls energy gap suppressed by impurities, i.e. in the ``dirty" limit: $\tau^{-1}\gg\Delta,T_p$ ($\tau$: relaxation time for band electrons). The opposite ``clean" case: $\tau^{-1}\ll\Delta,T_p$, i.e. with long electron relaxation times was addressed by Artemenko \textit{et al.} \cite{Artemenko87}. Due to long range correlations in this limit, it was not possible to formulate a local equation of motion. PSs should be treated on quantum level, as amplitude solitons with adiabatic dynamics. However, qualitatively, the same description of PS was obtained as in the Gor'kov limit. The Gor'kov model is only relevant for thin samples with a transverse dimension smaller than $\xi_\perp$. The order parameter vanishes in the entire plane located at a distance $x_0$ from the barrier.

The mechanical analogy, to be compared with eqs~(\ref{eq5-6}) and (\ref{eq5-7}), is that of a long elastic spring tightly fixed at $x=0$ which rotates in a viscous medium, the term with the electric field being the applied torque \cite{Gorkov84}. A steady-state behaviour is possible only if at some position $x=x_0$, because of the accumulation of stress, the coupling between the two parts of the spring periodically breaks, so that slippage of the phase occurs (typically $\varphi=2\pi$). Gor'kov \cite{Gorkov84} showed that each time there is slippage the phase relaxes over a distance
\begin{equation}
\vert x-x_0\vert\sim\xi_x\left(\frac{T_p}{eE\xi_x}\right)^{\frac{1}{2}}.
\label{eq5-10}
\end{equation}
In weak electric field, this distance is macroscopically large. The slipping process itself occurs over a very short time period: $\omega_0^{-1}$~= $t_0$.

The critical field needed for the formation of a 1D PS was estimated \cite{Gorkov84} and numerically evaluated \cite{Batistic84} as:
\begin{equation}
E_c\sim\frac{\Delta(T)}{e\xi_x}\,\left(\frac{\xi_x}{L}\right)^{1.23}.
\label{eq5-11}
\end{equation}
The difference between the $L^{-1}$ (eq.~(\ref{eq5-9})) and $L^{-1.23}$ results from the details of the boundary effects and amplitude variations.

Although the PS model was formulated for a 3D system, analysis were essentially limited to the 1D solution. However it was shown \cite{Jelcic91} that the Ginzburg-Landau diffusion equation, used in the Gor'kov model, has a special type of 3D solutions: dynamic phase vortices i.e. dislocation lines. These dislocation lines are the  3D generalisation of the 1D PS. In fact the core of the dislocation line is an array of simultaneous PS centres.

\subsubsection{Phase vortices}\label{sec5-2-3}

Rather than to have the amplitude of the order parameter to vanish in the entire cross-section, internal stresses can generate nucleation and growth of dislocation loops. Thus it was proposed \cite{Ong84,Ong85} that the conversion of the CDW current occurs through the formation of phase vortices in front of strong barriers, in particular at the electrodes. Ong and Maki used \cite{Ong84,Ong85} the vortex solution of the static Landau equation for the complex order parameter and imposed phenomenologically their transverse motion in order to ensure the evacuation of the accumulated charge. This model has some similarity with the model of static elastic medium considered by Feinberg and Friedel \cite{Feinberg88,Feinberg89} and  \cite{Ong84,Ong85}.

\begin{figure}[b]
\begin{center}
\includegraphics[width=7cm]{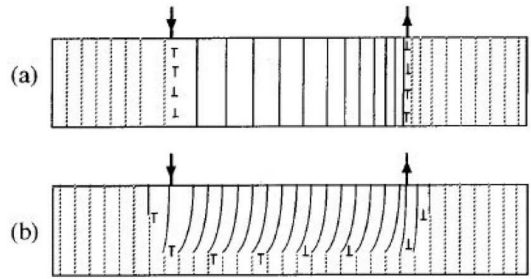}
\caption{Strain distributions, dislocation paths and Frank-Read sources. Wave fronts in moving and stationary regions of a CDW are shown as continuous and broken lines, respectively. The symbols $\perp$ and $T$ represent edge dislocations. in (a), motion (from left to right) of the CDW between the terminals is made possible by the climb to the surface of edge dislocations loops, generated by longitudinal stress. In (b), motion of a layer is made possible by the glide towards the terminals, and subsequent climb, of dislocations generated by a Frank-Read source on the boundary of the stationary layer (reprinted figure with permission from J.C. Gill, Physical Review B 63, p. 125125, 2001 \cite{Gill01}. Copyright (2001) by the American Physical Society).}
\label{fig5-3}
\end{center}
\end{figure}

Under the stress concentration, at each electrode, edge dislocation loops can nucleate, thus by climbing introduce or remove disk-like portions of a period $\lambda$ of the CDW as shown in figure~\ref{fig5-3}(a). Each time the phase advances by $\lambda$, a phase vortex (edge dislocation loop) disappears and a new one is created. The phase continuity imposes:
\begin{equation}
\frac{v_s}{\ell_s}=\frac{v_d}{\lambda},
\label{eq5-12}
\end{equation}
with $v_s$ the velocity of vortices in their climb to the surface separated by a distance $\ell_s$, $v_d$ the CDW velocity.

Equating the work $\sum_{xx}2\pi R\lambda dR$ of the Peach-Koehler force ($F_{\rm PK}$~= $\frac{1}{2}[\Sigma\times L]$, with the unit vector along the line $L$) due to the extrema stresses to the change in elastic energy: $d(2\pi RE)$ when the radius $R$ of the loop varies by $dR$, one obtains the equilibrium condition \cite{Feinberg88}:
\begin{equation}
F_x=\frac{2\pi\sqrt{K_\perp K_x}}{RL\lambda}\ln\left(\frac{\tilde{R}}{\xi}\right).
\label{eq5-13}
\end{equation}
The critical field \cite{Feinberg88} will correspond to the smallest nucleus i.e. for $\tilde{R}\cong\xi$. Thus:
\begin{equation}
F_x=\frac{2\pi\sqrt{K_\perp K_x}}{\xi_\perp L\lambda}=\frac{2\pi K_x}{\xi_xL\lambda}.
\label{eq5-14}
\end{equation}
The critical voltage for dislocation nucleation is then:
\begin{equation}
E_vL\cong\frac{2\pi K_x}{e\,n_s\,\xi_x\,L}.
\label{eq5-15}
\end{equation}
Equation~(\ref{eq5-15}) has to be compared with eq.~(\ref{eq5-9}). As stated by Feinberg and Friedel, both processes: either the total disrupt of the CDW amplitude or the nucleation of edge dislocations are not very different. Their ratio is:
\begin{equation}
\frac{E_v}{E_c}=\frac{1}{4\pi\xi_0}{\lambda}{\xi_x}.
\label{eq5-16}
\end{equation}
In both models, for long samples, the threshold field for CDW depinning is due to impurities and is a bulk effect, independent of the length $L$. For short samples, $E_T$ should vary as $L^{-1}$ or $L^{-1.23}$. This dependence is true either for longitudinal (contact) or lateral surface pinning when depinning resulting from critical stresses overcomes impurity pinning.

\subsubsection{Frank-Read sources}\label{FRsources}

There is another possible source of generation of dislocations initiated by the Frank-Read mechanism from pre-existing dislocations, as proposed by Lee and Rice \cite{Lee79}. Feinberg and Friedel \cite{Feinberg88,Feinberg89} considered this mechanism unlikely; essentially because the Burger vector having one single orientation parallel to the chains, the dislocation lines cannot build a stable 3D Frank network. However Gill \cite{Gill01} has noted that a macroscopic stress may arise when the mean velocity is spatially non-uniform. In crystals it may occur that the motion in the central segment does not extend over the entire cross-section and a layer remains stationary between current terminals. As shown in figure~\ref{fig5-3}(b), motion of a CDW layer is possible by the glide towards the terminals and subsequent climb of dislocations generated by a Frank-Read source on the boundary of the stationary layer. Although the surface pinning is contested \cite{Isakovic06}, bending of the CDW wave fronts (with a rotation of typically 0.03$^\circ$) may occur at steps in crystals with non-uniform thickness. The CDW shears from the thinner part of the sample which is more strongly pinned. Estimations of the shear strain were obtained from X-ray diffraction topography \cite{Li99}, X-ray microbeam diffraction \cite{Isakovic06}, as well as with transport measurements \cite{Neill04}.

\subsection{Direct observation of a single CDW dislocation}\label{sec5-3}

Up to now only indirect evidence of CDW dislocations close to contacts has been reported. It might be envisaged that high resolution STM measurements may reveal phase slip defects \cite{Brun09,Brazovskii11}. But a direct observation of CDW dislocation by diffraction techniques is very challenging. In fact, coherent X-ray diffraction measurements in the CDW state of K$_{0.3}$MoO$_3$ have been interpreted with the presence of a single CDW dislocation \cite{LeBolloch05}.

\subsubsection{Coherent X-ray diffraction and speckle}\label{sec5-3-1}

Improvement of the X-ray coherence of the beam is directly associated with the brightness of new (third generation) synchrotron sources. While in a ``classical" X-ray diffraction, the spatial coherence lengths of the beam are typically the atomic distances, in a coherent diffraction the diffraction resolution is dominated by the finite size of the beam (several $\umu$m \cite{Sutton91}). The small X-ray beam sufficiently coherent produces speckle patterns which characterise the specific structural arrangement of the part of the sample illuminated. The general conditions to get a coherent beam are the use of typically 10~$\umu$m$\times$10~$\umu$m entrance slits (playing the role of pinhole) in front of the sample. The beam quality and its intrinsic degree of coherence are then tested by using typically 2~$\umu$m$\times$2~$\umu$m entrance slits in order to observe their regular interference fringes in the Fraunhofer regime. The patterns are recorded on a direct illumination CCD camera.

\begin{figure}
\begin{center}
\subfigure[]{\label{fig5-4b}
\includegraphics[width=5.5cm]{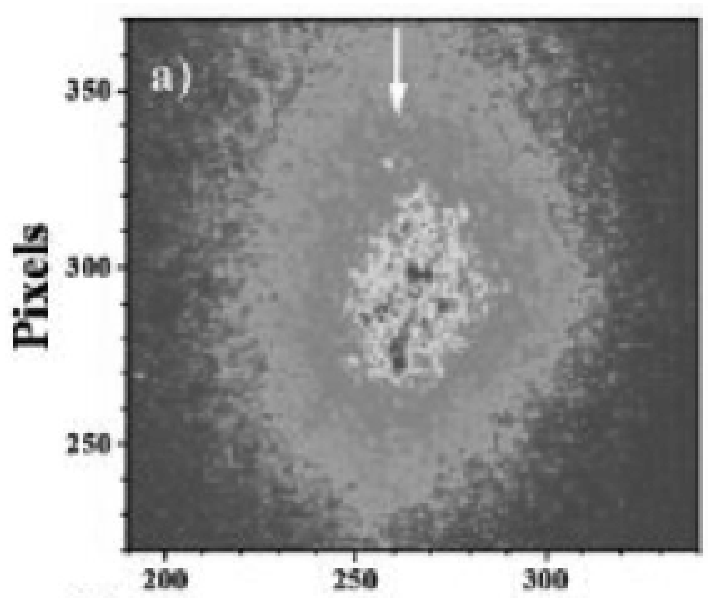}}
\subfigure[]{\label{fig5-4a}
\includegraphics[width=5.5cm]{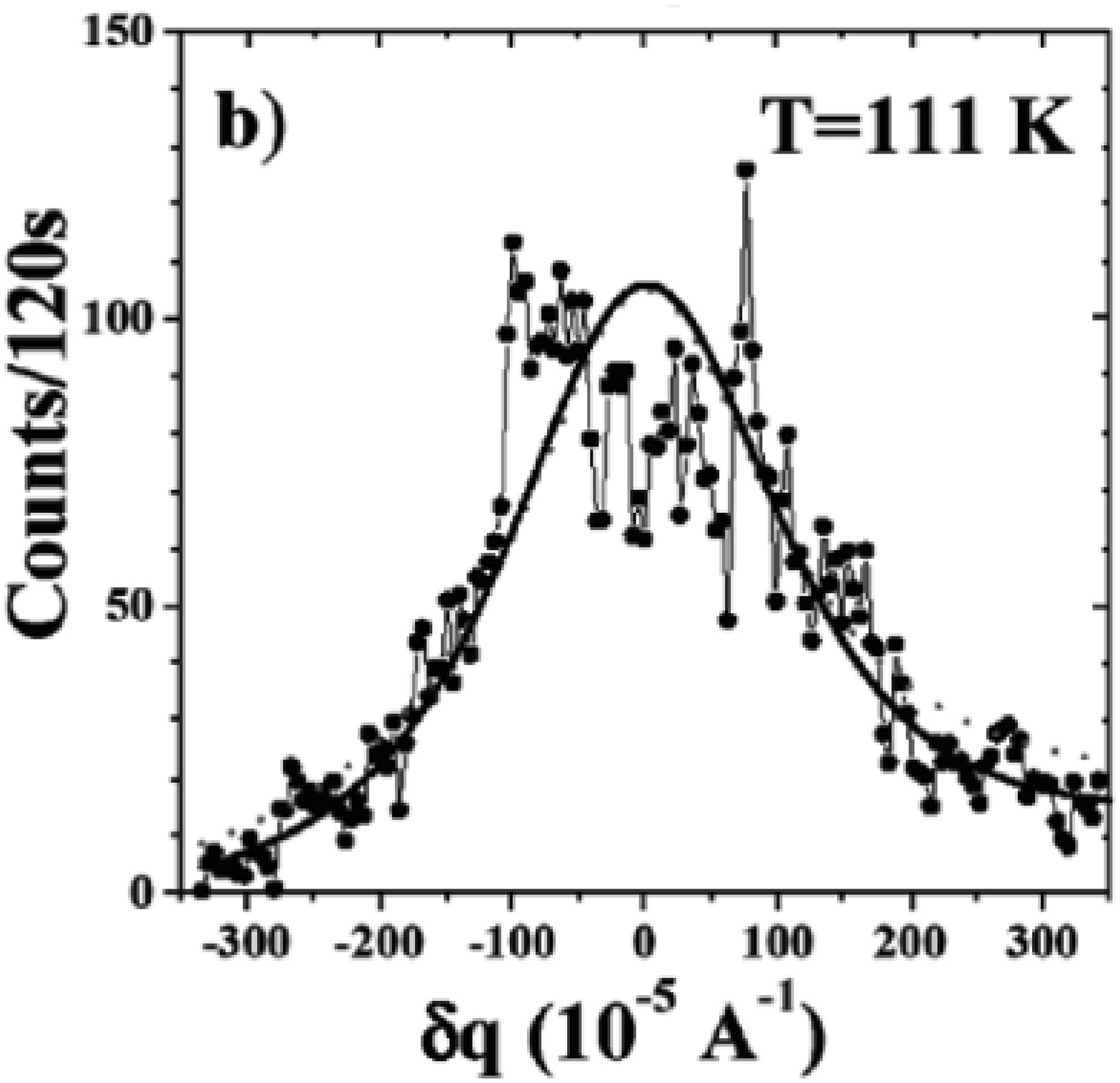}}
\vspace{-0.25cm}
\caption{(a) 2D speckle pattern (logarithmic colour scale) obtained at the $Q_s$ superstructure peak intensity resulting from the antiferrodistortive displacive transition of SrTiO$_3$. (b)~Display of the vertical profile indicated by the white arrow. The solid line is a fit to a Lagrangian line shape (reprinted figure with permission from S. Ravy \textit{et al.}, Physical Review Letters 98, p. 105501, 2007 \cite{Ravy07}. Copyright (2007) by the American Physical Society).}
\label{fig5-4}
\end{center}
\end{figure}
As an example, figure~\ref{fig5-4} shows the 2D pattern obtained \cite{Ravy07} at the position of the peak intensity of the $Q$ superstructure in the pre-transitional temperature range of the antiferrodistortive displacive transition of SrTiO$_3$. Coherent X-ray diffraction does not yield smooth averaged diffraction patterns but reveals speckled patterns which are related to the disordered distributed scatterers in the coherence volume. The pattern in classical diffraction would correspond to dotted line in figure~\ref{fig5-4}(b) fitted with a Lagrangian line shape.

This X-ray technique can be also used for studying intensity fluctuation spectroscopy (XIFS) \cite{Sutton02}. XIFS examines the temporal evolution of the speckle patterns to access the dynamics of the system. Difficulties however may arise concerning the limited time constants which can be detected (typically 1--10$^4$~s) when the integration time for data collection is compared with the intrinsic time constant of the sample. XIFS experiments were performed on NbSe$_3$ in the upper CDW state \cite{Sutton02}. Static speckles were observed but the CDW being to well ordered, the speckle pattern did not exhibit many peaks. With the applied current above threshold, the speckle went away.

\subsubsection{Single CDW dislocation}\label{sec5-3-2}

2D diffraction patterns were recorded at 75~K at the satellite reflection of K$_{0.3}$MoO$_3$ for different positions along the sample. For most of the positions the patterns exhibit only a sharp peak (no visible speckle reflecting disordered systems). At some beam positions, regular fringes appeared along the $t^\ast$ direction ($t^\ast$ is tilted by 19.5$^\circ$ away from the $2a^\ast$-$c^\ast$ direction) ascribed to interference effects induced by localised phase field deformations of the CDW \cite{LeBolloch05}. Detailed simulations of the X-ray coherent diffraction patterns in the presence of various types of dislocations were performed in ref.~\cite{Jacques09}. From these simulations, the doubling of the CDW diffraction peak along $t^\ast$ was interpreted as a screw-like static dislocation. That would be the first direct observation by a diffraction technique of a topological defect in an electronic crystal.

In the sliding regime, by the same X-ray coherent diffraction technique applied to the same compound, it was shown that secondary satellite reflections appear in addition to the $2k_{\rm F}$ one indicating correlations along the chain axis up to the micrometer range \cite{LeBolloch08}. Several explanations --soliton lattice, ordering of CDW dislocations-- have been suggested.

Similarly, a single dislocation in the SDW state of chromium \cite{Fawcett88}, embedded a few micrometers from the surface, was recently observed using coherent X-ray diffraction \cite{Jacques09}

\subsection{Phase slip voltage, $V_{\rm ps}$}\label{sec5-4}

The first evidences of the boundary conditions for initiating the CDW motion was brought experimentally by Gill \cite{Gill82} and theoretically by Gor'kov \cite{Gorkov83}.

\begin{figure}
\begin{center}
\includegraphics[width=7cm]{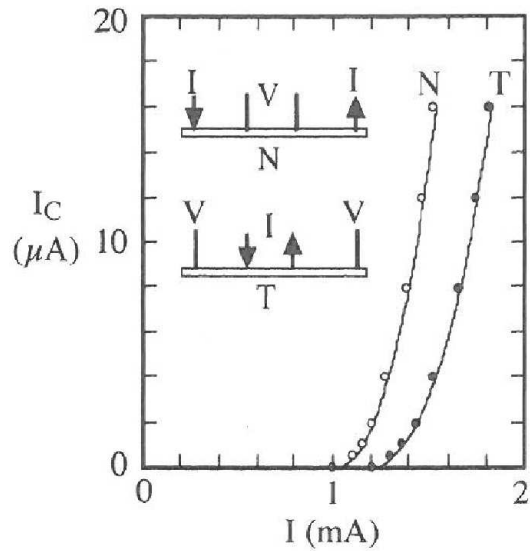}
\caption{Four-terminal measurement of the voltage required to induce phase-slip. The CDW current, $I_c$, recorded at 90~K using terminal configuration ``normal (N)" and ``transposed (T)" is shown as a function of total applied current $I$ (reprinted figure with permission from Solid State Communications 44, J.C. Gill, p. 1041, 1982 \cite{Gill82}. Copyright (1982) with permission from Elsevier).}
\label{fig5-6}
\end{center}
\end{figure}

In the usual (``normal") 4 probe measurement, a current $I$ is injected at the extremity of the sample and the voltage measured between 2 contacts separated by $L$ located between the current terminals. Gill \cite{Gill82} noticed that in the ``transposed" configuration, i.e. when current flows between the inner pair of contacts and voltage measured between the outer pair of contacts, the threshold voltage for depinning the CDW, $V_T$, was higher than in the ``normal" configuration, $V_N$ (see figure~\ref{fig5-6}), 
such as:
\begin{equation*}
V_T-V_N=V_{\rm PS}.
\end{equation*}
In the transposed configuration, the CDW has to be broken near the inner terminals in order to have the central part to move. That arises by the need to induce phase slippage at the boundaries between moving and stationary parts of the CDW \cite{Gill96,Brazovskii91a,Brazovskii91b} which requires the extra potential $V_{\rm PS}$.

Another type of measurements which reveal this phase slip voltage is the dependence of the threshold field when the distance between terminals is reduced.

\subsubsection{Shunting-non shunting electrodes}\label{sec5-4-1}

Measurement of the dependence of the threshold field as a function of the length of the sample requires to characterise the nature of electrodes. Contact at the electrode can be either metallic or resistive. If metallic, at the voltage electrodes the contact resistance is nearly zero and depending of the width of the electrode, it may occur that the whole current bypass the sample across the contact, creating a region with zero electric field below the electrode, thus disturbing strongly the current lines. This type of electrode is called shunting or injecting electrode. On the contrary, a resistive terminal will deviate only very little the current lines and then is called non-shunting electrode.

For instance, for NbSe$_3$, it was calculated and experimentally verified \cite{Saint-Lager88} that for a sample with a width $\sim 10~\umu$m, thickness $\sim~1~\umu$m, an electrical anisotropy between the chain direction and the direction perpendicular to the ($b,c$) plane of $\sim$~100, a contact resistance of a few $\Omega$, the electrodes are not perturbing if the contact width is less than $\sim 20~\umu$m.

\subsubsection{Length dependence of the threshold field}\label{sec5-4-2}

\begin{figure}
\begin{center}
\includegraphics[width=7.5cm]{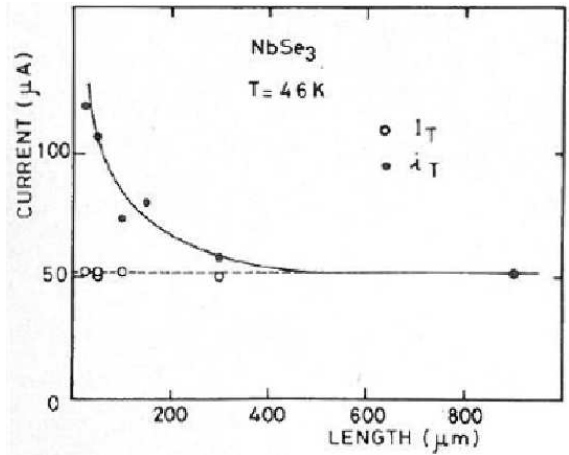}
\caption{Variation of the depinning current as a function of the length between electrodes. The current I is applied at the extremities of the NbSe$_3$ sample and the voltage electrodes are non-shunting; the current $i$ is applied through the voltage electrodes. The width of the electrodes is 8~$\umu$ (reprinted figure with permission from Physica B 143, P. Monceau \textit{et al.}, p. 64, 1986 \cite{Monceau86}. Copyright (1986) with permission from Elsevier).}
\label{fig5-7}
\end{center}
\end{figure}

A NbSe$_3$ sample was deposited and electrostatically tightened on a multicontact configuration with various distances between electrodes. Injecting the current $I$ far away from the voltage electrodes the threshold current, $I_T$, was measured between pairs of electrodes of 8~$\umu$m width separated by 25, 60, 100, 200~$\umu$m. As shown in figure~\ref{fig5-7}, 
$I_T$ is independent of the length.

If now, in the transposed configuration, the current is injected through the voltage leads and the voltage measured at the extremity of the sample, depinning occurs at a current $i_T>I_T$ ($I_T$ threshold current in the ``normal" configuration, $i_T$ threshold in the ``transposed" configuration) such as:
\begin{equation*}
i_T=I_T+i_0,
\end{equation*}
with $i_T$ increasing when $\ell$ is reduced (see figure~\ref{fig5-7}). The threshold voltage $V_T$~= $R\,i_T$ shows a linear variation as a function of the length with a finite intercept $V_0$ when $\ell\rightarrow 0$. $V_0$ is temperature dependent and typically $V_0\sim$~0.3~mV for NbSe$_3$ at 46~K \cite{Saint-Lager88}, 1~mV for o-TaS$_3$ \cite{Mihaly83}. The accuracy of measurements is not enough to discriminate the variation of $E_T\sim L^{-1}$ from $E_T\sim L^{-1.23}$ estimated numerically in ref.~\cite{Batistic84}. $V_0$ is thus the field for breaking the longitudinal CDW coherence.

The small value for $V_0$ discredits the total collapse of the amplitude of the order parameter in the whole cross-section; in that case $V_0$ should be $2\xi_x\,E_c\,\lambda$ ($E_c$: the condensation energy per electron) estimated for NbSe$_3$ to be $\sim$~20~mV, two orders of magnitude larger than the experimental value.

\subsection{Breakable CDW}\label{sec5-5}

\begin{figure}
\begin{center}
\includegraphics[width=12cm]{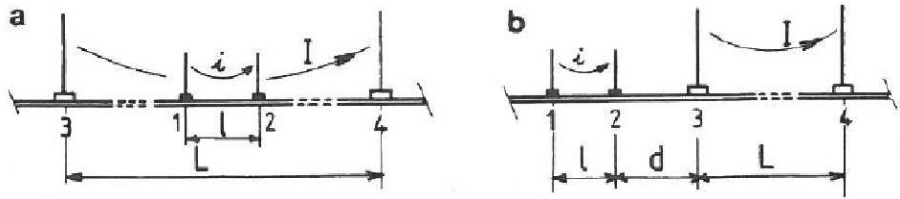}
\caption{Schema for studying CDW depinning when two independent current sources deliver current in different segments of the same crystal. Typically $L\sim 1$~mm, 20~$\umu$m$<\ell<80~\umu$m, the width of the electrodes~= 8~$\umu$m. In (a) segments $L$ and $\ell$ are overlapping, in (b) $L$ and $\ell$ are apart and separated by $d$ (reprinted figure with permission from M.-C. Saint-Lager \textit{et al.}, Europhysics Letters 9, p. 585, 1989 \cite{Saint-Lager89}. Copyright (1989) from EdpSciences.}
\label{fig5-8}
\end{center}
\end{figure}

The growth of dislocation loops should appear at any velocity discontinuity to release the charge accumulation, especially when depinning occurs under inhomogeneous conditions. That can be achieved by applying a thermal gradient along the sample length or when several independent sources deliver current in different segments of the same crystal. In the latter case, two types of configuration have been used as shown in figure~\ref{fig5-8}. 
In figure~\ref{fig5-8}(a), a current source supplies $i$ on a segment $\ell$ with typically 25~$\umu$m$<\ell<80~\umu$m whereas the whole crystal of length $L\sim$~1~mm is fed by a constant current delivering $I$ \cite{Saint-Lager89}. In figure~\ref{fig5-8}(b), segment $L$ and $\ell$ are now apart and separated by a distance $d$.

\subsubsection{Lateral current injection}\label{sec5-5-1}

\begin{figure}
\begin{center}
\includegraphics[width=7.5cm]{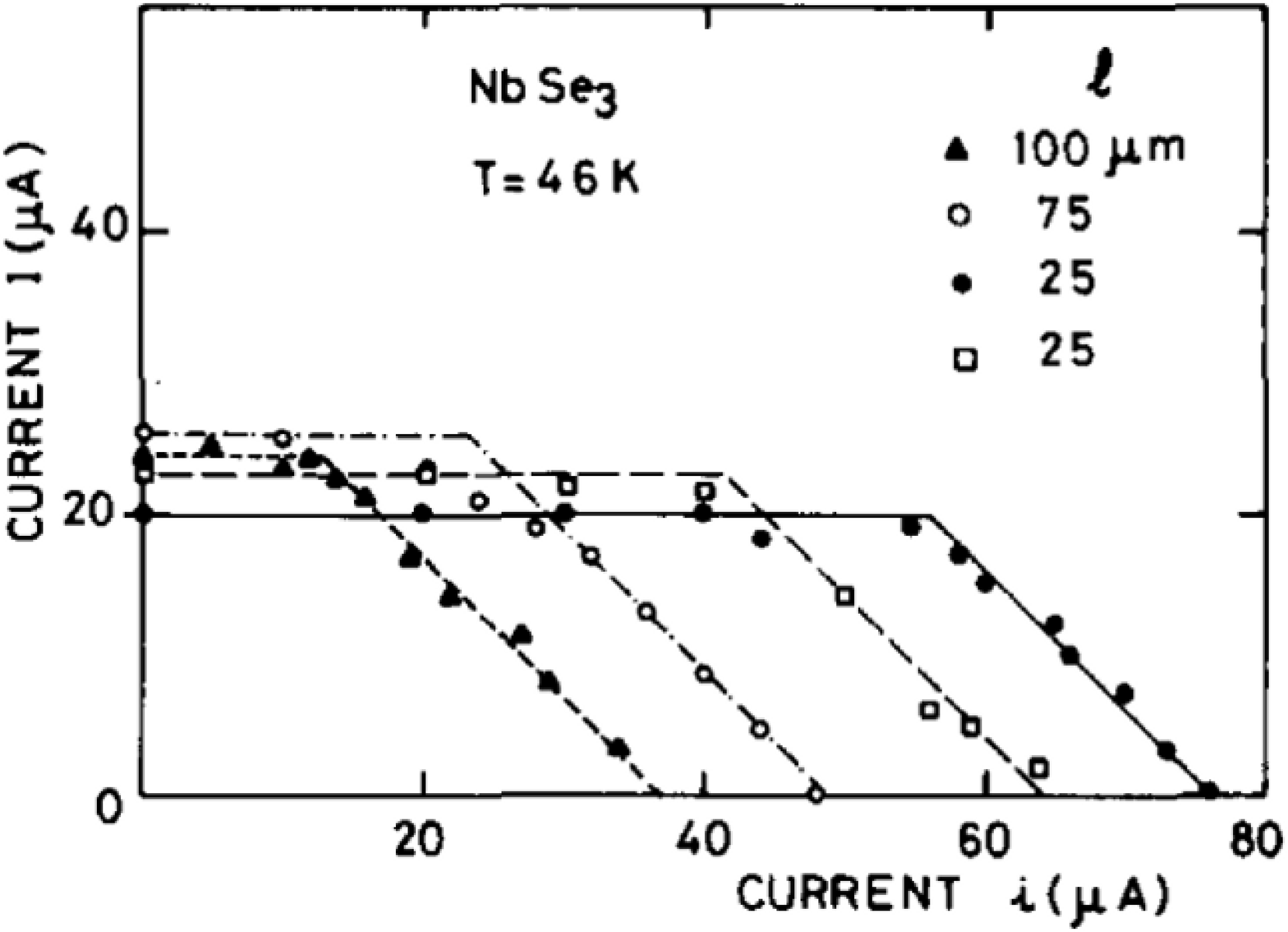}
\caption{($I$, $i$) diagram for depinning of a small segment of a NbSe$_3$ sample located between two electrodes apart of $\ell$ for several values of $\ell$. The width of electrodes is 8~$\umu$m. $I$ and $i$ have the same meaning as in figure~\ref{fig5-8}(a) (reprinted figure with permission from Physica B 143, P. Monceau \textit{et al.}, p. 64, 1986 \cite{Monceau86}. Copyright (1986) with permission from Elsevier).}
\label{fig5-9}
\end{center}
\end{figure}

When the segments $L$ and $\ell$ are overlapping, the depinning on segment $\ell$ can be studied with the conjunction of $I$ and $i$ currents. In figure~\ref{fig5-9}
it is shown, for different lengths $\ell$, that the threshold current $I_T$ is independent of $i$ up to a value $i_0$. Above $i_0$, $I$ and $i$ play the same role and depinning occurs along the line $I+i$~= $i_T$. It is thus possible to initiate CDW motion in a small part of the sample, totally decorrelated from the rest of the sample. That occurs because a vortex sheet has been created at the inner electrodes, requiring the voltage $V_0$. $V_0=r\,i_0$, $r$: the resistance of segment $\ell$. In the case where the voltage electrodes were large, they are shunting even when $i=0$, and then only $i_T$ can be defined.

\subsubsection{Long range CDW coherence}\label{sec5-5-2}

\begin{figure}
\begin{center}
\includegraphics[width=7.5cm]{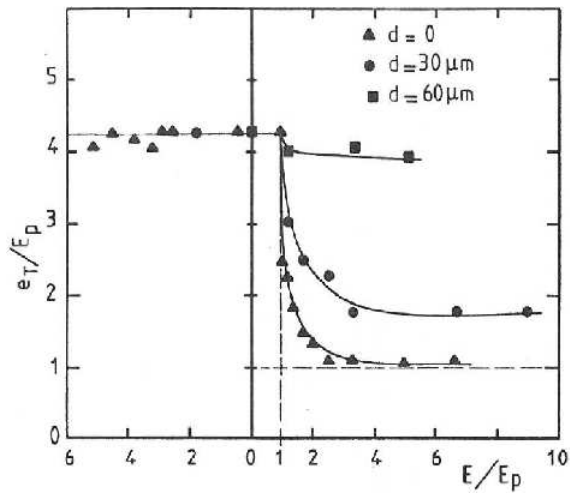}
\caption{Variation of the threshold field $e_T$ of a small segment $\ell$ (50~$\umu$m) of NbSe$_3$ at $T$~= 42~K as a function of the electric field on the neighbouring segment $L$ with $d$~= 0 ($\blacktriangle$), 30~$\umu$m) ({\large$\bullet$}) and 60~$\umu$m ({\tiny$\blacksquare$}), as schematically drawn in figure~\ref{fig5-8}(b). Electric fields are normalised with the threshold field for impurity pinning, $E_p$ (reprinted figure with permission from M.-C. Saint-Lager \textit{et al.}, Europhysics Letters 9, p. 585, 1989 \cite{Saint-Lager89}. Copyright (1989) from EdpSciences).}
\label{fig5-10}
\end{center}
\end{figure}

With $L$ and $\ell$ segments apart (figure~\ref{fig5-8}(b)), the depinning of the small segment $\ell$ can be studied according to the state --linear or non-linear-- of the segment $L$ \cite{Saint-Lager89}. The threshold of $\ell$ alone is $e_T$~= $E_p+V_0/\ell$ whereas the threshold of $L$ is $E_T\approx E_p$ since $L\gg \ell$ ($E_p$: impurity pinning). Figure~\ref{fig5-10} 
shows that $e_T$ decreases when the neighbouring segment $L$ starts to slide. This effect is only detectable for $d<100~\umu$m and for $I$ and $i$ with the same polarity. When $d$~= 0 (adjacent segments) $e_T$ decreases sharply to reach the volumic value $E_p$. Thus the depinning on the adjacent segment suppresses the boundary effects on the segment $\ell$ revealed by $V_0/\ell$.

\begin{figure}
\begin{center}
\includegraphics[width=7cm]{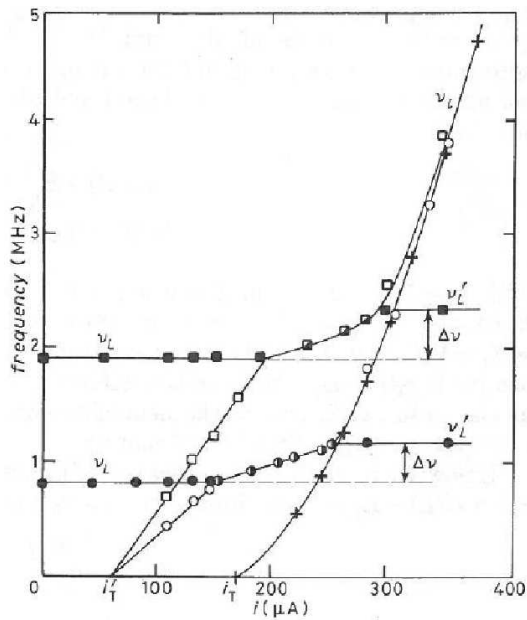}
\caption{Variation of the fundamental narrow-band noise frequency $\nu_\ell$ ($\circ$, $\square$) is a segment $\ell$ (50~$\umu$m) of a NbSe$_3$ crystal at 42~K as a function of the applied current $i$ for two different constant currents above threshold in a neighbouring segment $L$ (500~$\umu$m) inducing a narrow-band noise frequency $\nu_L$ ({\large$\bullet$}, {\tiny$\blacksquare$}) in $L$. The locking regime for each $\nu_L$ is shown by half-black circles and half-black squares (reprinted figure with permission from M.-C. Saint-Lager \textit{et al.}, Europhysics Letters 9, p. 585, 1989 \cite{Saint-Lager89}. Copyright (1989) from EdpSciences).}
\label{fig5-11}
\end{center}
\end{figure}

The CDW velocities have also been studied by NBN analysis. The current $I$ in $L$ is kept constant and the CDW velocity in $L$ is $v_L$. Figure~\ref{fig5-11} shows the CDW velocities in $L$ and $\ell$ as a function of the current $i$ in $\ell$. When $i$ is increased beyond $i^\prime_T$, the CDW starts to move in $\ell$ with a velocity $v_\ell$. When $v_\ell$ reaches $v_L$, the two segments lock together; the further increase of $i$ speeds up the unique velocity in the segment $L+\ell$. Then for higher value of $i$, coherence breaking occurs, the CDW velocity in $L$ staying constant at a value $v^\prime_L>v_L$, whereas $v_\ell$ continue to increase with $i$.

The very spectacular result in this experiment is the long range CDW coherence in motion as manifested by the variation of $v_L$ in the long segment: although $I$ has been kept constant, when the boundary conditions are changed at one electrode by a modification of the electrical condition of the adjacent segment of 50~$\umu$m long, the CDW velocity on a distance of 1~mm or more has been increased as a whole from $v_L$ to $v^\prime_L$. These experiments demonstrate that in absence of defects which may initiate carrier conversion, the CDW velocity does not suffer any small local variation along the sample length, the CDW velocity coherence being quasi infinite.

\subsubsection{Critical state model}\label{sec5-5-3}

These results have been interpreted using a phenomenological model for CDW transport in terms of motion of dislocation loops \cite{Saint-Lager89}. The loops can only grow under electrodes but the dislocation line suffers an irreversible short-range pinning force which hinders its growth. If $S$ is the oriented area of a loop projected perpendicular to the $b$-axis, then the equation of dislocation motion in a segment of length $X$ is given by:
\begin{equation}
X\left[E-E_p-\eta v\right]=V_0\left[{\rm sgn}\,S\right],
\label{eq5-17}
\end{equation}
with $\eta$ the phenomenological viscosity and $v$ the CDW velocity. The left part of eq.~(\ref{eq5-17}) represents the forces provided by external sources for enlarging or decreasing the dislocation loops and the right part corresponds to the irreversible forces provided by the dislocations pinning.

In the three-electrode configuration, let call 23 the common electrode (see figure~\ref{fig5-8}(b)). $I$ is defined positive when flowing from 1 to 23 as $I$ from 23 to 4. Loops growing under electrodes 1, 23, 4 are called $S_1$, $S_{23}$, $S_4$. The neutrality condition imposes that:
\begin{equation}
S_1+S_{23}+S_4=0.
\label{eq5-18}
\end{equation}

The forces on $L$ and $\ell$ following eqs~(\ref{eq5-17}) and (\ref{eq5-18}) can be written as a function only of $S_1$ and $S_{23}$. The equality of the work of these forces yields two equations with three determinations because the modulus of the sum $\vert S_1+S_{23}\vert$ depends on the sign of $S_1S_{23}$. The first determination corresponds to $I$ and $i$ in opposite direction. Then it is easy to see that $e_T(\ell)$ is not affected by $I$. In the case where $I$ and $i$ have the same polarity the force equations in $L$ and $\ell$ are
\begin{eqnarray}
\ell\left[E(\ell)-E_p-\eta v_\ell\right] & = & Z_\ell V_0,\label{eq5-19}\\
L\left[E(L)-E_p-\eta v_L\right] & = & (1-Z_\ell)V_0, \label{eq5-20}
\end{eqnarray}
with $Z_\ell$~= 1 if $v_\ell >v_L$ and $Z_\ell$~= 0 if $v_\ell <v_L$. If one first consider the threshold fields, when $E(L)<E_T(L)$, $Z_\ell$~= 1 and from eq.~(\ref{eq5-19}) one recovers the result for short samples: $e_T(\ell)$~= $E_p+V_0/\ell$. But if $L$ is in the non-linear state, then $Z_\ell$~= 0 and the new threshold for the short sample is $e_T(\ell)$~= $E_p$: the boundary effects are suppressed for $\ell$ as found experimentally and as shown in figure~\ref{fig5-10}. Due to the neutrality condition, $V_0$ is not divided by 2 in this case, but disappears completely of the $\ell$ motion.

If now one fix $v_L$ in $L$ by application of a constant $E(L)>E_T(L)$, when $v_\ell<v_L$, eq.~(\ref{eq5-20}) yields $\eta v_L$~= $E(L)-E_p-V_0/L$. But when $v_\ell>v_L$ the CDW velocity in $L$ is given by:
\begin{equation}
v^\prime_L=\frac{E(L)-E_p}{\eta}>v_L,
\label{eq5-21}
\end{equation}
with $\Delta v_L$ corresponding to the field $V_0/L$. This result has been explicitly found experimentally in figure~\ref{fig5-11}. The locking regime also shown in figure~\ref{fig5-11} corresponds to the change of the irreversible force acting on $L$ which goes from $-V_0$ when $v_\ell<v_L$ to 0 when $v_\ell>v_L$. In this regime $v_\ell$~= $v_L$ and the velocity is given by the average field on the total length $\ell+L$. Although $E(L)$ is fixed, $v_{L,\ell}$ increases slightly as also experimentally found. These experiments show that the irreversible forces due to the dislocation loop depinning are always paid by the segment with the faster velocity.

The same model can be applied if now contacts 2 and 3 are separated by a distance $d$. When $L$ is in the non-linear state, one can show that, until $E_p(d/\ell)<V_0$:
\begin{equation}
e_T(\ell)=E_p\left(1+\frac{d}{\ell}\right),
\label{eq5-22}
\end{equation}
in agreement with results shown in figure~\ref{fig5-10}. When $d$ is larger than 100~$\umu$m, one finds the result for isolated short samples, i.e. $e_T(\ell)$~= $E_0+V_0/\ell$. Clearly these measurements show that the phase perturbation penetrates outside the segment in the non-linear state. This effect will be directly demonstrated by X-ray diffraction (see figure~\ref{fig5-17}). This non-local effect takes place till a distance around 70~$\umu$m (for NbSe$_3$ at $T$~= 40~K).

\subsubsection{Thermal gradient}\label{sec5-5-4}

Inhomogeneous conditions for CDW depinning can also be achieved when  thermal gradient is applied through the sample. The local  nature of CDW oscillations has been thought to be proved by the observation of the splitting of the fundamental NBN frequency only into two frequencies \cite{Ong85,Verma84}: one corresponding to each contact. Contradictory interpretations of results have been reported. Some results show that the depinning occurs at a threshold given by the average temperature between the electrodes \cite{Zettl85,Mihaly84,Lyding86}. Other results indicate that the threshold is imposed by the minimum value possible in any given temperature configuration \cite{Zhang86}. It was also shown that voltage oscillations were absent if the sliding region was kept away from the contacts by, for instance, warming the injection contacts above the Peierls temperature transition \cite{Zhang86}. However experiments with a reentrant thermal gradient configuration with the sample in a cryogenic liquid \cite{Monceau86} were unable to detect any difference in the noise spectra when the contacts were warmed above $T_{\rm P}$. Numerical simulations using the Gor'kov model have shown \cite{Jelcic91} that, resulting from the competition between the long range CDW coherence and the local variation of the electric field variation along the sample induced by the applied thermal gradient, the sample can be divided into a finite number of dynamically coherent domains separated by phase slip centres. The density of phase slips depends on the $E(x)$ dependence, the length of the sample and the finite length on which phase slip occurs. For short distances, phase slips interfere and prevent the production of further phase slips even in strong thermal gradient.
 
The critical state model has also been applied in these conditions \cite{Saint-Lager89}. When a thermal gradient is applied, the electrodes are at temperature $T_A$ (for $x$~= 0) and $T_B$ for ($x$~= $L$). The electric field $E(x)$~= $\rho(x)I$ and the bulk pinning field $E_p(x)$ depend now on the distance $x$ between 0 and $L$. The threshold conditions can be obtained from the generalisation of eq.~(\ref{eq5-17}) as:
\begin{equation}
L\int^{x_0}_0\,\left[E(x)-E_p(x)\right]{\rm d}x=\frac{V_0(T_A)}{2}+\frac{V_0(T_{x_0})}{2},
\label{eq5-23}
\end{equation}
where $x_0$ is some distance along the sample. Two cases can occur: the first one, if eq.~(\ref{eq5-23}) implies $x_0>L$ (for instance for experiments with $L$ short or for a weak thermal gradient $\Delta T$, then, in spite of inhomogeneous conditions, the depinning will occur at a field given by the average value of $E$ along $L$. The second case will occur if $x_0<L$; then, at $x_0$, phase dislocation loops will grow, the CDW coherence will be broken and the threshold will be the smallest value for the two parts of the sample. Again this CDW phase breaking via dislocation loops results from the comparison between the temperature-dependent electric force $\sim(E-E_p)$ and the pinning force on dislocations given by $V_0$.

These studies with a thermal gradient were essentially aimed at discriminating the origin of the voltage oscillations in the sliding CDW state: from bulk impurity pinning or locally from phase slips or CDW dislocations at the normal $\leftrightarrow$ CDW interface. It is not so simple, because phase slippage may occur far away from contacts, at strong defects in the bulk as seen in X-ray scattering \cite{Rideau01} or at the surface where shear may occur with samples with irregular cross-section \cite{Li99}. In any attempt for reaching a conclusive answer, it appears essential, before any transport measurements under a thermal gradient, to characterise the samples under examination with high X-ray to detect any defect or bending and to evaluate the contact perturbation on the current profile.

\begin{figure}
\begin{center}
\includegraphics[width=12cm]{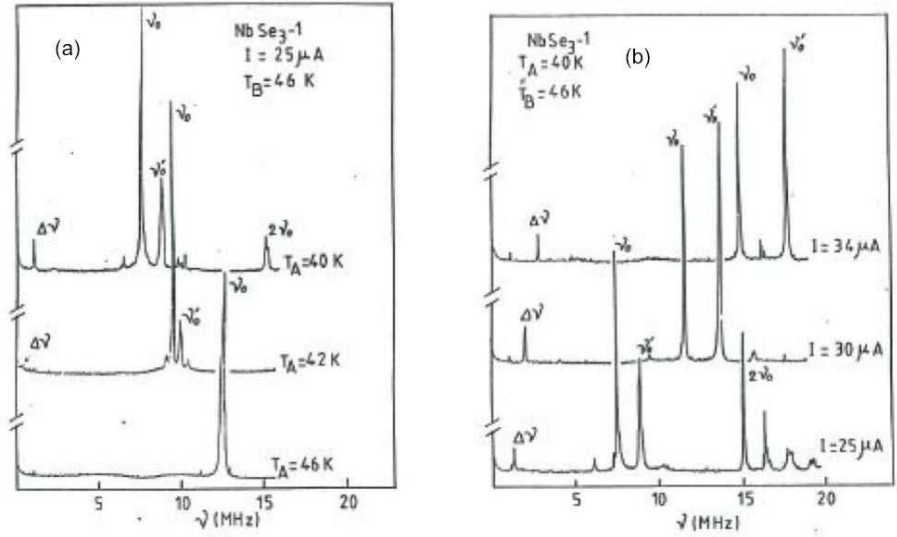}
\caption{Narrow-band noise (NBN) frequency of a NbSe$_3$ sample under thermal gradient between electrodes A and B (a) fixed applied current $I$, temperature of electrode B, $T_{\rm B}$, fixed = 46~K, variable temperature of electrode A, $T_{\rm A}$ from 46~K down to 40~K. Splitting of the fundamental NBN frequency $\nu_0$ in $\nu_0$ and $\nu'_0$ occurs for a finite $\Delta T$. (b) fixed gradient $\Delta T$~= $T_{\rm B}-T_{\rm A}$~= 6~K, variation of $\nu_0$ and $\nu'_0$ for different applied current $I$ and occurrence of the frequency $\Delta\nu$~= $\nu'_0-\nu_0$ (with permission from M.-C. Saint-Lager \cite{Saint-Lager88}).}
\label{fig5-12}
\end{center}
\end{figure}

Figure~\ref{fig5-12}(a) shows the variation of the voltage oscillation frequency of a NbSe$_3$ sample ($\ell$~= 0.4~mm, $s$~= 2~$\umu$m$^2$) at a given applied current $I$ when the temperature of one contact, $T_A$, is varied from 46~K to 40~K, the temperature of the second one, $T_B$, being fixed: $T_B$~= 46~K \cite{Saint-Lager88}. For $T_A$~= $T_B$~= 46~K one observes a single frequency, $\nu_0$. For a finite $\Delta T$~= $T_B-T_A$, $\nu_0$ is splitted in $\nu_0$ and $\nu^\prime_0$ with $\nu^\prime_0-\nu_0$, increasing for larger $\Delta T$. In figure~\ref{fig5-12}(b), $T_A$~=  40~K and $T_B$~= 46~K are kept fixed, but at different applied current. Remarkably in this experiment (although that is not the general case) is detected the appearance of the additional line, $\Delta\nu$~= $\nu_0-\nu'_0$, in the low-frequency part of the voltage spectra which accommodates the difference of velocity of the CDW in the two coherent domains induced by the thermal gradient \cite{Saint-Lager88}. It appears to be the first experimental evidence of the phase continuity (see eq.~(\ref{eq5-12})) at the interface between two sliding regions, away from contacts. Thus, Anderson \cite{Anderson66} showed that, in the situation when different phase winding rates of the order parameter $\phi_1$ and $\phi_2$ are imposed in two regions 1 and 2 of a sample, the matching between $\phi_1$ and $\phi_2$ is realised by a train of vortices crossing the line joining 1 and 2 at a rate $\dot{N}_v$ given by: $2\pi\,\dot{N}_v$~= $\dot{\phi}_1-\dot{\phi}_2$.

\subsection{Thermally activated phase slippage} \label{sec5-6}

Experiments have shown that $V_{\rm ps}$, for a given $I_c$, increases rapidly when temperature is decreased, indicating that the nucleation of phase vortices is induced thermally. Ramakrishna \textit{et al.} \cite{Ramakrishna92} have developed a theory of the thermal nucleation process. They assumed that the critical (maximum) strain (see eq.~(\ref{eq5-8})) which drives CDW phase slippage near the electrodes depends only on $V_{\rm ps}$ dropped uniformly between the current contacts separated by the distance $L$. The CDW strain profile derived from eq.~(\ref{eq5-7}) is given by:
\begin{equation}
e(x)=\frac{1}{Q}\,\frac{\upartial\phi}{\upartial x}=\frac{e\,n_c}{Q^2K}\,V_{\rm ps}\left(\frac{x}{L}-\frac{1}{2}\right),
\label{eq5-24}
\end{equation}
with $x$ the distance measured from one of the contacts.

\subsubsection{Phase slip rate}\label{sec5-6-1}

In this model \cite{Ramakrishna92}, dislocation loops are formed in the presence of the CDW strain by homogeneous thermal nucleation and not by the subsequent growth and motion of dislocation loops. Assuming that the local phase-slip rate is simply proportional to the local nucleation rate, they predicted:
\begin{equation}
r_{\rm ps}(x)=r_0\exp\left[-\left(\frac{e\,n_c}{Q}\right)\frac{V_a}{2QKe(x)}\right].
\label{eq5-25}
\end{equation}
The net CDW current is obtained by integration the nucleation rate over the strain profile with the result:
\begin{equation}
I_c=I_0\left[\frac{V_{\rm ps}}{V_a}\right]\exp\left[-\frac{V_a}{V_{\rm ps}}\right],
\label{eq5-26}
\end{equation}
with $\qquad\displaystyle V_a \sim \frac{\pi^2Q}{e\,n_c}\,\frac{\overline{K}^2}{k_{\rm B}T}\sim\frac{1}{T}\left(\frac{\Delta}{\Delta_0}\right)^3,\quad$ and $\displaystyle\quad I_0 \sim \omega\,\frac{e\,n_c}{\pi Q}\,\frac{A^2}{4\xi^3}\,L\sim L\left(\frac{\Delta}{\Delta_0}\right)^4,$

\medskip
\noindent with $A$: the cross-section of the sample, $\xi^3$: the BCS amplitude coherence volume, $\omega$: attempt frequency, $\overline{K}$~= $-\frac{1}{2}[(K_xK_y)^{1/2}+(K_xK_z)^{1/2}]$ the normalised transverse elastic constant.

The general qualitative form of the measured (4-probe technique) $I_c-V_{\rm ps}$ curves are relatively well described by eqs~(\ref{eq5-25}) and (\ref{eq5-26}). The temperature dependence of $V_a$ and $I_0$ are roughly consistent with experimental results \cite{Maher95}. But quantitatively the predicted value of $V_A$ is many times larger than experimentally derived ($\sim 50$ for the upper CDW in NbSe$_3$, several hundreds for the lower one); the predicted value of $I_0$ is ten orders of magnitude larger than experimentally measured \cite{Maher95}.

\subsubsection{Phase-slip and strain coupling}\label{sec5-6-2}

Subsequent spatially resolved experiments using bipolar current pulses have allowed to determine the CDW current profile \cite{Adelman96,Lemay98}. It was shown that the CDW current, uniform in the middle of long enough samples, decays as the current contacts are approached. Thus, contrary to the previous assumptions of a linear CDW strain profile between current contacts, CDW current and strain are coupled. This coupling between current and phase-slip profiles enhances the strain near contacts and consequently appreciable phase slippage occurs in a smaller region than if the strain profile were linear. In this context, phase slippage has been studied by analysis the local relation between the phase-slip rate $r_{\rm ps}(x)$ and the CDW strain $e(x)$ \cite{Lemay98} and not $V_{\rm  ps}$ as in ref.~\cite{Adelman96}. It was shown \cite{Lemay98} that the exponential variation of $r_{\rm ps}$ (phase-slip rate) versus $1/\bar{K}e$ (CDW strain) is well followed but data for various positions along the sample did not collapse onto a unique straight line as predicted in phase-slip models \textit{via} homogeneous dislocation line nucleation \cite{Ramakrishna92}. It was supposed that additional phase slippage may be due to dislocation motion in the direction of current flow, as suggested also in refs~\cite{Gill90a,Gill90b}.

\subsubsection{CDW elastic constant}\label{sec5-6-3}

As shown above, conductivity measurements on multicontacted samples can indirectly yield the CDW phase gradient under application of an electric field. But phase gradient is directly observable by means of X diffraction as a longitudinal shift:
\begin{equation*}
q^\prime(x)\propto\frac{\upartial\phi}{\upartial x},
\end{equation*}
of the CDW satellite peak position in the reciprocal space in the non-linear state. In the frame of the thermally nucleation model \cite{Ramakrishna92} the CDW wave vector varies between the contacts according to:
\begin{equation}
\Delta Q(x)=\frac{\upartial\phi}{\upartial x}=\frac{e\,n_c}{Q\,K_x}\,V_{\rm ps}\left(\frac{x}{L}\right),
\label{eq5-29}
\end{equation}
with here $x$ measured from the midpoint between the contacts.

In the first experiment, CDW deformations were monitored using a 0.8~mm wide X-ray beam on a 4--5~mm long sample. The data \cite{DiCarlo93b} suggest an approximately linear variation of the satellite shift with $x$ in the central part of the sample, but experimental limitations did not allow the contact regions to be investigated. Nevertheless, from this linear dependence, the longitudinal CDW elastic constant
\begin{equation}
K_x=\left(\frac{e\,n_c}{Q}\,\frac{V_{\rm ps}}{L}\right)\left(\frac{\upartial\Delta Q}{\upartial x}\right)^{-1}
\end{equation}
was estimated to be $\sim(1.7\pm 0.25)\times 10^{-2}$~eV$\AA^{-1}$ \cite{DiCarlo93b}. However if one assume a)~that the CDW strain varies only linearly in the middle of the sample and is strongly enhanced near the current contacts, b)~that the strain at the current contacts to initiate phase slippage is $V_{\rm ps}$, then for a given $V_{\rm ps}$ and $K_x$ the mid-sample strain gradient will be smaller than if the strain profile was linear. Consequently the assumption of a linear profile overestimated $K_x$ and makes the CDW to appear stiffer than it is \cite{Adelman96}.

\subsection{X-ray spatially resolved studies of current conversion}\label{sec5-7}

The previous models of phase slippage at current contacts resulting from conversion of normal current into CDW current were all assuming that the phase slip stress was distributed over the length $L$ between the contacts, leading to a constant wave number gradient $q^\prime_x$~= $\upartial Q/\upartial x\neq 0$ is the bulk of the sample.

The conclusions of the model developed by Brazovskii \textit{et al.} \cite{Brazovskii00,Requardt98a} is opposite to these statements. The model shows that for a long enough sample there must be no linear gradient like $q$~= $q^\prime_{\rm blk}x$ due to an applied or contact voltage. Instead there are a strong dependence of $q^\prime(x)$ near contacts  that flattens (exponentially in the simplest case) towards the sample centre.

The high brilliance synchrotron source (ESRF, Grenoble, France) providing a high photon density on the sample was used to obtain a high spatial resolution allowing thus the investigation of the near-contact region.

\begin{figure}[h!]
\begin{center}
\includegraphics[width=8cm]{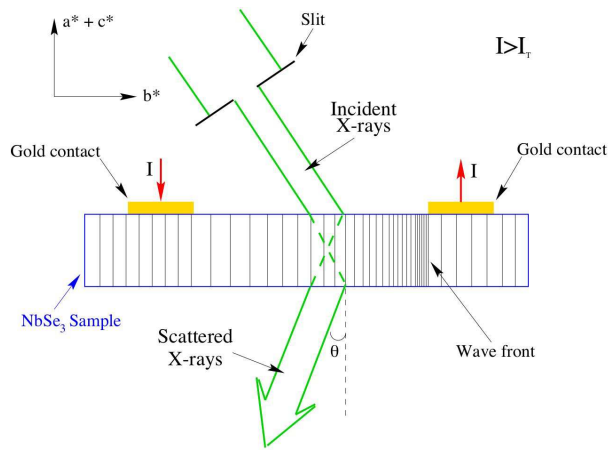}
\caption{Schematic view of the X-ray scattering geometry. The CDW wave-front spacing between electrodes is purposely exaggerated in the figure.}
\label{fig5-13}
\end{center}
\end{figure}

The measurements were essentially carried out on NbSe$_3$ samples with a very low mosaicity and in the upper CDW phase. The spatial definition was controlled by a slit (from 10 to 30~$\umu$m) placed before the sample. The sample position was shifted with respect to the X-ray beam with a positional reproductivity of a few $\umu$m. A schematic view of the X-ray scattering geometry is shown in figure~\ref{fig5-13}. The NbSe$_3$ samples, of typical cross-sections $10\times 2~\umu$m$^2$, were mounted on sapphire substrates of 100~$\umu$m thickness to provide homogeneous sample cooling together with suitable beam transmission (50\%). Electrical contacts were prepared by evaporation of wide 2~$\umu$m thick gold layers on the sample, leaving a section of a few mm of free sample length between electrodes.

\subsubsection{Stationary state}\label{sec5-7-1}

\begin{figure}[b]
\begin{center}
\subfigure[]{\label{fig5-14a}
\includegraphics[width=7.25cm]{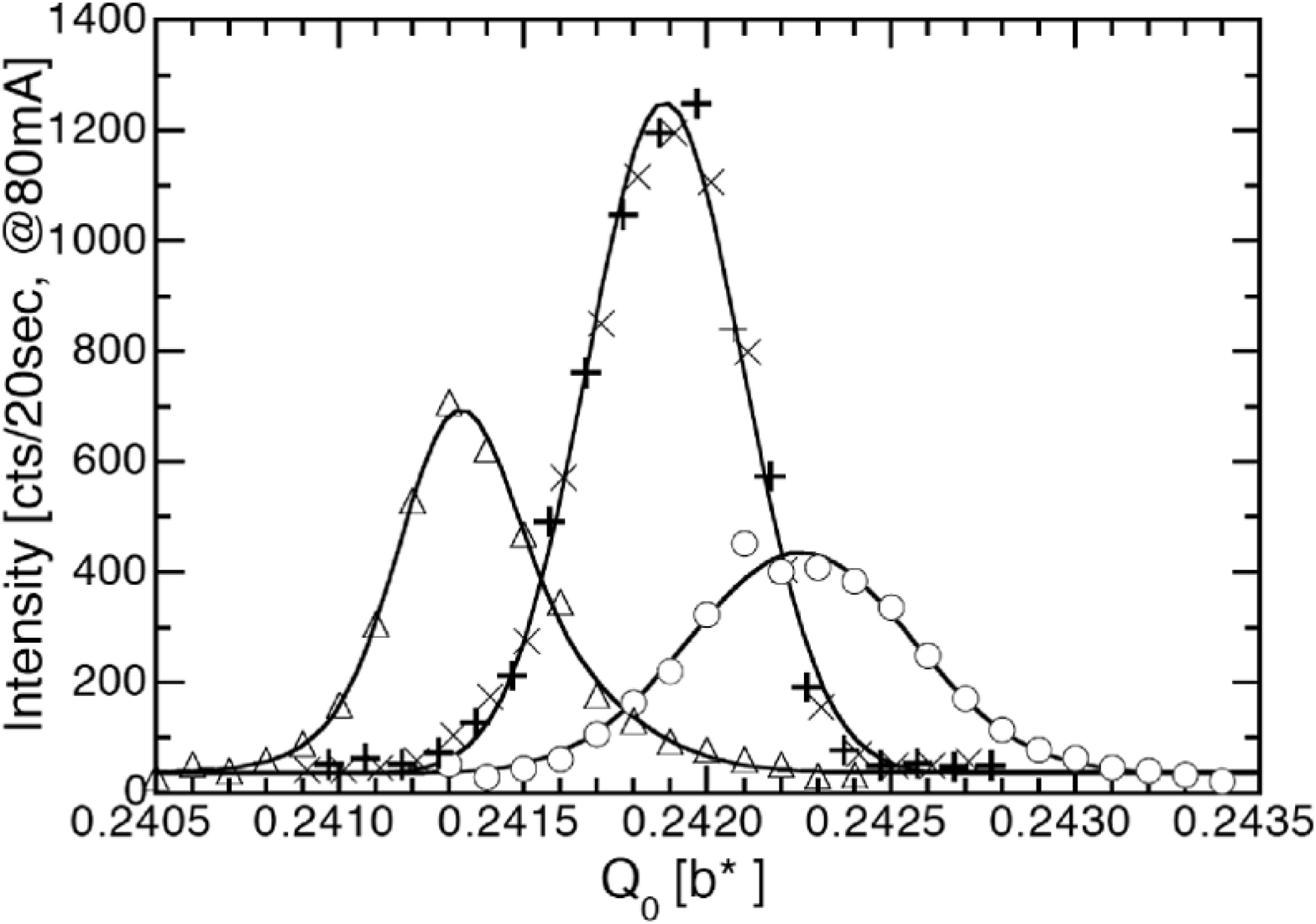}}
\subfigure[]{\label{fig5-14b}
\includegraphics[width=6cm]{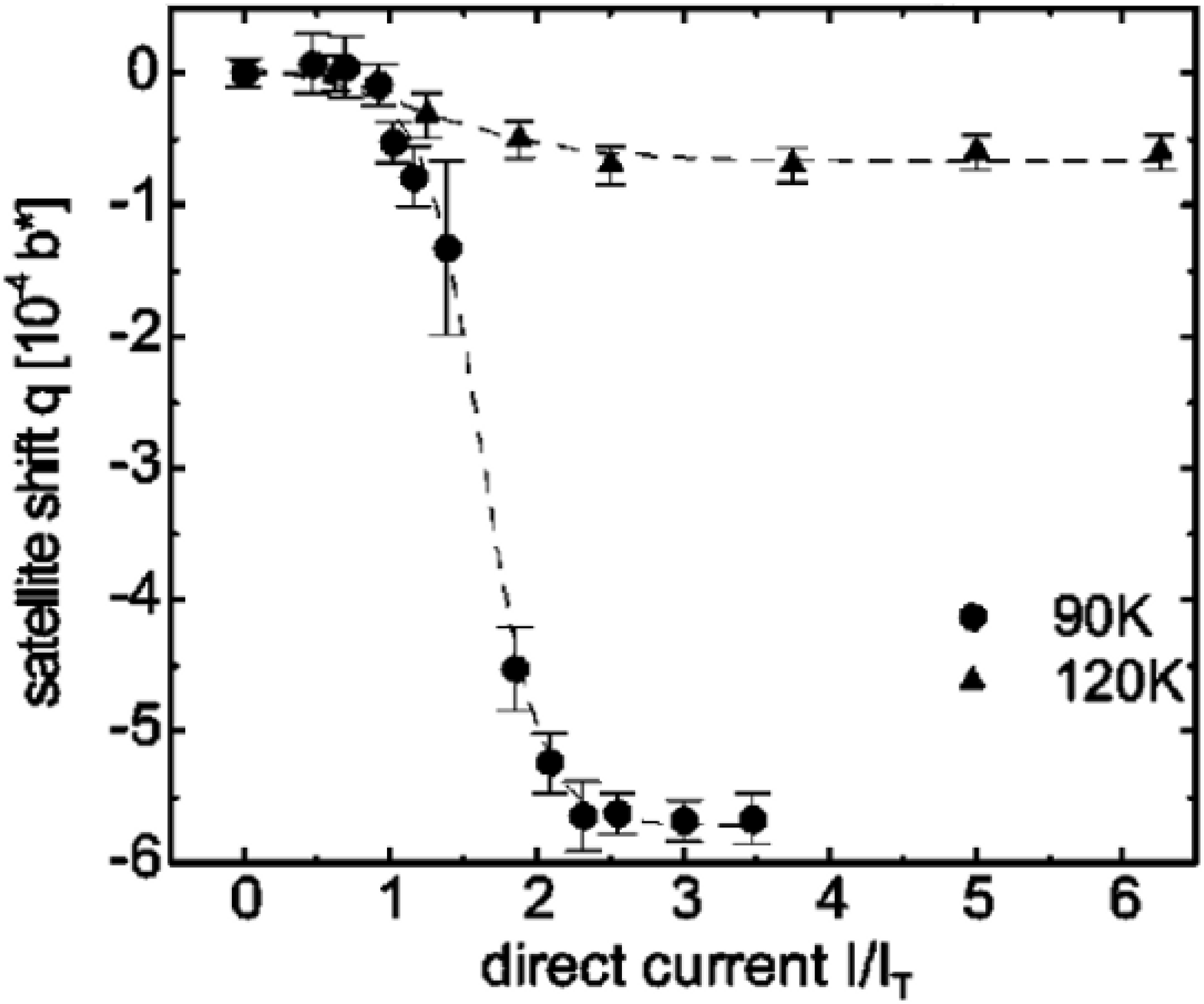}}
\caption{(a) Longitudinal satellite profiles at $x$~= 100~$\umu$m (+) $I_{\rm dc}$~= 0, ($\circ$) $I_{\rm dc}$~= 2.13~$I_T$, ($\times$) $I_{\rm dc}$~= 0 after depolarisation, and ($\vartriangle$) $I_{\rm dc}$~= -2.13~$I_T$; NbSe$_3$, $T$~= 90~K (after ref.~\protect\cite{Requardt98a}). (b)~Change of satellite position $q$ for direct current of varying intensities normalised to the threshold value $I_T$ at the contact boundary; {\large$\bullet$}, $T$~= 90~K, $\blacktriangle$, $T$~= 120~K; dashed lines are guides for the eye (reprinted figure with permission from S. Brazovskii \textit{et al.}, Physical Review B 61, p. 10640, 2000 \cite{Brazovskii00}. Copyright (2000) by the American Physical Society).}
\label{fig5-14}
\end{center}
\end{figure}

The profile of the (0, 1+$Q_0$, 0) CDW is shown in figure~\ref{fig5-14}(a)
at zero field as well as at a dc current of such $I/I_T$~= 2.13, with a positive and negative polarity. The data are taken at a position 100~$\umu$m away from the positive electrode with a beam of 30~$\umu$m wide. When the current exceeds the threshold value $I_T$, the satellite position along $b^\ast$ shifts according to the current polarity and becomes broader and deformed which indicates an inhomogeneously stretched or compressed sliding state. Figure~\ref{fig5-14}(b) shows the satellite shift at the contact boundary, as a function of the applied current normalised to $I_T$ for temperatures $T$~= 90~K and 120~k. At 90~K the satellite shift increases strongly at $I>I_T$ and saturates for currents $I>2I_T$ at a maximum shift of $5.6\times 10^{-4}b^\ast$. The satellite shift is much smaller, namely $6\times 10^{-5}b^\ast$, at 120~K.

\begin{figure}
\begin{center}
\subfigure[]{\label{fig5-15a}
\includegraphics[width=6.7cm]{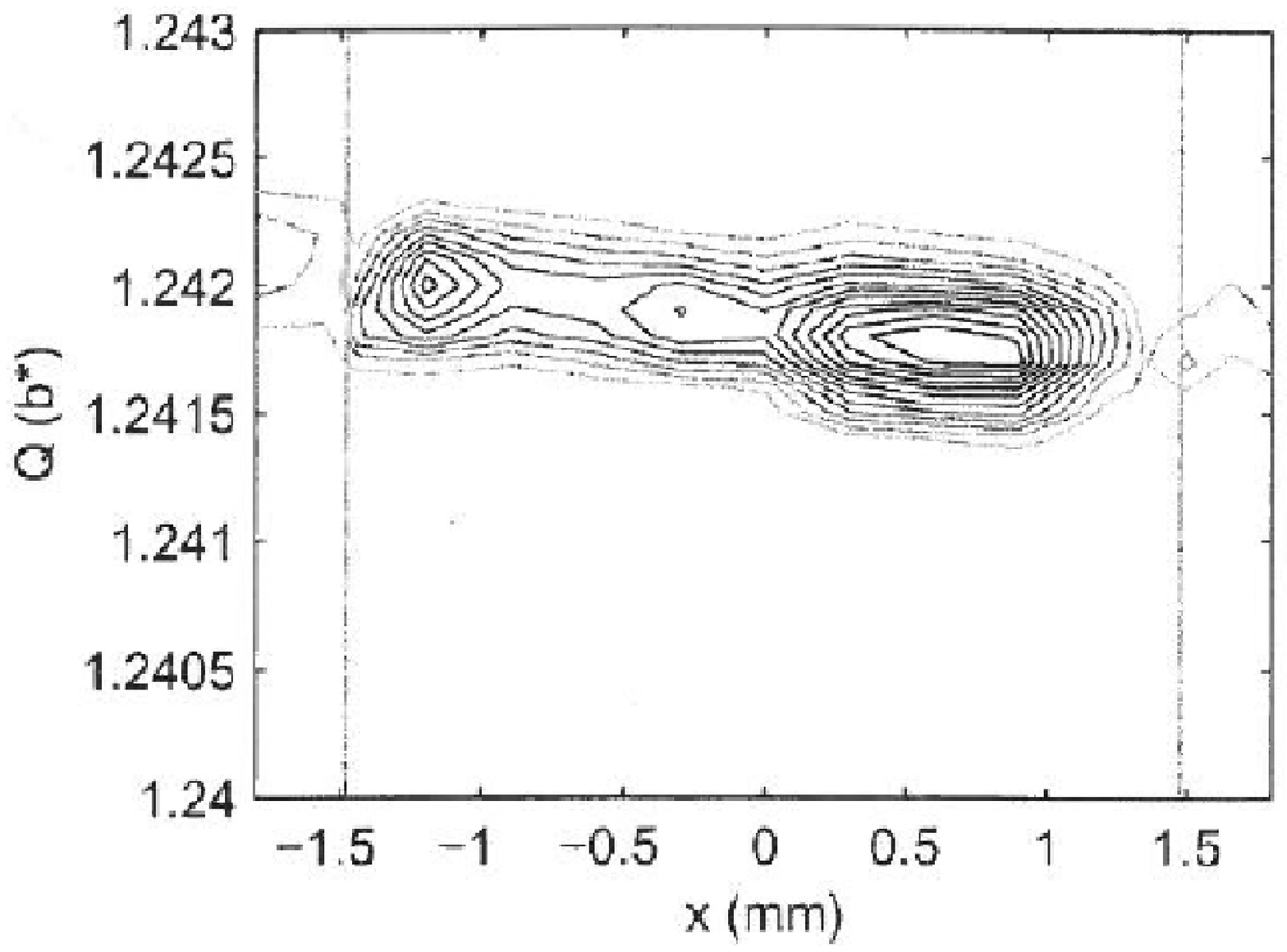}}
\subfigure[]{\label{fig5-15b}
\includegraphics[width=6.7cm]{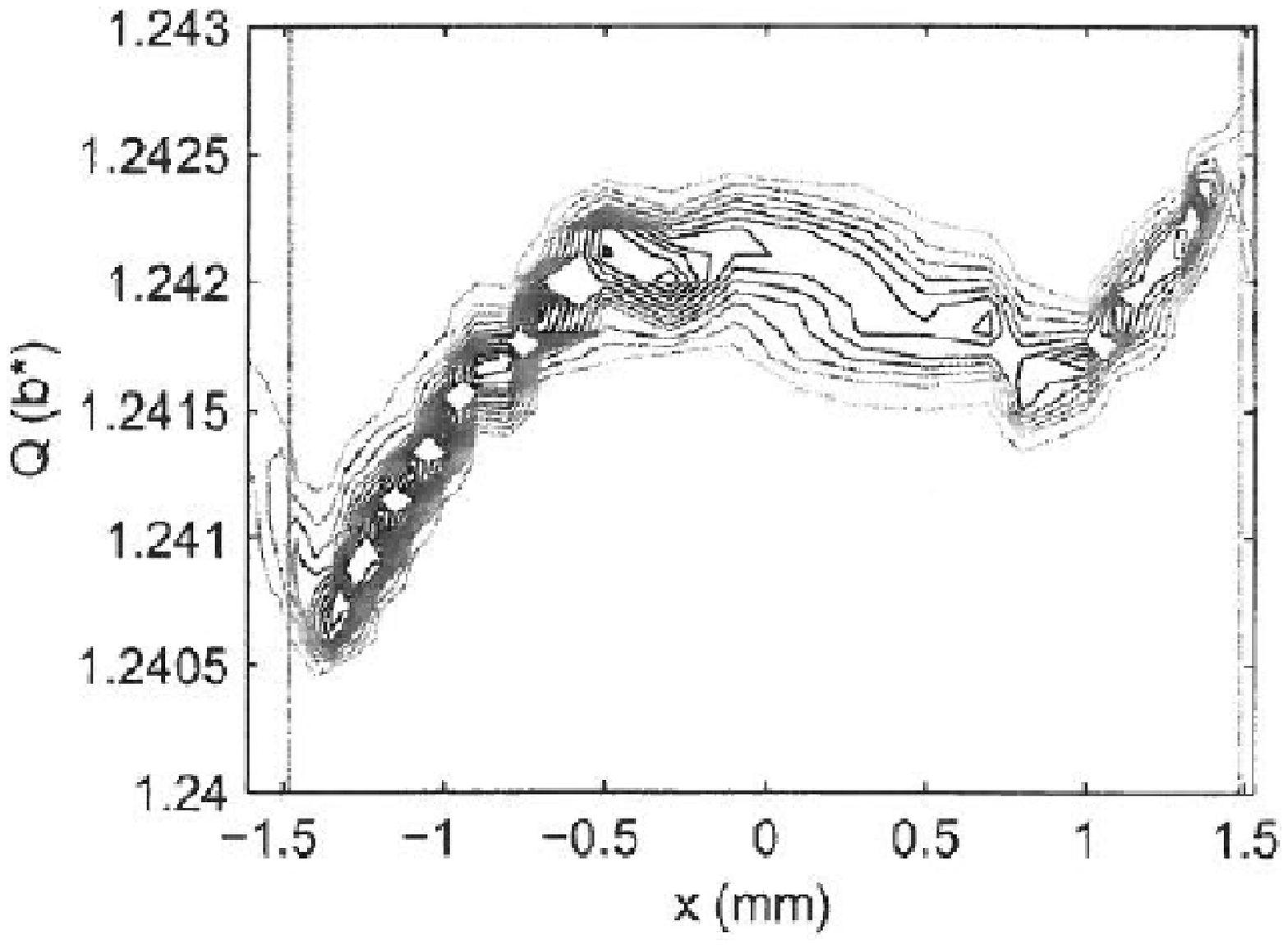}}
\caption{Intensity maps $I(Q,x)$ (arbitrary limits) of the (0, 1+$Q_0$, 0) satellite peak as a function of the position, $x$, and coordinate, $Q$, along the $b^\ast$-direction in reciprocal space: in the pinned state (a), for $I$~= 0~mA; in the sliding state, for $I$~= 3~mA ($I/I_T$~= 3) (b); beam width: 30~$\umu$m, NbSe$_3$, $T$~= 90~K (reprinted figure with permission from Nuclear Instruments and Methods, Physics Research A 467-468, D. Rideau \textit{et al.}, p. 1010 \cite{Rideau01b}. Copyright (2001) with permission from Elsevier).}
\label{fig5-15}
\end{center}
\end{figure}

Figure~\ref{fig5-15}
shows the intensity map $I(x,Q)$ of the (0, 1$\pm Q_0$, 0) CDW satellite peak as a function of beam position along the $b^\ast$-direction in the pinned state $I=0$ (a) and in the sliding state ($I/T$~= 3) (b) at $T$~= 90~K. Only a few samples have a defect-free section and a constant mosaicity  (typically below 0.013$^\circ$) all along several mm between current contacts as evidenced in figure~\ref{fig5-15}(a). Defects which appear to be related to clusters of impurities or grain boundaries provide local misorientations of the $b$-axis. Typically this type of defect occurs once per mm along the sample length \cite{Rideau01a}.

The vertical lines show the contact boundaries. Clearly the intensity map in figure~\ref{fig5-15}(b) reveals the asymmetric shift of the satellite between both polarities and a very small, if any, distortion in the middle part of this 3~mm long sample.
\begin{figure}
\begin{center}
\includegraphics[width=7.5cm]{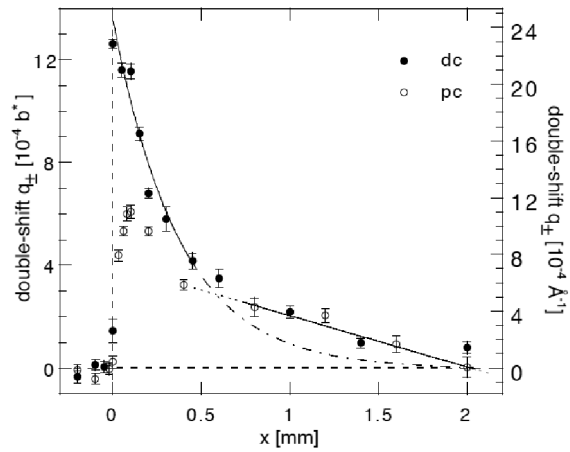}
\caption{Double-shift $q_\pm(x)$~= $Q(+I)-Q(-I)$ (in units of b$^\ast$) for direct ({\large$\bullet$}) and pulsed ({\large$\circ$}) current ($I/I_T$~= 2.13). The full line shows the exponential fit from eq.~(\ref{eq5-33}) near the contact ($0<x<0.5$) and a linear dependence for $0.7<x<2$. The dash-dotted line extrapolates the exponential fit into the central section. The vertical dashed line represents the contact boundary; the horizontal dashed line, the line of zero sift. NbSe$_3$, $T$~= 90~K (reprinted figure with permission from H. Requardt \textit{et al.}, Physical Review B 80, p. 5631, 1998 \cite{Requardt98a}. Copyright (1998) by the American Physical Society).}
\label{fig5-16}
\end{center}
\end{figure}

Figure \ref{fig5-16} shows the spatial dependence of the (double) shift $q_+(x)$~= $Q(+I)-Q(-I)$ of the (0, 1+$Q_0$, 0) CDW satellite both at direct (dc) as well as pulsed (pc) current of magnitude $I/I_T$~= 2.13. With dc current the satellite shift $q$ will reflect the stationary state of the CDW strain, The shift $q(x)$ is monitored as a function of beam position $x$ along one half of the sample-length ($0<x<2$). The spatial variation of $q_+(x)$ can be fitted with an  exponential decay near the electrodes ($0<x<0.7$~mm) with a characteristic length of $\lambda$~= $375\pm 50~\umu$m and a linear variation for 0.7~mm~$<x<2$~mm with a slope $\upartial q_+/\upartial x$~= $-(2.0\pm 0.1)10^{-4}b^\ast$~mm$^{-1}$ (for another sample at 90~K, $\lambda$ for $q_+$~= 290$\pm$~40~$\umu$m and $\lambda$~= $230\pm40~\umu$m for $q_-$).

It has to be noted that in the middle part of the sample there is no observed difference between direct and pulsed currents. On the contrary the pc shift is nearly zero at the electrode position and reaches a maximum at $\sim 100~\umu$m away from the contact boundary. These differences indicate a spatially dependent relaxational behaviour for the CDW deformations, the fastest relaxation occurring at the contact position. In the pulsed current experiments the observed decay of the satellite shift $q$ near the contacts, as a function of the frequency $f$~= $1/\tau$, $\tau_i$: the time between current pulses), varied from $10^{-3}$~s to 10~s, resembles the Kim-Anderson law ($-\ln t$) for the supercurrent decay in superconductors \textit{via} the creep of pinned vortices.

\subsubsection{Normal $\leftrightarrow$ CDW current conversion model}\label{currentconversionmod}\label{sec5-7-2}

The semi-microscopic model by Brazovskii \textit{et al.} \cite{Brazovskii00} describing the normal $\leftrightarrow$ condensed carrier conversion by nucleation and growth of phase dislocation loops in a highly rigid CDW electronic crystal is based on the assumption of the local equilibrium between the electrochemical potentials of the phase dislocations, $U$, and of the free carriers, $\mu_n$. The quantity $\eta\equiv U-\mu_n\propto q$, which is directly measured by X-ray diffraction, measures the electrochemical potential imbalance between $U$ and $\mu_n$ and hence the excess or lack of normal carriers. The stationary distribution of the CDW deformation obeys the equation (see eq.~(18) in ref.~\cite{Brazovskii00}).
\begin{equation}
\frac{\upartial\eta}{\upartial x}=F_r(J_c)-\frac{e(J_n-J_T)}{\sigma_n}.
\label{eq5-31}
\end{equation}
Here $J_n$, $J_c$, $\sigma_n$ and $J_T$ are, respectively, the normal carrier and CDW current densities, the normal carrier conductivity, and the current density at threshold. $F_r(J_c)$, the friction force, is approximated for high current values ($J_{\rm tot}\leq 2J_T$) as $F_r(J_c)\approx eJ_c/\sigma_c$, where $\sigma_c$ is the high-field CDW conductivity.

A boundary condition for eq.~(\ref{eq5-31}) is that all partial currents are stationary in the bulk of the sample which implies the absence of current conversion, all types of carriers being in equilibrium with equal chemical potentials.

The balance between the different types of carriers is controlled by the injection and extraction rates from the electrodes and by the conversion rate $R$:
\begin{equation}
\frac{\upartial J_c}{\upartial x}\propto R(\eta)
\label{eq5-32}
\end{equation}
between normal carriers and CDW condensate.

They are two extreme scenarios for $R(\eta,J_c)$. The first refers to an ideal host crystal, where only homogeneous thermal nucleation is present. Another extreme refers to samples with a sufficiently large density of defects acting as nucleation centres for DLs (heterogeneous nucleation). Both scenarios are of the ``passive" type, when the CDW motion itself plays no role.

A plausible ``active" scenario emerges for a fast enough CDW motion when the DLs are created by the CDW sliding through bulk or surface defects with $R\propto\eta\,J_c$ as the simplest version.

Assuming heterogeneous passive nucleation of dislocation lines, for which the conversion rate can be approximated as $R\propto\tau^{-1}_{\rm conv}\eta$, the solution of eqs~(\ref{eq5-31}) and (\ref{eq5-32}) yields \cite{Brazovskii00}: $q(x)\propto\sin h(x/\lambda_0)$ which can be written as:
\begin{equation}
q(x)\propto\exp\left(-\frac{|x-a|}{\lambda_0}\right)-\exp\left(-\frac{|x+a|}{\lambda_0}\right)
\label{eq5-33}
\end{equation}
with electrodes at $x$~= $\pm a$. Thus, $q$ decreases exponentially near the contacts as shown in figure~\ref{fig5-16} and vanishes in the central part of the sample. $\lambda_0\propto\sqrt{\tau_{\rm conv}}$, typically a few hundred $\umu$m, characterises the length scale of the phase slip distribution. $\tau_{\rm conv}$ is the lifetime of an excess carrier with respect to its conversion to the condensate, i.e. the mean free carrier lifetime before adsorption by a dislocation line.

As shown above the dc data of the $q$ shift shown in figure~\ref{fig5-16} follows satisfactorily this exponential variation in the region near the electrodes. Furthermore, the observed linear variation of $q$ in the central part of the sample (figure~\ref{fig5-16}) suggests that the conversion rate may be suppressed by pinning of the DLs below some finite non-equilibrium threshold $\eta_t$: $R=0$ if $|\eta |<\eta_t$. In this latter case, the carrier conversion is blocked, and the normal and collective current densities, $J_n$ and $J_c$, are fixed at values above ($J_n$) or below ($J_c$), their equilibrium values. Thus, this bulk gradient results from the residual quenched distortion.

By modelling experimental conditions, it was shown \cite{Brazovskii00} that the large gradients observed over the entire bulk in multicontact studies \cite{Adelman96} might be due to an extreme size effect: distances between contacts were so short that only a small part (estimated to $\sim$~16\%) of the normal current was converted to the CDW one, thus, the partial currents being far from their mutual equilibrium values.

\subsubsection{Bulk phase slippage}\label{sec5-7-3}

CDW phase discontinuities may also occur along the sample, away from any contact, due to defects which obstruct the CDW current.
\begin{figure}
\begin{center}
\subfigure[]{\label{fig5-17a}
\includegraphics[width=6.7cm]{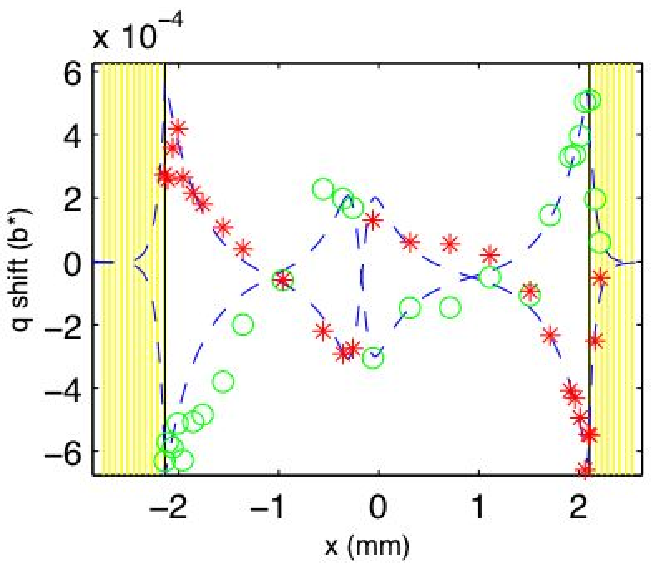}}
\subfigure[]{\label{fig5-17b}
\includegraphics[width=6.7cm]{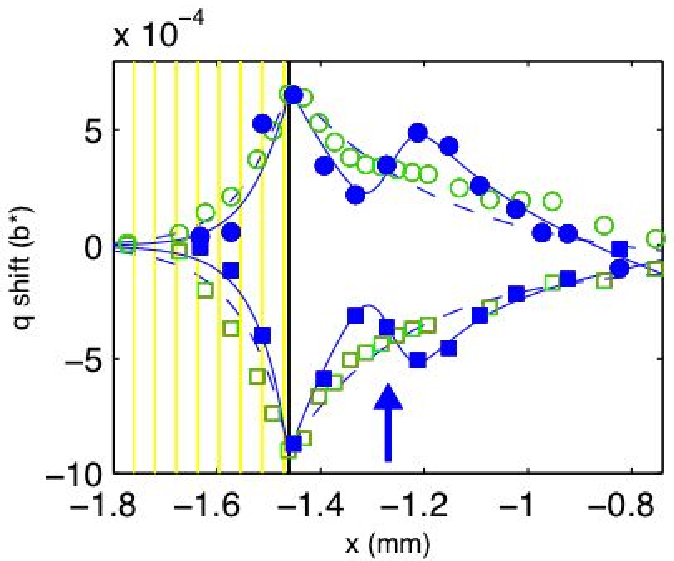}}
\caption{(a) Shift $q(x)$~= $Q_I(x)-Q_0$ of the CDW satellite peak position as a function of beam for positive (red $\ast$) and negative (green $\circ$) polarities. $I$~= $\pm$4.6~mA ($I/I_T$~= 2.13). Phase slippage occurs at the defect position $x$~= -0.15~mm. The vertical lines show the boundaries of the gold-covered contacts. Beam width: 30~$\umu$m, $T$~= 90~K. (b) Shift $q(x)$, along the $\ell hs$ part of a NbSe$_3$ sample for positive (green $\circ$) and negative (green {\tiny $\square$}) polarities; ($I/I_T$~=  3). Full symbols show the shift $q(\pm I)$ after a local irradiation at $x_d$~= -1.27~mm (arrow). NbSe$_3$, $T$~= 90~K (reprinted figure with permission from D. Rideau \textit{et al.}, Europhysics Letters 56, p. 289, 2001 \cite{Rideau01}. Copyright (2001) from EdsSciences).}
\label{fig5-17}
\end{center}
\end{figure}

Figure~\ref{fig5-17}(a) shows \cite{Rideau01} the shift $q(x)$~= $Q(I)-Q(0)$ for positive and negative current polarities as a function of beam position, $x$, along the sample with $I/I_T$~= 2.13. This sample exhibits between electrodes one localised defect at the position $x_d\approx$~-0.15~mm. The sliding CDW satellite shift changes sign abruptly at this position, with maxima on either sites of the defect position, as well as at the electrical contacts.

A damaged region can also be created by exposing a small part of a sample ($\sim 30~\umu$m) to the X-ray beam over  long period of time (4 hours). Figure~\ref{fig5-17}(b) shows the shift $q(x)$ for both current polarities, as a function of $x$ before and after irradiation. As seen, the value of the CDW satellite position changes abruptly at the irradiated position ($x_d$~= -1.27~mm, the arrow in figure~\ref{fig5-17}(b)). Using the Brazovskii model with an enhanced pinning force at the defect position, it was possible to give a coherent description of the spatial dependence of $q(x)$ near both types of defects \cite{Rideau01,Rideau01a}.

On the NbSe$_3$ sample on which irradiation was performed, one current electrode was 1~mm wide, 15~$\umu$m wide for the other. Data shown in figure~\ref{fig5-17}(b) were performed near this electrode. One can note that the CDW distortions extend over an appreciable distance ($\approx 100~\umu$m beyond the contact boundary, as discussed in section~\ref{sec5-2-2}).

\subsubsection{Transient structure of sliding CDW}\label{sec5-7-4}

The difference in the $q(x)$ deformation between dc and pc currents near electrodes (figure~\ref{fig5-16}) reveals strong spatially dependent CDW relaxation during the pauses between current pulses. Time dependence of the longitudinal deformation, $q(t)$, with a high spatial resolution (beam width: 30~$\umu$m) at several distances from the current contact was studied with application of unipolar pulses \cite{Requardt99,Requardt02}. The relaxation of the satellite shift after the pulse is best fitted by a profile of the stretched exponential type: $q(t)$~= $q_0\exp[(-t/\tau)^\mu]$, a time dependence found in scaling solutions of statistical models. $\mu$ was found in the range of 0.4~$\sim$~0.7; the relaxation time $\tau$ becoming faster with increasing $T$ and slowing down with increasing distance from the current contact.

After the pulse, the $q(x)$ shift relaxes towards the non-deformed state, but does not reach the initial position corresponding to zero current, indicating that remnant CDW deformations remain frozen. To erase such remnant deformations due to sample history and in order to have reproducible results, a procedure analogous to that used in demagnetisation or depolarisation techniques, i.e. application of current of alternating polarities and decreasing amplitude, was used. This technique allows to recover the original zero-field satellite position as well as the corresponding profile widths, along b$^\ast$ and perpendicular to the chain direction (rocking width) (see reproducible profile of the satellite with this depolarisation procedure in figure~\ref{fig5-14}(a)).

Other X-ray work on CDW time dependence have been also performed with relaxed spatial resolution either by application of bipolar pulsed currents \cite{Sweetland94} (which include effects of sample history in the observable CDW deformations --see above), or studying the transverse CDW deformations in the centre of the sample where the longitudinal $q$ shift is zero \cite{Ringland99}. It is then difficult to compare directly these results which do not address the same objects of study.

\subsubsection{Phase slip in narrow superconducting strips}\label{sec5-7-5}

Very similarly the non-equilibrium region in the vicinity of a phase slip centre in narrow superconductivity strip was described by the imbalance between the electrochemical potentials $\mu_{\rm P}$ for the paired electrons and $\mu_{\rm Q}$ for quasi particles \cite{Skocpol74}. While $\mu_{\rm P}$ varies spatially only over a distance of the order of the coherence length, it was shown that $\mu_{\rm Q}$ decays exponentially towards $\mu_{\rm P}$ with a characteristic length $\lambda_{\rm Q}$. The gradient in $\mu_{\rm Q}$ corresponds to a quasi particle current. It was possible to access directly $\mu_{\rm Q}/e$ and $\mu_{\rm P}/e$ as potentials by I-V measurements using probes made of a normal metal or a superconductor one respectively \cite{Dolan77}.

\subsubsection{Controllable phase slip}\label{sec5-7-6}

\begin{figure}
\begin{center}
\includegraphics[width=7.5cm]{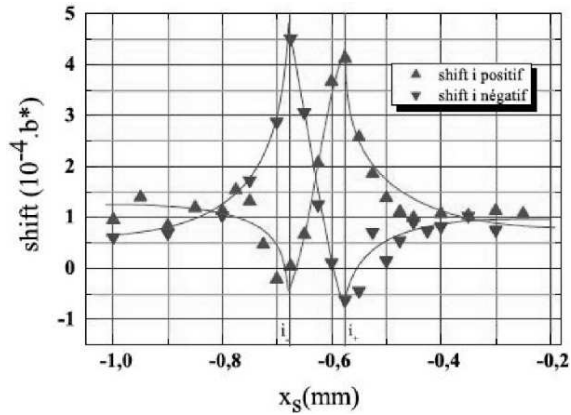}
\caption{Satellite shift near the closely-spaced (95~$\umu$m) electrodes for $i$~= $\pm$5.6~$i_{\rm th}$ ($I$~= 0). The origin of the vertical scale refers to the satellite position at $i$~= 0 in the depolarised state. The lateral beam width is 20~$\umu$m. NbSe$_3$, $T$~= 100~K (after ref.~\protect\cite{ESRF04}, P. Monceau, unpublished).}
\label{fig5-19}
\end{center}
\end{figure}

In order to get rid of slippage at an uncontrolled defect as shown in figure~\ref{fig5-17}(a), it has been proposed to perform a detailed study of the local CDW deformations by X-ray when the CDW coherence is broken by an additional current applied is a very short segment of the sample \cite{ESRF04} (corresponding to the schema shown in figure~\ref{fig5-8}(a)). Figure~\ref{fig5-19} 
shows the observed CDW satellite shift $q(x)$ in the vicinity of electrode pair (1-2, see figure~\ref{fig5-8}(a)) corresponding to respectively: $x$~= -0.675 and -0.570~mm (i.e. for a distance between electrodes of 95~$\umu$m, width of electrodes: 5~$\umu$m) for $i$~= $\pm 5.6\,i_{\rm th}$ at $T$~= 100~K and $I$~= 0. One can note that the CDW deformation peaks precisely at the electrode position and that the CDW deformation extends far ($\sim 300~\umu$m) outside the inter-electrode region. The linear variation of $q(x)$ between electrodes results from the strong finite-size effect as discussed in \ref{currentconversionmod}. Unfortunately technical problems have made impossible the second part of experiment with simultaneous application of $I$ (between electrodes 3-4) and $i$.

\subsubsection{CDW deformations on compounds with a semiconducting ground state}\label{sec5-7-7}

- o-TaS$_3$

\begin{figure}
\begin{center}
\subfigure[]{\label{fig5-20a}
\includegraphics[width=6.25cm]{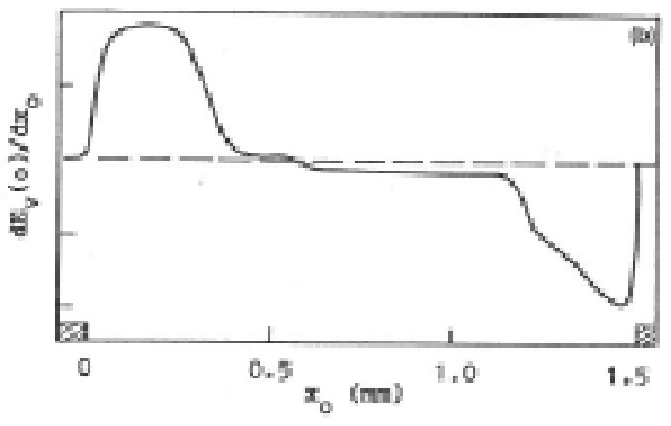}}
\subfigure[]{\label{fig5-20b}
\includegraphics[width=6.5cm]{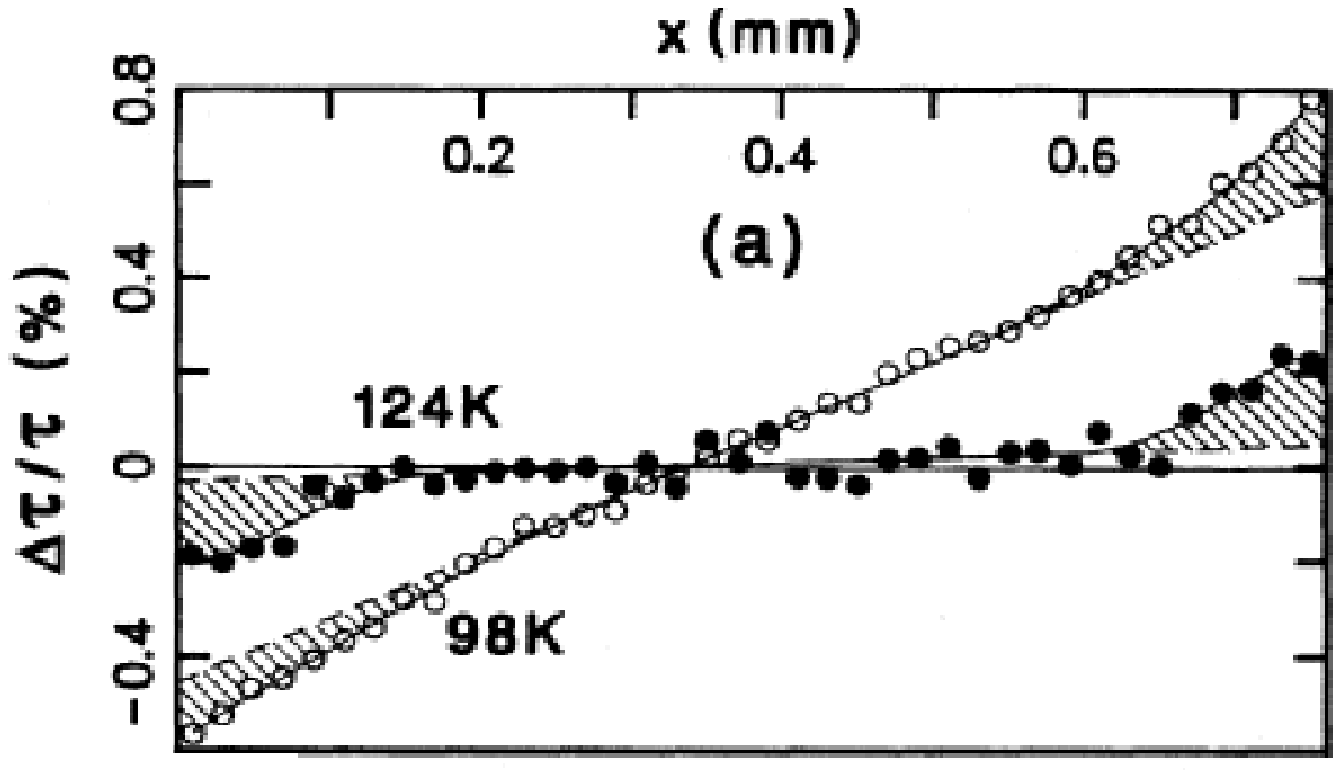}}
\caption{(a) Dependence of the derivative of the half-width of the field-induced hysteretic loop as a function of the position $x_0$ at which the sample is illuminated with a laser beam (laser beam width 10~$\umu$m); TaS$_3$, $T$~= 125~K (reprinted figure with permission from Soviet Physics JETP 63, M.E. Itkis \textit{et al.}, p. 177, 1986  \cite{Itkis86}. Copyright (1986) from Springer Science and Business media). (b)~Spatial dependence of electromodulated infrared transmission of K$_{0.3}$MoO$_3$ at two temperatures with the same non-linear (CDW) current (reprinted figure with permission from M.E. Itkis \textit{et al.}, Physical Review B 52, R11545, 1995 \cite{Itkis95}. Copyright (1995) by the American Physical Society).}
\label{fig5-20}
\end{center}
\end{figure}

It was shown that laser illumination with a narrow laser beam on a o-TaS$_3$ sample gives rise to a thermopower voltage, which depends on the position on the sample which is illuminated and on the magnitude and sign of the applied electric field \cite{Itkis86}. This thermopower voltage is caused by the non-uniform heating of the sample by the electromagnetic radiation with a temperature distribution decreasing from the lighted spot towards the contacts. By sweeping the applied voltage from positive to negative values above the threshold, the observed hysteretic loop reflects the remnant deformations of the CDW. The dependence of the magnitude of this hysteretic loop on the position $x_0$ of the illumination was measured by moving the laser beam along the sample \cite{Itkis86}. Figure~\ref{fig5-20}(a) 
shows the spatial derivative of the half-width of the hysteretic loop ${\rm d}E_v/{\rm d}x_0$ on the coordinate, $x_0$, at the point at which the sample is illuminated (laser beam diameter $\sim 10~\umu$m, positional precision $\sim 10~\umu$m, sample length $\sim 1.5$~mm). ${\rm d}E_v/{\rm d}x_0$ determines the distribution of the residual deformation of the CDW. The deformation is maximal in the vicinity of the contacts and extends within the sample over a macroscopic distance of $\sim 300~\umu$m. But no deformation occurs in the middle part of the sample, at least at this given temperature ($T\sim 100$~K).\\

- K$_{0.3}$MoO$_3$

In blue bronze the spatial distribution of the CDW deformations was studied by electromodulated infrared transmission \cite{Itkis95}. A difference in the transmitted intensities is observed between the pinned and the sliding state. That suggests that the infrared absorption comes from excitation of quasi particles the local density of which compensates the excess (or the lack) of charges $\delta\rho$ resulting from the deformations of the sliding CDW. Thus the phase gradient $\upartial_x\phi$ is proportional to the relative difference in IR transmitted intensity:
\begin{equation*}
\frac{\Delta\tau}{\tau}\propto\upartial_x\phi .
\end{equation*}

Spatial dependence of $\Delta\tau/\tau$ at different applied voltages (symmetric square waves) exhibit some differences between samples which may reflect some hidden defects along the sample. However, results indicate a contact strain which appears within $\sim 100~\umu$m of the contacts. Figure~\ref{fig5-20}(b) shows the spatial dependence of $\Delta\tau/\tau$ for a given sample at two temperatures for the same non-linear (CDW) current. At $T$~= 124~K, the profile shows only strains at the contacts but no bulk strain, very similarly to the data on o-TaS$_3$ (figure~\ref{fig5-20}(a)), while at $T$~= 98~K, there is a linear bulk strain.

No systematic temperature dependence of the length $\lambda$ on which the quasi particle conversion takes place was presented. One can envisage some differences of $\lambda(T)$ depending if the ground state is (semi)-metallic (NbSe$_3$) or semiconducting (o-TaS$_3$, K$_{0.3}$MoO$_3$, \ldots) due to different screening effects. It may also occur that $\lambda$ in K$_{0.3}$MoO$_3$ is strongly $T$-dependent and that, at $T$~= 98~K, there are interferences between contacts leading to size effects as discussed above.

\subsection{$Q_1-Q_2$ coupling in NbSe$_3$}\label{sec5-8}

A recurrent question arisen in NbSe$_3$ concerns the apparent independence of both CDWs which, however, occur on adjacent chains in the same unit cell. Thus in (TTF-TCNQ) the coupling between the CDW which occurs first on TCNQ chains at 54~K and that on TTF chains at 49~K is manifested by a temperature dependence of the CDW wave-vector along $a^\ast$ from $0.5a^\ast$ to $0.3a^\ast$ with a discontinuous locking at $0.25a^\ast$ at 38~K \cite{Comes79}.

Below $T_{\rm P_2}$,  phase locking of the coexisting CDWs was anticipated \cite{Bruisma80} from the fact that the two modulation vectors nearly satisfy the relation $2(Q_1+Q_2)\approx$~$(111)~= 0$ which  suggest the possibility of a joint commensurability between the lattice and the two CDWs. However no anomaly in the temperature dependence of $Q_1$ was detected in the vicinity of $T_{\rm P_2}$ \cite{Moudden90}. However high resolution STM images, in addition to the modulation on chains III corresponding to $Q_1$ and on chains I corresponding to $Q_2$ have revealed a beating phenomena between the $Q_1$ and $Q_2$ periodicities \cite{Brun09}. That is manifested by a new domain superstructure with a periodicity $u$~= $2(0.26-0.24)b^\ast$, i.e. twice the difference in commensurability between the $Q_1$ and $Q_2$ CDW components along the chains.

Using high-resolution X-ray scattering in the presence of an applied current, a dynamical decoupling between the CDWs in NbSe$_3$ was reported \cite{Ayari04}. It was shown that, when both CDWs are set in motion, the $Q_1$ and $Q_2$ satellites undergo simultaneous and opposite shifts along $b^\ast$, away from their static positions. This sliding-induced structural change implies a dynamical electronic charge transfer between the $Q_1$ and $Q_2$ condensates.

This effect should not be confused to the satellite shifts that are current polarised, i.e. which change sign with current polarity as discussed in section~\ref{sec5-7} (see figure~\ref{fig5-17}), ascribed to CDW current conversion at the electrodes.
\begin{figure}[h!]
\begin{center}
\includegraphics[width=7.5cm]{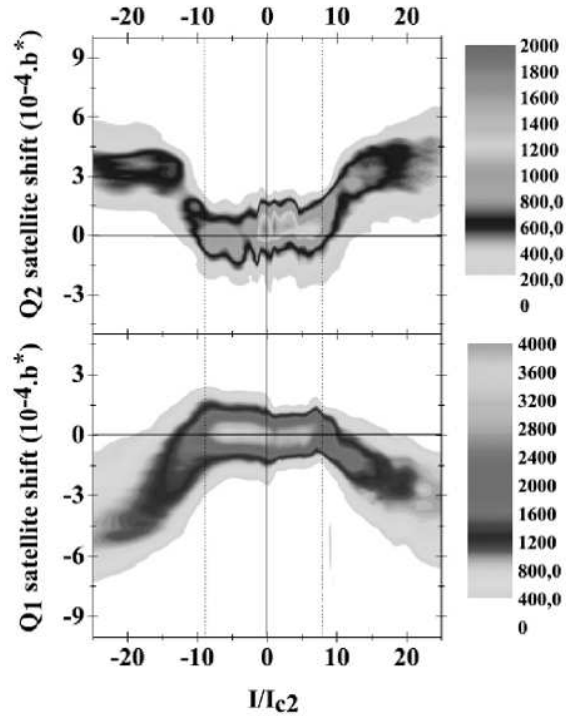}
\caption{NbSe$_3$: longitudinal (bottom) $Q_1$- and (top) $Q_2$-satellite profiles versus normalised current at $T$~= 45~K. The origin on the vertical scale coincides with the unshifted (zero-current) peak center and the two vertical dotted lines mark the position of the upper threshold current, $I_{C1}$ (reprinted figure with permission from A. Ayari \textit{et al.}, Physical Review Letters 93, p. 106404, 2004 \cite{Ayari04}. Copyright (2004) by the American Physical Society).}
\label{fig5-21}
\end{center}
\end{figure}

The present effect is not current polarised and not related to electrode conversion. Figure~\ref{fig5-21} shows the longitudinal $Q_1$ (bottom trace) and $Q_2$ (top trace) satellite profiles versus normalised current at 45~K. Measurements show opposite shifts for the two satellites at $(0,\,1+Q_{1\parallel}\,0)$ and $(0.5,1+Q_{2\parallel},0.5)$. Within experimental error, the decrease of $Q_{1\parallel}$ compensates for the increase of $Q_{2\parallel}$, the deviation from commensurability of the sum $Q_{1\parallel}+Q_{2\parallel}$ remaining unchanged.

From interlayer tunnelling technique (see sec.~\ref{sec11-6}), it was also shown that the formation of the low-$T$ CDW gap in NbSe$_3$ is accompanied by an increase of the high-$T$ CDW below $T_{\rm P_2}$ \cite{Orlov06}.

\section{Screening effects}\label{sec6}
\setcounter{figure}{0}
\setcounter{equation}{0}

It was shown in section \ref{sec4} that the existence of a threshold field for sliding requires that the CDW is deformable. The lattice and impurities distort the moving CDW. These distortions resulting from the competition between the elastic energy and the random pinning energy accumulate locally charges which are screened by the remaining normal electrons. These screening effects will be more important for C/S DW with a semiconducting ground state with a decrease of normal conductivity by many orders of magnitude below the CDW or SDW phase transition.

\subsection{Relaxation of polarisation}\label{sec6-1}

One way to study the effect of temperature on screening is the measurement of relaxation of the electrical polarisation. The CDW is polarised by application of an electric field (typically with amplitude higher than $E_T$) and the depolarisation current is measured as a function of time from the switch-off of the polarisation pulse field. For blue bronze it was shown that a good fit \cite{Kriza86} for the time decay of the CDW polarisation has the form of a stretched exponential function:
\begin{equation}
P(t)=P_0\exp\left\{-\left[\frac{t}{\tau(T)}\right]^{1-n}\right\},
\label{eq6-1}
\end{equation}
with the temperature dependence included in $\tau(T)$.

\begin{figure}
\begin{center}
\subfigure[]{\label{fig6-1a}
\includegraphics[width=6.5cm]{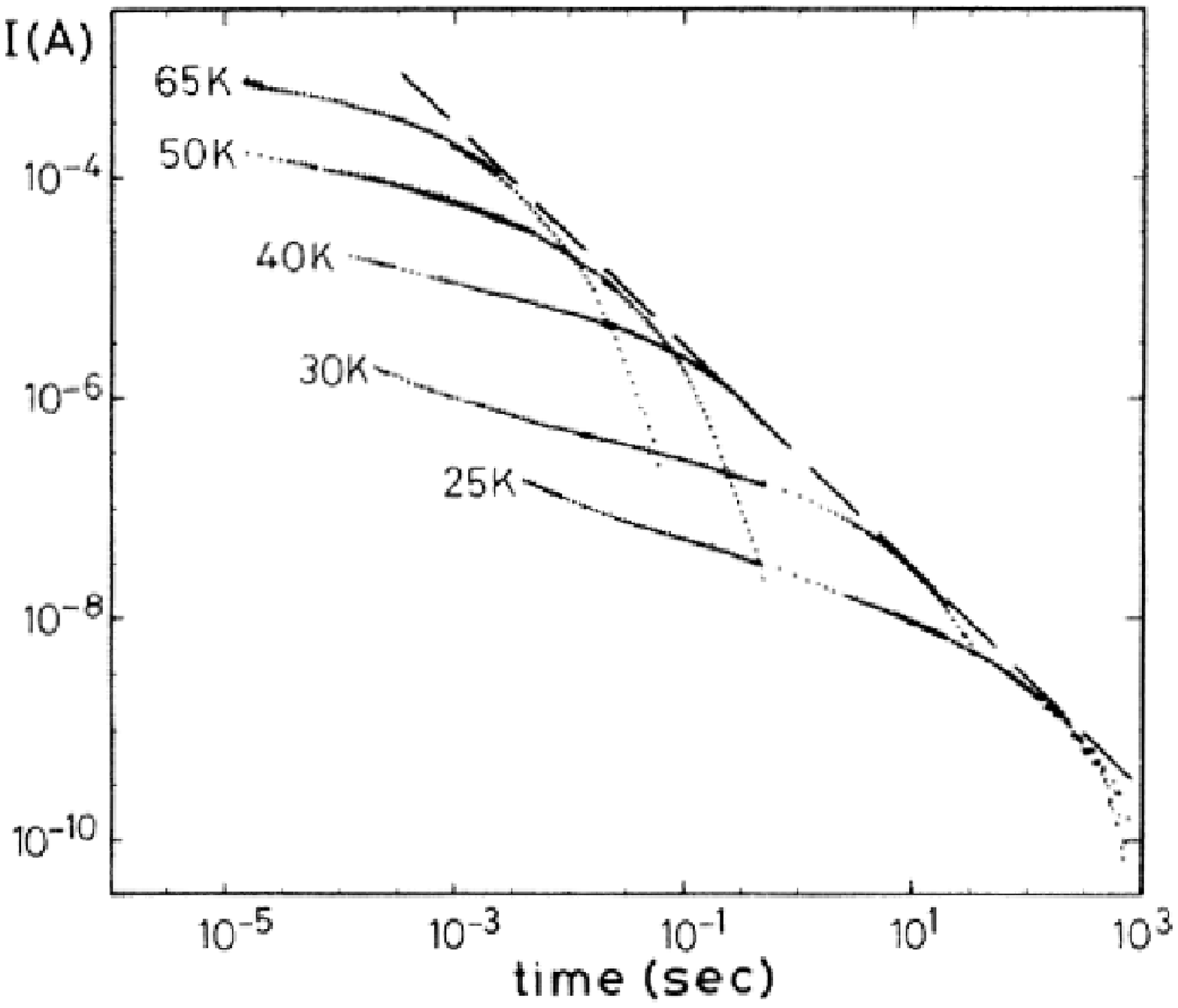}}
\subfigure[]{\label{fig6-1b}
\includegraphics[width=6.5cm]{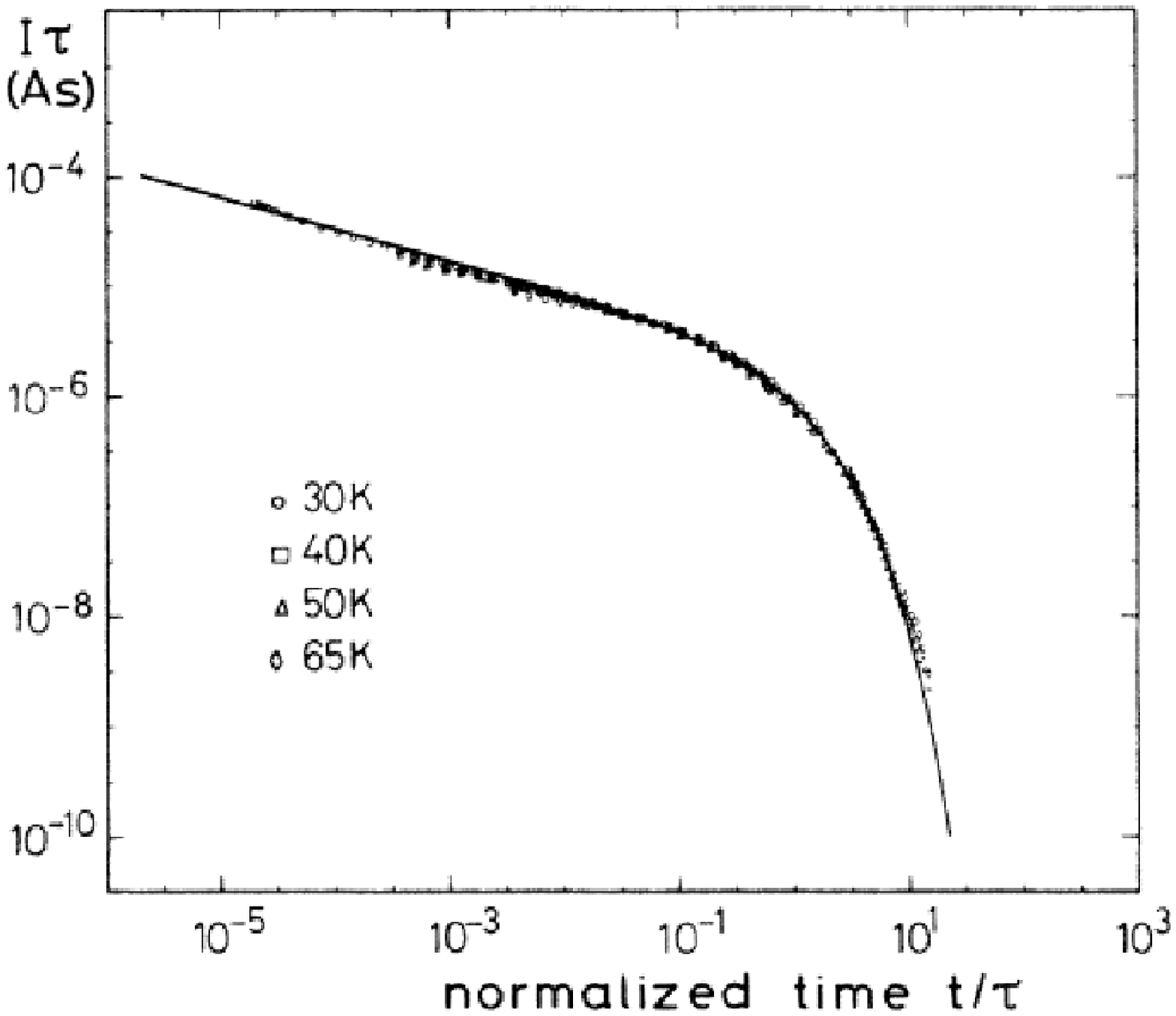}}
\subfigure[]{\label{fig6-1c}
\includegraphics[width=6.5cm]{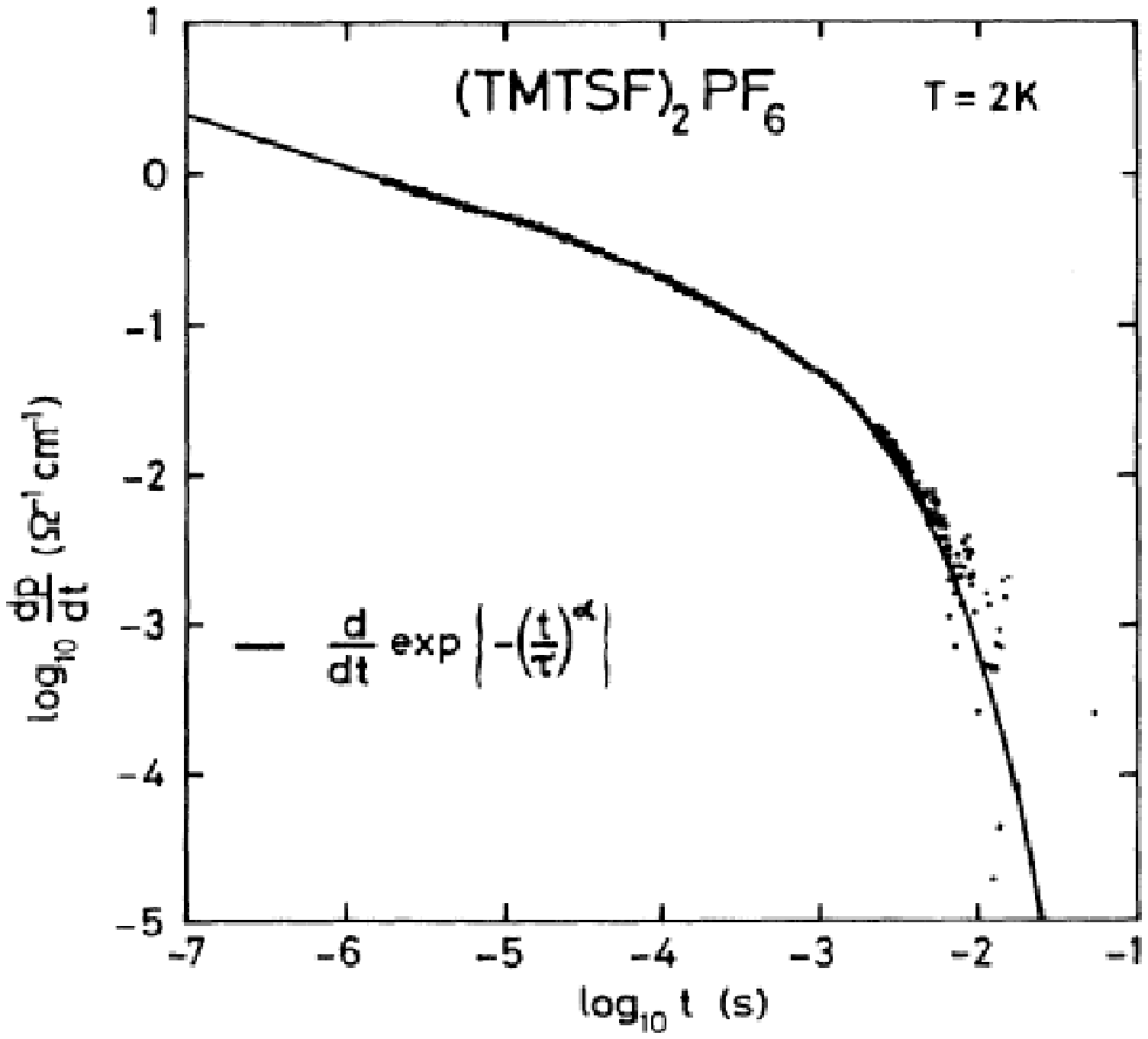}}
\vspace{-0.25cm}
\caption{Time dependence of the depolarisation current (i.e. derivative of the polarisability) after switch off of a polarisation pulse field with $E>E_T$ (a)~for K$_{0.3}$MoO$_3$ at different temperatures. (b)~The same as in (a) plotted on normalised time and current scales. (c)~For (TMTSF)$_2$PF$_6$. The solid lines in (b) and (c) correspond to a stretched exponential behaviour (reprinted figure with permission from G. Kriza and G. Mihaly, Physical Review Letters 66, p. 2806, 1991 \cite{Kriza86}. Copyright (1991) by the American Physical Society for (a) and (b) and reprinted figure with permission from G. Mihaly \textit{et al.}, Physical Review Letters 67, p. 2713, 1991 \cite{Mihaly91}. Copyright (1991) by the American Physical Society for (c)).}
\label{fig6-1}
\end{center}
\end{figure}

Figure~\ref{fig6-1}(a) 
shows the time dependence of depolarisation current (derivative of the polarisation) at different temperatures on K$_{0.3}$MoO$_3$. When time and current are normalised with $\tau(T)$, one obtains a master curve as shown in figure~\ref{fig6-1}(b) in which the solid line correspond to the stretched exponential dependence with $n$~= 0.3. Thus, the only temperature-dependent parameter involved in the decay of the polarisation is relaxation time which scales with temperature following an activated behaviour: $\tau(T)$~= $\tau_0\,e^{\Delta/T}$ as the ohmic resistance. This broad relaxation process, larger than it would be expected for single-particle excitations, reveal internal density wave excitations indicating a wide distribution of metastable pinned CDW configurations.

A similar behaviour is observed \cite{Mihaly91a,Mihaly91} for the dielectric relaxation of (TMTSF)$_2$PF$_6$ in the pinned state (see figure~\ref{fig6-1}(c)), indicating the very similar role of internal modes with a broad distribution of relaxation times in the CDW and SDW condensate.

Huge CDW polarisation can be directly investigated by the thermally stimulated discharging current technique largely used for many electric  materials such as polymers, dipolar glasses \cite{Turnhout75}. Experimentally a blue bronze sample was quenched to 4.2~K from 77~K in presence of an electric field large enough to initiate sliding. Then the electric field being removed, the current released was measured as a function of the heating rate. This method extends the dielectric measurements down to very low frequencies $\sim 10^{-4}$~Hz. Depolarisation current of the order of $10^{-9}$~A for K$_{0.3}$MoO$_3$ \cite{Cava84a}, $10^{-10}$~A for (TaSe$_4$)$_2$I and around $10^{-13}$~A for o-TaS$_3$ \cite{Staresinic99} were released on heating with a maximum which moves at higher temperature for larger heating rates.

\subsection{Elastic hardening due to Coulomb interaction}\label{sec6-2}

But the ac method is the most intensively used for measurements of the dielectric susceptibility in the frequency range below $10^7$~Hz. From the measurement of the complex conductivity $\sigma(\omega)$, the calculation of the real and imaginary parts of the dielectric susceptibility are obtained for the equations:
\begin{eqnarray}
\varepsilon^\prime= & \displaystyle\frac{\Im \sigma(\omega)}{\omega}\label{eq6-2}\\
\varepsilon^{\prime\prime}= & \displaystyle\frac{[\Re \sigma(\omega)-\sigma_{\rm dc}]}{\omega}, \label{eq6-3}
\end{eqnarray}
where $\sigma_{\rm dc}$ is the conductivity for dc current ($\omega\rightarrow 0$). In addition to the well defined peak of the dielectric constant close to the pinning frequency $\Omega_0$ in the range of GHz ascribed to the collective mode of the pinned CDW as calculated by Lee, Rice and Anderson \cite{Lee74}, it exists an extra loss peak at lower frequency (see figure~\ref{fig4-2}). The dielectric response was shown to follow the empirical law \cite{Fleming89,Cava84b}:
\begin{equation}
\varepsilon(\omega)=\frac{\varepsilon_0}{\left[1+(i\omega\tau)^{(1-\alpha)}\right]^\beta},
\label{eq6-5}
\end{equation}
which describes the overdamped CDW motion within a single metastable well ($\alpha$~= 0.25, $(1-\alpha)\beta$~= 0.7) which can be interpreted in terms of a broad distribution $n(E)$ of localised relaxational modes.

Measurements of $\varepsilon(\omega)$ have been first performed in a limited temperature range on K$_{0.3}$MoO$_3$ ($60<T<101$~K) \cite{Cava84b,Fleming89}, o-TaS$_3$ ($73<T<128$~K) \cite{Cava85,Wu86}, (TaSe$_4$)$_2$I ($90<T<180$~K) \cite{Cava86,Wu86}. It was shown that the main relaxation time, typically between $10^{-4}\sim 10^{-7}$~s, of the broad mode increases when $T$ is decreased, with a thermally activated law, with an activation energy of the same order of magnitude than the ohmic resistivity, but not equivalent, except for (TaSe$_4$)$_2$I where both are the same.

Littlewood \cite{Littlewood87b} showed that the low-frequency dielectric behaviour is the response of longitudinal screened collective modes of the CDW (which couples to the electrostatic potential). These longitudinal modes are strongly damped by the interaction with free carriers. At frequencies close to the pinning frequency, the coupling is essentially to transverse CDW modes which, unscreened, are much less damped. However, because the non-uniform nature of the pinning, it was shown that a ac current with zero wave-vector $q$ excites finite $q$-modes which are never purely longitudinal nor purely transverse which makes possible the coupling to longitudinal modes.

The effect of Coulomb interaction on the collective modes of the CDW was investigated by many authors \cite{Barisic87,Barisic89,Baier90,Virosztek93,Wong87,Nakane85}. In particular Sneddon \cite{Sneddon84} studied effects of the screening of CDW distortions resulting from pinning on the non-linear I-V characteristics. He showed that led to an enhancement of the effective damping and consequently to a reduction of the non-linear CDW conductivity, that was experimentally observed \cite{Fleming86}.

When taking into account screening effects of CDW deformations and long range Coulomb interactions \cite{Sneddon84,Littlewood87b}, the charged CDW deformations are electrostatically coupled to normal electrons. The total current should be now written \cite{Littlewood87b} as:
\begin{eqnarray}
j=\varepsilon
\begin{array}{c}
\mbox{\tiny$\circ$}\vspace{-3pt} \\
E\\ \vspace{-4pt}
\\
\end{array}
+\sigma E+j_{\rm CDW},
\end{eqnarray}
where the first term is the displacement current with $\varepsilon$ the dielectric constant, the second one the linear ohmic current where $\sigma$ is thermally activated as $\exp(-\Delta/kT)$ and the last term the extra CDW current. The relevant parameter is now $\omega\,\varepsilon/\sigma$ or $\omega/\omega_1$ ($\omega_1$~= $\sigma/\varepsilon$).

Two regimes are distinguished:
\vspace{-6pt}
\begin{itemize}
\item[-] at low frequencies or at relatively high temperature, the conduction electrons are able to screen the CDW deformations. This black flow current induces a Ohmic dissipation which accounts for the enhanced damping. The effective damping was found \cite{Littlewood87b} to be
\begin{equation}
\gamma_{\rm eff}=\gamma_0+\gamma_1\left[1+\left(\omega/\omega_1\right)^2\right]^{-1}.
\label{eq6-7}
\end{equation}
$\gamma_0$ is the damping which comes from phason-phason or phason-phonon scattering \cite{Takada85}. This damping varies as $T^2$ and should vanish at $T\rightarrow 0$.

\item[-] At high frequencies, for $\omega\varepsilon/\sigma\gg 1$, and at low $T$, the few remaining normal carriers are no more able to screen the CDW deformations and there are Coulomb interactions of the CDW with itself. These long range Coulomb interactions introduce a gap for excitations in the phason mode (section~\ref{sec6-4}), believed to led to switching effects in the I-V characteristics (see below section~\ref{sec6-2}).
\end{itemize}

\subsection{Low frequency dielectric relaxation at low temperature}\label{sec6-3}

If the elastic deformations of the CDW can explain the gradual screened dielectric response as experimental results have shown in a limited temperature range, one can raise the following questions: are only elastic properties of CDW have to be taken into account? What would be the contribution of plastic deformations of the CDW superstructure, essentially phase slip processes, nucleation and propagation of CDW dislocation loops \cite{Tucker88,Brazovskii91a}. What link between dielectric relaxation at relatively high temperature and the very slow relaxational dynamics with very low activation energies (1--2~K) in the temperature range $T<1$~K, where ergodicity breaking with ``ageing" effects, have been established \cite{Biljakovic91a,Biljakovic93} (see section~\ref{sec8-1}).

In fact investigation of the dielectric response of K$_{0.3}$MoO$_3$ at lower temperature \cite{Yang91,Kriza91c} (data for $10<T<40$~K down to frequencies in the range of Hertz) have shown that an essential ingredient was missing in the characterisation of the longitudinal mode \cite{Yang91}.

\begin{figure}
\begin{center}
\includegraphics[width=7.5cm]{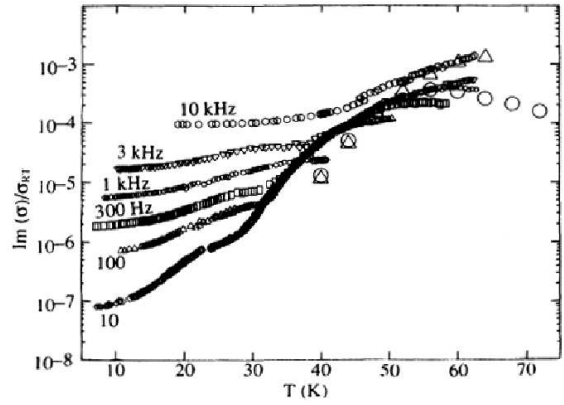}
\caption{Temperature dependence of the imaginary part of the ac conductivity normalised to the room temperature value, $\sigma_0$, of K$_{0.3}$MoO$_3$ at different frequencies. $\circ$ ($\vartriangle$) are data at 1~kHz (10~kHz) from Cava \textit{et al.} \protect\cite{Cava84b} (reprinted figure with permission from J. Yang and N.P. Ong, Physical Review B 44, p. 7912, 1991 \cite{Yang91}. Copyright (1991) by the American Physical Society).}
\label{fig6-2}
\end{center}
\end{figure}

Figure~\ref{fig6-2} 
shows the temperature dependence of the imaginary part of the ac conductivity (normalised to the room temperature conductivity) for different low frequencies \cite{Yang91}. On the same figure, data at 1~kHz and 10~kHz from ref.~\cite{Cava84b} have been added (reproduced from figure~3 in ref.~\cite{Yang91}). Clearly a dielectric relaxation occurs at much lower temperature (down to a few K) and lower frequencies than derived from the elastic treatment discussed above.

Then measurements of the low frequency dielectric susceptibility of o-TaS$_3$ \cite{Nad93a,Nad95a}, monoclinic TaS$_3$ \cite{Nad97}, K$_{0.3}$MoO$_3$ \cite{Nad93b,Staresinic04} have been extended in the low $T$ range 4.2--50~K and in the frequency range $10^{-2}-10^7$~Hz. In these experiments, a great care should be taken concerning the non-ideal orthogonality of channels for measurements of the real part and of the imaginary part of the conductivity which always exists in real devices. This non-orthogonality is usually characterised by a coefficient which, for the best devices, has a typical magnitude of about $10^{-2}-10^{-3}$~degree. This coefficient determines how many times one component of the conductivity can be different of the other with keeping an appropriate accuracy in the measurements.

\subsubsection{Dielectric relaxation of o-TaS$_3$}\label{sec6-3-1}

\begin{figure}
\begin{center}
\includegraphics[width=7.5cm]{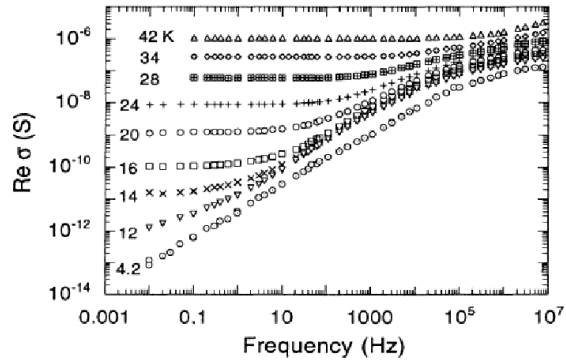}
\caption{Frequency dependence of the real part of the ac conductivity $\Re\sigma$ of o-TaS$_3$ at temperatures 4.2--42~K (reprinted figure with permission from F.Ya. Nad and P. Monceau, Physical Review B 51, p. 2052, 1995 \cite{Nad95a}. Copyright (1995) by the American Physical Society).}
\label{fig6-3}
\end{center}
\end{figure}
\begin{figure}
\begin{center}
\includegraphics[width=7cm]{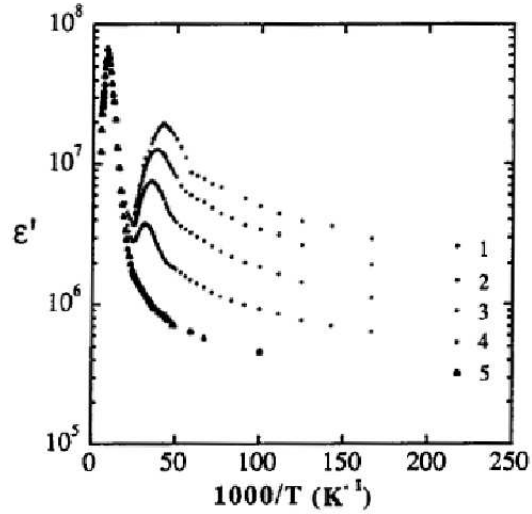}
\caption{Dependence of the real part, $\varepsilon^\prime$, of the dielectric susceptibility of o-TaS$_3$ versus the inverse of temperature at frequencies (Hz): 1--11, 2--111, 3--1.1 10$^3$, 4--10$^4$, 5--10$^5$ (reprinted figure with permission from F.Ya. Nad and P. Monceau, Journal de Physique IV, Colloque C2 3, p. 343, 1993 \cite{Nad93b}. Copyright (1993) from EdpSciences).}
\label{fig6-4}
\end{center}
\end{figure}

Figure~\ref{fig6-3} shows the $\Re\sigma(\omega)$ frequency dependences of o-TaS$_3$ at various temperatures. Below 14~K $\Re\sigma$ does not reach the dc value $\sigma_{\rm dc}$ even down to the lowest frequency of $10^{-2}$~Hz. In this temperature range, the $\Re\sigma(\omega)$ dependences proceed gradually to a power-law. Thus at 4.2~K in the frequency range $10^{-2}-10^5$~Hz $\Re\sigma(\omega)$ and $\Im\sigma(\omega)$ vary $\sim\omega^s$ with $s\simeq 0.8$ indicating a variable range hopping conductivity. The temperature dependence of $\varepsilon^\prime$, the real part of the complex permittivity of o-TaS$_3$ is plotted in figure~\ref{fig6-4} 
at fixed frequencies. Below the Peierls transition temperature ($T_{\rm P}$~= 220~K), the magnitude of $\varepsilon^\prime$ becomes very large and reaches a maximum $\sim 7\times 10^7$ near 120~K. Afterwards in the temperature range between 100~K and 50~K, $\varepsilon^\prime$ decreases exponentially with an activation energy of the order of that of the Peierls gap as resulting from screening of elastic deformations as discussed in section~\ref{sec6-2}. Below 50~K, $\varepsilon^\prime$ measured at frequencies $f\geq 100$~kHz continues to decrease monotonously. However for frequencies $f<100$~kHz, the $\varepsilon^\prime(T,f)$ dependences show pronounced peaks. The magnitude and the position of the peak maxima on the temperature scale are dependent on frequency. The $\varepsilon^\prime$ peak is shift at lower temperature when the frequency is decreased.

The frequency dependence of $\varepsilon^\prime(\omega)$ and $\varepsilon^{\prime\prime}(\omega)$ in the frequency range between 10$^{-2}$-10$^7$~Hz was analysed in detail below 40~K. The high frequency branches (above $10^4$~Hz) are very similar and they are well described by a power law $\varepsilon^\prime\sim\varepsilon^{\prime\prime}\sim\omega^{-n}$. On the contrary the dependences of $\varepsilon^\prime(\omega)$ and $\varepsilon^{\prime\prime}(\omega)$ exhibit a different behaviour in the low frequency range. It is worth noting that the loss function ($\varepsilon^{\prime\prime}$) changes from practically symmetrical in the temperature range around 30~K to more and more non symmetrical at lower temperature \cite{Nad95a}. All these features of $\varepsilon^\prime(\omega,T)$ and $\varepsilon^{\prime\prime}(\omega,T)$ dependences are typical for many different types of disordered materials with non-exponential relaxation \cite{Jonscher83}.

Dynamic effects in disordered glass materials have often be analysed in the frame of dynamic scaling
\cite{Castaing91,Souletie94,Ma81}. In particular, it was shown \cite{Souletie94} that, in enough correlated systems with some disorder, the distribution of logarithms of the relaxation times is a Gaussian:
\begin{equation}
P(\ln\tau)=\frac{1}{\sqrt{2\pi\lambda}}\exp -\,\frac{\left[\ln^2(\tau/\tau^\ast)\right]}{2\lambda^2},
\label{eq6-8}
\end{equation}
where $\tau^\ast$ is the most probable value of relaxation time and $\lambda$ corresponds to the width of the distribution. Taking into account this suggestion it was shown that in the frequency domain $\varepsilon^\prime(\log_{10}\omega)$ dependence corresponds to an error function. In this case, the relation between $\varepsilon^\prime$ and $\varepsilon^{\prime\prime}$, which in the general case is determined by Kramers-Kronig relations, is reduced to the so-called ``$\pi/2$ rule" \cite{Pytte87}
\begin{equation}
\varepsilon^{\prime\prime}(\log_{10}\omega)=-\frac{\pi}{2}\,\frac{{\rm d}\varepsilon^\prime(\omega)}{{\rm d}\log_{10}\omega}.
\label{eq6-9}
\end{equation}

Experimental data $\varepsilon^{\prime\prime}(\log_{10}\omega)$ and data calculated by relation~(\ref{eq6-9}) are in a satisfactory agreement  for $T\geq 30$~K, but a quantitative difference up to $\sim 30\%$ occurs below 24~K. Additionally, in the frame of this approach, the effective width $\lambda$ of the loss peak $\varepsilon^{\prime\prime}(\omega)$ should be proportional to ($\log_{10}\omega_p)^{0.5}$, where the frequency $\omega_p$ corresponds to the maximum of the loss peak. The relation $\lambda^2\sim\log_{10}\omega_p$ has been, indeed, effectively observed \cite{Nad95a} in o-TaS$_3$ in the temperature range above $\sim 30$~K. Thus, in this temperature range, the form of the loss function is very similar to Gaussian. However, with decreasing temperature, it deviates from Gaussian and the distribution below 25~K becomes more and more wider.

\subsubsection{Dielectric relaxation in the SDW state of (TMTSF)$_2$PF$_6$}\label{sec6-3-2}

Similar ac conductivity measurements have been performed on the SDW compound (TMTSF)$_2$PF$_6$ in the temperature range below half the SDW transition temperature ($T_{\rm P}$~= 12.5~K) and in the frequency range $10^2-10^7$~Hz \cite{Nad95b}.

\begin{figure}
\begin{center}
\includegraphics[width=6.5cm]{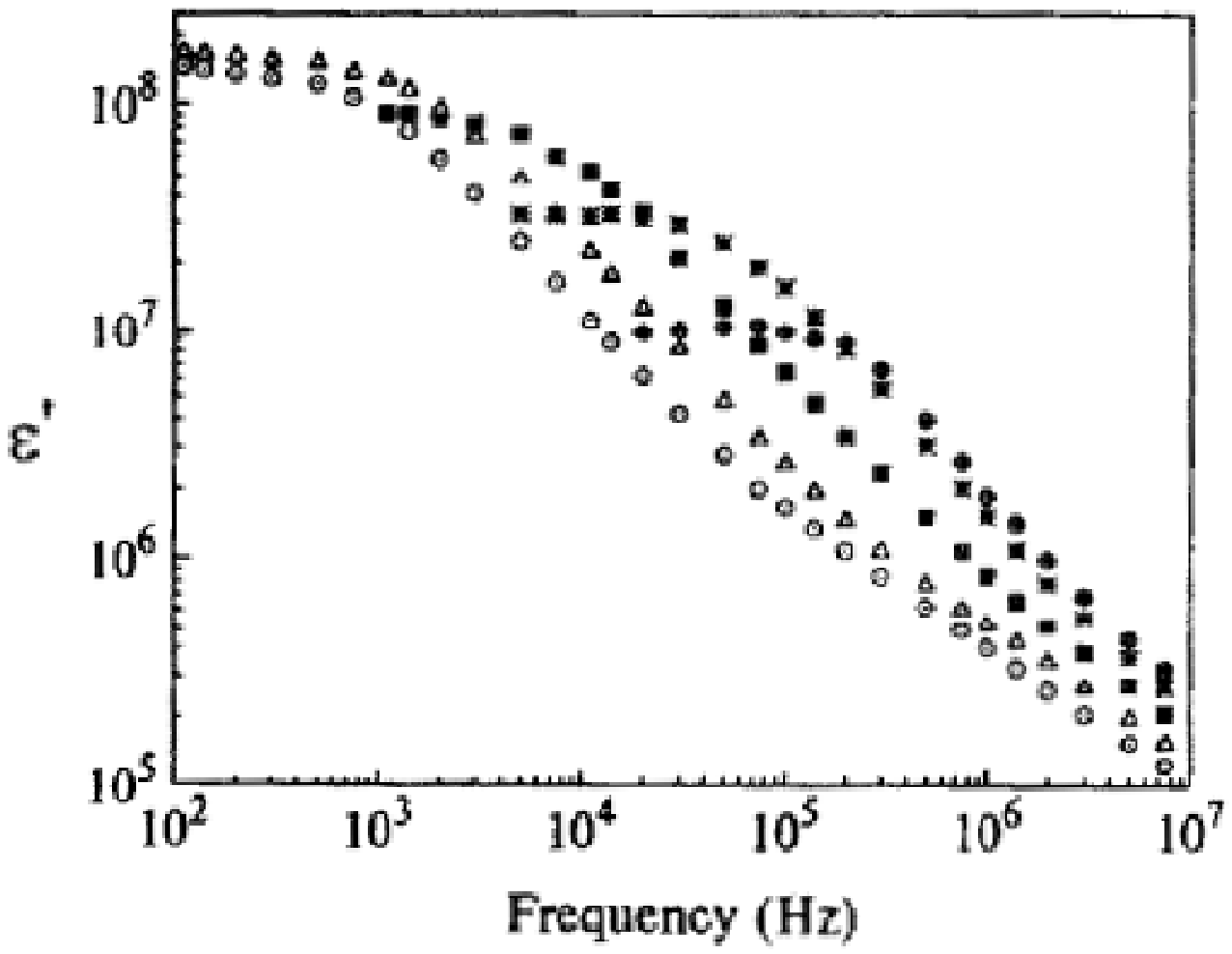}
\includegraphics[width=6.5cm]{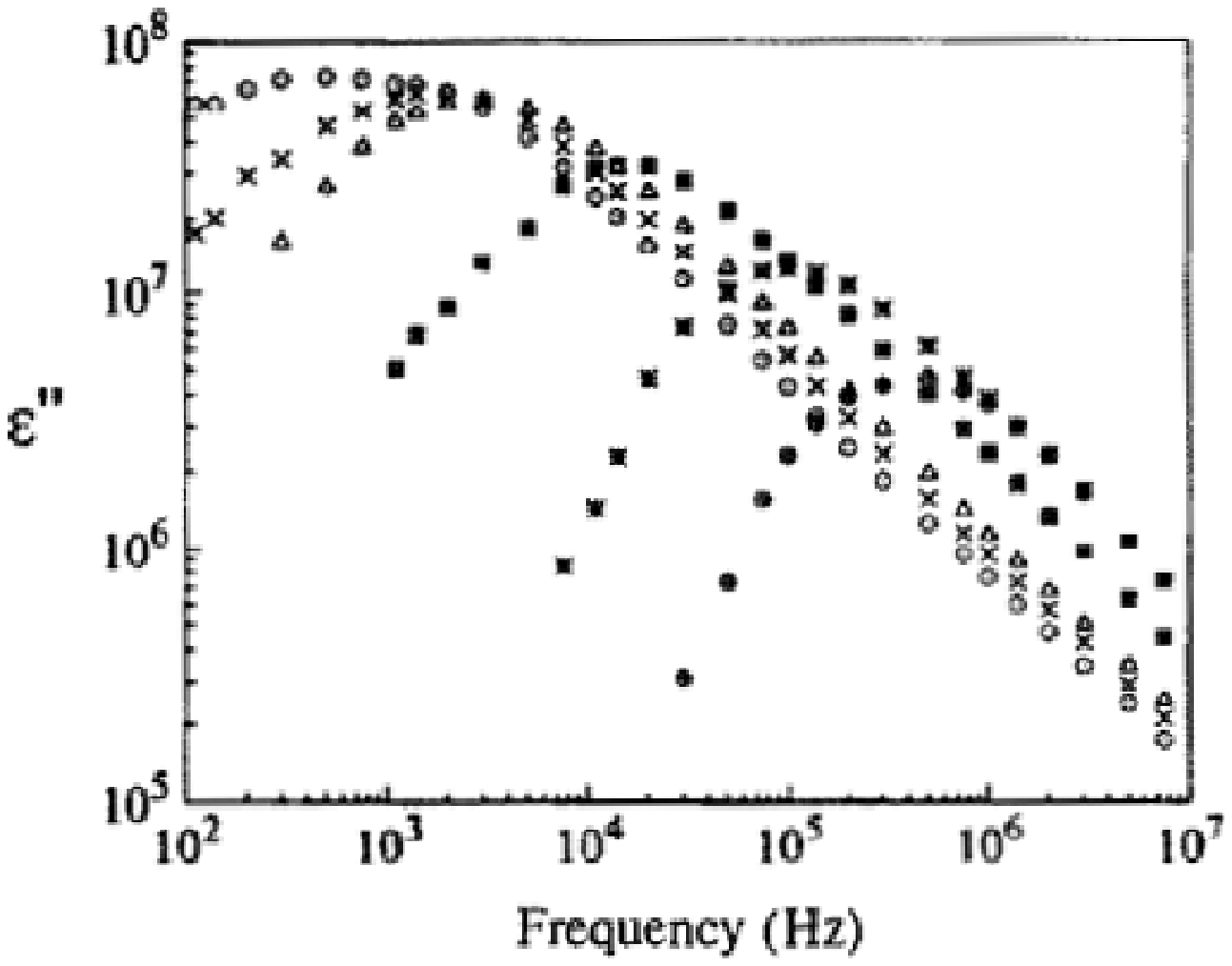}
\caption{Frequency dependence of the real part $\varepsilon^\prime$ and imaginary part $\varepsilon^{\prime\prime}$ of the dielectric permittivity of (TMTSF)$_2$PF$6$ at given temperatures (in K): $\odot$: 1.3, $\times$: 1.6, $\vartriangle$: 1.8, {\tiny$\square$}: 2.4, $\ast$: 3.2, $\oplus$: 4 (reprinted figure with permission from Solid State Communications 95, F. Nad \textit{et al.}, p. 655, 1995 \cite{Nad95b}. Copyright (1995) with permission from Elsevier).}
\label{fig6-11}
\end{center}
\end{figure}

Figure~\ref{fig6-11} 
shows $\varepsilon^\prime(\omega)$ and $\varepsilon^{\prime\prime}(\omega)$ dependences at several temperatures. In the high frequency range ($10^5-10^7$~Hz) $\varepsilon^\prime(\omega)$ and $\varepsilon^{\prime\prime}(\omega)$ dependences are similar such as $\varepsilon^\prime(\omega)\sim\varepsilon^{\prime\prime}(\omega)\sim\omega^{-\alpha}$ with $\alpha$~= 0.6--0.7. With decreasing frequency $\varepsilon^{\prime\prime}(\omega)$ shows a pronounced maximum the position of which shifts at lower frequency when $T$ is reduced. Simultaneously the width of $\varepsilon^{\prime\prime}(\omega)$ grows.

The temperature dependence of $\varepsilon^\prime(T)$ at fixed frequencies between 111~Hz and 1~MHz is plotted in figure~\ref{fig6-7}(b). For a given frequency $\varepsilon^\prime(T,\omega)$ shows a pronounced maximum. The position of this maximum is shifted to lower $T$ with decreasing frequency. It should be noted that the high temperature parts of these curves (on the right of the maximum) at different frequencies merge practically into a single master curve.

\subsubsection{Glassy state}\label{sec6-3-3}

\begin{figure}[b]
\begin{center}
\subfigure[o-TaS$_3$]{\label{fig6-7a}
\includegraphics[width=7.5cm]{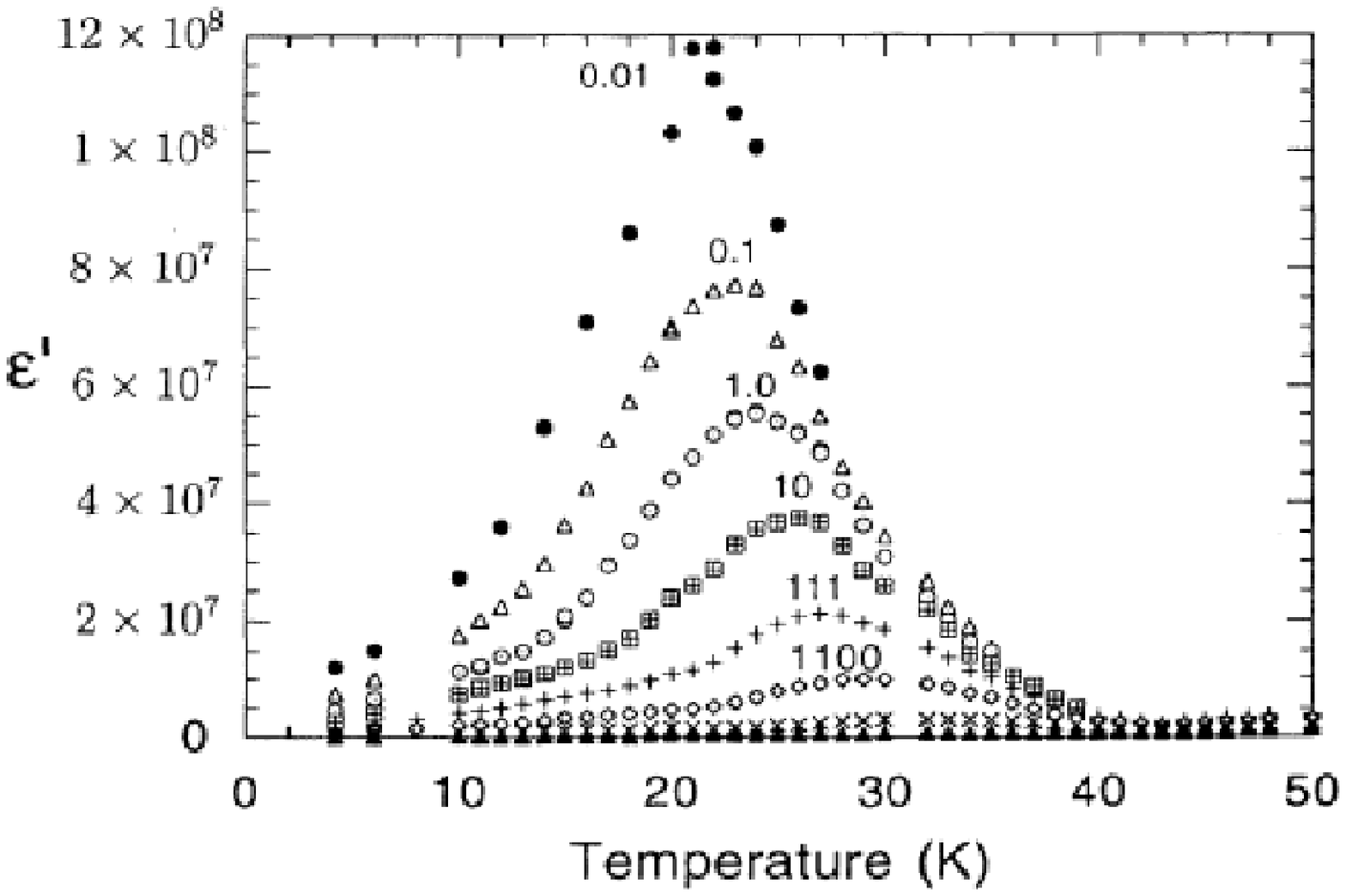}}
\subfigure[(TMTSF)$_2$PF$_6$]{\label{fig6-7b}
\includegraphics[width=5.5cm]{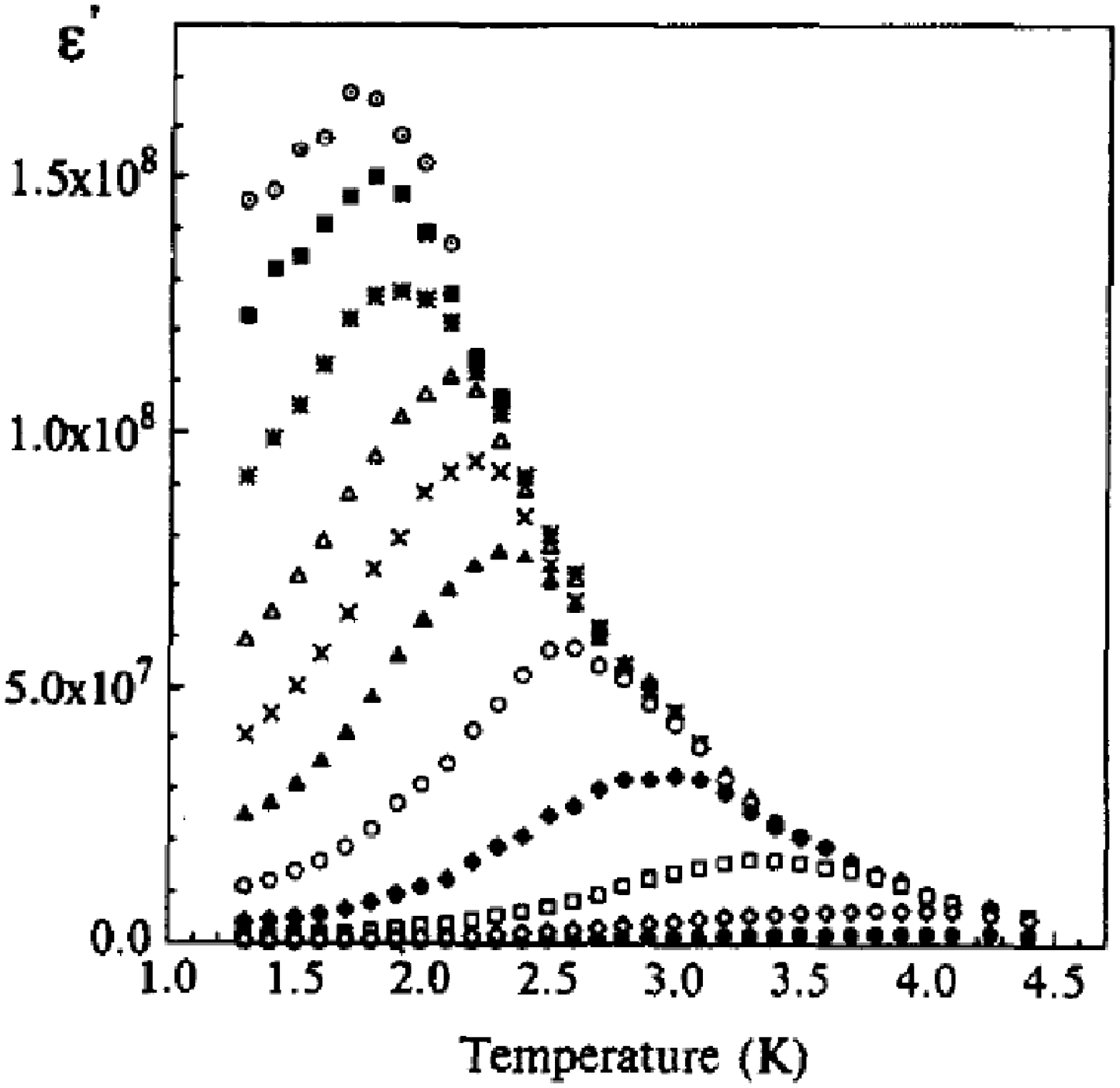}}
\caption{Temperature dependence of the real part $\varepsilon^\prime$ of the dielectric susceptibility (a)~of o-TaS$_3$ at frequencies (in Hz) indicated in the figure (reprinted figure with permission from F.Ya. Nad and P. Monceau, Physical Review B 51, p. 2052, 1995 \cite{Nad95a}. Copyright (1995) by the American Physical Society); (b)~of (TMTSF)$_2$PF$_6$ at frequencies (in kHz) $\odot$ -- 0.111, {\tiny$\square$} -- 0.5, {\large$\ast$} -- 1.1, $\vartriangle$ -- 2, $\times$ -- 3, $\blacktriangle$ -- 5, {\large$\circ$} -- 11, $\lozenge$ -- 30, {\tiny$\square$} -- 100, $\lozenge$ -- 300, {\large$\bullet$} -- 1000 (reprinted figure with permission from Solid State Communications 95, F. Nad \textit{et al.}, p. 655, 1995 \cite{Nad95b}. Copyright (1995) with permission from Elsevier).}
\label{fig6-7}
\end{center}
\end{figure}

\begin{figure}
\begin{center}
\subfigure[o-TaS$_3$]{\label{fig6-8a}
\includegraphics[width=7.25cm]{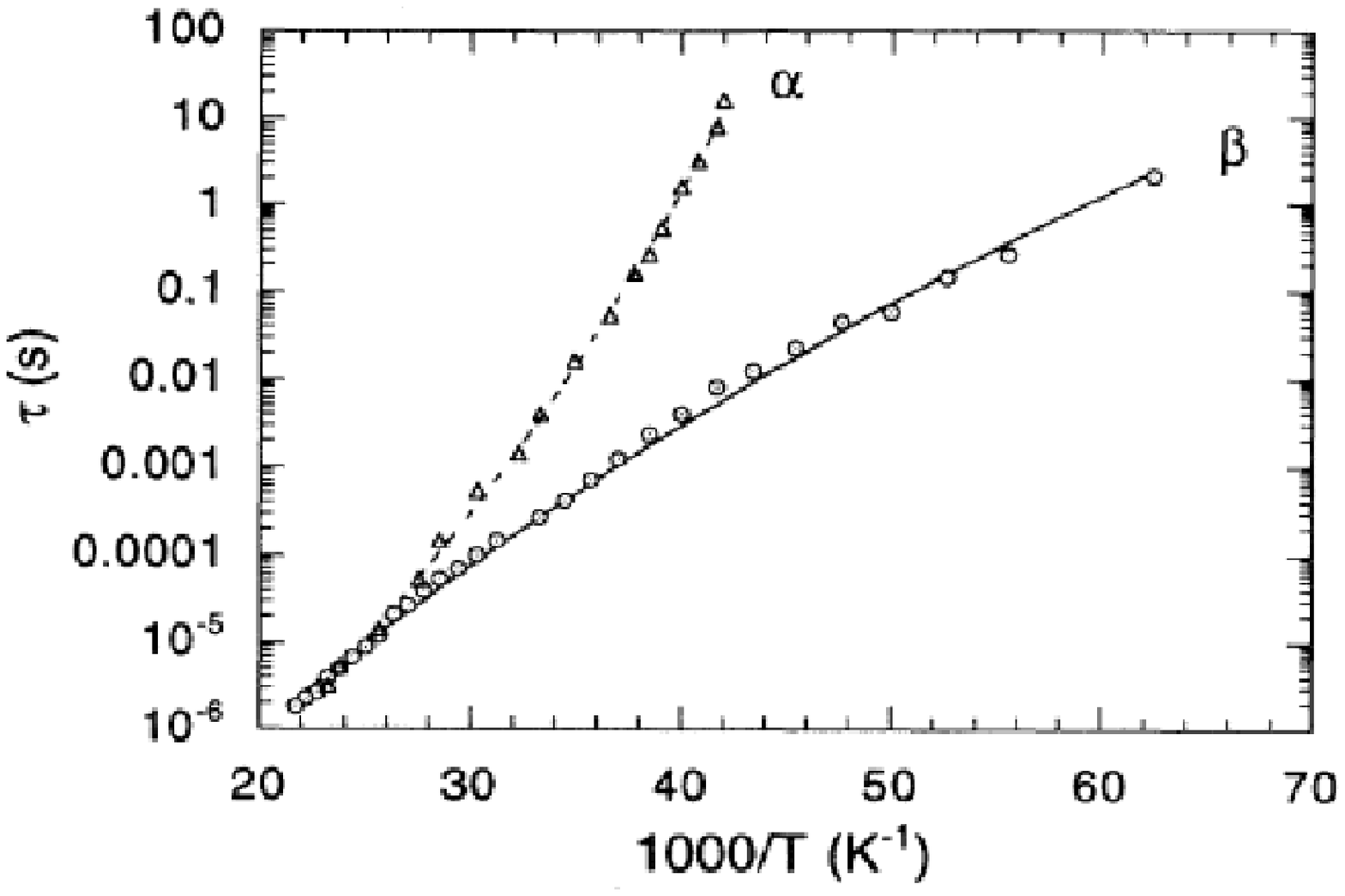}}
\subfigure[(TMTSF)$_2$PF$_6$]{\label{fig6-8b}
\includegraphics[width=6cm]{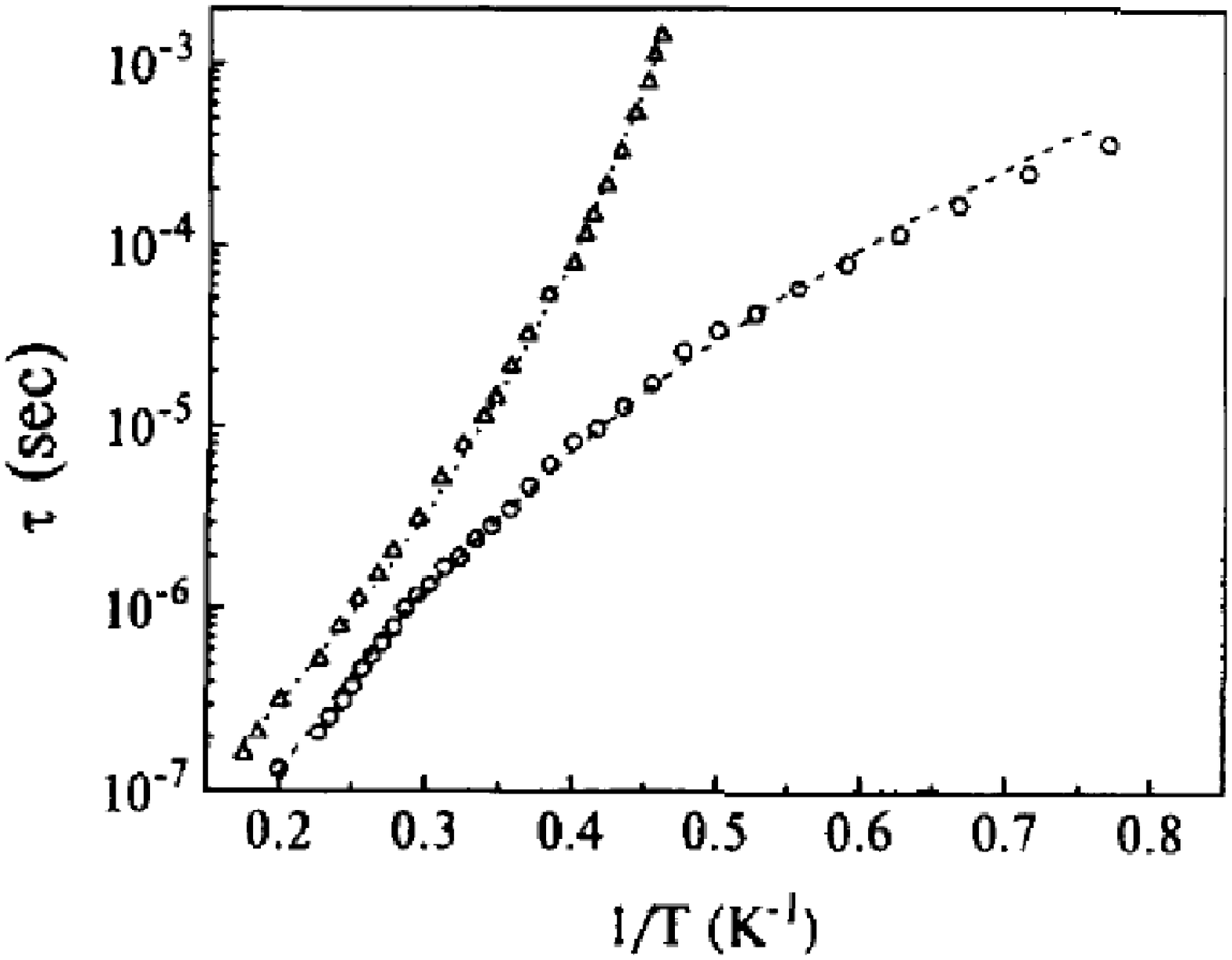}}
\caption{Temperature dependence of relaxation time $\tau$ (a)~of o-TaS$_3$, (b)~of (TMTSF)$_2$PF$_6$ $\vartriangle$: $\tau$ value determined by deflection points of $\varepsilon^{\prime\prime}(T,\omega)$ dependences; $\circ$: $\tau$ value determined from position of maximum of $\varepsilon^{\prime\prime}$($\log_{10}\omega$) dependences (reprinted figure with permission from F.Ya. Nad and P. Monceau, Physical Review B 51, p. 2052, 1995 \cite{Nad95a}. Copyright (1995) by the American Physical Society for (a), reprinted figure with permission from Solid State Communications 95, F. Nad \textit{et al.}, p. 655, 1995 \cite{Nad95b}. Copyright (1995) with permission from Elsevier for (b)).}
\label{fig6-8}
\end{center}
\end{figure}

The temperature dependence of $\varepsilon^\prime$ at fixed frequencies is shown in figure~\ref{fig6-7}(a) 
for o-TaS$_3$. $\varepsilon^\prime$ exhibits a sharp peak.  The magnitude of this peak increases when the frequency is decreased. The average relaxation time, $\tau^\ast$, can be determined at the deflection point on $\varepsilon^\prime(T)$ curve. $\tau^\ast(T)$ can be fitted either by a Fulcher law: $\tau^\ast$~= $\tau_0\exp(B/T-T_c)$ with $\tau_0$~= $10^{-9}$~s, $B$~= 200~K and $T_c\simeq 13$~K and $zv\simeq 24$. The latter dependence is drawn in figure~\ref{fig6-8}(a) 
by a solid line. In the same figure is plotted the dependence of $\tau_p\equiv(\omega_p)^{-1}$ where $\omega_p$ corresponds to the maximum of the loss function $\varepsilon^{\prime\prime}$. Above 32~K, $\tau^\ast(1/T)$ and $\tau_p(1/T)$ show a similar thermally activated dependence with an activation energy of 780~K which is close to the Peierls energy gap in o-TaS$_3$. These data are also in good agreement with those obtained at higher temperature \cite{Cava85}. However below 32~K a branching between two relaxation processes appears: $\tau^\ast(1/T)$ exhibits an upward curvature fitted by a slowing down equation and $\tau_p(1/T)$ a downward curvature.

As in the case of CDW systems, figure~\ref{fig6-8}(b) shows the temperature dependence of two relaxation time processes: the mean relaxation time $\tau^\ast$ from the deflection points in $\varepsilon^\prime(T,\omega)$ dependences, and the relaxation  time from the maximum in $\varepsilon^{\prime\prime}(T,\omega)$.

There are great similarities between the data of CDW systems as o-TaS$_3$, K$_{0.3}$MoO$_3$ and those of the SDW (TMTSF)$_2$PF$_6$ (see figures~\ref{fig6-7}(a) and \ref{fig6-7}(b), \ref{fig6-8}(a) and \ref{fig6-8}(b)).

The dielectric susceptibility results from the summation of polarisation effects and dipole interactions between these randomly distributed solitons. The main cause of the $\epsilon^\prime(T)$ growth comes from the increase of the CDW rigidity with decreasing screening which tends to make more rigid the CDW and more homogeneous. Then the CDW coherence length and the dielectric constant will increase. However at low temperature dynamic retarding effects occur and the CDW has no enough time to respond to the ac perturbation and its response (i.e. the $\varepsilon^\prime$ magnitude) decreases.

The divergence of $\varepsilon^{\prime}(T)$, the slowing down relation with a finite $T_c$ for the average relaxation time, the change of the distribution of relaxation times (from $\varepsilon^{\prime\prime}$) from Gaussian to non-Gaussian, all are features suggestive of a phase transition into a glassy state \cite{Littlewood88a,Erzan90,Biljakovic98}. This approach is strengthened by the observation of two relaxation times as in structural glasses \cite{Angell00}: the slower one (or $\alpha$ relaxation) may correspond to relaxation of large CDW regions and the faster one ($\beta$ relaxation) may concern smaller scales, probably phase slip regions with a typical size of the order of one unit cell of the CDW superstructure. Below 20~K, in o-TaS$_3$ for instance, the relaxation time becomes so long that the system is frozen in some glassy state. In this state the CDW collective excitations - solitons - are pinned on impurities. The conductivity is carried by rarely jumps of these solitons by tunnelling  transitions \cite{Nad93c,ZZ93}. At microscopic scale these residual excitations can be responsible for the enthalpy relaxation with ageing effects observed at very low temperature \cite{Biljakovic91a} (see section~\ref{sec7}).

\begin{figure}[h!]
\begin{center}
\includegraphics[width=6.5cm]{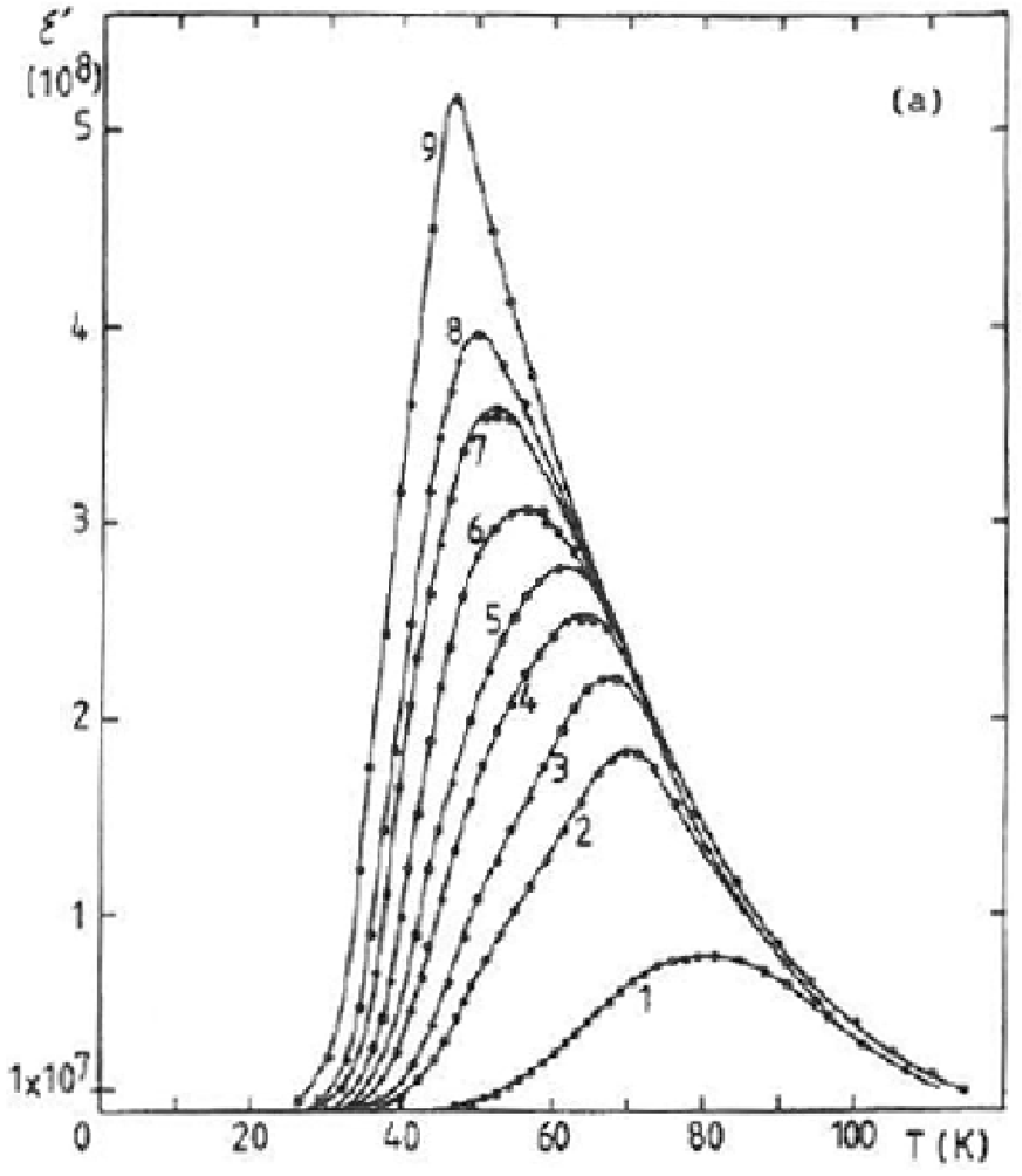}
\includegraphics[width=6.5cm]{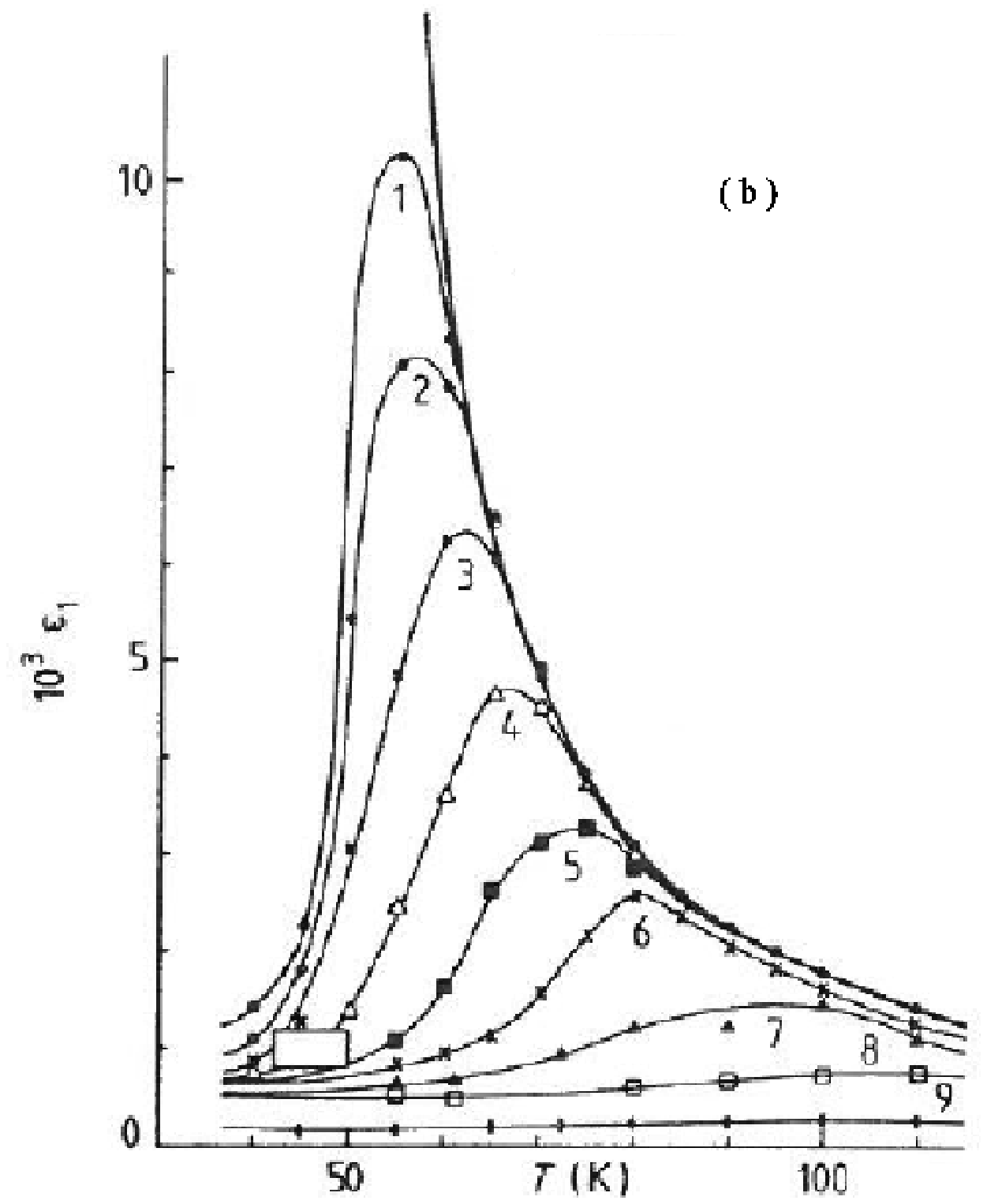}
\caption{Variation of the real part $\varepsilon^\prime$ of the dielectric susceptibility as a function of temperature (a)~of K$_{0.3}$MoO$_3$ at given frequencies: 1 -- 100~kHz, 2 -- 10~kHz, 3 -- 5~kHz, 4 -- 2~kHz, 5 -- 1.1~kHz, 6 -- 500~Hz, 7 -- 200~Hz, 8 -- 111~Hz, 9 -- 31~Hz (reprinted figure with permission from F.Ya. Nad and P. Monceau, Journal de Physique IV (France), Colloque C2 3, p. 343, 1993 \cite{Nad93b}. Copyright (1993) from EdpSciences); (b)~of the dipolar glass K$_{0.974}$Li$_{0.026}$TaO$_3$ (reprinted figure with permission from U.T. H\"ochli and M. Maglione, Journal of Physics: Condensed Matter 1, p. 2241, 1989 \cite{Hochli89}. Copyright (1989) by the Institute of Physics).}
\label{fig6-9}
\end{center}
\end{figure}

Similar results than those on o-TaS$_3$ were obtained \cite{Nad93b} on K$_{0.3}$MoO$_3$ as shown in figure~\ref{fig6-9}(a) 
(see figure~\ref{fig6-2} in semi-logarithmic scale from ref.~\cite{Yang91} for comparison). In figure~\ref{fig6-9}(b) is also drawn the variation of the real part of the dielectric constant for the dipolar glass K$_{1-x}$Li$_x$TaO$_3$ \cite{Hochli89}.

\begin{figure}
\begin{center}
\includegraphics[width=7.5cm]{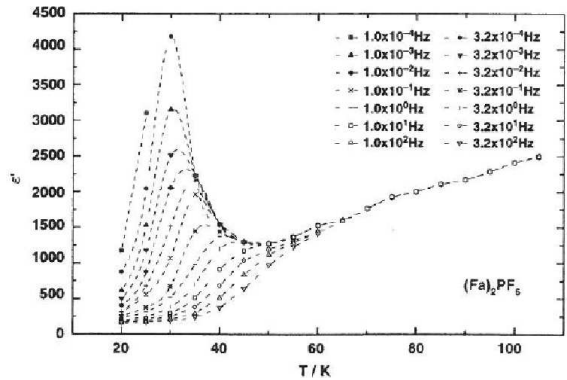}
\caption{Variation of the real part $\varepsilon^{\prime}$ of the dielectric susceptibility of (Fluoranthene)PF$_6$ for different frequencies (reprinted figure with permission from C. Math \textit{et al.}, Europhysics Letters 35, p. 221, 1996 \cite{Math96}. Copyright (1996) from EdpSciences).}
\label{fig6-10}
\end{center}
\end{figure}

Dielectric spectroscopy has been also studied \cite{Math96} in the CDW organic (fluoranth\`ene)$_2$PF$_6$. Similarly to o-TaS$_3$ and K$_{0.3}$MoO$_3$ a high frequency mode at $7\times 10^{11}$~Hz was interpreted as the pinning mode; this mode appears at much higher frequency than in o-TaS$_3$ or K$_{0.3}$MoO$_3$, probably due to the commensurate CDW. Below the pinned mode a lower frequency mode develops shifting a low frequency where $T$ is reduced. The main relaxation time $\tau$ of this mode is thermally activated in the temperature range 100~K--40~K with the same activation energy than the dc resistivity. Below 40~K, as seen in figure~\ref{fig6-10}, 
below 10~Hz a peak in the dielectric function $\varepsilon^\prime$ occurs, which seems to indicate a divergence around 30~K. Thus, as o-TaS$_3$ and K$_{0.3}$MoO$_3$ the dielectric function of (fluoranth\`ene)$_2$PF$_6$ shows features of a glassy state \cite{Math96} down to a very low frequency ($\sim 10^{-4}$~Hz).

A very wide-temperature and frequency range dielectric response has been studied in o-TaS$_3$ \cite{Staresinic02} and K$_{0.3}$MoO$_3$ \cite{Staresinic04}. For o-TaS$_3$ three relaxation processes have been identified: the $\alpha$ process which splits off from the high $T$ pinning resonance, the $\beta$ process which splits off from the $\alpha$ process below $\sim 100$~K, and the $\beta_0$ process which emerges from the high frequency side of the $\beta$ process below $\sim 25$~K. This $\beta_0$ process is detectable down to the lowest temperature. A criterion for freezing of the $\alpha$ process has been presented \cite{Staresinic02} for which freezing occurs when there is less than one carrier per CDW domain. The $\beta_0$ process is almost temperature independent and thus can be linked to some kind of thermally assisted quantum tunnelling as proposed \cite{ZZ93} for the non-linear conductivity of o-TaS$_3$ at very low temperatures. The $\beta$ process was also observed \cite{Staresinic04} in K$_{0.30}$MoO$_3$ with the temperature dependence of the relaxation time following an Arrhenius behaviour. By contrast, for o-TaS$_3$, the $T$ dependence of the $\beta$ relaxation time deviates from an activated law. This difference may reflect the behaviour between fragile (o-TaS$_3$) and strong (K$_{0.3}$MoO$_3$) glasses which is related to a different topography of the phase space \cite{Staresinic04}. Finally the comparison of the  amplitude of relaxation processes is very appealing \cite{Staresinic04}: it shows that the amplitude of the $\beta$ process is two orders of magnitude lower in K$_{0.3}$MoO$_3$ than in o-TaS$_3$ which may indicate a smaller number of topological defects active at low temperature.

\subsubsection{Competition between weak and strong pinning}\label{sec6-3-4}

In the temperature range $T_{\rm P}/2<T<T_{\rm P}$, the interaction of the CDW with impurities has essentially a weak pinning collective character. With decreasing temperature below $\sim T_{\rm P}/2$, because of the reduction of screening and the resulting hardening of the CDW due to the exponentially vanishing of the number of free carriers, the weak pinning is exhausted and the pinning becomes essentially local induced by strong pinning impurities. Local deformations at these pinning centres are nucleated in the CDW superstructure which dominates the kinetic properties of the system.

At variance with the glass transition phenomenology described above, the low frequency peak in $\epsilon^\prime$ experimentally measured in o-TaS$_3$, K$_{0.3}$MoO$_3$, (TMTSF)$_2$AsF$_6$ was explained as resulting from the competition between strong local pinning and collective pinning \cite{Larkin95}. At very low temperature, only strong pinning sites are effective for initiating plastic deformations and metastable states \cite{Larkin94,Larkin95}. Effect of local and collective pinning are additive to the pinning force: $f_{\rm pin}$~= $f_{\rm local}+f_{\rm coll}$. Then:
\begin{equation}
\varepsilon^{-1}=\varepsilon^{-1}_{\rm loc}+\varepsilon^{-1}_{\rm coll},
\label{eq6-10}
\end{equation}
which was calculated as:
\begin{equation}
\varepsilon^{-1}={\rm cst}\,n_i\,\omega^{T/V_0}+{\rm cst}\,n_i^2\exp(-\Delta/T).
\label{eq6-11}
\end{equation}
The first term originates from the decay of metastable states around strong pinning impurities with concentration $n_i$ due to a thermal excitation over barriers with a distribution (assumed exponential):
\begin{equation}
P=\frac{1}{V_0}\exp(-V/V_0).
\label{eq6-12}
\end{equation}
The exponential $-\Delta/T$ in the second term of eq.~(\ref{eq6-11}) comes from the effect of Coulomb hardening. The function $\varepsilon(\omega,T)$ is not monotonic and demonstrates a peak at $T_{\rm max}(\omega_{\rm max})$ defined as:
\begin{equation}
\omega_{\rm max}\sim\frac{1}{\tau_0}\exp\left(-\frac{E_aV_0}{T^2_{\rm max}} + \frac{V_0}{T_{\rm max}}\ln n_i\right).
\label{eq6-13}
\end{equation}
A good fit using eq.~(\ref{eq6-13}) with the data of o-TaS$_3$ was reported in ref.~\cite{Brazovskii04}.

Following \cite{Larkin95,Brazovskii04}, near $T_{\rm max}$ the pinning force is minimum. When increasing $T$ from the lowest temperature, the local metastable states approach thermal equilibrium and $f_{\rm loc}\rightarrow 0$; $\varepsilon(\omega,T)$ grows until the collective pinning force becomes dominant.

Finally the peak in $\varepsilon^\prime$ was also obtained from a model of defects forming a period lattice interacting with a regular lattice of phase solitons with the same periodicity \cite{Volkov93}. A model with weakening of collective pinning was also proposed  \cite{Wonnenberger96}.

\subsection{Second threshold at low temperature}\label{sec6-4}

I-V characteristics in K$_{0.3}$MoO$_3$ in the low temperature range have shown a switching from the insulating state (typically $\sim 10^{10}-10^{12}~\Omega$cm) to a conducting state with a nearly zero differential resistance above a very large threshold $V^\ast_T$ \cite{Mihaly87,Mihaly88c}.

$V^\ast_T$ was first considered to be a contact threshold voltage \cite{Mihaly88a,Mihaly87} but found later to vary linearly with the length of the sample.

Abrupt change in noise spectra with jumps between distinct current levels revealed that the conducting cross-section of the sample may vary. It was also envisaged that the sharpness of the I-V curves could result from a breakdown phenomenon due to a filamentary instability. Real time current oscillation measurements have led to values of $\nu/J_{\rm CDW}$ ($\nu$: frequency of the current oscillation, $J_{\rm CDW}$ current density carried by the CDW) with nearly identical value than measured at high temperature, indicating a bulk phenomenon \cite{Mihaly88a,Tessema87}; but for some other experiments, it was shown that only a fraction of the cross-section was involved in the sliding state (which has probably to be related to contact effects and current injection phenomena): typically $\sim 14\%$ in ref.~\cite{Martin88}.

Switching itself is influenced by the electric circuit used to study the I-V characteristics. At the onset of the initial switching, the voltage jumps down to a lower value revealing an hysteresis for repetitive further field cycles; this hysteresis for initiating switching has been shown to depend on the giant remnant dielectric polarisation which relaxes very slowly with a logarithmic or stretched time dependence, associated with metastable CDW configurations \cite{Mihaly86}. A smaller time allowing to change the metastable states below $V^\ast_T$ results in a smaller hysteretic effect \cite{Kim89}.

Switching in K$_{0.3}$MoO$_3$ is not always so abrupt as sometimes reported \cite{Maeda90,Ogawa05,Martin88}. It was shown that the breakdown to the high conducting state shows always a finite slope, even down to 1.25~K. Calculating the differential conductivity $\sigma$~= ${\rm d}I/{\rm d}V$ in the breakdown region from the I-V characteristics, a fit \cite{Martin88} of data over 2--3 orders of magnitude yields $\sigma\propto I^\beta$ with $\beta$~= 0.8--1.1. A region of negative differential conductivity was also found in a short current range.

The current response to a step voltage pulse with an amplitude above $V^\ast_T$ is very quick and in the range of some dozens of nanoseconds. Similar switching I-V characteristics have also been measured in o-TaS$_3$ and (TaSe$_4$)$_2$I \cite{Mihaly88a}.

\subsubsection{Rigid CDW motion}\label{sec6-4-1}

A crude phenomenological explanation of switching lays on a single degree-of-freedom for the CDW motion equation. The low frequency low field dynamics \cite{Littlewood87b} are frozen out at low temperatures. The phase mode develops a gap because the long range Coulomb forces. The resulting increased rigidity suppresses the possibility of creating local CDW deformations. The equation of motion is reduced to:
\begin{equation}
\frac{1}{\tau}\,\frac{{\rm d}x}{{\rm d}t}=F-F_T.
\label{eq6-14}
\end{equation}
The absence of the dissipative term results from the gap in the phason spectrum \cite{Mihaly88b}.

\begin{figure}[h!]
\begin{center}
\includegraphics[width=7cm]{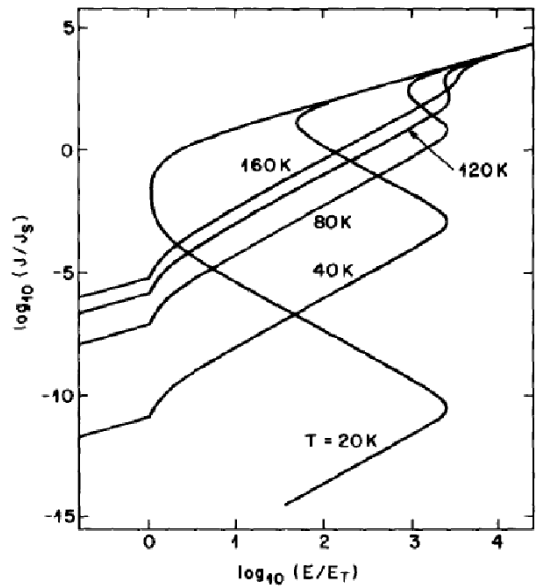}
\caption{Total current normalised to the CDW current as a function of the normalised electric field showing bistability at low temperatures (reprinted figure with permission from Solid State Communications 65, P.B. Littlewood, p. 1347, 1988 \cite{Littlewood88b}. Copyright (1988) with permission from Elsevier).}
\label{fig6-12}
\end{center}
\end{figure}

However the rigidity of the condensate does not exclude the configurational rearrangement in the pinned state as seen \cite{Chen88} by the huge dielectric polarisation below $V^\ast_T$.  Littlewood \cite{Littlewood88b,Littlewood89} has extended in the non-linear CDW state his previous analysis of the FLR equation of motion with Coulomb interaction. It is shown that the I-V characteristic becomes bistable. When the CDW moves with the velocity $v$, the local ac voltage (NBN) generated at the frequency $\omega_0$~=  $Qv$ acts on the CDW and produce local distortions with the same periodicity which are screened by a back-flow current. Then, the effective damping parameter, $\gamma_{\rm eff}$, becomes velocity-dependent. There is a cross-over in the effective damping parameter when the CDW velocity exceeds the value $\sigma/\varepsilon Q$. There is, then, a transition from low velocity screened CDW motion to a high velocity where the effective damping is reduced to $\gamma_0$ and where the CDW motion is rigid. The form of  this transition appears as a ``S-shaped'' bistability in (I-V) curves with a large region of a negative differential conductivity as shown in figure~\ref{fig6-12}. In the lower branch the CDW current follows the same dependence than at higher temperature: $J_{\rm CDW}\propto(E-E_T)^\alpha$. In the higher branch $J_{\rm CDW}$ was evaluated as: $J_{\rm CDW}$~= $(\rho^2_0/\gamma_0)E^\ast_T$. As $\gamma_0$ originating from phason-phason or phason-phonon scattering \cite{Takada85} vanishes when $T\rightarrow 0$, in this model, without other sources of dissipation, the CDW can move undeformed indefinitely.

Experimentally it may be difficult to record such a negative differential conductivity (NDC) especially because the regime may change from voltage controlled to current controlled in the instability region, although such NDC was shown in a limited current range; also one can expect at the instability point a sudden jump from the lower branch to the upper branch, corresponding to the observed switching.

\subsubsection{Phase slip processes}\label{sec6-4-2}

However a different description of the sharp increase in conductivity at liquid helium temperatures on m-TaS$_3$ and o-TaS$_3$ has been brought, based on phase-slip processes.

Non-linear conductivity of o-TaS$_3$, m-TaS$_3$ have been measured in the wide range of electric field from $10^{-2}$ to 500~V/cm and temperatures down to helium temperatures. As already seen in section~\ref{sec3} the case of o-TaS$_3$ is particular: indeed from $T_{\rm P}$~= 200~K down to 100~K, the dependence of the conductivity is such that $\log\sigma$ is close to linear as a function of $1/T$ with an activation energy $\Delta$ of about 800~K. Below 100~K, the variation of $\log\sigma$ is smaller with activation energy of about 220--240~K. This variation has been attributed to the thermal excitation of soliton-like excitations \cite{Zhilinskii83}. On the other hand the $\log\sigma(1/T)$ of m-TaS$_3$ is a linear function of $1/T$ with an activation energy $\Delta$ of 950~K down to 25~K.

\begin{figure}
\begin{center}
\includegraphics[width=12.5cm]{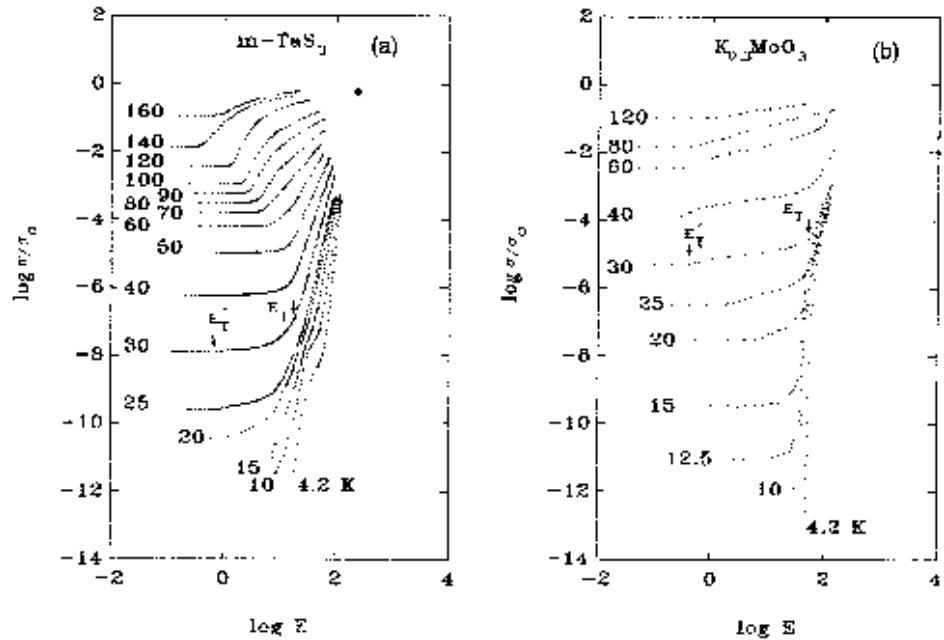}
\caption{Variation of the conductivity normalised to its room temperature value as a function of the electric field at the indicated temperatures (a)~for m-TaS$_3$, (b)~for K$_{0.3}$MoO$_3$ (reprinted figure with permission from Synthetic Metals 41-43, M.E. Itkis \textit{et al.}, p. 4037, 1991 \cite{Itkis91}. Copyright (1991) with permission from Elsevier).}
\label{fig6-13}
\end{center}
\end{figure}

Figure~\ref{fig6-13}(a) 
shows in a log-log plot the conductivity of m-TaS$_3$ normalised to the conductivity at room temperature as a function of electric field and similarly in figure~\ref{fig6-13}(b) for K$_{0.3}$MoO$_3$, with a variation in conductivity on 12 orders of magnitude. In the temperature range $T>20$~K for m-TaS$_3$ \cite{Itkis90,Nad92}, $T>12.5$~K for K$_{0.3}$MoO$_3$ \cite{Itkis91} it was possible to measure the initial linear part of the I-V curve (ohmic conductivity) up to the electric field $E^\prime_T$. With the subsequent increase of $E$, one observe a smooth deviation of linearity with a sharp growth of conductivity at higher field $E>E_T$. Above $\sim 70$~K, the $\sigma(E)$ dependence has the well-known form with a single threshold $E_T$.

\begin{figure}
\begin{center}
\includegraphics[width=7cm]{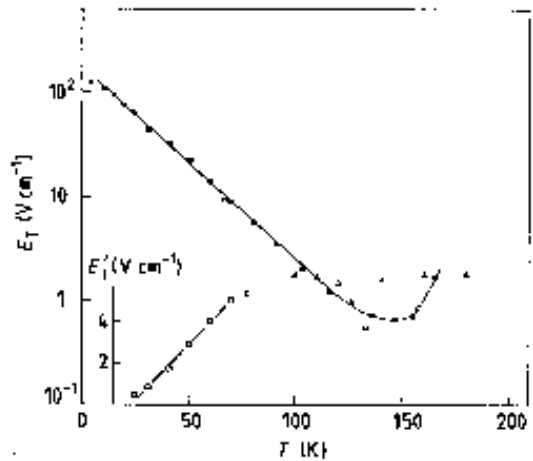}
\caption{Variation of the thresholds $E_T$ and $E^\prime_T$ for m-TaS$_3$ as a function of temperature (reprinted figure with permission from M.E. Itkis \textit{et al.}, Journal of Physics: Condensed Matter 2, p. 8327, 1990 \cite{Itkis90}. Copyright (1990) by the Institute of Physics).}
\label{fig6-14}
\end{center}
\end{figure}

For m-TaS$_3$, in figure~\ref{fig6-14}, 
it is shown that when $T$ is increased from 4.2~K, $E^\prime_T$ increases linearly with $T$ while $E_T$ decreases exponentially as $\exp(-T/T_0)$ with $T_0\simeq 20$~K (as shown in section~\ref{sec4}).
\begin{figure}
\begin{center}
\includegraphics[width=8.5cm]{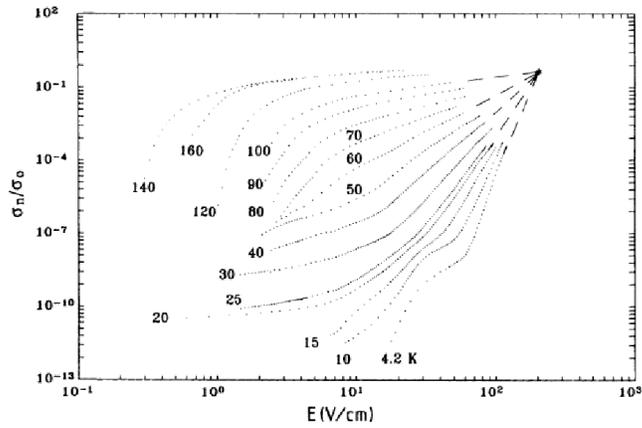}
\caption{Variation of the non-linear (CDW) conductivity $\sigma_{\rm n}$ normalised to the room temperature conductivity as a function of the electric field at the indicated temperatures (reprinted figure with permission from F.Ya. Nad and P. Monceau, Physical Review B 46, 7413, 1992 \cite{Nad92}. Copyright (1992) by the American Physical Society). When extrapolated to a very large $E$, all the curves will cross at the asterisk point.}
\label{fig6-15}
\end{center}
\end{figure}

\begin{figure}
\begin{center}
\includegraphics[width=7.5cm]{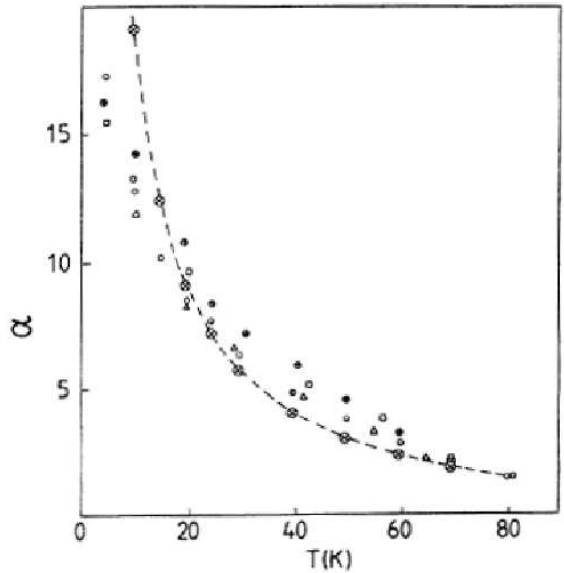}
\caption{Temperature dependence of the $\alpha$ exponent of the experimental $\log_{10}(\sigma_{\rm n}/\sigma_0)$ versus $\log_{10}(E)$ dependences for several m-TaS$_3$ samples. The dashed curve is calculated from the relaxation $\alpha$~= $(F-T)/T$ with $F$~= 200~K (reprinted figure with permission from F.Ya. Nad and P. Monceau, Physical Review B 46, 7413, 1992 \cite{Nad92}. Copyright (1992) by the American Physical Society).}
\label{fig6-16}
\end{center}
\end{figure}

From the same set of (I-V) curves shown in figure~\ref{fig6-13}(a), are drawn in figure~\ref{fig6-15} the electric field dependences of the non-linear (CDW) conductivity $\sigma_n$. This conductivity is equal to the difference between the total conductivity $\sigma$ and its linear part $\sigma_1$ (well defined in figure~\ref{fig6-13}(a) for $T>25$~K): $[\sigma_n=\sigma-\sigma_1]$. One can note in figure~\ref{fig6-15} a change of convexity of the $\sigma_n(E)$ curves around the curve at $T\sim 60$~K. However the variation of $\log_{10}(\sigma_n)$ versus $\log_{10}(E)$ shows a linear dependence for all the curves at different temperatures. Being prolonged, all the curves will cross in a small region marked by an asterisk. Thus, $\sigma_n/\sigma_0\sim\beta(E/E_0)^\alpha(T)$ where $\beta$ and $E_0$ are the coordinates of the asterisk point. $\alpha(T)$ grows from $\simeq 1$ to $\simeq 17$ with decreasing temperature as shown in figure~\ref{fig6-16} for different m-Ta samples.

Similar dependences were observed in o-TaS$_3$ \cite{Itkis90,Itkis91} and in blue bronze at higher temperature \cite{Mihaly88d}. The observation of $E^\prime_T$ in K$_{0.3}$MoO$_3$ down to 12.5~K as seen in figure~\ref{fig6-13}(b) results from the large diapason in the change of conductivity to be compared with other data \cite{Mihaly88b}.

At variance with models only based on elasticity and screening as described above, phase slippage was considered as a possible mechanism to explain the sharp increase of conductivity above $E_T$. Thus the description of the CDW dynamics at $E>E_T$ was tentatively undertaken using theoretical \cite{Langer67,Ivlev84} formulas derived for phase-slip processes in superconductivity which have some common features with CDWs.

The dependence on temperature and electric field for the time derivative of the phase difference $\varphi$ resulting from the phase-slip process in superconductors \cite{Langer67,Ivlev84} derived such as:
\begin{equation}
\frac{{\rm d}\varphi}{{\rm d}t}=\Omega\exp\left[-\frac{E}{k_T}\right]\sinh\left[\frac{\delta F}{2kT}\right],
\label{eq6-15}
\end{equation}
where $F$ is the energy barrier between two states before and after phase slip (when the phase difference changes by $2\pi$), $\delta F$ the difference between barrier height resulting from the application of voltage $V$, $\Omega(T)$~= $N(T)/\tau(T)$ with $N(T)$: the number of electrons involved in phase slips in the sample and $\tau(T)$ the average relaxation time of excitations, was applied for CDWs \cite{Nad92}. In that case, the current on a single chain is equal to $I$~= $e/\pi\,{\rm d}\varphi/{\rm d}t$. But in real systems, the phase slip process is nucleated in a finite minimum volume estimated to the volume of one unit cell in the CDW superstructure. So, as a result of a change of phase by $2\pi$,
\begin{equation}
\delta F=\int IV\,{\rm d}t=\int\,V(en/\pi)({\rm d}\varphi/{\rm d}t){\rm d}t=2neV,
\label{eq6-16}
\end{equation}
where $n$ is the number of electrons in the unit-cell volume involved in the phase-slip process. Then the CDW current can be written in the form \cite{Borodin87,Gill86}
\begin{equation}
I_c=I_0\exp\left[-\frac{F}{kT}\right]\sinh\left[\frac{neV}{kT}\right],
\label{eq6-18}
\end{equation}
\medskip
\noindent with $\displaystyle I_0=\frac{e}{\pi}\,\Omega=\frac{e}{\pi}\,\frac{N}{\tau}$.

\medskip
\noindent It was shown earlier that the CDW current near $I_T$ ($V\gtrsim V_T$) is well described by eq.~(\ref{eq6-18}) for o-TaS$_3$ at temperatures 100--150~K \cite{Borodin87} as well as for NbSe$_3$ \cite{Gill86}.

In the high electric-field range, when $neV>kT$, eq.~(\ref{eq6-18}) becomes:
\begin{equation}
I_c=I_0\exp\left[\frac{neV-F}{kT}\right].
\label{eq6-19}
\end{equation}

A more simple equation for the high electric-field range can be obtained from eq.~(\ref{eq6-19}). Let denote $F$~= $nF_0$~= $neE_0\ell$, where $F_0$ is the energy spent by a single quasiparticle for overcoming the barrier. This energy corresponds to  that gained by this quasiparticle in the electric field $E_0$ on a length $\ell$. This length corresponds to the region of a strong CDW deformation due to the external electric field $E$~= $V/\ell$ \cite{Tucker88}. The $E_0$ value corresponds to the total suppression of the barrier $F$ and consequently $E_0\gg E_T$. Then one derive from eq.~(\ref{eq6-19}) (supposing that $k=1$)
\begin{equation}
I_c=I_0\exp\left\{(F/T)[(E/E_0)-1]\right\}.
\label{eq6-20}
\end{equation}
Using the approximate formula $(E/E_0)-1\cong\ln(E/E_0)$, one obtains
\begin{equation}
I_c/I_0=\exp[(F/T)\ln(E/E_0)]=(E/E_0)^{F/T}.
\label{eq6-21}
\end{equation}
Then, denoting $J_c=\sigma_nE$ and $J_0=\sigma_0E^\prime_0$, we can write
\begin{equation}
\sigma_n/\sigma_0=\gamma(E/E^\prime_0)^{(F-T)/T},
\label{eq6-22}
\end{equation}
where $\gamma$ ($E_0/E^\prime_0$) is a constant. Equation~(\ref{eq6-22}) provides a good fit to the experimental dependences at $E\gg E_T$ obtained for m-TaS$_3$ samples as shown in figure~\ref{fig6-15}. It should be noted that in the simplest case the exponent in $\log_{10}\sigma_n(\log_{10}E)$ dependences is equal to $\alpha(T)$~= $(F-T)/T$.

Figure~\ref{fig6-16} shows the experimental $\alpha(T)$ dependence for several m-TaS$_3$ samples. The dependence $\alpha(T)$~= $(F-T)/T$ is also drawn with a dashed line with $F$~= 200~K. This value is approximately equal to the CDW condensation energy for one quasi particle $\sim(T_{\rm P})^2/E_F$ multiplied by the number of quasiparticles in a unit cell of the CDW superlattice \cite{Nad92,Itkis90,Itkis91}. This energy is that needed for the microscopic primary act of momentary collapse of a unit cell in the CDW superlattice and thus for the creation of a $2\pi$ soliton. Afterwards macroscopic dislocation lines formed by aggregation of these solitons can develop practically without any energy barrier \cite{Brazovskii91a,Brazovskii91b}. The energy $F$ experimentally determined in figure~\ref{fig6-16} also corresponds to the theoretical estimation of the energy needed for a CDW phase slip by $2\pi$ at a strong-pinning centre estimated \cite{Tucker88} as $E_{\rm P}\simeq\Delta/4\sim 230$~K (for m-TaS$_3$, the CDW gap $\Delta\simeq 900$~K).

The $\alpha(T)$ value depends on the variation of the effective barrier as a function of temperature, electric field and on dissipation processes for the CDW motion. At temperatures lower than $T_0$, no thermally activated solitons are available and tunnelling through the barrier becomes predominant in a weak electric field.

Measurements on o-TaS$_3$ with a very small cross-section ($\sim 10^{-2}~\umu$m$^2$) down to 30~mK found that rather the power law dependence $I\alpha E^\alpha$ as presented above, I-V curves can alternatively be fitted by the exponential law: $I\alpha\exp[-(V_0/V)^\beta]$ with $\beta\sim 1-2$ \cite{ZZ93}, which can be accounted for as CDW quantum creep at very low temperatures \cite{ZZ93,ZZ97}.

\subsubsection{Low temperature transport properties of (TMTSF)$_2$X salts}\label{sec6-4-3}

\begin{figure}
\begin{center}
\includegraphics[width=7.5cm]{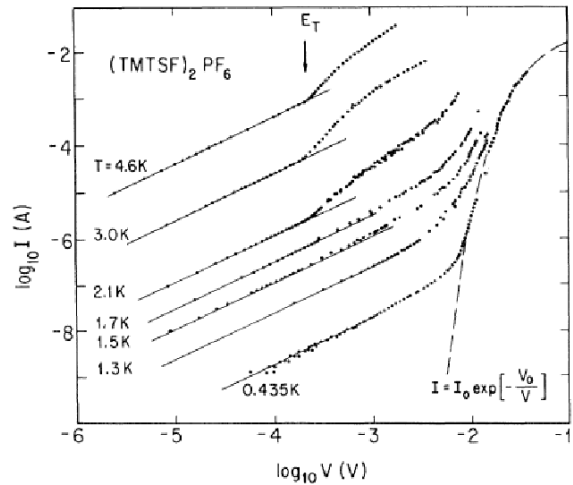}
\caption{I-V characteristics of (TMTSF)$_2$PF$_6$ measured at different temperatures. The solid lines represent the ohmic conductivity of normal carriers. At $T$ not too far below $T_{\rm SDW}$, the threshold field $E_T$ is well defined. At low $T$, the I-V curve follows a Zener-type expression (reprinted figure with permission from G. Mihaly \textit{et al.}, Physical Review Letters 67, p. 2713, 1991 \cite{Mihaly91a}. Copyright (1991) by the American Physical Society).}
\label{fig6-17}
\end{center}
\end{figure}

I-V characteristics of (TMTSF)$_2$PF$_6$ were performed from $T_{\rm SDW}$ temperature transition down to very low temperature. For $T$ not too far from $T_{\rm SDW}$, the non-linear response with a well-defined threshold field $E_T$ (see figure~\ref{fig6-17}) 
is determined by the internal SDW deformations, with the freezing of the SDW current between 5 and 2~K, as previously observed for CDWs. The scaling between normal and collective conductivity holds down to $\sim 2.5$~K indicating that damping arises from the dynamic screening of the SDW deformations by normal carriers. But below 1.5~K the shape of the (I-V) characteristics changes with a huge increase of current at higher electric field \cite{Mihaly91a}, found to follow the law: $j=j_0\,e^{-E_0/E}$, with $E_0$: a few V/cm. It was suggested that a new type of collective mode excitation occurs at low $T$ through Zener tunnelling. The characteristic electric field $E_0$ can be expressed as: $E_0=\alpha\frac{W^2}{e\hbar v_F}$, with $\alpha$: a constant and $W$: the barrier height for tunnelling \cite{Traetteberg92}. The small value of $E_0$ excludes Zener tunnelling of single carriers through the CDW gap ($W=2\Delta$).

\subsubsection{Macroscopic quantum tunnelling}\label{sec6-4-4}

\begin{figure}[h!]
\begin{center}
\subfigure[]{\label{fig6-18a}
\includegraphics[width=6.5cm]{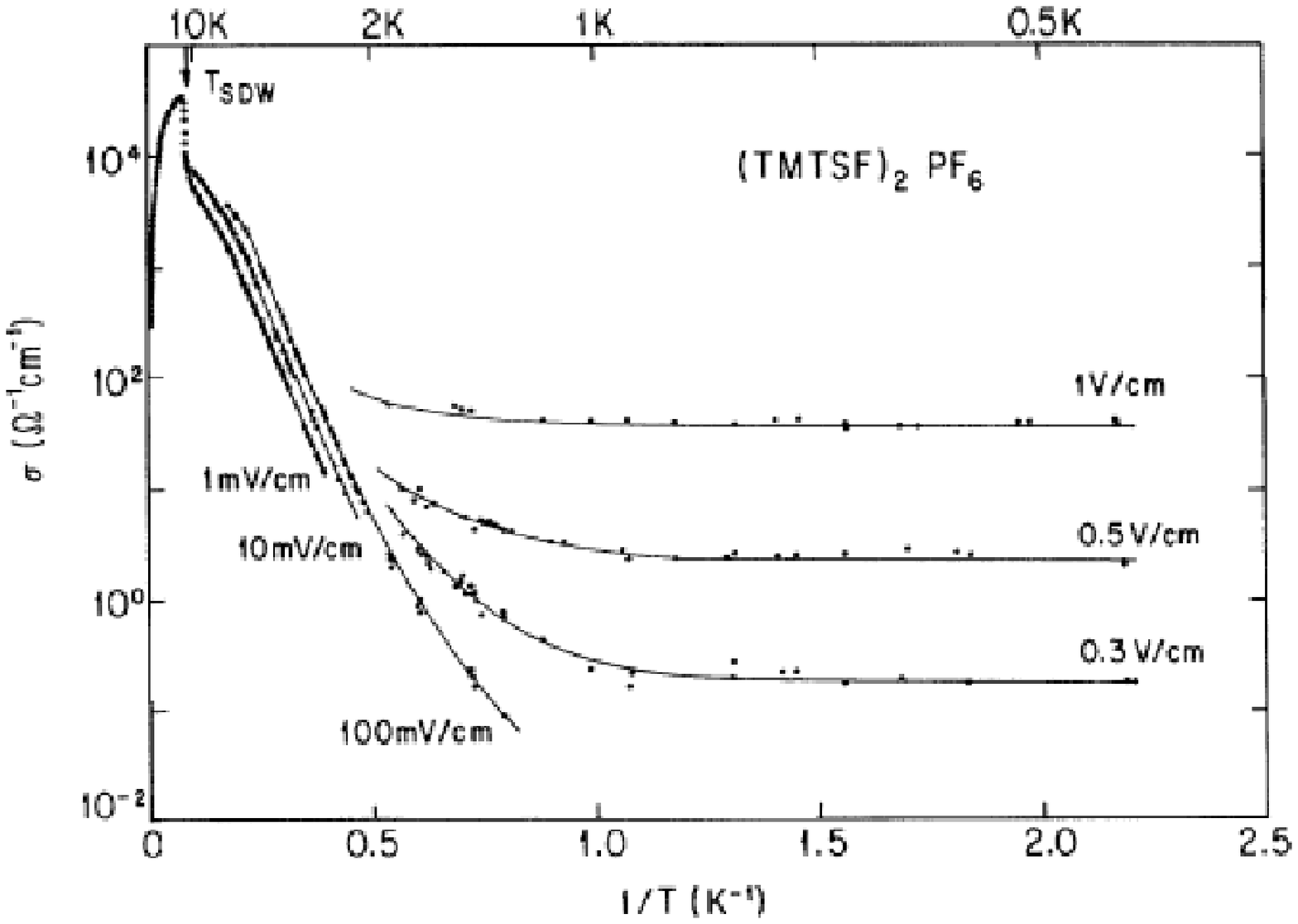}}
\subfigure[]{\label{fig6-18b}
\includegraphics[width=6.5cm]{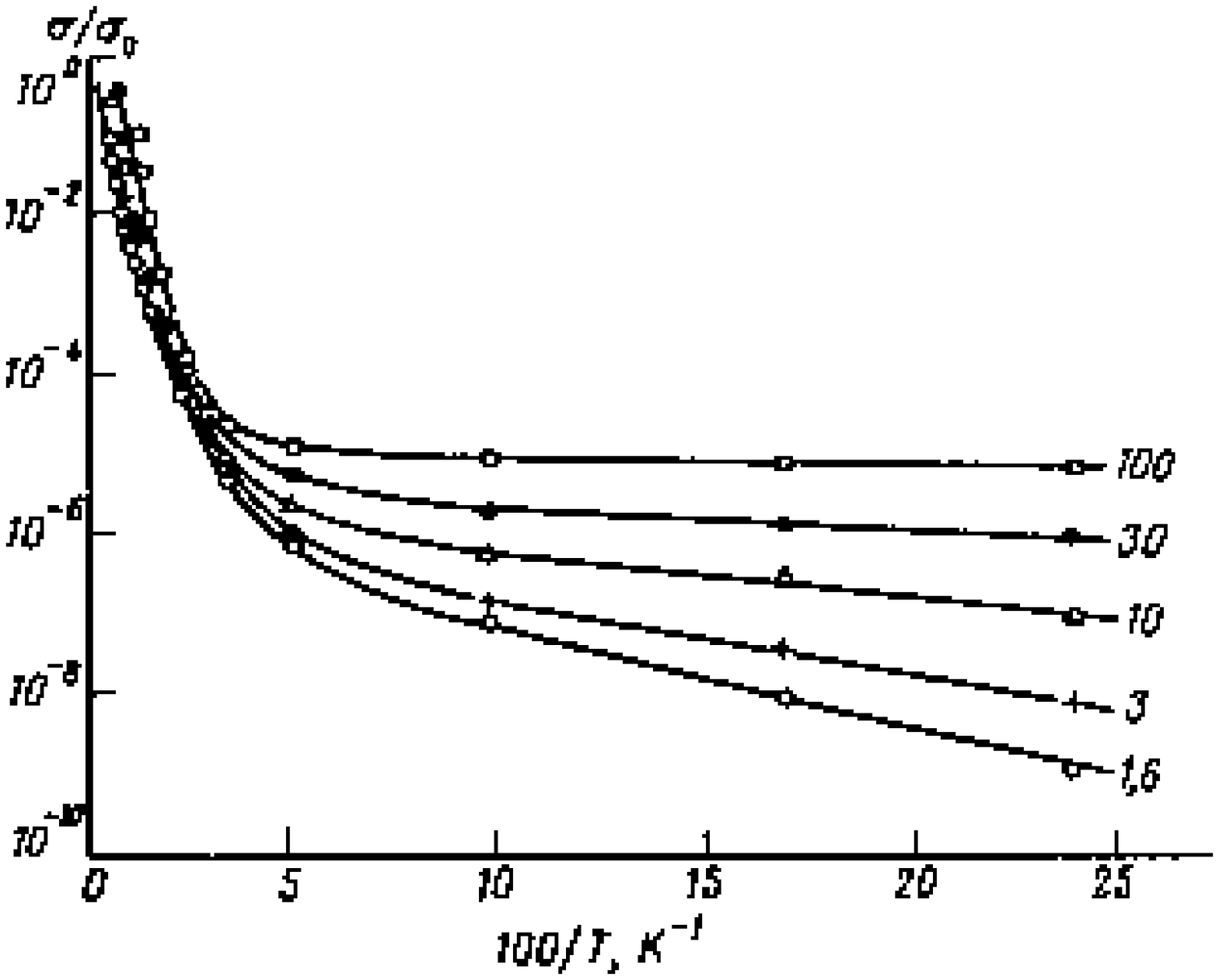}}
\caption{Temperature dependence of the conductivity at various electric fields at low temperatures (a) in (TMTSF)$_2$PF$_6$ (reprinted figure with permission from G. Mihaly \textit{et al.}, Physical Review Letters 67, p. 2713, 1991 \cite{Mihaly91a}. Copyright (1991) by the American Physical Society), (b)~in o-TaS$_3$ (reprinted figure with permission from JETP Letters 58, F.Ya. Nad, p. 111, 1993 \cite{Nad93c}. Copyright (1993) from Springer Science and Business media).}
\label{fig6-18}
\end{center}
\end{figure}

When the data of figure~\ref{fig6-17} are replotted as a function of $1/T$ for given values of the electric field (see figure~\ref{fig6-18}(a)), 
the conductivity is found to be independent of temperature at very low temperatures. The same behaviour is, in fact, also observed for CDWs as shown in figure~\ref{fig6-18}(b) for o-TaS$_3$ \cite{Nad93b,Nad93c}, as well as, at $T$~= 2~K, for a o-TaS$_3$ sample with a very small cross-section \cite{ZZ97}.

A common explanation for CDWs and SDWs may be proposed involving interaction between strong pinning centres and $2\pi$ solitons. The detachment of a pinned $2\pi$ soliton from an impurity site results from an instantaneous local disruption of the order parameter and a change of $2\pi$ in the phase difference. This phase slippage at very low temperature occurs as a transition between two potential minima with an energy barrier. The disruption of the order parameter occurs in a minimum volume corresponding to a unit cell of the DW superstructure. Such a unit cell contains a number of electrons typically $\sim 10$, thus the phase slippage corresponds to a macroscopic quantum tunnelling. This latter tunnelling was studied in the case of Josephson junctions \cite{Caldeira81,Larkin83}. The tunnelling of a soliton through a potential barrier $V$ set up by a strong pinning centre was evaluated in ref.~\cite{Larkin78}. Production of soliton-antisoliton pairs by the external field was estimated to be the main contribution to the electrical conductivity at very low temperatures when the energy soliton $E_\varphi$ is very small compared with $kT$: $E_\varphi/kT\ll 1$. In the same context, a theory (with numerical simulations) of nucleation of quantum solitons was recently proposed \cite{Miller12}.

The electrical conductivity dominated by quantum tunnelling was then calculated \cite{Maki96} and found to have the $\exp(-(E_0/E)$ dependence as experimentally found \cite{Mihaly91a,Nad93b,Nad93c}.

\subsubsection{Metastable plastic deformations}\label{sec6-4-5}

\begin{figure}
\begin{center}
\includegraphics[width=7.5cm]{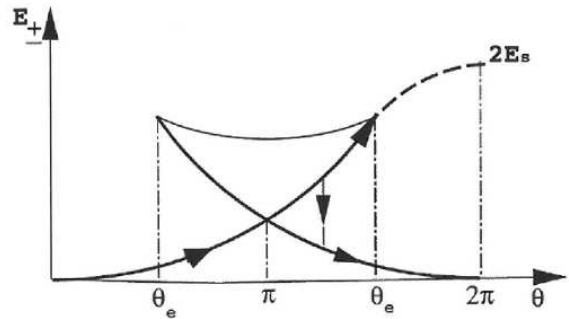}
\caption{Energy of local metastable plastic deformations after a polarisation $\delta\theta$ of the original equilibrium density wave showing the bistability with two branches $E_+(\theta)$ and $E_-(\theta)$ with a transition over a barrier between them at $\theta_e$. The difference at given $\theta>\pi$ between the ascending and the descending lines gives the dissipated energy (reprinted figure with permission from S. Brazovskii and A. Larkin, Journal de Physique IV (France) 9, p. Pr-77, 1999 \cite{Larkin95}. Copyright (1999) from EdpSciences).}
\label{fig6-19}
\end{center}
\end{figure}

A local pinning approach in terms of local metastable states created by solitons or dislocation loops induced by pinning has been developed in \cite{Brazovskii04,Larkin95}. In this model the CDW is considered as an elastic periodic medium with topological defects interacting with impurities. Following \cite{Brazovskii04}, let consider an isolated impurity at some point $r_i$ interacting with the CDW $\sim\cos(Q\bm{r}+\varphi)$. Introduce the positionally random phase $\theta$~= $-\varphi_0-Qr_i$ and the CDW phase at the impurity site $\psi$~= $\varphi(r_i)-\varphi_0$, $\varphi_0$ being the reference value within the large correlated volume of the collective pinning. In the sliding state $\varphi_0$~= vt. The effective Hamiltonian showing the competition between the cost of soliton generation and elastic deformation has been expressed \cite{Brazovskii04} as:
\begin{equation*}
H(\psi,\theta)=W(\psi)+V_{\rm pin}(\psi-\theta),
\end{equation*}
where $W(\psi)$~= $E_S[1-\cos(\psi/2)]$ is the  energy deformation and $V_{\rm pin}(\psi-\theta)$~= $V_{\rm pin}[1-\cos(\psi-\theta)]$ is the pinning potential. $2E_S$ is the energy absorbed in the nucleation of a pair of solitons each time $\psi$ acquires a $2\pi$ increment. $H(\psi,\theta)$ should be minimised with respect to $\psi$ at a given $\theta$.
For large impurity potential it was seen \cite{Larkin94,Larkin95,Brazovskii04} that, at each strong pinning site, a bistability occurs with two branches $E_+$ and $E_-$ with a transition over a barrier between them at a given value of $\theta$ as shown in figure~\ref{fig6-19}. This transition releases the energy $\Delta E(\theta)$~= $E_+(\theta)-E_-(\theta)$. At very low temperatures and high CDW velocity, the metastability is maintained over the whole period till $\theta$~= $2\pi$. The maximum energy, $\Delta E_{\rm max}$~= $2E_S$ ($E_S$: the soliton energy), is then absorbed after each cycle and it is this energy which creates the pair of solitons (or dislocation loops). In these conditions the pinning force reaches an asymptotic value. The second threshold field $E^\ast_T$ has been identified by the comparison between the dissipated energy $\Delta E_{\rm max}$ times the linear concentration of strong pinning impurities and the energy $2eE^\ast_T$ gained by the applied electric field:
\begin{equation}
2eE^\ast_T=\Delta E_{\rm max}\,n_i=2E_Sn_i.
\label{eq6-25}
\end{equation}
It should be noted that the soliton proliferation at each $\theta$ cycle occurs when the pinning force $f$ (which is the tangent $\upartial E/\upartial\theta$) reaches its asymptotic value (see the schematic plot of pinning force versus velocity in figure 6.17 in \cite{Brazovskii04}, 
recovering at low velocity the collective sliding regime and the divergent velocity when $f$ reaches its maximum value $f_m$).

Thermodynamical properties of C/S DW below 1~K have revealed the contribution of low-energy excitations (LEE), seen as an additional contribution to the specific heat to the regular phonon term. These LEE were interpreted \cite{Larkin95,Ovchinnikov96} as metastable states and analysed as two-level systems resulting from local deformations of the DW at strong pinning impurity sites (see section~\ref{sec7}).

\subsection{Switching in NbSe$_3$}\label{sec6-6}

\begin{figure}
\begin{center}
\includegraphics[width=7.5cm]{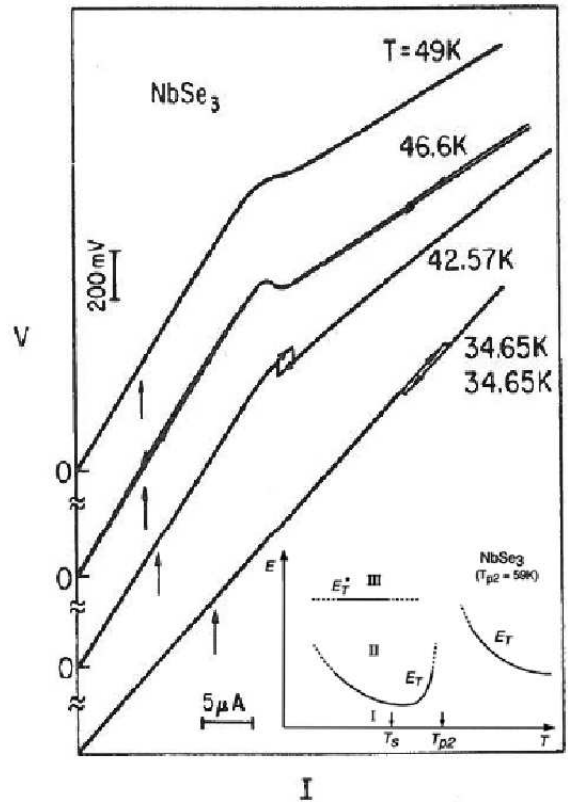}
\caption{I-V characteristics of NbSe$_3$ at several temperatures showing switching. Arrows indicate the value at which the differential resistance ${\rm d}V/{\rm d}I$ starts slightly to decrease (with permission from M.-C. Saint-Lager \cite{Saint-Lager83}). Inset shows the schematic phase diagram of NbSe$_3$ showing the two thresholds $E_T$ and $E^\ast_T$ for the lower CDW of switching samples (from ref.~\protect\cite{Li01}).}
\label{fig6-21}
\end{center}
\end{figure}

It has been observed that, in the low temperature CDW phase of NbSe$_3$, switching (in the I-V curves) occurs for some samples \cite{Hall88}. An example is drawn in figure~\ref{fig6-21} 
showing the evolution when $T$ is decreased of the shape of the I-V characteristics \cite{Saint-Lager83}. Two threshold fields are determined: the lower one $E_c$ (marked by an arrow in figure~\ref{fig6-21} 
at which a very low increase of conductivity occurs and a sharp higher field $E^\ast_T$ at which the non-linear current jump by several orders of magnitude. Hysteresis in switching is also observed between increasing and decreasing the applied current. It was shown that at variance with $E_T$, $E^\ast_T$ is nearly temperature independent \cite{Li01} (see inset of figure~\ref{fig6-21}). It was noted also that, after application of a current pulse, some delay was needed for the abrupt transition to the high conductivity state \cite{Levy93}. Not all NbSe$_3$ samples present switching and the ratio between switching samples to non-switching samples is largely dependent on different batches. Even more surprisingly some switching samples were shown to loose its property after a few  months \cite{Hall88}.

Switching in NbSe$_3$ was ascribed to the presence inside crystals of extended strong pinning centres which generates phase slippage \cite{Inui88}. Location of these pinning centres was determined by measuring the I-V characteristics with movable non-pertubating voltage electrodes \cite{Hall86}. On the contrary, in ref.~\cite{Adelman93}, it was first stated that $E^\ast_T$ is not determined by isolated strong pinning impurities but by a bulk pinning. Later the central role due to the competition between local and collective pinning was recognised \cite{Lemay99,Thorne02a}. There is a striking similarity between switching in NbSe$_3$ and in o-TaS$_3$, K$_{0.3}$MoO$_3$ \cite{Maeda90} in spite of the different nature of their ground states. Then the same models as developed above \cite{Brazovskii04} seem to be  applicable for all these DW systems.

\section{Excitations}\label{sec7}
\setcounter{figure}{0}
\setcounter{equation}{0}

As discussed in section~\ref{sec2}, the Peierls transition in quasi-1D systems, which occurs at a finite $T_{\rm P}$ due to a finite interchain coupling, arises from the first-order coupling between the 1D electron gas and the phonon field, inducing a static modulated distortion below $T_c$ of the lattice positions such:
\begin{equation*}
u(r)=A\cos(Qx+\phi).
\end{equation*}
The low energy excitations of the incommensurate modulation are those of the amplitude $A$ and of the phase $\phi$ which give rise to two branches: the amplitudon and the phason branches in the phonon spectrum.

\subsection{Amplitudons and phasons}\label{sec7-1}

\subsubsection{Incommensurate dielectrics}\label{sec7-1-2}

These two new collective excitations are peculiar to aperiodic phases which include incommensurate modulated phases, composites, quasi crystals (for reviews see \cite{Janssen07} and \cite{Currat88}) and CDW compounds. A unified description of phason modes from the hydrodynamic theory of aperiodic crystals is given in ref.~\cite{deBoissieu08}. In displacively modulated crystals, the infinite-wavelength phason --so-called sliding mode-- corresponds, as in CDWs, to an overall phase shift of the modulation with respect to the average lattice. It may be gapless, as in Goldstone theory, or pinned depending of the analytic or non-analytic behaviour of the superspace functions representing the modulation profile \cite{Aubry78}.

In composite binary systems as the alkane-urea insertion compounds \cite{Lefort96}, the phason branch corresponds to the anti-translation of the two intermodulated sublattices along the direction of incommensurability \cite{Ollivier98,Currat02}.

Dynamical studies of incommensurate systems have been carried out by means of light, neutron and resonance spectroscopy. Raman scattering has been intensively used in the study of amplitude modes \cite{Poulet86}. Phason branches were clearly observed in several incommensurate dielectrics by high resolution cold 3-axis neutron scattering \cite{Currat89,Currat00}.

The molecular compound of bis(4-Chlorophenyl)sulphone (ClC$_6$D$_4$)$_2$SO$_2$, in short BCPS) presents a displacive incommensurate transition at $T_I$~= 150~K with a modulated wave vector $q_s$~= $0.78b^\ast$. The low soft-mode damping has permitted \cite{Ollivier98} a detailed analysis of the lineshape of the soft-mode above $T_I$, allowing to disentangle its contribution from the quasi-elastic scattering or central peak one (common to displacive phase transition and plausibly resulting from a linear coupling of the soft-mode with a slowly relaxing defect), as well as the phason lineshape below $T_I$.

ThBr$_4$ undergoes a phase transition at $T_I$~= 95~K from a tetragonal phase to a 1D incommensurate modulated phase resulting from the condensation of a soft optic mode near the (0, 0, 1/3) wave vector. Between 300~K and 120~K, the entire optic branch softens while the transverse acoustic (LA) slope remains unchanged \cite{Bernard83}. Double-peak structures were observed in constant-$Q$ scans with the origin of wave vectors corresponding to the (2, 3, 0.69) satellite reflection at $T$~= 81~K ($T_c$-14~K); the lower branch was identified to the propagation phase mode and the upper branch to the corresponding amplitude mode seen also in Raman experiment \cite{Hubert81}. For $\xi$~= 0, the data cannot ascertain the existence of a zero-wave vector gap for the phase mode from the extrapolation of the linear phason dispersion. Due to the experimental uncertainties, a upper limit of 70~GHz was estimated \cite{Bernard83}.

Phason dispersion was also measured \cite{Hlinka02} in the incommensurate phase of (deuterated) thioura [SC(ND)$_2$)$_2$] \cite{Denoyer80,Denoyer86} which occurs between the high temperature paraelectric phase ($T>T_I$~= 202~K) and the low temperature ferroelectric phase ($T<T_c$~= 169~K). In the intermediate temperature interval $T_c<T<T_I$ the modulation wave vector $q_I$ displays a monotonous devil' staircase-like temperature dependence with $q_I(T_I)\approx 0.14$b$^\ast$ and $q_I(T_c)$~= b$^\ast/9$ \cite{Denoyer80}.

The same collective modes were also determined in biphenyl (C$_{12}$H$_{10}$). Biphenyl undergoes two successive phase transitions: the first one of second order type structural phase transition occurs at $T_I$~= 37~K driven by a soft mode related to the torsional motion of the molecules. Below $T_I$ (phase II), this mode is frozen into a static incommensurate modulation wave characterised by four $\pm q_{s_1}$ and $\pm q_{s_2}$ wave vectors. The order parameter has four components ($n$~= 4), directly related to the modulation amplitudes $A_1$ and $A_2$, and phases $\phi_1$ and $\phi_2$. At $T_{II}=20$~K, a first order phase transition occurs with a partial lock-in. Below $T_{II}$ (phase III) the two modulation wave vectors lie along the twofold screw axis, b$^\ast$ ($n$~= 2). Propagating  phase mode and amplitude mode branches were measured in both phase III \cite{Cailleau80,Cailleau86} and phase II \cite{Moussa87,Launois89}.

In all these systems the determination of the phase mode by inelastic neutron scattering requires an incommensurate system with a small damping, so that the phason becomes underdamped away from the satellite position.

\subsubsection{Screening of the phason mode}\label{sec6-5}

The frequency dependences of the phason and amplitudon mode were given in eqs.~(\ref{eq2-13}) and (\ref{eq2-14}) and the velocity of the phason mode is given as:
\begin{equation}
v_\varphi=\left(\frac{m}{M^\ast}\right)^{1/2}\,v_{\rm F}.
\label{eq6-28}
\end{equation}

In the case where the effects of impurities are neglected ($\omega_p$~= 0), the dispersion law of the phase mode in the case of screening has been calculated by several authors \cite{Lee78,Kurihara80,Barisic87,Nakane85,Wong87,Artemenko89,Virosztek93}. Longitudinal and transverse modes have to be analysed separately: a longitudinal compression of the CDW implies a charge redistribution leading to long range Coulomb forces which consequently at $T$~= 0 raises the $\omega$-phase mode to a finite frequency calculated \cite{Lee74} to be ($1.5\lambda)^{1/2}\omega_A$, $\omega_A$: the frequency of the amplitude mode. On the contrary a transverse shear only implies a dephasing of the CDW between adjacent chains but not a charge redistribution which preserves the acoustic character of the phase mode. Following \cite{Wong87} with Coulomb force interactions, the dispersion relation eq.~(\ref{eq2-14}) becomes, with $x$ the longitudinal chain axis:
\begin{equation}
\omega_-^2=\left(v_\varphi^2+\frac{\Omega_p^2}{\varepsilon_zq_x^2+q_0^2}\right)q_x^2,
\label{eq6-29}
\end{equation}
with two limits
\begin{equation}
\omega_-^2=\left(v_\varphi^2+\frac{\Omega_p^2}{q_0^2}\right)q_x^2\quad\mbox{for}\quad q_x^2\ll\frac{q_0^2}{\varepsilon_z},
\label{eq6-30}
\end{equation}
and
\begin{equation}
\omega_-^2=\frac{3}{2}\lambda\omega^2_A+v_\varphi^2q_x^2\quad\mbox{for}\quad q_x^2\gg\frac{q_0^2}{\varepsilon_z},
\label{eq6-31}
\end{equation}
where $\varepsilon_z$ is the screening dielectric constant: $\varepsilon_z$~= $1+\omega^2_{pe}/6\Delta^2$ with $\omega_{pe}^2$~= $4\pi\,ne^2/m$. $\Omega_p^2$~= $4\pi ne^2/M^\ast$ ($\Omega_p$: the plasma frequency) and $q_0^2$~= $4\pi e^2n_{qp}/T$ the square of the Thomas-Fermi screening wave number of the quasiparticle gas with the concentration $n_{qp}$. This quasiparticle concentration is thermally activated. The cross-over between both regimes  (eqs.~(\ref{eq6-30}) and (\ref{eq6-31})) results from the comparison between the phase mode length and the Thomas-Fermi screening length. When $\vert q_x\vert\ll q_0\varepsilon^{-1/2}_z$ the phase mode length is much larger than the Thomas-Fermi screening length, there is no local charge density in the scale of a phase mode length and the excitations remain acoustic. However the phason velocity is seen to become stiffer. On the other hand, when $\vert q_x\vert\gg q_0\varepsilon_z^{-1/2}$, quasi particles are unable to screen any charge fluctuation in the scale of the phase mode length and the spectrum approaches the optical frequency $\frac{3}{2}\lambda\omega^2_A+v^2_\varphi q^2_x$ which in the limit $q\rightarrow 0$ agrees with the result from ref.~\cite{Lee74}. In the temperature range where the phase mode is still acoustic, the temperature dependence of the phase mode velocity was given by \cite{Wong87,Nakane85}:
\begin{equation}
v_\varphi=v_{\rm  F}\left(\frac{m}{M^\ast}\right)^{1/2}\left[1+\frac{\exp(\Delta/T)}{(2\pi\Delta/T)^{1/2}}\right]^{\frac{1}{2}}.
\label{eq6-32}
\end{equation}
It shows that the longitudinal phason velocity increases as temperature decreases and ultimately the phason itself disappears completely and the CDW becomes rigid.

Experimentally neutron-scattering experiments in blue bronze have reported the observation of the Kohn anomaly at $Q$~= $2k_{\rm F}$ and below $T_{\rm P}$, the amplitudon and phason branches \cite{Pouget91,Hennion92}. The linear dispersion measured from the satellite reflection corresponds  to the phase mode with a very anisotropic slope, leading at $T$~= 175~K to the phason velocity $v_\varphi$ along the chains of $3.3\pm 0.5\times 10^3$~m/s and perpendicular to the chains of $1.8\pm 10^3$~m/s. When $T$ was decreased down to 100~K, the phason mode remains acoustic but with a velocity $\sim 2.5$ larger at 100~K \cite{Hennion92}, confirming the importance of Coulomb forces. On the other hand, there is no significant variation of the velocity along the transverse ($2a^\ast$-$c^\ast$) direction.

The counting rate in inelastic neutron scattering is very limited when $T$ is decreased, preventing thus any measurement at temperatures $\sim$ below 100~K. However, inelastic X-ray scattering (IXS) due to the high brilliance of third generation synchrotron sources is a very powerful recent technique for studying low-energy excitations in solids. This technique was used especially on R$_{0.30}$MoO$_3$ down to 40~K. Energy scans were measured \cite{Ravy04} at positions $Q_S+0.05(2a^\ast-c^\ast)+\delta kb^\ast$ near the strongest satellite reflection $Q_S$~= (15, 0.75, -7.5). Fits were made with the sum of four damped harmonic oscillators: two high-energy modes called optics I and II at about 10 and 20~meV, the amplitudon and the phason mode. Energy and damping for the amplitude mode were taken from previous inelastic neutron scattering, Raman and ultrafast reflectivity measurements. From 170~K down to 100~K, at different positions along the chains $0.75+\delta k$, phason modes were well identified with the slope of the dispersion in good agreement with results from inelastic neutron scattering. But at $T$~= 40~K, there is no intensity between the elastic and the amplitudon peaks, i.e. there is no evidence of the phason mode. At this temperature, the phason mode becomes optic-like with a frequency given by eq.~(\ref{eq6-31}) which can be estimated to be $\sim 8.5$~meV. This optic phason mode could not be disentangled from the amplitude mode.

\subsection{Excess heat capacity}\label{sec7-3}

The lattice specific heat $C_{\rm P}$ of crystalline matter is well described by the Debye model at $T\ll\theta_{\rm D}$ where $k_{\rm B}\theta_{\rm D}$ is a cut-off energy. But this model does not describe the low temperature specific heat if additional degrees of freedom have to be considered; thus excitations in the order of 10~GHz contribute to $C_{\rm P}$ in the temperature range 100~mK. CDW phasons are a special type of lattice vibrations with a linear $q$ dependence of their frequency. Their contribution to the low temperature specific heat was evaluated \cite{Boriack78} using a Debye model. Hereafter very low temperature thermodynamic experiments are presented, performed with the aim to detect the possible contribution of phasons to the specific heat.

\begin{figure}
\begin{center}
\includegraphics[width=6.5cm]{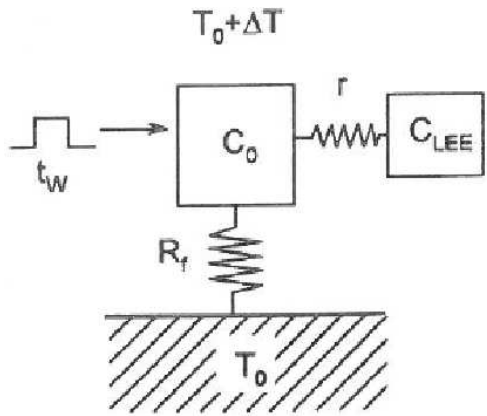}
\caption{Schematic representation of the heat method for measuring specific heat: $R_f$ is the thermal link of the sample to the thermal bath, $C_0$ the regular part of the specific heat weakly coupled through internal thermal link $r$ to the low energy excitations (LEEs). Their specific heat contribution $C_{\rm LEE}$ shows strong time dependence.}
\label{fig7-13}
\end{center}
\end{figure}

Specific heat was measured with a transient-heat-pulse technique schematically displayed in figure~\ref{fig7-13} \cite{Kis99}. The specific heat is calculated from the increment of temperature $\Delta T$~= $T-T_0$ that occurs during a heat pulse by using the exponential decay of the temperature: $T(t)-T_0$~= $(T_{\rm P}-T_0)\exp -t/\tau$ with the relaxation time $\tau$~= $C_{\rm P}R_\ell$, where $C_{\rm P}$ is the specific heat and $R_\ell$ the thermal resistivity of the thermal link to the regulated cold sink.

\begin{figure}
\begin{center}
\subfigure[]{\label{fig7-14a}
\includegraphics[width=6.5cm]{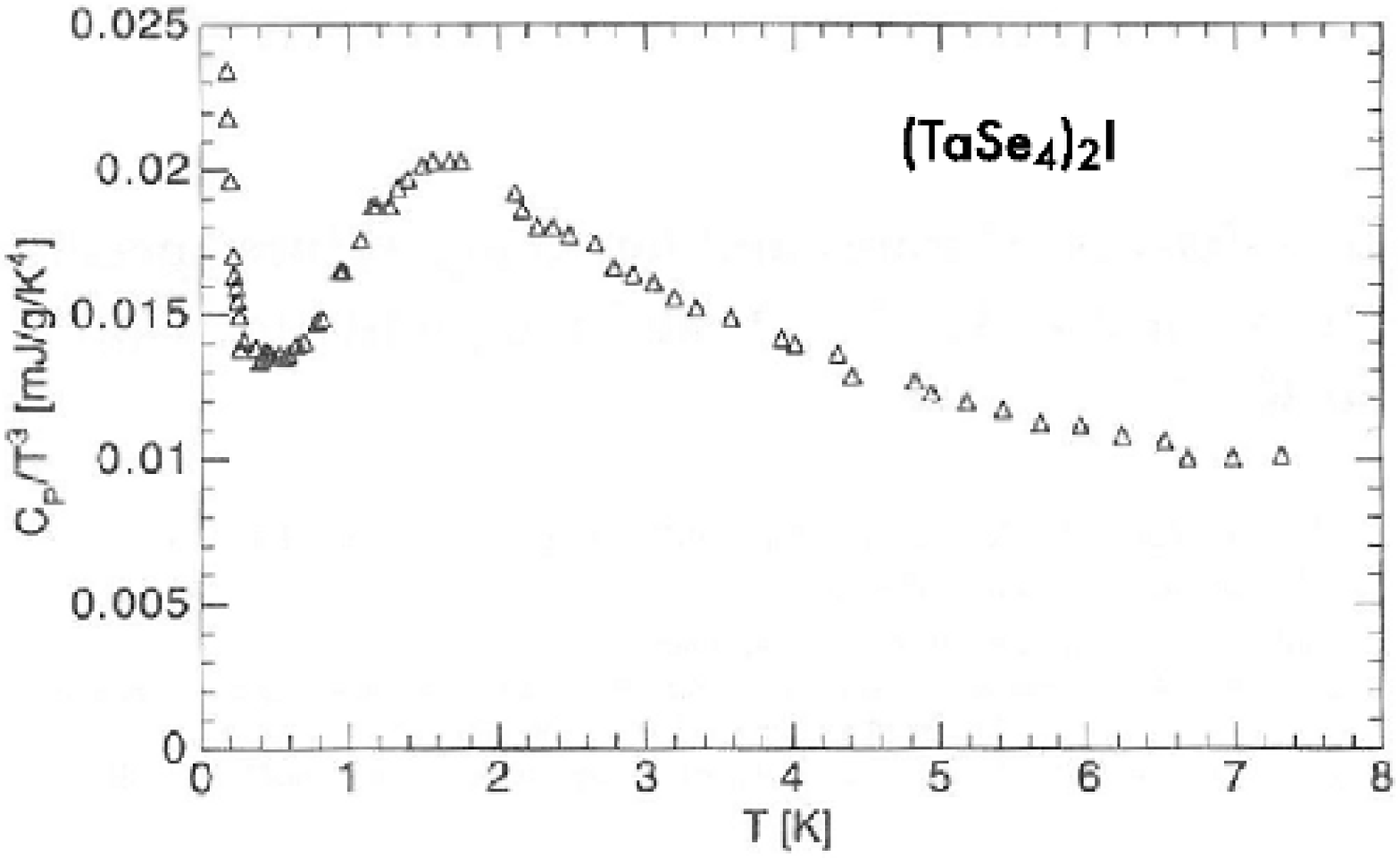}}
\subfigure[]{\label{fig7-14b}
\includegraphics[width=6.5cm]{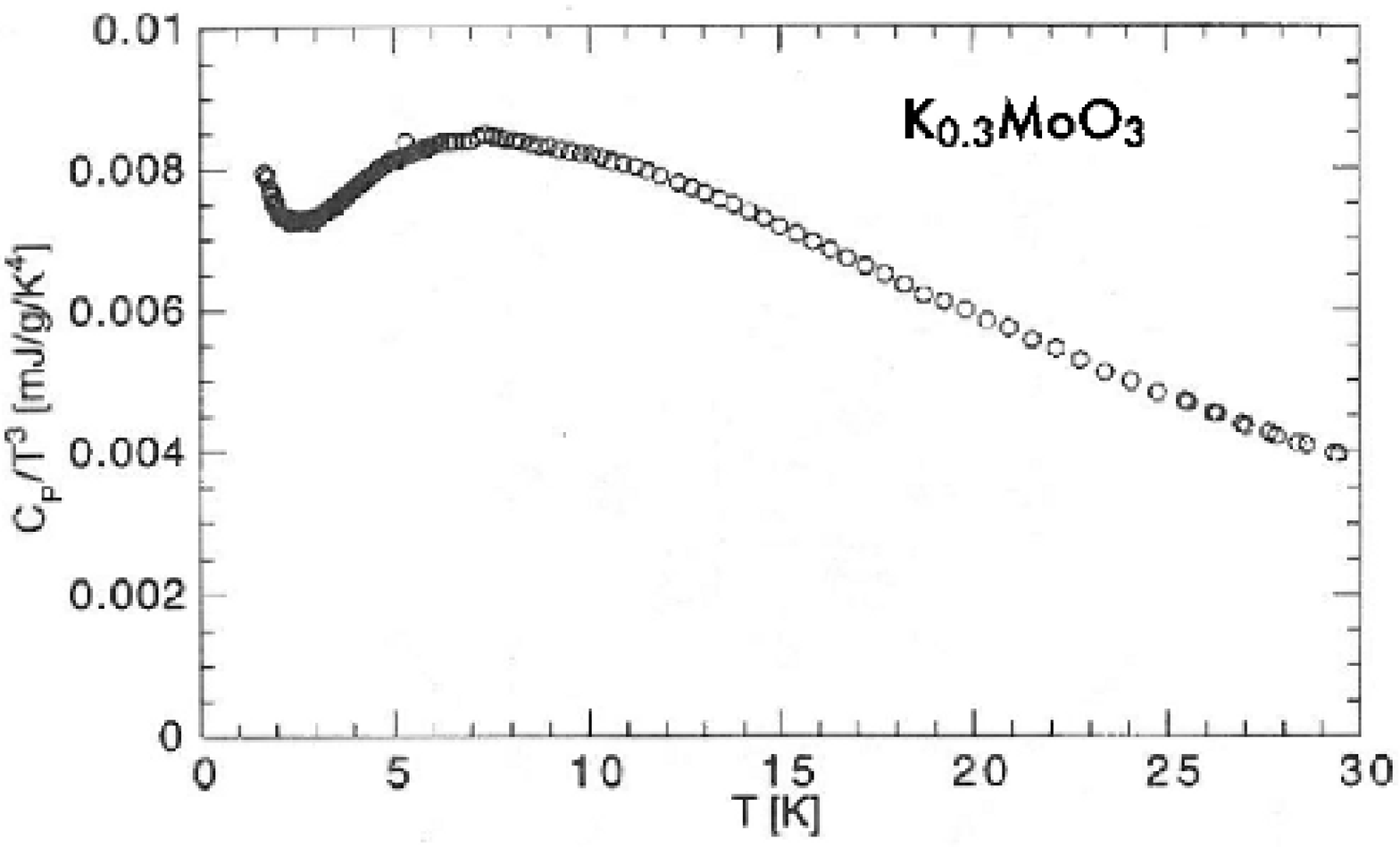}}
\subfigure[]{\label{fig7-14c}
\includegraphics[width=6.5cm]{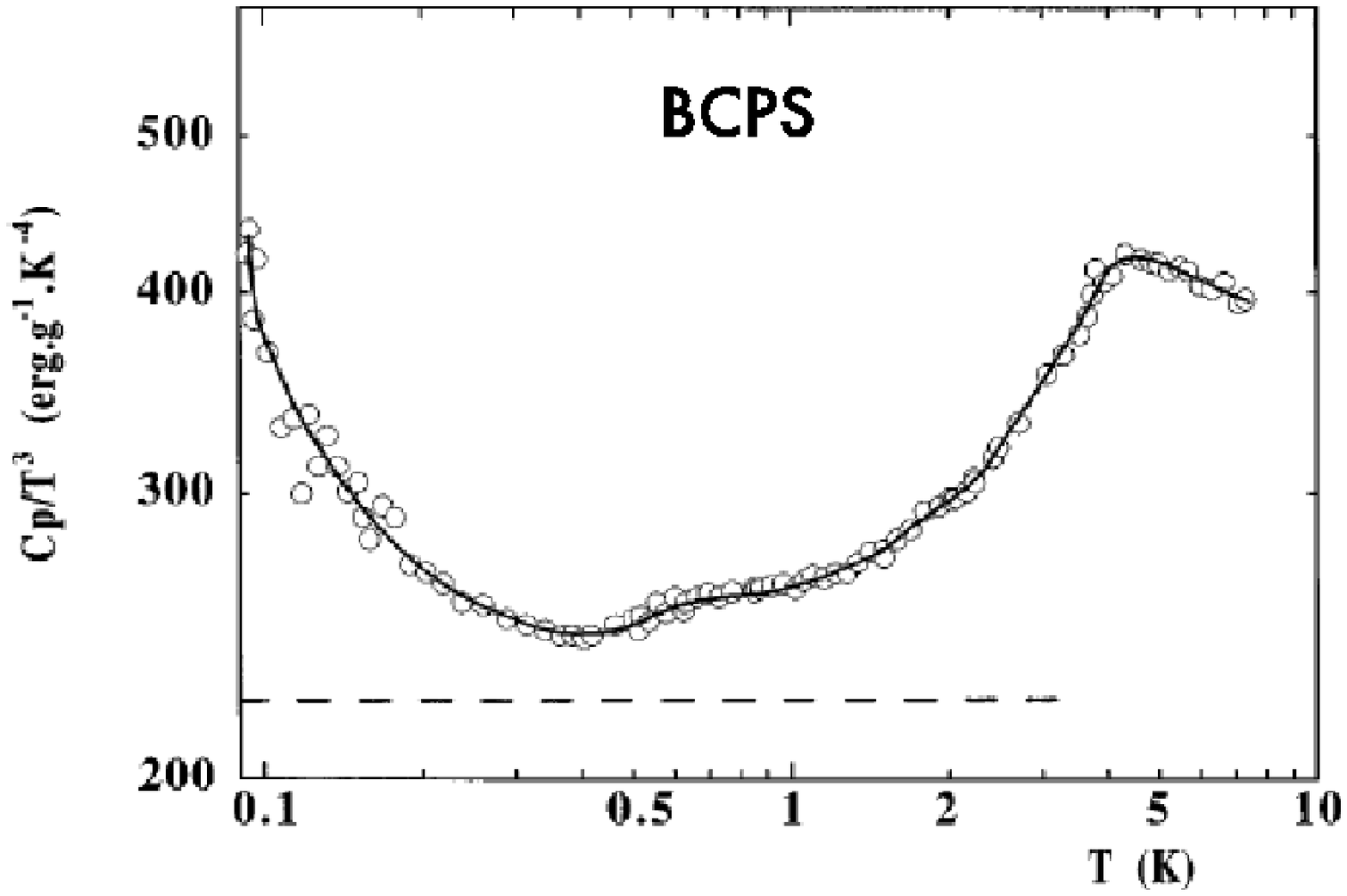}}
\subfigure[]{\label{fig7-14d}
\includegraphics[width=6.5cm]{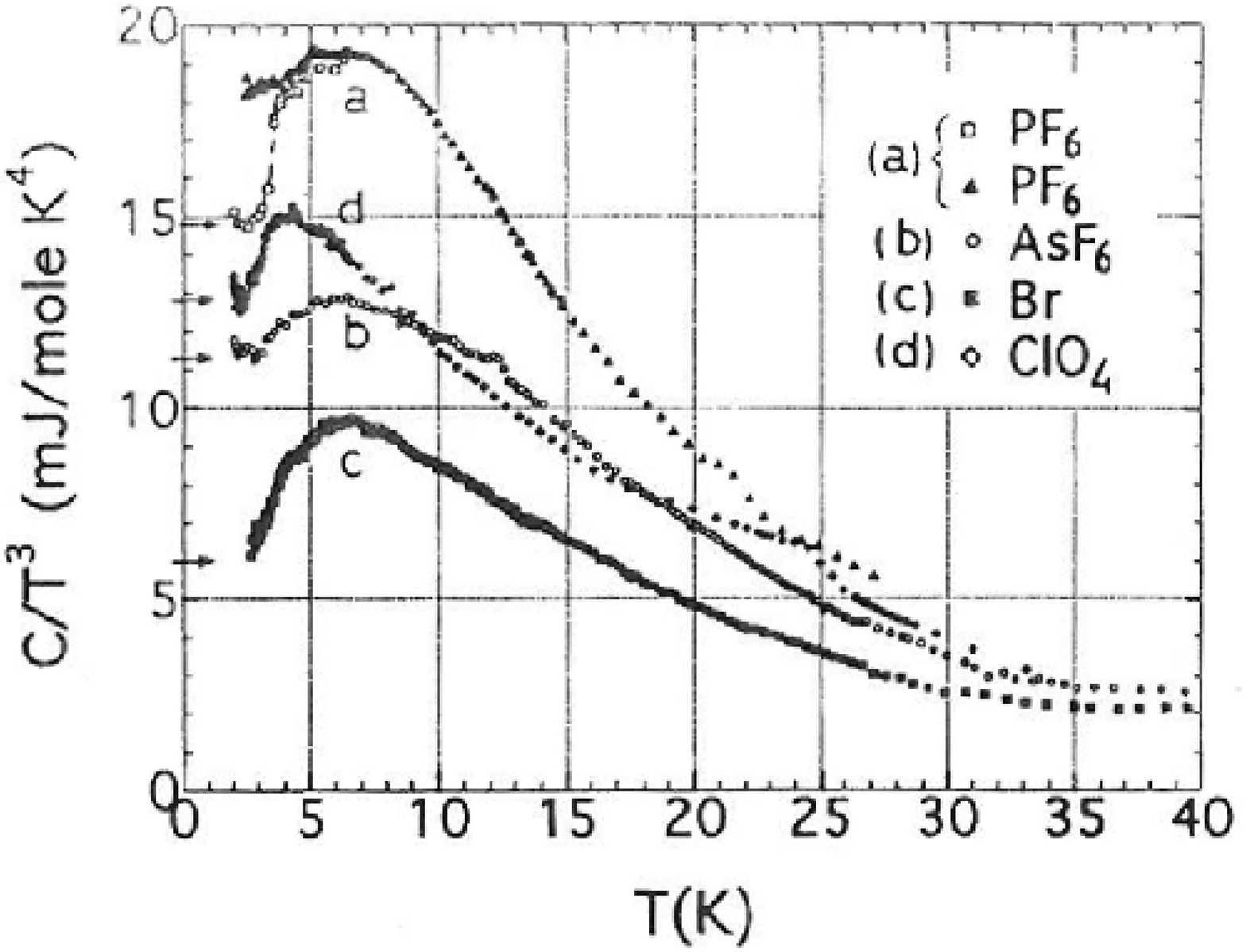}}
\caption{Specific heat in $C_{\rm P}/T^3$. a)~(TaSe$_4$)$_2$I (reprinted figure with permission from K. Biljakovi\'c \textit{et al.}, Physical Review Letters 57, p. 1907, 1986 \cite{Biljakovic86}. Copyright (1986) by the American Physical Society). b)~KCP (reprinted figure with permission from J. Odin \textit{et al.}, Physical Review B 46, p. 1326, 1992 \cite{Odin92}. Copyright (1992) by the American Physical Society). c)~(ClC$_6$D$_4$)$_2$SO$_2$(BCPS) (reprinted figure with permission from J. Etrillard \textit{et al.}, Physical Review Letters 76, p. 2334, 1996 \cite{Etrillard96}. Copyright (1996) by the American Physical Society). d)~(TMTSF)$_2$PF$_6$, (TMTSF)$_2$AsF$_6$, (TMTTF)$_2$Br, (TMTSF)$_2$ClO$_4$ ((reprinted figure with permission from H. Yang \textit{et al.}, Journal of Physics: Condensed Matter 11, p. 5083, 1999 \cite{Yang99}. Copyright (1999) by the Institute of Physics).}
\label{fig7-14}
\end{center}
\end{figure}

\begin{figure}
\begin{center}
\includegraphics[width=7cm]{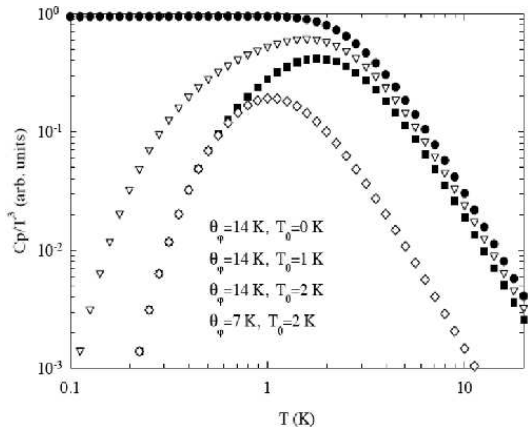}
\caption{Schematic presentation of the phason contribution to the specific heat in $C_{\rm P}/T^3$ calculated from eq.~(\ref{eq7-2}): $T_0$ represents the lower cut-off, $T_0$~= $h\nu_0/k_{\rm B}$, $\nu_0$: phason gap which dominates the low-$T$ range; $\theta_\phi$ is the equivalent of the Debye cut-off frequency for phasons which influences the maximum position (reprinted figure with permission from the European Physical Journal B - Condensed Matter and Complex Systems 28, J.-C. Lasjaunias \textit{et al.}, p. 187, 2002 \cite{Lasjaunias02c}. Copyright (2002) from Springer Science and Business media).}
\label{fig7-15}
\end{center}
\end{figure}

\begin{figure}
\begin{center}
\includegraphics[width=8cm]{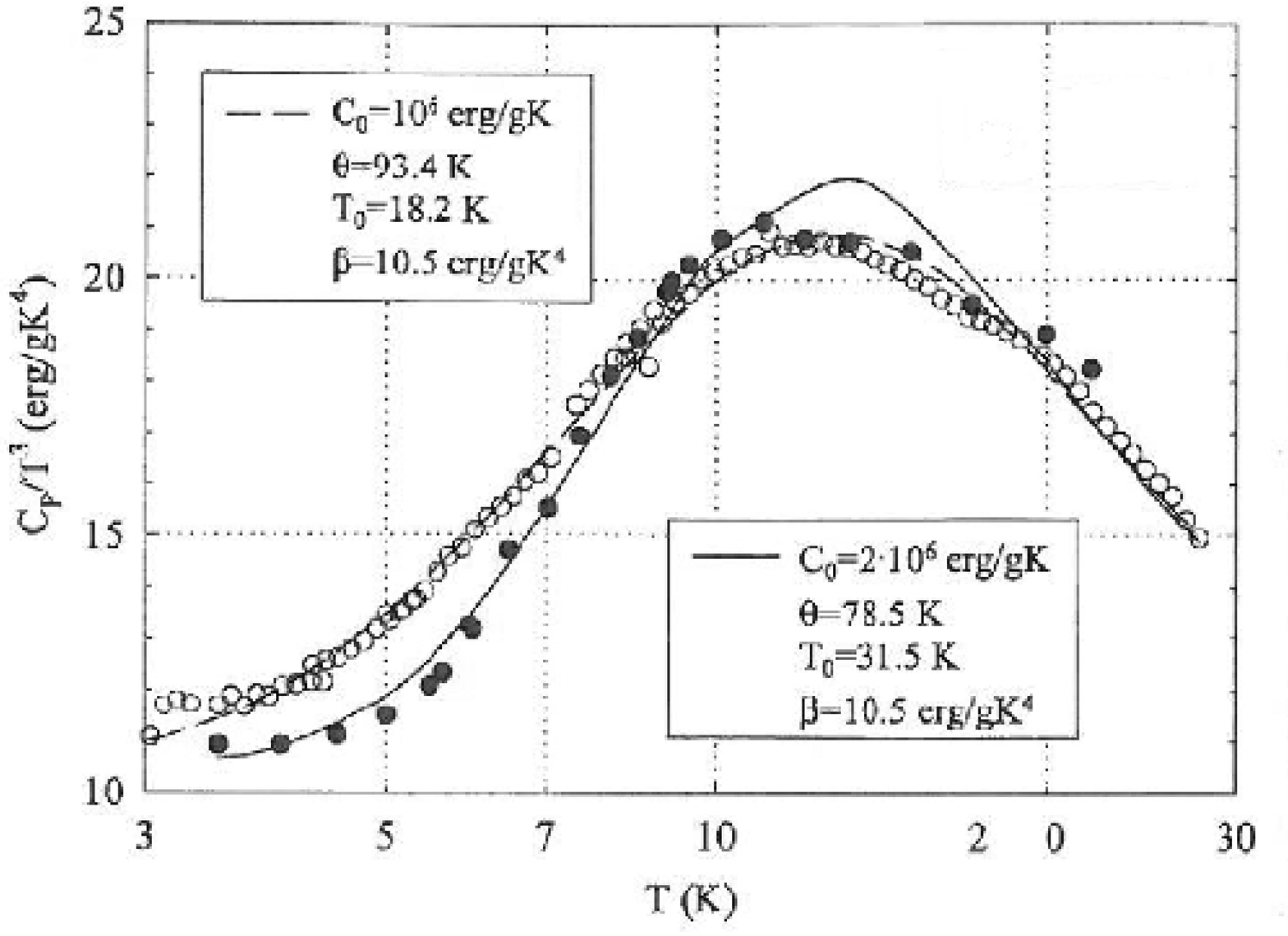}
\caption{Specific heat in $C/T^3$ of a pure sample of K$_{0.3}$MoO$_3$ ({\large$\circ$}) and of a sample of a lower purity or structural quality ({\large$\bullet$}). Fits of the bump by the phason model (eq.~(\ref{eq7-2})) are indicated by a continuous line for data ({\large$\bullet$}) and  a dashed line for ({\large$\circ$}) with parameters reported in the insets (reprinted figure with permission from European Physical Journal B - Condensed Matter and Complex Systems 24, J. Odin \textit{et al.}, p. 315, 2001 \cite{Odin01}. Copyright (2001) from Springer Science and Business media). Other data on the low temperature $C_{\rm P}$ of K$_{0.3}$MoO$_3$ were reported in \cite{Dalhauser86} and \cite{Konate84}.}
\label{fig7-17}
\end{center}
\end{figure}

For many C/S DW systems $C_{\rm P}$ is characterised by four generic features:\\
a)~a sub-cubic regime for the acoustic phonon contribution which may originate from the strong anisotropy of force constants along and between the adjacent chains (sec.~\ref{secbendingforces}),\\
b)~a bump in $C_{\rm P}/T^3$ as shown in figure~\ref{fig7-14} for different types of systems: K$_{0.3}$MoO$_3$, (TaSe$_4$)$_2$I, BCPS and Bechgaard-Fabre salts,\\
c)~a contribution of low-energy excitations (LEE) additional to that of regular phonons following the law: $C_{\rm LEE}\sim T^\nu$ with $\nu<1$ (sec.~\ref{sec8-1})\\
d)~metastability and non-exponential relaxation with ageing effects at very low temperatures ($T<0.5$~K), indicating that $C_{\rm P}$ becomes strongly time-dependent (sec.~\ref{sec8-4}).

\subsubsection{Phason heat capacity}\label{sec7-3-1}

Following ref.~\cite{Boriack78}, the phason contribution vanishes at a temperature $\theta_\phi$ defined by $h\nu_\phi$~= $k_{\rm B}\theta_\phi$ where $\nu_\phi$ is the frequency at which the phason modes cannot be distinguished from the phonon modes. The excess heat capacity due to phasons was analysed by a Debye excitation spectrum \cite{Boriack78}, modified to take into account two cut-off frequencies: the upper one analogous to a Debye temperature for phasons, $\theta_\phi$, and the lower one corresponding to the pinning frequency $h\nu_0$~= $k_{\rm B}T_0$ if the phason mode is not gapless, such:
\begin{equation}
C_\phi=3N_\phi k_{\rm B}(T/\theta_\phi)^3\int^{\theta_\phi/T}_{T_0/T}(x-x_0)^2\left[\frac{x^2e^x}{(e^x-1)^2}\right]dx\,
\label{eq7-4}
\end{equation}
with $x$~= $h\nu/kT$, $N_\phi$ being the number of phason excitations, $\eta$ the anisotropy of the phason velocity.

Figure \ref{fig7-15} illustrates the effect of the low frequency cut-off on the $C_{\rm P}/T^3$ variation. It appears that the low part of the $C_{\rm P}/T^3$ bump is strongly depressed when the low-frequency cut-off $T_0$, i.e. the pinning frequency, is increased, keeping the upper part above the maximum nearly unchanged \cite{Lasjaunias02c}. 

Good fits were reported from eq.~(\ref{eq7-4}) for (TaSe$_4$)$_2$I with \cite{Biljakovic86}, K$_{0.3}$MoO$_3$ \cite{Odin01} platinum chain compound KCP \cite{Odin92} with, when available, acceptable agreement with pinning frequency obtained from microwave conductivity experiments.

Similarly the two bumps in $C_{\rm P}/T^3$ of BCPS have been identified \cite{Etrillard96} as resulting from the phason contribution with a pinning frequency of $\sim 80$~GHz and from an Einstein mode at $\sim 500$~GHz which corresponds to the low temperature amplitudon mode as determined by neutron scattering and shown in figure~\ref{fig7-14}(c). The excess specific heat to the phonon contribution in ThBr$_4$ can be analysed by taking into account a phason contribution \cite{Biljakovic11}. However in the case of molecular nonadecane/urea inclusion composite, no phason mode contribution to $C_{\rm P}$ at low temperature was detected, the excess specific heat was assigned to the dynamical disorder of the guest molecules \cite{Etrillard00}.

\subsubsection{Contribution from low energy modes}\label{seccontriblowenergymodes}

The interpretation of the bump in $C_{\rm P}/T^3$ in terms of a phason contribution has been contested in refs~\cite{Requardt97,Lorenzo96,Lorenzo02}. A generalised phonon density-of-states (PDOS) in (TaSe$_4$)$_2$I and (NbSe$_4$)$_3$I was derived from neutron time-of-flight measurements. A low-frequency step-like structure in the PDOS of (TaSe$_4$)$_2$I was observed, assigned to the flat transverse-acoustic sheet with in-chain polarisation (see figure~\ref{fig7-6}). The observed specific heat anomaly can then be reproduced from this step in the PDOS \cite{Lorenzo96}.

A PDOS was also measured in K$_{0.3}$MoO$_3$ \cite{Requardt97}. Data show a Debye-like behaviour up to about 3~meV; when the parabolic $a\omega^2$ is extrapolated to higher energies, the PDOS between 4 and 7~meV exceeds a Debye behaviour. When calculating the specific heat from the generalised PDOS, the $C_{\rm P}/T^3$ variation reveals qualitatively the same behaviour as the measured specific heat: a flat part -- the ``Debye plateau"-- below about 4~K, corresponding to a pure Debye behaviour in this $T$-range (the $\omega^2$ variation in the PDOS curve) and a bump in $C_{\rm P}/T^3$ for temperatures above 4~K with a maximum at around 11~K revealing a non-Debye behaviour. The temperature of this maximum corresponds to an energy of about 5--6~meV in the PDOS, i.e. to the energy range of the excess over a Debye law in the generalised PDOS.

The bump in $C_{\rm P}/T^3$ in (TaSe$_4$)$_2$I, K$_{0.3}$MoO$_3$ and KCP can also be obtained from calculations of the phonon density of states derived from the low energy phonon branches, especially low optic modes with no (or a small) dispersion in the Brillouin zone and flat sections of acoustic branches near the Brillouin zone boundary \cite{Requardt97,Lorenzo96}.

However the interpretation of solely a lattice phonon origin for explaining the anomalous specific heat behaviour observed in (TaSe$_4$)$_2$I and K$_{0.3}$MoO$_3$ should be considered with respect to results on doped samples. The frequency of the TA$_z$ sheet in (TaSe$_4$)$_2$I was shown by neutron scattering to be not sensitive to Nb-doping (0.4\%). However the $C_{\rm P}/T^3$ variation is steeper below the maximum peak for the doped sample. The maximum in $C_{\rm P}/T^3$ was shown to be affected (reduced) by substitutional impurities \cite{Brown88}. Similarly figure~\ref{fig7-17} displays the specific heat of two K$_{0.3}$MoO$_3$ samples of different purity: ({\large$\circ$}) data for a purer sample, ({\large$\bullet$}) for a dirtier one \cite{Odin01}. While there is an excellent agreement for temperatures above the maximum in $C_{\rm P}/T^3$, below the maximum the variation is much sharper for the more impure sample. Fits by the phason model are indicated in figure~\ref{fig7-17} with the parameters of fitting. As shown in figure~\ref{fig7-15}, in the phason model, a sharper $C_{\rm P}/T^3$ variation at low $T$ is expected from a higher pinning frequency.

While in quasi 1D compounds, peculiar low energy phonon modes contribute to the non-Debye behaviour of the specific heat, one also are led to consider that, at least, a part of the $C_{\rm P}/T^3$ bump may originate from a phason contribution.

\subsubsection{Low-energy ``intra"-molecular phonon modes in Bechgaard-Fabre salts}\label{sec7-3-2}

The specific heat of (TMTSF)$_2$PF$_6$, (TMTSF)$_2$AsF$_6$, (TMTSF)$_2$ClO$_4$ (in the quenched-SDW state) and (TMTTF)$_2$Br are presented in figure~\ref{fig7-14}(d) in the form of the temperature dependences of $C_{\rm P}/T^3$ from 1.8 to 40~K. The SDW transitions appear as a shallow anomaly in the total specific heat at $T_c$~= 12.2, 12.4, 4.5 and 11.7 for the four compounds, respectively. The anion ordering for quenched (TMTSF)$_2$ClO$_4$ is also clearly detectable from the anomaly at 24~K. In the case of (TMTSF)$_2$PF$_6$ an abrupt jump at $T$~= 3.5~K is observed by the transient-heat pulse technique whereas it disappears under quasi-adiabatic conditions \cite{Odin94}. This jump was ascribed as resulting from a glass-like transition due to the freezing-in of internal degrees of freedom of the disordered SDW \cite{Lasjaunias94} (see section~\ref{sec7-3-2.a} for a detailed discussion).

From figure~\ref{fig7-14}(d), one can see that $C_{\rm P}/T^3$ presents a bump with a maximum located at about 7.0~K for (TMTSF)$_2$PF$_6$, (TMTSF)$_2$AsF$_6$ and (TMTTF)$_2$Br and 4.0~K for (TMTSF)$_2$ClO$_4$ in the quenched-SDW state. Such an upward deviation of the specific heat from the usual Debye $T^3$-law was ascribed to low-energy phonon modes \cite{Yang99,Yang00}. In the case of organic compounds, the low-energy ``intra"-molecular phonon modes have been revealed by far-infrared spectroscopy. In the case of (TMTSF)$_2$PF$_6$, two sharp peaks were measured \cite{Eldridge85} at 18~cm$^{-1}$ and 45$\pm$5~cm$^{-1}$ (equivalent to 26~K and 64$\pm$7~K).

Independently of the contribution of the Einstein modes the substantial decrease of $C_{\rm P}/T^3$ in the high $T$-range cannot be explained by the usual assumption of deviation from the $T^3$-law which starts at $T\approx\theta_{\rm D}$/20, at least ($\theta_{\rm D}$: Debye temperature~= 200~K for (TMTSF)$_2$PF$_6$). Then, the variation of $C_{\rm P}/T^3$ was described using a temperature dependent phonon background $T^\alpha$, with $\alpha<3$. A self-consistent fit for (TMTSF)$_2$PF$_6$ is based \cite{Yang99} on the assumption of two main contributions: two additional Einstein modes, the frequency of which are based on far-IR measurements plus an acoustic background which changes from a cubic regime below a cross-over temperature $T^\ast$ towards a $T^\alpha$ dependence above. The best fit was found with $T^\ast$~= 2.5~K, $\alpha$~= 2.4 and the Einstein temperatures $\theta_1$~=  35~K and $\theta_2$~= 65~K. The shift  of the maximum of $C_{\rm P}/T^3$ from 7~K for (TMTSF)$_2$PF$_6$ to 4~K for (TMTSF)$_2$ClO$_4$ \cite{Yang00} is also in agreement with the related shift to 7 and 25~cm$^{-1}$ of the two first IR modes \cite{Ng83,Challener83}.

\subsubsection{Bending forces}\label{secbendingforces}

The sub-cubic regime for the lattice contribution can be interpreted within the model of Genensky and Newell \cite{Genensky57} made for a chain polymer crystal. Materials with restricted dimensionality are characterised by dominant valence forces between atoms in planes (as for graphite \cite{Nicklow72}) or along chains. They act as strong restoring forces for bending. Let assume $z$ the chain direction and $x$ and $y$ the transverse directions. The low frequency modes along the chains  are exactly the same as for a tridimensional lattice. The transverse modes  are abnormal, and the vibrations out the chains  can be evaluated \cite{Genensky57} such as:
\begin{equation}
\omega_x^2=\mu q_x^2+\frac{\lambda}{2}q^2_y+\frac{\lambda}{2}q_z^2+\frac{\kappa}{4}q^4_z\,
\label{eq7-5}
\end{equation}
where $\lambda$ is the force constant between second neighbours on adjacent chains, $\mu$ a combination of force constant between first and second neighbours and $\kappa$ the bending force constant along the chains. It is supposed that $\kappa\gg\lambda$. At low frequencies $0<\omega^2_x<\lambda^2/4\kappa$, the $q^4_z$ contribution is negligible and the dispersion law gives a frequency distribution law, $q_x(\omega)$, proportional to $\omega^2$ and therefore a specific heat in $T^3$. At moderate frequency for $\lambda^2/4\kappa<\omega_x^2\ll\lambda$, the $(\kappa/4)q_z^4$ becomes predominant. The frequency distribution is therefore proportional to $\omega^{3/2}$ and the corresponding specific heat follows a $T^{2.5}$ law. The total specific heat $C_{\rm P}$~= $(C_x+C_y+C_z)/3$ is the sum of the transverse and longitudinal contribution.

For the transverse contribution at very low temperatures $T\ll T_m\,\lambda/(4\kappa)^{1/2}$, the specific heat follows a cubic law; then in the range $T_m/(4\kappa)^{1/2}\ll T\ll T_m\sqrt{\lambda}$, it follows a $T^{5/2}$ law ($T_m$ is a limit temperature analogous to a Debye temperature). Then, in the same temperature range, one have simultaneously a $T^3$ variation for $C_z$ and a $T^{2.5}$ for $C_{x,y}$. As the measurements cannot separate the longitudinal and transverse contributions, one observes a linear combination of $T^3$ and $T^{2.5}$ terms corresponding to a power law with an intermediate coefficient between 2.5 and 3. The crossover temperature $T^\ast$ between a pure $T^3$ variation and the $T^{2.5}$ contribution from the bending force term is:
\begin{equation}
T^\ast=T_m\,\lambda/(4\kappa)^{1/2}.
\label{eq7-6}
\end{equation}

The chain bending force was clearly detected from the dispersion of transverse acoustic modes propagating along the chain direction at room temperature for (NbSe$_4$)$_3$I \cite{Monceau89} (see figure~\ref{fig7-6} and \ref{fig7-9}), for (TaSe$_4$)$_2$I \cite{Lorenzo98} (see figure~\ref{fig7-6}) and in NbSe$_3$ \cite{Requardt02a} (see figure~\ref{fig7-10}).

The anisotropic force constant appears to be a general property of Q-1D materials leading to a $T^\alpha$ dependence of the specific heat with $\alpha$~= 2.8 in NbSe$_3$ \cite{Lasjaunias82} corrected to 2.9 \cite{Biljakovic91b}, 2.8 in KCP \cite{Odin92}. The value 2.4$\pm$0.1 for (TMTSF)$_2$PF$_6$, (TMTSF)$_2$AsF$_6$, and (TMTTF)$_2$Br \cite{Yang99}, close to the limit case in the Genensky-Newell model, implies that the contribution of the transverse modes is predominant in the specific heat of these organic salts. Concerning (TMTSF)$_2$ClO$_4$, either in the metallic state or in the SDW-quenched state, the same value $\alpha$~= 2.7 was estimated for both states \cite{Yang00} indicating a reduced anisotropy of the lattice-force constants in this salt.

\subsubsection{A possible phason contribution in the specific heat of {\rm\bf (TMTSF)$_2$AsF$_6$}}\label{sec8-7}

\begin{figure}
\begin{center}
\includegraphics[width=7.5cm]{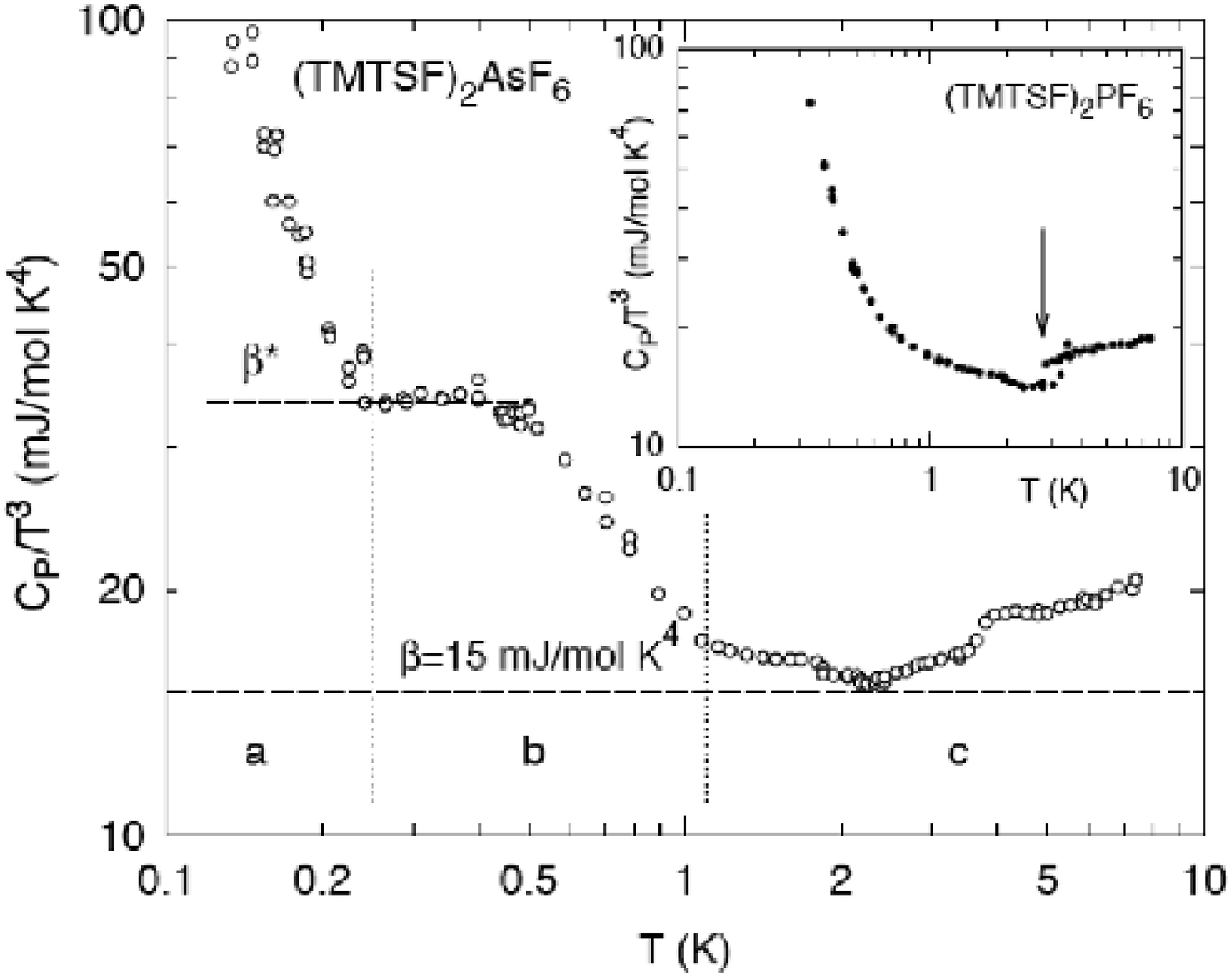}
\caption{Temperature dependence of the specific heat, $C_{\rm P}$ of (TMTSF)$_2$AsF$_6$, divided by $T^3$, between 0.15 and 7.5~K. At variance to (TMTSF)$_2$PF$_6$ (in the inset), $C_{\rm P}$ shows an intermediate cubic regime between 0.25 and 0.5~K. The arrow at $T$~= 3~K for (TMTSF)$_2$PF$_6$ indicates the onset of hysteresis phenomena ascribed to a glassy-like transition (see figure~\ref{fig8-6} for details). The horizontal dashed line $\beta$~= 15~mJ/mol$\,$K$^4$ is the estimation of the regular phonon background, similar to (TMTSF)$_2$PF$_6$ (reprinted figure with permission from European Physical Journal B - Condensed Matter and Complex Systems 7, J.-C. Lasjaunias \textit{et al.}, p. 541, 1999 \cite{Lasjaunias99}. Copyright (1999) from Springer Science and Business media).}
\label{fig8-14}
\end{center}
\end{figure}

The comparison between the specific heat of (TMTSF)$_2$AsF$_6$ and (TMTSF)$_2$PF$_6$ in a $C_{\rm P}/T^3$ plot versus $T$ is displayed in figure~\ref{fig8-14}. In (TMTSF)$_2$AsF$_6$, $C_{\rm P}/T^3$ exhibits two plateaux: the first one ascribed to the regular phonon contribution with $\beta$~= 15~mJ/mol\,K$^4$, a common value to (TMTSF)$_2$AsF$_6$ and (TMTSF)$_2$PF$_6$ (as shown in figure~\ref{fig8-5}) and a second one below 0.5~K with $\beta$ twice as large: $\beta^\ast\sim 34$~mJ/mol\,K$^4$. This contribution was possible to be observed in (TMTSF)$_2$AsF$_6$ only because the very small LEE contribution.

In section~\ref{sec7-3-2} it has been shown that bumps in the $C_{\rm P}/T^3$ contribution occur in all the Q-1D organic salts in the temperature range between 3--8~K. Low energy intramolecular modes were, then, involved for explaining these bumps. Therefore one may think that a phonon origin can be ruled out and that this low temperature $C_{\rm P}/T^3$ contribution around 0.2~K in (TMTSF)$_2$AsF$_6$ results from the phason mode.

Following the analysis performed in section~\ref{sec7-3-1} (eq.~(\ref{eq7-4})) one use a modified Debye mode with two cut-off frequencies: the lower one, $\nu_0$, corresponding to the phason gap given by the pinning frequency, the upper one being the ``phason Debye frequency", $\nu_\varphi$, which can be the amplitude mode frequency (ref.~\cite{Boriack78}). A good fit was reported in figure~3 in ref.~\cite{Lasjaunias99}, with $\nu_0\approx 0$ and $\nu_\varphi$~= 60~GHz~= 2.8~K. In the absence of pinned mode ($\nu_0$~= 0), the phason contribution to $C_{\rm P}$ appear as a step in excess to the phonon contribution in the $C/T^3$ versus $T$ plot, as experimentally shown in figure~\ref{fig8-14}. For (TMTSF)$_2$PF$_6$ the pinning frequency was found to be $\sim$~5--6~GHz at $T$~= 2~K \cite{Donovan94}. With such a pinning frequency the low-$T$ contribution  to $C_{\rm P}$ is suppressed and the overall phason contribution appears as a bump (see figure~\ref{fig7-15}). The impurity content in (TMTSF)$_2$AsF$_6$ is much less than in (TMTTF)$_2$PF$_6$ that yields a very low value of the pinning frequency and consequently the phason contribution to the specific heat may appear as a step in the $C_{\rm P}/T^3$ versus $T$ plot.

\subsection{Very low temperature energy relaxation}\label{sec8-2}

In section~\ref{sec7} the temperature dependence of the specific heat of quasi 1-D materials was presented in a large temperature range between 1--2~K up to $\sim 40$~K. This section is essentially devoted to thermodynamical properties at very low temperatures of C/S DW systems down to 70~mK which exhibit characteristics of glassy materials with slow non exponential heat relaxation indicating that $C_{\rm P}$ becomes strongly time-dependent \cite{Phillips96} as in disordered systems.

It is then necessary to precise the time-scale conditions of $C_{\rm P}$ measurements. In the usual heat pulse technique the heat capacity at ``short time" is determined from the response to a brief energy pulse from the exponential $\Delta T(t)$ decay. But as shown below, at temperatures below 1~K, deviations to the exponential relaxation develop progressively as a tail at long time. In addition, the thermal transient $\Delta T(t)$ depends on the duration of the heat supplied to the sample, $t_{\rm w}$ (named ``waiting time" and later ``pumping time"). The specific heat at equilibrium is obtained when  the time dependence of $C_{\rm P}$ is exhausted when $t_{\rm w}\geq t_{\rm eq}$. In the intermediate time scale the heat capacity is defined by integration of the total heat release through the heat link.

In addition, to ascertain that the LEE contribution and time dependence effects take their origin to the C/S superstructure, measurements of $C_{\rm P}$ in similar experiments as for C/S DW systems have been performed \cite{Lasjaunias02b} on a single crystal of Si. Crystalline Si exhibits low-$T$ mechanical and thermal transport properties due to structural defects \cite{Kleiman87,Liu98} which are treated as two-level systems (TLS) in glasses \cite{Phillips81,Hunklinger86}. Data of a undoped Si sample for different heat links corresponding rather well to the extreme values of $R_\ell$ measured in various experiments in Q1D materials give the same heat capacity \cite{Lasjaunias02b}.

\subsubsection{Low energy excitations}\label{sec8-1}

\begin{figure}
\begin{center}
\subfigure[o-TaS$_3$]{\label{fig8-1a}
\includegraphics[width=7.5cm]{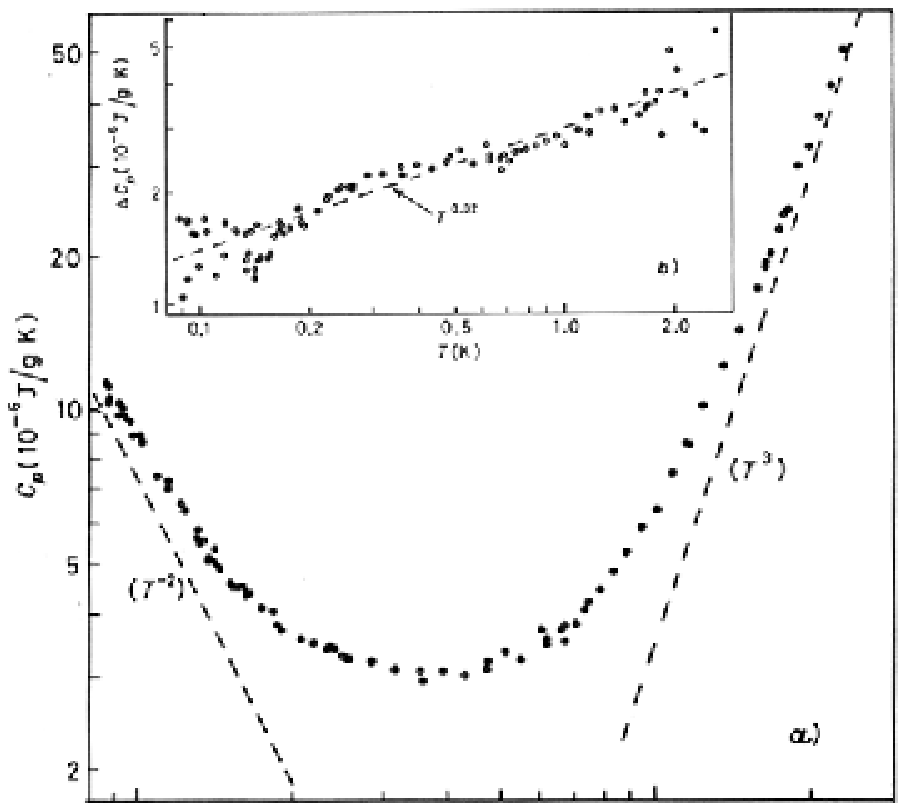}}
\subfigure[(TMTTF)$_2$Br]{\label{fig8-1b}
\includegraphics[width=5.75cm]{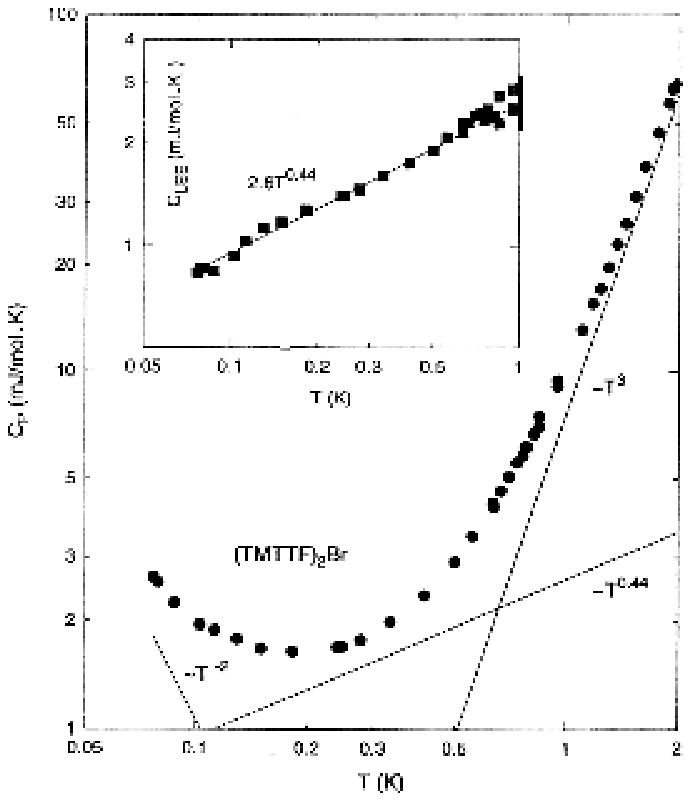}}
\caption{Low temperature specific heat, obtained by the short pulse technique, in a log-log plot of (a)~orthorhombic TaS$_3$ (reprinted figure with permission from K. Biljakovi\'c \textit{et al.}, EuroPhysics Letters 8, p. 771, 1989 \cite{Biljakovic89c}. Copyright (1989) from EdpSciences), (b)~(TMTTF)$_2$Br (reprinted figure with permission from J.-C. Lasjaunias, Journal de Physique I (France) 7, p. 1417, 1997 \cite{Lasjaunias97}. Copyright (1997) from EdpSciences). The phonon $T^3$ contribution, the ``hyperfine" $T^{-2}$ term are indicated by lines. Insets show the respective low energy excitation (LEE) $T^\nu$ contribution.}
\label{fig8-1}
\end{center}
\end{figure}

As prototypes of CDW and SDW systems, the temperature dependence of $C_{\rm P}$ of o-TaS$_3$ and (TMTTF)$_2$Br are shown in figure~\ref{fig8-1}(a) and \ref{fig8-1}(b). Below 2~K, $C_{\rm P}$ at ``short time" deviates progressively from a $T^3$ law on decreasing $T$, reflecting the additional contribution of low energy excitations (LEEs).

Short time $C_{\rm P}$ data are well described, below 3~K to the lowest temperature, by:
\begin{equation}
C_{\rm P}=C_hT^{-2}+AT^\nu+\beta T^3.
\label{eq8-1}
\end{equation}

The two first terms are the ``hyperfine" (tail of a Schottky anomaly) and the power law contributions of the LEEs, the third one being the usual phonon contribution. For (TMTTF)$_2$Br, the best fit gives $\nu$~= 0.44 as shown in the inset of figure~\ref{fig8-1}(b), and $\nu$~= 0.32 for o-TaS$_3$ (inset of figure~\ref{fig8-1}(a)).

\begin{figure}
\begin{center}
\includegraphics[width=6.5cm]{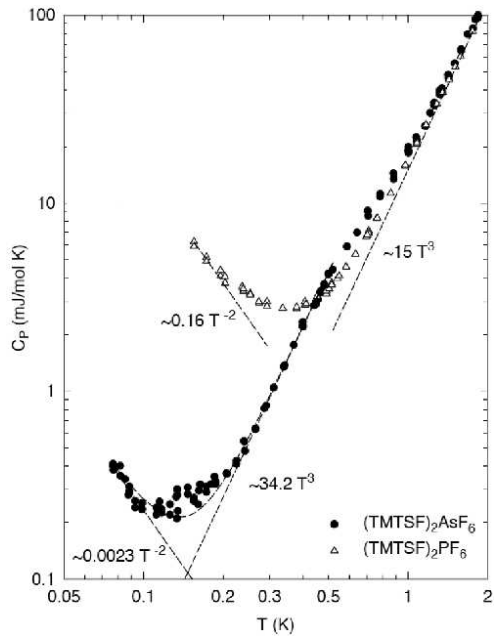}
\caption{Temperature dependence of the low temperature specific heat on the short time scale below 2~K in a log-log plot of (TMTSF)$_2$PF$_6$ and (TMTSF)$_2$AsF$_6$ (reprinted figure with permission from European Physical Journal B - Condensed Matter and Complex Systems 7, J.-C. Lasjaunias \textit{et al.}, p. 541, 1999 \cite{Lasjaunias99}. Copyright (1999) from Springer Science and Business media).}
\label{fig8-2}
\end{center}
\end{figure}

Comparison between $C_{\rm P}$ of (TMTSF)$_2$PF$_6$ and (TMTSF)$_2$AsF$_6$ on the short-time scale below 2~K in a log-log plot is presented in figure~\ref{fig8-2} \cite{Lasjaunias99}. It is seen that $C_{\rm P}$ for (TMTSF)$_2$AsF$_6$ is considerably smaller than for (TMTSF)$_2$PF$_6$, that can be ascribed to a large reduction of  the LEE contribution. While for (TMTSF)$_2$PF$_6$, $C_{\rm P}$ data are analysed with the three contributions of eq.~(\ref{eq8-1}) with a $AT^\nu$ contribution with $\nu\simeq$~1.2, for (TMTSF)$_2$AsF$_6$ $C_{\rm P}$ follows a cubic variation down to 0.25~K, which rapidly turn to a $T^{-2}$ regime obeyed between 100 and 75~mK; that means that the amplitude of the $T^\nu$ contribution is negligible or extremely weak.

\begin{figure}[h!]
\begin{center}
\subfigure[]{\label{fig8-3a}
\includegraphics[width=5cm]{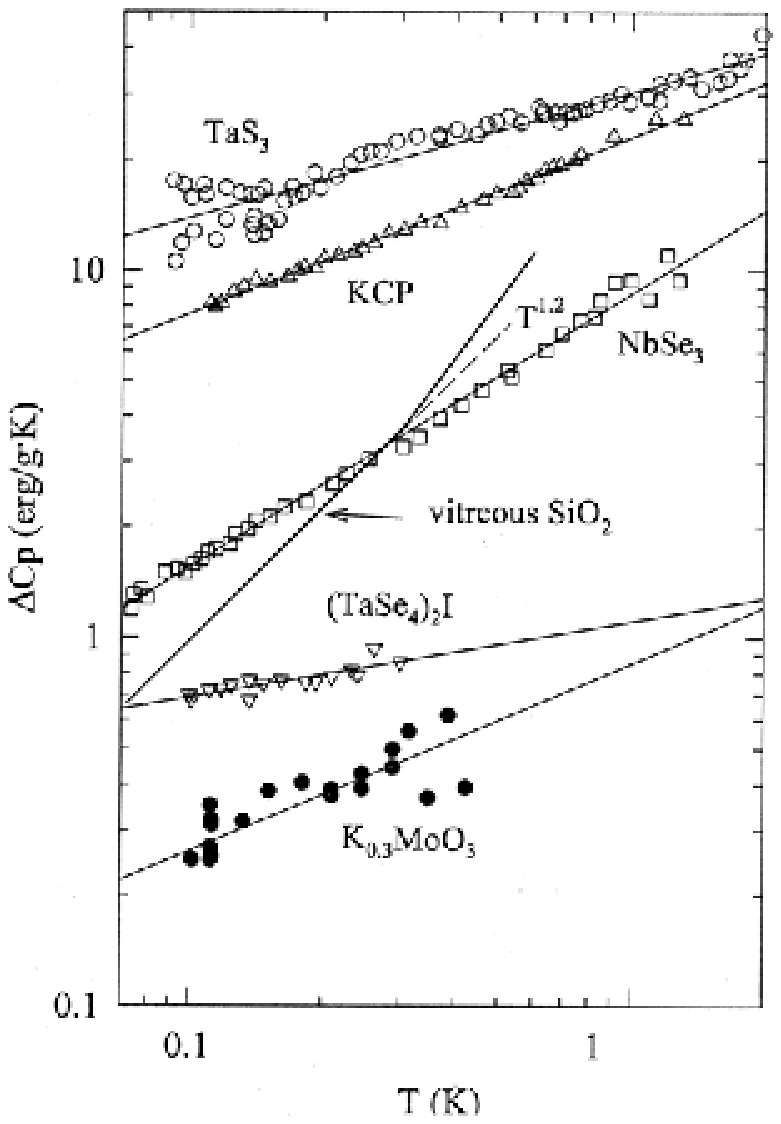}}
\subfigure[]{\label{fig8-3b}
\includegraphics[width=6.5cm]{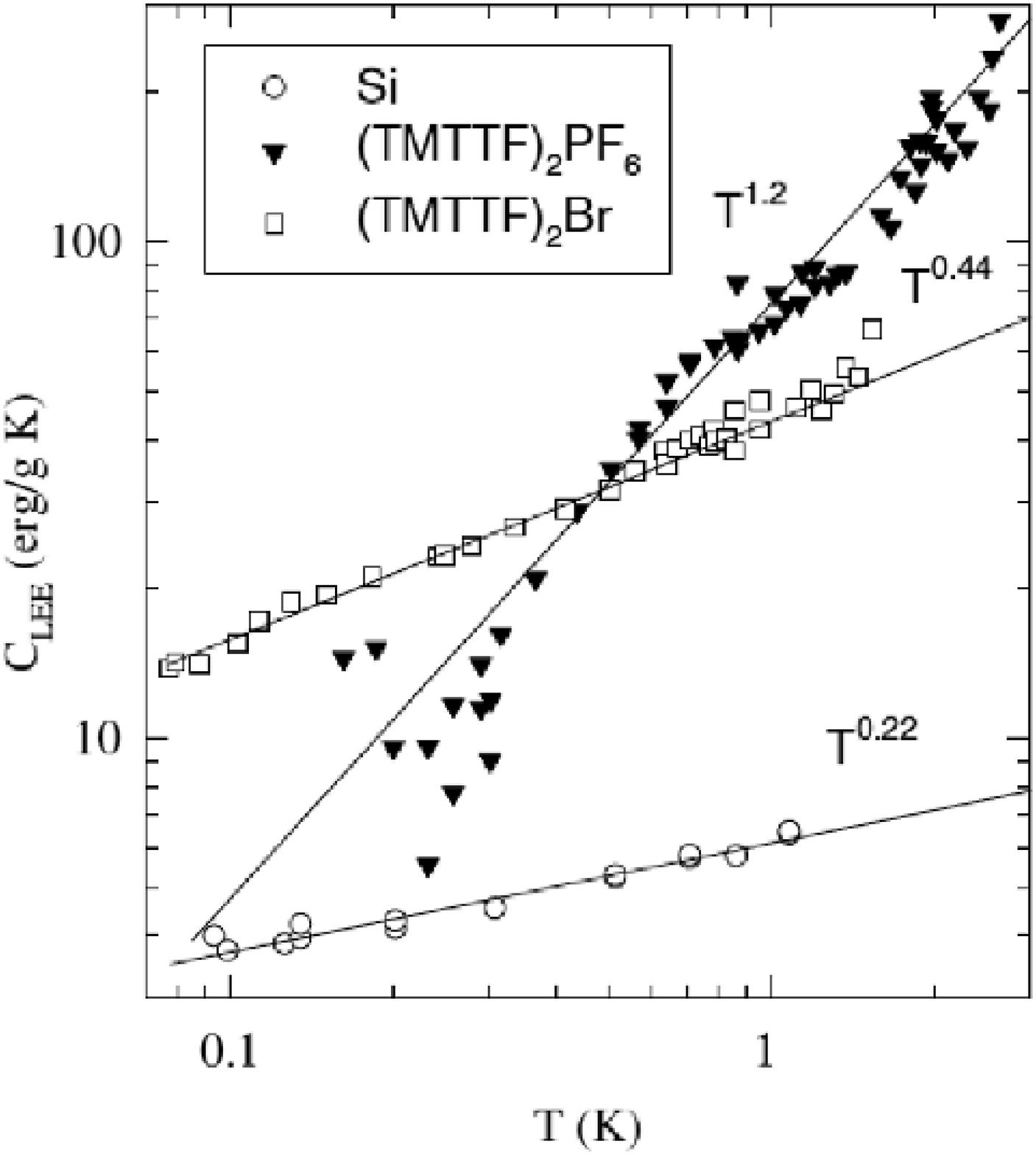}}
\caption{Residual specific heat $\Delta C(T)$ or low energy excitation $C_{\rm LEE}$ excitations exhibiting a power law dependence $T^\nu$ after subtraction of the phonon $T^3$ term and the ``hyperfine" $T^{-2}$ term (a)~for various quasi 1D materials in the CDW state (reprinted figure with permission from European Physical Journal B - Condensed Matter and Complex Systems 24, J. Odin \textit{et al.}, p. 315, 2001 \cite{Odin01}. Copyright (2001) from Springer Science and Business media), (b)~for two Fabre salts (reprinted figure with permission from J.-C. Lasjaunias \textit{et al.}, Journal of Physics: Condensed Matter 14, p. 8583, 2002 \cite{Lasjaunias02b}. Copyright (2002) by the Institute of Physics). For comparison data for vitreous silica are reported in figure~\ref{fig8-3}(a), and for crystalline silicon in figure~\ref{fig8-3}(b).}
\label{fig8-3}
\end{center}
\end{figure}

The contribution of LEE to the short-time scale of the residual specific heat --the quasi-linear term-- $\Delta C_{\rm P}$ or $C_{\rm LEE}$ (after subtraction of the $T^{-2}$ and $T^3$ terms) is plotted in figure~\ref{fig8-3}(a) for CDW systems and in figure~\ref{fig8-3}(b) for organic Fabre salts. Heat capacity has also been measured on Nb-doped (0.5\% nominal) o-TaS$_3$ \cite{Biljakovic03} and on titanium-doped (TaSe$_4$)$_2$I \cite{Biljakovic89b}. It was found that the amplitude of the $AT^\nu$ LEE contribution was increased by doping, but that the exponent $\nu$ was not depending of doping (at least in the small concentration range considered).

\subsubsection{Sub-SDW phase transitions}\label{sec8-3}

Anomalies in the $^1$H-NMR relaxation rate, well below the SDW transition, in (TMTSF)$_2$PF$_6$, (TMTSF)$_2$ClO$_4$ with different SDW temperature transition depending on the quenching conditions have been reported \cite{Takahashi86,Nomura93,Nomura95}. It was proposed the appearance of different SDW phases: SDW$_1$ between $T_{\rm SDW}\simeq 12$~K (the main metal-semiconducting phase transition) and 3.5~K, SDW$_2$ between 3.5~K and $\sim 1.9$~K and SDW$_3$ at low $T$ below 1.9~K. There are also some changes in the collective transport in the same $T$ range. Thus, the ratio $\sigma^\prime_{\rm SDW}/\sigma_n$ between the SDW conductivity $\sigma^\prime_{\rm SDW}$ normalised to the normal conductivity $\sigma_n$ drops shortly \cite{Kriza91d} at a temperature $T^\ast$ such $T_{\rm SDW}/T^\ast\sim 3.5$, which may indicate a different qualitative sliding mode \cite{Mihaly91}. It was proposed \cite{Wonnenberger91} that an incommensurate-commensurate transition with $N$~= 4 occurs at $T^\ast$, but no significant change in $^1$H-NMR lineshapes were observed which are very sensitive to the ratio $Q_{\rm SDW}/Q_{\rm lattice}$.

\begin{figure}
\begin{center}
\includegraphics[width=7cm]{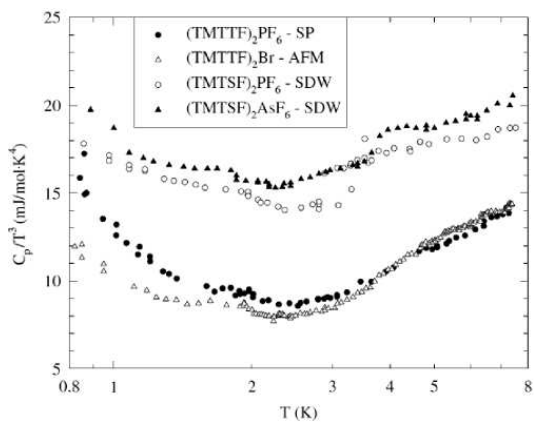}
\caption{Temperature dependence of the specific heat defined on the short time scale divided by $T^3$ between 0.8~K and 8~K of Bechgaard (TMTSF)$_2$PF$_6$ and (TMTSF)$_2$AsF$_6$ salts (SDW ground state) and Fabre (TMTTF)$_2$PF$_6$ (spin-Peierls ground state) and (TMTTF)$_2$Br (antiferromagnetic  ground state) salts (reprinted figure with permission from J.-C. Lasjaunias \textit{et al.}, Journal of Physics: Condensed Matter 14, p. 8583, 2002 \cite{Lasjaunias02b}. Copyright (2002) by the Institute of Physics).}
\label{fig8-5}
\end{center}
\end{figure}

What is the nature of the transition at $T^\ast$ and is it coupled to a lattice instability ? The best way to represent a lattice contribution is a $C_{\rm P}/T^3$ plot versus $T$. Comparison between two selenides (TMTSF)$_2$PF$_6$ and (TMTSF)$_2$AsF$_6$, and two sulphides (TMTTF)$_2$PF$_6$ and (TMTTF)Br is shown in figure~\ref{fig8-5}.

The lattice contribution follows a $T^3$ law only in a small $T$-interval below 3~K. There is a strong similarity between both selenide salts on one hand and both sulphide ones on the other. Therefore it can be concluded that the lattice term is mainly determined by the (TMTSF) or (TMTTF) molecules forming the stacks, the role of anions being minor. Analysis of $C_{\rm P}$ yields a common $\beta$~= $C_{\rm P}/T^3$~= 14.5~mJ\,mol$^{-1}$K$^{-4}$ for both (TMTSF)$_2$PF$_6$ and (TMTSF)$_2$AsF$_6$ and $\beta$~= 7.8~mJ\,mol$^{-1}$K$^{-4}$ for both sulphides.

From figure~\ref{fig8-5} one can see that a discontinuity in specific heat is observed at very likely the same temperature $\sim 1.9$~K for the selenides and sulphides salts. Then, the sub-phase transition detected at this temperature by NMR is \cite{Takahashi86} appears to not be related to any (electronic) superstructure induced in these salts but probably results from some (unknown) re-arrangement in the stacks along the chains.

For $T$ higher than 2.5~K, there is a rapid increase of $C/T^3$ up to 7~K for sulphides salts without the sharp discontinuity detected in (TMTSF)$_2$PF$_6$ around 3--3.5~K.

\medskip
\noindent \textit{7.3.2.a. Glassy-like transition at 3.5~K in (TMTSF)$_2$PF$_6$}\label{sec7-3-2.a}
\medskip

\begin{figure}
\begin{center}
\includegraphics[width=8.5cm]{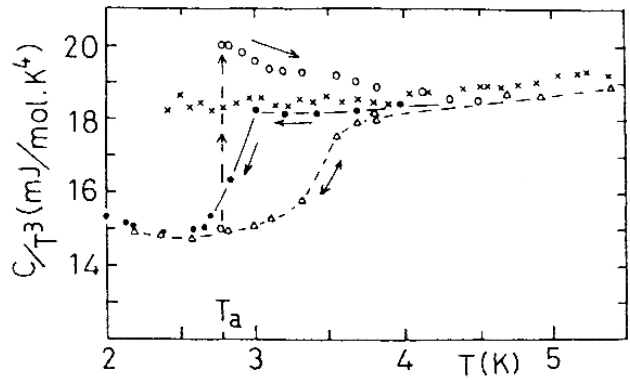}
\caption{Temperature dependence of the total specific heat of (TMTSF)$_2$PF$_6$ in a $C_{\rm P}/T^3$ plot in the vicinity of the glass transition, obtained in very different kinetic conditions: Data ($\times$) are obtained with the quasi-adiabatic technique. Other data are obtained with the transient-heat-pulse technique: ($\vartriangle$) correspond to a regular heating or cooling at a rate of 0.5~Kh$^{-1}$ (reversible cycle) between 2.9 and 4.5~K for an unannealed sample. After a long annealing at $T_a$~= 2.77~K for 40h, a large endothermic maximum appears on reheating ({\large$\circ$}), whereas on subsequent cooling ({\large$\bullet$}), $C_{\rm P}$ follows the equilibrium value down to 3~K (reprinted figure with permission from J.-C. Lasjaunias \textit{et al.}, Physical Review Letters 72, p. 1283, 1994 \cite{Lasjaunias94}. Copyright (1994) by the American Physical Society).}
\label{fig8-6}
\end{center}
\end{figure}

Indeed, for (TMTSF)$_2$PF$_6$, a well-defined jump of $C_{\rm P}$ with a characteristic hysteretic behaviour has been measured in the vicinity of 3.5~K, in relation with a strong dependence on previous thermal history, allowed to ascribe this discontinuity to a glassy-like transition \cite{Odin94,Lasjaunias94}. The $C_{\rm P}/T^3$ variation of the total specific heat of (TMTSF)$_2$PF$_6$ is drawn in figure~\ref{fig8-6} 
under different time scale conditions. Data ($\vartriangle$) and ($\times$) correspond to two extreme kinetic conditions: the former obtained with the transient-short time heat  pulse technique which shows a reversible cycle obtained between 1.9~K and 6.3~K when the data are taken regularly by heating or by cooling at a rate of 0.5~K/h; the latter obtained by a quasi-adiabatic (QA) technique \cite{Odin94}. In this technique, after cooling down to the lowest temperature, the sample was isolated from the thermal bath and heated adiabatically step by step. While an excellent agreement between the two techniques (transient pulse and QA) is obtained between 4 and 7~K, the characteristic jump which occurs around 3--3.5~K with the transient heat pulse technique, does not appear down to 2.5~K with the QA technique.

Hysteresis loops occur as soon as isothermal treatments (or stabilisations in glass terminology) are performed. After annealing at any temperature around 3.5~K, the specific heat spontaneously increases during the stabilisation towards the ``supercooled" equilibrium value. Open circles in figure~\ref{fig8-6} correspond to heating after a long annealing at $T_a$~= 2.77~K for 40~h, exhibiting thus a large endothermic effect, often observed in DSC for a glass relaxed very close to the structural equilibrium. This overshoot on heating results in a highly relaxed state; indeed, on the cooling part of the cycle, $C_{\rm P}$ follows the equilibrium value $C_{\rm P_e}$ down to $T$~= 3.0~K (dark circles) below which the system freezes rapidly into the glassy state, as $C_{\rm P}$ drops to the lower vibrational value \cite{Lasjaunias94,Odin94}.

With respect to glass transition of supercooled liquids, the jump in specific  heat represents the freezing-in of the configurational degrees of freedom over the experimental time scale \cite{Jones71}. A characteristic property of the glass transition is its strong sensitivity on the thermal history of the system and of the kinetics of the measurement as demonstrated in figure~\ref{fig8-6} by the isothermal annealing process performed in the vicinity of the glass transition. It is also known for glasses that the temperature of the jump is sensitive to the frequency of the measurement. Compared to the dc adiabatic technique, the pulse heat technique is similar to a low frequency ac technique (0.1--0.001~Hz).

Above $T_g$ all degrees of freedom --vibrational and configurational-- can be excited during the experimental time span. This corresponds to the specific heat $C_{\rm P_e}$ of the supercooled liquid in the thermodynamical equilibrium. For (TMTSF)$_2$PF$_6$ $C_{\rm P_e}$ varies very closely to a cubic law up to 8~K ($C_{\rm P_e}$~= $\beta T^3$ with $\beta$~= 18.5--19~mJ\,mol$^{-1}$K$^{-4}$). Below $T_g$, there remain only the vibrational contribution of the ``glass", $C_{\rm P_g}$, which corresponds here to the lattice term following the Debye law: $C_{\rm P_e}$~= $\alpha T^3$ with $\alpha$~= 15~mJ\,mol$^{-1}$K$^{-4}$. It is worth to note that $C_{\rm P_e}$ obeys a cubic law like the vibrational specific heat. In comparison with usual glass forming systems where $T_g$ occurs at several 100~K, in the present case $T_g$ appears at very low $T$, in the liquid helium $T$-range. In this region, $C_{\rm P}$ is dominated by the lattice specific heat ($\alpha T^3$) and the transition appears as a jump in a $C_{\rm P}/T^3$ diagram \cite{Lasjaunias94}.

As  seen in section~\ref{sec7-3}, the SDW transition appears as a very small bump superimposed to the continuous background. The relative amplitude $\Delta C_{\rm P}/C_{\rm P}$ at $T_{\rm SDW}$ is very small, at maximum $\sim 1.5\%$ of $C_{\rm P}$ \cite{Odin94,Coroneus93}. From this jump, the electronic entropy in the condensate state below $T_{\rm SDW}$ was estimated to be $\sim 25$~mJ/mol\,K$^2$, giving the electronic entropy $\gamma T_{\rm SDW}\sim 300$~mJ/mol\,K.

In ref.~\cite{Odin94}, identifying the supercooled liquid with the SDW entity, the value of the configurational entropy of the ``supercooled" state was estimated (it was supposed that the configurational contribution ceases at $T_{\rm SDW}$) to be at least five times larger than the electronic one. This analysis raised the question of the role of the electron-phonon interaction in the SDW stabilisation  of (TMTSF)$_2$PF$_6$.

Contrary to (TMTSF)$_2$PF$_6$, (TMTSF)$_2$AsF$_6$ does not show any annealing effects or hysteretic behaviour. However it exhibits important dynamical effects \cite{Lasjaunias99}.

\medskip
\noindent \textit{7.3.2.b. Coexistence of CDW with SDW}
\medskip

X-ray diffuse scattering experiments are another way for detecting a possible coupling between the SDW and the lattice. $2k_{\rm F}$ and $4k_{\rm F}$ superlattice  spots were, indeed, found in the SDW state of (TMTSF)$_2$PF$_6$ in the restricted $T$ range between 13 and 11~K \cite{Pouget96,Pouget97}. These satellite reflections were only observed near the origin of the reciprocal lattice suggesting a purely electronic origin of the CDW without  involving lattice distortion. These X-ray diffuse scattering experiments have been extended \cite{Kagoshima99} at lower temperature down to 1.6~K, performed on (TMTSF)$_2$PF$_6$ as well as on (TMTSF)$_2$AsF$_6$. Satellite reflections were observed in (TMTSF)$_2$PF$_6$ only below $T_{\rm SDW}$, confirming the coexistence of CDW and SDW in the SDW phase. The $2k_{\rm F}$ CDW spots were found to agree well with the $Q$ vector evaluated by NMR \cite{Takahashi86}. But the more intriguing result was the disappearance of the $2k_{\rm F}$ and $4k_{\rm F}$ satellites below 3.5~K (or at least a strong reduction of their intensity, satellites being not detectable below 2~K). Moreover no satellite spots, at any temperature, were detected in (TMTSF)$_2$AsF$_6$, except two very weak, possibly $4k_{\rm F}$, spots. That might be related to the intrinsic better quality of (TMTSF)$_2$AsF$_6$ crystals with regards to (TMTSF)$_2$PF$_6$.

This CDW-SDW coexistence has raised an intense theoretical activity, sometimes controversial, using mean field calculations  \cite{Kobayashi98}, renormalisation group \cite{Yoshioka01} and numerical studies \cite{Mazumdar99}. It was shown that it is essential for the coexistence between CDW and SDW to consider the internal degrees of freedom in the dimers forming the (TMTSF)$_2$PF$_6$ unit cell. A mean field calculation was performed \cite{Kobayashi98} using a one-dimensional extended Hubbard model with a 1/4 filled band, taking into account the effect of the next-nearest-neighbour Coulomb repulsion, $V_2$. The effect of dimerisation due to anion chains results in unequal value of the intra- and interdimer transfer integral, $t_a$ and $t_b$, and of the intra- and interdimer neighbour Coulomb interaction, $V_a$ and $V_b$. The on-site Coulomb energy is the same on each of the molecule forming the dimer. A phase diagram was derived \cite{Kobayashi98} from this model where, depending on the parameters $t_a$, $t_b$, $V_a$, $V_b$, $U$, $V_2$, there is, indeed, coexistence of the $2k_{\rm F}$ SDW --$4k_{\rm F}$ CDW as well as $2k_{\rm F}$ SDW-- $2k_{\rm F}$ CDW. It appears thus that dimerisation ($t_a\neq t_b$) is necessary to observe the $2k_{\rm F}$ SDW and $2k_{\rm F}$ CDW coexistent phases. Similarly, if $V_a$~= $V_b$, CDW and SDW coexistence will require a considerable large $V_2$ value.

However, none of these models can interpret the disappearance of $2k_{\rm F}$ and $4k_{\rm F}$ CDW below 3.5~K, as well as the difference between (TMTSF)$_2$PF$_6$ and (TMTSF)$_2$AsF$_6$. There are, however, no doubt that the appearance and the disappearance of the CDW in (TMTSF)$_2$PF$_6$ as well as the absence of satellite spots in (TMTSF)$_2$AsF$_6$ are deeply connected with the nature of the glassy-like transition observed in thermodynamical measurements.

\subsubsection{Non-equilibrium dynamics}\label{sec8-4}

In the conditions where the contribution of the LEE to the specific heat dominates the phonon one, a highly non exponential relaxation is observed in almost all the C/S DW systems. It was also shown that $C_{\rm P}$ depends upon the time delivery of energy (or ``waiting" time $t_w$) \cite{Biljakovic89a}. This expression borrowed from non-ergodic phenomena in spin glass, however, recovering a different experimental situation: in spin glasses, the system is quenched under a small magnetic field from a temperature above $T_g$ at a temperature below $T_g$, kept during a variable time $t_w$ after which the field is switched off \cite{Lundgren83,Campbell90,Vincent72}; in C/S DW the system, at equilibrium at $T_0$, is slightly heated at $T_1$ such $\Delta T/T_0$~= $T_1-T_0/T_0\sim$ a few \% and this thermal difference is kept constant during $t_w$ where the power supplied is switched off.

In o-TaS$_3$, the logarithmic derivative of $\Delta T(t)$ shows a peak in the relaxation rate at $\tau_p$. The distribution of relaxation rate is smeared out over a few orders of magnitude in time, reflecting the wide spectrum of relaxation times $\tau$ and its correlation with $t_w$. $\tau_p$ increases with $t_w$ but saturates at a maximum value $\tau_{\rm max}$ for longer $t_w$ indicating that the system has reached its thermodynamical equilibrium \cite{Biljakovic91a} (thus, for o-TaS$_3$, $\tau_{\rm max}$ is of the order of $\sim 10$~h).

\begin{figure}
\begin{center}
\subfigure[]{\label{fig8-9a}
\includegraphics[width=6.5cm]{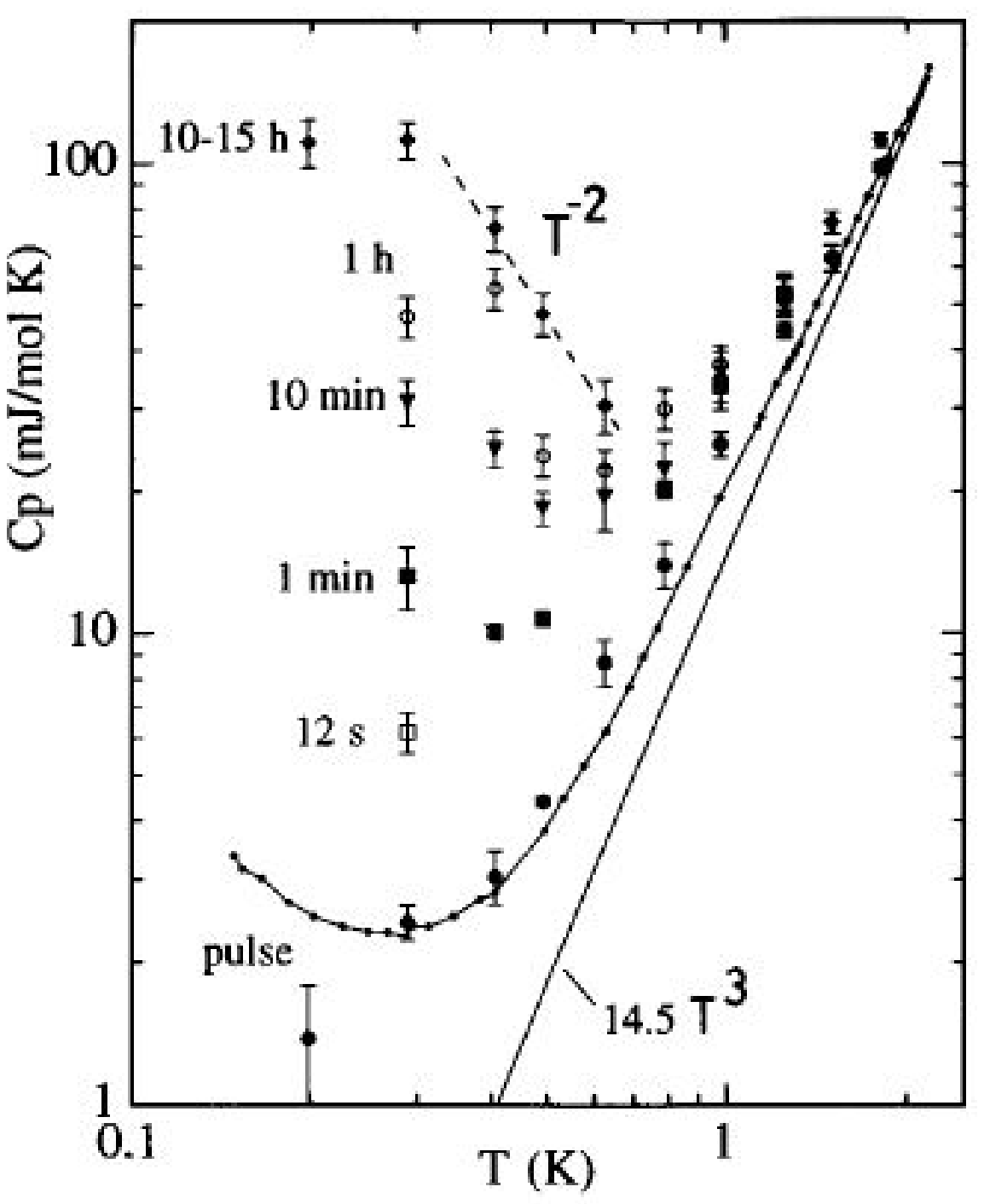}}
\subfigure[]{\label{fig8-9b}
\includegraphics[width=6cm]{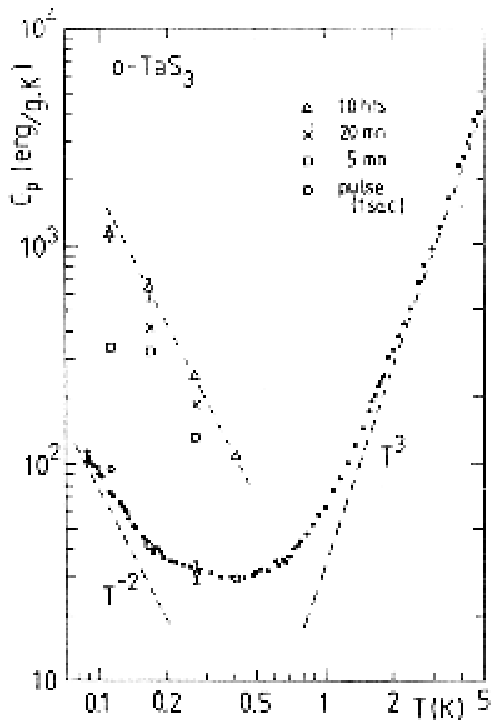}}
\caption{Dependence of the specific heat on the duration of energy delivery ($t_w$) (a)~for (TMTSF)$_2$PF$_6$ (reprinted figure with permission from J.-C. Lasjaunias \textit{et al.}, Physical Review B 53, p. 7699, 1996 \cite{Lasjaunias96}. Copyright (1996) by the American Physical Society), (b)~for o-TaS$_3$ (reprinted figure with permission from P. Monceau \textit{et al.}, Physica B: Condensed Matter 194-196, p. 403, 1994 \cite{Monceau94}. Copyright (1994) with permission from Elsevier). $C_{\rm P}$ is calculated by the integration of total energy release through the heat link.}
\label{fig8-9}
\end{center}
\end{figure}

Figure~\ref{fig8-9} shows the temperature dependence of the specific heat determined by the integration technique, as explained above, of o-TaS$_3$ \cite{Monceau94} and (TMTSF)$_2$PF$_6$ \cite{Lasjaunias96} with different duration of energy delivery $t_w$. The amplitude of the Schottky anomaly tail in $T^{-2}$ is largely increased for large $t_w$. In a time scale of $\sim 10^4$~s, at $T$~= 0.2~K, the time dependence effect on $C_{\rm P}$ for both the CDW o-TaS$_3$ compound or the SDW (TMTTF)$_2$PF$_6$ salt is similar and tremendously large. At or nearby the thermodynamical equilibrium, one can see that $C_{\rm P}$ for $t_w\sim 10$~h is $\sim 70$ times larger than the value defined on a short-time scale.

\begin{figure}
\begin{center}
\includegraphics[width=8cm]{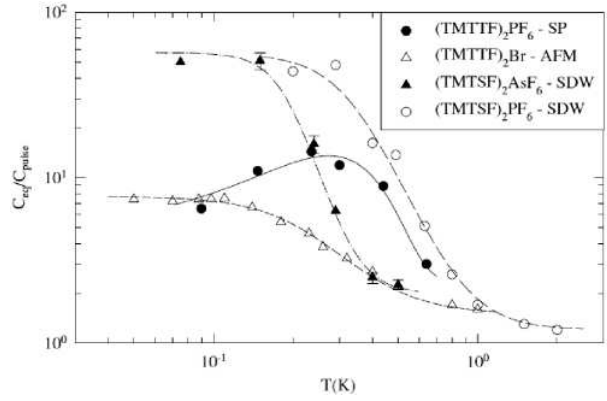}
\caption{Comparison of the amplitude of the time dependence of the specific heat in Bechgaard (TMTSF)$_2$PF$_6$ and (TMTSF)$_2$AsF$_6$ salts and Fabre (TMTTF)$_2$Br and (TMTTF)$_2$PF$_6$ salts measured as the ratio between the specific heat obtained under equilibrium conditions, $C_{\rm eq}$ and the short-time scale specific heat, $C_{\rm P}$ (reprinted figure with permission from J.-C. Lasjaunias \textit{et al.}, Journal of Physics: Condensed Matter 14, p. 8583, 2002 \cite{Lasjaunias02b}. Copyright (2002) by the Institute of Physics).}
\label{fig8-10}
\end{center}
\end{figure}

Similarly to (TMTSF)$_2$PF$_6$, the specific heat of (TMTTF)$_2$PF$_6$ and (TMTTF)$_2$Br are strongly dependent on the time scale \cite{Lasjaunias02a,Lasjaunias02b}. It was also noted that $C_{\rm P}$ for (TMTTF)$_2$PF$_6$ is much larger than that for (TMTTF)$_2$Br for similar dynamical conditions. For comparison, the amplitude of the time dependence of $C_{\rm P}$ --measured as the ratio between $C_{\rm eq}$ at the thermodynamical equilibrium and the short-time scale (pulse) $C_{\rm P}$-- is plotted in figure~\ref{fig8-10} in a log-log plot for Bechgaard salts: (TMTSF)$_2$PF$_6$ and (TMTSF)$_2$AsF$_6$ and Fabre salts: (TMTTF)$_2$PF$_6$ and (TMTTF)$_2$Br.

\begin{figure}
\begin{center}
\includegraphics[width=7.5cm]{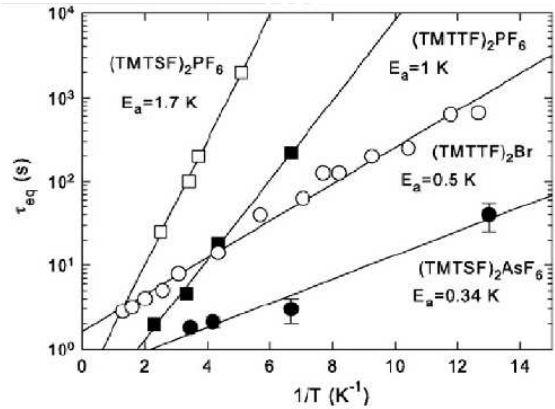}
\caption{Variation of the characteristic heat relaxation times versus $1/T$ in the equilibrium thermodynamical conditions for temperatures below 1~K for Bechgaard (TMTSF)$_2$PF$_6$ and (TMTSF)$_2$AsF$_6$ salts and Fabre (TMTTF)$_2$PF$_6$ and (TMTTF)$_2$Br salts (reprinted figure with permission from Synthetic Metals 159, K. Biljakovi\'c \textit{et al.}, p. 2402, 2009 \cite{Biljakovic09}. Copyright (2009) with permission from Elsevier).}
\label{fig8-11}
\end{center}
\end{figure}

The characteristic relaxation times, $\tau_{\rm max}$, in the thermodynamical equilibrium below 1~K are plotted in figure~\ref{fig8-11} for the same four organic salts. It is shown that $\tau_{\rm max}$ varies as $\exp(-E_a/kT)$ with $E_a$ in the range of 1~K or below. It is worth to note that while for (TMTSF)$_2$PF$_6$ thermal equilibrium will be reached at 150~mK after a few days, it will take only 1~mn for (TMTSF)$_2$AsF$_6$. That reflects again the strong difference between very ``similar" compounds as far as the low $T$ properties are concerned, very likely because a strong difference in the defect concentration.

\medskip
\noindent \textit{7.3.3.a. Bimodal energy relaxation in commensurate {\rm\bf (TMTTF)$_2$PF$_6$}}
\medskip

\begin{figure}
\begin{center}
\includegraphics[width=6.5cm]{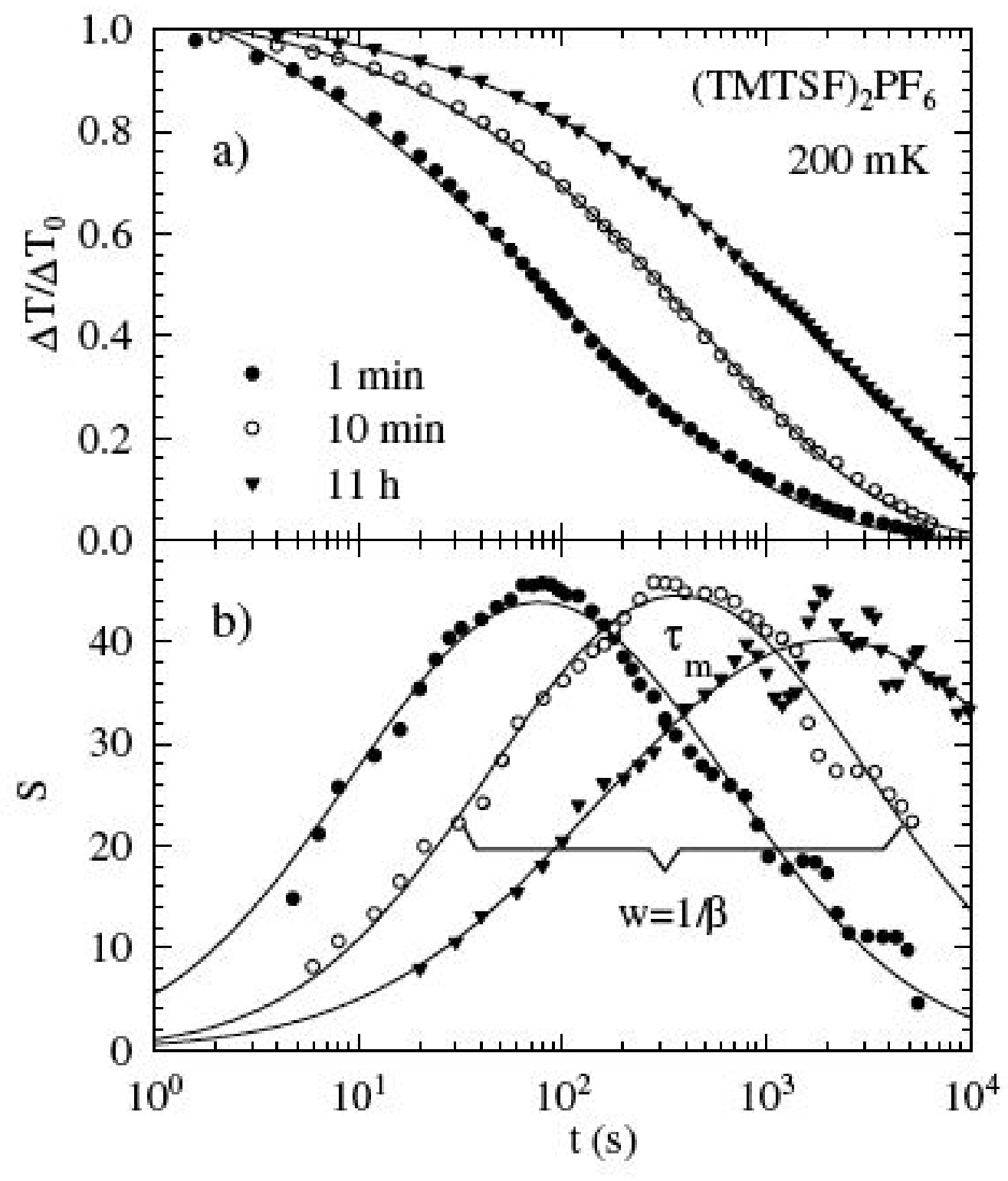}
\includegraphics[width=6.5cm]{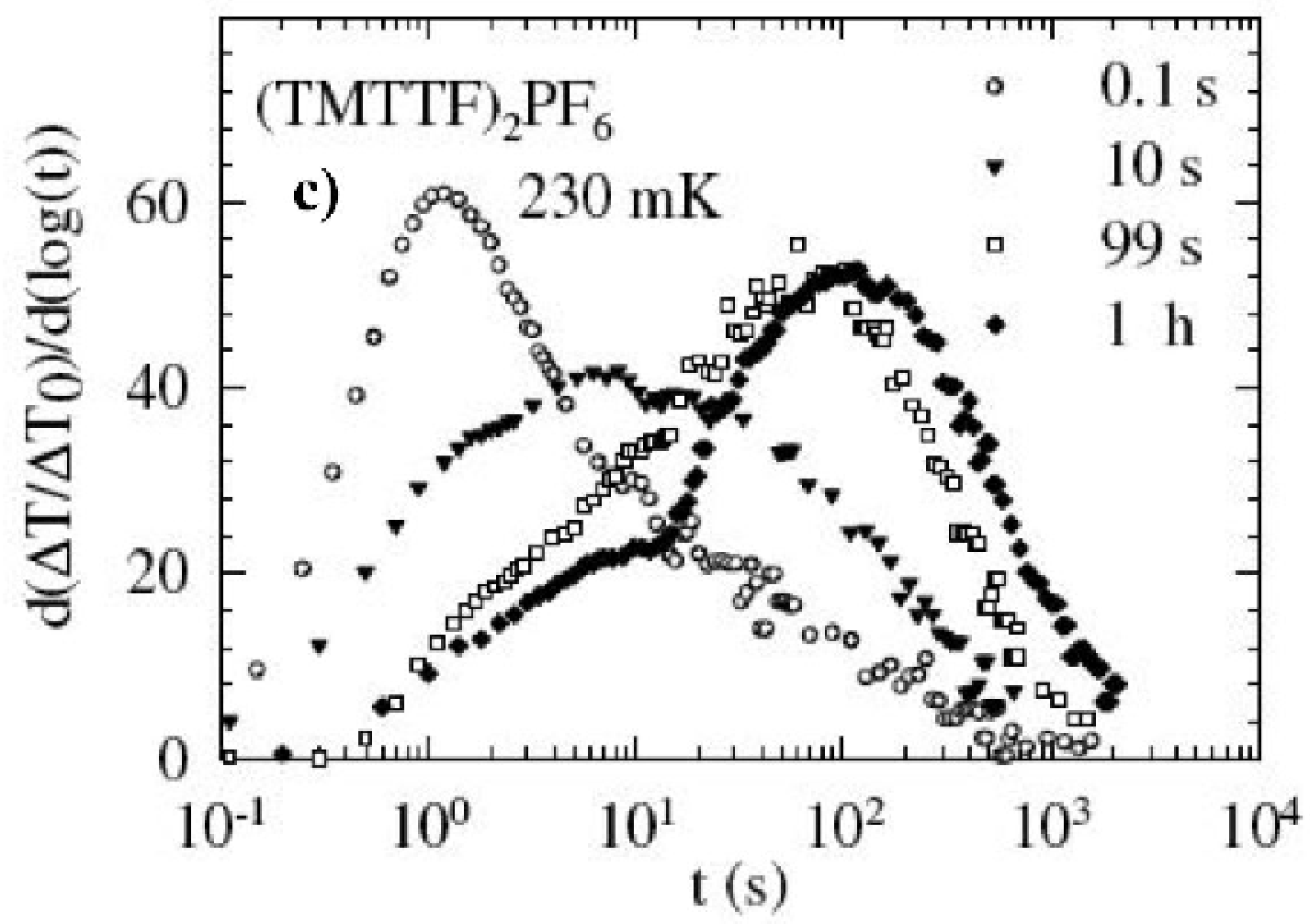}
\caption{a) and b) Homogeneous broadening and shift of the heat relaxation with increasing the pumping time, $t_w$, for (TMTSF)$_2$PF$_6$. a)~Relaxation $\Delta T/\Delta T_0$ versus $\log t$; curves are fits to a stretched  exponential $\exp(t/\tau)^\beta$, b)~Variation of the relaxation rate $S(t)$~= ${\rm d}(\Delta T/\Delta T_0)/{\rm d}(\ln t)$ fitted to a Gaussian function $G(\ln t)$ with a width $w$~= $1/\beta$. c)~bimodal relaxation with redistribution of the spectral weight and saturation for $t_w\sim 10^2$~s for the spin-Peierls Fabre (TMTTF)$_2$PF$_6$ salt (reprinted figure with permission from J.-C. Lasjaunias, Physical Review Letters 94, p. 245701, 2005 \cite{Lasjaunias05a}. Copyright (2005) by the American Physical Society).}
\label{fig8-12}
\end{center}
\end{figure}

The time-dependent non exponential energy relaxation exhibits quite different properties according to the nature of the modulated ground state \cite{Lasjaunias05a,Lasjaunias05b,Lasjaunias05c}: figure~\ref{fig8-12}(a) shows the time dependence of the relaxation $\Delta T/\Delta T_0$ for (TMTSF)$_2$PF$_6$ at 200~mK. This relaxation can be satisfactory fitted with a phenomenological stretched-exponential function  $\frac{\Delta T(t)}{\Delta T_0}=\exp(-t/\tau)^\beta$, with $\beta$ the stretching parameter. In the high $T$ limit the relaxation recovers an exponential decay, such as $\beta(T)$ ($T>1$~K)~= 1. However $\beta(T)$ decreases when $T$ is decreased, leading to a very slow relaxation. A better characterisation of the non-exponential relaxation can be obtained by the relaxation rate: $S(t)=\frac{{\rm d}(\Delta T/\Delta T_0)}{{\rm d}\ln t}$.

Noting that \cite{Lasjaunias05a} the logarithmic derivative of a stretched exponential is a Gaussian centred on $\ln\tau_m$ and with a width $w\sim 1/\beta$, the width $w$ can be extracted directly from $S(t)$ curves as shown in figure~\ref{fig8-12}(b) (width at half height).

However, in the case of (TMTTF)$_2$PF$_6$ with a commensurate spin-Peierls ground state, while still non-exponential, the heat relaxation is quite different. Instead of the homogeneous broadening and shift of the $S(t)$ spectra in (TMTSF)$_2$PF$_6$ with $t_w$, relaxation times in (TMTTF)$_2$PF$_6$ exhibit ``discrete bands" with a narrower distribution and smaller time for reaching thermal equilibrium (a few tens of mn to be compared with days in the selenium salt). Figure~\ref{fig8-12}(c) shows a bimodal redistribution of the relaxation spectrum indicating a transfer between two different relaxation processes when the ``waiting" or pumping time is increased: a contribution of fast modes dominant in dynamics at short $t_w$ while the relative weight of slow modes takes the advantage at large $t_w$ \cite{Lasjaunias05a}.

Concerning dynamics of heat relaxation, the CDW o-TaS$_3$ shares the same properties than the SDW (TMTSF)$_2$PF$_6$; there are both incommensurate systems. On the other hand, (TMTTF)$_2$PF$_6$ and the commensurate antiferromagnet (TMTTF)$_2$Br behave similarly with a multimodal dynamics.

\medskip
\noindent \textit{7.3.3.b. Collective dynamics in a strong pinning model}
\medskip

As shown in previous sections, metastability in C/S DW systems results from the compromise/competition between elastic and pinning energies. At intermediate temperature below $T_{\rm P}$, collective pinning is frozen below a glass transition as shown by dielectric susceptibility experiments (section~\ref{sec6-3}) which involves activation energy of the order of the C/S DW gap. It is obvious that the relaxation processes below 1~K, described above, concern a totally different type of entities with activation energy of the order of 1~K. In the context of glasses, that might be related to $\beta$ process relaxation resulting from the branching of $\alpha$ and $\beta$ processes at $T_g$ \cite{Angell00}. As presented below these entities were described as local defects in the local model of strong pinning \cite{Larkin94}.

\begin{figure}[h!]
\begin{center}
\includegraphics[width=8cm]{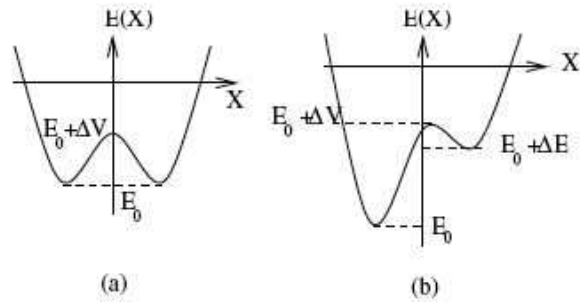}
\caption{Schematic representation of the energy landscape (a)~in the commensurate case, (b)~in the incommensurate case. The ground state is at energy $E_0$. The metastable state is at energy $E_0+\Delta E$ where $\Delta E$ is the splitting. The unstable ``bounce" state is at energy $E_0+\Delta V$ where $\Delta V$ is the energy barrier (reprinted figure with permission from the European Physical Journal B: Condensed Matter and Complex Systems 43, R. Melin \textit{et al.}, p. 489, 2005 \cite{Melin05}. Copyright (2005) from Springer Science and Business media).}
\label{fig8-13}
\end{center}
\end{figure}

The low temperature heat capacity anomalies were first explained using a model of independent strong impurities \cite{Larkin94,Ovchinnikov96}. Bisolitons (superposition of a $2\pi$-soliton and a $2\pi$-antisoliton) are generated at an impurity by a sufficient strong pinning potential. It is then possible to define, for an incommensurate material, an effective two-level system with a ground state $E_0$ separated from a metastable state $E_0+\Delta E$ by an unstable ``bounce" state at energy $E_0+\Delta V$ where $\Delta V$ is the energy barrier as schematically drawn in figure~\ref{fig8-13}(b). 
This model explains the $T^{-2}$ contribution to the specific heat as the high $T$ tail of a Schottky anomaly, as well as its dependence on the waiting or pumping time. Increasing the waiting time increases the energy transferred to the two-level system (TLS) and therefore increases the amplitude of the $1/T^2$ term. A maximum in $C_{\rm P}$ is expected at $T_{\rm max}$ related to the level splitting $\Delta E$ of the TLS such: $\Delta E\simeq 2.5~k_BT_{\rm max}$. $T_{\rm max}$ is a temperature lower than experimentally accessible, that means typically at $T\lesssim 30$~mK \cite{Lasjaunias02a}. It was noted \cite{Melin05} that the two-level splitting, $\Delta E$, is a universal property, valid for commensurate and incommensurate C/SDW with a large spread on their critical temperature. In fact, $\Delta E$ is only related \cite{Melin05} to the strength of the interchain interactions and of the Fermi velocity $v_{\rm F}$ which are nearly identical for all the compounds considered.

However the amplitude of the $T^{-2}$ term requires a very large concentration of defects, $\sim 1.4\times 10^{23}$/mole, that means one of four molecules must be a defect, which is not really compatible with independent rare strong pinning impurities.

Moreover the $T^\nu$ contribution to $C_{\rm P}$ cannot be explained in a dynamical model of independent bisolitons. Collective effects resulting from interaction between bisolitons were then taken into account using two successive approaches: one \cite{Melin02} using the dynamical renormalisation group where the collective behaviour is similar to domain growth dynamics in presence of disorder, the second \cite{Melin05} using a random energy-like trap model inspired from trap models developed for glasses and spin glasses \cite{Bouchaud92}.

It was then shown that heat relaxation experiments can be described by assuming two types of defects: strong pinning impurities and substitutional impurities. The $T^\nu$ contribution was attributed to these substitutional impurities by noting that doping o-TaS$_3$ by niobium changes only the amplitude of $T^\nu$ contribution but keeps unchanged either the exponent $\nu$ and the $1/T^2$ contribution \cite{Biljakovic03}. The substitutional disorder in C/S DW was studied with the extended model \cite{Melin05} of doped spin-Peierls system \cite{Fabrizio97} which provides the explanation of the power-law, $T^\nu$ for the specific heat, and the power $T^{-1+\nu}$ to the magnetic susceptibility \cite{Biljakovic03} measured in o-TaS$_3$.

It was also noted \cite{Melin05} that the energy landscape, in the commensurate case, is symmetric as shown in figure~\ref{fig8-13}(a). It is then not possible to transfer energy over long time scales to the effective two-level systems by applying the increment $\Delta T$ in temperature in $C_{\rm P}$ measurements. Thus, the classical model predicts no slow relaxation although observed in experiments. For restoring a finite heat response, quantum tunnelling between the two energy minima of the energy landscape has to be added \cite{Melin05}. Thus, from these models \cite{Melin02,Melin05} ``the final picture for Q1D C/S DWs at very low temperature is a coexistence between strong pinning and substitutional disorder as well as a coexistence between classical and quantum effects". However a direct search of two-level systems in K$_{0.3}$MoO$_3$ through their characteristic non-linear behaviour in an electron-echo experiment was unconclusive \cite{Mozurkewich87}.

A different theoretical study was aimed to the formation of a CDW with long range order in a system of repulsive 1D electrons coupled to 3D phonons \cite{Artemenko07}. It was found that the coupling of 1D electrons on CDW chains to 3D phonons leads to stabilisation of the Luttinger Liquid (LL) state and results to the formation of a long range order CDW \cite{Artemenko07}. In presence of defects, electrons are localised near impurities and the system is broken into independent segments with bounded LL \cite{Artemenko07,Artemenko05} which yields violation of the  spin-charge separation. The energy spectrum of these segments consists of zero modes the lowest energy of them corresponding to the number of extra electrons. The smallest energy is achieved by zero-mode states with the number of extra electrons localised near the impurity $\delta_n$= 0, $\pm 1$, with their energies in both the charge sector and the spin sector evaluated in ref.~\cite{Artemenko09}. The energy of the states with $\pm 1$ extra spin is shifted by the applied magnetic field by $\pm\mu_{\rm B}H$ which leads to the magnetic field dependence of the specific heat \cite{Artemenko09} (see below).

In this model of CDW in a Luttinger Liquid, spins degrees of freedom are coupled to $H$ by Zeeman coupling both in CDW systems as well as in SDW ones. It results that the specific heat is magnetic field dependent in CDWs and in SDWs, as experimentally found \cite{Lasjaunias05b,Lasjaunias05c,Sahling07}. On the other hand, in the model developed in refs~\cite{Melin05,Melin06,Sahling07} solely the spin degrees of freedom of bisolitons in SDWs are coupled to $H$.

\subsection{Effect of a magnetic field}\label{sec8-8}

The strong sensitivity of the low-$T$ specific heat to moderate magnetic fields was pointed out either for CDWs (o-TaS$_3$ and RbMoO$_3$) \cite{Lasjaunias05b,Lasjaunias05c} and SDWs ((TMTTF)$_2$Br and (TMTTF)$_2$PF$_6$) \cite{Lasjaunias03,Melin06,Sahling07,Lasjaunias02d} systems.

It was shown \cite{Sahling03} that in (TMTSF)$_2$PF$_6$ with the short-pulse technique, the amplitude of the tail of the Schottky anomaly, the $A/T^2$ term, exhibits a sharp maximum at $H\sim 0.2$~T, independently on the orientation of the magnetic field $H\parallel a$ or $H\parallel b^\prime$. A different magnetic field dependence was found for the amplitude of the quasi-linear term in $C_{\rm P}$ which increases up to 0.2~T and saturates at higher field.
\begin{figure}[b]
\begin{center}
\includegraphics[width=7.5cm]{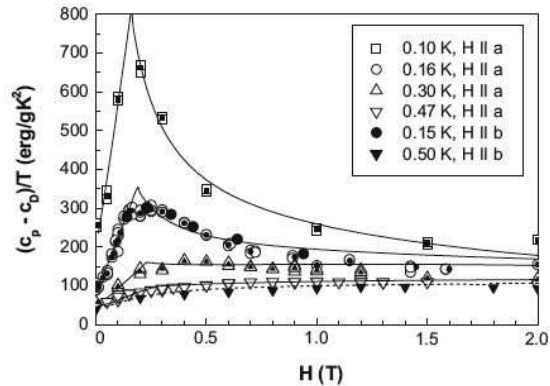}
\caption{Contribution of the low energy excitations to the specific heat (total specific heat minus the phonon $\beta T^3$ term) divided by $T$ of (TMTSF)$_2$PF$_6$ as a function of magnetic field applied parallel to $a$ and $b'$-axis at different temperatures. The dashed curve (for $T$~= 0.5~K) shows the contribution of the quasi-linear term (reprinted figure with permission from Journal of Low Temperature Physics 133, S. Sahling \textit{et al.}, p. 273, 2003 \cite{Sahling03}. Copyright (2003) from Springer Science and Business media).}
\label{fig8-15}
\end{center}
\end{figure}

From waiting time experiments, it was shown above that the maximum in the relaxation time spectrum is much shorter in the commensurate antiferromagnetic (TMTTF)$_2$Br than in the incommensurate (TMTSF)$_2$PF$_6$ which allows for the former salt to investigate the equilibrium specific heat. 

\begin{figure}
\begin{center}
\subfigure[]{\label{fig8-16a}
\includegraphics[width=6.5cm]{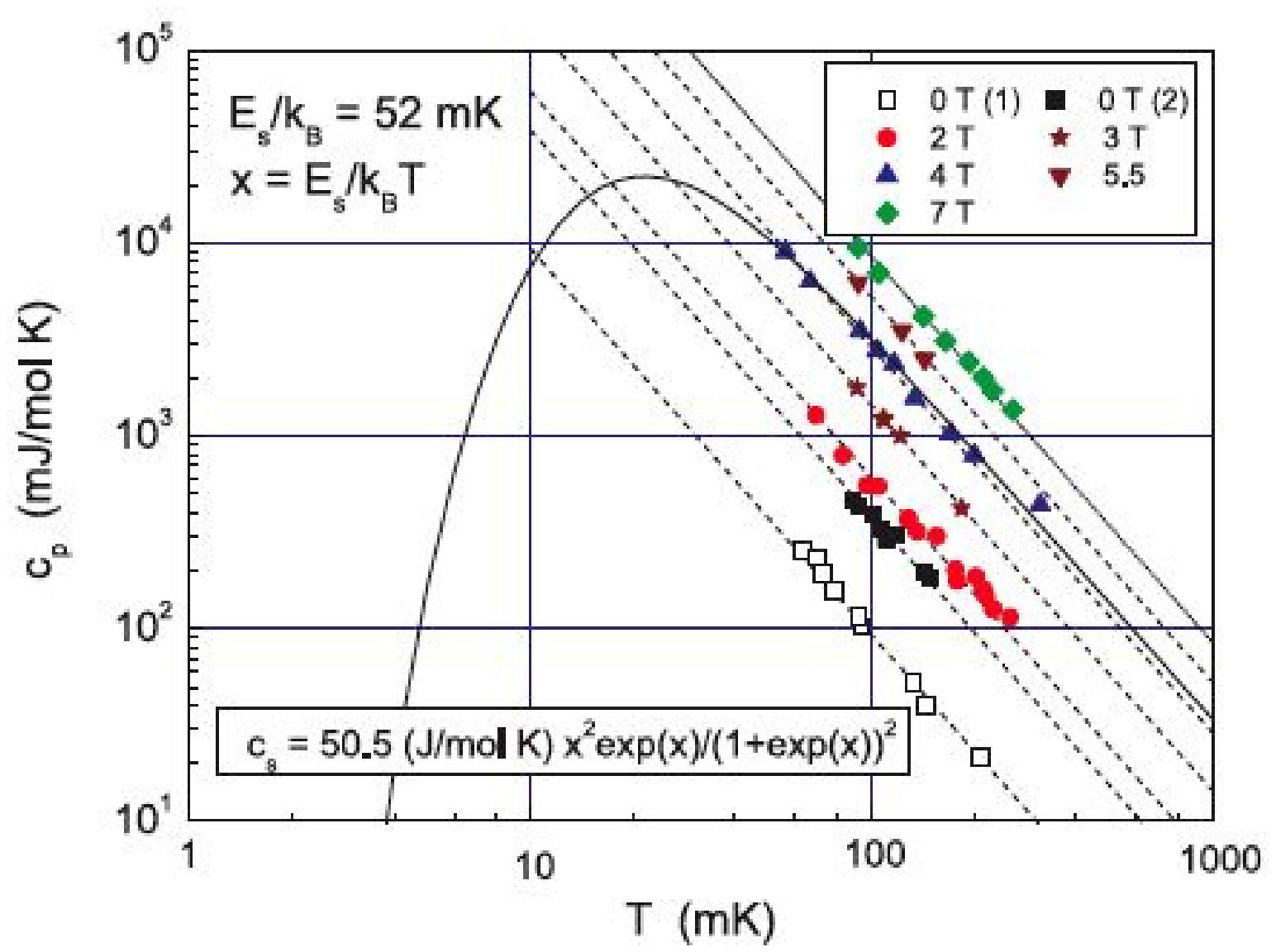}}
\subfigure[]{\label{fig8-16b}
\includegraphics[width=6.5cm]{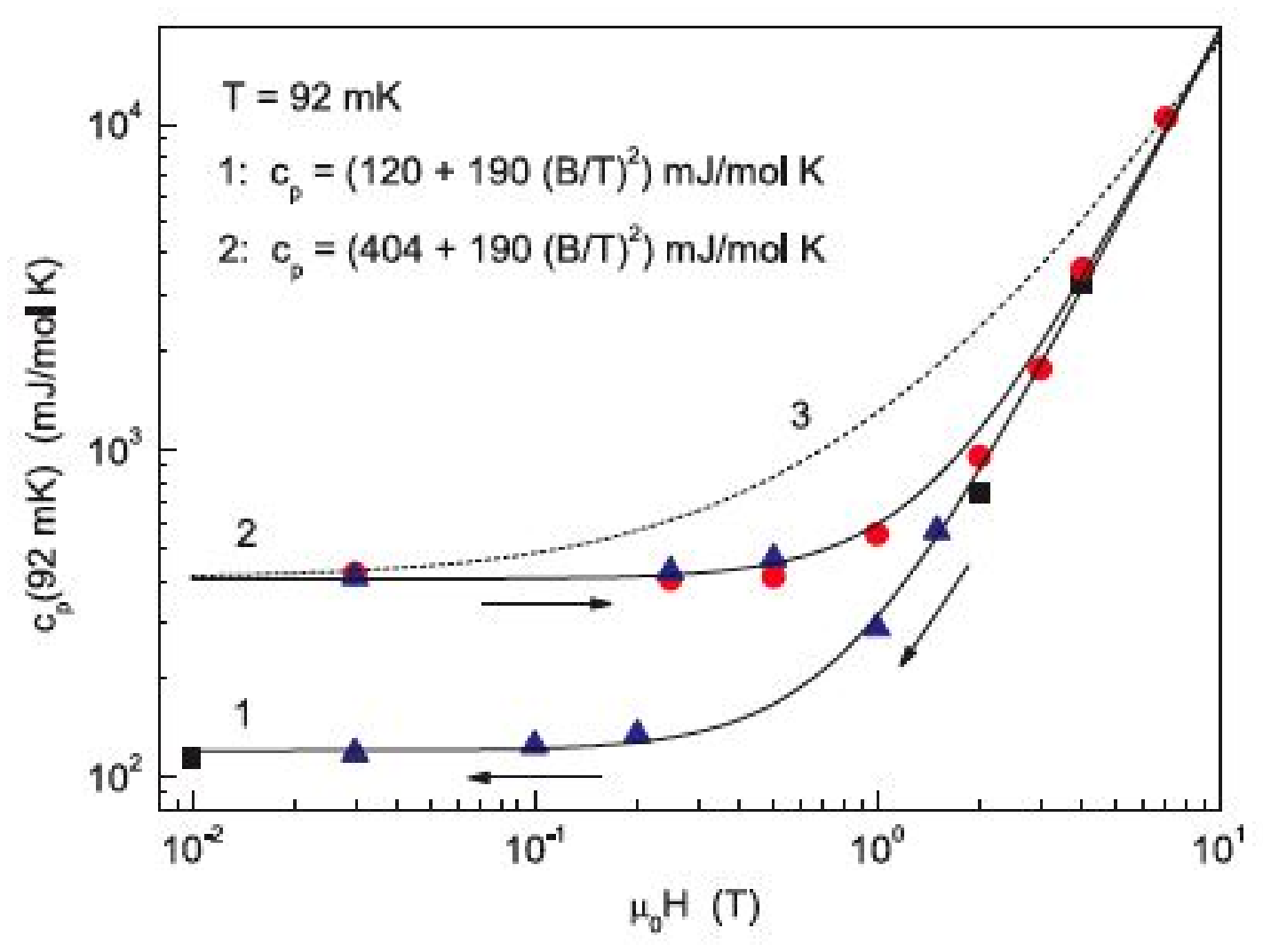}}
\caption{(a)~Temperature dependence of the heat capacity at the equilibrium of (TMTTF)$_2$Br for different magnetic fields. All data in this temperature range can be fitted by $T^{-2}$ laws (straight lines) representing the high-$T$ tail of the Schottky specific heat. (b)~Equilibrium heat capacity of (TMTTF)$_2$Br at $T$~= 92~mK as a function of magnetic field with two metastable branches. Starting from the initial lower branch 1, one recovers the upper branch 2 after exposure to high field ($>$~5~T) at low $T$ ($<$~100~mK). Solid lines 1 and 2 are fits with $C_{\rm P}$~= $C_0+\alpha B^2$ (reprinted figure with permission from the European Physical Journal B - Condensed Matter and Complex Systems 59, S. Sahling \textit{et al.}, p. 9, 2007 \cite{Sahling07}. Copyright (2007) from Springer Science and Business media).}
\label{fig8-16}
\end{center}
\end{figure}

\begin{figure}[h!]
\begin{center}
\includegraphics[width=7.5cm]{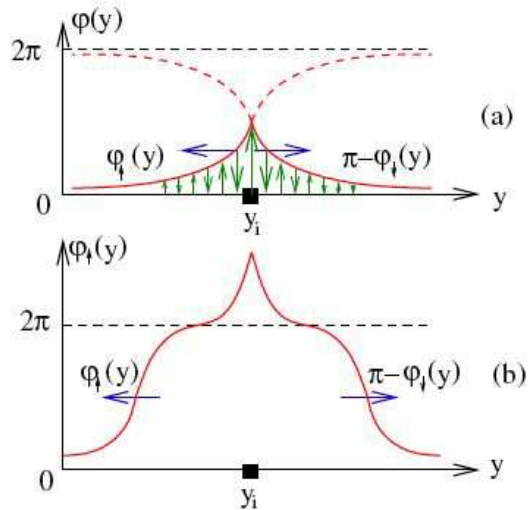}
\caption{a)~Deconfinement of a spin-polarised bisoliton made up of the superposition of a spin-up electron-like soliton and a spin-down hole-like soliton. b)~Generation of a new bisoliton (reprinted figure with permission from R. Melin \textit{et al.}, Physical Review Letters 97, p. 227203, 2006 \cite{Melin06}. Copyright (2006) by the American Physica Society).}
\label{fig8-17}
\end{center}
\end{figure}

The temperature dependence of the specific heat at equilibrium of (TMTTF)$_2$Br is shown \cite{Sahling07} in figure~\ref{fig8-16}(a) for different values of $H$. It is seen that the $1/T^2$ contribution increases by almost 2 orders of magnitude when $H$ increases from $B$~= 0 to $B$~= 7~T. As a function of $H$, the specific heat varies as shown in figure~\ref{fig8-16}(b):
\begin{equation*}
C_{\rm P}=C_0+\alpha B^2.
\end{equation*}
Moreover metastability occurs depending on the field excursion. After the first cooling without magnetic field, $C_0$ is found to be 120~mJ/mol\,K. Increasing the magnetic field up to $B_c$~= 3--4~T does not change $C_0$ (branch 1 in figure~\ref{fig8-16}(a)). If the magnetic field increases above $B_c$, the specific heat follows the branch 2 with a $B^2$ dependence when the field is decreased with $C_0$~= 404~mJ/mol\,K (a value $3\sim 4$ times larger than the initial value) and remains on this branch 2 if $H$ is further cycled above $B_c$. Branch 2 is very robust and the sample has to be reheated in zero field up to 20~K above the AF phase transition to recover its initial state (above the Peierls transition temperature for CDWs). The model of bisolitons generated at the strong pinning centres \cite{Larkin94,Ovchinnikov96} has been extended \cite{Melin05,Melin06} to the case of an applied magnetic field. A mechanism of deconfinement of bi-solitons resulting in the creation of new solitons under field was proposed as shown schematically in figure~\ref{fig8-17}. 
In the absence of impurities, the SDW ground state is unstable against the nucleation of pairs of solitons and antisolitons carrying a net spin if the gain in the Zeeman energy exceeds the soliton energy $E_S$, a condition \cite{Melin06} equivalent to $B>B_c$. The consequence of it yields the characteristic $B^2$ dependence of $C_{\rm P}$ at large field, for $B>B_c$, where spin-polarised solitons appear, in addition to initial spin-less bisolitons. At large field above $B_c$, the proliferation of solitons generates a random phase configuration (which can be called a density wave glass) carrying spin polarised magnetic moments \cite{Sahling07}. When H is reduced to zero a remnant phase configuration occurs due to solitons carrying magnetic moments with a random orientation. These remnant solitonic defects contribute to the specific heat (branch 2 in figure~\ref{fig8-16}(b) and are vey difficult to anneal.

\subsection{Electronic excitations}\label{sec7-4}

The complete electrodynamic spectrum from optical reflectivity data together with conductivity measurements over a broad frequency range was performed on C/S DW systems, namely on K$_{0.3}$MoO$_3$ \cite{Degiorgi91a,Degiorgi91b,Degiorgi95,Schwartz95}, (TaSe$_4$)$_2$I \cite{Degiorgi91a,Degiorgi91b,Schwartz95,Berner93}, o-TaS$_3$ \cite{Degiorgi91a,Creager91}, and on Bechgaard-Fabre salts \cite{Donovan94}.

For these systems the following common features were observed:

\noindent a. a relaxational contribution with a broad temperature-dependent frequency distribution in the MHz range;

\noindent b. an oscillator contribution in the 10--100~GHz range associated with the pinned collective oscillation of the C/S DW condensate;

\noindent c. for CDWs, a contribution from infrared active lattice modes in the 1--10~THz range. The lowest such frequency, $\omega_1$, was interpreted as a ``bound'' collective mode arising from the presence of polarisable impurities \cite{Degiorgi91a,Degiorgi91b,Degiorgi95};

\noindent d. a high frequency peak corresponding to single excitations across the C/S DW gap.

\subsubsection{Optical conductivity in CDWs}\label{sec7-4-1}

\begin{figure}[h!]
\begin{center}
\includegraphics[width=7.5cm]{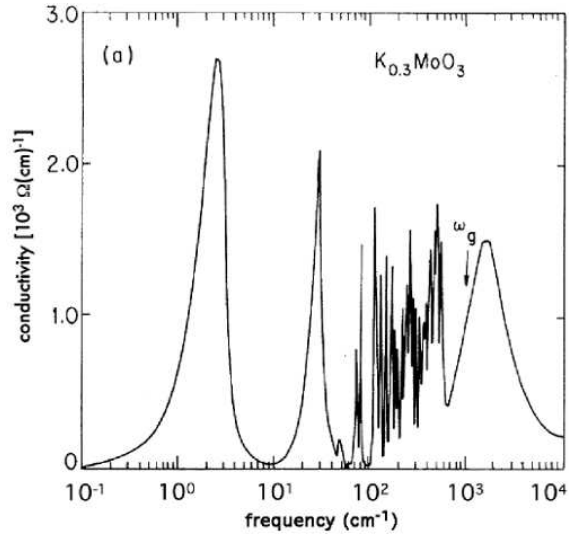}
\caption{Optical conductivity of K$_{0.3}$MoO$_3$ at 10~K as evaluated from the Kramers-Kronig analysis (reprinted figure with permission from L. Degiorgi \textit{et al.}, Physical Review B 44, p. 7808, 1991 \cite{Degiorgi91a}. Copyright (1991) by the American Physical Society).}
\label{fig7-19}
\end{center}
\end{figure}

The b, c, d features are well visible in figure~\ref{fig7-19} which displays the experimental optical conductivity of K$_{0.3}$MoO$_3$ on a large frequency range: the CDW pinned mode at 3.33~cm$^{-1}$, the ``band" collective mode at $\omega_1$~= 40~cm$^{-1}$ and the absorption peak through the single-particle gap, $\omega_g$ \cite{Degiorgi91b}.

\medskip
\noindent \textit{7.5.1.a. ``Bound" collective mode}
\medskip

The $\omega_1$ mode in fir experiments was also observed at 10~cm$^{-1}$ in o-TaS$_3$ and 38~cm$^{-1}$ in (TaSe$_4$)$_2$I (see figure~5 in refs~\cite{Degiorgi91a,Degiorgi91b}).

By analogy with the theory of lattice modes in doped semiconductors, an effective-medium dielectric function, $\varepsilon_{\rm eff}$, following a Closius-Mossati dielectric function was derived \cite{Degiorgi91a,Degiorgi91b} from the bulk medium with the dielectric function $\varepsilon_m(\omega)$ containing spherical regions due to impurities randomly distributed with a different dielectric function $\varepsilon_s(\omega)$. The extra polarisability around impurities locally modifies the dielectric function and $\varepsilon_{\rm eff}(\omega)$ presents two zero crossings: one at the pole of $\varepsilon_m(\omega)$ as in the pure case, the other one defining a localised resonance centred on impurities.

It has been searched the analogue of the infrared $\omega_1$ resonance from inelastic neutron scattering in the 1-THz range. Indeed a mode at 1.1~THz (36~cm$^{-1}$) at the $\Gamma$-point in the Brillouin zone was identified in (TaSe$_4$)$_2$I \cite{Lorenzo93}. This optic mode disperses with $q\parallel c^\ast$ and shows anticrossing with the longitudinal acoustic mode; it is observed up to the zone boundary for $q$ along $a^\ast+b^\ast$. The two modes, at the same frequency $\omega_1$, in fir and in neutron scattering appear to have related symmetry and polarisation characteristics. However the fir mode was observed at temperatures well below the CDW temperature transition, while the $\Gamma$-point phonon in neutron scattering was measured at room temperature above $T_{\rm P}$. It would be worth to know if, in the infrared case, the mode which corresponds to the $\Gamma$-point phonon becomes unscreened below $T_{\rm P}$, due to the progressive disappearance of free carriers, or whether it corresponds to a ``new mode" in the same sense as the phason.

Apparently the observation of the 1.1 - THz mode above the Peierls transition does not seem compatible with the interpretation of the $\omega_1$ mode as  a bound CDW collective mode. However as seen below, in (TaSe$_4$)$_2$I, all the gap structure is already visible in optical conductivity above $T_{\rm P}$ \cite{Berner93}.

\medskip
\noindent \textit{7.5.1.b. Phase phonons}
\medskip

As seen in figure~\ref{fig7-19} several intense resonance peaks dominate the optical conductivity in the 40--700~cm$^{-1}$ range. These results have led to apply the many-phonon coupling model \cite{Rice76,Rice77,Rice78} developed for organic CDWs (in particular TTF-TCNQ \cite{Rice75}) to inorganic K$_{0.3}$MoO$_3$ and (TaSe$_4$)$_2$I compounds \cite{Degiorgi91b}. This model considers that, beyond the Peierls-Fr\"ohlich CDW state where conduction electrons are coupled to a single acoustic phonon band, in organic linear chain conductors there are additional coupling of the conduction electrons to the large number of intramolecular phonon bands which arise from the internal vibrations of the organic molecules. These phonon bands which, in absence of el-ph coupling $\lambda$, are infrared non-active, renormalised as collective modes they become infrared active in the chain direction, and are called phase phonons \cite{Rice75,Rice76,Rice78}. In the limit $\omega\rightarrow 0$ the static dielectric constant along the chain direction can be written as \cite{Rice76}:
\begin{equation}
\varepsilon_0=1+\left[\frac{\omega_p}{2\Delta}\right]^2\,\left[\frac{2}{3}\,+\,\frac{\lambda\Delta}{V}\right]\,
\label{eq7-7}
\end{equation}
where $\omega_p$ is the plasma frequency, $\omega_p^2$~= $4\pi ne^2/m$. The first two terms describe the usual electron transition across the energy gap $2\Delta$, $V$: periodic potential of wave vector $Q$~= $2k_{\rm F}$. The third term describes the contribution of the phase phonons. Each phase phonon may be considered as having a particular el-ph coupling constant $\lambda_n$.

Transposed to inorganic compounds such K$_{0.3}$MoO$_3$, and (TaSe$_4$)$_2$I, among many distinct phonons that contains the unit cell, some of them may couple to the electrons \cite{Degiorgi91a,Degiorgi91b}. Fits of $\sigma(\omega)$ shown in figure~\ref{fig7-19}, using the multiphonon model were performed delivering values of $\lambda_n$, $\omega_n$ and damping $\Gamma_n$ for a series of phase phonons. Summing all the el-ph coupling constants $\lambda_n$ the total electron-phonon constant was estimated \cite{Degiorgi91a,Degiorgi91b,Degiorgi95} to be 1.16 for K$_{0.3}$MoO$_3$ and 2.22 for (TaSe$_4$)$_2$I (table~2 in ref.~\cite{Degiorgi95}) implying a strong el-ph coupling for this last compound.

\medskip
\noindent \textit{7.5.1.c. Gap structure}
\medskip

\begin{figure}[b]
\begin{center}
\subfigure[]{\label{fig7-20a}
\includegraphics[width=5.5cm]{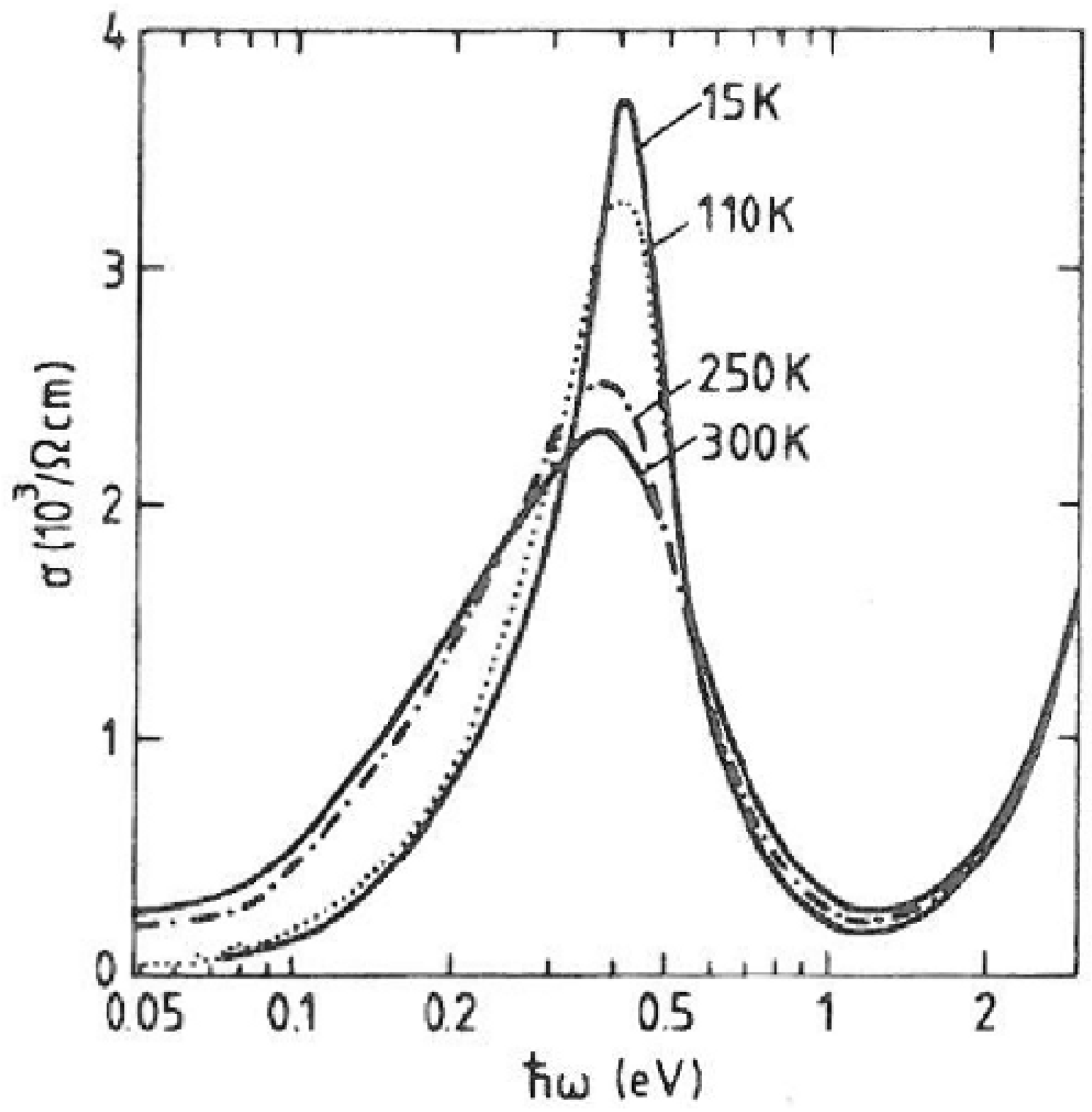}}
\subfigure[]{\label{fig7-20b}
\includegraphics[width=6.5cm]{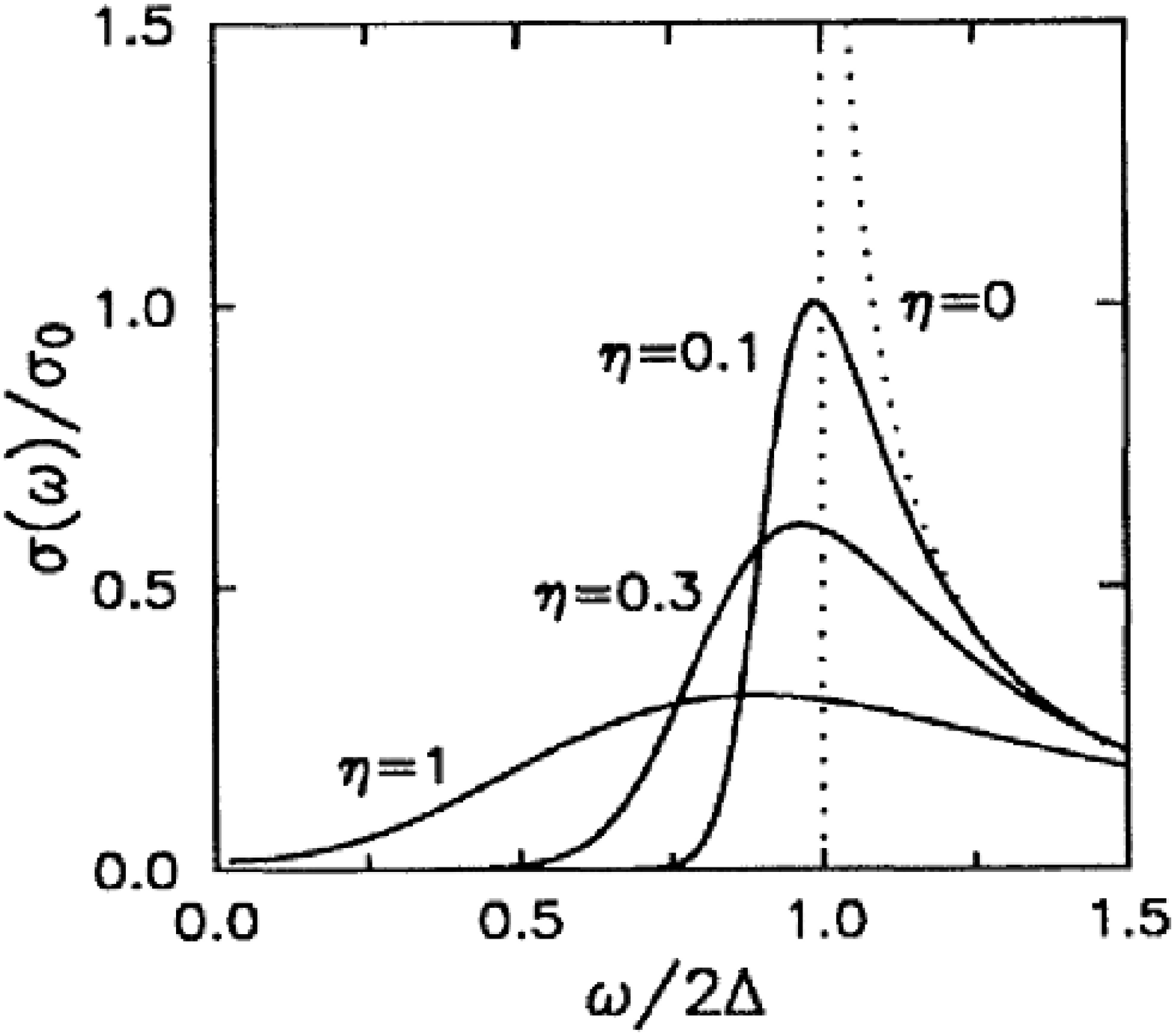}}
\vspace{-0.30cm}
\caption{a)~Frequency dependence of the longitudinal optical conductivity of (TaSe$_4$)$_2$I at different temperatures (reprinted figure with permission from D. Berner \textit{et al.}, Journal de Physique IV (France) C2, 3, p. 255, 1993 \cite{Berner93}. Copyright (1993) from EdpSciences). b)~Frequency dependence of the optical conductivity spectrum for different values of the disorder parameter $\eta$ in the model taking into account zero-point and thermal lattice motion (reprinted figure with permission from K. Kim \textit{et al.}, Physical Review Letters 71, p. 4015, 1993 \cite{Kim93}. Copyright (1993) by the American Physical Society).}
\label{fig7-20}
\end{center}
\end{figure}

A sharp structure in optical conductivity appears at $\omega_g$ close to the CDW gap $2\Delta$, as seen in figure~\ref{fig7-19} for K$_{0.3}$MoO$_3$. The temperature dependence of the optical conductivity of (TaSe$_4$)$_2$I, exhibiting the strong absorption due to the CDW gap, is shown in figure~\ref{fig7-20}(a). 
It is remarkable to note \cite{Berner93} that a gap is already present in the electronic excitation spectrum above the Peierls transition, manifesting that, even at high temperature, free carriers are condensed into a CDW ground state. However, above $T_c$~= 263~K, as shown by neutron or X-ray scattering, the CDW does not exhibit long-range order, but has the physical properties of a liquid. Below $T_c$, this ``liquid" condensate crystallises into a superlattice structure. Optical spectra show a nearly continuous change between 15~K and 300~K. The shape of the longitudinal optical conductivity spectrum exhibits just a continuous narrowing through the Peierls transition \cite{Berner93}. Moreover the temperature dependence of the optical conductivity is quite different of that of a rigid lattice, namely an inverse-square-root singularity at $\omega$~= $2\Delta$ and to be zero for $\omega\ll 2\Delta$. Considering the effect of zero-point and thermal lattice motions (as discussed in the introduction of this section) it was demonstrated \cite{McKenzie92,Kim93} that the density of states contains a tail of strongly localised states below the gap, the contribution of it increasing significantly with temperature. Figure~\ref{fig7-20}(b) shows \cite{Kim93} the computed optical conductivity following this model for different values of the disorder parameter $\eta$ (see eq.~(\ref{eq7-2})). $\eta$ is temperature dependent (see eq.~(\ref{eq7-2})) and increases with $T$. The model accounts well for the experimental data.

\medskip
\vbox{\noindent \textit{7.5.1.d. Time-resolved optical spectroscopy}
\medskip

Ultra fast optical spectroscopy can measure the dynamics of CDW collective excitations with a very high accuracy \cite{Sagar07,Shimatake06,Demsar99,Tomeljak09,Schafer10}. An ultrashort laser pump-pulse excites electron-hole pairs into higher energy states far above the CDW gap. Hot carriers quickly release their energy through a combination of exponential responses and damped sinusoidal oscillations. The exponential part is assigned to the transient response of single particles (SPs) and the oscillation part results from coherent motion of phonon and collective CDW modes, essentially amplitudons.}

\begin{figure}
\begin{center}
\includegraphics[width=8cm]{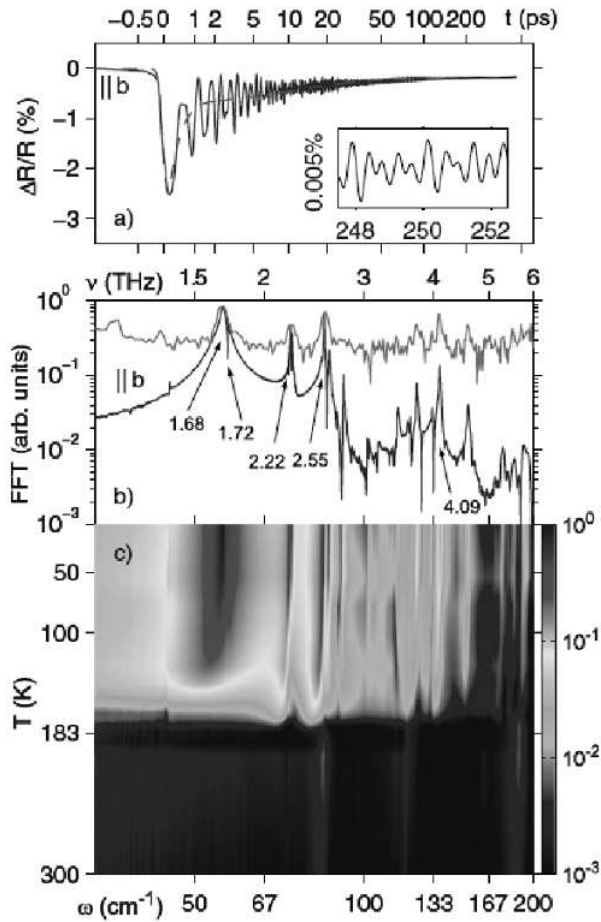}
\caption{a)~Transient change in reflectivity of K$_{0.3}$MoO$_3$ with $\vec{E}\parallel\vec{b}$ at 10~K following photoexcitation with a 40 fs laser pulse. Dashed line is the fitted electronic transient. b)~The FFT spectrum (amplitude) of the coherent parts of the signals, compared to recent Raman data (grey) \cite{Sagar08}. c)~T-dependence of the corresponding FFT spectrum in the range 1-6~THz (reprinted figure with permission from H. Sch\"afer \textit{et al.}, Physical Review Letters 105, p. 066402, 2010 \cite{Schafer10}. Copyright (2010) by the American Physical Society).}
\label{fig7-21}
\end{center}
\end{figure}

The SP transient response occurs via two exponential decays: first the relaxation to states near the band edge which is detected by an abrupt change in the reflectivity $\Delta R$, and then a slower relaxation involving the recombination time, $\tau_s$, of carriers which were accumulated on the upper edge of the CDW gap, as seen in figure~\ref{fig7-21}(a) for K$_{0.3}$MoO$_3$. Experimentally, $\tau_s$ was shown to diverge at the CDW temperature transition and to drop sharply above, which reflects the formation of a collective gap. The dominant recombination mechanism across the gap is phonon emission via phonons whose energy $\hbar\omega>2\Delta$. With the gap closing near $T_c$ more low-energy phonons become available for reabsorption and the recombination mechanism becomes less and less efficient \cite{Demsar99}. Similar dynamics of the gap closing followed by the recovery of the SDW gap was observed in the far-infrared spectrum of (TMTSF)$_2$PF$_6$ \cite{Watanabe09}.

It  was first shown that the photo-induced non thermal CDW-metal phase transition takes place on the 100~fs time scale. The adsorbed energy density required to optically induces the phase transition was experimentally found to be comparable with the CDW condensation energy \cite{Tomeljak09}. During the process of melting and sub-ps recovery of the CDW electronic modulation, the lattice remains frozen as determined by the observation of zone-folded phonons.

\begin{figure}
\begin{center}
\includegraphics[width=7cm]{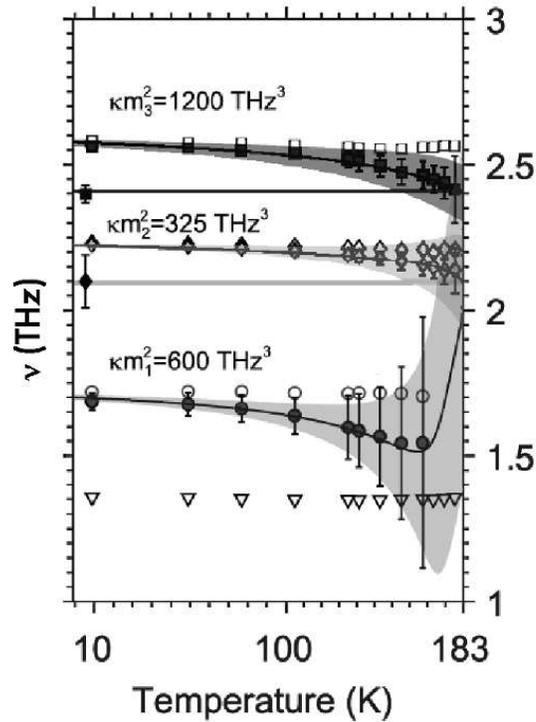}
\caption{Temperature dependence of selected phonon modes of K$_{0.30}$MoO$_3$ (from figure \ref{fig7-21}(b)): frequency $\nu$ and damping (solid symbols and bars) fitted with the TDGL model (solid lines and shaded areas. $m_i$ describes the strength of the coupling between the phonon mode and the electronic order parameter. The frequencies (dampings) of the infrared modes of 6~K (from refs~\cite{Degiorgi91a,Degiorgi91b}) are shown by solid black symbols (reprinted figure with permission from H. Sch\"afer \textit{et al.}, Physical Review Letters 105, p. 066402, 2010 \cite{Schafer10}. Copyright (2010) by the American Physical Society).}
\label{fig7-22}
\end{center}
\end{figure}

Fourier transformation of the modulated transient response yields the characteristic frequencies of phonons and collective CDW excitations in K$_{0.3}$MoO$_3$ as shown in figure~\ref{fig7-21}(b). Numerous frequency components are detected which are attributed to the coherently excited phonon modes. However these modes are not simple phonon modes but some phase modes take part in the oscillatory behaviour of the transient. Most of them are also seen in Raman scattering experiments \cite{Sagar07}. However, the high frequency resolution ($\sim$0.1~cm$^{-1}$) of the data obtained by the time-resolved technique allowed a detailed study of the temperature dependence of their amplitude, frequency and damping. The temperature dependence of the frequency and the damping of three most intense phonon modes were fitted using this model. Not only the 1.68~THz mode currently assigned to the amplitude mode, but all three modes show a pronounced softening in the whole temperature range except near $T_c$ (where the TGDL equation is no more valid) as seen in figure~\ref{fig7-22}. At very low temperatures, the frequencies of the photo-excited modes are in excellent agreement with those measured by fir spectroscopy \cite{Degiorgi91a,Degiorgi91b} and analysed as phase phonon modes \cite{Rice76} (see section~\ref{sec7-4-1}.b.).

The same ultrafast optical spectroscopy technique was used to study the recovery of several CDW phases including 2H-TaSe$_2$, DyTe$_3$ and K$_{0.3}$MoO$_3$, after having been destroyed by a high intensity pulse \cite{Yusupov10,Averitt10}. A destruction pulse quenches the system in a high temperature high-symmetry phase while a pump-probe sequence probes the change in reflectivity at a later time $\Delta t$. The evolution of the recovery of the CDW order is then obtained from changes in the transient electronic response (as discussed above). It has been shown \cite{Yusupov10} that the recovery takes place through multistep events leading to formation of topological defects or CDW domain walls in a 1--3 ps time range, those being annihilated at longer time (3--10~ps). The dynamics in the creation and annihilation of topological defects, as revealed in this experiment, place these CDW systems into the general properties of non-equilibrium dynamics of symmetry-breaking phase transitions.

\subsubsection{Optical conductivity in Bechgaard-Fabre salts}\label{sec7-4-2}

Similarly to the CDWs, the electrodynamical response of the Bechgaard-Fabre salts has been investigated over a very broad spectral range \cite{Degiorgi04c,Dressel96}. The optical conductivity of (TMTTF)$_2$X salts is quite different of that of (TMTSF)$_2$X ones \cite{Vescoli00}; it displays several transitions with large absorption features due to lattice vibrations, namely inter- and intra-molecular vibrations of the (TMTTF) molecules. Because the large phonon contribution, the evaluation of the charge gap in (TMTTF)$_2$X salts is not straightforward.

\medskip
\noindent \textit{7.5.2.a. Charge gap}
\medskip

\begin{figure}[b]
\begin{center}
\includegraphics[width=8.5cm]{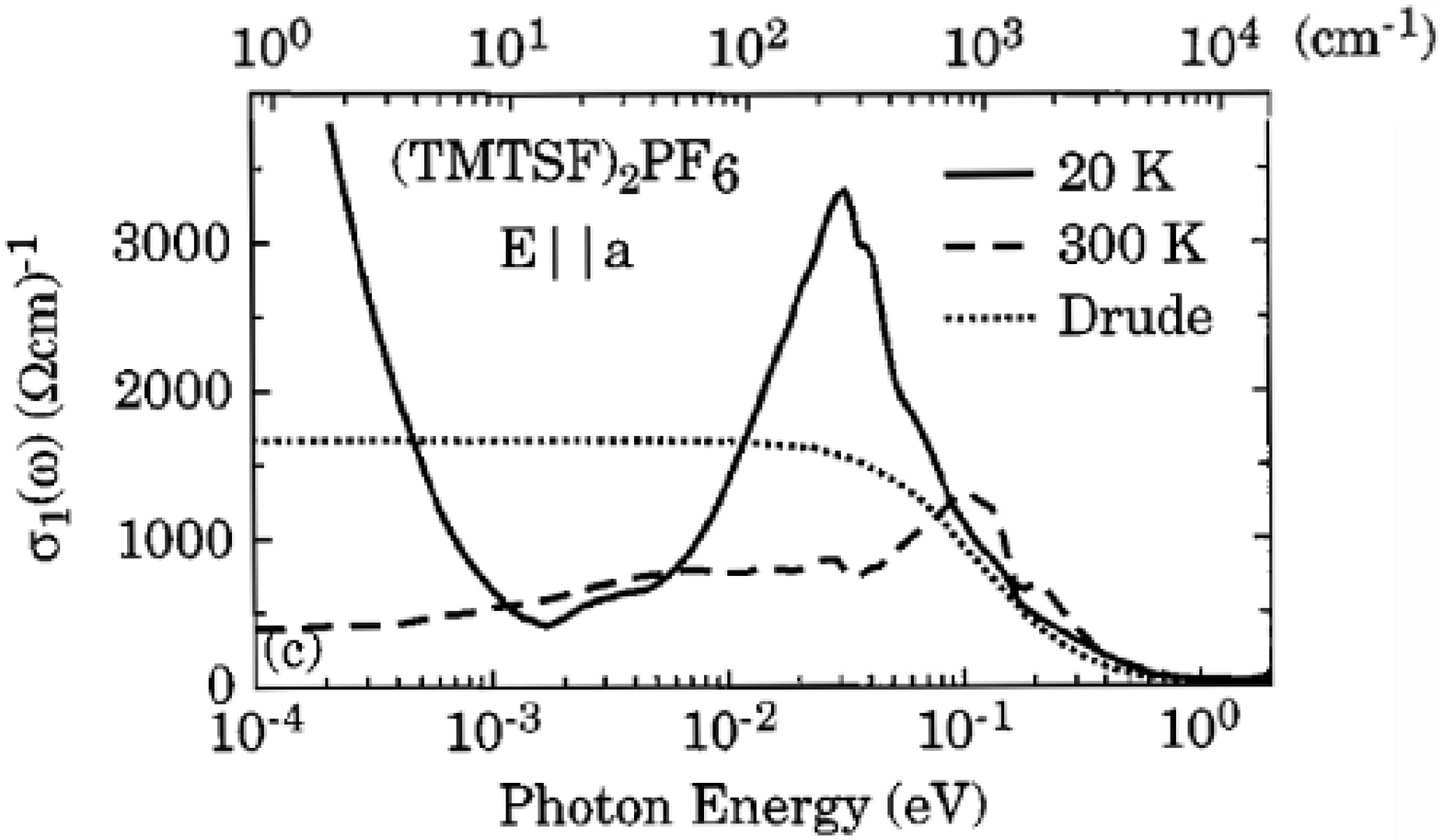}
\caption{Frequency dependence of the on-chain optical conductivity of (TMTSF)$_2$PF$_6$ at temperatures above the spin-density wave phase transition. The gap feature around 25~meV is ascribed to the Mott-Hubbard charge gap (reprinted figure with permission from the European Physical Journal B - Condensed Matter and Complex Systems 13, V. Vescoli \textit{et al.}, 2000 \cite{Vescoli00}. Copyright (2000) from Springer Science and Business media).}
\label{fig7-23}
\end{center}
\end{figure}

Figure~\ref{fig7-23} displays the on-chain optical conductivity of (TMTSF)$_2$PF$_6$ at several temperatures above the SDW phase transition \cite{Schwartz98,Vescoli00}. Two features are predominant: a well-defined absorption gap around 25~meV and a zero frequency mode at low temperature with a very small spectral weight of the order of 1\% of the total. This mode is responsible for the metallic dc conductivity.

Depending on the importance of the dimerisation along the (TMTTF) and (TMTSF) states, the electronic band can be described either as half-filled (strong dimerisation) or quarter-filled (weak dimerisation). Due to the commensurate filling (full charge transfer from the organic molecules to the counter ions) a strictly 1D model leads to a Mott insulator with a charge gap, as measured in (TMTTF)$_2$X salts.

\medskip
\noindent \textit{7.5.2.b. Non-Fermi liquid}
\medskip

In such a 1D model the effects of electron-electron interactions lead to the formation of a non-Fermi-liquid state, the so-called Tomonaga-Luttinger liquid (TLL) (see section~\ref{sec2-6}) \cite{R18Giamarchi03,Voit95,Schonhammer02}. Such a state is characterised by the absence of single electron-like quasi particles and by a power law dependence for the decay of all correlation functions controlled by the Luttinger-liquid exponent $\kappa_\rho$. The Mott insulating state, described in the TLL model, results from Umklapp processes which yields the charge gap such  as \cite{Schwartz98}:
\begin{equation*}
\Delta\rho\sim W\left(\frac{g_{1/2\,n}}{W}\right)^{1/(2-2n^2\kappa_\rho)}.
\end{equation*}
The coupling constant $g_{1/2\,n}$ is the Umklapp process corresponding to the commensurability $n$ ($n$~= 1 for half filling, $n$~= 2 for  quarter filling), $W$ the bandwidth. The peak in optical conductivity corresponds to the Mott gap and will contain the whole spectral weight in the finite frequency range for a strictly 1D system. However, in (TMTSF)$_2$X, there are competition between the Mott gap and the interchain hopping, $t_\perp$. A finite interchain hopping can be viewed as an effective doping leading to deviations from the commensurate filling due to the warping of the Fermi surface; and in this case, the physical properties are those of a doped-1D Mott insulator. As long as the correlation gap $\Delta\rho$ related to the charge localisation is larger than $t_\perp$, which occurs at high temperatures and high frequencies in (TMTSF)$_2$X salts, the interchain coupling is inoperative and the system is in the 1D regime. Theoretical models have calculated \cite{Giamarchi97} the dynamic conductivity in such 1D Mott insulator and found its functional power-law dependence for $\omega >\Delta\rho$ as:
\begin{equation*}
\sigma(\omega)\sim\omega^{-\nu}\sim\omega^{4n^2\kappa_\rho-5}.
\end{equation*}
Experimentally this law was found \cite{Schwartz98} to be obeyed for (TMTSF)PF$_6$, (TMTSF)AsF$_6$ and (TMTSF)ClO$_4$ with the same exponent $\nu$~= 1.3$\pm$0.1. Assuming quarter-filled band ($n$~= 2) $\kappa_\rho$ was estimated to be $\approx 0.23$. Such a small value of $\kappa_\rho$ corresponds to a relatively large repulsion.

\begin{figure}[h!]
\begin{center}
\includegraphics[width=7.5cm]{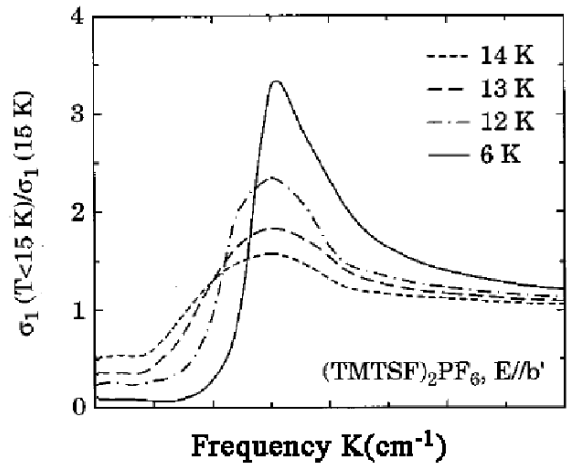}
\caption{Frequency dependence of the optical conductivity of (TMTSF)$_2$PF$_6$ in the SDW state normalised to its value at $T$~= 15 K ($T>T_{\rm SDW}$), as obtained from Kramers-Kroning transformation of the reflectivity data (reprinted figure with permission from V. Vescoli \textit{et al.}, Physical Review B 60, p. 8019, 1999 \cite{Vescoli99}. Copyright (1999) by the American Physical Society).}
\label{fig7-24}
\end{center}
\end{figure}

\medskip
\vbox{
\noindent \textit{7.5.2.c. SDW gap}
\medskip

When $T$ is reduced below $T_{\rm  SDW}$ no clear indication of the SDW gap was observed \cite{Donovan94} in the optical conductivity with electric field polarised along the chain, $a$-axis. It was speculated that, in these conditions, the SDW gap should develop in the low-frequency tail of the Mott-Hubbard gap feature displayed in figure~\ref{fig7-23}. Nevertheless, performing optical experiments on large crystals of Bechgaard salts that enhanced the experimental accuracy, optical measurements with polarisations along the perpendicular direction, $b'$-axis, allowed \cite{Degiorgi96,Vescoli99} to identify a gap feature developing at about 70~cm$^{-1}$ at $T<T_{\rm SDW}$. The gap spectrum is rather sharp at low temperatures but decreases in intensity and broadens as the temperature increases as shown in figure~\ref{fig7-24} for (TMTSF)$_2$PF$_6$.}

\subsubsection{Photoemission}\label{sec7-4-3}

In section~\ref{sec3}, ARPES measurements were essentially presented in the scheme of independent particles with the aim of reconstruction of the electronic dispersion relation $E(k)$. However, strong electronic correlations deeply modify the nature of electronic states \cite{Grioni04,Grioni09}.

\medskip
\noindent \textit{7.5.3.a. Electronic states near $E_{\rm F}$}
\medskip

\begin{figure}
\begin{center}
\subfigure[]{\label{fig7-25a}
\includegraphics[width=5.5cm]{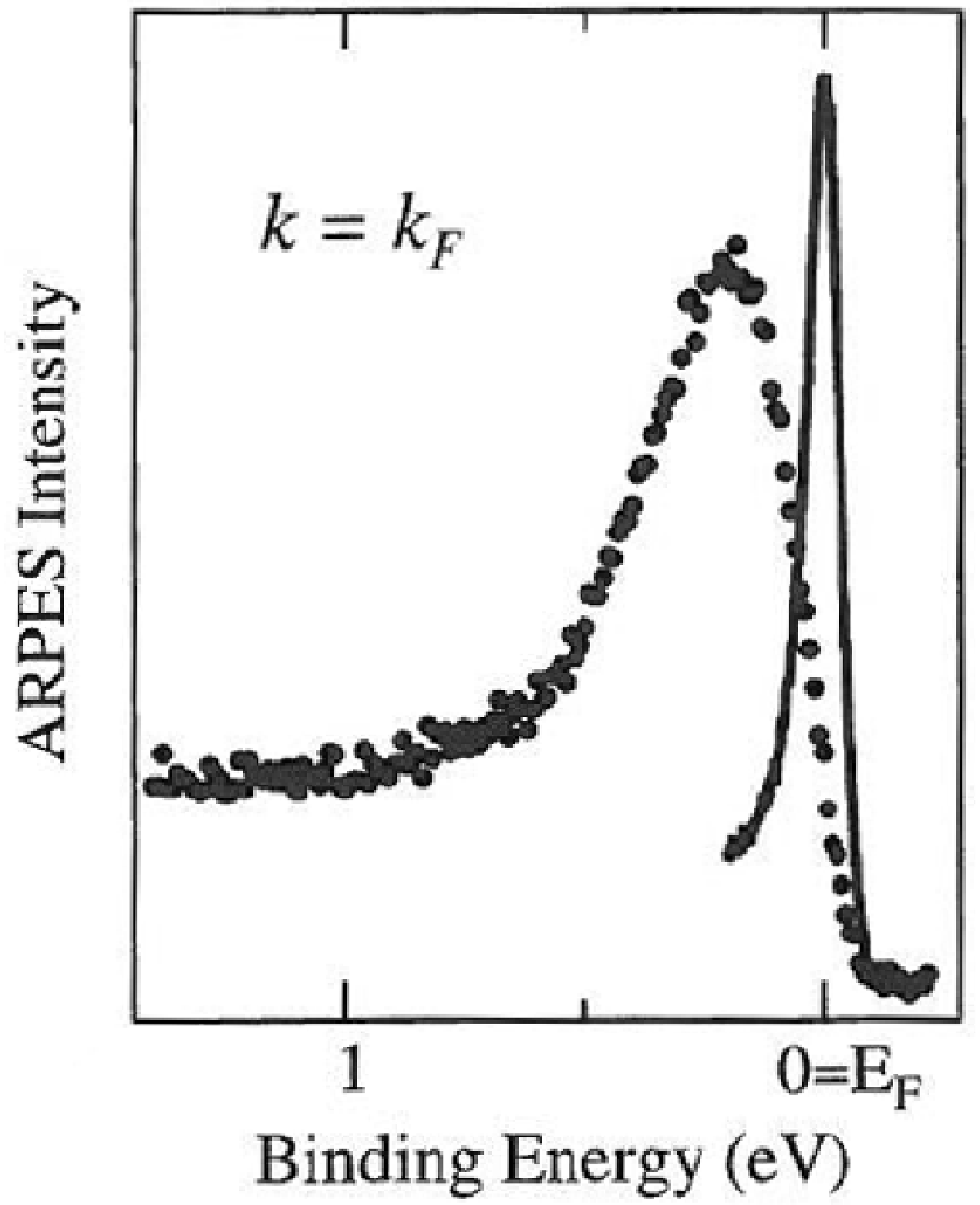}}
\subfigure[]{\label{fig7-25b}
\includegraphics[width=6cm]{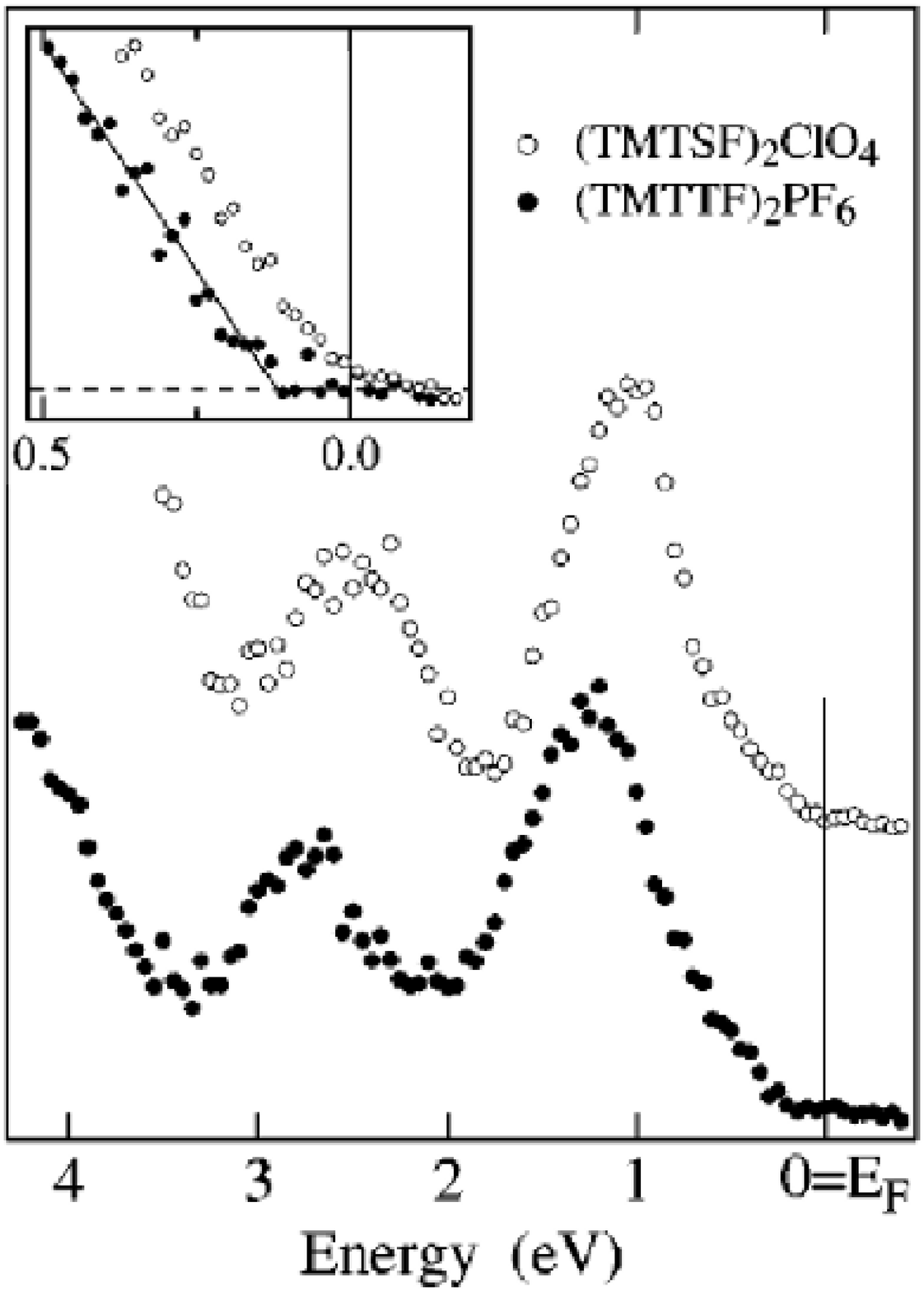}}
\caption{ARPES spectrum along the 1D-chain direction of a)~K$_{0.3}$MoO$_3$ at $T$~= 200~K. Solid line corresponds to the quasi particle spectrum of the Fermi liquid reference 1T-TiTe$_2$ (reprinted figure with permission from Journal of Electron Spectroscopy and Related Phenomena 127, L. Perfetti \textit{et al.}, p. 77, 2002 \cite{Perfetti02b}. Copyright (2002) from Springer Science and Business media). b)~Bechgaard-Fabre salts (TMTSF)$_2$ClO$_4$ and (TMTTF)$_2$PF$_6$ at the $\Gamma$ point in the Brillouin zone. Inset shows the linear fit of the leading edges of both compounds; the energy shift of 100$\pm$20~meV in (TMTTF)$_2$PF$_6$ is the indication of the Mott-Hubbard charge gap of this compound (reprinted figure with permission from F. Zwick \textit{et al.}, Physical Review Letters 79, p. 3982, 1997 \cite{Zwick97}. Copyright (1997) by the American Physical Society).}
\label{fig7-25}
\end{center}
\end{figure}

Thus, photoelectron spectroscopy (PES) with angular (momentum) resolution (ARPES) on Q-1D systems have revealed the lack of intensity in the vicinity of $E_{\rm F}$ \cite{Dardel91,Dardel93,Hwu92,Gweon01}. In figure~\ref{fig7-25}(a) is displayed the ARPES spectrum of K$_{0.3}$MoO$_3$ at $T$~= 100~K. Two significant features are remarkable: the absence of a quasi particle peak in the vicinity of $E_{\rm F}$ indicating a Fermi level crossing and a peak with a very broad lineshape. The comparison with the ARPES spectrum of the model Fermi liquid system 1T-TiTe$_2$ exhibiting the sharp quasi particle lineshape in the vicinity of $E_{\rm F}$ with a large spectral weight at $E_{\rm F}$ vividly illustrates the non Fermi liquid character of the Q-1D systems. Similarly no Fermi edge was detectable in Bechgaard-Fabre salts \cite{Dardel93,Zwick97,Vescoli00}.

Figure~\ref{fig7-25}(b) illustrates \cite{Zwick97} ARPES spectra at 150~K along the 1D direction of the Bechgaard salt (TMTSF)$_2$ClO$_4$ and of the Fabre salt (TMTTF)$_2$PF$_6$ at the $\Gamma$ point of the Brillouin zone. The two compounds exhibit similar spectra with non-dispersive features at 4, 2.5~eV and a prominent one at 1~eV. From band calculations, the feature at 1~eV was tentatively identified with the lower Hubbard subband of the 3/4 filled system \cite{Zwick97} followed by an almost linear tail. In the case of metallic (TMTSF)$_2$ClO$_4$, this linear tail extends to the Fermi energy $E_{\rm F}$, demonstrating the absence of a metallic Fermi step. The spectrum of (TMTTF)$_2$PF$_6$ is almost rigidly shifted to higher binding energy, manifesting a gap evaluated by the extrapolated leading edge of the spectrum to be $\sim 100$~meV as shown in the inset of figure~\ref{fig7-25}(b). This value agrees well with the charge gap observed in optical conductivity (see section~\ref{sec7-4-2}.a.).

\medskip
\noindent \textit{7.5.3.b. Luttinger liquid modelling}
\medskip

The form of the spectral function near $E_{\rm F}$ was modelled with the properties of the Luttinger liquid in which the momentum-integrated spectral function $\rho(\omega)$ vanishes at $E_{\rm F}$ following the law: $\rho(\omega)\sim\omega^\alpha$ ($\omega=E_{\rm F}-E$), the exponent $\alpha$~= $\frac{1}{8}(\kappa_\rho+\kappa^{-1}_\rho-2)$ reflecting the nature and the strength of interactions. However this interpretation has been questioned; it was shown that the possible signatures of electronic correlations were strongly affected by surface effects \cite{Claessen02}. Moreover specific 1D characters in the Luttinger liquid scheme should be totally washed out \cite{Castellani94} with a finite interchain coupling, i.e. Luttinger liquid properties in Bechgaard salts can only be searched in the high temperature range.

Nevertheless, in the case of K$_{0.3}$MoO$_3$, from the temperature dependence of the leading edges of spectra, it was possible to extract \cite{Perfetti02a,Perfetti02b} a phenomenological order parameter consistent with the Peierls-(half) gap value measured in optics or from transport.

\medskip
\vbox{\noindent \textit{7.5.3.c. Polaron model}
\medskip

Electronic correlations can transfer spectral weight from the coherent quasi-particle peak to the incoherent part of the spectral function at higher binding energy, the nature of them depending on materials. While electron-electron interactions are strong in organic Bechgaard-Fabre salts, electron-phonon interactions are more appropriate for inorganic CDW systems such (TaSe$_4$)$_2$I and K$_{0.3}$MoO$_3$.}

\begin{figure}
\begin{center}
\includegraphics[width=7.5cm]{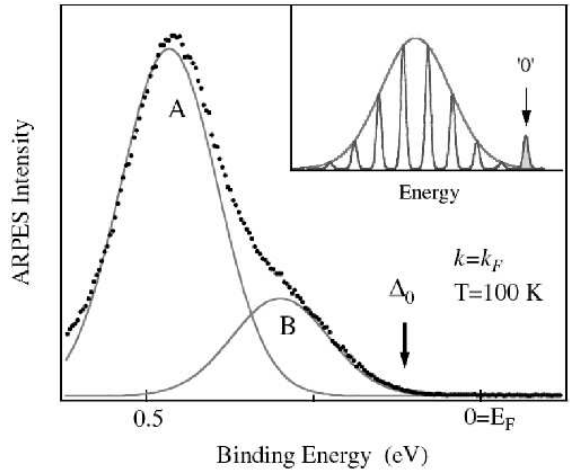}
\caption{High-resolution ARPES spectrum of (TaSe$_4$)$_2$I at $T$~= 100~K and $k$~= $k_{\rm F}$. The line shape is well fitted with two Gaussian lines. Inset: ARPES spectral function of an electron coupled to an harmonic oscillator in the strong coupling model.  The quasi particle peak (`0') should lie under the general Gaussian envelop (reprinted figure with permission from L. Perfetti \textit{et al.}, Physical Review Letters 87, p. 216404, 2001 \cite{Perfetti01}. Copyright (2001) by the American Physical Society).}
\label{fig7-26}
\end{center}
\end{figure}

The absence of any Fermi surface crossing and the shift of the spectral weight into an incoherent high frequency feature (as shown in figure~\ref{fig7-25}(a)) were interpreted \cite{Perfetti01,Perfetti02a,Perfetti02b} as resulting from the formation of a polaron liquid with mass $\sim 10\,m_e$. These polaronic quasi particles condense in the CDW state below $T_c$ with long range order. A qualitative model \cite{Perfetti01,Perfetti02a,Perfetti02b} was proposed: it consists of one electron coupled to a single harmonic oscillator of frequency $\Omega$ (see inset figure~\ref{fig7-26}). The electron removal spectrum presents a ``zero phonon" (0) peak, followed by a series of equally spaced satellites with a Poissonian envelop, corresponding to the excited vibrational final states of the oscillator. The (0) peak represents the coherent quasi particle peak of the model, while the satellites are the ``incoherent" part of the spectrum. In the strong coupling adiabatic limit (when $\hbar\Omega\ll t$, $t$: the electron hopping integral) the satellites merge into a single Gaussian peak at an energy $E\sim\langle n\rangle\hbar\Omega$ below the (0) peak where $\langle n\rangle$ is the number of vibrations. The heavily renormalised carriers are small mobile polarons; i.e. electrons moving coherently with the distortion they induce in the lattice. Figure~\ref{fig7-26} shows the spectrum of (TaSe$_4$)$_2$I at $k_{\rm F}$ and  at $T$~= 100~K, the spectrum exhibits a main peak (A) and a shoulder (B) which were fitted with two polaronic  line shapes accounting for the bonding and antibonding bands predicted by band theory \cite{Gressier84a}.

\subsection{Strong coupling model}\label{sec7-5}

The ARPES and optical spectral properties in CDWs were interpreted by taking into account strong electron-phonon interactions and the formation of small polarons with a large mass. Large values of the electron-phonon coupling were deduced from optical measurements. The large value of $2\Delta/kT_c\approx 5-15$ beyond the mean-field BCS value, 3.52, although commonly considered to result from fluctuations, was also derived by taking into account zero-point and lattice motions involving again a large electron-phonon coupling. Finally the absence of a Kohn anomaly in the phonon spectrum at $Q=2k_{\rm F}$ in some compounds ranks the Peierls transitions among order-disorder phase transitions rather than displacive ones.

Indeed, as discussed in ref.~\cite{Lorenzo98}, the existence of Peierls transitions with order-disorder dynamics has been predicted by Aubry and co-workers \cite{Aubry89,Aubry92,Raimbault95,Aubry93} in the context of strong-coupling theory. Beyond a critical value $k_c$ of the electron-phonon coupling parameter $k$, the ground state of the interacting electron-phonon system cannot be calculated perturbatively, starting from the unperturbed metallic state. This `non-analytical' regime is characterised by the existence of localised electronic states (bipolarons) strongly pinned to the lattice. However, this bipolaronic state should not be confused with the bipolaronic regime discussed above. In this latter case, the bipolarons are isolated defects with small or negligible overlapping of the corresponding wave functions. Aubry's bipolarons form a dense ensemble which in the limit $k\rightarrow\infty$ gives rise to a bond ordering wave. The normalised electronic eigenstate of the bipolaron is given by:
\begin{equation*}
\Psi_{(x)}=\frac{k}{2\sqrt{2}}\,\frac{1}{\cosh(k^2x/4)},
\label{eq7-8}
\end{equation*}
where the parameter $k$ is defined as:
\begin{equation*}
k=\lambda\sqrt{\frac{2}{tM\omega^2_0}}.
\label{eq7-9}
\end{equation*}
$\omega_0$ and $M$ are the constant frequencies and masses of identical oscillators located at each lattice site, $t$ is the electronic exchange coupling between neighbouring sites and $\lambda$ is the usual dimensionless electron-phonon coupling constant.

In such a picture, the energy scale corresponding to the Peierls gap $2\Delta(0)$ is the bipolaron formation energy. This latter energy is closely related to the energy of pinning of a single bipolaron to the lattice (the Peierls-Nabarro potential $E_{\rm PN}$). The activated nature of the electrical conductivity below as well as above $T_{\rm P}$ (the `pseudo-gap') is accounted for in terms of bipolaron hopping. The Peierls transition corresponds to the ordering temperature of the bipolaronic fluid. The order of magnitude of $k_{\rm B}T_{\rm P}$ is fixed by bipolaron interaction energies which, consistently, may be assumed to be much lower than the bipolaron formation energy; hence $k_{\rm B}T_{\rm P}\ll 2\Delta(0)$ such as $2\Delta(0)$~= $11.4k_{\rm B}T_{\rm P}$ for (TaSe$_4$)$_2$I).

The above predictions are in qualitative agreement with the results obtained for pure (TaSe$_4$)$_2$I (see sec.~\ref{sec7-2-1}). However it was found difficult to account for the strong influence of dilute isoelectronic impurities within a localised-electron framework \cite{Lorenzo98}. The primary effect of dilute isoelectronic impurities should be to localise the conduction electron wave functions over chain segments with an average length of the order of the average distance between impurities along a given chain. If, however, the electronic wave functions are already localised over a much smaller distance, as implied in Aubry's strong-coupling picture, one would not expect impurities to play such a major role.

\section{Field-induced density waves}\label{sec8}
\setcounter{figure}{0}
\setcounter{equation}{0}

Charge and spin density wave states resulting from delicate nesting properties of the Fermi surface with the formation of electron-hole pairs and a gap in the electronic excitations can be affected by external parameters. In section~\ref{sec3-2-5}, it was seen that under pressure C/SDW ground states can disappear due to this denesting mechanism. Application of magnetic fields, acting on the electron spin, can also be detrimental on the stability of the DW condensate. However the effect of $H$ is different for CDW and SDW \cite{Brooks08}.
\begin{figure}
\begin{center}
\subfigure[]{\label{fig9-1a}                    
\includegraphics[width=6.5cm]{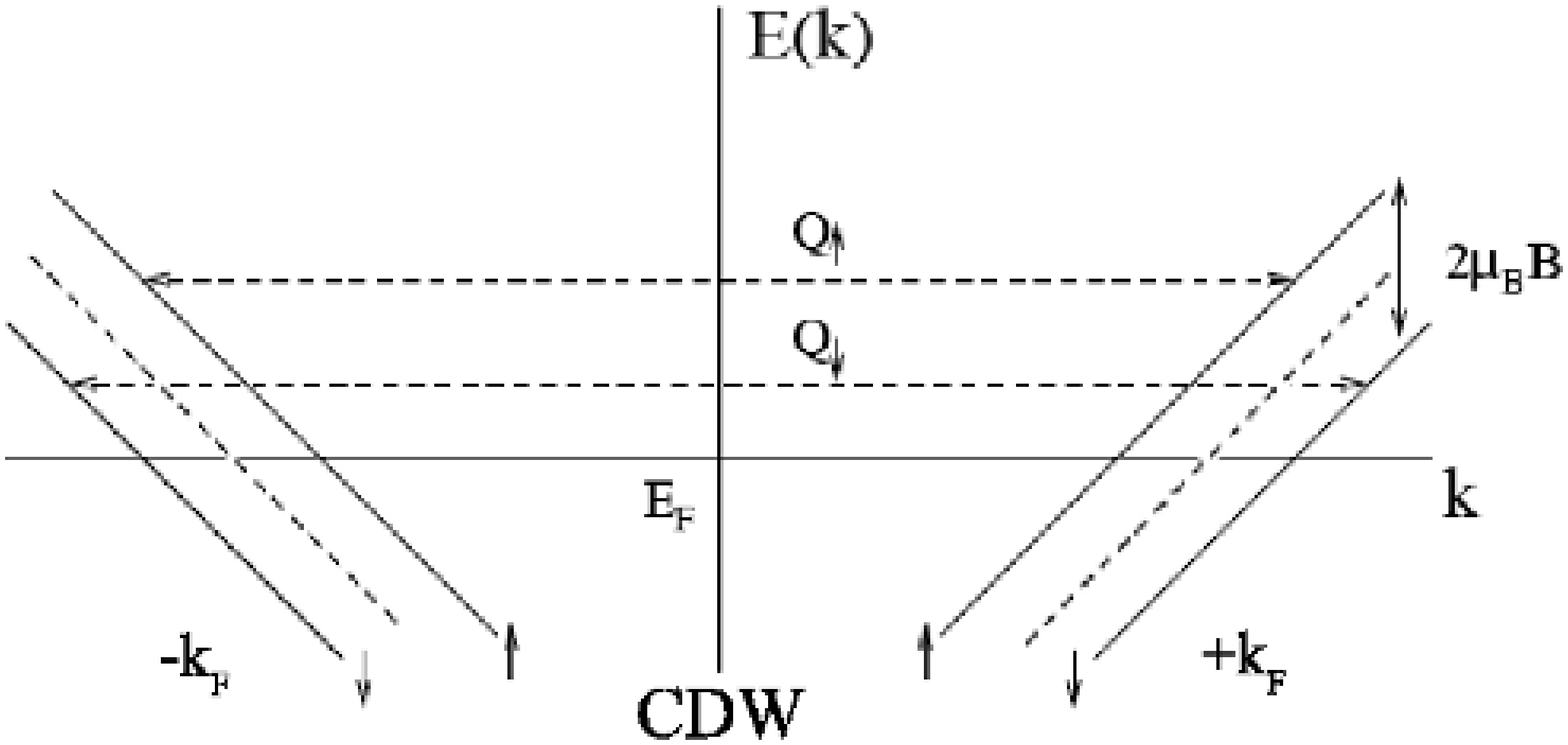}}
\subfigure[]{\label{fig9-1b}
\includegraphics[width=6.5cm]{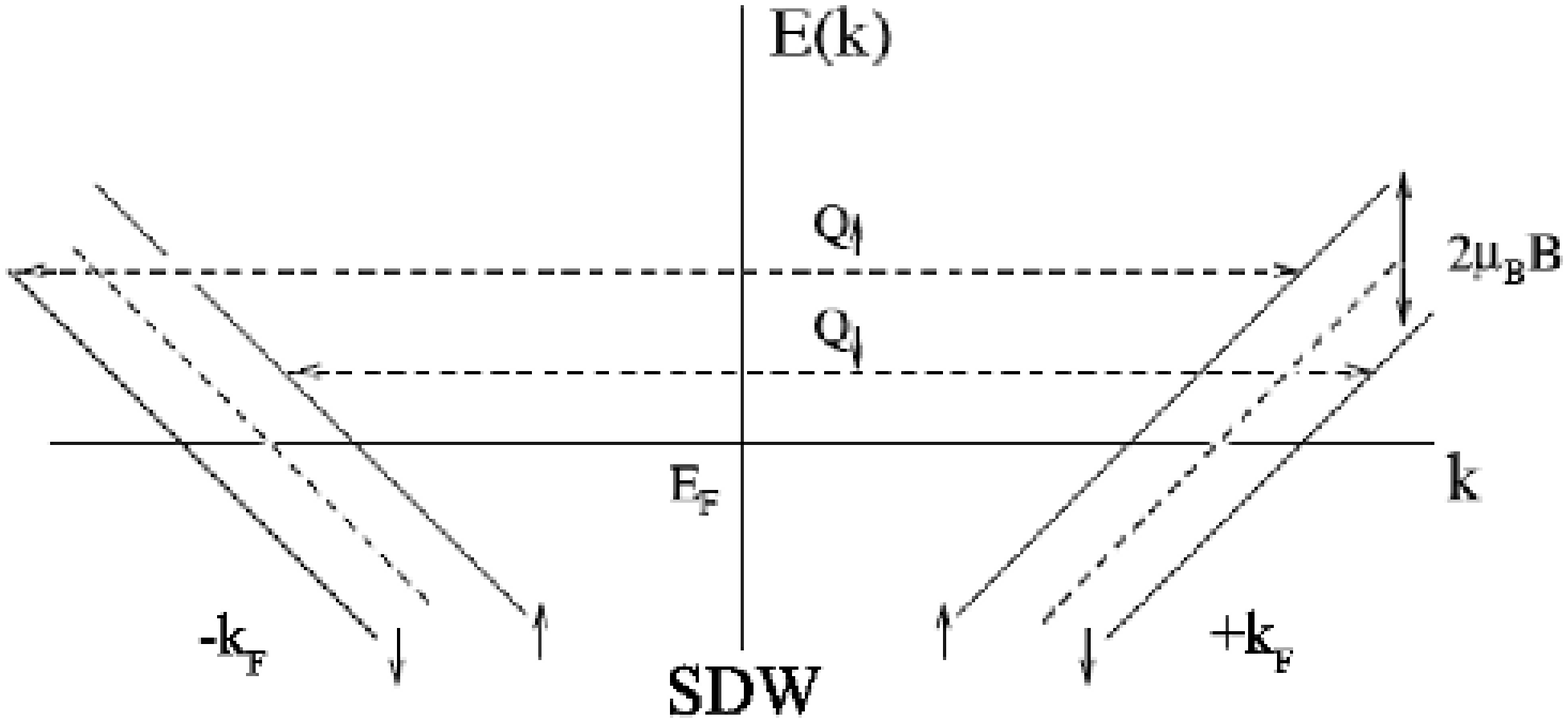}}
\caption{Linearly energy dispersion $E(k)$ in the vicinity of $E_{\rm F}$ with application of a magnetic field. Different spin bands separated by $2\mu_{\rm B}H$ result from the Zeeman effect: a)~in CDW the nesting vector is different from spin up ($\uparrow$) and spin ($\downarrow$) sub-bands, b)~in SDW the nesting vector is unaffected by $H$.}
\label{fig9-1}
\end{center}
\end{figure}

Thus, under a magnetic field, the Zeeman (Pauli) effect leads to the split of the electronic bands (represented linearised in the vicinity of $E_{\rm F}$ in figure~\ref{fig9-1} into two sub-bands for electrons with opposite spins. The energy difference between the bands in $g\mu_{\rm B}B$ with $g\simeq 2$ the Land\'e factor and $\mu_{\rm B}$: the Bohr magneton. A CDW connects apart the Fermi surface carriers with the same spin. The nesting vector, $Q$, is now different for spin up ($\uparrow$) and spin down ($\downarrow$). Consequently, a perfect nesting cannot be assumed for both sub-bands, making the CDW system energetically less favourable, resulting in a decrease of $T_{\rm CDW}$. The stability of the CDW will be exhausted at the Pauli limit, when $\mu_{\rm B}H_{\rm P}\approx\Delta(0)$ with $\Delta(0)$: the CDW gap at $T$~= 0. The Zeeman splitting being independent of the magnetic field orientation, the Pauli effect is isotropic.

On the other hand, for a SDW, the condensate is formed by electron-pair with opposite spins. The $Q$ vector is then unaffected by $H$ as schematically shown in figure~\ref{fig9-1}(b), and therefore the SDW is not suppressed in magnetic fields.

\subsection{FISDW}\label{sec9-1}

It is well known that (TMTSF)$_2$PF$_6$ becomes superconducting at $T\sim 1.1$~K when a pressure of $\sim 5$~kbars is applied, which suppresses the SDW state \cite{Jerome80}. Then, above a threshold magnetic field around 5~T (the normal state is restored for a much weaker field), a large increase in resistance was observed, the manifestation of a phase transition, followed by a cascade of different first order phase transitions at higher fields \cite{Gorkov84b,Virosztek86,Heritier84}. Exhaustive reviews of these spectacular field-induced spin density wave (FISDW) transitions are found in refs~\cite{Lebed08,Chaikin96,Montambaux91} as well as angular magnetoresistance oscillations (AMRO) \cite{Lebed04}. Then, hereafter a minimum of the properties of the FISDW states will be presented, essentially for comparison with the effect of $H$ in CDW systems.

This recovery of a SDW state with $H$ is specifically an orbital effect. When $H$ is applied perpendicularly to the ($a,b$); $a$: chain direction) conducting planes ($c$-direction) in Bechgaard salts, the conduction electrons moving along the open sheets of the Fermi surface experience a Lorentz force $ev_{\rm F}\wedge B$. One can define a frequency, $\omega_c$~= $ev_{\rm F}bH/\hbar$ corresponding to the frequency of the electron motion along the open sheets of the Fermi surface. Since the electron velocity is perpendicular to the Fermi-surface, it will result an oscillatory motion in real space, with a wavelength:
\begin{equation}
\lambda= h/eBb\,
\label{eq9-2}
\end{equation}
which will become more and more restricted to a single chain when $H$ is increased. Then, this one-dimensionalisation of the electron motion by a magnetic field stabilises the imperfect nested SDW state \cite{Gorkov84,Montambaux85,Chaikin96}.

\begin{figure}
\begin{center}
\includegraphics[width=7.5cm]{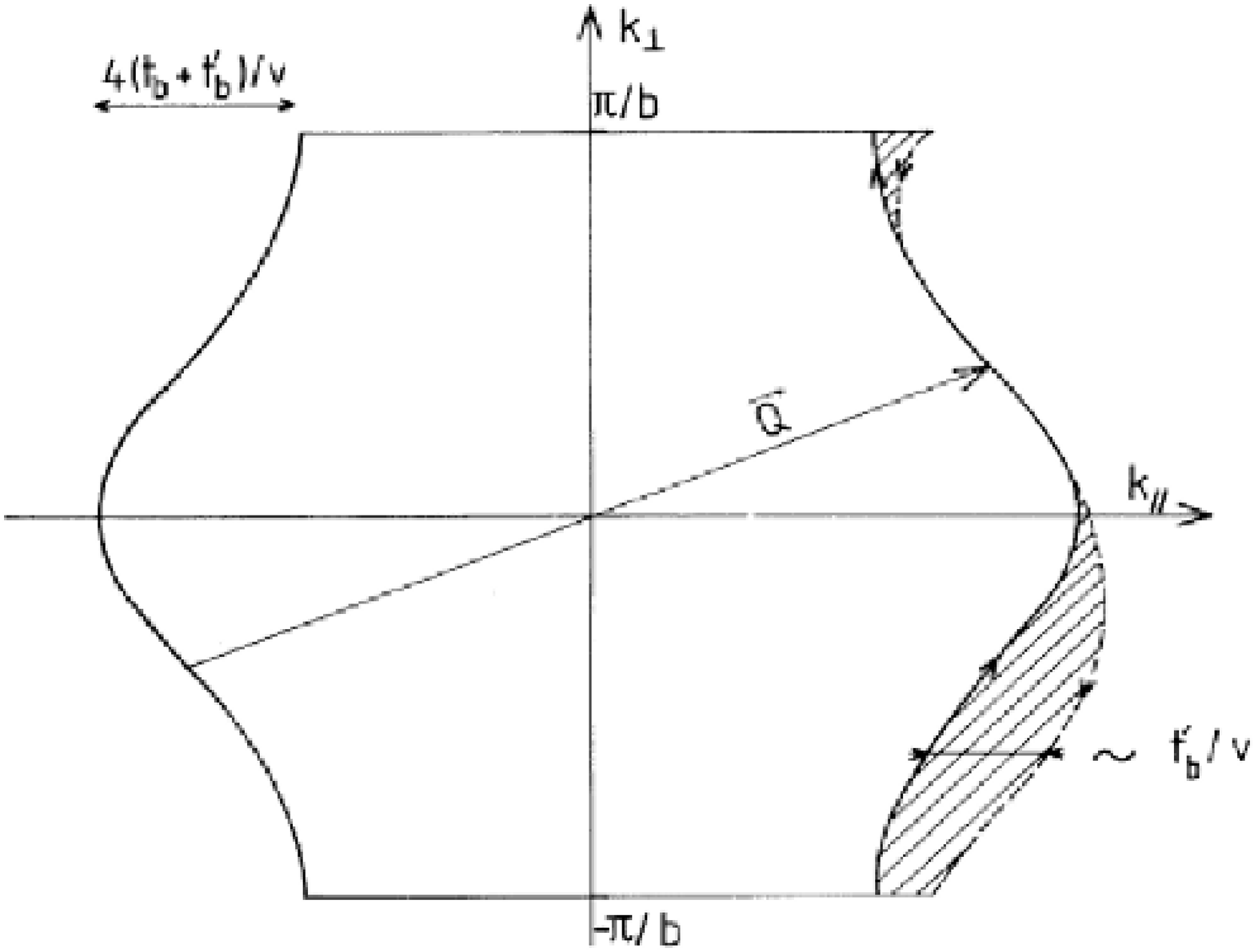}
\caption{Two-dimensional open Fermi surface of quasi-one-dimensional systems with imperfect nesting conditions; $t_b$: the transfer integral along the $b$-axis, $t'_b$: the imperfect nesting parameter, $Q_0$~= $(2k_{\rm F}$, $\pi/b)$: the nesting wave vector. Under magnetic field the electron and hole pockets with size $t'_b/v_{\rm F}$ are quantified in Landau levels (reprinted figure with permission from G. Montambaux \textit{et al.}, Physical Review Letters 55, p. 2078, 1985 \cite{Montambaux85}. Copyright (1985) by the American Physical Society).}
\label{fig9-2}
\end{center}
\end{figure}

Recall the energy dispersion (section~\ref{sec2-12}), linearised around $k_{\rm F}$, neglecting the transfer integral perpendicular to the ($a,b$) planes (2D approximation):
\begin{equation}
\varepsilon_k=v_{\rm F}\hbar(|k_x|-k_{\rm F})+2t_b\cos(bk_y)+2t'_b\cos(2bk_y).
\label{eq9-1}
\end{equation}
In zero field, the best nesting vector $Q$ connects the inflexion points of the two sheet forming the Fermi surface. Due to imperfect nesting, that leaves electron and hole pockets between one sheet of the FS and the other translated by $Q$ (figure~\ref{fig9-2}). 
The size of these pockets is characterised by $t'_b/v_{\rm F}$. At low temperatures and with application of $H$, electron and hole pockets are quantised in Landau levels with spacing $\hbar\omega$~= $eH/m^\ast c$, which yield maxima in calculated 2D spin susceptibility. The distortion SDW wave vector adjusts itself such as $\varepsilon_{\rm F}$ lies between Landau levels. $\varepsilon_{\rm F}$ being completely located in a gap, only filled Landau levels occur resulting in quantum Hall effect in transport properties. When $H$ is varied, $Q_{\rm SDW}$ changes so as to keep $\varepsilon_{\rm F}$ in the gap between Landau levels. At some field, it may be energetically favourable for $Q_{\rm SDW}$ to jump to a different value. Then the series of SDW distortion vectors are given \cite{Heritier84} by:
\begin{equation}
Q_{\rm FISDW}=\left(2k_{\rm F}\pm\frac{2n\pi}{\lambda},\pi/b\right).
\label{eq9-3}
\end{equation}

\begin{figure}[h!]
\begin{center}
\subfigure[]{\label{fig9-3a}                    
\includegraphics[width=6.5cm]{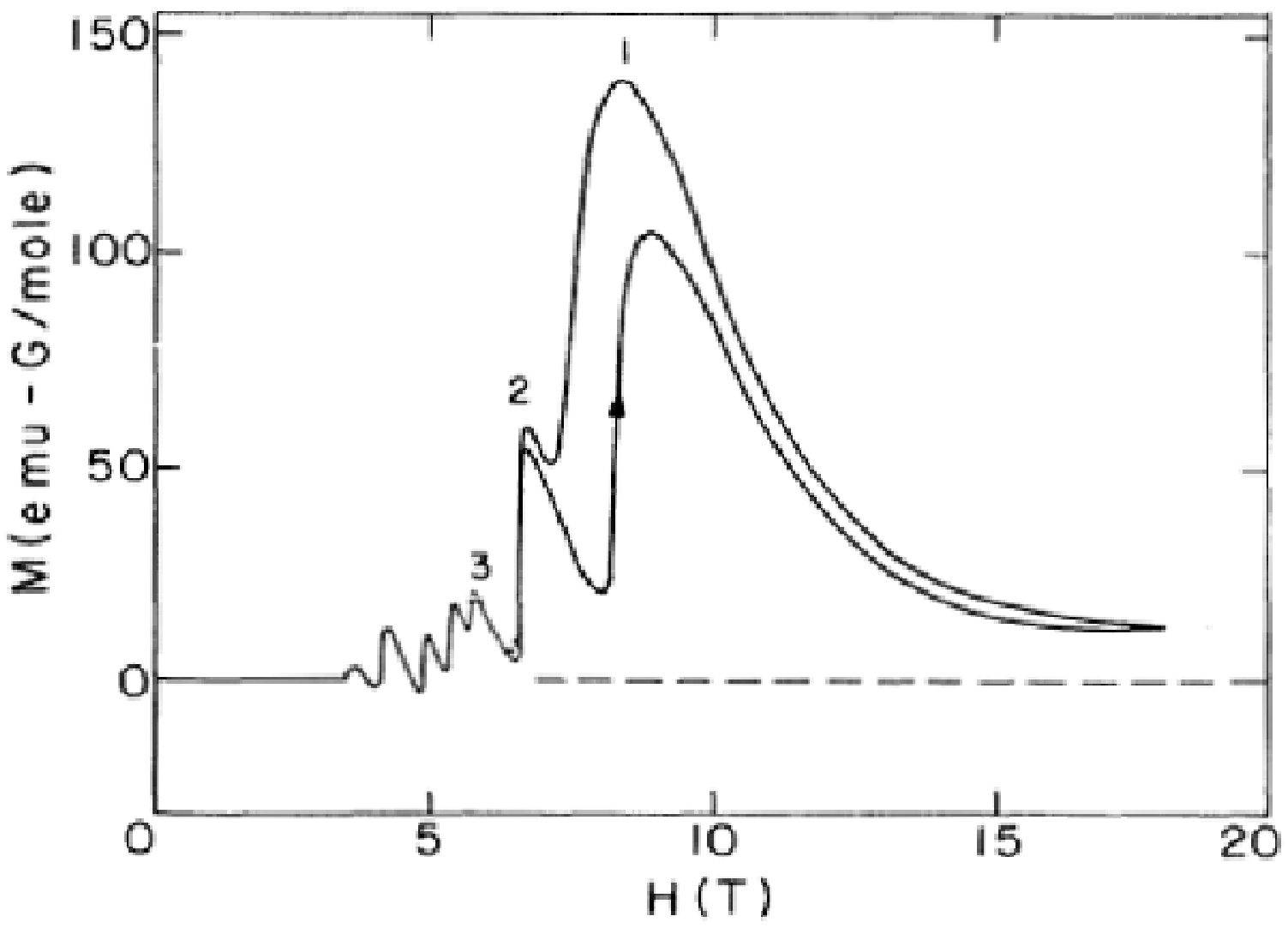}}
\subfigure[]{\label{fig9-3b}                    
\includegraphics[width=6.5cm]{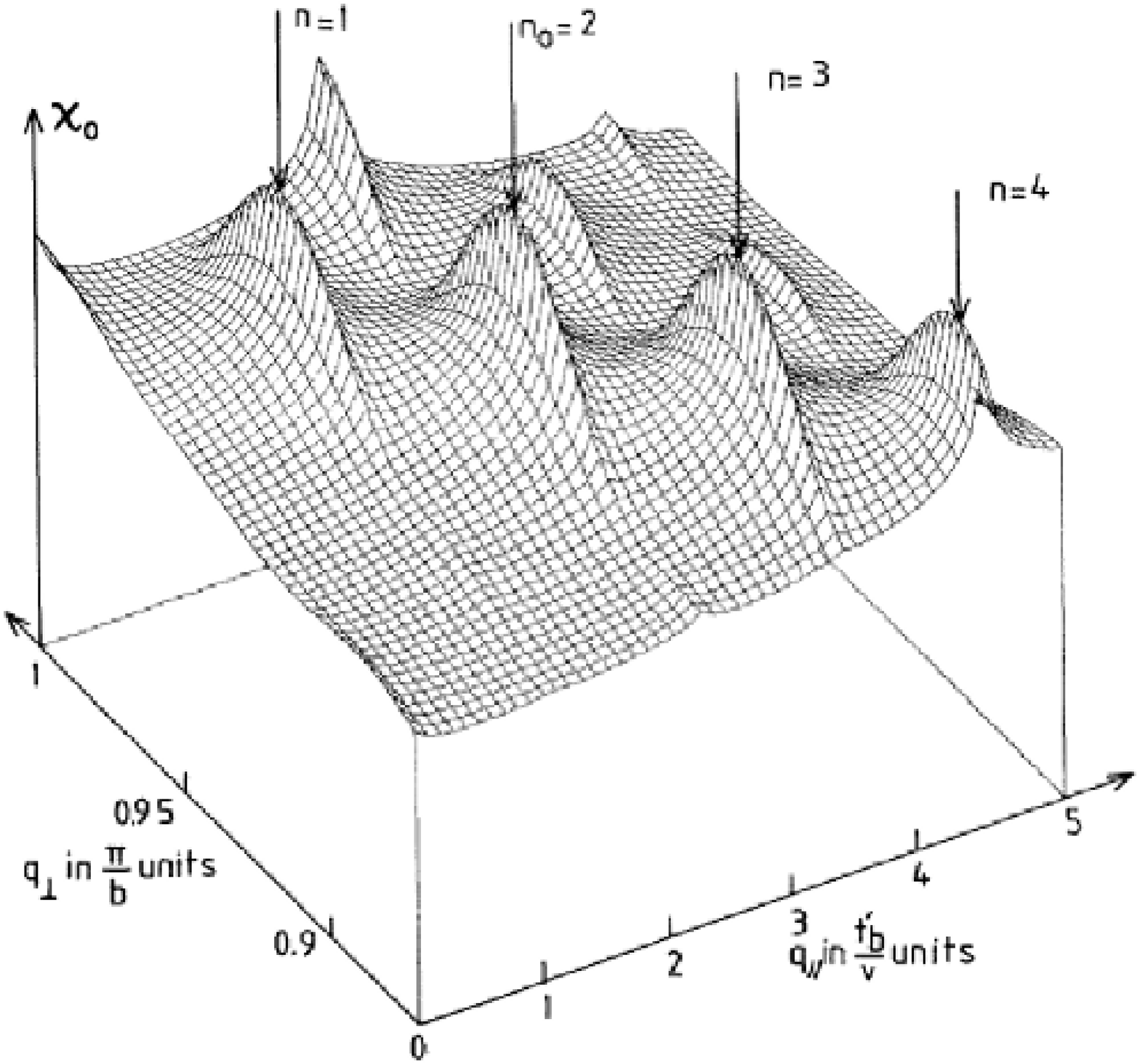}}
\caption{a)~Magnetic field dependence of the magnetisation of (TMTSF)$_2$ClO$_4$ in the relaxed state at $T$~= 60~mK (from ref.~\cite{Naughton85}); b)~Calculated spin susceptibility $\chi_0(Q)$ as  function of $Q$~= $(2k_{\rm F}+q_\parallel,\,q_\perp)$ in the case where $t'_b/t_b$~= 0.1. It exhibits peaks at quantised values of $q_\parallel$= $neHb/\hbar$ (reprinted figure with permission from G. Montambaux \textit{et al.}, Physical Review Letters 55, p. 2078, 1985 \cite{Montambaux85}. Copyright (1985) by the American Physical Society).}
\label{fig9-3}
\end{center}
\end{figure}

For an example, the magnetisation of (TMTSF)$_2$ClO$_4$ in the relaxed (slow cooling) state with $H$ applied nearly along $c^\ast$ direction is shown \cite{Naughton85} in figure~\ref{fig9-3}(a). The calculated \cite{Montambaux85} spin susceptibility, showing a series of peaks at wave vectors which obey a quantised nesting condition is drawn in figure~\ref{fig9-3}(b) in the case where $t'_b/t_b$~= 0.1. Similar cascade of phase transitions was also measured in thermodynamic experiments \cite{Pesty85}.

The increase of the SDW phase transition with $H$ evidences also the role of orbital effects in imperfect nesting conditions. The dependence of $T_{\rm SDW}$ on $H$ and the imperfect nesting parameter $t'_b$ was calculated \cite{Danner96} as:
\begin{equation}
T_c(H)=T_c(0)+f(\beta)\,\frac{\omega^2_c}{t^{'\ast}_b}.
\label{eq9-4}
\end{equation}
$f(\beta)$: a dimensionless function of the ratio $\beta$ between $t'_b$ and the value of the ``critical" imperfect nesting sufficient to destroy the SDW: $t^{'\ast}_b$.  Then, the field dependence of $T_{\rm SDW}$ of (TMTSF)$_2$PF$_6$ under applied pressure of 5.2~kbars ($H$ applied along $c$-axis) which reduces $T_{\rm SDW}(H=0)$ at around 3.2~K was shown \cite{Danner96} to follow a $H^2$ dependence (at 8~T, $T_{\rm SDW}$ reaches 4.5~K). No field dependence of $T_{\rm SDW}$ was observed with $H$ applied along $a$ and $b$-axis).

\subsection{Pauli paramagnetic limit}\label{sec9-2}

When a magnetic field is applied, the limit of stability of a condensate formed of pairs, either a superconductor or a density wave, occurs at a field at which the gain in energy in the metallic state due to the paramagnetism of electrons, $-\chi H^2$ with $\chi$~= $\mu^2_{\rm B}N(E_{\rm F})$ being the Pauli spin susceptibility, $N(E_{\rm F})$ the density of electron states of the normal state at $E_{\rm F}$, becomes equivalent to the DW condensate gap: $N(E_{\rm F})\Delta^2_0$. $T_{\rm CDW}$ is reduced to zero at the Pauli paramagnetic limit, or Chandrasekhar-Clogston limit \cite{Chandrasekhar62,Clogston62} without any orbital effect defined by:
\begin{equation}
B_{\rm P}=\frac{\Delta_0}{\sqrt{2}gs\mu_{\rm B}}=\frac{1.2~k_{\rm B}T_{\rm CDW}(0)}{\mu_{\rm B}}
\label{eq9-5}
\end{equation}
$\Delta_0$: the CDW gap at $T=0$, $g\cong 2$: the Land\'e factor, $\mu_{\rm B}$: the Bohr magneton, $s$: the electron spin.

The Zeeman splitting of the bands at $E_{\rm F}$ reduces the pairing interaction. The similarity between BCS and CDW ground states has led to predict \cite{Dieterich73} the field dependence of the CDW such as:
\begin{equation}
\frac{\Delta T_{\rm CDW}}{T_{\rm CDW}(0)}=-\frac{\gamma}{4}\left[\frac{\mu_{\rm B}B}{k_{\rm B}T_{\rm CDW}(0)}\right]^2.
\label{eq9-6}
\end{equation}
The Pauli reduction of $T_{\rm CDW}$ has been well demonstrated on CDW systems with a low CDW transition temperature such as $\mu_BB/k_{\rm B}T_{\rm CDW}\sim 1$ can be in the range of the accessible high magnetic fields. That is the case for the class of organic (Per)$_2$M(mnt)$_2$ systems, with $M$: Au, Pt. The structure of these compounds was described in section~\ref{sec3-4-2}. Let recall that a CDW distortion occurs on perylene chains at $T_{\rm CDW}$~= 12~K for M~= Au, and 8~K for M~= Pt. For Pt compounds, in addition to the CDW distortion, a spin-Peierls transition occurs on the paramagnetic anions in the M(mnt)$_2$ anion structure at the same temperature as the CDW.

\begin{figure}
\begin{center}
\includegraphics[width=7.5cm]{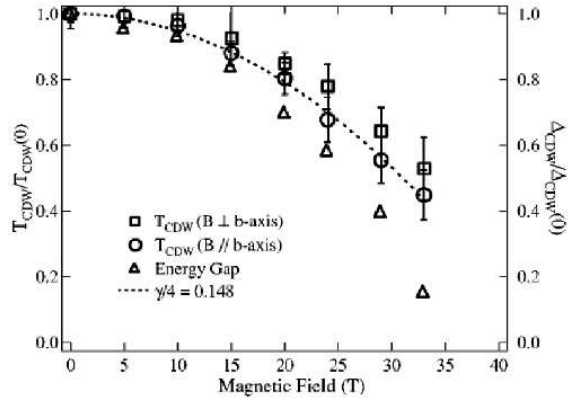}
\caption{Magnetic field dependence of $T_{\rm CDW}(B)/T_{\rm CDW}(0)$ for field parallel and perpendicular to the chain direction and of $\Delta_{\rm CDW}(B)/\Delta_{\rm CDW}(0)$ for field parallel to the chain axis for (Per)$_2$Au(mnt)$_2$ $T_{\rm CDW}(0)$~= 11.3~K, $\Delta_{\rm CDW}(0)$~= 88~K (reprinted figure with permission from D. Graf \textit{et al.}, Physical Review Letters 93, p. 076406, 2004 \cite{Graf04b}. Copyright (2004) by the American Physical Sociey).}
\label{fig9-5}
\end{center}
\end{figure}

The field dependence of $T_{\rm CDW}$ for $H$ applied parallel and perpendicular to the $b$-chain axis and $\Delta_{\rm CDW}$ for $H\parallel b$ is shown in figure~\ref{fig9-5} for (Per)$_2$Au(mnt)$_2$ \cite{Graf04a,Bonfait91,Graf04b,Matos96}. Up to $\sim 30$~T, the $B^2$ dependence, as expected from eq.~(\ref{eq9-5}), is well followed with $\gamma/4\simeq 0.148$. However at higher field (up to 45~K), exceeding the Pauli paramagnetic limit which can be estimated around 37~T from figure~\ref{fig9-5}, an inhomogeneous CDW phase was anticipated \cite{McDonald04} from the survival of CDW electrodynamics (non-linear transport properties, with a reduced threshold field), in agreement with theoretical models as discussed below.

\subsection{FICDW for perfectly nested Fermi surfaces}\label{sec9-3}

\begin{figure}[h!]
\begin{center}
\includegraphics[width=8.5cm]{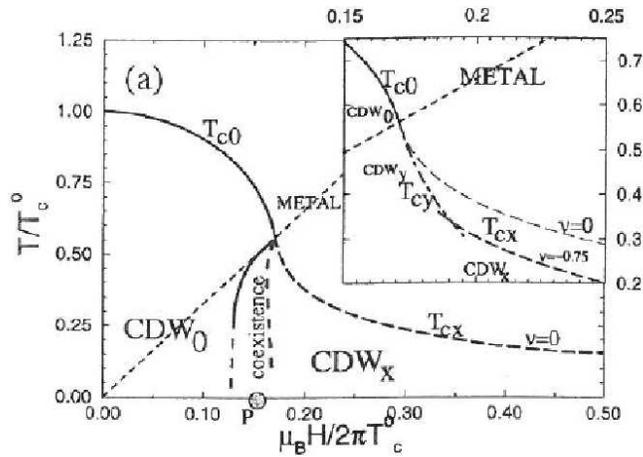}
\caption{$(H,T)$ phase diagram for a CDW in the perfect nesting case: $T$ normalised to the CDW temperature transition at $H=0$. Inset shows the stability of the (CDW)$_y$ state for negative values of SDW/CDW coupling constant ratio (reprinted figure with permission from A. Bjelis \textit{et al.}, Journal de Physique IV (France) 9, p. Pr10-203, 1999 \cite{Bjelis99}. Copyright (1999) from EdpSciences).}
\label{fig9-6}
\end{center}
\end{figure}

The phase diagram of a CDW under a magnetic field was derived \cite{Zanchi96,Osada06} from electron susceptibility calculations based on metallic-state electron functions in a mean field theory. Ignoring orbital effects, and thus considering the perfect nested case with $t'_b$~= 0 in eq.~(\ref{eq9-1}), it was shown that at enough low fields $h\equiv\mu_BH/2kT\leq h_c$~= 0.304, the CDW, called then CDW$_0$, keeps the perfect nesting wave vector $Q_0$~= $(2k_{\rm F},\,\pi/b)$, although the critical transition temperature $T_{\rm CDW}$ decreases. For $h>h_c$, the $Q$ vector shifts to an incommensurate value $Q$~= $Q_0\pm q_x(H)$ denoting a CDW$_x$ order with:
\begin{equation}
q_x=\frac{2\mu_BH}{v_{\rm F}}.
\label{eq9-7}
\end{equation}
Having taken into account the finite coupling between the CDW and the component of the SDW parallel to $H$ in the susceptibility response, the CDW$_x$ is considered \cite{Zanchi96} as a hybridised CDW-SDW state. The $(H,T)$ phase diagram \cite{Bjelis99} drawn in figure~\ref{fig9-6} defines a tricritical point when the phases metal, CDW$_0$ and CDW$_x$ joint together.

Inset of figure~\ref{fig9-6} shows, that under some conditions, a phase CDW$_y$ with a transverse shift of the wave vector may be stabilised in a very tiny part of the phase diagram. This appears only for negative values of the parameter $\nu$ which is the ratio between the SDW and CDW coupling constant. Near $T_{\rm CDW}$, the transition between CDW$_0$ and CDW$_x$ occurs near $T_c$ along the line $h_c$~= 0.34, but it deviates at low temperature towards the point $P$ defined as $\mu H\propto T_c(0)$. The transition between CDW$_0$ and CDW$_x$ is of the first order.

The incommensurate modulation of the CDW in the CDW$_x$ phase, with $q_x$ varying with $H$, shows some similarity with the superconducting state above the paramagnetic Pauli limit, labelled as the Fulde-Ferrel-Larkin-Ovchinnikov (FFLO) state \cite{Fulde64,Larkin64}. As noted \cite{Grigoriev05} there are, however, differences between both condensates. In particular, the lowest energy excitations in the CDW$_x$ phase may be soliton kinks \cite{Brazovskii81,Brazovskii84b}. A detailed study \cite{Grigoriev05} of the CDW phase diagram below the phase transition has shown that the CDW$_x$ in high magnetic field is the sum of $\cos [(Q_0+q_x)x+\phi_1]$ and $\cos [(Q_0-q_x)x+\phi_2]$ distortions.

The problem of ordering of two CDWs with nearly equivalent wave vectors has been previously studied either in the limit of very strongly coupled CDWs \cite{Brazovskii84,Brazovskii81} or in the limit of a weak coupling between them \cite{Bjelis86}. It was shown that, in the strong coupling limit, a single lattice deformation was associated with the CDWs in both electron bands, this deformation being viewed as a lattice of amplitude kinks, each band having two gaps with a mid-gap sub-band built by soliton electronic states. In the weak coupling limit, the CDW ordering is characterised by the stabilisation of a soliton lattice \cite{Bjelis86}. In the present case of Pauli splitting of the bands, the increase of the magnetic field increases the density of amplitude kinks and the order parameter at very high field tends continuously to a single phase with the same $Q_0$ as in zero magnetic field.

\begin{figure}
\begin{center}
\includegraphics[width=6cm]{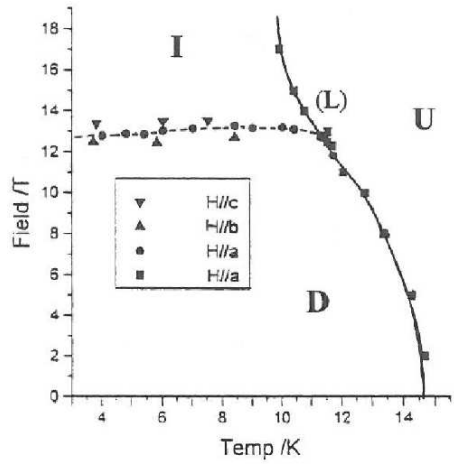}
\caption{Phase diagram of the spin-Peierls compound CuGeO$_3$ showing the three different phases: uniform (U), dimerised (D) and incommensurate (I), which meet at the Lifschift point L (reprinted figure with permission from W. Palme, G. Ambert, J.-P. Boucher, G. Dhalenne, and A. Revcoleski, Journal of Applied Physics 79, p. 5384, 1996 \cite{Palme96}. Copyright (1996) from American Institute of Physics).}
\label{fig9-7}
\end{center}
\end{figure}

Very similar phase diagram occur in the magnetic field dependence in spin-Peierls systems \cite{Bray83}, and in particular in CuGeO$_3$ \cite{Boucher96}. The application of a magnetic field induces a transition from the commensurate spin-Peierls state to an incommensurate modulated phase. This commensurate-incommensurate transition results from the competition between the spin-lattice coupling which favours the formation of a non-magnetic spin-Peierls state and the Zeeman energy which is minimised for states with a magnetic moment; it occurs at a critical field $H_c$ related to the spin-Peierls temperature at zero field and to the spin-gap between the ground state singlet and spin excitation triplets, identical to eq.~(\ref{eq9-5}) such as $H_c$~= $1.2k_{\rm B}T_{\rm SP}/g\mu_{\rm B}$. $T_{\rm SP}$ decreases with $H$ with the same $H^2$ dependence as in eq.~(\ref{eq9-5}). The experimental phase diagram obtained for CuGeO$_3$ is shown in figure~\ref{fig9-7} \cite{Palme96}. The transition lines between uniform (U)-dimerised (D) as well that between uniform (U)-incommensurate (I) are second order, while that between dimerised (D) and incommensurate (I) is first order with hysteresis.

The incommensurate phase has been expected to be formed by a stacking of dimerised regions regularly spaced by a 3D array of domain walls (solitons) carrying each one a spin 1/2 \cite{Meurdesoif99}. This soliton lattice has been experimentally demonstrated by the line shape of NMR spectra of Cu in the incommensurate phase \cite{Fagot96,Horvatic99}, as well by observation \cite{Kiryukhin96} of harmonics of the incommensurate Bragg reflections by X-ray scattering. The anharmonic modulation is directly related to the density of solitons, which at very large $H$ are overlapping, rending the incommensurate modulation harmonic.

\subsection{Fermi surface deformation by the pinned CDW structure}\label{sec9-4}

\begin{figure}
\begin{center}
\includegraphics[width=4.5cm]{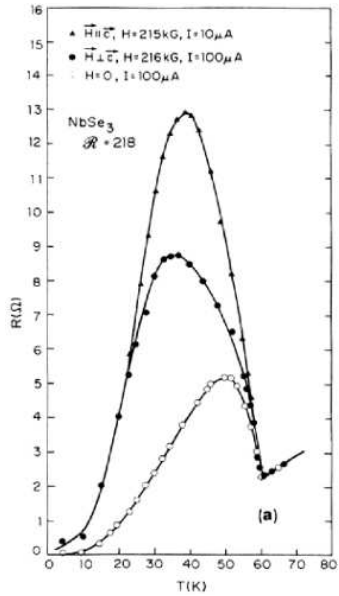}
\caption{Resistance enhancement below the low temperature CDW transition in NbSe$_3$ for the transverse magnetic field of $\sim 22$~T applied parallel and perpendicular to the $c$-axis (reprinted figure with permission from R.V. Coleman \textit{et al.}, Physical Review B 41, p. 460, 1990 \cite{Coleman90}. Copyright (1990) by the American Physical Society).}
\label{fig9-8}
\end{center}
\end{figure}

Effect of a magnetic field has been studied in NbSe$_3$. A strong resistance enhancement by $H$ was observed \cite{Coleman90} for all field directions applied in the ($a-c$) plane, with the largest effect with $H\parallel c$, as shown in figure~\ref{fig9-8}. 
This resistance enhancement does not change significantly in magnitude with impurity doping which reduces the resistance ratio RRR from $\sim 200$ to 20. It was suggested \cite{Balseiro85} that field-induced modifications of the electronic energy spectrum drive the Fermi surface in a more perfect nesting condition.
\begin{figure}
\begin{center}
\includegraphics[width=7cm]{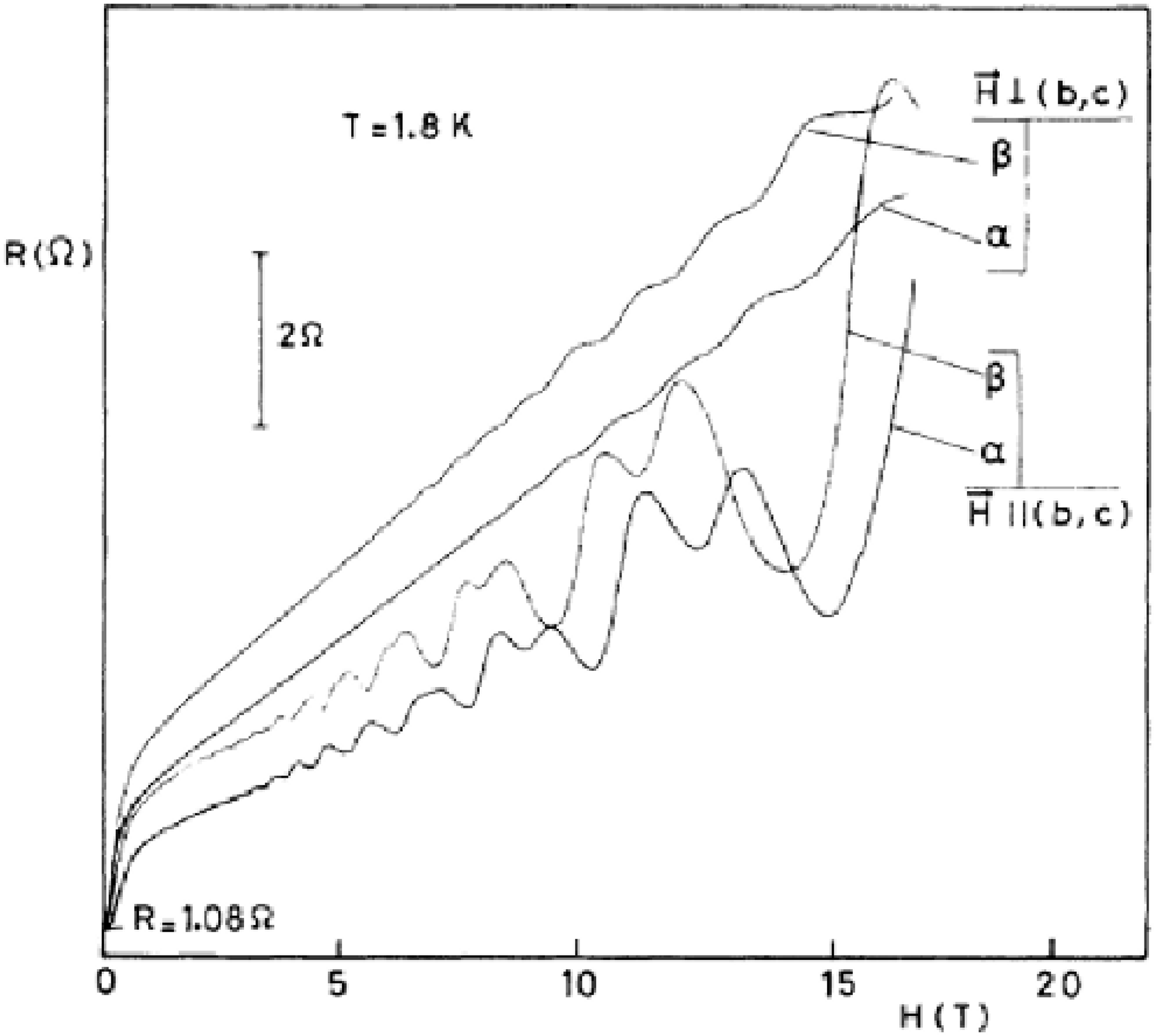}
\caption{Magnetoresistance of NbSe$_3$ at $T$~= 1.8~K with $H$ parallel and perpendicular to the ($b,c$) plane. Curves $\alpha$ correspond to the virgin state when the sample has just been cooled from room temperature. Curves $\beta$ are recorded after a current large enough to depin the CDW has been swept up and reduced to zero (reprinted figure with permission from J. Richard \textit{et al.}, Physical Review B 35, p. 4533, 1987 \cite{Richard87}. Copyright (1987) by the American Physical Society).}
\label{fig9-9}
\end{center}
\end{figure}

The existence of electron and hole pockets arising from the imperfect nesting at the low-$T$ CDW in NbSe$_3$ was revealed by magneto-quantum oscillations in magnetoresistance at low $T$ \cite{Coleman90,Richard87}, shown in figure~\ref{fig9-9}. These oscillations for $H\parallel c$-axis are dominated by a single frequency of $\sim 0.3$~MG. This frequency increases smoothly as $H$ is rotated is the $a-c$ plane and reaches $\sim 1$~MG for $H\perp c$  \cite{Monceau78,Coleman90}. The volume of this ellipsoid FS pocked is $\sim 1.2\times 10^{-3}$ the Brillouin zone volume with an effective mass of the carriers in the pocket deduced from the temperature dependence of the oscillation amplitude $m^\ast\sim 0.24m_e$. A huge oscillation in magnetoresistance is also observed at 38~T in the ultra quantum limit \cite{Audouard93}.

It was also found that the fundamental frequency with $H\parallel c$ and $I\parallel b$ can vary from 0.28 to 0.32~MG depending on the metastable states in the pinned CDW. Thus, this frequency measured after a first cooling is shifted to a lower value when the CDW has been depinned and repinned. Shifts also occur by reversing either the electric field or the magnetic field without depinning the CDW \cite{Everson87}. But the lowest value in the most homogeneous conditions is 0.28~MG.

The quantum oscillations show a large harmonic content at high field caused by spin splitting; but it should be noted that 80\% of the change in the total magnetoresistance comes from the oscillatory component, that indicates the presence of magnetic breakdown (MB) \cite{Everson87}. MB is extremely sensitive to changes in the Fermi surface cross-section. A beat structure at $\sim 2.5$~T was clearly detected in the $ac$ transverse magnetoresistance. This beat is produced by interference effects generated through MB. The beat structure disappears after depinning and repinning, rending the CDW ground state more homogeneous \cite{Everson87}.

It was proposed \cite{Coleman90} that, although the frequency of the magneto-oscillations is determined by small pockets of normal electrons, the main mechanism contributing to the oscillation amplitude is magnetic breakdown to open orbits existing on the Fermi surface sheets that nest to form the low-$T$ CDW phase. The normal electron pockets are closely coupled to the pinned CDW structure through the MB process, and this coupling is extremely sensitive to the pinned configuration of the CDW. Deformations of the CDW can shift the local Fermi level and change the extremal FS area of the pockets and the MB gaps. From the calculated component of the oscillatory conductivity of a single open orbit coupled to small lens-shaped closed orbits by MB \cite{Sowa85}, detailed configurations of metastable states in the pinned CDW were obtained \cite{Coleman90}.

Non-linear transport properties were measured under $H$ at low temperature \cite{Monceau88}. It was shown that the lower threshold field at which noise first occurs without a significant decrease in ${\rm d}V/{\rm d}I$ decreases above a threshold magnetic field. On the other hand, the second higher threshold field at which ${\rm d}V/{\rm d}I$ exhibits a strong discontinuity is independent of $H$. It was suggested a critical $T(H)$ line from this reduction of $E_c$ with $H$. Speculation was made on a possible variation of the nesting vector across this transition line. High resolution X-ray diffraction in magnetic field up to 10~T with an upper bound resolution of $\Delta Q/Q\leq 2.5\times 10^{-3}$ did not detect any shift of the $Q_2$ wave-vector \cite{Kiryukhin98}. From a preliminary experiment at the ESRF at $T$~= 2.5~K, $H$~= 8~T a relative change of the $Q_2$ satellite with respect to the (0 2 0) main Bragg peak was found of the order of 1.5--2$\times 10^{-4}$, slightly above the resolution of the measurements \cite{ESRF07}.

Finally the reduction of the threshold field $E_c$ is angular dependent, and can be relatively well fitted \cite{Monceau87b} with the relation:
\begin{equation*}
\frac{E_c[H_\parallel(b,c)]}{E_c[H_\perp(b,c)]}=(\cos^2\theta+\varepsilon^2\sin^2\theta)^{-1/2},
\end{equation*}
with $\varepsilon\sim 3$ at $H$~= 17~T. This anisotropic behaviour of $E_c$ indicates the role of orbital effect due to the magnetic field  on the spin susceptibility of the CDW condensate, as discussed hereafter.

\subsection{Interplay between Zeeman and orbital effects}\label{sec9-5}

\begin{figure}
\begin{center}
\includegraphics[width=7.5cm]{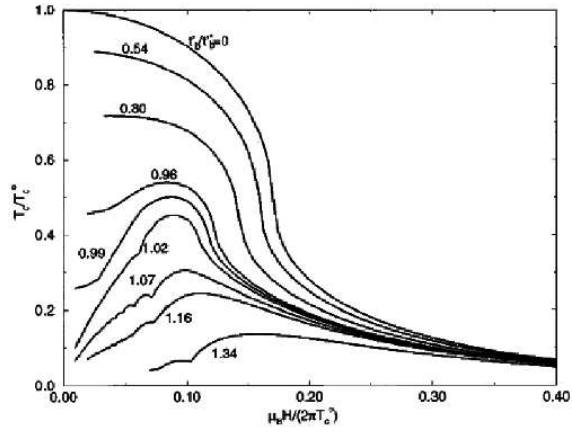}
\caption{Phase diagram showing interplay between Pauli effect which tends to suppress the CDW and orbital effects which enhances the CDW for different values of the imperfect nesting parameter $t'_b$ (reprinted figure with permission from D. Zanchi \textit{et al.}, Physical Review B 53, p. 1240, 1996 \cite{Zanchi96}. Copyright (1996) by the American Physical Society).}
\label{fig9-13}
\end{center}
\end{figure}

With imperfect nesting, $t'_b\neq 0$ in eq.~(\ref{eq9-1}), and as in the case of FISDW, orbital effects are implied. Phase diagram in these conditions is shown in figure~\ref{fig9-13} showing the magnetic field dependence of the critical temperature at different values of the imperfect nesting normalised to $t'^*_b$ (at which $T_{\rm CDW}$ is zero at zero field). The curve at $t'_b/t'^*_b$~= 0 is the same as that plotted in figure~\ref{fig9-6} for the parameter $\nu$~= 0. For low values of $t'_b$, the results obtained are similar to those in the perfect nesting case, the critical temperature being shift to lower temperatures.

The orbital effects enter into play at large values of $t'_b$ enhancing first $T_c$ at low field and suppressing it at high field. In the bad nesting conditions, namely at $t'_b/t'^*_b\sim 1$, a cascade of transitions associated with the FICDW phases occur \cite{Lebed03}. Taking into account both Pauli and orbital effects, the $Q(H)$ wave-vector is written as:
\begin{equation}
Q_x=2k_{\rm F}\pm\frac{2\mu_{\rm B}B}{\hbar v_{\rm F}}\pm N\frac{ebB\cos\theta}{\hbar}.
\label{eq9-8}
\end{equation}
The second term in eq.~(\ref{eq9-8}) is due to the Pauli effect (cf. eq.~(\ref{eq9-7})) and the last one due to the orbital effect (cf. eq.~(\ref{eq9-3})) where $B\cos\theta$ is the field component with $\theta$ the tilt angle between the field and the normal to the conducting plane.

\begin{figure}[h!]
\begin{center}
\includegraphics[width=6cm]{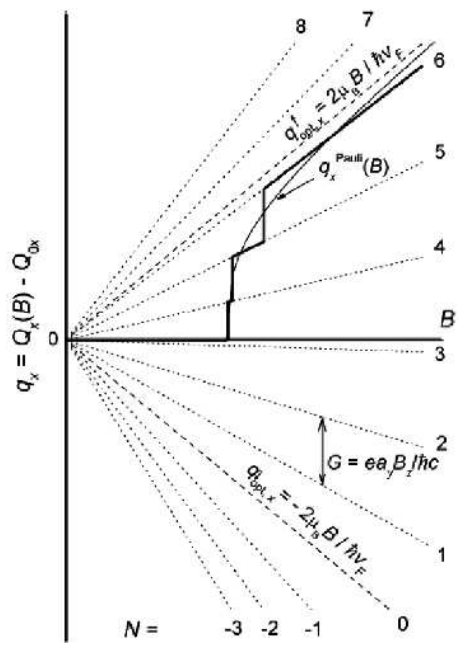}
\caption{Schematic illustration of the superposition of the Pauli effect thin solid line) and orbital quantisation (dotted lines) on the CDW nesting vector (thick line) (reprinted figure with permission from D. Andres \textit{et al.}, Physical Review B 68, p. 201101, 2003 \cite{Andres03}. Copyright (2003) by the American Physical Society).}
\label{fig9-14}
\end{center}
\end{figure}
\begin{figure}
\begin{center}
\includegraphics[width=6cm]{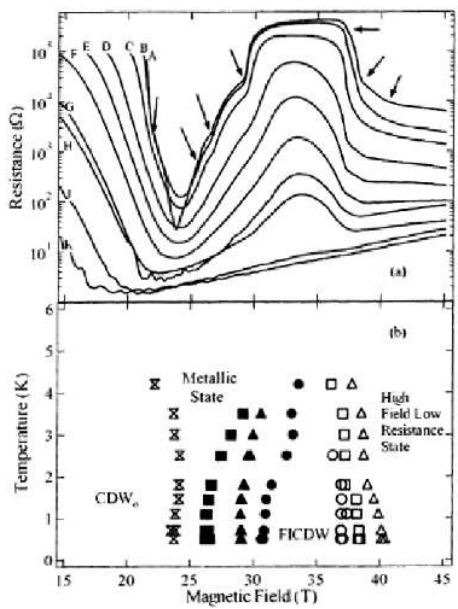}
\caption{a) Magnetoresistance of (Per)$_2$Pt(mnt)$_2$ with $B$ parallel to $c$-axis at temperatures 0.5, 1.1, 1.8, 2.5, 3.0, 3.5, 4.2, 5.0, 6.0 and 7.0~K respectively (following labels A-K). Arrows indicate phase boundaries and cascade-like anomalies. b)~proposed $T-B$ phase diagram based on magnetoresistance in a) showing the suppression of the low field CDW state CDW$_0$, the onset and termination of FICDW states (reprinted figure with permission from Synthetic Metals 153, D. Graf \textit{et al.}, p. 361, 2005 \cite{Graf05}. Copyright (2005) with permission from Elsevier).}
\label{fig9-15}
\end{center}
\end{figure}

A schematic illustration \cite{Andres03} of the superposition of the Pauli effect and orbital quantisation on the CDW nesting vector is shown in figure~\ref{fig9-14}. The most favourable values of $Q$ in the CDW$_x$ state above $B_{\rm P}$ are the intersection between the Pauli $q_x^{\rm Pauli}(B)$ line (which asymptotically approaches the $2\mu_{\rm B}B/\hbar v_{\rm F}$ value) and the straight lines defining the different quantised levels. A cascade of CDW subphases characterised by different quantised values of the wave-vector is obtained by varying $H$. To maximise the FICDW phase transition temperature, it was proposed that deviation of the FICDW nesting vector should be taken into account for both $x$ and $y$ axes \cite{Lebed09}.

While orbital effects arise in tilted magnetic field in (Per)$_2$Au(mnt)$_2$, the interplay between Zeeman and orbital effects is more apparent in (Per)$_2$Pt(mnt)$_2$ \cite{Graf04a,Graf04b,Graf05}. The magnetoresistance up to 45~T with $B$ parallel to the $c$-axis shows that, after the suppression of low field CDW phase at $B\sim 23$~T, a sharp upturn is observed in higher field exhibiting subphases indicated by arrows in figure~\ref{fig9-15}(a) before a plateau around 30~T. Reentrance to a low resistance state is also visible above 40~T. The resulting phase diagram is shown in figure~\ref{fig9-15}(b). These data are in qualitative agreement with the theoretical predictions. However the interaction between magnetic effects on the M(mnt)$_2$ anion chains exhibiting a spin-Peierls transition with the CDW perylene chains would need to be more precised.

A model was proposed \cite{Lebed07} for explaining the high resistance plateau at high field as shown in figure~\ref{fig9-15}(a). Due to 4 donors in the unit cell, the Fermi surface of (Per)$_2$Pt(mnt)$_2$ consists of 4 sheet \cite{Canadell04}. At relatively low field, due to Zeeman effect, a soliton superlattice state formed of $Q$~= $2k_{\rm F}\pm q_x$, two nesting vectors and two gaps, is stable. At a given higher field, a spin up of one branch can overlap with the spin down of another branch, resulting in  new nesting condition called spin improved nesting in ref.~\cite{Lebed07}. At higher field, there is restoration of the Peierls CDW phase with $Q$~= $2k_{\rm F}$ \cite{Lebed07}, supposed to be responsible for the stabilisation of the high resistance state. At even higher field, Zeeman effects destroy this CDW state.

Orbital and Zeeman effects have been also intensively studied in the 2D layered organic $\alpha$-(BEDT-TTF)$_2$KHg(SCN)$_4$ formed with alternating layers of (BEDT-TTF) donor molecules separated by acceptor anion layers \cite{Harrison00,Andres03,Brooks08,Kartsovnik09}. Phase diagram between CDW$_0$ and CDW$_x$ was traced \cite{Harrison00}. Increase of the CDW transition temperature with magnetic field was found in a certain range of pressure and magnetic field \cite{Andres01}. At ambient pressure a cascade-like transitions were observed in the magnetoresistance at different angles $\theta$ of $H$ with the $a$, $c$ plane \cite{Andres03}. By varying $\theta$, the orbital term (proportional to $\cos\theta$) changes while the Zeeman term is independent of $\theta$. It was also shown \cite{Kartsovnik09} that FICDW features are more pronounced if the spin splitting and the distance between the orbitally quantised levels become commensurate \cite{Lebed03}. This commensurability can be achieved by a specific angle defined from eq.~(\ref{eq9-8}) as
\begin{equation*}
\cos\theta_c=\frac{1}{N}\,\frac{2\mu_{\rm B}}{ev_{\rm F}b}.
\end{equation*}

\subsection{CDW gap enhancement induced by magnetic field in NbSe$_3$}\label{sec9-5-2}

The spectroscopy of CDW gaps and intragrap states have been studied in NbSe$_3$ using the interlayer tunnelling technique on mesa type layered nanostructures \cite{Latyshev07}, as it will be presented in detail in section~\ref{sec11-6}. Hereafter the magnetic effect on CDW gaps is presented. The mesa structures prepared by double-side etching have typically lateral dimensions 1x1~$\umu m$ and contains 20--30 elementary conducting layers.

Energy gaps, $\Delta_1$ and $\Delta_2$, of both CDWs in NbSe$_3$ manifest themselves \cite{Orlov08} as maxima in the ${\rm d}I/{\rm d}V(V)$ spectra measured in a low magnetic field ($2T$). 

When $T$ approaches $T_{{\rm P}_2}$, the peaks corresponding to $\pm 2\Delta_2$ merge into a wide maximum at zero bias, that is attributed to CDW fluctuations.

\begin{figure}
\begin{center}
\subfigure[]{\label{fig9-17a}         
\includegraphics[width=6.5cm]{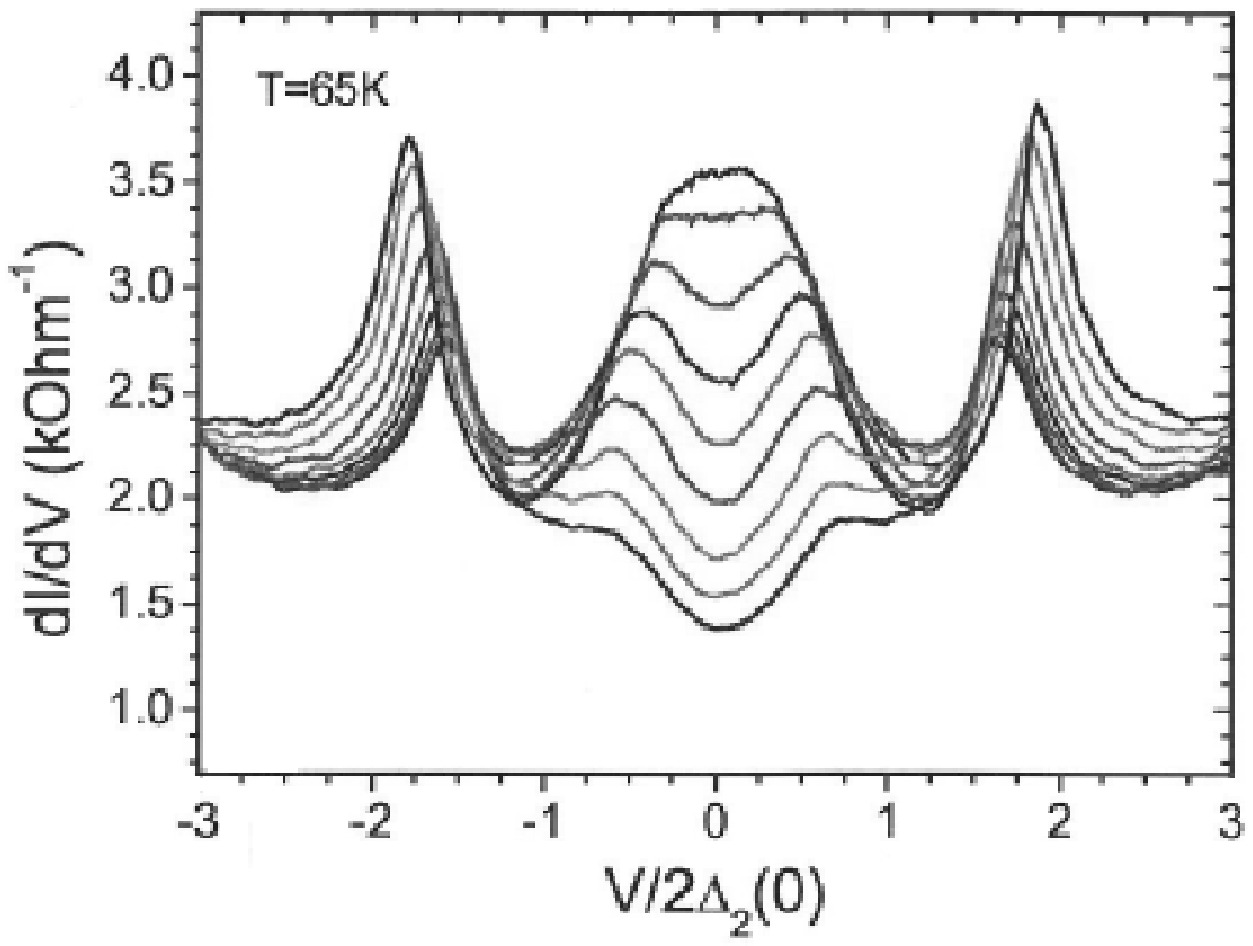}}
\subfigure[]{\label{fig9-17b}         
\includegraphics[width=6.75cm]{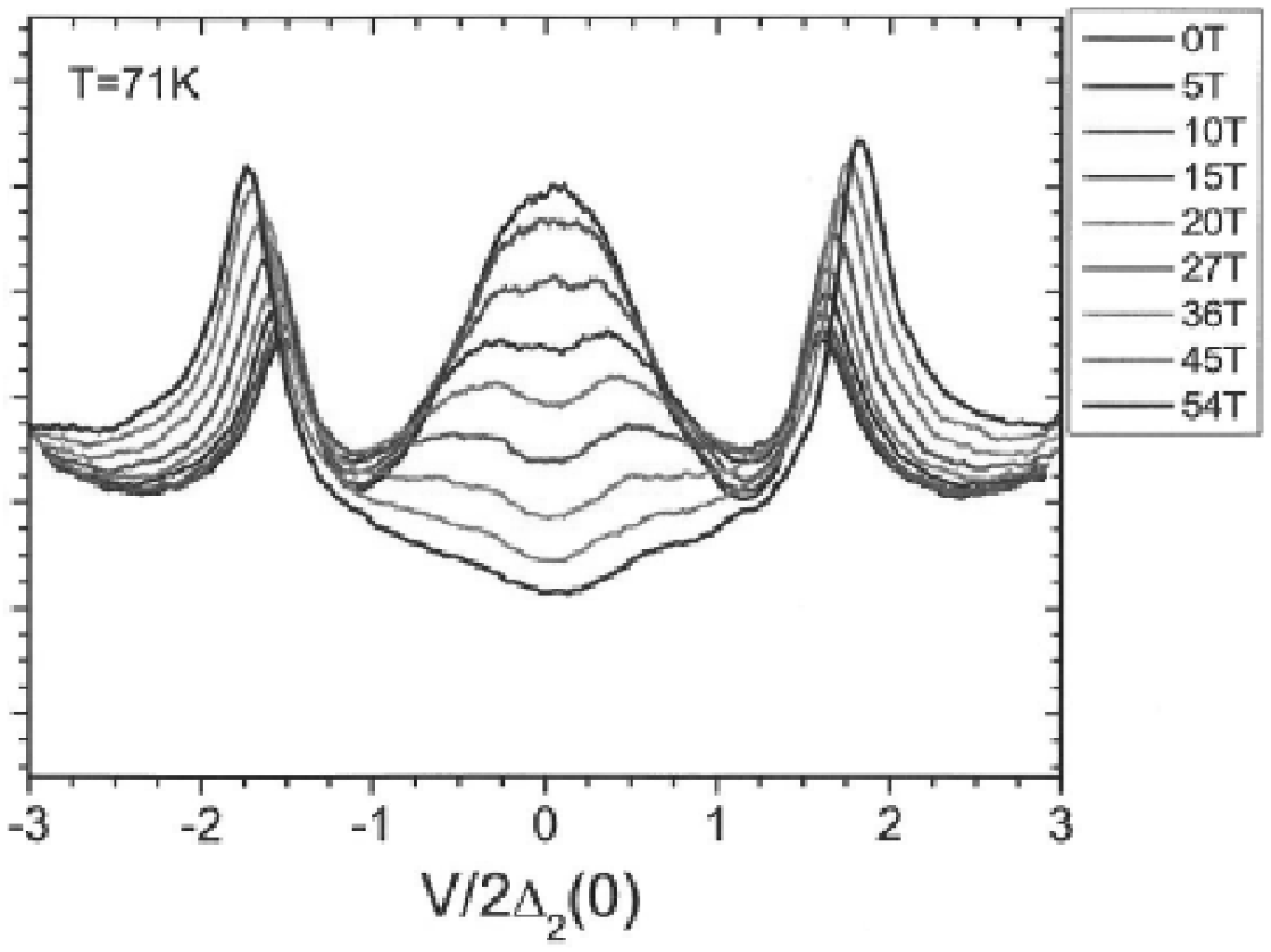}}
\caption{Interlayer tunnelling spectra in NbSe$_3$ in various magnetic fields $H\parallel a^\ast$. a)~$T$~= 65~K, b)~$T$~= 71~K (reprinted figure with permission from JETP Letters 87, A.P. Orlov \textit{et al.}, p. 433, 2008 \cite{Orlov08}. Copyright (2008) from Springer Science and Business media).}
\label{fig9-17}
\end{center}
\end{figure}
\begin{figure}
\begin{center}
\subfigure[]{\label{fig9-18a}         
\includegraphics[width=6.15cm]{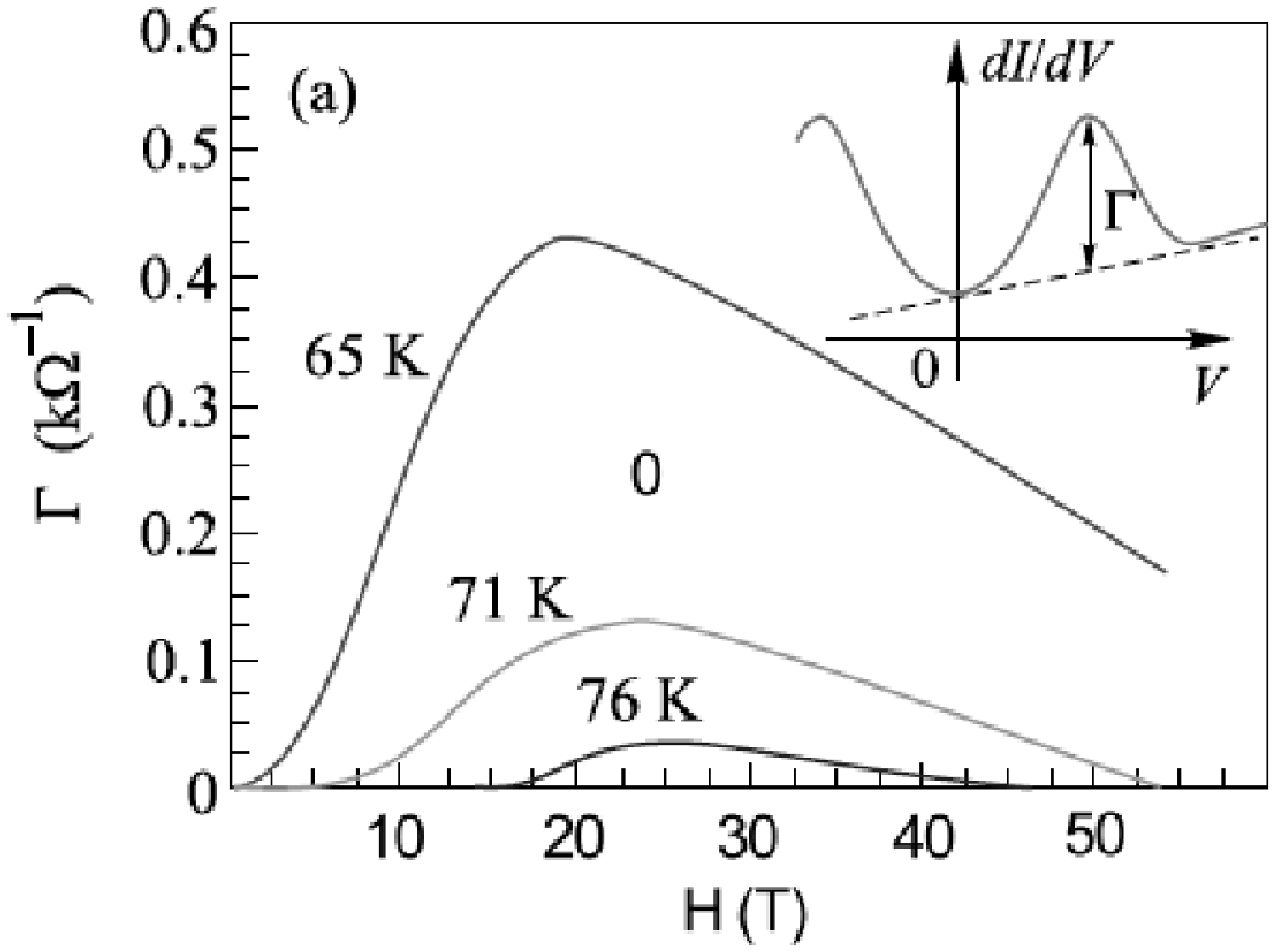}}
\subfigure[]{\label{fig9-18b}         
\includegraphics[width=7.25cm]{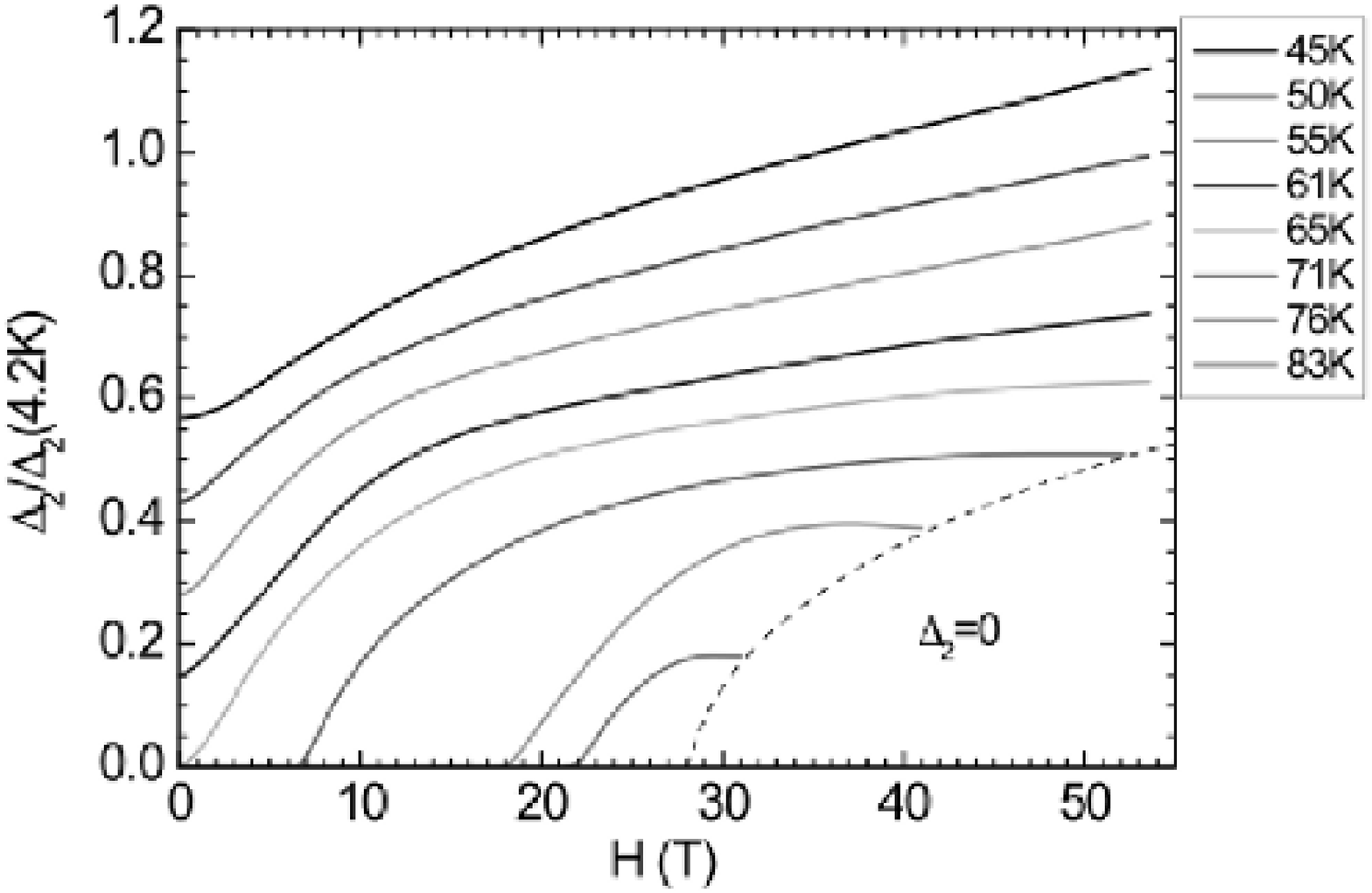}}
\caption{Magnetic field dependence of a)~amplitude, $\Gamma$, of the field-induced CDW gap singularity as defined in inset; b)~voltage at the gap peak at several temperatures below and above $T_{\rm P}$~= 59~K (at $H$~= 0). The dashed line marks the region where the gap singularity amplitude $\Gamma$ vanishes (reprinted figure with permission from JETP Letters 87, A.P. Orlov \textit{et al.}, p. 433, 2008 \cite{Orlov08}. Copyright (2008) from Springer Science and Business media).}
\label{fig9-18}
\end{center}
\end{figure}
\begin{figure}
\begin{center}
\includegraphics[width=7cm]{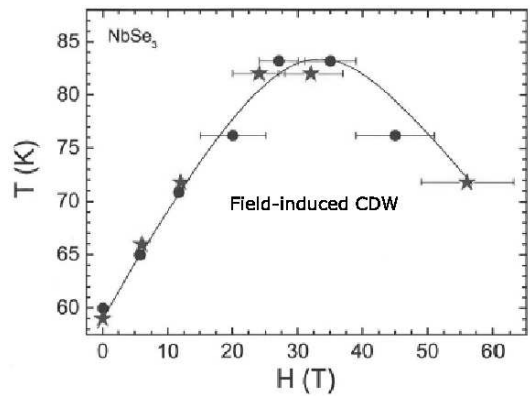}
\caption{Phase diagram of the field-induced CDW state in NbSe$_3$ (reprinted figure with permission from Yu.I. Latyshev \textit{et al.}, Physica B: Condensed Matter 404, p. 399, 2009 \cite{Latyshev09}. Copyright (2009) with permission from Elsevier).}
\label{fig9-19}
\end{center}
\end{figure}

Figure~\ref{fig9-17} shows the effect of a high magnetic field up to 55~T applied parallel to $a^*$ on the interlayer spectra ($I\parallel a^*$) at two temperatures above $T_{{\rm P}_2}(H=0)$~= 59~K. The magnetic field suppresses the density of states at zero bias, increasing the amplitude and the voltage position of the gap singularities. The restoration of the $\Delta_2$ CDW gap induced by $H$ is observed up to temperatures exceeding $T_{{\rm P}_2}$ by 20~K. However the gap peaks occur only within a certain magnetic field range. At $T$~= 71~K, as shown in figure~\ref{fig9-17}(b). $\Delta_2$ gap singularities appear at $H$~= 7~T and almost disappear at $H$~= 52~T.

This non-monotonous variation of the gap induced by a large field is demonstrated \cite{Orlov08} in figure~\ref{fig9-18}(a), which shows the field dependence of the amplitude $\Gamma$ of the field-induced gap singularity (as defined in inset of figure~\ref{fig9-18}(a): $\Gamma(H)$ first increases, has a maximum at about 25~T and then decreases at higher field. From figure~\ref{fig9-17} and data obtained at different temperatures, the amplitude of the voltage $V$ at the gap singularity position can be traced as a function of $H$, as drawn in figure~\ref{fig9-18}(b).

Finally the phase diagram of the magnetic field dependence of the CDW state of the lower CDW in NbSe$_3$ is plotted \cite{Orlov08} in figure~\ref{fig9-19}. One can see that at a given temperature above $T_{{\rm P}_2}$ in zero field, when $H$ is increased, a field induced CDW state occurs. This phase diagram is very similar to that shown in figure~\ref{fig9-13}, in the case of imperfect nesting conditions \cite{Zanchi96}. Due to orbital effects which improves the nesting between Fermi surface sheets, the lower CDW in NbSe$_3$ under a magnetic field occurs at higher temperature than in zero field. This effect is huge, since the CDW is still nearly stabilised at 80~K with 30~T. At higher field, the Zeeman effect destabilises the CDW, reducing the temperature at which it appears. The cross-over between orbital and Zeeman effects at the field $B_{\rm P}$ defined as $2\mu_{\rm B}B_{\rm P}\approx k_{\rm B}T_{\rm P}$  \cite{Zanchi96} agrees roughly with the experimental maximum $\sim 30$~T, as seen in figure~\ref{fig9-19}.

\begin{figure}
\begin{center}
\includegraphics[width=13.5cm]{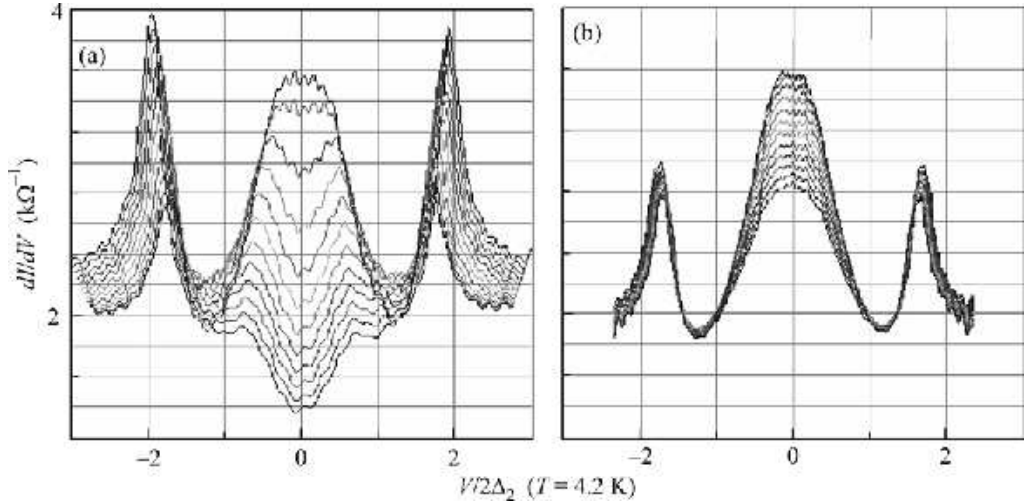}
\caption{Interlayer tunnelling spectra in NbSe$_3$ in various magnetic fields. a)~$H\parallel a^\ast$, b)~$H\parallel b$ (chain axis), demonstrating the orbital effect within enhances the CDW gap only $H$ perpendicular to the chain axis from top to bottom 0, 4.9, 9.8, 14.7, 19.7, 24.4, 29.3, 34.4, 39.1, 44, 48.9 and 53.7~T (reprinted figure from JETP Letters 94, Yu.I. Latyshev and A.P. Orlov, p. 481 (2011) \cite{Latyshev11}. Copyright (2011) from Springer Science and Business media).}
\label{fig9-20}
\end{center}
\end{figure}

The orbital origin of this interplay between orbital and Zeeman effects in NbSe$_3$ is even more emphasised by the influence of the field orientation on the FICDW. Gap enhancement above $T_{{\rm P}_2}(0)$ is only effective with a perpendicular component of $H$ along the chains and it is not observed with $H$ parallel to the $b$-axis \cite{Latyshev11} as demonstrated in figure~\ref{fig9-20}. These experiments confirm the imperfect nesting conditions for the low temperature CDW in NbSe$_3$.

Anomalies in the interlayer magnetoresistance at high field above 28~T have also been reported in the 2D KMo$_6$O$_{17}$ compound which exhibits a commensurable CDW transition at $T$~= 110~K. First order transitions to smaller gap states take place at low temperature above 30~T \cite{Guyot05,Balaska05}.

\section{Mesoscopy}\label{sec9}
\setcounter{figure}{0}
\setcounter{equation}{0}

The development of modern technologies has given the possibility of fabrication of DW structures in the submicron and nanometer scale. That has opened the way of studying quantum coherent properties of CDW conductors on a length scale comparable to or smaller than the phase and the amplitude ($\xi=\hbar v_{\rm F}/\pi\Delta$) correlation lengths of the CDW order parameter.

\subsection{Shaping}\label{sec11-1}

\subsubsection{Growth of thin films of Rb$_{0.30}$MoO$_3$}\label{sec11-1-1}

Thin films growth of Rb$_{0.3}$MoO$_3$, first reported in ref.~\cite{Zant96}, were fabricated using the pulsed-laser-deposition (PLD) method, which was developed earlier for the epitaxial growth of high $T_c$ copper oxides \cite{Chrisey94}. The stoichiometry and morphology of the deposited thin films was shown \cite{Zant96,Mantel97,Steinfort98} to depend on the thermodynamic conditions during growth (temperature of substrat, nature of substrat, oxygen pressure and deposition rate). The thin films are granular with the largest grains  when grown at high $T$ (350$^\circ$C$~<T<~$500$^\circ$C) and low deposition rate. The majority of grains of $\umu$-length has the chain $b$-axis parallel to the substrate plane. According to substrates --sapphire (Al$_2$O$_3$) or SrTiO$_3$-- the  morphology of films were studied \cite{Mantel97,Steinfort98} using X-ray diffraction (XRD), energy-dispersive analysis of X-rays (EDX), scanning electron microscopy (EM), transmission electron microscopy (TEM), atome-force microscopy (AFM).

Patterning of Rb$_{0.3}$MoO$_3$ films was performed by photolithography with the process described in ref.~\cite{Mantel99a}. Wires with widths down to 1~$\umu$m could be fabricated. Au contacts were realised \cite{Mantel99a,Mantel99b} either by optical lithography or for smaller contact spacings by electron-beam lithography. Voltage probes with a width as small as 400~nm and a spacing between adjacent voltage probes as small as 0.1~$\umu$m were realised \cite{Mantel99a}.

Characteristics of patterned Rb$_{0.3}$MoO$_3$ wires were studied by electric transport measurements. The resistivity of wires at room temperature was found to vary in a large extent; that was explained by the random-in-plane orientation of the grains on the Al$_2$O$_3$ substrate with, in addition, the presence of several grain boundaries.

\begin{figure}
\begin{center}
\subfigure[]{\label{fig11-1a}         
\includegraphics[width=6.5cm]{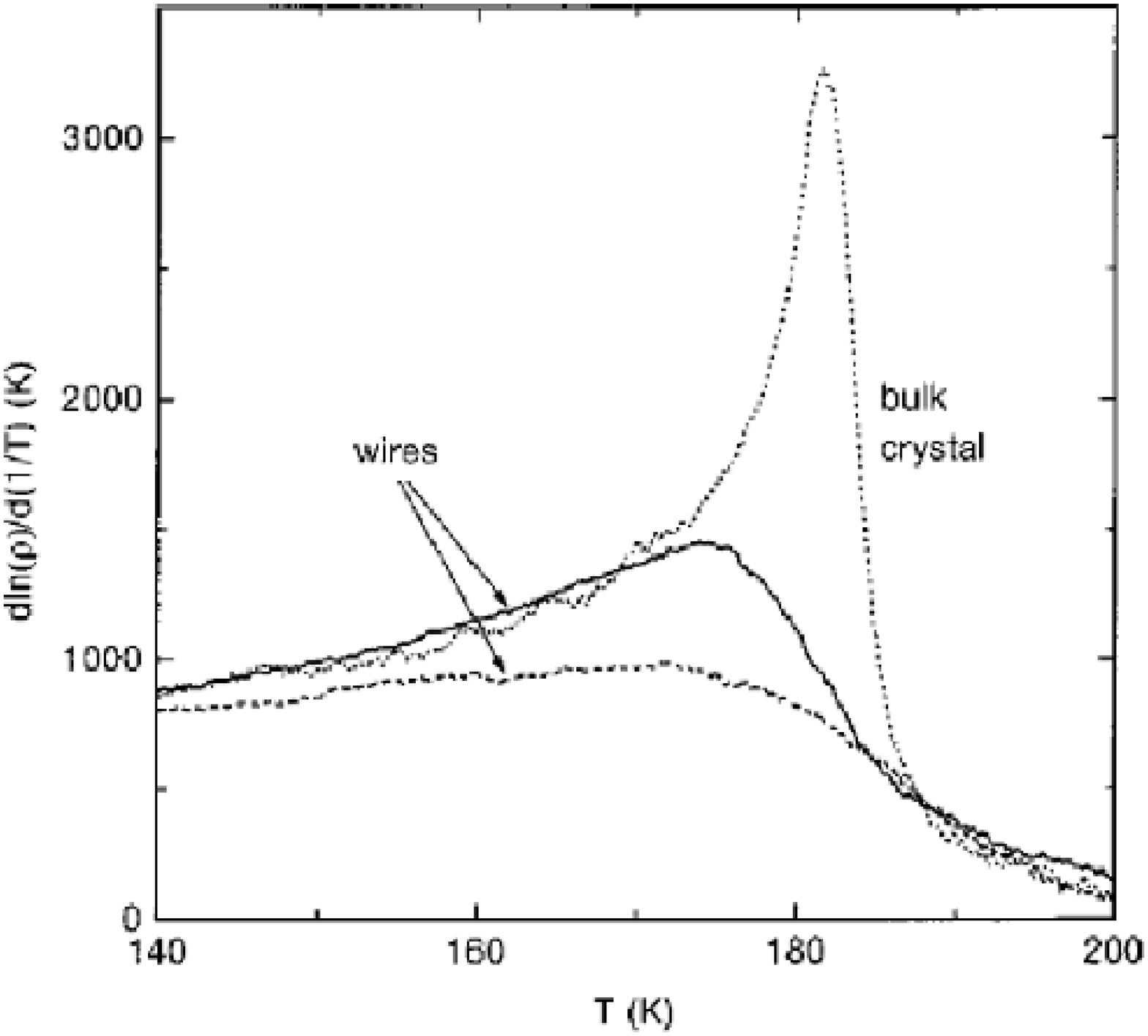}}
\subfigure[]{\label{fig11-1b}         
\includegraphics[width=6.5cm]{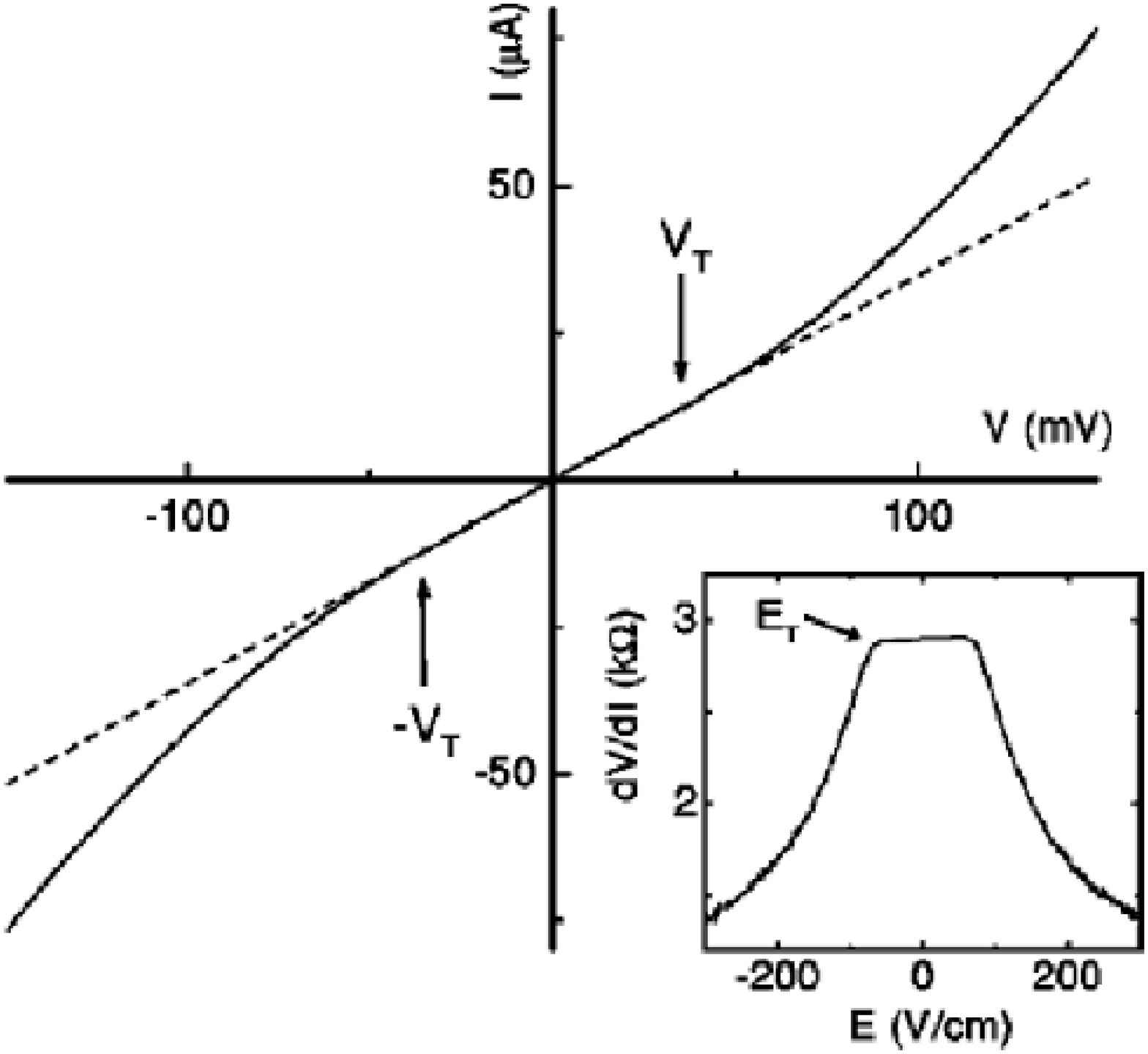}}
\caption{a) Temperature dependence of the logarithmic derivative ${\rm d}\ln(\rho)/{\rm d}(1/T)$ of the resistivity of patterned thin wires of Rb$_{0.3}$MoO$_3$, 2~$\umu$m wide and 0.3~$\umu$m thick with spacing, $d$, between voltage electrodes: 5~$\umu$m (solid line) and 1.5~$\umu$m (dashed line), and of a 0.2~mm thick single crystal with $d$~= 1~mm (dotted line) (reprinted figures with permission from O.C. Mantel \textit{et al.}, Physical Review B 60, p. 5287, 1999 \cite{Mantel99b}. Copyright (1999) by the American Physical Society). b)~I-V characteristics at 110~K of a patterned wire of Rb$_{0.3}$MoO$_3$ (2~$\umu$m wide, 0.3~$\umu$m thick, 5~$\umu$m interelectrode spacing). Inset shows the differential resistance, defining $E_T$~= 60~V/cm. $E_T$ increases with decreasing temperature such as $E_T$~= $E_T(0)e^{-T/T_0}$, with $T_0$~= 77~K and $E_T(0)$~= 300~V/cm (reprinted figures with permission from O.C. Mantel \textit{et al.}, Physical Review B 60, p. 5287, 1999 \cite{Mantel99b}. Copyright (1999) by the American Physical Society).}
\label{fig11-1}
\end{center}
\end{figure}

Films selected with a room temperature resistivity nearly equivalent to that of bulk crystals clearly exhibit a transition, although broadened, in the CDW state. Figure~\ref{fig11-1}(a) 
shows \cite{Mantel99a,Mantel99b} the temperature variation of the logarithmic derivative of the resistivity for two Rb$_{0.3}$MoO$_3$ wires of 2~$\umu$m wide, 0.3~$\umu$m thick and 5 and 1.5~$\umu$m voltage probe spacing respectively; and that of a 0.2~mm thick single crystal. Instead of a sharp peak at $T_{\rm CDW}$~= 182~K in the wires, the CDW transition occurs more gradually and at a slightly lower temperature, this behaviour being often found in disordered single crystals.

CDW sliding in these thin films was also observed \cite{Mantel99a,Mantel99b} as shown in figure~\ref{fig11-1}(b). However, the threshold field, $E_T$, for CDW depinning was found 2 at 3 orders of magnitude larger than in bulk crystals, typically in the range of 100~V/cm. Between 60 and 180~K, the $T$-dependence of $E_T$ can be fitted \cite{Mantel99a,Mantel99b} by an exponential variation $E_T$~= $E_T(0)e^{-T/T_0}$. This dependence is radically different to that measured in bulk crystals (see sec.~\ref{sec4-3-1}.c. and figure~\ref{fig4-21}), but with the same functional dependence than for NbSe$_3$, o-TaS$_3$, (TaSe$_4$)$_2$I. This large amplitude of $E_T$ in the patterned wires can be due to size effects as discussed in sections~\ref{sec4-3-1}.d. and \ref{sec5-4-2}  and also to surface roughness, resulting from the granular nature of the thin films.

Recently, using the same PLD technique, the growth of K$_{0.3}$MoO$_3$ 100~nm thick films has been developed \cite{Dominko11}. The morphology consists of nanocrystalline grains (200--300~nm) rending the room temperature resistivity very large; consequently no direct sign of the CDW state was visible in the temperature variation of the resistivity. However preliminary femtosecond time-resolved spectroscopy measurements \cite{Dominko11} on thin films exhibit features of the CDW ground state in nanocrystalline grains. Improvement in the CDW thin film quality would be of a great interest for time-resolved THz conductivity dynamics in transmission configuration as well as for femtosecond electron diffraction studies \cite{Demsar11}.

\subsubsection{Patterning of whisker crystals} \label{sec11-1-2}

The huge threshold field for CDW sliding in thin films is a difficulty which, at the present time, largely prevents the use of these films in mesoscopic devices. Another route with this aim was to take advantage of the whisker morphology of some CDW compounds and to pattern well selected whiskers. Transition metal trichalcogenides such as NbSe$_3$, o-TaS$_3$ are very well suitable with this respect. Thus, it is relatively easy to extract from the growth batch NbSe$_3$ whiskers 20--50~$\umu$m wide and down to 0.3~$\umu$m thick. Reduction of the thickness of whiskers can also be achieved using reacting ion etching.

The first patterning on a NbSe$_3$ single crystal was aimed to produce a periodic array of submicronic antidots \cite{Lorenzo98}. The technique was further developed \cite{Mantel99c}. By use of electron beam lithography, a given pattern is etched into the crystal with a SF$_6$ plasma. In a second step, contact Au pads are also patterned in a second e-beam lithography step.

\begin{figure}
\begin{center}
\subfigure[]{\label{fig11-3a}
\includegraphics[width=6.5cm]{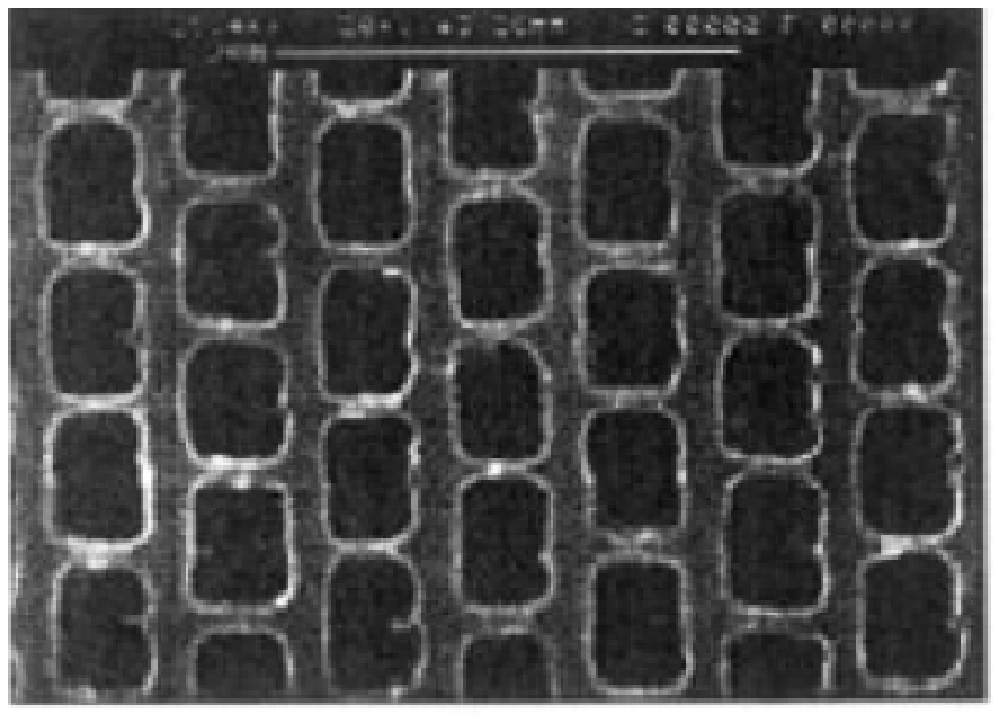}}
\subfigure[]{\label{fig11-3b}
\includegraphics[width=6.5cm]{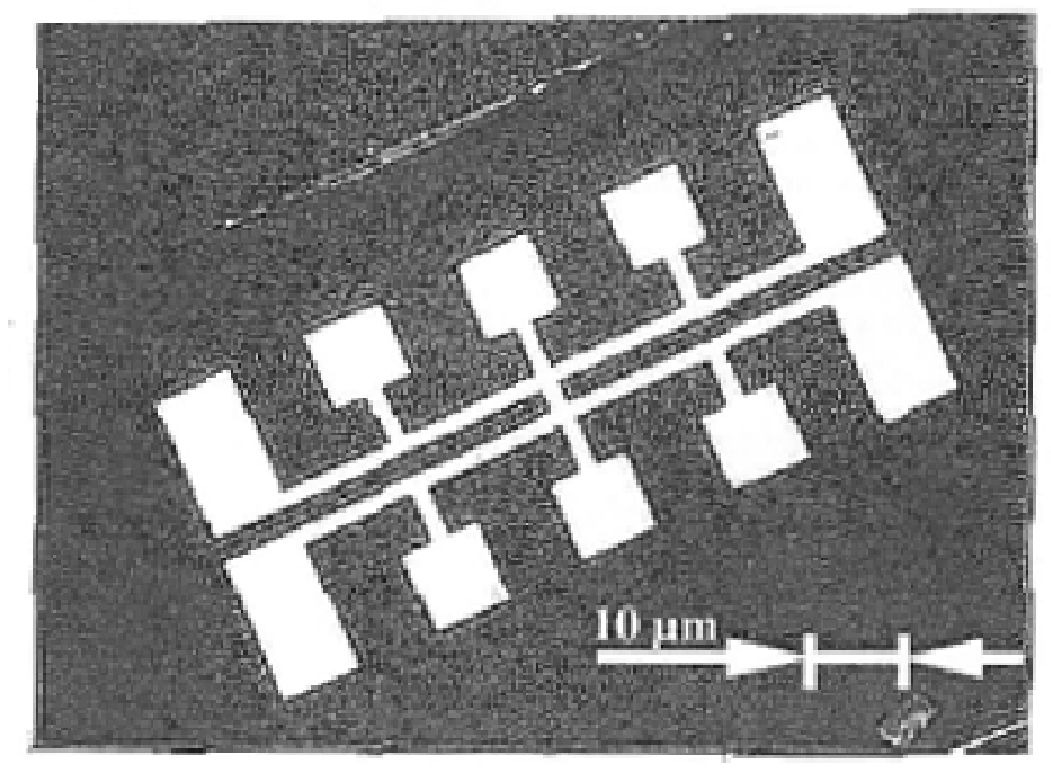}}
\subfigure[]{\label{fig11-3c}
\includegraphics[width=6.5cm]{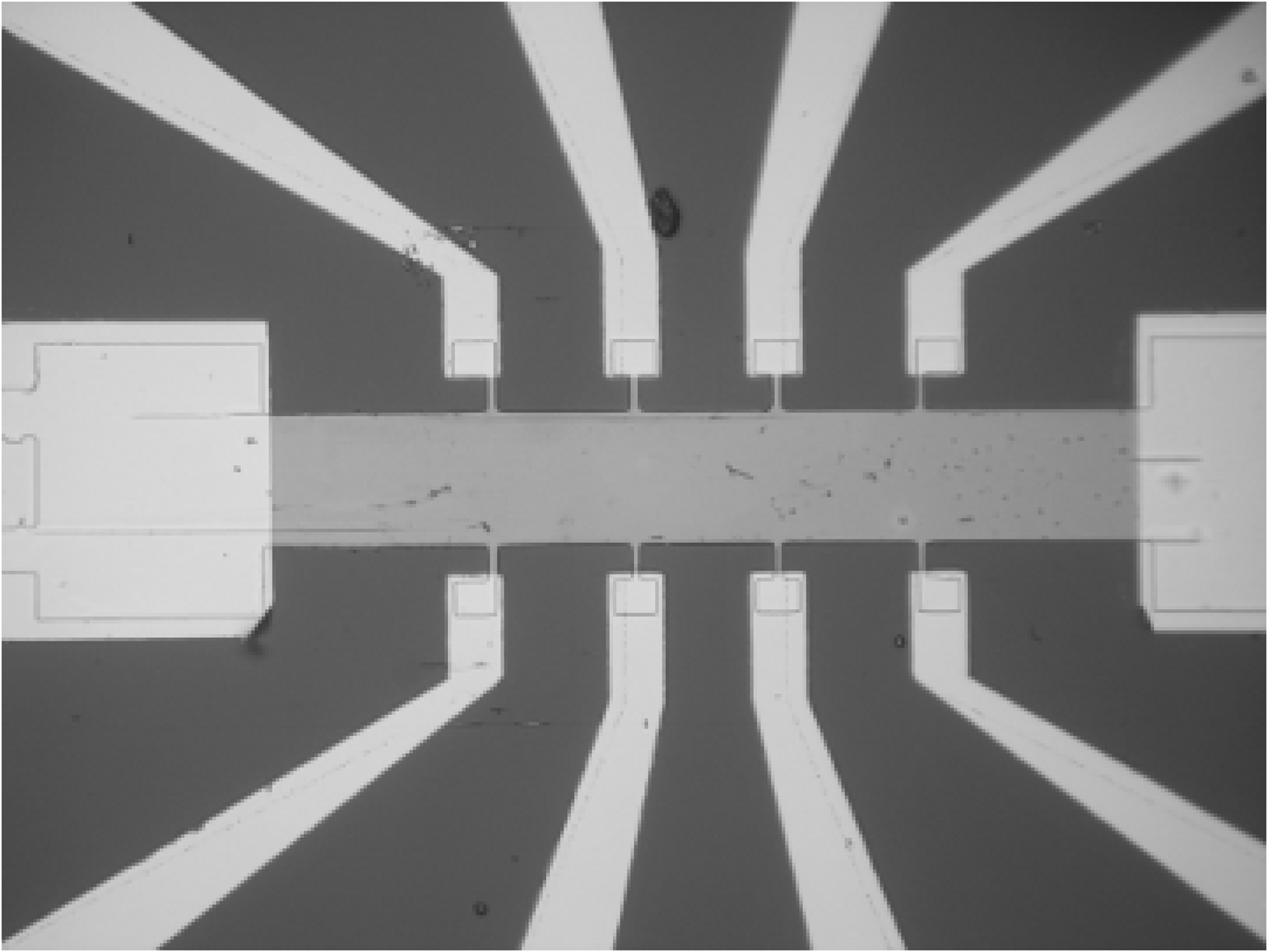}}
\subfigure[]{\label{fig11-3d}
\includegraphics[width=6.5cm]{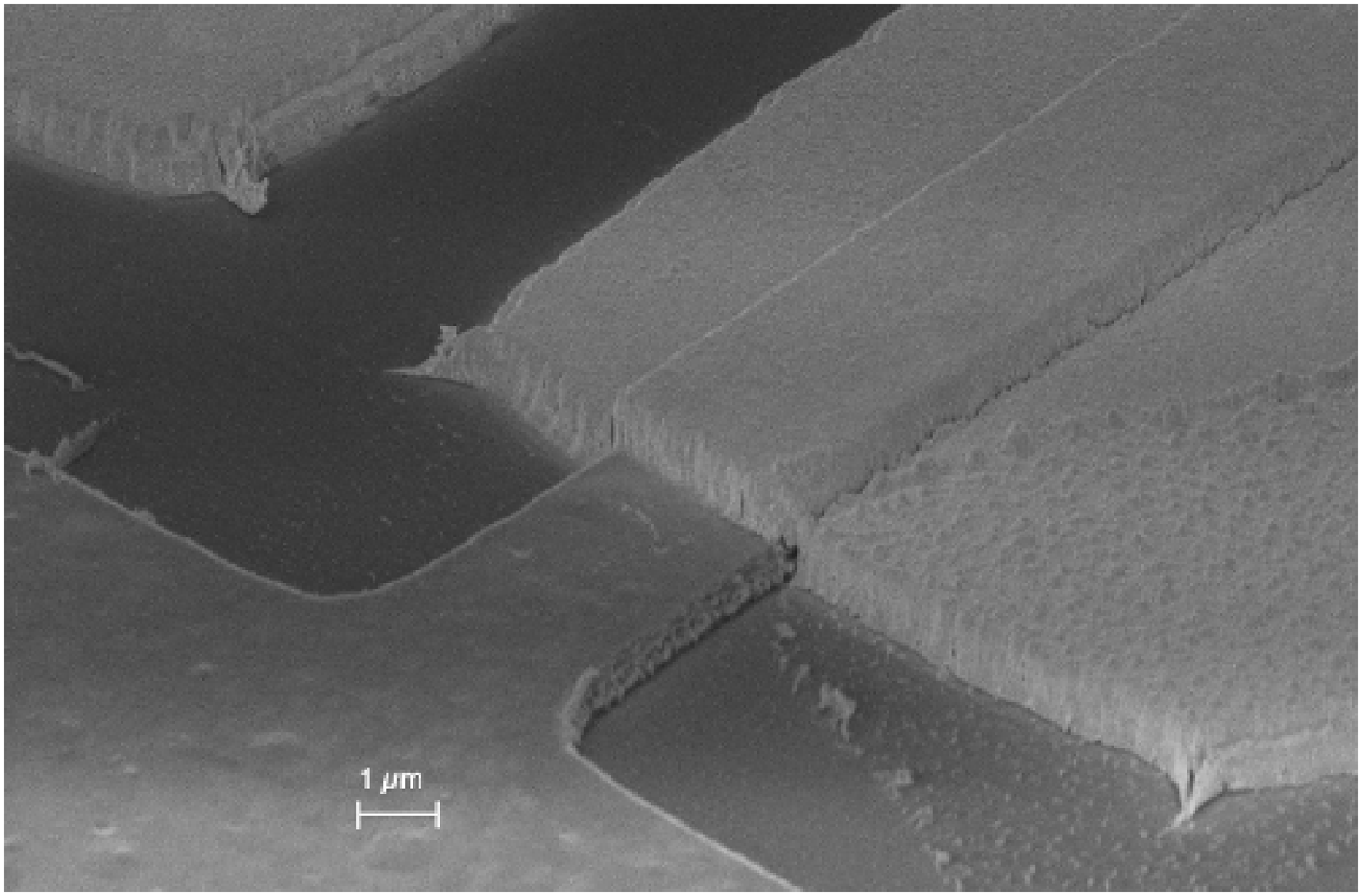}}
\caption{Selected structures patterned on NbSe$_3$ single crystals: a)~Antidot lattice with hole cross-section of 0.3~$\umu$m along $b$-axis and 0.4~$\umu$m along $c$-axis (reprinted figure with permission from the European Physical Journal B  - Condensed Matter and Complex Systems 3, Yu.I. Latyshev \textit{et al.}, p. 421, 1998 \cite{Latyshev98}. Copyright (1998) from Springer Science and Business media); b)~Structure for longitudinal and transverse current injection (reprinted figure with permission from A. Ayari and P. Monceau, Physical Review B 66, p. 23511, 2002 \cite{Ayari02}. Copyright (2002) by the American Physical Society); c)~Hall configuration on a crystal 0.265~$\umu$m thick, 20~$\umu$m wide, width of electrodes: 2~$\umu$m; d)~Scanning electron microscope image of the Au contact on the same sample as in c) (reprinted figure with permission from JETP Letters 93, A.A. Sinchenko \textit{et al.}, p. 56, 2011 \cite{Sinchenko11}. Copyright (2011) from Springer Science and Business media).}
\label{fig11-3}
\end{center}
\end{figure}

Figure~\ref{fig11-3} shows some typical structures patterned on NbSe$_3$: a triangular lattice of antidots with a rectangular hole cross-section of 0.3~$\umu$m along the $b$-axis and 0.4~$\umu$m along the $c$-axis \cite{Lorenzo98}, a sample with ten contacts for longitudinal and transverse current injection \cite{Ayari02}, a Hall configuration on a 0.265~$\umu$m thick and 20~$\umu$m wide crystal with the SEM image of one of the Au contacts shown in figure~\ref{fig11-3}(d) \cite{Sinchenko11}. The advantage of the layouts b, c, d, where etching side contacts are made in the crystal itself, is the negligible shunting of current through the voltage probes.

\begin{figure}
\begin{center}
\subfigure[]{\label{fig11-4a}
\includegraphics[width=7.5cm]{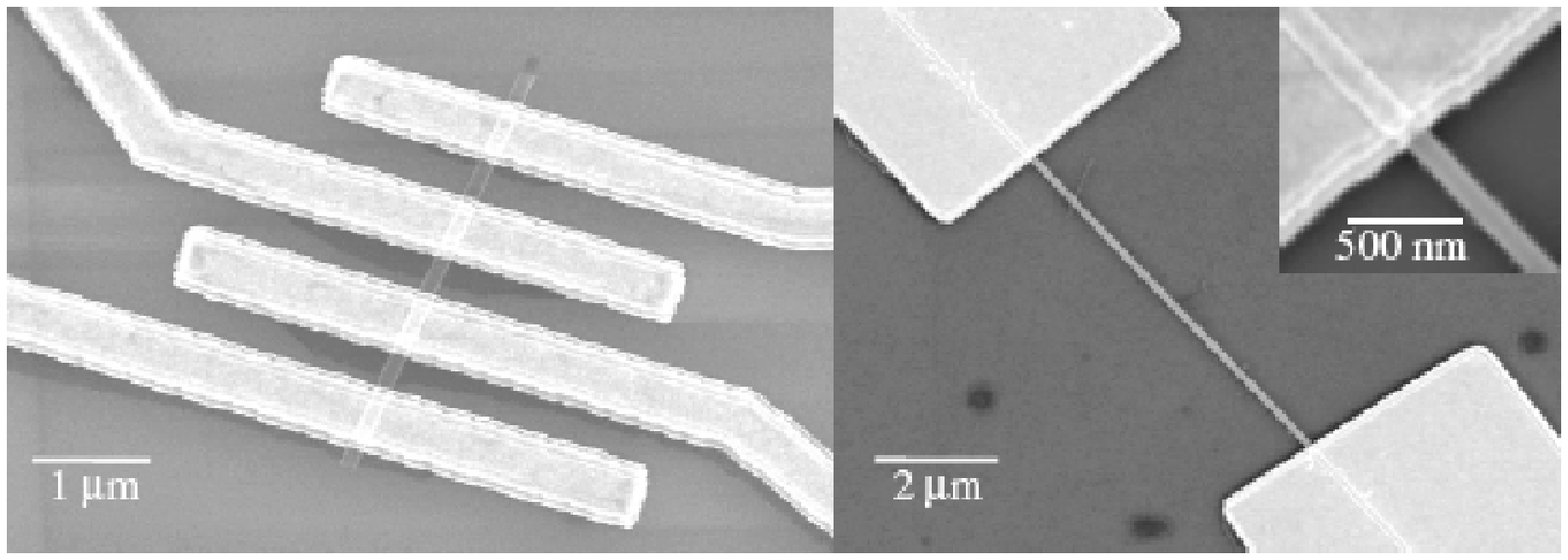}}
\subfigure[]{\label{fig11-4b}
\includegraphics[width=5cm]{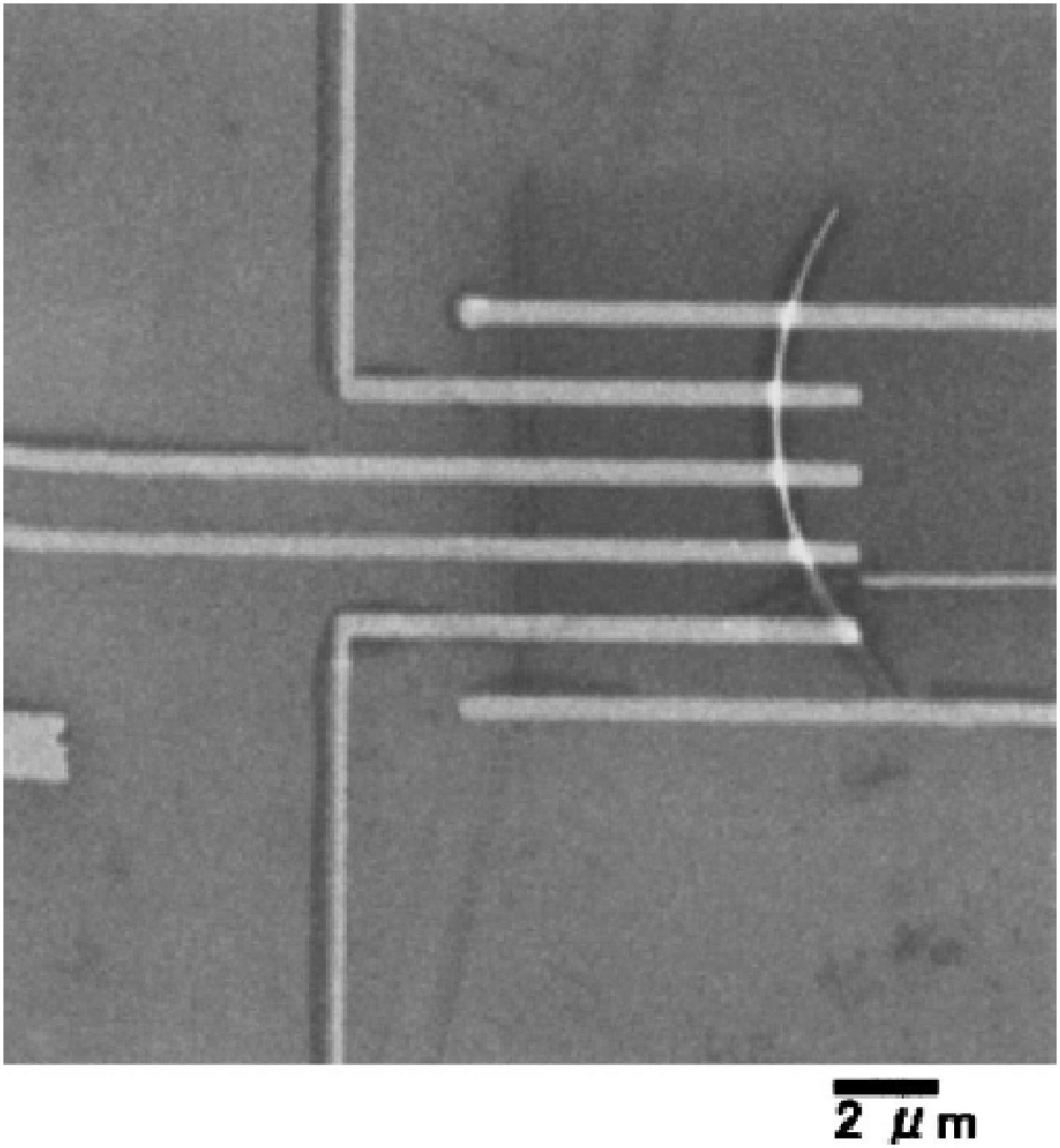}}
\vspace{-0.25cm}
\caption{Scanning electron microscope images of : a)~NbSe$_3$ nanowires with 4 and 2 probes (reprinted figure with permission from E. Slot \textit{et al.}, Physical Review Letters 93, p. 176602, 2004 \cite{Slot04}. Copyright (2004) from the American Physical Society); b)~o-TaS$_3$ nanowire (reprinted figure with permission from K. Inagaki, T. Toshima, S. Tanda, and K. Yamaya, Applied Physics Letters 86, p. 073161, 2005 \cite{Inagaki05}. Copyright (2005) from American Institute of Physics).}
\label{fig11-4}
\end{center}
\end{figure}

\subsubsection{Nanowires}\label{sec11-1-3}

Nanowires were obtained from bulk NbSe$_3$ and o-TaS$_3$ bulk crystals by ultrasonically cleaving in either a pyridine solution for NbSe$_3$ \cite{Slot04} or toluene for o-TaS$_3$ \cite{Inagaki05}. Size of nanowires ranges from 30~nm to 300~nm and length from 2 to 20~$\umu$m. A drop of the suspension is deposited onto a silicon substrate with predefined markers. Contacting is made by e-beam lithography on selected nanowires with respect to these markers. Scanning electron microscope images of NbSe$_3$ nanowires \cite{Slot04} and o-TaS$_3$ nanowires \cite{Inagaki05} are shown in figure~\ref{fig11-4}.

\subsubsection{Focus-ion beam (FIB) technique} \label{sec11-1-4}

\begin{figure}
\begin{center}
\includegraphics[width=12cm]{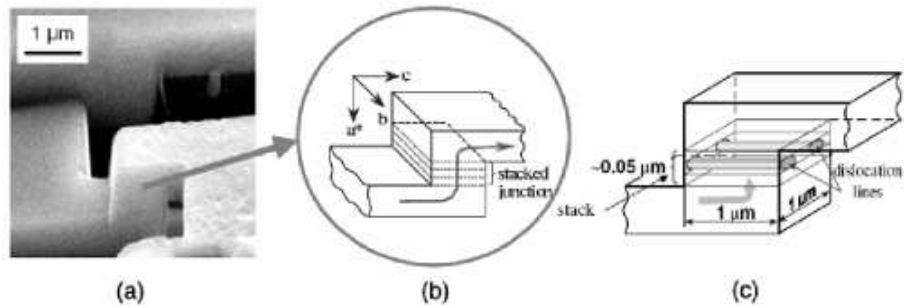}
\caption{Scanning electron microscope image of a junction tailored in a NbSe$_3$ whisker (a) and its scheme (b). The stepwise arrowed line shows the current in the junction area. (c)~Schema of the tunnelling device with the current injected along $c$-axis. $a^\ast$, $b$, $c$ are the crystallographic axes of NbSe$_3$, $b$:~the chain direction (reprinted figure with permission from Yu.I. Latyshev \textit{et al.}, (a)+(b): Physical Review Letters 95, p. 266402, 2005 \cite{Latyshev05}; (c): Physical Review Letters 96, p. 116402, 2006 \cite{Latyshev06}. Copyright (2005,2006) by the American Physical Society).}
\label{fig11-5}
\end{center}
\end{figure}

The FIB technique, very accurate and less-time consuming than e-lithography was also used. That is specifically the case for studying transport properties across the ($b,c$) plane of NbSe$_3$ with submicronic overlap junctions. Technically, a square trench of typical size $6\times 6~\umu$m$^2$ is etched on one side of a thin whisker by FIB to the depth of $d/2+\delta$ where $d$ (typically less than 1~$\umu$m) is the thickness of the crystal and $\delta$ a small excess depth (typically $\delta$~= (0.05--0.1)$d$. At the second stage, the crystal is turned over and a second trench is etched on the opposite side of the crystal near the first one to the same depth. The junction, called a mesa or an overlap junction, with thickness of $2\delta$ is thus formed between the trenches. Two types of devices were fabricated: a)~one with a longitudinal geometry where the current enters the junction via the in-plane direction $b$ along the chain axis (see figure~\ref{fig11-5}(b)) \cite{Latyshev05}; 
b)~the other with a transverse geometry where the current, before entering the junction, changes in the plane direction to the $c$-axis transverse to $b$ \cite{Latyshev06}. This latter configuration shown in figure~\ref{fig11-5}(c) prevents the interference between the interlayer tunnelling and a possible CDW sliding within connecting channels; the typical lateral sizes are 1~$\umu$m by 1~$\umu$m and the thickness along $a^\ast$ between 0.05--0.2~$\umu$m (self-heating is largely reduced with a decrease of lateral sizes down to micron scale).

\subsubsection{Topological crystals}\label{sec11-1-5}

\begin{figure}
\begin{center}
\includegraphics[width=12cm]{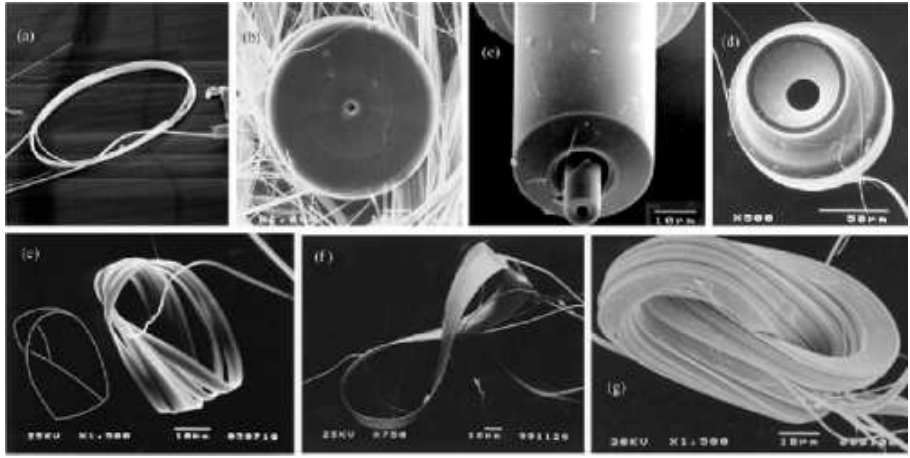}
\caption{Scanning electron microscope images of NbSe$_3$ topological crystals with typical diameters between 10--200~$\umu$m: a)~a basic ring or a thin short cylinder; b)~a cylinder that grew radially; c)~a cylinder grew longitudinally; d)~concentric composite cylinders with a continuous or abrupt change in diameter; e)~a M\"obius strip with a $\pi$-twist; f) and g)~a $2\pi$ twisted strip (reprinted figure with permission from T. Tsuneta and S. Tanda, Journal of Crystal Growth 264, p. 223, 2004 \cite{Tsuneta04}. Copyright (2004) with permission from Elsevier).}
\label{fig11-6}
\end{center}
\end{figure}

A topological deformation such as rolling a planar sheet into a cylinder, or forming a spherical structure (such as for instance for carbon-based materials : fullerenes \cite{Kroto85} or carbon nanotubes \cite{Ijima91}) modifies significantly the physical properties. By varying the typical growth conditions, topological crystals of NbSe$_3$ (and later o-TaS$_3$, TaSe$_3$) with different shapes, as shown in figure~\ref{fig11-6}, were prepared \cite{Tanda02}. A large temperature gradient creates a non-equilibrium state inside the quartz tube, which yields a strong convection of selenium with different-vapour and liquid (droplet) phases.

The different forms (rings, disks, and tubes) were explained from a growth model such as the NbSe$_3$ whisker is bent by the surface tension of the selenium droplet on which it is grown \cite{Tanda02,Tsuneta04}. The monoclinic structure promotes also bending with twisting. The different morphology can be classified into three classes according to the degree of twist: $0$: a single ring, $\pi$ (M\"obius crystal), $2\pi$ twist formed by a double encirclement around a Se droplet. For a ring, typical dimensions are 30--50~$\umu$m in diameter and 1~$\umu$m in width. For a M\"obius crystal: 50~$\umu$m in diameter and 1~$\umu$m in width. For a double twist: 200~$\umu$m in diameter. X-ray diffraction and transmission microscope diffraction have confirmed that the ring and M\"obius structure are equivalent to that of a crystalline ribbon, but with a large strain. The ring crystals are plastically deformed by mechanical compression due to the bending. That is reflected by significant variation of the lattice spacings with respect to those of whiskers. Disorder in the structure is also revealed by a relatively low resistance ratio between room temperature and He temperature: typically 20--30 compared to 100 for a whisker grown in the same conditions \cite{Tsuneta04}. It was also shown \cite{Hayashi07}  that the bending of the lattice causes frustration in the CDW order. That may give rise to the formation of a mixed CDW state analogous to that formed in a type II superconductor under a magnetic field \cite{Hayashi07}.

\subsection{Aharonov effect} \label{sec11-2}

The quantum nature of the CDW was considered \cite{Bogachek90} to be revealed by the possible observation of quantum interference effects [Aharonov-Bohm (AB) effects] in a ring formed by a CDW conductor with a diameter smaller or comparable to the CDW coherence length. As a result, oscillations of the magnetic susceptibility and of the CDW electric conductivity due to large scale quantum fluctuations of the CDW order parameter phase, the instantons, can occur with a period corresponding to a flux change equal to that of a single superconducting quantum flux. The occurrence of this period results from the equivalence of a $2\pi$ change in phase along the ring to a $2e$ change in the charge. Thus, the global $2\pi$ symmetry imposes oscillations with the period $hc/2e$ \cite{Bogachek90}.

\subsubsection{Columnar defects} \label{sec11-2-1}

\begin{figure}
\begin{center}
\includegraphics[width=7.5cm]{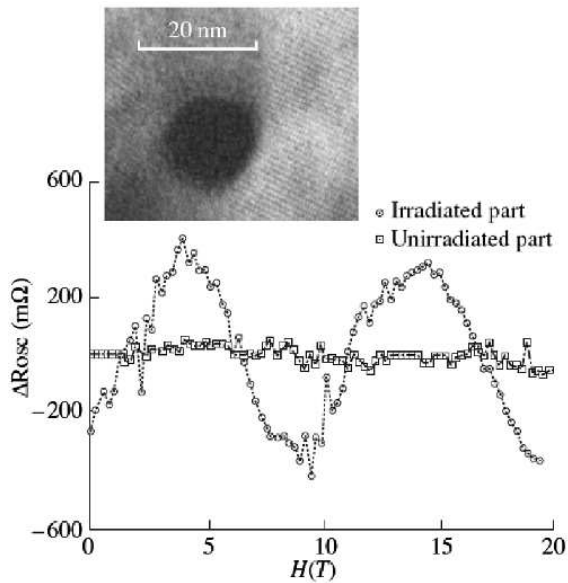}
\caption{Variation of the oscillatory part of the magnetoresistance of NbSe$_3$ in the sliding state, $\Delta R_{\rm osc}(H)$ as a function of the magnetic field $H\parallel a^\ast\parallel$ columnar defect axis at $T$~= 52~K. Columnar defect concentration: $4\times 10^9$~defect/cm$^2$. No oscillation in $\Delta R$ is observed on the unirradiated part of the sample (reprinted figures with permission from Yu.I. Latyshev \textit{et al.}, Physical Review Letters 78, p. 919, 1997 \cite{Latyshev97}. Inset shows the high resolution electron microscopy image of a columnar defect (reprinted figure with permission from Physical Review B 60, p. 14019, 1999 \cite{Latyshev99}. Copyright (1997,1999) by the American Physical Society).}
\label{fig11-7}
\end{center}
\end{figure}

The search of this AB effect was undertaken \cite{Latyshev97,Latyshev99} in a planar geometry. A NbSe$_3$ crystal, with a thickness less than 1~$\umu$m, on which an array of columnar defects was pierced from irradiation with heavy ions (Xe, Pb, U) with energy in the GeV range with a density from $2\times 10^9$ to $2\times 10^{10}$ defects/cm$^2$. A columnar defect is an homogeneous amorphous cylinder formed in the crystal along the particle track as a result of melting and quench of the material. Each columnar defect being formed by a single ion, all of them have the same size. The direction of the heavy ion beam was parallel to the $a^\ast$-axis. The diameter of defects was determined using transmission electron microscopy and was found to be $\sim$~16~nm (as show in inset of figure~\ref{fig11-7}).

From studies of differential current-voltage characteristics and Shapiro-step spectra, it was shown that, for defect concentration less than $10^{10}$ defect/cm$^2$, no significant change in CDW transport characteristics was detected between irradiated et non-irradiated samples, and that the coherent CDW motion persists over the entire length of the sample \cite{Latyshev99}. Magnetoresistance in the sliding state was measured simultaneously on two sections of the same crystal, one with columnar defects, the other defect-free. Oscillation of the magnetoresistance with a period of 10~T was observed in the section containing columnar defects, whereas, in the same conditions, no oscillations were detected in the defect-free section (figure~\ref{fig11-7}) \cite{Latyshev97,Latyshev99}.

Detailed studies \cite{Latyshev99} have shown that this effect is reproducible on several samples, although interdiffusion processes between the amorphous core of the columnar defect and the crystalline matrix can yield change in the local size of the amorphous region with time (that may explain the lost of the AB effect when samples kept at room temperature for several months were remeasured).

The period of oscillations was shown to correspond, with an experimental accuracy of $\approx 15\%$, to a flux change of half a single quantum $hc/2e$ in the defect, ($\phi_0$~= $hc/e$) independent of the defect concentration ($3\times 10^9-10^{10}$~defect/cm$^2$) and of temperature (36--52~K). The oscillation amplitude is maximum at $2\sim 3\,I_T$, the threshold current; the oscillations disappear with $H$ applied perpendicular to the defect axis.

It was proposed \cite{Latyshev97} that these magnetoresistance oscillations are the collective response of the moving CDW to the Aharonov-Bohm flux trapped inside the columnar defect. Taking the hole diameter of the columnar defect as the size for the scattering of the sliding CDW, one expects the period of the magnetoresistance oscillations to be:
\begin{equation}
\Delta H=\frac{hc}{e^\ast}\,/\,\frac{\pi D^2}{4}.
\label{eq11-1}
\end{equation}
From the measurement of $D$ and the value of $\Delta H$ (in figure~\ref{fig11-7}), within experimental uncertainty, the flux quantification corresponds to $e^\ast=2e$.

This interpretation was challenged in ref.~\cite{Visscher98} and \cite{Duhot07}. It was considered \cite{Visscher98} that nanoholes are surrounded by a normal region where the CDW order is destroyed. In this normal region, electrons can encircle the hole. For a single columnar defect, the threshold field was found periodic in the flux quantum but containing higher harmonics from weak-localisation paths. Averaging over an ensemble of columnar defects retains only \cite{Altshuler81} the half-flux quantum periodicity $\phi_0/2$, and consequently yields oscillation in the magnetoresistance with the $\phi_0/2$ periodicity.

Similarly, the mechanism of transport around a nanohole of diameter comparable to the coherence length was elucidated \cite{Duhot07}. Weak localisation-like subgap tunnelling was shown to be coupled to the sliding motion by charge accumulation in the interrupted chains at the nanohole position. Thus, modulation of the magnetoresistance in this planar geometry with array of holes may not be the result of interference effect associated to the collective quantum mechanical CDW ground state. Experiments with a single hole would be of great interest.

\subsubsection{CDW ring}\label{sec11-2-2}

\begin{figure}
\begin{center}
\includegraphics[width=8.5cm]{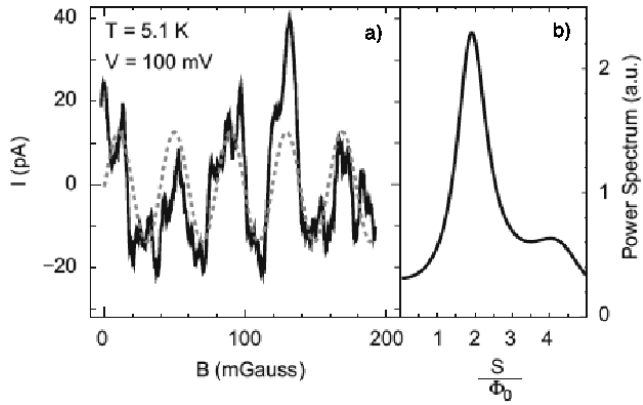}
\caption{a) Magnetic field dependence of the CDW current at $T$~= 5.1~K at fixed applied voltage 300~mV of a o-TaS$_3$ ring the period of oscillation is 39.7~mG. b)~Power spectrum of the observed oscillations (reprinted figure with permission from M. Tsubota \textit{et al.}, Physica B: Condensed Matter 404, p. 416, 2009 \cite{Tsubota09}. Copyright (2009) with permission from Elsevier).}
\label{fig11-8}
\end{center}
\end{figure}

However crystals in the form of rings can provide CDW loops, without having the CDW chains interrupted by holes. AB oscillations were measured on a o-TaS$_3$ ring with a diameter of 27~$\umu$m and a cross-section $S$~= $1\times 0.1$~$\umu$m$^2$. Figure~\ref{fig11-8}(a) shows \cite{Tsubota09} periodic oscillations ($\Delta H$~= 39.7~mG) in the CDW current in the sliding state with a fixed 300~mV field applied at 5.1~K with the magnetic field perpendicular to the ring cross-section. The power spectrum of the oscillations shown in figure~\ref{fig11-8}(b) exhibits a peak for $S/\phi_0=2$.

Assuming the origin of oscillations to result from the AB effect, the estimated charge using eq.~(\ref{eq11-1}) is $e^\ast=2e$ within 10\% of accuracy. The dashed line in figure~\ref{fig11-8}(a) shows the sinusoidal oscillation with the period and the amplitude corresponding to the peak in the power spectrum. The AB effect with 2e charge was confirmed on rings with different diameters \cite{Tsubota12}. It was also suggested \cite{Tsubota09} that interference effect originates from solitons in the low temperature CDW ground state of o-TaS$_3$.

With a different approach, the stability of the CDW ground state with a  magnetic flux threading the centre of a mesoscopic ring was theoretically studied \cite{Nathanson92,Yi97}.

\subsubsection{Circulating CDW current} \label{sec11-2-3}

\begin{figure}[h!]
\begin{center}
\includegraphics[width=7.5cm]{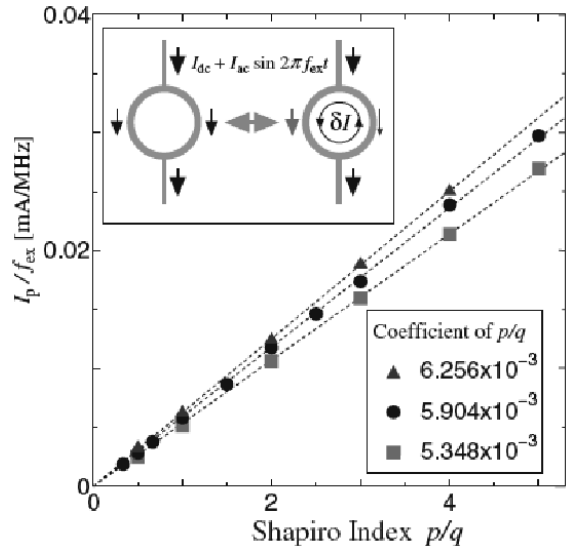}
\caption{(sub)harmonic content $p/q$ of Shapiro steps represented by $I_{\rm CDW}/f_{\rm ext}$ on a NbSe$_3$ ring at $T$~= 120~K, showing a beat structure (reprinted figure with permission from T. Matsuura \textit{et al.}, Physical Review B 79, p. 014304, 2009 \cite{Matsuura09}. Copyright (2009) by the American Physical Society).}
\label{fig11-9}
\end{center}
\end{figure}

The ring topology of crystals allowed the possibility of searching long range CDW coherence without being hampered by contact effects. Thus, measurements of Shapiro steps were performed \cite{Matsuura09} in the sliding state of a NbSe$_3$ ring with an outer diameter of 120~$\umu$m and an inner diameter of 10~$\umu$m. Small beat peaks were found on both sides of the main Shapiro peaks (see figure~\ref{fig11-9}). This result was interpreted \cite{Matsuura09} as an interference effect associated with the ring topology, resulting from a circulating CDW current along a path inside the ring as shown in the inset of figure~\ref{fig11-9}. 

In fact, this interpretation can be only understood if one considers that, in addition to the CDW current, $I_{\rm CDW}$, between the two electrodes at the outer part of the ring, there is an additional CDW current, $\delta I$, circulating around the ring. In these conditions the resonance condition for Shapiro step is modified such as $I_{\rm CDW}\pm\delta I/2e=(p/q)f_{\rm ext}$ which may yield the beat structure as shown in figure~\ref{fig11-9}.

\subsection{Mesoscopic CDW properties}\label{sec11-3}

\subsubsection{CDW transport}\label{sec11-3-1}

Transport properties were studied as a function of the number of parallel chains for NbSe$_3$ nanowires ultrasonically cleaved (see figure~\ref{fig11-4}(a)). A remarkable change from metallic to non-metallic behaviour at low temperatures occurs for a cross-section of $\sim$~500~nm$^2$ corresponding to a total of 2000 chains \cite{Slot04}. At low temperature the resistance at zero bias was found to follow a power law $R\alpha T^{-\alpha}$ and the (I,V) characteristics at large bias such $I\alpha V^\beta$. All the (I-V) characteristics collapse in a single master curve when plotted $I/T^{1+\alpha}$ versus $eV/k_{\rm B}T$. These power law dependences are characteristic for one-dimensional systems. However taking into account the interaction between chains, comparison to a single Luttinger liquid channel without disorder such as single-well carbon tubes, was considered inappropriate \cite{Slot04}. A model for explaining these results is still to be developed.

Studies of CDW dynamics on a submicron length of o-TaS$_3$ and NbSe$_3$ have revealed regions of negative absolute resistance \cite{Zant01}. I-V curves were recorded with probe spacing in the submicron range exhibiting a large variation in shape from segment to segment. In some segments, the I-V characteristics showed a negative absolute resistance. This effect is a local effect occurring only on a micron scale. The proposed explanation considers \cite{Zant01} that an increased scattering in the region with this negative absolute resistance --due to a macroscopic defect, for instance, a line dislocation-- leads to a quasiparticle current that flows in a direction opposite to that in the rest of the sample.

Using NbSe$_3$ wire structures of mesoscopic dimensions, it was shown \cite{Mantel99c} that the phase slip voltage associated to the CDW current conversion (see section~\ref{sec5-4}) is strongly reduced when the spacing between contacts is smaller than a few $\umu$m: a reduction of more than a factor 2 for a 0.5~$\umu$m spaced contacts with respect to a 3~$\umu$m spacing. Interaction between phase-slips for short distances was discussed in section~\ref{sec5-7-2}. It was proposed \cite{Mantel99c,Artemenko03} that phase-slip events are correlated, the addition of one wave front at one contact being synchronously associated with the removal of a wave front at the other contact, this coherent phase slippage requiring less deformation of the CDW.

\subsubsection{Quantised CDW wave vector variation} \label{sec11-3-2}

\begin{figure}[b]
\begin{center}
\includegraphics[width=7cm]{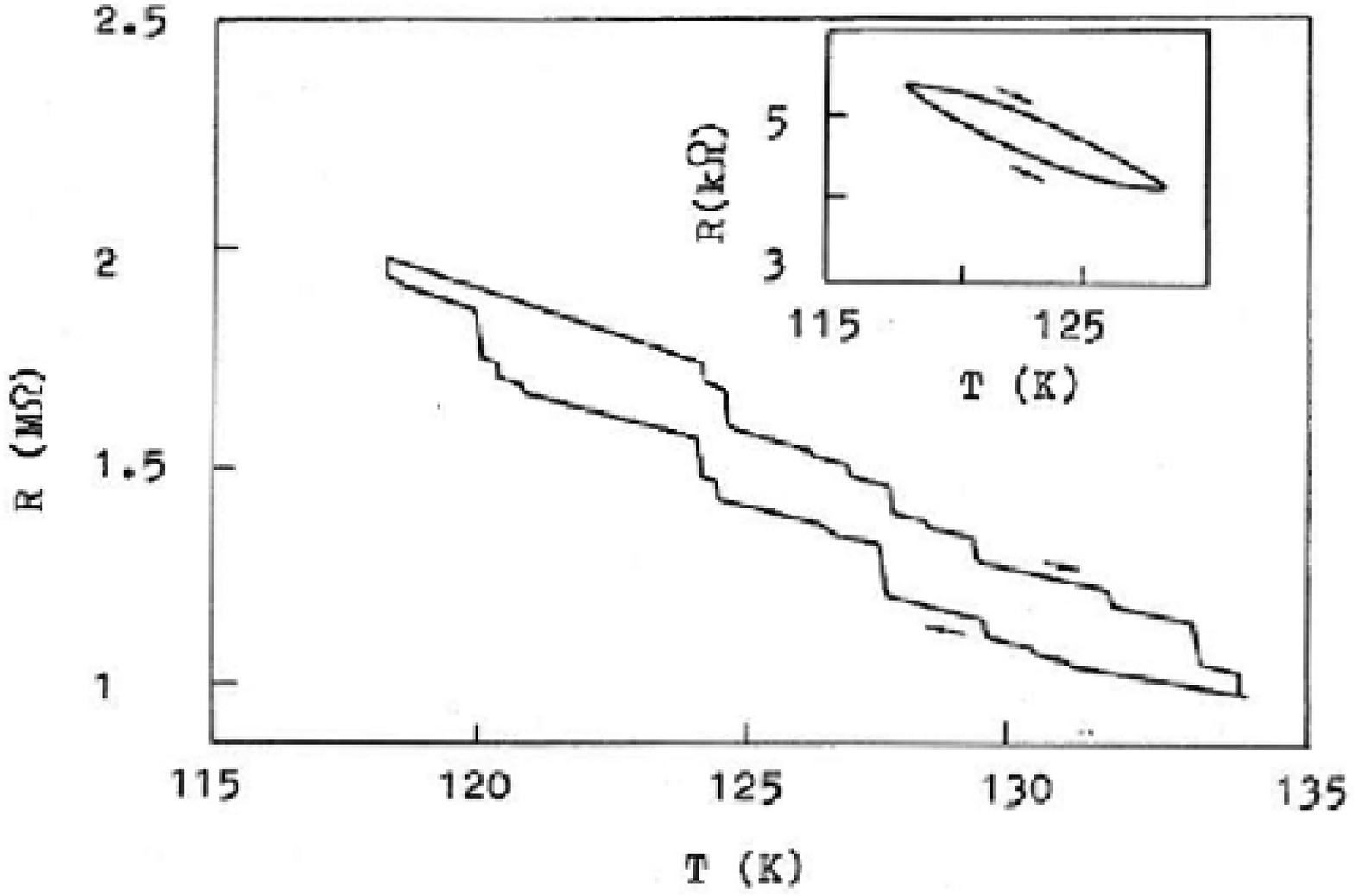}
\caption{Temperature dependence of the resistance of a o-TaS$_3$ sample with cross-section area $S$~= $10^{-2}~\umu$m$^2$ and distance between contacts $L\simeq 20~\umu$m. Inset shows the similar dependence for a sample with $S$~= 40~$\umu$m$^2$ and $L\simeq 1$~mm (reprinted figure with permission from Soviet Physics JETP 66, D.V. Borodin \textit{et al.}, p. 793, 1987 \cite{Borodin87}. Copyright (1987) from Springer Science and Business media).}
\label{fig11-10}
\end{center}
\end{figure}

The CDW $Q$-vector of o-TaS$_3$ (section~\ref{sec3-1-2}) and K$_{0.3}$MoO$_3$ (section~\ref{sec3-3}) is temperature dependent below $\sim 100$~K with hysteresis between temperature cycling. In the case where the CDW phase is strongly pinned at the contacts, the number of CDW wavelengths, $N$, in the sample should be integer. In mesoscopic samples, the change of $Q$ can be realised by a set of discrete states with different values of $N$. Transition between these states correspond to a single phase slip process with addition or subtraction of a single CDW period. This step-like form with jumps was observed \cite{Borodin87} below 150~K in the temperature dependence of the resistance of o-TaS$_3$ samples with a cross-section of $\sim$~0.3--0.4~$\umu$m$^2$ and a length of 20~$\umu$m as shown in figure~\ref{fig11-10}. For thick samples, the $R(T)$ is smooth.

\begin{figure}[h!]
\begin{center}
\includegraphics[width=7cm]{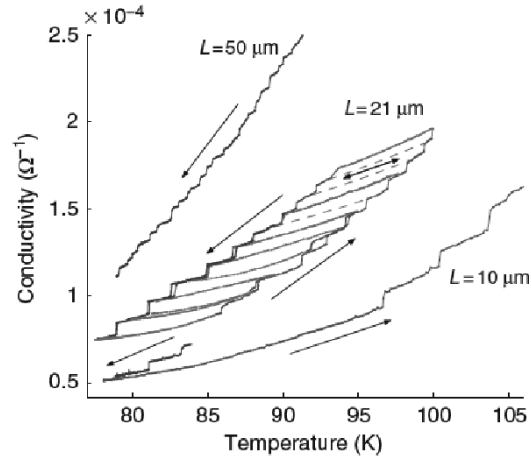}
\caption{Temperature dependence of the conductivity of K$_{0.3}$MoO$_3$ samples with dimensions 21$\times$5$\times 0.3~\umu$m$^3$ (central curve), 50$\times$7$\times 0.3~\umu$m$^3$ (upper curve: $\sigma$ is multiplied by 1.5), 10~$\umu~\times~5~\umu$m$^2$ (lower curve: $\sigma$ is divided by 15.). The arrows indicate the direction of temperature sweep (reprinted figure with permission from S.G. Zybtsev \textit{et al.}, Physica B: Condensed Matter 407, p. 1810, 2012 \cite{Zybtsev10}. Copyright (2012) with permission from Elsevier).}
\label{fig11-11}
\end{center}
\end{figure}
\begin{figure}[h!]
\begin{center}
\includegraphics[width=8cm]{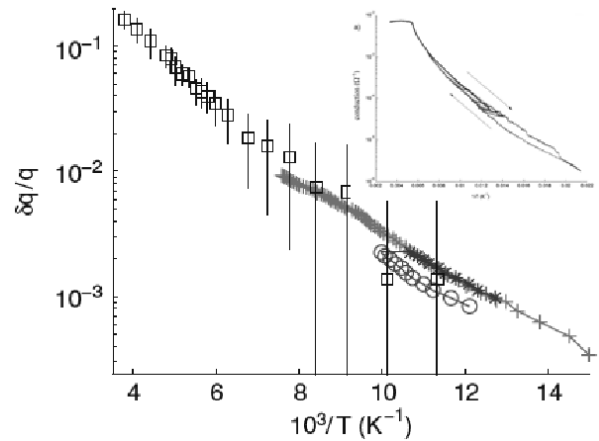}
\caption{Variation $\Delta Q/Q$ versus the inverse of temperature of the CDW wave vector of K$_{0.3}$MoO$_3$ derived from the $\sigma(T)$ curve in figure~\ref{fig11-11} (central curve) assuming that $\Delta Q$ changes by $2\pi/L$ at each step (squares represent data from X-ray diffraction results \cite{Girault88}) (reprinted figure with permission from S.G. Zybtsev \textit{et al.}, Physica B: Condensed Matter 407, p. 1810, 2012 \cite{Zybtsev10}. Copyright (2012) with permission from Elsevier). Inset: hysteresis loop of the conductivity between cooling and heating temperature variation.}
\label{fig11-12}
\end{center}
\end{figure}

Better resolved steps (due to better quality crystals) have been measured \cite{Zybtsev10} on K$_{0.3}$MoO$_3$ samples with submicron thickness and contact separation on the order of tens of microns. Figure~\ref{fig11-11} shows the temperature variation of the conductivity in a limited temperature range for a sample with contact separation $L$~= 21~$\umu$m. Steps between discrete states of the CDW are regular. The $\sigma(T)$ segments connecting the heating and the cooling curves are reversible. The number of steps is directly coupled to $L$ as seen in figure~\ref{fig11-11} for samples with $L$~= 50~$\umu$m.

Each step reveals a single phase slip act providing the variation in $Q$ such $\delta Q$~= $\pm 2\pi/L$. By counting the number of steps over a temperature range, the $Q(T)$ variation can be obtained \cite{Zybtsev10} as shown in figure~\ref{fig11-12} where the results of diffraction studies \cite{Girault88} are also shown. The hysteresis loop between warming and cooling a K$_{0.3}$MoO$_3$ sample measured from the $\sigma(1/T)$ dependence in the full temperature range between room temperature and 45~K is shown in inset of  figure~\ref{fig11-12}.

The $\Delta Q/Q$ for a single phase-slip can be estimated to be $\sim 10^{-4}$, in agreement with the X-ray studies for current conversion presented in section~\ref{sec5-7}.

\subsection{Mesoscopic CDW junctions} \label{sec11-4}

The recent experimental capability of structuring mesoscopic CDW devices and the inspiration from the formal similarity between theories for CDW and superconductivity have given an impetus for studying heterostructures based on CDW.

\subsubsection{Carrier reflection at the N/CDW interface}\label{sec11-4-1}

The question of the possibility of observing the subgap carrier reflection at the N/CDW interface (N: normal metal) was first raised in ref.~\cite{Kasatkin84}.

\begin{figure}
\begin{center}
\includegraphics[width=5.cm]{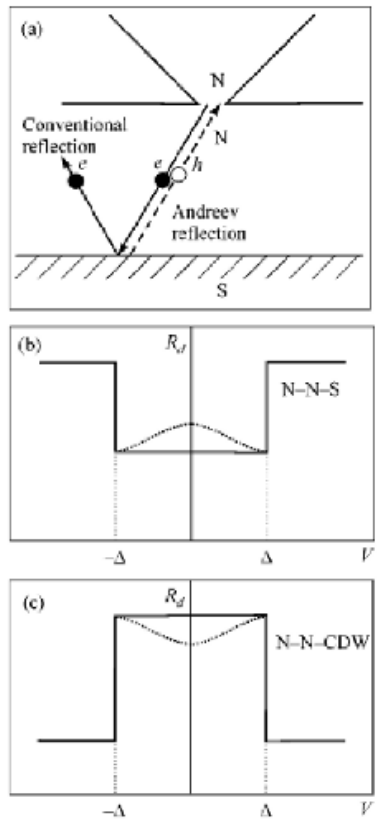}
\caption{Schematic representations of a) Andreev quasiparticle reflection at the metal normal - superconductor interface with carrier injection from a point contact. b)~Differential resistance as a function of bias $V$ for a metal normal-superconductor interface in presence of Andreev reflection. c)~Differential resistance as a function of $V$ for a metal normal - CDW interface in presence of Andreev-type reflection (reprinted figure with permission from JETP Letters 75(11), Yu.I. Latyshev and A.A. Sinchenko, p. 593, 2002  \cite{Latyshev02a}. Copyright (2002) from Springer Science).}
\label{fig11-13}
\end{center}
\end{figure}

In the case of a metal normal - superconductor (N/S) interface, an Andreev reflection (AR) may occur, in which a spin up electron from the normal side is reflected as a hole in the spin-down band and a Cooper pair is transmitted in the superconductor \cite{Andreev64}. Such a quantum transformation of charge is possible only if the energy of the incident particle is lower than the superconducting gap \cite{Son87}. Experimentally a thin film in normal metal with a thickness $d$ less than the mean free path $\ell$ is deposited on the superconducting surface. Electrons are injected into the normal film from a point contact with diameter $a\ll 1$ and reach ballistically the N-S interface. Due to Andreev reflection, most of them return as holes along the incident trajectories to the point contact (as schematically drawn in figure~\ref{fig11-13}(a)), 
thereby reducing the contact resistance by a factor $\sim 2$. Consequently the differential resistance $R_d(V)$ of the contact as a function of bias $V$ should have (see figure~\ref{fig11-13}(b)) the form of a step function:
\begin{eqnarray*}
R_dV=\left\{
\begin{array}{ll}
A & \mbox{ for $|V|<\Delta$}\\
2A & \mbox{ for $|V|>\Delta$}
\end{array}\right.
\end{eqnarray*}
An non-ideal N-S interface reduces the Andreev reflection probability at small biases, giving rise to a local maximum of $R_d(V)$ at $V=0$ (the dotted line in figure~\ref{fig11-13}(b)).

In the case of a N/CDW interface, by analogy with Andreev reflection in superconductivity, due to the CDW gap, an electron near the Fermi surface cannot propagate through the CDW region unless it simultaneously drags a hole from the Fermi sea forming a new electron-hole pair. This pair is transferred into the condensate whereas an extra electron is reflected back into the Fermi sea. The momentum of the reflected electron is slightly less ($-k_{\rm F}-\delta k$) than that of the incident electron, $k_{\rm F}$, the momentum transfer to the CDW being $2k_{\rm F}+\delta k$. Thus, only a momentum transfer occurs through the N-CDW interface, and not a charge as in the Andreev reflection \cite{Kasatkin84,Rejaei96}.

The expected $R_d(V)$ for a N-CDW interface is then mirror-symmetric about the $V$ axis: $R_d(|V|<\Delta)$ should be $> R_d(|V|>\Delta)$ (figure~\ref{fig11-13}(c) because most of the particles reflected along the incident trajectories return to the point contact without changing the charge sign \cite{Latyshev02a}.

However it was stated \cite{Duhot07b} that the motion of reflected particles cannot occur at a normal metal/CDW interface, since the pair in the CDW is a superposition of two electrons with momenta differing by the CDW wave vector and not with opposite momenta. Then, the momentum component parallel to the interface can either be conserved or change by an amount equal to the component of the CDW wave vector parallel to the interface.

\subsubsection{CDW heterostructures}\label{sec11-4-2}

Theoretical models were developed for heterostructures formed with, from one hand, a CDW and from the other hand, either a normal metal, a superconductor or another CDW.

\medskip
\noindent \textit{9.4.2.a. N/CDW/N junction}
\medskip

The N/CDW/N junction was modelled \cite{Rejaei96} by parallel 1-D CDW chains of length $L$ ($L$~= a few coherence length $\xi_0$) perpendicular to the interfaces with two normal, low impedance, normal leads connected to two large reservoirs at the chemical potentials $\mu_L$ and $\mu_R$. The quasiparticle transport was calculated with the Andreev-like type of reflection. The transmission probability is extremely small below the gap and exhibits oscillations above the gap with a period proportional to $\xi_0/L$. A perfect interface does not pin the CDW and therefore allows the possibility for collective CDW motion through a N/CDW interface \cite{Rejaei96}.

\medskip
\vbox{\noindent \textit{9.4.2.b. S/CDW/S}
\medskip

With the same scheme as described in section~10.4.2.a, but with the CDW sandwiched between two superconducting leads, the question was raised \cite{Visscher97} of the consequence on the Josephson effect in a S/CDW/S heterostructure when the CDW is sliding. It was shown that the Josephson current oscillates as a function of the velocity of the sliding CDW, and that an additional dynamic phase resulting of the CDW motion is added to the conventional superconductor phase difference. Internal mode locking between Josephson and CDW frequencies were thus predicted with plateaux in the CDW conductance at fractions $m/n$ of the Josephson and CDW frequencies \cite{Visscher97}.}

This approach was challenged in ref.~\cite{Duhot07} where it was stated that it is impossible to observe a Josephson current through a CDW. It was shown that an Andreev pair entering a CDW (with electron-hole pairing) is dephased on the Fermi wavelength $\lambda_{\rm F}$ (a factor 2 with the CDW wavelength). This dephasing results \cite{Duhot07} from the non-cancellation of the random phase factors acquired by a spin-up electron visiting different impurities with the phase of a spin-down hole visiting the same sequence of impurities.

In contrast, for non magnetic impurities, random phases of spin-up electrons and spin-down holes cancel with each other in the total phase of the Andreev pair propagating through a SDW. Thus a Josephson current through a SDW is predicted \cite{Duhot07} to be observable in a structure where the SDW width is a few coherence length $\xi_{\rm SDW}$.

\medskip
\noindent \textit{9.4.2.c. CDW/N/CDW junctions}
\medskip

The mechanism of the coupling of two phase-coherent ground states with phase $\varphi_1$ and $\varphi_2$ spatially separated by a normal metal was investigated in ref.~\cite{Visscher96}. The phase coupling was understood \cite{Visscher96} as a phase locking of charge oscillations in the quasiparticle transmission probability (discussed for a single N/CDW interface section~\ref{sec11-4-1}.a.) arising from both interfaces. Differently to the case of a Josephson S/N/S junction in which the resonant states correspond to the transfer of a Cooper pair, in a CDW/N/CDW junction there is no charge transfer, only a momentum transfer. In the case of perfect interfaces ($Z$~= 0), the time evolution of the phase $\varphi_1$ follows linearly that of $\varphi_2$. When impurities are included ($Z\neq 0$), if $\varphi_1$ is changed adiabatically in time, the change of $\varphi_2$ in time is non linear and shows a phase jump \cite{Visscher96} which can be detectable in the transport properties of the CDW/N/CDW junction in the non-linear state.

A similar problem was previously considered \cite{Artemenko84b,Artemenko97a} in which two CDWs are weakly linked (either through a tunnel junctions or a direct contact). It was shown that the current flowing between the weak link contains a component which depends on the difference between the phases of each CDW: $I\alpha\Delta_1\Delta_2\cos(\varphi_1-\varphi_2)$. There is a similarity with the time dependent Josephson effect in weakly linked superconductors. However, in the superconductivity case, the phase difference $\varphi_1-\varphi_2$ is determined by the applied voltage, $\dot{\varphi}_1-\dot{\varphi}_2$~= $2eV/\hbar$ according to the Josephson relation. For CDW, the phase difference does not depend on the current through the contact. However $\varphi_1$ and $\varphi_2$ can vary over time if the CDW is sliding. It was also noted that this oscillatory current will vanish if CDW chains across the weak link are mis-oriented. It was then suggested \cite{Artemenko84b} to study this effect between weak linked conductors fabricated from a common crystal, which can be technically feasible now with FIB technology.

\subsection{Point contact spectroscopy}\label{sec11-5}

The practical realisation of heterostructures considered in the theoretical calculations presented above in section~\ref{sec11-4}, when the length of the CDW (possibly sliding) in the chain direction is a few coherence lengths (50--100~nm) is still out of reach. However properties of a N/CDW interface can be studied using point contact spectroscopy.

\subsubsection{Point contact with a semiconducting CDW}\label{sec11-5-1}

The first observation \cite{Sinchenko96,Sinchenko98} of the subgap reflection of carriers of a N/CDW interface was performed on Au-K$_{0.3}$MoO$_3$ and Cu-K$_{0.3}$MoO$_3$ structures. The N/CDW interface showed an excess resistance for incident carriers with energy $E<\Delta_{\rm CDW}$, consistent with the fact that these carriers are reflected from the interface (see figure~\ref{fig11-13}(c)).

However, at low temperatures ($T\lesssim 77$~K), the (I-V) characteristics of Cu-K$_{0.3}$MoO$_3$ point contacts are asymmetric. The electric field in the tiny region where the point contact is formed lead to a deformation of the CDW and to a change in the quasiparticle density and consequently to a shift of the chemical potential \cite{Sinchenko98}. The energy bands of the semiconducting CDW are strongly bent near the boundary. As a consequence of the contact with the metal, a $p-n$ junction is formed in a sub-surface layer of K$_{0.3}$MoO$_3$. Band bending should be substantially irrelevant in NbSe$_3$ where screening by the remaining normal carriers is effective.

\subsubsection{Point contact with NbSe$_3$}\label{sec11-5-2}

\medskip
\noindent \textit{9.5.2.a. N/CDW interface}
\medskip

I-V characteristics of direct-type junctions (without any insulating layer) formed on NbSe$_3$ were measured along the three crystallographic orientations, $b$, $c$, and $a^\ast$ \cite{Sinchenko03a}. The small dimensions of any NbSe$_3$ whisker along $b$-axis and $c$-axis (typically thickness $\sim 1~\umu$m, and width: 10--50~$\umu$m) made impossible to use the metallic needle-type tip method to form point-contact along $b$ and $c$-axis. Therefore a thin gold strip of 50~$\umu$m wide and 4~$\umu$m thick was used. Thus, the maximum possible area of the contact is about 4~$\umu$m$^2$. But, taking into account the real relief of the contacting surfaces, more likely the contact consists of multishorts with a net contacting area much less than 4~$\umu$m$^2$, as it can be evaluated from the value of the contact resistance (see below). Along the $a^\ast$ direction, the counter electrode was an electrochemically etched thin metal (Au, Cu, In) wire with the tip diameter less than 1~$\umu$m. The appropriate contact geometry is shown in inset of figure~\ref{fig11-16}.

Energy gap spectroscopy by means of point contact is reliable only if the mean free path, $\ell$, of conduction electrons is larger than the size of the contact (ballistic regime). The resistance of the point contact in that regime is given by the Sharvin formula \cite{Sharvin65}:
\begin{equation*}
R_{\rm contact}=\frac{\rho\ell}{d^2},
\end{equation*}
where $\rho$ is the resistivity, and $d$ the contact diameter. With point contact resistance between 20 and 1000~$\Omega$, $\rho$~= $10^{-5}~\Omega$cm and $\ell$~= $10^{-4}$~cm, the diameter of point contact was estimated to be $d\leq 10^{-5}$~cm, less than $\ell$.

\begin{figure}
\begin{center}
\includegraphics[width=7.5cm]{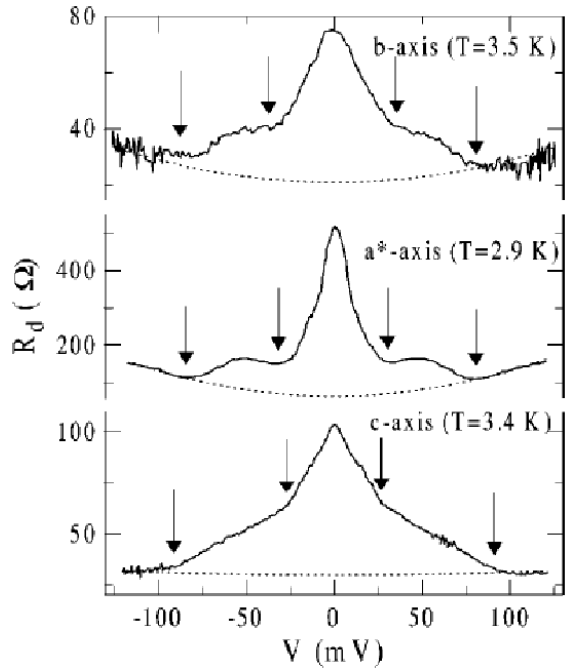}
\caption{Bias dependences of the differential resistance $R_d(V)$ for point contacts Au-NbSe$_3$ oriented along $b$-, $a^\ast$-, and $c$-axis directions measured at $T<4.2$~K. Energy gap positions are indicated by arrows. Dotted line is the normal state background (reprinted figure with permission from A.A. Sinchenko and P. Monceau, Physical Review B 67, p. 125117, 2003 \cite{Sinchenko03a}. Copyright (2003) by the American Physical Society).}
\label{fig11-14}
\end{center}
\end{figure}

Bias dependences of the differential resistance, $R_d(V)$, for point contacts oriented along $a^\ast$, $b$ and $c$ axis at $T\leq 4.2$~K are shown in figure~\ref{fig11-14}. 
\begin{figure}
\begin{center}
\includegraphics[width=7.5cm]{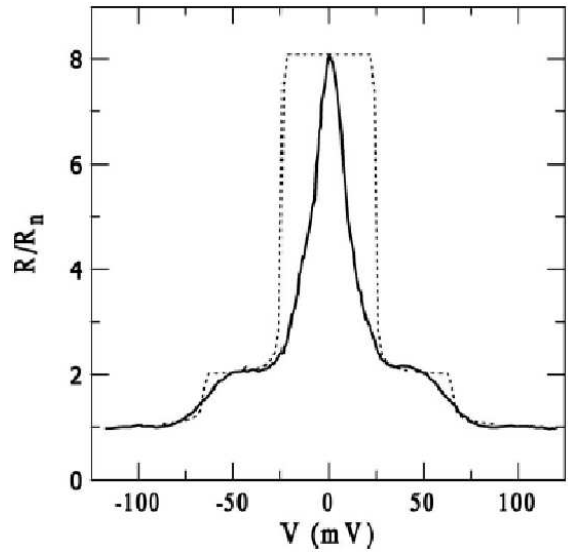}
\caption{Modelling of the bias dependence of the normalised differential resistance $R_d(V)/R_{dN}(V)$ measured at $T$~= 3.8~K for a point contact Au-NbSe$_3$ oriented along the a$^\ast$-axis. The dotted line is the fit with $\Delta_1(0)$~= 65.0~mV and $\Delta_2(0)$~= 24.5~mV (reprinted figure with permission from A.A. Sinchenko and P. Monceau, Physical Review B 67, p. 125117, 2003 \cite{Sinchenko03a}. Copyright (2003) by the American Physical Society).}
\label{fig11-15}
\end{center}
\end{figure}
\begin{figure}
\begin{center}
\includegraphics[width=7.5cm]{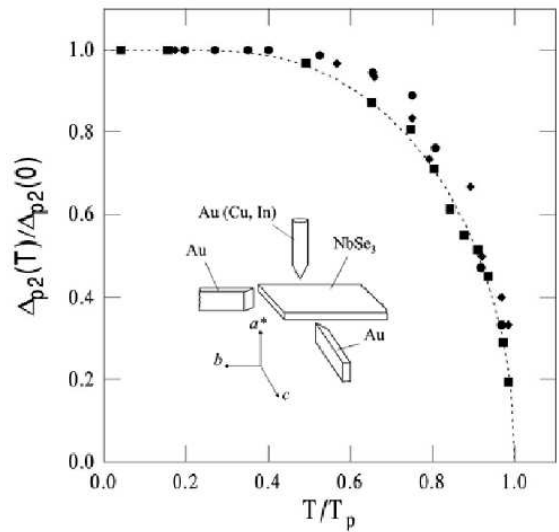}
\caption{Temperature dependence of the low temperature CDW energy gap of NbSe$_3$ normalised to the value at $T\rightarrow 0$ obtained from different contacts oriented along a$^\ast$-axis. The dashed curve corresponds to the BCS theory: inset shows the schema of the experimental set-up (reprinted figure with permission from A.A. Sinchenko and P. Monceau, Physical Review B 67, p. 125117, 2003 \cite{Sinchenko03a}. Copyright (2003) by the American Physical Society).}
\label{fig11-16}
\end{center}
\end{figure}
At high voltage ($V>100$~mV) the increment of $R_d$ is proportional to the square of the bias voltage (dotted lines in figure~\ref{fig11-14}), that is typical for Joule heating. For the three curves shown in figure~\ref{fig11-14} at $|V|<100$~mV, the differential resistance increases in two steps and is maximum at $V$~= 0. These two steps maxima were associated with the excess resistance arisen from the reflection of injected carriers on each Peierls gap, respectively. Thus, the upper CDW gap $\Delta_1$ is 75--80~mV along $b$, 85--95~mV along $c$ and 65-80~mV along $a^\ast$, while the lower CDW gap, $\Delta_2$, is 28--34~mV along $b$ and $c$-axis and 24--30~mV along $a^\ast$.

The current flowing through the contact was calculated \cite{Sinchenko03a} by modelling each energy gap as a potential step with a height $\Delta_1$ and $\Delta_2$ respectively (perfect N-CDW boundary). The estimated differential resistance normalised to the background resistance $R_{dN}$ is compared with the experimental data in figure~\ref{fig11-15} for contact along $a^\ast$-axis. The temperature dependence of the  $\Delta_2$ gap is shown in figure~\ref{fig11-16}; it follows a BCS type variation, although the ratio $(2\Delta_2(0)/kT_{\rm P_2})\simeq 8.2-14$ is far beyond the value from the mean field theory.

For contacts oriented along $a^\ast$-axis (for which the bias dependence of $R_d(V)$ is very similar to that for contacts along $b$-axis (see figure~\ref{fig11-14}), it was considered \cite{Sinchenko03a} that, because of the very large anisotropy $\sigma_b/\sigma_a^\ast$, in fact, the density of states along $b$ was probed.

\medskip
\noindent \textit{9.5.2.b. S/CDW interface}
\medskip

\begin{figure}
\begin{center}
\subfigure[]{\label{fig11-17a}
\includegraphics[width=6.5cm]{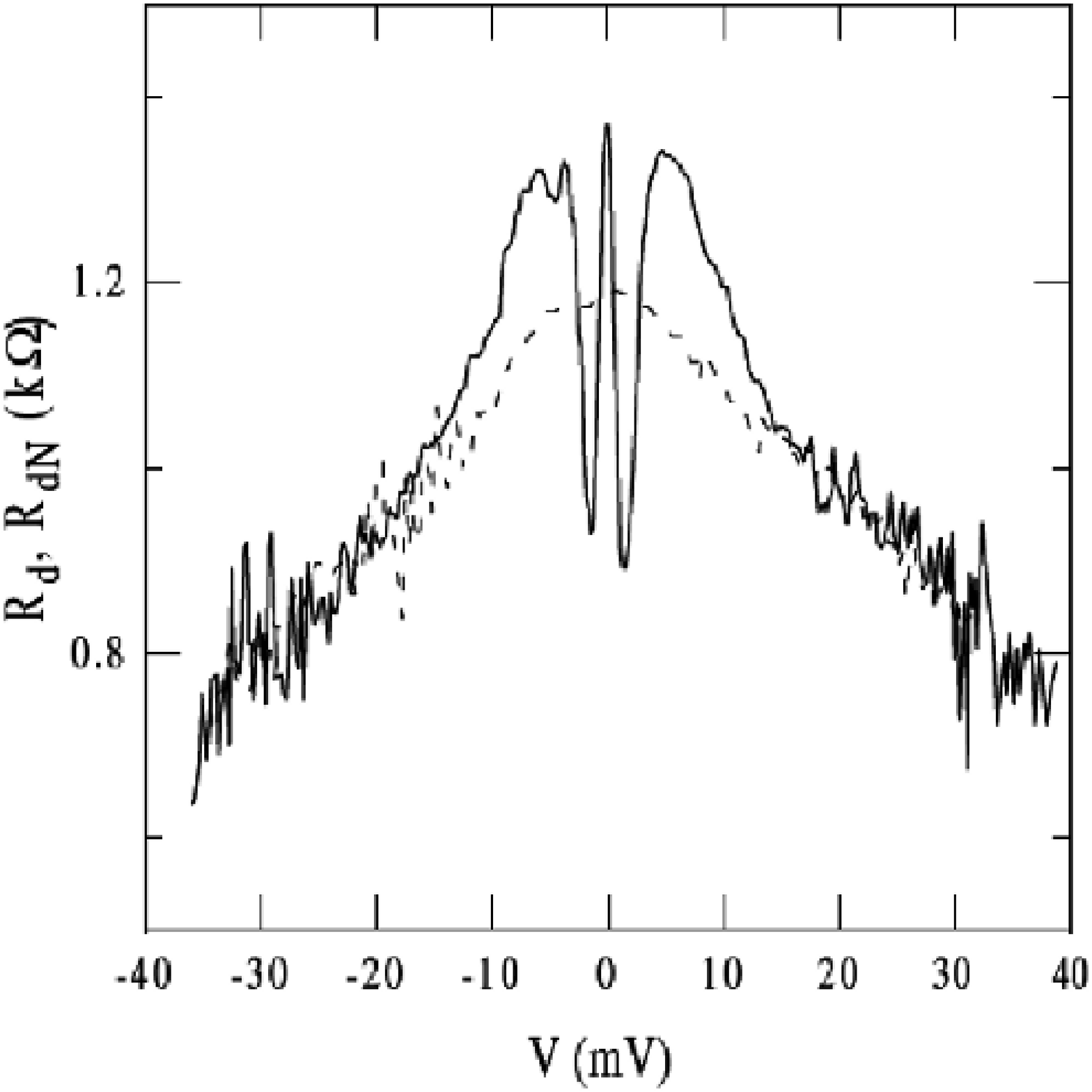}}
\subfigure[]{\label{fig11-17b}
\includegraphics[width=6.5cm]{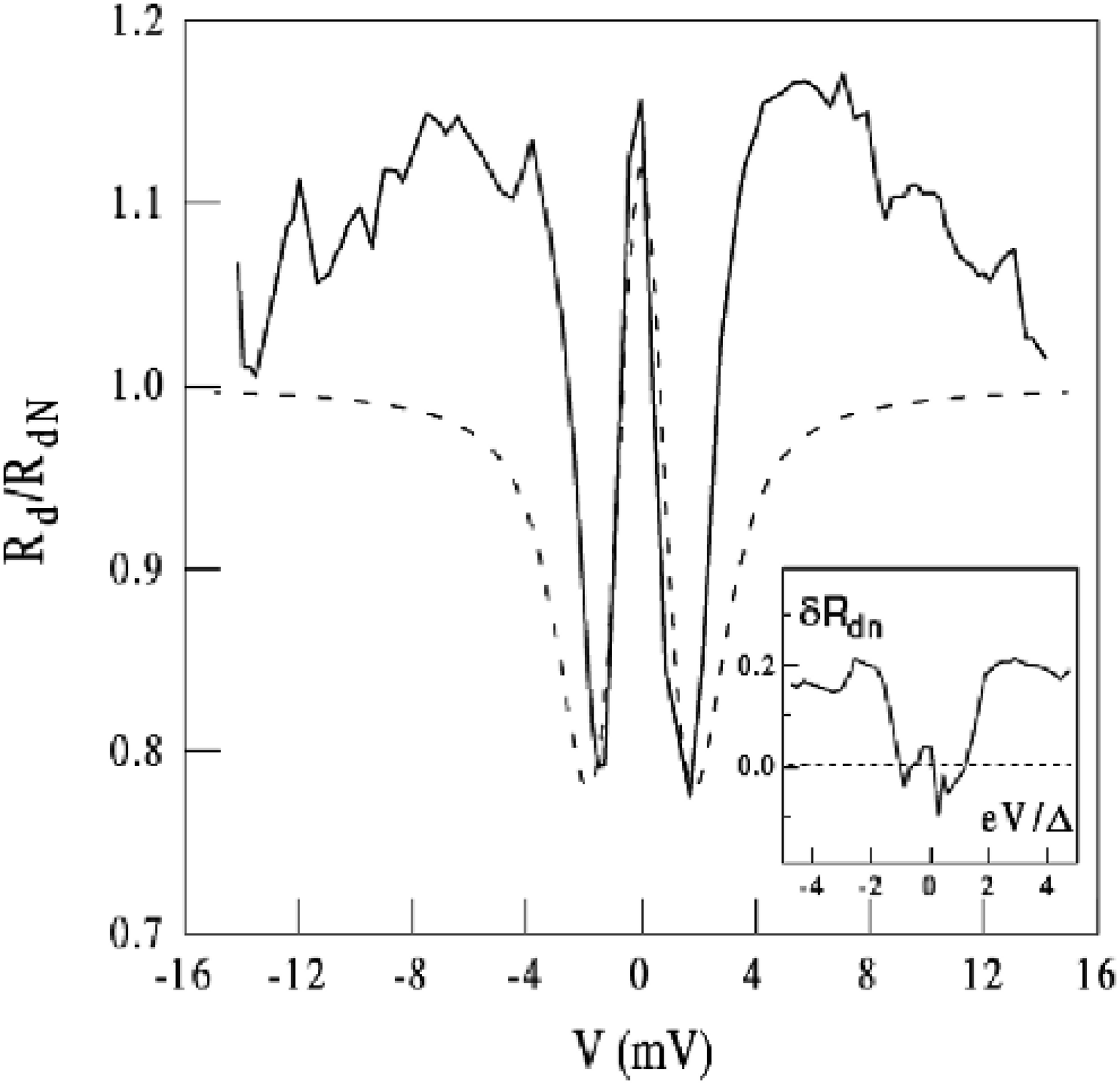}}
\caption{a) Bias dependence of the differential resistance of a Nb/NbSe$_3$ contact revealing the Andreev reflection singularity: solid curve $R_d(V)$: $T$~= 3.6~K; dotted curve $R_{dN}(V)$: $T$~= 10.1~K. Superconducting transition temperature of niobium: $T$~= 9.2~K. b)~Bias dependence of the normalised differential resistance $R_d(V)/R_{dN}(V)$ for the same Nb/NbSe$_3$ contact as in a). The dashed curve is the Blonder, Tinkham, and Klapwijk \cite{Blonder82} fit to the experimental dynamical resistance. Inset shows the difference between $R_d(V)/R_{dN}(V)$ and the BTK fit (reprinted figure with permission from A.A. Sinchenko and P. Monceau, Journal of Physics: Condensed Matter 15, p. 4153, 2003 \cite{Sinchenko03}. Copyright (2003) by the Institute of Physics).}
\label{fig11-17}
\end{center}
\end{figure}

The direct possible interaction between a superconductor and a CDW was tested by studying Nb/NbSe$_3$ point contacts \cite{Sinchenko03} with the same configuration as that shown in inset of figure~\ref{fig11-16}. The bias dependences of the differential resistance of a Nb-NbSe$_3$ contact in the superconducting state of Nb at $T$~= 3.6~K, and in the normal state ($T$~= 10.1~K) are shown in figure~\ref{fig11-17}(a). 
\begin{figure}
\begin{center}
\includegraphics[width=7.5cm]{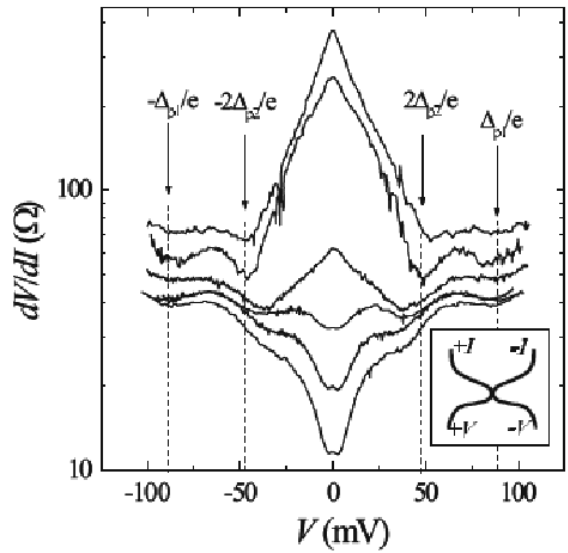}
\caption{Bias dependences of the differential resistance $R_d(V)$ (in a logarithmic scale) measured at $T$~= 4.2~K for several NbSe$_3$/NbSe$_3$ point contacts with different zero bias resistances $R_{d_0}$ (from 10 to 400~$\Omega$) oriented along the $a^\ast$-axis. Arrows indicate gap singularities. Inset illustrates schematically the contact geometry (reprinted figure with permission A.A. Sinchenko and P. Monceau, Physical Review B 76, p. 115129, 2007 from \cite{Sinchenko07}. Copyright (2007) by the American Physical Society).}
\label{fig11-18}
\end{center}
\end{figure}

At $T$~= 10.1~K $R_{dN}(V)$ strongly decreases with the voltage is increased; that is related to the excess resistance arising from the reflection of normal quasiparticles injected from Nb on the Peierls CDW gap.

At temperature below the superconducting transition temperature of niobium, and at low bias voltage the I-V spectra exhibit a double-minimum structure, characteristic of Andreev reflection at a N-S point contact. The observed Andreev reflection has been attributed only to the remaining electrons of NbSe$_3$ at low temperature below both CDW transitions, which are gapless.

The point contact was modelled \cite{Sinchenko03} using the Blonder, Tinkham, and Klapwijk (BTK) formulation \cite{Blonder82}. The fit of the normalised $R_d(V)/R_{dN}(V)$ with the energy gap of Nb, $\Delta_s$~= 1.5~meV, and the barrier transmission $Z$~= 0.83 is drawn in figure~\ref{fig11-17}(b), to be compared with the experimental data. There is a good agreement between them at $|V|\leq\Delta_s/e$. But at $|V|\geq\Delta_s/e$ the experimental dependence deviates from the theoretical BTK fit, as illustrated in the inset of figure~\ref{fig11-17}(b) showing the difference between them. It was assumed \cite{Sinchenko03} that the deviation from the BTK theory, in this specific case of a S/CDW junction, results from the suppression of the superconductivity of Nb near the contact at $|V|>\Delta_s/e$. This effect may originate from non-equilibrium effects resulting from quasi particles reflected on the Peierls energy gap; then, when $|V|$ is increased, the increase in the quasiparticle concentration from reflection on the CDW gap induces a decrease of the superconducting energy gap of Nb (till a critical concentration at which $\Delta_s\rightarrow 0$) in a region at the surface of the Nb electrode.

Thus, current conversion at the Nb/NbSe$_3$ point contact boundary proceeds, at $|V|\leq\Delta_s/e$, by means of Andreev reflection of gapless electrons of NbSe$_3$ and at $|V|\geq\Delta_s/e$, by suppression of  the superconductivity near the S/CDW interface, by means of the N/CDW scenario.

\medskip
\noindent \textit{9.5.2.c. CDW/CDW junctions}
\medskip

I-V characteristics were measured on NbSe$_3$-NbSe$_3$ contacts (point contacts) formed along the $a^\ast$ axis. To form a NbSe$_3$-NbSe$_3$ contact \cite{Sinchenko99,Sinchenko07}, two bent crystals with parallel $b$-axis were brought together (see inset of figure~\ref{fig11-18}) with the help of a precise mechanical motion transfer system. Two different behaviours in the I-V curves were observed depending on the contact resistance of the contact. Figure~\ref{fig11-18} shows the bias dependence of the differential resistance $R_d(V)$ measured at $T$~= 4.2~K for several NbSe$_3$-NbSe$_3$ point contacts with different zero bias resistance $R_{d_0}$ from 10~$\Omega$ to 400~$\Omega$. Curves with a large contact resistance $R_{d_0}>100~\Omega$ exhibit a behaviour with a large zero bias resistance peak while, on the contrary, those with a low contact resistance are characterised by a deep minimum of $R_{d_0}$. Independently of their contact resistances all the curves show gap singularities \cite{Sinchenko07} for both CDWs. Contacts with high and low zero bias resistances have a different temperature dependence. 
Junctions with large $R_{d_0}$ values exhibit a typical tunnelling behaviour, for which $R_{d_0}$ increases when $T$ is reduced below $T_{\rm p_2}$~= 59~K, similarly as in S/I/NbSe$_3$ tunnel junctions \cite{Fournel86,Sorbier96} or in longitudinal nanoconstrictions \cite{Neill06} described as NbSe$_3$/I/NbSe$_3$ tunnel junctions. In the opposite, for low contact resistances, $R_{d_0}$ decreases when $T$ is reduced. The low resistance point contacts were considered \cite{Sinchenko07} to correspond to coherent interlayer tunnelling similar to properties of mesa-type junctions as described in the following section~\ref{sec11-6}.

\medskip
\noindent \textit{9.5.2.d. Superconducting NbSe$_3$-NbSe$_3$ point contact junction}
\medskip

Junctions were also formed \cite{Escudero01} with a NbSe$_3$ ribbon laid along a cylindrical glass rod of about 100~$\umu$m in diameter and the counter-electrode (another NbSe$_3$ whisker) stretched perpendicularly across the NbSe$_3$, pressing on it to make electrical contact \cite{Escudero01}. It occurred that the pressure on the edges of the NbSe$_3$ whiskers was strong enough to make superconducting the NbSe$_3$-NbSe$_3$ junction. In section~\ref{sec3-2-5} it was shown \cite{Nunez92} that NbSe$_3$ becomes superconducting at $T_c$~= 2.5~K above a pressure of 0.6~GPa which suppresses the lower CDW and at $T_c\sim 5$~K when the upper CDW is completely destroyed. The superconducting critical current of the junction was shown to disappear when $T\rightarrow 4.18$~K. The amplitude of the critical current at $T\rightarrow 0$ has been calculated in different models considering either short weak links \cite{Kulik75} in the dirty and clean limits or superconducting tunnel junctions \cite{Ambegaokar63} where:
\begin{equation*}
I_c(0)=A\,\frac{\pi\Delta_S(0)}{2eR_N},
\end{equation*}
where $\Delta_S(0)$ is the superconducting gap at $T$~= 0, $R_N$: the normal resistance of the contact, and $A$ between 1.32 and 2 according to the models. $2\Delta_S(0)/k_{\rm B}T_c$ for the superconducting state was measured to range between 5.3 and 7.3 much above the mean field 3.52 value, indicating a strong electron-phonon coupling for the superconducting state, that may be related to the deviation of the mean field theory for the CDW gaps (section~\ref{sec11-6-2}).
\begin{figure}
\begin{center}
\includegraphics[width=7.5cm]{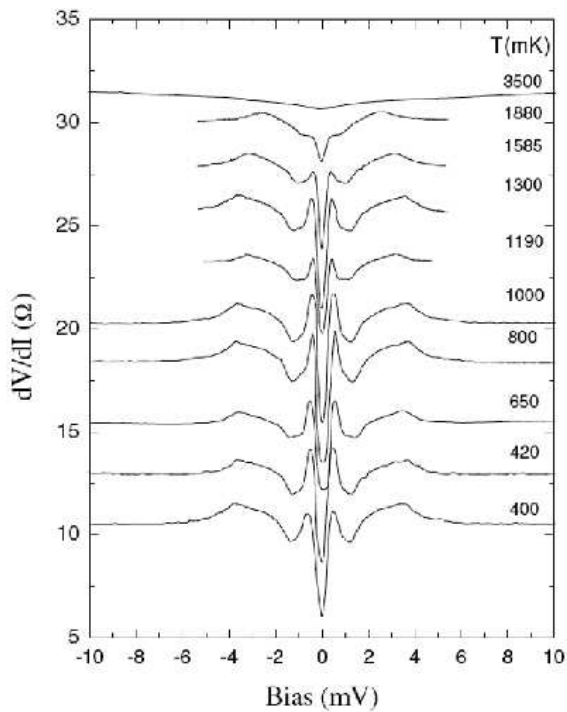}
\caption{Bias dependence of the differential resistance, ${\rm d}V/{\rm d}I$ of a superconducting NbSe$_3$/NbSe$_3$ junction from 0.4~K to 3.5~K. The curves are vertically displaced for clarify (reprinted figure with permission from R. Escudero \textit{et al.}, Journal of Physics: Condensed Matter 13, p. 6285, 2001 \cite{Escudero01}. Copyright (2001) by the Institute of Physics).}
\label{fig11-19}
\end{center}
\end{figure}
Differential resistances versus the bias voltage for this type of NbSe$_3$-NbSe$_3$ junction are shown in figure~\ref{fig11-19} from 0.4~K to 3.5~K. The central peak minimum is related to the superconducting state. This peak disappears at 2.2~K defining a superconducting transition temperature. The broad maximum around $\pm 4$~meV evolves slowing at low temperature and is not detectable above 3.5~K. These two peaks were assigned \cite{Escudero01} to the two superconducting states which appear successively under pressure after destruction of first the lower CDW, then the second one.

\subsection{Intrinsic interlayer tunnelling spectroscopy}\label{sec11-6}

\subsubsection{Conductivity anisotropy along the a$^\ast$ axis}\label{sec11-6-1}

\begin{figure}[b]
\begin{center}
\includegraphics[width=6cm]{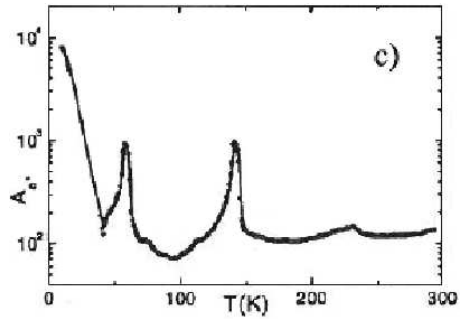}
\caption{Temperature dependence of the $a^\ast$ axis anisotropy $A_{a^\ast}$~= $\rho_{a^\ast}/\rho_b$ of NbSe$_3$ (reprinted figure with permission from E. Slot and H.S.J. van~der~Zant, Journal de Physique IV (France) 12, p. Pr9-103, 2002 \cite{Slot02}. Copyright (2002) from EdpSciences).}
\label{fig11-20}
\end{center}
\end{figure}

A characteristic feature of NbSe$_3$ is its large conductivity anisotropy. Evaluated from the chain conductivity along the $b$-axis, the conductivity anisotropy in the $b$-$c$ plane was estimated as $\sigma_b/\sigma_c\sim 10$, whereas the conductivity ratio perpendicular to the $b,c$ plane, $\sigma_b/\sigma_{a^\ast}$ may reach the value of $\sim 10^4$ a low temperatures \cite{Slot02}. This value was obtained \cite{Slot02} from measurements of the spreading resistance of a narrow trench across the width of a NbSe$_3$ crystal, with a depth $\sim 50$\% of its thickness (figure~\ref{fig11-20}).

From the crystallographic structure shown in figure~\ref{fig3-1}, it was seen that (in the $a,c$ plane) the conducting chains are assembled in elementary conducting layers in which the selenium prisms (the cross-section of chains perpendicular to $b$, the chain direction) are rotated and shifted with their edges toward each other. Each elementary conducting layer is, thus, separated for the adjacent ones by an insulating layer formed by a double barrier of the bases of selenium prisms. This type of layered structure combined with the large conductivity anisotropy allows to consider, as in layered high $T_c$ superconductors, that the CDW order parameter is modulated along the $a^\ast$ axis and  that the transport across the layers is determined by the intrinsic interlayer tunnelling between the CDW elementary layers.

\subsubsection{CDW gaps}\label{sec11-6-2}

\begin{figure}
\begin{center}
\includegraphics[width=7cm]{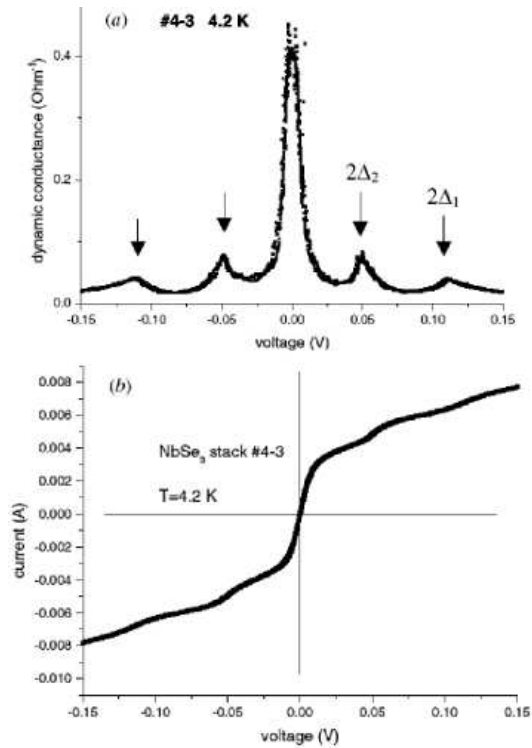}
\caption{Bias dependence of I-V characteristic and dynamical conductance, ${\rm d}I/{\rm d}V$ of a NbSe$_3$ mesa of 1~$\umu$m along $b$-axis, 1~$\umu$m along $c$-axis, and 0.03~$\umu$m along $a^\ast$-axis measured at $T$~= 4.2~K (reprinted figure with permission from Yu.I. Latyshev \textit{et al.}, Journal of Physics A: Mathematical and General 36, p. 9323, 2003 \cite{Latyshev03}. Copyright (2003) by the Institute of Physics).}
\label{fig11-21}
\end{center}
\end{figure}

This method has been intensively used to study intrinsic Josephson effect, superconducting energy gap and pseudogap, symmetry of the order parameter, ... of layered cuprate superconductors \cite{Kleiner94}. Figure~\ref{fig11-21} shows the (I-V) characteristic and the dynamic conductance ${\rm d}I/{\rm d}V$ of a NbSe$_3$ mesa with a thickness of approximately 30 individual layers at 4.2~K. In addition to the zero bias peak, i.e. ZBCP, the symmetric structure with peaks at $V_1$~= $\pm 120$~mV and $V_2$~= $\pm 50$~mV is very well resolved. The position of these peaks determines the double gap values $2\Delta_1$, and $2\Delta_2$ for the upper and the lower CDW gap of NbSe$_3$, respectively as expected for CDW/I/CDW tunnelling junction, namely 60~mV for CDW$_1$ and 25~mV for CDW$_2$.

CDW gaps in NbSe$_3$ have been measured revealing a relatively broad distribution in their amplitude depending on the techniques used; STM crossed junctions, point contacts, optics, angle-resolved photo-emission (ARPES), interlayer tunnelling. Concerning the lower CDW gap, $\Delta_2(0)$ was estimated 35~meV \cite{Dai92}, 36~meV \cite{Ekino94}, 36~meV \cite{Fournel86,Sorbier96}, 45~meV \cite{Schafer03}, 52~meV \cite{Neill06}, from optical measurements 70~meV which the light polarised along the $b$-axis and 136~mV for the light polarised along the $c$-direction \cite{Perruchi04} and 25~meV from point contact and interlayer tunnelling spectroscopy \cite{Latyshev03,Sinchenko03}. The distribution of the upper CDW gap amplitude seems to be less broad, except for optics measurements with $\Delta_1(0)$~= 101~meV \cite{Dai92}, 100~meV \cite{Ekino94}, 110~meV \cite{Schafer03}, 60~meV \cite{Sinchenko03,Latyshev03}, 281~meV \cite{Perruchi04}.

Spectroscopy by STM on individual chains in the unit cell of NbSe$_3$ showed \cite{Dai92} changes in the relative CDW  modulation depending of doping which was correlated with large changes in the amplitude of the CDW gaps. Thus, with Fe or Co doping, which does not alterate  either the CDW low range order or the transition temperature, $\Delta_2$ was found to be 25~mV (30\% reduction) for a Fe concentration of less than 1\% and 48~mV (37\% increase) for a Co concentration less than 3\% with respect to the 35~mV value for a ``pure" NbSe$_3$ crystal.

The tunnelling conductance ${\rm d}I/{\rm d}V$ spectra were also found to be substantially broadened as compared with those from the BCS theory. A probability distribution of BCS gaps resulting from either the spatial variation of the CDW modulation amplitude or the CDW gap anisotropy such as:
\begin{equation*}
P(\Delta,\Delta_0)=\frac{1}{(2\pi\delta\Delta)^{1/2}}\exp\,-\,\left(\frac{\Delta-\Delta_0}{2\delta\Delta}\right)^2,
\end{equation*}
where $\Delta_0$ is the mean gap value, and $\delta\Delta$ the standard deviation, should be taken into account \cite{Ekino94}. While the conductance peak energy yielded $\Delta^{\rm peak}_2$~= 37~meV and $\Delta_1^{\rm peak}$~= 100~meV, the fits to the raw data including $P(\Delta,\Delta_0)$ gave  the values of $\Delta_{0,2}$~= 23.4~meV and $\Delta_{0,1}$~= 51.3~meV which were considered \cite{Ekino94} to be taken as the CDW gap values.

In addition, it was shown in section~\ref{sec3-1-1} that the band structure of NbSe$_3$ is anisotropic and that the transfer integral perpendicular to the chains is not negligible; that can be reflected in the tunnelling experiments. In that case it was calculated \cite{Huang90} that singularities in tunnelling spectra do not occur at the gap $\pm\Delta_0$ value but at $\pm(\Delta_0+\epsilon_0)$ when $\epsilon_0$ is the transverse corrugation of bands. $\epsilon_0$ was estimated \cite{Sorbier96,Neill06} to be of the same order of magnitude than $\Delta_0$ (which makes a factor two in the gap value).

\begin{figure}
\begin{center}
\includegraphics[width=6cm]{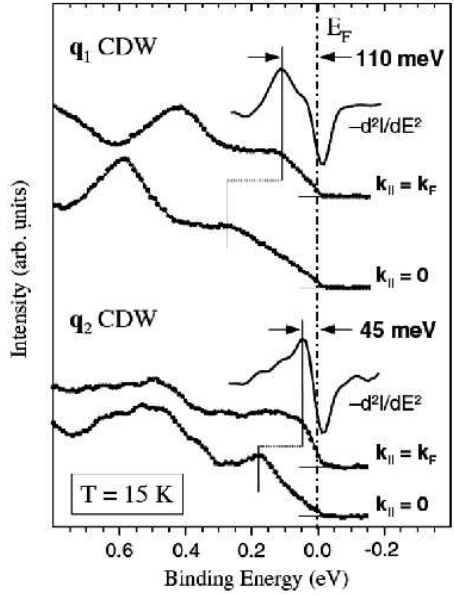}
\caption{Spectral functions from ARPES line scans  at $k_\parallel=0$ and $k_\parallel=k_{\rm F}$ measured in NbSe$_3$ at $T$~= 15~K; gap spectra are exhibited in second derivatives $-{\rm d}^2I/{\rm d}E^2$ (reprinted figure with permission from J. Sch\"afer \textit{et al.}, Physical Review Letters 91, p. 066401, 2003 \cite{Schafer03}. Copyright (2003) by the American Physical Society).}
\label{fig11-22}
\end{center}
\end{figure}

Figure~\ref{fig11-22} 
shows the spectral functions derived from angle-resolved photoemission spectroscopy (ARPES) line scans at two positions $k_\parallel=0$ and $k_\parallel=k_{\rm F}$ measured \cite{Schafer03} in NbSe$_3$ at $T$~= 15~K, the gap singularities, $\Delta_1$~= 110~meV and $\Delta_2$~= 45~meV, being better resolved in the ${\rm d}^2I/{\rm d}E^2$ spectra. It was noted \cite{Schafer03} that tunnelling averages over $k$ space (as in ARPES) lead to an underestimation of the gap, that appear to be even more important in optics \cite{Perruchi04}.

The absolute determination of CDW gaps in NbSe$_3$ is impeded by the effects of impurity content, anisotropy, distribution of CDW modulation amplitude, imperfect nesting, ... . Nevertheless taking the lowest values determined by point contact and interlayer tunnelling, namely $\Delta_1$~= 25~meV and $\Delta_2$~= 60~meV the ratio $2\Delta_1/kT_{\rm P_1}$~= 9.59 and $2\Delta_2/kT_{\rm P_2}$~= 9.82 is, in any case, much larger than the mean BCS value: 3.52.

It is important to note  that the voltage at which the dynamic conductance of NbSe$_3$ mesas show peaks are close to that measured in traditional point contact or tunnelling experiments. That differs of the behaviour for Josephson effects in high $T_c$ superconducting mesas \cite{Kleiner94}. It was shown  that, if a voltage $V_n$ drops across the $n^{\rm th}$ junction in the mesa, the Josephson current will oscillate at the frequency $f_n$~= $(2e/\hbar)V_n$. For synchronised $N$ junctions, the maximum emitted power occurs at $V$~= $N(h/2e)f$, with $V=\sum^N_{n=1}V_n$, with $N$ the number of junctions in the mesa. Such a behaviour does not occur in NbSe$_3$ mesas. Thus, only one elementary tunnel junction in the mesa is operating at high bias voltage; a single junction was considered \cite{Latyshev03} to be the weakest elementary junction where the CDW phase decoupling occurs in increasing $V$.

\subsubsection{Zero bias conductance peak (ZBCP)}\label{sec11-6-3}

In figure~\ref{fig11-21} the large amplitude conductance peak at zero bias voltage appears as the most prominent feature in the I-V characteristics. This conductivity peak has been ascribed \cite{Latyshev02a,Latyshev02b,Latyshev03} to the almost coherent interlayer tunnelling of charge carriers non condensed in the CDW ground state, localised at low temperature in the pockets of the Fermi surface. Both the height and the width of ZBCP characterise the mesa quality. For the best samples, the ratio of the ZBCP height to the background value at $V>2\Delta$ can reach $\sim 30$ while the ZBCP half-width is $\sim 10$~mV. For poor quality stack junctions, or for point contact along $a^\ast$, this ratio is strongly reduced.

\begin{figure}
\begin{center}
\includegraphics[width=7.5cm]{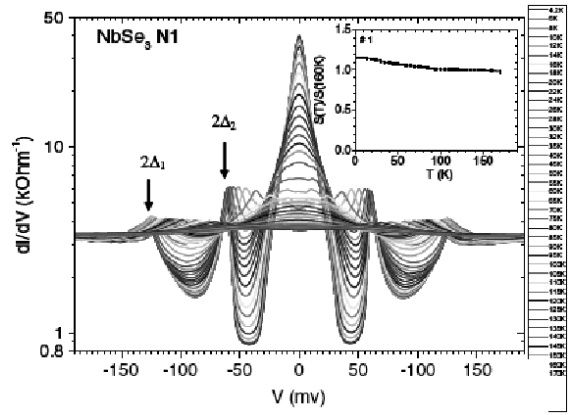}
\caption{Temperature evolution of interlayer tunnelling spectra of NbSe$_3$ in the 4.2~K--170~K temperature range. Inset shows the temperature dependence of the integral $S=\int^{+V_0}_{-V_0}{\rm d}I/{\rm d}V(V)dV$ (reprinted figure with permission from Yu.I. Latyshev \textit{et al.}, Superconductor Science and Technology 20, p. 587, 2007 \cite{Latyshev07}. Copyright (2007) by the Institute of Physics).}
\label{fig11-23}
\end{center}
\end{figure}

\begin{figure}
\begin{center}
\includegraphics[width=6.5cm]{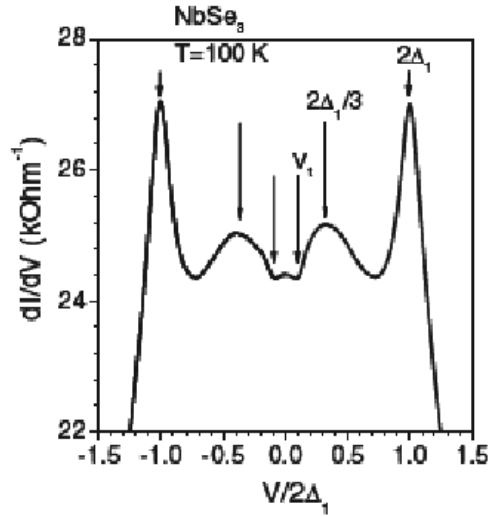}
\caption{Schematic interlayer tunnelling spectrum of NbSe$_3$ at $T$~= 100~K.}
\label{fig11-24}
\end{center}
\end{figure}

It was shown \cite{Latyshev07} also  that, in spite of very sharp changes of ${\rm d}I/{\rm d}V(V)$ below the Peierls transitions, the total spectral weight $S$ is nearly constant in the whole temperature range 4~K--170~K (inset figure~\ref{fig11-23}). $S$ was defined as  the integral at each temperature of the interlayer tunnelling spectrum (as shown in figure~\ref{fig11-23}) between extrema voltages from $-V_0$ to $+V_0$, $V_0$~= 200~mV being much above the highest CDW singularity. That result indicates that the opening of CDW gaps does not affect the total number of single-particle states.As also shown in figure~\ref{fig11-23} and in more detail in figure~\ref{fig11-25} the conductance peak at $V=0$ decreases with $T$ is increased and is totally suppressed above $\sim 25$~K. The interlayer tunnelling spectrum for the upper CDW is schematically drawn in figure~\ref{fig11-24}. It emphasises the CDW gap, a sub-gap singularity: at $2\Delta/3$ ascribed to amplitude solitons and the threshold $V_t$ for CDW phase decoupling as discussed below.

\subsubsection{Intragap CDW states}\label{sec11-6-4}

\begin{figure}
\begin{center}
\includegraphics[width=7.5cm]{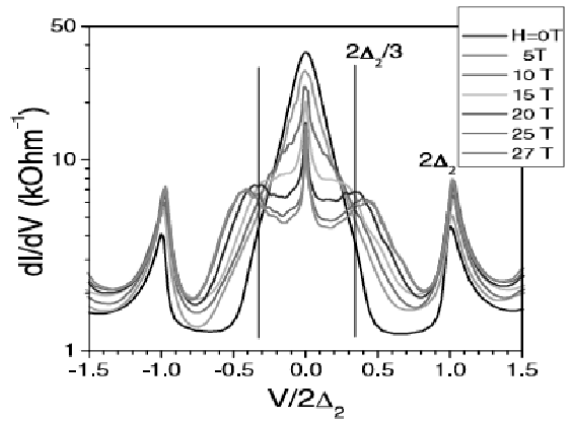}
\caption{Magnetic field dependence spectra of the tunnelling conductivity ${\rm d}I/{\rm d}V$ of a NbSe$_3$ stack with the voltage $V$ normalised to the CDW gap $2\Delta_2$ measured at $T$~= 4.2~K. $H\parallel c$ (reprinted figure with permission from Yu.I. Latyshev \textit{et al.}, Physical Review Letters 95, p. 266402, 2005 \cite{Latyshev05}. Copyright (2005) by the American Physical Society).}
\label{fig11-25}
\end{center}
\end{figure}

It was found that the application of a high magnetic field applied along $c$- or $a^\ast$-direction drastically narrow or suppress the ZBCP, recovering and sharpening the structures at higher $V$. Figure~\ref{fig11-25} 
shows the tunnelling spectra of NbSe$_3$ at 4.2~K for the lower CDW at several magnetic fields $B$ up to 27~T applied in the junction plane \cite{Latyshev05}. In addition to the gap singularities at $2\Delta_2$, surprisingly sharp features appear inside the $\Delta_2$ gap as an additional singularity at $V$~= $V_{AS}\simeq 2\Delta_2/3$. Similar singularity is also visible for the upper CDW \cite{Latyshev05}. This unexpected peak was interpreted \cite{Latyshev05} in the picture of solitons as special elementary excitations of the CDW state \cite{Brazovskii84,Brazovskii05}.

\begin{figure}
\begin{center}
\includegraphics[width=6.5cm]{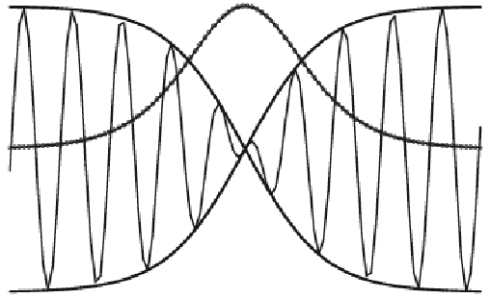}
\caption{Profile of the oscillating electronic density (thin line) with its envelop (thick line) and of the distribution of the spin-density (dotted line) of an amplitude soliton (AS) (reprinted figure with permission from Yu.I. Latyshev \textit{et al.}, Physical Review Letters 95, p. 266402, 2005 \cite{Latyshev05}. Copyright (2005) by the American Physical Society).}
\label{fig11-26}
\end{center}
\end{figure}

The incommensurate CDW order parameter in the uniform ground state, $\Delta_0$~= $A\cos(Qx+\varphi)$, with $\varphi$ the arbitrary phase and $A$~= cst, is degenerated with respect to the transformation $A\leftrightarrow -A$. That leads the possibility of a non-uniform ground state with local phase change of $\pi$ and the simultaneous acceptance of one electron from the free band, excitation called amplitude soliton ($\Delta S)$. In that case $A$~= $\tanh(x/\xi_0)$, $\xi_0$: the amplitude CDW order parameter. The $AS$ is a self-localised state with an energy $E_s$~= $2\Delta_0/\pi$~= $0.65\Delta_0$ \cite{Brazovskii84}. This state is more stable because the energy $0.35\Delta_0$ is gained by converting the electron into the $AS$. The peak at $V_{AS}$ in tunnelling spectra results from the tunnelling between the Fermi level (middle of the gap) of the remnant electronic pocket to the $AS$ level below the $\Delta_0$ gap. Profile of the $AS$ is shown in figure~\ref{fig11-26} showing the oscillating electronic density and the spin density distribution of a impaired electron. The $AS$ carries the electronic spin $s$~= 1/2 but its charge is zero, being the realisation of a spinon \cite{Brazovskii89}.

\subsubsection{Phase decoupling}\label{sec11-6-5}

\begin{figure}[h!]
\begin{center}
\includegraphics[width=7.5cm]{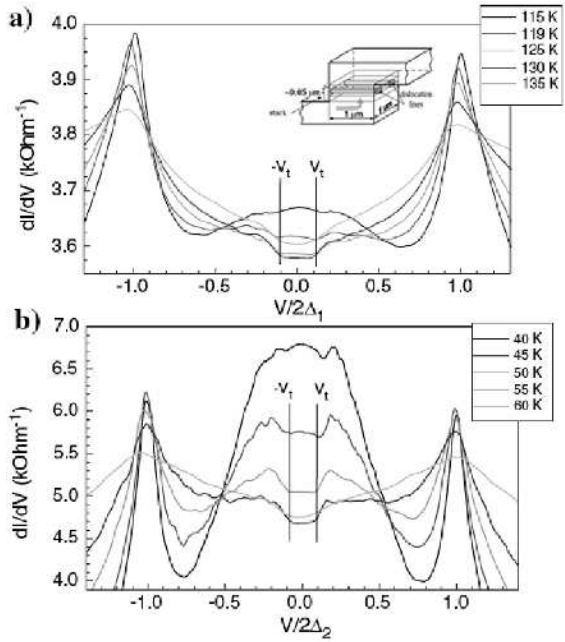}
\caption{Tunnelling spectra ${\rm d}I/{\rm d}V$ in NbSe$_3$ as a function of the voltage $V$ normalised to the CDW gap $2\Delta$ at different $T$: a)~upper CDW, b)~lower CDW (reprinted figure with permission from Yu.I. Latyshev \textit{et al.}, Physical Review Letters 96, p. 116402, 2006 \cite{Latyshev06}. Copyright (2006) by the American Physical Society).}
\label{fig11-27}
\end{center}
\end{figure}

Another remarkable effect that appears at lower energies within the CDW gap is a sharp voltage threshold $V_t$ for onset of the interlayer tunnelling conductivity (as seen in figure~\ref{fig11-27} 
for both CDWs in NbSe$_3$ when the ZBCP is suppressed and schematically shown in figure~\ref{fig11-24}. The connecting electrodes to  the mesa are along the $c$-axis, preventing the interference between interlayer tunnelling and a possible CDW sliding within connecting channels (figure~\ref{fig11-5}(c)). $V_t$ was found to be $V_t\sim 1.3k_{\rm B}T_{\rm P}$ for both CDWs in NbSe$_3$ and in o-TaS$_3$, energy corresponding to the CDW phase decoupling between neighbouring layers. A model has described \cite{Latyshev06} this phase decoupling via the formation, in the single weakest junction, of dislocation lines (DLs), oriented across the chains as represented schematically in inset of figure~\ref{fig11-27}(a). DLs appear as the result of the shear stress of the CDW induced by the electric field. The circulation around the DL core gives a phase variation of $2\pi$. The excess charge of the DL accumulates the electric field within the DL core. In the vertical direction, the DL core has an atomic size, the field being concentrated within one elementary junction; but the in-plane size of the DL core is larger, about $\sim 100~\AA$ \cite{Brazovskii91a,Brazovskii91b}.
\begin{figure}
\begin{center}
\subfigure[]{\label{fig11-28a}
\includegraphics[width=6.5cm]{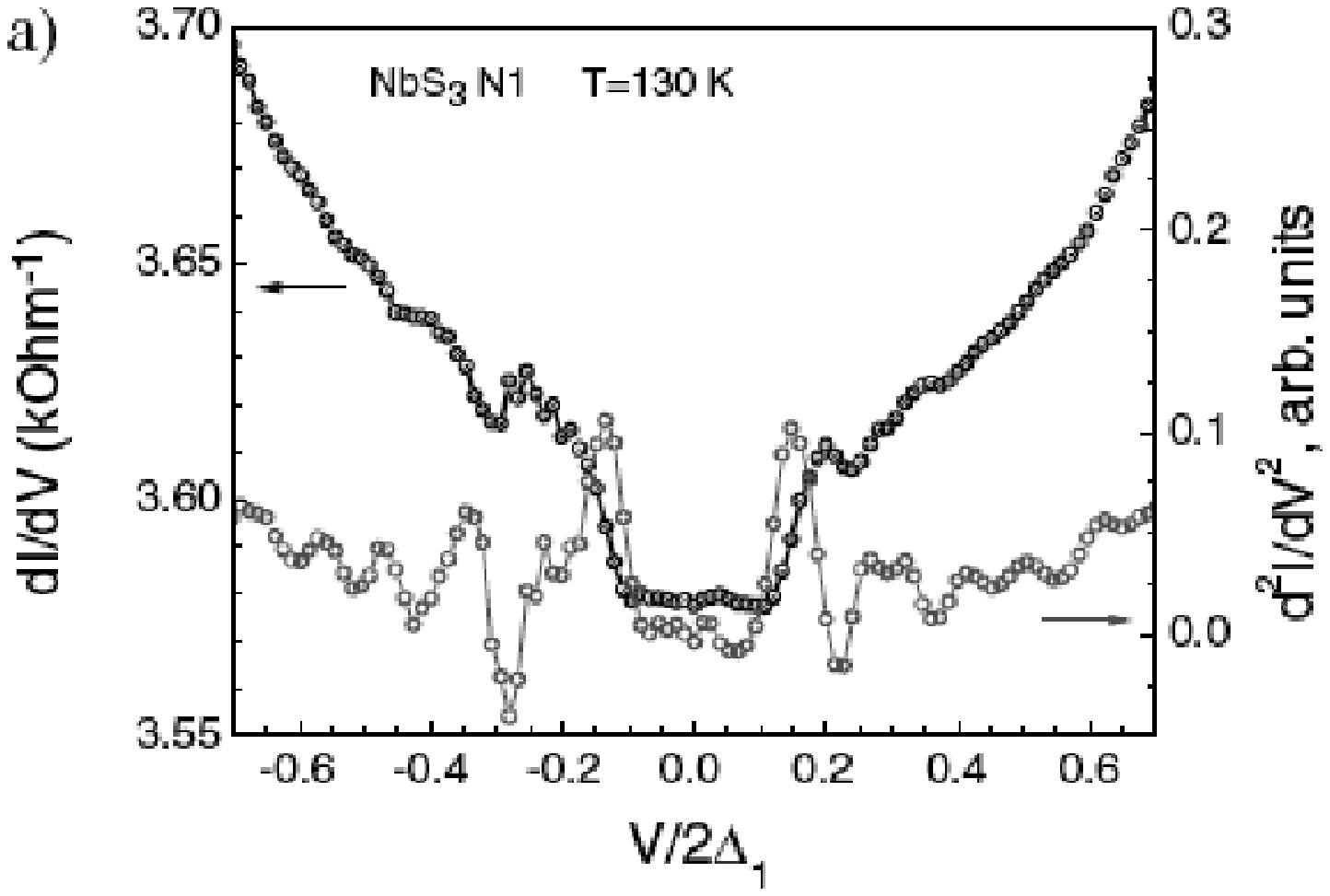}}
\subfigure[]{\label{fig11-28b}
\includegraphics[width=6.25cm]{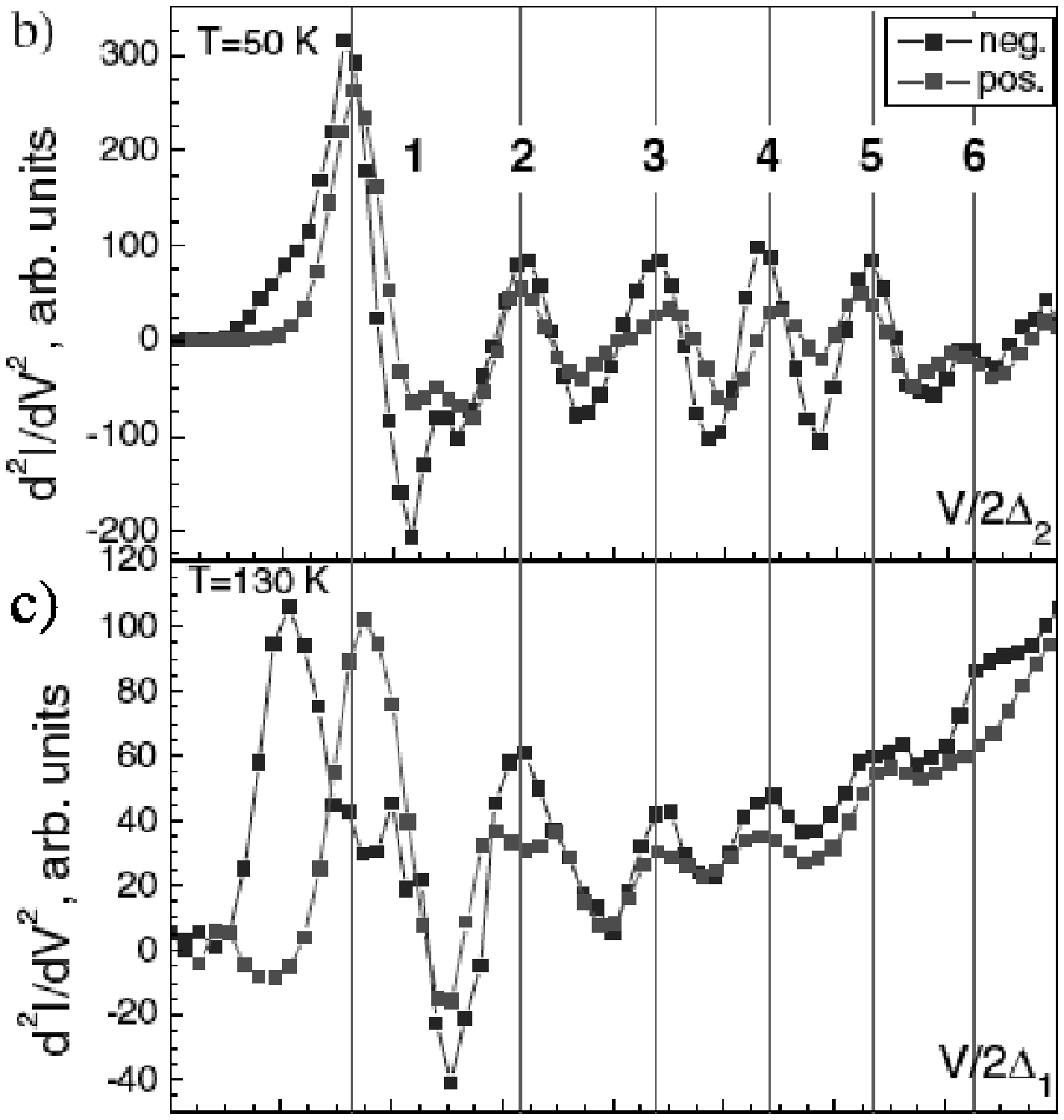}}
\caption{Tunnelling spectra in NbSe$_3$ within the magnified threshold region with the voltage $V$ normalised to the CDW gap $2\Delta$. a)~${\rm d}I/{\rm d}V$ and its derivative ${\rm d}^2I/{\rm d}V^2$ at $T$~= 130~K as a function of the voltage $V$ normalised to the CDW gap; b)~comparison of ${\rm d}^2I/{\rm d}V^2$ for both polarities at $T$~= 50~K (lower CDW) and at $T$~= 130~K (upper CDW). Peaks are interpreted as a sequential entering of dislocation lines into the junction area (reprinted figure with permission from Yu.I. Latyshev \textit{et al.}, Physical Review Letters 96, p. 116402, 2006 \cite{Latyshev06}. Copyright (2006) by the American Physical Society).}
\label{fig11-28}
\end{center}
\end{figure}

Figure~\ref{fig11-28} shows \cite{Latyshev06} the fine structure of interlayer tunnelling spectra in NbSe$_3$ at $T$~= 130~K. Steps in ${\rm d}I/{\rm d}V$ and corresponding sharp peaks in ${\rm d}^2I/{\rm d}V^2$ are clearly observed. For each CDW, the structure is well reproduced for both bias polarities. It is also shown that for the normalised $V/2\Delta_{1,2}$, the peak positions coincide for both CDWs. $V_t$ was thus identified \cite{Latyshev06} as the DL entry energy. The periodic structure above the threshold voltage corresponds to the appearance of the first, second, ... till six dislocation lines.

There is a remarkable similarity between layered superconductors and CDW systems that manifest themselves in similar mechanisms of phase decoupling via the formation of phase vortices. In both cases, a threshold energy for phase decoupling associated with $H_{c_1}$ for superconductors or $V_t$ is much less than the value of the energy gap. The duality of phase topological defects in layered high $T_c$ superconducting Josephson vortices \cite{Clem90} and CDW systems \cite{Latyshev07} is summarised in table~\ref{tab11-1}.
\begin{table}[h!]
  \tbl{Duality of phase topological defects in layered superconductors (high $T_c$) and CDW systems (from \cite{Latyshev07}).}
{\begin{tabular}{@{}cc}\toprule
$\begin{array}{c}
   \mbox{Layered superconductors}\\
   \mbox{(high $T_c$)}
   \end{array}$
   & CDW\\
   \colrule
   flux & charge \\
   $\phi_0=hc/2e$ & $Q_0=2e$ \\
   $H\parallel\,$layers & $E\perp\,$layers \\
   $H_{c_1}$ & $E_t$ \\
   \botrule
  \end{tabular}}
      \label{tab11-1}
\end{table}

However, as noted in \cite{Coppersmith91} there are fundamental differences between CDWs and flux lattices in type II superconductors when defects are accounted for, essentially because CDW wavelengths can be created freely at phase slips while magnetic flux lines cannot be created freely at grain boundaries of the flux lattice.

\subsection{Current-effect transistor and gate effect} \label{sec11-7}

A few attempts have been performed to modulate the threshold field for depinning the CDW.

\subsubsection{Current effect transistor} \label{sec11-7-1}

The damping in moving periodic structures has been intensively studied theoretically \cite{Balents98,Giamarchi98}. A normal (single current) current density $J_x$ transverse to the direction of the CDW motion was predicted \cite{Radzihovsky98} to have dramatic effects on the longitudinal depinning field for $J_c$ above a threshold value. For a fixed applied $J_x$, the threshold longitudinal $E_T$ was calculated as:
\begin{eqnarray*}\begin{array}{c}
E_T(0) \mbox{ for } J_x<J_c\\
E_T(J_x)   =  E_T(0) \,\displaystyle\frac{J_x}{J_c}\,\exp [-2(J_x-J_c)/2]\hspace*{0.5cm}\mbox{for $J_x>J_c$},\end{array}
\end{eqnarray*}
with $J_c$~= $\sigma_\infty E_T(0) (k_{\rm F}\xi_L)\,\rho_n/\rho_{\rm CDW}$ (with $\sigma_\infty$ is the large field limit of the longitudinal CDW conductivity, $\rho_n$ and $\rho_{\rm CDW}$ the normal and the CDW electron density, $\xi_L$ the Larkin coherence length). The exponential decay of $E_T$ above a critical transverse current $J_c$ was reported in ref.~\cite{Markovic00}. The conduction in the CDW channel appeared to be modulated by a current in the single-particle channel, making the device to operate in principle as a transistor. This non-equilibrium dynamics was taking its origin \cite{Radzihovsky98} by the introduction in the Fukuyama-Lee-Rice model of an additional ``convective" term in the damping such as $\gamma\rightarrow\gamma[\partial_t\varphi+v\partial_x\varphi]$. Physically it means that the CDW becomes longitudinally more ordered with transversely moving normal carriers \cite{Radzihovsky98}. The roughness of the CDW wave fronts in the motion direction, due to the presence of defects or impurities in the crystal, is reduced (or ``ironed") \cite{Markovic00} by the transverse current. The CDW transport being more coherent, the pinning is less effective and the longitudinal threshold for depinning is reduced.

However a microscopic derivation \cite{Artemenko00} of the damping showed that the effect of convective terms is orders of magnitude smaller than evaluated in ref.~\cite{Radzihovsky98}. Contribution from transverse currents of electron- and hole-like quasiparticles to the force exerted on the moving CDW were shown to act in opposite directions. The convective term is still operating \cite{Artemenko00} but at a $J_c$ several orders of magnitude larger as previously estimated \cite{Radzihovsky98}.

The apparent reduction of the threshold field \cite{Markovic00,Yue01} was ascribed \cite{Ayari02,Ayari02a} to electric field inhomogeneities near lateral current leads. Using an appropriate design reducing considerably field inhomogeneity, no reduction of the threshold field was observed up to $J_\perp$~= $2\times 10^4$~A/cm$^2$ compared to $J_c$~= 750~A/cm$^2$ in ref.~\cite{Markovic00}.

\subsubsection{Gate effect} \label{sec11-7-2}

Field-effect modulation of CDW transport can also be achieved by the realisation of MOSFET-like devices. Both single particle and collective transport in NbSe$_3$ were modulated by application of a gate voltage \cite{Adelman95,Slot05,Latyshev09}. For a sample with 5.7$\times 10^{-3}$~$\umu$m$^2$ cross section, the induced quasiparticle was of the order of 0.1\% of the total carrier density \cite{Adelman95} for a gate voltage of 15~V. For samples with a thickness down to 20~nm at temperatures between 30 and 50~K, the increase of the single particle conduction was up to 2.5 times \cite{Latyshev09} with a gate of a few $V$. This effect was associated with the penetration of the electric field in a large part of the sample volume, which shifts the Fermi level increasing the number of carriers. The gate effect also was shown to decrease rapidly above the threshold for CDW depinning, suggesting that the CDW sliding screens much more efficiently the transverse field than in the pinned state \cite{Latyshev09}.

Similar attempts were made on organic charge order/Mott insulators \cite{Yamamoto09}. Injecting carriers by electrostatic doping into molecular conductors would allow to study the strength of electron correlations depending on the band filling. Two kinds of procedure were used to fabricate field-effect transistor structures, either by a direct growth of the organic single crystals by electrochemical reaction on platinum electrodes deposited on the gate dielectric or by using very thin crystals obtained by lamination. Field effect was obtained \cite{Yamamoto09} on the charge-ordered phase of $\alpha$-(BEDT-TTF)$_2$I$_3$ which a maximum in the gain of $\sim 2$ in the $T$ range between 20--50~K.

\section{Conclusions}\label{sec12}
\setcounter{figure}{0}
\setcounter{equation}{0}

The preceding sections have reviewed the general properties of quasi-one-dimensional electronic inorganic as well as organic crystals. In spite of the variety of systems, their properties are quite universal: a depinning transition above a threshold, collective transport properties associated to non-linear and non-stationary effects due to the sliding phenomena, screening effects, glassy properties due to numerous metastable states, ... These systems exhibit the collective Fr\"ohlich conductivity, with the current carried by the motion of the C/S DW corresponding to the total electron concentration in the bands affected by the density wave gaps or to 2e per chain. Quite exciting, this coherent motion of a macroscopic number of electrons was even observed above the room temperature. Although similar to superconductivity however, damping prevents superconductivity.

Phase slippage is ubiquitous in many properties and may correspond to the microscopic mechanism for sliding. Elementary events were found in the temperature dependence of the CDW $Q$ vector in mesoscopic CDW systems, in phase dislocations in planar junctions similar to $H_{c_1}$ in superconductors, and along a single chain in NbSe$_3$ by STM.

Advanced techniques often developed for other purposes have been used for electronic crystals: atomic resolution by STM, X-ray coherent scattering, femtosecond spectroscopy, photoemission, spatially-resolved X-ray diffraction, focused ion beam for mesoscopic devices, ... they have then brought a better microscopic understanding of their properties.

In the following, some additional topics are shortly reviewed.

\section{Epilogue}\label{secepilogue}
\setcounter{figure}{0}
\setcounter{equation}{0}

However, the present review can only be a part of the much wider research activity in low dimensional systems. In this epilogue several lines of researchs will be shortly presented such as the instability of the Fermi liquid picture in 1D, unconventional density wave, search of the sliding mode in 2D systems. Phase separation with stripe formation is often observed in perovskite oxides. In the underdoped regime of 2D conducting oxides exhibiting superconductivity at high temperature, different symmetry-broken ground states lead to various forms of spontaneous electronic superstructures; that extend and enrich the studies of static and dynamic properties of superstructures reported essentially in quasi-1D materials in this review.

\subsection{Potassium}\label{sec12-2}

First of all, it is worth to remind that historically the very early theoretical development of instabilities among itinerant electrons presented by A. W. Overhauser \cite{Overhauser68} suggested that a CDW may be the ground state of simple metals such as the alkali metals. The argument was essentially based on the weak Born-Mayer ion-ion interactions, known to be extremely weak in the alkali metals; thus, the equilibrium position of the positive ions could be displaced from the ideal cubic sites, cancelling most of the Coulomb correlation energy; the positive ion charge modulation will just cancel the electronic charge density wave.

Over many years, predictions were made on the effect of a CDW in potassium in magnetoresistance, optical adsorption, Haas-van~Alphen periodicity, specific heat (\cite{Amarasekara82} and references therein). The observation of satellites in potassium was reported \cite{Giebultowicz86} from neutron scattering with the CDW wave vector $Q$~= [0.995,~0.995,~0.015] in the extreme vicinity of the (110) Bragg reflection. The intensity of these satellites was $\sim 10^{-5}$ that of the nearby Bragg peak. This result was thought to be the end of a long search over several decades \cite{Giebultowicz86}. However the identification of the peaks observed in inelastic scans in potassium to CDW satellites was challenged in ref.~\cite{Pintschovius87}; these peaks being analysed as resulting from double-scattering processes. It was shown that no satellite-like structures appear in the limit of $\sim 10^{-7}$ of the intensity of the (110) Bragg peak.

However the mosaic width of the Bragg peak of K single crystal in the [110] direction with polarisation $[1\bar{1}0]$ was shown to soften a few \% between 100~K and 4~K, near the zone boundary \cite{Blaschko88}. The softening in potassium is similar in size to that observed in Na which undergoes a martensitic phase transition at 36~K \cite{Blaschko84,Smith87} but smaller to that at the martinsitic phase transition in bcc lithium which occurs near 70~K \cite{Ernst86}.

\subsection{Luttinger liquid}\label{sec12-5}

As seen in section~\ref{sec2}, when dimensionality is reduced to one, the Fermi liquid state becomes unstable to Coulomb interactions and the conduction electrons behave according to the Tomonaga Luttinger liquid theory. In this context, the charge density is large enough to ignore the strong correlations between electron positions. The single particle spectral function $\rho(\omega)$ is, then, predicted to have a power law dependence on the binding energy $\rho(\omega)\alpha |\omega |^\alpha$ with $\alpha$ depending on the strength and range of interactions related to the parameter $\kappa_\rho$ of the Luttinger model \cite{Voit95}: $\kappa_\rho$~= $\frac{1}{8}[\kappa_\rho+\kappa^{-1}_\rho-2]$. The dispersion of the quasi particle peak of the Fermi liquid is replaced by two distinct singularities representing the spinon and the holon excitations; these peaks are degenerate at $k_{\rm F}$ but disperse with different velocities reflecting the spin-charge separation \cite{Giamarchi04,Dressel03,Deshpande10,Fiete07}.

For Q-1D organic Bechgaard-Fabre salts, Luttinger liquid fingerprints have to be searched at high temperature when $k_{\rm B}T>t_\perp$, with $t_\perp$ is the transverse coupling energy, or at high energy, because any transverse coupling between 1D chains will destroy inevitably the Luttinger liquid state \cite{Castellani94}. $\kappa_\rho$ was determined by transport properties ($\kappa_\rho\sim 0.2$) \cite{Moser98}, photoemission ($\kappa_\rho\sim 0.2$) (see review in ref.~\cite{Grioni04} and ref.~\cite{Grioni09}, also in  optical conductivity ($\kappa_\rho\sim 0.23$) \cite{Vescoli00,Schwartz98}, in relative agreement with theoretical predictions for quarter-filled band systems \cite{Giamarchi97}. However a clear observation of spin-charge separation which will establish the Luttinger liquid scenario in Q-1D materials is more elusive (see discussion in ref.~\cite{Grioni04,Grioni09} with the exception of the double Cu-O chain compound SrCuO$_2$ \cite{Kim06}, or in the anisotropic metal Li$_{0.9}$Mo$_6$O$_{17}$ \cite{Wang09}.

Power law dependences in transport properties of many Q-1D \cite{Bockrath99,Aleskin04,Zaitsev-Zotov00,Slot04} were ascribed to Luttinger liquid behaviour namely for low applied voltage $I\alpha VT^\alpha$ and for high voltage $I\alpha V^{\beta+1}$.  Thus, in the case of NbSe$_3$ with a cross-section of a few thousand of chains, the non-linear I-V characteristics when plotted $I/T^{1+\alpha}$ versus $eV/k_{\rm B}T$ collapse in a master curve \cite{Slot04}. However, cautions should be taken concerning the relevance of the Luttinger liquid effects in these experiments. Indeed it was shown that, in such Q-1D systems, the conventional mechanism of transport, which is the variable range hopping, can also lead to the same power law dependences \cite{Rodin10}.

Then, Tomonaga-Luttinger states were searched in other crystalline 1D systems, specifically carbon nanotubes considered to be ideal 1D systems. Indeed, photoemission \cite{Ishii03} measurements have shown power law dependence on binding energy as expected for Luttinger liquids, and a similar behaviour was also observed in conductance measurements \cite{Bockrath99}. However caution should be taken because the intertube interaction, as well as the distribution in nanotube diameters and in chirality should affect the spectral profiles in photoemission.

Finally artificial 1D structures were fabricated at metal interfaces, such as metallic vicinal surfaces Au111 or Cu111 self-assembled atomic chains at silicon single crystal surfaces. These 1D objects have been studied conjointly by a $k$-resolved probe like ARPES and a local probe by STM (for a review see \cite{Grioni09}).

\subsection{Quantum wires}\label{sec12-6}

It was shown \cite{Schulz93} that for dilute systems of electrons in 1D gas, the long range Coulomb force leads to a $4k_{\rm F}$ state, expected for a 1D Wigner crystal, and that all unusual dynamical properties associated with non-linear transport in classical CDWs should occur in 1D electron system. Quantum wires are usually made by confining a 2D electron gas to a 1D channel (for a review see \cite{Meyer09}). The electron density can be controlled by means of the gate voltage. For a stronger gate-induced depletion of the channel, when $n\;a_{\rm B}\ll 1$ ($a_{\rm B}$~= $\epsilon\hbar^2/me^2$: Bohr radius of the semiconductor material, $n$: electron density), Coulomb interactions become dominant. Non-linearity and oscillations in conductance in quantum wires in GaAs heterostructures \cite{Meirav89} and in Si-MOSFET \cite{Scott-Thomas89,Field90} were explained as suggestive of pinned charge density wave, although an alternative model of Coulomb blockade was also presented \cite{Field90}. Conductance of a quantum wire was calculated \cite{Matseev04} in the Wigner crystal regime taking into account the perturbation of the leads. Studying the pinning of a 1D Wigner crystal in the presence of weak disorder, it was shown \cite{Glazman92} that quantum fluctuations soften the pinning barrier and charge transfer occurs due to thermally assisted tunnelling. Conducting polymer nanowires (polypyrrole) with low electron densities were shown \cite{Rahman07} to exhibit a power law dependence of the I-V characteristics, a switching to highly conducting state above a threshold voltage with a negative differential resistance and enhancement of noise that may indicate a Wigner crystallisation.

Progresses in sample growth for carbon nanotubes have yield the production of very clean and low-disorder semiconducting 1D systems which a large gap ($\sim 100$~meV) which reveal to be Wigner crystals \cite{Deshpande08}.

\subsection{Unconventional density waves}\label{sec2-17}

In the preceeding sections the density wave order parameter $\Delta$ was implicitly considered to be constant ($\Delta$~= cst) or slightly anisotropic. However the wave vector dependence of the order parameter $\Delta(k)$ taking different values at different positions on the Fermi surface was shown to have a crucial effect in theories of superconductivity, including high $T_c$ materials. In these compounds, the superconductivity is unconventional (in the sense of the BCS model) or nodal, the quasi particle spectrum having no finite gap along specific directions.

As for superconductivity, a similar situation was suggested for density waves, then called unconventional density waves (UDW) exhibiting unconventional properties due to the $k$-dependent gap \cite{Nayak00}: $\Delta(k)$~= $\Delta_0\,f(k)$. $f(k)$~= 1 corresponds to the ``classical" Peierls state and thus to a singlet s-wave density wave, $f(k)$~= $\sin k_xa$ for a singlet $P_x$ state and $f(k)$~= $\cos(k_xa-\cos k_ya)$ for a singlet $d_{x^2-y^2}$ state. The average of the gap function over the Fermi surface is zero, causing a lack of the periodic modulation of the C/S DW. It was theoretically determined \cite{Dora04a,Dora04b} that the two hallmarks of U-DW would be the angle-dependent magnetoresistance and giant Nernst effect experimentally measured \cite{Basletic02,Basletic07}.

Due to the vector nature of the order parameter, the SDW may have a chiral character. Although the CDW order parameter is a scalar, however a CDW chiral state has been recently discovered in layered dichalcogenide 1T-TiSe$_2$ \cite{Ishioka10,Ishioka11} and 2H-TaS$_2$ \cite{Guillamon11} by STM microscopy. Topographic images show hexagonal atomic lattice and CDW with clockwise and counterclockwise charge modulations. It was explained \cite{Wezel12} that the helical symmetry of TiSe$_2$ can be understood as the consequence of the combined presence of charge and orbital order and results from the combinations of at least three differently polarised CDW components with no zero relative phase differences. These stringent presequisites may restrict chiral charge order to a very few systems \cite{Wezel12}

\subsection{Electronic phase separation}

Systems with strong electronic correlations exhibit a wide variety of properties due to strong coupling and competition among spin, charge, orbital and lattice degrees of freedom. One fingerprint of the resulting ground states is often phase separation taking the form of stripes. Charge order in 2D organic compounds reveals such inhomogeneous ground states. Phase separation is magnificently demonstrated in underdoped perovskite oxides leading to high temperature superconductivity (HTSC) and in manganites in the colossal magnetoresistance regime \cite{Tokura00}.

\subsubsection{Charge-order stripe in 2D (BEDT-TTF)$_2$X salts}

Two-dimensional organic compounds, such as those from the bis(ethylenenedithio)-tetrathiafulvalene) (BEDT-TTF)$_2$X family (see figure~\ref{fig3-26} for molecule structure) are strongly correlated systems in which the ground state resulting from the interplay between the on-site Coulomb energy $U$ and the intersite Coulomb $V_{ij}$ may result in a CO state (for a review \cite{Seo00,Seo06}.

A large variety in the 2D arrangement of (BEDT-TTF) molecules yields different polytypes classified by Greek characters: $\alpha$, $\beta$, $\kappa$, $\theta$, $\lambda$, ... . In the case of $\alpha$ and $\theta$ polytypes, the (BEDT-TTF) molecules are disposed in the conducting layers along a triangular lattice with one hole for every two molecules. Not only the anisotropy in the transfer integral $t_{ij}$ has to be taken into account but also the anisotropy of the intersite $V_{ij}$: namely, in the case of the $\theta$-structure, $V_c$ along the stacking $c$ direction and $V_p$ along the bonds in the transverse direction.

Mean field calculations with the extended Hubbard model on this anisotropic triangular lattice were performed on materials with $\theta$, $\theta_d$ and $\alpha$-type structures \cite{Seo00}. It was found that CO occurs on the form of charge rich (0.5+$\delta$) and charge poor (0.5-$\delta$) stripes which, depending on the relative value of $V_c$ and $V_p$ can be horizontal, vertical or diagonal \cite{Seo00,Seo06}. The $\theta_d$-structure corresponds to the case where CO is accompanied by a dimerised structure. Mean-field calculations have shown that when $V_c\gtrsim V_p$, the horizontal stripe solution is the more stable for $\theta_d$ and $\alpha$-type structures.

Metal-insulator transitions observed in $\theta$-(BEDT-TTF)$_2$RbZn(SCN)$_4$ around 190--200~K \cite{Mori98} and $\alpha$-(BEDT-TTF)$_2$I$_3$ \cite{Bender84} were ascribed as being due to CO \cite{Takahashi06}. A drastic change in the NMR line-shape was found below $T_{\rm CO}$ in $\theta$-(BEDT-TTF)$_2$RbZn(SCN)$_4$ with evidence of charge-rich and charge-poor sites \cite{Miyagawa00,Chiba00}. Raman spectroscopy has also revealed the split of the charge sensitive $\nu_2$ $C=C$ stretching mode, indicating a charge difference among (BEDT-TTF) molecules at non-equivalence sites \cite{Yamamoto02}. The charge distribution confirms the appearance of charge disproportionation (CD) below $T_{\rm CO}$ with an average CD rates 0.2--0.8. The loss of inversion in the low-$T$ phase, as concluded from the selection rule, suggest that CO occurs along horizontal stripes. Structural data have yielded ``charge-rich" and ``charge-poor" sites from intramolecular bond-length distribution with ionicities  0--0.2 and 0.8--1 with a spatial pattern with hole-rich and hole-poor molecules aligned alternatively along the $c$-axis \cite{Watanabe04}. The CO transition is coupled to a first-order structural transition with doubling of the unit cell along the $c$-axis \cite{Watanabe04}, transforming the $\theta$-phase into the $\theta_d$-phase. It was also shown that large fluctuations in space already appear well above $T_{\rm CO}$ in the so-called metallic state \cite{Takahashi06}. The transition at $T_{\rm CO}$ may be the formation of a 3D long-range ordering assisted by lattice dimerisation along the $c$-axis.

Similarly CO was established by NMR studies in $\alpha$-(BEDT-TTF)$_2$I$_3$ \cite{Takano01} and by Raman spectroscopy \cite{Wojciechowski03}. CD also exists above $T_{\rm CO}$ \cite{Moroto04} with the degree of CD increasing when $T$ is decreased from room temperature.

At the CO transition in $\theta$-(BEDT-TTF)$_2$RbZn(SCN)$_4$ and in $\alpha$-(BEDT-TTF)$_2$I$_3$ the space group changes breaking some inversion centre, and thus allowing the possibility of a ferroelectric character to the CO phase transition.

Indeed generation of a ferroelectric polarisation was demonstrated in $\alpha$-(BEDT-TTF)$_2$I$_3$ by time-resolved pump and probe measurements of the optical second harmonic which appears below $T_{\rm CO}$ \cite{Yamamoto08}. Optical second harmonic generation was also detected in (TMTTF)$_2$SbF$_6$ \cite{Yamamoto11}. In addition images of ferroelectric domains were obtained by means of a second harmonic generation interferometry \cite{Yamamoto10}.

While in 1D-(TMTTF)$_2$X compounds, CO is stabilised by the shift of the anion sublattice as a whole, in 2D organic compounds, the interaction between electronic and lattice subsystems, being stronger, CO is stabilised by dimerisation as observed in $\theta$-(BEDT-TTF)$_2$RbZn(SCN)$_4$ and in $\alpha$-(BEDT-TTF)$_2$I$_3$.

However the horizontal CO stripes in $\theta$-(BEDT-TTF)$_2$RbZn(SCN)$_4$ and $\alpha$-(BEDT-TTF)$_2$I$_3$ are different. For the former compound the bonds between charge rich sites are all equivalent along one kind of bonds, resulting in a uniform Heisenberg coupling between spins along these stripes \cite{Seo06}. There is no abrupt change in magnetic susceptibility at $T_{\rm CO}$, which is well described by the Bonner-Fisher model 1D Heisenberg antiferromagnet, known to be unstable at lower temperature towards a spin-Peierls state. On the other hand, the horizontal stripe pattern of $\alpha$-(BEDT-TTF)$_2$I$_3$ results in an alternation of bonds (a BOW) along the charge rich sites \cite{Seo06}. In that case the spin susceptibility drops abruptly below $T_{\rm CO}$, reflecting the opening of a spin gap \cite{Rothaemel86}.

Finally, it was shown that the band structure of $\alpha$-(BEDT-TTF)$_2$I$_3$ when CO is suppressed by pressure is characterised by a linear energy dispersion \cite{Kobayashi07} expressed as $E=\pm v_\phi\hbar k$ where plus and minus signs correspond to electron and hole bands respectively. The Fermi energy is equal to zero when the electron and hole bands are in contact. This bulk Dirac cone type linear dispersion is, by some aspects, akin to the Dirac cone system in graphene \cite{Novoselov05,Novoselov05b}.

\subsubsection{Oxides}

Phase separation has been intensively studied in manganese oxides \cite{Mori98Nature} where inhomogeneity involves ferromagnetic metallic and antiferromagnetic charge and orbital ordered insulating domains. The role of disorder on the first order metal-insulator transition was shown to lead to formation of randomly distributed clusters of the competing phases (for a review see \cite{Dagotto01,Dagotto05}).

When antiferromagnetic layered perovskite oxides are doped with charge carriers at a small doping level, incommensurate magnetic superstructure reflections are observed. They were analysed in the stripe model in which charge carriers in the CuO$_2$ layers segregate into hole rich stripes that alternate with intermediate spin stripes with locally antiferromagnetic correlations (for a recent review on stripes, see \cite{Vojta09}, \cite{Tranquada12}).

Stripe modulation describes a uni-dimensional density wave with a single wave vector $Q$, but alternatively a checkboard state when two density waves with equal modulation occur in perpendicular directions.

STM with atomic resolution reveals also the nature of the spatial charge ordered symmetry either a site-centred or a bond-centred, the last symmetry effectively realised \cite{Kohsaka07}.

\begin{figure}
\begin{center}
\includegraphics[width=7.5cm]{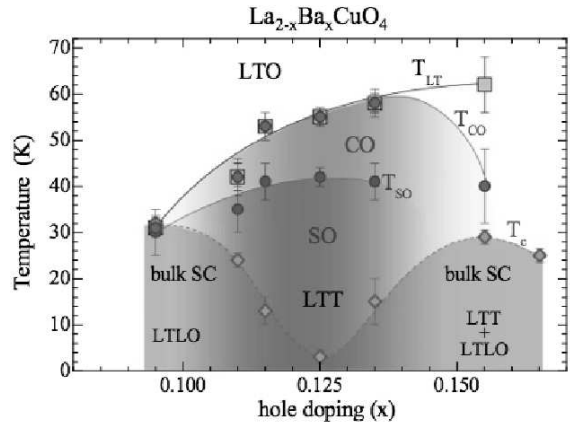}
\caption{Temperature versus hole-doping phase diagram of La$_{2-x}$Ba$_x$CuO$_4$ single crystals. Onset temperatures: $T_c$: bulk superconductivity (sc), $T_{\rm CO}$: charge stripe order (CO), $T_{\rm SO}$: spin stripe order (SO), $T_{\rm LT}$: structural transition LTO$\rightarrow$LTT and LTLO (reprinted figure with permission from M. H\"ucker \textit{et al.}, Physical Review B 83, p. 104506, 2011 \cite{Hucker11}. Copyright (2011) by the American Physical Society).}
\label{fig10-1new}
\end{center}
\end{figure}

Static stripe order with large correlation lengths was essentially observed in the single layer of doped La$_2$CuO$_4$ (214) cuprates. To illustrate the richness of various possible ground states, phase diagram of La$_{2-x}$Ba$_x$CuO$_4$ as a function of doping $x$ is shown in figure~\ref{fig10-1new}. 

Many transitions are observed: a structural one from HTT$\rightarrow$LTO, charge and spin stripes. $T_c$ is strongly reduced in a narrow region around $x=1/8$ associated with charge and spin stripe order. However competition between superconductivity and charge/spin order involves phase coherence of the superconducting state perpendicular to CuO$_2$ planes rather the local pairing amplitude \cite{Berg09}.

The nature of the order parameter of the pseudo-gap state is a major challenge for the last years \cite{Chakravarty01,Varma06}. Static stripe order implies that translation and rotation symmetries of CuO$_2$ planes are spontaneously broken (smectic order in the liquid crystal terminology). Another possible state-nematic-breaks the rotational symmetry of the lattice preserving its translational symmetry \cite{Vojta09,Fradkin10}. Such a nematic broken symmetry has been observed in underdoped YBa$_2$Cu$_3$O$_6$ \cite{Hinkov08} by neutron scattering, in Bi$_2$Sr$_2$CaCu$_2$O$_8$ by STM \cite{Lawler10}, in iron pnictides by ARPES \cite{Zhang12} and by in-plane resistivity anisotropy \cite{Chu10}. A coupling between both -smectic and nematic- locally broken electronic symmetries was also demonstrated \cite{Mesaros11}. Finally a broken-time-reversal symmetry was proposed due to non zero local charge current circulating around each elementary plaquette of the square lattice, identified by the ordered magnetic moments pointing perpendicular to the CuO$_2$ planes, and detectable by polarised neutron diffraction \cite{Fauque06,Li08} or inelastic neutron scattering \cite{Li10} experiments.

\subsection{Search for sliding mode in higher dimensionality}

Non-linearity in transport properties associated with broad band noise were found in many systems. Considered as similar to those in 1D systems, as in NbSe$_3$, they were ascribed as resulting from a CDW ground state.

\subsubsection{Misfit ladder Sr$_{14}$Cu$_{24}$O$_{18}$ compound}

Sr$_{14}$Cu$_{24}$O$_{18}$ is formed by ($a,c$) planes of weakly coupled Cu$_2$O$_3$ two-leg ladders stacked along the $b$-axis and separated by 1D CuO$_2$ edge sharing chain sheets, the Cu-Cu distances in the two sub-units being incommensurate with the relation $10c_{\rm chain}\simeq 7c_{\rm ladder}$. The insulating state of Sr$_{14}$Cu$_{24}$O$_{18}$ was ascribed to be charge ordering in the ladder planes. But non-linearity in the $c$-axis conductivity and a giant dielectric response \cite{Blumberg02,Gorshunov02} has led to identify the low  $T$ state to a pinned CDW which can slide above a threshold field. This interpretation was challenged in ref.~\cite{Vuletic06}. Non-linear transport properties were also found in the high magnetic field-induced phase of graphite \cite{Iye85}.

\subsubsection{2D(BEDT-TTF)$_2$X salts}

The current-voltage characteristics along the $a$ (in-plane) axis of $\theta$-(BEDT-TTF)$_2$RbZn(SCN)$_4$ and of $\theta$-(BEDT-TTF)$_2$CsZn(SCN)$_4$ measured at temperatures below 1~K have shown high non-linearity especially at lower temperatures, but without any clear threshold for higher conductivity as in C/S DWs. At low $T$, the I-V curves follow a power law $I\propto V^\alpha$ with $8<\alpha<10$. This feature was attributed to electric field induced unbinding of pairs of an electron and a hole that are thermally excited and attracted to each other due to the 2D long range logarithmic Coulomb interaction \cite{Takahide10}. However this description seems to not depend on the different nature of the low $T$ \cite{Takahide06} ground state: spin-Peierls for $\theta$-(BEDT-TTF)$_2$RbZn(SCN)$_4$ and a glass-like short range CO state for $\theta$-(BEDT-TTF)$_2$CsZn(SCN)$_4$ \cite{Nad08}.

A totally different model was proposed for explaining non-linear conductivity of organic conductors providing a macroscopic energy-balance based on a thermal (hot electron) model \cite{Mori09,Mori07b}.

\subsubsection{Manganites}

Manganese perovskites of general formula RE$^{3+}_{1-x}$AE$^{2+}_x$MnO$_3$ with RE: rare earths, AE: alkaline earth, exhibit a very rich variety of phases resulting from the competition between delocalised effects of the electron kinetic energy and localisation effects due to Coulomb repulsion (for a review see \cite{Tokura00}). Charge order between Mn$^{3+}$ and Mn$^{4+}$ was found with a modulation given by the average wave vector $Q\simeq (1-x)a^\ast$ ($a^\ast$: the reciprocal unit cell). Possible commensurate phases may occur for $x$~= 1/2, 2/3 or 3/4. But the strong electron lattice coupling that localises the Mn valence charge in the form of stripes \cite{Chen97} was challenged \cite{Loudon05}, and the superlattice being found to be periodically uniform. That led to considered the charge modulation as a charge density wave. To ascertain this statement, anisotropic transport properties \cite{Cox08}, non-linearity in I-V \cite{Wahl03,Cox08} curves with history-dependent features, broad-band noise \cite{Cox08,Barone09} spectra were reported.

In fact very similar non-linear (I-V) curves in manganites  (figure~2 in ref.~\cite{Wahl03}) and in organic conductors (figure~2 in ref.~\cite{Mori09}, figure~3 in ref.~\cite{Niizeki08}, ...) have been tentatively interpreted on the basis of a sliding  CDW. However, noise generation cannot by itself be the signature of CDW motion as discussed at length in section~\ref{sec3}. Moreover the signature of the collective transport is the generation of an ac voltage with the frequency proportional to the C/S DW velocity, which is the most easily identified in mode-locking experiments (Shapiro steps in the I-V characteristics).

\subsubsection{2D electronic solids}

\medskip
\noindent  \textit{9.6.4.a. 2D electrons on liquid helium substrate}
\medskip 

2D array of electrons above a superfluid helium film adsorbed on a dielectric substrate undergoes a metal-insulator transition at $T_m$, ascribed as the realisation of a 2D electronic solid, a Wigner crystal in which the electron form a triangular lattice (for a review see \cite{Andrei97}). Coulomb interactions play the main dominant role. Wigner crystallisation in this 2D system was first detected by the onset of coupled plasmon-ripplon resonances below $T_m$ \cite{Grimes79}. These modes arise from interaction between electrons and ripplons which produce a coherent periodic surface deformation beneath each electron. This crystal is also a model system for the study of 2D melting and the Kosterlitz-Thouless transition.

Non-linearity, either in ac Corbino conductivity \cite{Jiang89} in magneto-conductivity \cite{Kristensen96} or in a confined one-dimensional channel geometry \cite{Glasson01}, occurs below $T_m$. That was ascribed to the collective sliding of electrons out of the periodic deformation of the He surface \cite{Jiang89,Shirahama95}. Frequency and field dependence of the ac conductivity and noise characteristics (broad band noise) were considered \cite{Jiang89} to be analogous to those of sliding charge density waves. The ac force-velocity characteristic, $F(v)$, for electrons in a quasi-1D channel deviates \cite{Glasson01} from a linear Ohmic regime when the drift velocity of electrons approaches the critical value, $v_1$, of the radiation of coherent ripplons (whose wavelength equals the electron lattice spacing), the value depending on the electron current density. Above $v_1$, the ordered electrons decouple from the ripplons and the drag force decreasing produces a negative differential conductivity which may generate bistability and inhomogeneous current distribution through current filaments.

A second threshold occurs at much higher drift velocity where velocity increases at constant force. It is interesting to note the similarity of the ac force-velocity characteristic of this Wigner crystal with that derived for Q1D systems \cite{Brazovskii04}.

\medskip
\noindent  \textit{9.6.4.b. 2D GaAs/AlGaAs heterostructures}
\medskip 

Non-linearity behaviour is also a crucial feature of the metal-insulator transition in the 2D GaAs/AlGaAs heterostructures, also in high mobility Si-metal-oxide-semiconductor field-effect transistors (Si-MOSFET) at zero magnetic field \cite{Pudalov93}. A reentrant insulating phase was observed at high magnetic field in the vicinity of fractional (mainly 1/5 for GaAs, but also 1/3) quantum Hall states (FQHE) (for a review see \cite{Shayegan97}) in the regime of low filling factor $\nu=nh/eB$. The insulating nature of this state was ascribed to a Wigner crystal pinned by disorder, with non-linear I-V characteristics, noise generation, resonance in the microwave range \cite{Goldman90,Jiang91,Williams91,Ye02}. To disentangle the combined effect of carrier-carrier Coulomb interactions and disorder, measurements of the conductivity in the microwave range were performed \cite{Li00} by changing in situ at low temperature the carrier density, $n_s$, keeping the disorder potential unchanged. A microwave resonance peak in the real part of the frequency dependence of the diagonal conductivity at a fixed $T$ ($T$~= 25~mK) and fixed $H$ ($H$~= 13~T) was found to depend on $n_s$ as $n_s^{-\alpha}$, with $\alpha$~= 1/2 or 3/2. This resonance has been interpreted as caused by the pinned mode of Wigner crystal domains oscillating in the disorder potential, in agreement with the theory of a weakly pinned Wigner crystal \cite{Chitra98}. Microwave resonance  were also predicted for Wigner crystals in zero magnetic field  \cite{Chitra05}.

In clean 2D electron systems, electronic solid phases were shown to exist in presence of a moderate perpendicular field, which are distinct from the FQHE states which dominates at high field and low $\nu$. These new phases with a spatial modulation of charge form when a high Landau level is partially filled by electrons, close to integer fillings $\nu$~= 1, 2, 3. Within the integer quantum Hall effect (IQHE), a Wigner crystal is stabilised by inter-electron repulsion, that is similar to the Wigner crystal found at low $\nu$ of the lowest Landau level. Microwave resonances in the real part of the frequency dependent diagonal conductivity were found \cite{Chen03}, interpreted as the pinned mode of the Wigner crystals phase formed in the top Landau level, around the corresponding integer fillings.

A striking anisotropy in dc diagonal resistivity with the magnetic field applied perpendicular near half-integer Landau fillings $\nu$~= $9/2,11/2,13/2$ with Landau levels $N\geq 1$ completely filled has been reported \cite{Lilly99}. No analogous anisotropy was observed at very low $H$ or with $N=1$ Landau level. The anisotropy was initially analysed as ``stripe states'' with a spatial charge modulation along a given direction \cite{Lilly99}. The recent description refers this strongly temperature transport anisotropy to an electronic nematic phase (for a review see \cite{Fradkin10} in which the rotational invariance in the ($x-y$) plane is broken spontaneously. The signature of a nematic phase in thus the in-plane anisotropy contrary to otherwise essentially isotropic electronic systems.

Striped phases with a strong anisotropy in the conductivity occur near half integer fillings $\nu$~= 9/2, 11/2, 13/2. On either side of these striped phases, near 1/4 or 3/4 filling, additional phases exist, ``bubble" phases which are made up of multiple-electron (hole) bubbles that arrange into a ``super" Wigner crystal \cite{Fogler96}. The bubble phase exhibits non-linear properties similar to those of conventional CDWs; threshold onset for non-linearity \cite{Cooper03}, resonances in the microwave conductivity suggesting that they may reflect the pinned mode of the bubble crystal. A periodic time dependence of voltage was observed in the non-linear (I-V) characteristics of the bubble phase \cite{Cooper03}. However the frequency of this periodic voltage is much lower than that originating from the sliding CDW picture. Anisotropy of the stripe phases was well established by the observation of a microwave resonance only for $rf$ electric field across the stripe orientation, whereas the rf field parallel to the stripes yields a flat spectrum \cite{Sambandamurphy08}.  This resonance is then interpreted as the pinned mode of the striped phase.

2D atomic crystals can be obtained by mechanical cleaving of bulk layered crystals. That is one of the techniques used \cite{Novoselov05} to obtained graphene or a few layer thick sample from graphite. Few/single layer samples were obtained, especially those having a gap in the electronic structure as MoS$_2$ \cite{Radisavljevic11,Mak10}, a superconducting and/or a CDW gap such as NbSe$_2$ \cite{Novoselov05}. The Fermi level of these atomic crystals can be tuned by electric gate. A new route towards  2D crystals is thus opened.

\subsubsection{Sliding mode in RTe$_3$ compounds}

However a new class of layered compounds namely rare-earch tritellurides RTe$_3$ (R~= Y, La, Ce, Nd, Sm, Gd, Tb, Ho, Dy, Er, Tm) \cite{DiMasi95,Brouet08,Ru08} has recently raised an intense research activity on CDW. These systems exhibit an incommensurate CDW through the whole $R$ series with a wave vector $Q_{\rm  CDW_1}$~= $(0,0,\sim 2/7c^\ast)$ with a Peierls transition temperature above 300~K for the light atoms (La, Ce, Nd). For the heavier R (Dy, Ho, Er, Tm) a second CDW occurs with the wave vector $Q_{\rm CDW_2}$~= $(\sim 2/7 a^\ast,0,0)$. Non-linear transport properties were recently reported in DyTe$_3$ below $T_{\rm CDW_1}$~= 302~K: conductivity is increasing sharply above a threshold field and under application of a 
$rf$ field Shapiro steps are clearly observed, features demonstrating for the first time CDW sliding in 2D compounds \cite{Sinchenko12}.

Although band structure is two dimensional, RTe$_3$ compounds have a ``hidden'' 1D character \cite{Yao06}. That may explain the observation of CDW sliding at variance with experiments on transition metal dichalcogenides. In that case, absence of non-linearity in current-voltage characteristics was reported in 2H-TaSe$_2$ and 1T-TaSe$_2$ up to electric field 1~V/cm and 10~V/cm respectively \cite{DiSalvo80}. These values are much higher that typical $E_T$ in 1D compounds (section \ref{sec3}). One possible reason might be a commensurability pinning resulting from the triple-$Q$ structure with three wave vectors of equal amplitude, 120� apart.  Very compelling is the occurrence of two orthogonal CDWs with nearly identical wave vectors for the heaviest rare earth which may lead to checkboard type of superstructures.

\section*{Acknowledgements}
I acknowledge Georges Waysand for having introduced me to physics of low dimensional materials. I am very grateful for having worked closely with Katica Biljakovic, Roland Currat, Jean-Claude Lasjaunias, Yuri Latyshev, Fran{\c c}ois L\'evy, Jos\'e-Emilio Lorenzo-Dias, Felix Nad, Michel Renard, Michel Saint-Paul and Alexander Sinchenko for a lasting, fruitful collaboration on the different topics of this review. I also thank many others, students, colleagues and visitors for their important contribution. I am very thoughtful of the late Madeleine Monceau who was a permanent help to me for so many years. I am deeply indebted to Dani\`ele Devillers for her continuous, endless assistance in shaping the manuscript. Finally, I am grateful to Professor, David Sherrington, the Editor, for giving me the opportunity to write this review, and for having been so patient in waiting for its submission.

\vfill{\eject}

\bibliographystyle{tADP}
\bibliography{MonceauAP}

\begin{thebibliography}{1000}
\newcommand{\noopsort}[1]{}
\newcommand{\printfirst}[2]{#1}
\newcommand{\singleletter}[1]{#1}
\newcommand{\switchargs}[2]{#2#1}
\providecommand{\url}[1]{\normalfont{#1}}
\providecommand{\urlprefix}{Available at }

\bibitem{Peierls55}
{\rm R.E. Peierls, in} (1955).

\bibitem{Frohlich54}
{\rm H. Fr{\"o}hlich} (1954).

\bibitem{BCS57}
J.~Bardeen, L.N. Cooper, and J.R. Schrieffer, Phys. Rev. 108 (1957), p. 1175.

\bibitem{Kohn59}
W.~Kohn, Phys. Rev. Lett. 2 (1959), p. 393.

\bibitem{Overhauser59}
A.W. Overhauser and A.~Arrott, Phys. Rev. Lett. 3 (1959), p. 414.

\bibitem{Overhauser62}
A.W. Overhauser, Phys. Rev. 128 (1962), p. 1437.

\bibitem{Lomer62}
W.M. Lomer, Proc. Phys. Soc. London 80 (1962), p. 489.

\bibitem{Little64}
W.A. Little, Phys. Rev. 134 (1964), p. A1416.

\bibitem{R20Brazovskii99}
\emph{International Workshop on Electronic Crystals {\rm[ECRYS-1999]}}, J.
  Phys. IV (France), Vol.~9, S. Brazovskii and P. Monceau eds, 1999.

\bibitem{R21Brazovskii02}
\emph{International Workshop on Electronic Crystals {\rm[ECRYS-2002]}}, J.
  Phys. IV (France), Vol.~12, S. Brazovskii, N. Kirova and P. Monceau eds,
  2002.

\bibitem{R22Brazovskii05}
\emph{International Workshop on Electronic Crystals {\rm[ECRYS-2005]}}, J.
  Phys. IV (France), Vol. 131, S. Brazovskii, P. Monceau and N. Kirova eds,
  2005.

\bibitem{R23Brazovskii09}
\emph{International Workshop on Electronic Crystals {\rm[ECRYS-2008]}}, Physica
  B - Condensed Matter, Vol. 404, Issues 3-4, Guest editors: S. Brazovskii, N.
  Kirova and P. Monceau, 2009.

\bibitem{R24Brazovskii12}
\emph{International Workshop on Electronic Crystals {\rm[ECRYS-2011]}}, Physica
  B - Condensed Matter, Vol. 407, Issue 11, Guest editors: S. Brazovskii, N.
  Kirova and P. Monceau, 2012.

\bibitem{R1Jerome77}
\emph{Low-Dimensional Conductors and Superconductors}, NATO ASI Series, Series
  B: Physics, Vol. 155, D. J\'erome and L.G. Caron eds, Plenum Press, New York,
  1977.

\bibitem{R3Devresse79}
\emph{Highly Conducting One-Dimensional Solids}, J.T. Devresse, R.P. Evrard and
  V.E. van Doren eds, Plenum Press, New York, 1979.

\bibitem{R5Gruner94}
\emph{Density Waves in Solids}, G. Gr\"uner, edited by Addison-Vesley
  Publishing Company, Reading, Massachusetts, USA, 1994.

\bibitem{R6Ishiguro98}
\emph{Organic Superconductors}, T. Ishiguro, K. Yamaji and G. Saito, edited by
  Springer, Berlin, 2nd edition, 1998.

\bibitem{R7Gorkov89}
\emph{Charge Density Waves in Solids}, Modern Problems in Condensed Matter
  Sciences, Vol.~25, L.P. Gor'kov and G. Gr\"uner eds, North-Holland,
  Amsterdam, 1989.

\bibitem{R9Schlenker89}
\emph{Low-Dimensional Electronic Properties of Molybdenum Bronzes and Oxides},
  C. Schlenker ed., Kluwer Academic Publishers, 1989.

\bibitem{R10Schlenker96}
\emph{Physics and Chemistry of Low-Dimensional Inorganic Conductors}, NATO ASI
  Series B: Physics,, Vol. 354, C. Schlenker and J. Dumas and M. Greenblatt and
  \mbox{S. van Smaalen} eds, Plenum Press, New York, 1996.

\bibitem{R11Baeriswyl04}
\emph{Strong Interactions in Low Dimensions}, D. Baeriswyl and L. Degiorgi eds,
  Kluwer Academic Publishers, 2004.

\bibitem{R12Boswell99}
\emph{Advances in the Crystallographic and Microstructural Analysis of Charge
  Density Wave Modulated Crystals}, F.W. Boswell and J.C. Bennett eds, Kluwer
  Academic Publishers, 1999.

\bibitem{R14Lebed08}
\emph{The Physics of Organic Superconductors and Conductors}, A. Lebed ed.,
  Springer, Berlin, 2008.

\bibitem{R16Monceau85}
\emph{Electronic Properties of Inorganic Quasi-One-Dimensional Materials}, P.
  Monceau ed., Reidel, Dordrecht, Part I and Part II, 1985.

\bibitem{R17Rouxel86}
\emph{Crystal Chemistry and Properties of Materials with One-Dimensional
  Structures}, J. Rouxel ed., Reidel, Dordrecht, 1986.

\bibitem{R18Giamarchi03}
\emph{Quantum Physics in One Dimension}, T. Giamarchi, Clarendon Press, Oxford,
  2003.

\bibitem{Overhauser60}
A.W. Overhauser, Phys. Rev. Lett. 4 (1960), p. 462.

\bibitem{Kagoshima88}
S.~Kagoshima, H.~Nagasawa, and T.~Sambongi, \emph{One-Dimensional Conductors},
  Springer-Verlag, Berlin, 1988.

\bibitem{Chan73}
S.K. Chan and V.~Heine, J. Phys. F: Metal Phys. 3 (1973), p. 795.

\bibitem{Fazekas99}
\mbox{\rm P. Fazekas,}, \emph{Lecture Notes on Electron Correlation and
  Magnetism}, Series in Modern Condensed Matter Physics, Vol.~3, World
  Scientific, Singapore, 1999.

\bibitem{Friedel77}
\mbox{J. Friedel in \textit{Electron-Phonon Interactions and Phase
  Transitions}}, NATO ASI Series, Series B: Physics, Plenum Press, New York,
  Vol. 29  (1977), p.~1.

\bibitem{Rice73}
M.J. Rice and S.~Strassler, Solid State Commun. 13 (1973), p. 1931.

\bibitem{Bishop81}
M.F. Bishop and A.W. Overhauser, Phys. Rev. B 23 (1981), p. 3638.

\bibitem{Tutis91}
E.~Tuti{\u s} and S.~Bari{\u s}i{\'c}, Phys. Rev. B 43 (1991), p. 8431.

\bibitem{Pouget91}
\mbox{J.-P. Pouget}, B.~Hennion, \mbox{C. Escribe-Filippini}, and M.~Sato,
  Phys. Rev. B 43 (1991), p. 8421.

\bibitem{Lee73}
P.A. Lee, T.M. Rice, and P.W. Anderson, Phys. Rev. Lett. 31 (1973), p. 462.

\bibitem{Lee74}
P.A. Lee, T.M. Rice, and P.W. Anderson, Solid State Commun. 14 (1973), p. 703.

\bibitem{Solyom79}
J.~S{\'o}lyom, Adv. Phys. 28 (1979), p. 209.

\bibitem{Firsov85}
\mbox{Yu.A. Firsov}, V.N. Prigodin, and C.~Seidel, Phys. Rep. 126 (1985), p.
  245.

\bibitem{Voit95}
J.~Voit, Rep. Prog. Phys. 58 (1995), p. 977.

\bibitem{Emery79}
\mbox{\rm V.J. Emery, in \cite{R3Devresse79}} , p. 247.

\bibitem{Schulz91}
H.J. Schulz, Int. Mod. Phys. B 5 (1991), p.~57.

\bibitem{Hubbard78}
J.~Hubbard, Phys. Rev. B 17 (1978), p. 494.

\bibitem{Schulz77}
\mbox{H.J. Schulz, in ref.~\cite{R1Jerome77}} , p.~95.

\bibitem{Pytte74}
E.~Pytte, Phys. Rev. B 10 (1974), p. 4637.

\bibitem{Bray83}
\mbox{\rm J.W. Bray, L.V. Interrante, I.S. Jacobs, and J.C. Bonner,},
  \emph{Extend Linear Chain Compounds}, Vol.~3, J.S. Miller ed., Plenum Press,
  New York, \mbox{1983, p 353}.

\bibitem{Cross79}
M.C. Cross and D.S. Fisher, Phys. Rev. B 19 (1979), pp. 402; M. C. Cross, Phys.
  Rev. B 20 (1979) p. 4606.

\bibitem{Lieb68}
H.H. Lieb and F.Y. Lu, Phys. Rev. Lett. 20 (1968), p. 1445.

\bibitem{Haldane81}
F.D.M. Haldane, J. Phys. C: Solid State Phys. 14 (1981), p. 2585.

\bibitem{Hirsch83}
J.F. Hirsch and D.J. Scalapino, Phys. Rev. Lett. 50 (1983), p. 1168; Phys. Rev.
  B 27 (1983) p. 7169; Phys.Rev. B 29 (1984) p. 5554.

\bibitem{Mila93}
F.~Mila and X.~Zotos, Europhys. Lett. 24 (1993), p. 133.

\bibitem{Ejima05}
S.~Ejima, F.~Gebhard, and S.~Nishimoto, Europhys. Lett. 70 (2005), p. 492.

\bibitem{Seo06}
\mbox{For a review : H. Seo}, J.~Merino, H.~Yoshioka, and M.~Ogata, J. Phys.
  Soc. Jpn 75 (2006), p. 051009.

\bibitem{Seo04}
H.~Seo, C.~Hotta, and H.~Fukuyama, Chem. Rev. 104 (2004), p. 5005.

\bibitem{Seo97}
H.~Seo and H.~Fukuyama, J. Phys. Soc. Jpn 66 (1997), p. 1249.

\bibitem{Nishimoto00}
S.~Nishimoto, M.~Takahashi, and Y.~Ohta, J. Phys. Soc. Jpn 69 (2000), p. 1594.

\bibitem{Yoshioka00}
H.~Yoshioka, M.~Tsuchiizu, and Y.~Suzumura, J. Phys. Soc. Jpn 69 (2000), p.
  651.

\bibitem{Shibata01}
Y.~Shibata, S.~Nishimoto, and Y.~Ohta, Phys. Rev. B 64 (2001), p. 235107.

\bibitem{Tsuchiizu01}
M.~Tsuchiizu, H.~Yoshioka, and Y.~Suzumura, J. Phys. Soc. Jpn 70 (2001), p.
  1460.

\bibitem{Tanaka05}
Y.~Tanaka and M.~Ogata, J. Phys. Soc. Jpn 74 (2005), p. 3283.

\bibitem{Ung94}
K.C. Ung, S.~Mazumdar, and D.~Toussaint, Phys. Rev. Lett. 73 (1994), p. 2603.

\bibitem{Mazumdar99}
S.~Mazumdar, S.~Ramasesha, R.~Clay, and D.K. Campbell, Phys. Rev. Lett. 82
  (1999), p. 1522; Phys. Rev. B 62 (2000) p. 13400.

\bibitem{Clay07}
R.T. Clay, R.P. Hardikar, and S.~Mazumdar, Phys. Rev. B 76 (2007), p. 205118.

\bibitem{Su79}
W.P. Su, J.R. Schrieffer, and A.J. Heeger, Phys. Rev. Lett. 42 (1979), p. 1698;
  Phys. Rev. B 22 (1980) p. 2099.

\bibitem{Dixit84}
S.N. Dixit and S.~Mazumdar, Phys. Rev. B 29 (1984), p. 1824.

\bibitem{Riera00}
J.~Riera and D.~Poilblanc, Phys. Rev. B 62 (2000), p. R16243.

\bibitem{Riera01}
J.~Riera and D.~Poilblanc, Phys. Rev. B 63 (2001), p. 241102.

\bibitem{Yamaji82}
K.~Yamaji, J. Phys. Soc. Jpn 51 (1982), p. 2787.

\bibitem{Montambaux88}
G.~Montambaux, Phys. Rev. B 38 (1988), p. 4788.

\bibitem{Huang92}
X.~Huang and K.~Maki, Phys. Rev. B 46 (1992), p. 162.

\bibitem{Mihaly97}
G.~Mih{\'a}ly, A.~Virosztek, and G.~Gr{\"u}ner, Phys. Rev. B 55 (1997), p.
  R13456.

\bibitem{Sham79}
\mbox{L.J. Sham in ref. \cite{R3Devresse79}}  (1979), p. 227.

\bibitem{Scalapino72}
D.J. Scalapino, M.~Sears, and R.A. Ferrell, Phys. Rev. B 6 (1972), p. 3409.

\bibitem{McKenzie92}
R.H. McKenzie and J.W. Wilkins, Phys. Rev. Lett. 69 (1992), p. 1085.

\bibitem{Allender74}
D.~Allender, J.W. Bray, and J.~Bardeen, Phys. Rev. B 9 (1974), p. 119.

\bibitem{Blinc86}
\mbox{For a review, }, \emph{Incommensurate Phases in Dielectrics Modern
  Physics in Condensed Matter Sciences}, Vol. 14.1 Fundamentals and 14.2
  Materials, R. Blinc and A.P. Levanyuk eds, North Holland, Amsterdam, 1986.

\bibitem{Berge84}
B.~Berge, G.~Dolino, M.~Vallade, M.~Boissier, and R.~Vacher, J. Physique 45
  (1984), p. 715.

\bibitem{MacMillan76}
\mbox{W.L. Mac Millan}, Phys. Rev. B 14 (1976), p. 1496; Phys. Rev. B 16 (1977)
  p. 4655.

\bibitem{Brink08}
\mbox{J. van den Brink} and D.I. Khomskii, J. Phys.: Cond. Matter 20 (2008), p.
  434217.

\bibitem{Ishihara10}
S.~Ishihara, J. Phys. Soc. Jpn 79 (2010), p. 011010.

\bibitem{Naka10}
M.~Naka and S.~Ishihara, J. Phys. Soc. Jpn 79 (2010), p. 063707.

\bibitem{Arima11}
T.~Arima, J. Phys. Soc. Jpn 80 (2011), p. 052001.

\bibitem{Brazovskii08}
\mbox{S. Brazovskii in Ref.~\cite{R14Lebed08}} , p. 313.

\bibitem{Torrance81}
J.B. Torrance, A.~Girlando, J.J. Mayerle, J.I. Crowley, V.Y. Lee, and
  P.~Batail, Phys. Rev. Lett. 47 (1981), p. 1747.

\bibitem{Torrance81b}
J.B. Torrance, J.E. Vazquez, J.J. Mayerle, and V.Y. Lee, Phys. Rev. Lett. 46
  (1981), p. 253.

\bibitem{Buron03}
\mbox{M. Buron-Le Cointe}, \mbox{M. H. Lemee-Cailleau}, H.~Cailleau,
  B.~Toudi{\'c}, A.~Mor{\'e}ac, F.~Moussa, C.~Ayache, and N.~Karl, Phys. Rev. B
  68 (2003), p. 064103.

\bibitem{Lecointe95}
\mbox{M. Le Cointe}, \mbox{M.-H. Lemee-Cailleau}, H.~Cailleau, B.~Toudi{\'c},
  L.~Toupet, G.~Heger, F.~Moussa, P.~Schweiss, K.H. Kraft, and N.~Karl, Phys.
  Rev. B 51 (1995), p. 3374.

\bibitem{Tokura89}
Y.~Tokura, S.~Koshihara, Y.~Iwasa, H.~Okamoto, T.~Komatsu, T.~Koda, N.~Iwasawa,
  and G.~Saito, Phys. Rev. Lett. 63 (1989), p. 2405.

\bibitem{Collet03}
\mbox{E. Collet, M.-H. Lemee-Cailleau, M. Buron-Le Cointe, H. Cailleau, M.
  Wulff, T. Luty, S-Ya Koshihara}, M.~Meyer, L.~Toupet, P.~Rabiller, and
  S.~Techert, Science 300 (2003), p. 612.

\bibitem{Kagawa10}
F.~Kagawa, S.~Horiuchi, M.~Tokunaga, J.~Fujioka, and Y.~Tokura, Nature Phys. 6
  (2010), p. 169.

\bibitem{Rouxel89}
\mbox{J. Rouxel and C. Schlenker, in ref. \cite{R7Gorkov89}} , p.~15.

\bibitem{Meerschaut75}
A.~Meerschaut and J.~Rouxel, J. Less Comm. Metals 39 (1975), p. 197.

\bibitem{Hodeau78}
\mbox{J.-L. Hodeau}, M.~Marezio, C.~Roucau, R.~Ayrolles, A.~Meerschaut,
  J.~Rouxel, and P.~Monceau, J. Phys. C 11 (1978), p. 4117.

\bibitem{Chaussy76}
J.~Chaussy, P.~Haen, \mbox{J.-C. Lasjaunias}, P.~Monceau, G.~Waysand,
  A.~Waintal, A.~Meerschaut, P.~Molini{\'e}, and J.~Rouxel, Solid State Commun.
  20 (1976), p. 759.

\bibitem{Fleming78}
R.M. Fleming, D.E. Moncton, and \mbox{D.B. McWhan}, Phys. Rev. B 18 (1978), p.
  5560.

\bibitem{Brun06}
C.~Brun, Thesis, University of Paris-Sud, unpublished,  2006.

\bibitem{Brun09}
C.~Brun, Z.Z. Wang, and P.~Monceau, Phys. Rev. B 80 (2009), p. 045423.

\bibitem{vanSmaalen92}
\mbox{S. van Smaalen}, \mbox{J.L. de Boer}, A.~Meetsma, H.~Graafsma,
  \mbox{H.-S. Sheu}, A.~Darovskikh, P.~Coppens, and F.~Levy, Phys. Rev. B 45
  (1992), p. 3103.

\bibitem{Brun10}
C.~Brun, Z.Z. Wang, P.~Monceau, and S.~Brazovskii, Phys. Rev. Lett. 104 (2010),
  p. 256403.

\bibitem{Brazovskii11}
S.~Brazovskii, C.~Brun, Z.Z. Wang, and P.~Monceau, Phys. Rev. Lett. 108 (2012),
  p. 096801.

\bibitem{Bruinsma80}
R.~Bruinsma and S.E. Trullinger, Phys. Rev. B 22 (1980), p. 4543.

\bibitem{Moudden90}
A.H. Moudden, J.D. Axe, P.~Monceau, and F.~Levy, Phys. Rev. Lett. 65 (1990), p.
  223.

\bibitem{Ayari04}
A.~Ayari, D.~Danneau, H.~Requardt, L.~Ortega, J.E. Lorenzo, P.~Monceau,
  R.~Currat, S.~Brazovskii, and G.~Gr{\"u}bel, Phys. Rev. Lett. 93 (2004), p.
  106404.

\bibitem{Prodan01}
A.~Prodan, N.~Jug, \mbox{H. J.-P. van Midden}, H.~B{\"o}hm, F.W. Boswell, and
  J.C. Bennett, Phys. Rev. B 64 (2001), p. 115423.

\bibitem{Prodan10}
A.~Prodan, \mbox{H. J.-P. van Midden}, R.~Zitko, E.~Zupanic, J.C. Bennett, and
  H.~B{\"o}hm, Solid State Commun. 150 (2010), p. 2134.

\bibitem{Rouziere96}
S.~Rouzi{\`e}re, S.~Ravy, \mbox{J.-P. Pouget}, and R.E. Thorne, Solid State
  Commun. 97 (1996), p. 1073.

\bibitem{Yamaya83}
K.~Yamaya and G.~Oomi, J. Phys. Soc. Jpn 52 (1983), p. 1886.

\bibitem{Bullett79}
D.W. Bullett, J. Phys. C 12 (1979), p. 277; 15 (1982) p. 3069.

\bibitem{Shima82}
N.~Shima, J. Phys. Soc. Jpn 51 (1982), p. 11; 52 (1983) p. 578.

\bibitem{Schafer01}
J.~Sch{\"a}fer, E.~Rotenberg, S.D. Kevan, P.~Blaha, R.~Claessen, and R.~Thorne,
  Phys. Rev. Lett. 87 (2001), p. 196403.

\bibitem{Canadell90}
E.~Canadell, \mbox{I. E.-I. Rachidi}, \mbox{J.-P. Pouget}, P.~Gressier,
  A.~Meerschaut, J.~Rouxel, D.~Jung, M.~Evain, and \mbox{M.-H. Whangbo}, Inorg.
  Chem. 29 (1990), p. 1401.

\bibitem{Pouget83}
\mbox{J.-P. Pouget}, R.~Maret, A.~Meerschaut, L.~Guemas, and J.~Rouxel, J.
  Phys. (Paris) 44 (1983), pp. C3--1729.

\bibitem{Cava81}
R.J. Cava, V.L. Himes, A.D. Mighell, and R.S. Roth, Phys. Rev. B 24 (1981), p.
  3634.

\bibitem{Meerschaut81a}
A.~Meerschaut, L.~Guemas, and J.~Rouxel, J. Solid State Chem. 36 (1981), p.
  118.

\bibitem{Hillenius81}
S.J. Hillenius, R.V. Coleman, R.M. Fleming, and R.J. Cava, Phys. Rev. B 23
  (1981), p. 1567.

\bibitem{Meerschaut79}
A.~Meerschaut, J.~Rouxel, P.~Haen, P.~Monceau, and M.~N{\'u}{\~n}ez-Regueiro,
  J. Phys. Lett. (Paris) 40 (1979), p. L157.

\bibitem{Roucau80}
C.~Roucau, R.~Ayrolles, P.~Monceau, L.~Guemas, A.~Meerschaut, and J.~Rouxel,
  Phys. Stat. Solidi a 62 (1980), p. 483.

\bibitem{Sambongi77}
T.~Sambongi, K.~Tsutsumi, Y.~Shiozaki, M.~Yamamoto, K.~Yamaya, and Y.~Abe,
  Solid State Commun. 22 (1977), p. 729.

\bibitem{Roucau83}
C.~Roucau, J. Phys. (France) C3 44 (1983), pp. C3--1725.

\bibitem{Inagaki08}
\mbox{K. Inagaki, \;M. Tsubota, \;K. Higashiyama, K. Ichimura, S. Tanda, K.
  Yamamoto, N. Hanasaki}, N.~Ikeda, Y.~Nogami, T.~Ito, and H.~Toyokawa, J.
  Phys. Soc. Jpn 77 (2008), p. 093708.

\bibitem{Higgs83}
A.W. Higgs and J.C. Gill, Solid State Commun. 47 (1983), p. 737.

\bibitem{Nad85}
{\rm F. Ya Nad, in {\it ``Charge Density Waves in Solids"}, Lecture Notes in
  Physics} (1985).

\bibitem{Rijnsdorp78}
J.~Rijnsdorp and F.~Jellinek, J. Solid State Chem. 25 (1978), p. 325.

\bibitem{Cornelissens78}
T.~Cornelissens, \mbox{G. van Tendeloo}, \mbox{J. van Landuyt}, and
  S.~Amelinckx, Phys. Status Solidi A 48 (1978), p.~K5.

\bibitem{Wang89}
Z.Z. Wang, P.~Monceau, H.~Salva, C.~Roucau, L.~Guemas, and A.~Meerschaut, Phys.
  Rev. B 40 (1989), p. 11589.

\bibitem{Zettl82}
A.~Zettl, C.M. Jackson, A.~Janossy, G.~Gr{\"u}ner, A.~Jacobsen, and A.H.
  Thompson, Solid State Commun. 43 (1982), p. 345.

\bibitem{Zybtsev09}
S.G. Zybtsev, \mbox{V.Ya. Pokrovskii}, V.F. Nosretdinova, and S.V.
  Zaitsev-Zotov, Applied Physics Lett. 94 (2009), p. 152112.

\bibitem{Takahashi84}
S.~Takahashi, T.~Sambongi, J.W. Brill, and W.~Roark, Solid State Commun. 49
  (1984), p. 1031.

\bibitem{Eaglesham84}
D.~Eaglesham, J.W. Steeds, and J.~Wilson, J. Phys. C 17 (1984), p. L697.

\bibitem{Tokoya05}
T.~Tokoya, T.~Kiss, A.~Chainani, S.~Shin, and K.~Yamaya, Phys. Rev. B 71
  (2005), p. 140504.

\bibitem{Hoesch09}
M.~Hoesch, A.~Bosak, D.~Chernyshov, H.~Berger, and M.~Krisch, Phys. Rev. Lett.
  102 (2009), p. 086402.

\bibitem{Gressier85b}
{\rm P. Gressier and A. Meerschaut and J. Rouxel and M.-H. Whangbo in {\it
  Lecture Notes in Physics ``Charge Density Waves in Solids"}} (1985).

\bibitem{Wang83a}
Z.Z. Wang, \mbox{M.-C. Saint-Lager}, P.~Monceau, M.~Renard, P.~Gressier,
  A.~Meerschaut, L.~Guemas, and J.~Rouxel, Solid State Commun. 46 (1983), p.
  325.

\bibitem{Fujishita84}
H.~Fujishita, M.~Soto, and S.~Hoshino, Solid State Commun. 49 (1984), p. 313.

\bibitem{Wang83b}
Z.Z. Wang, P.~Monceau, M.~Renard, P.~Gressier, L.~Guemas, and A.~Meerschaut,
  Solid State Commun. 47 (1983), p. 439.

\bibitem{Gressier84a}
P.~Gressier, M.H. Wangbo, A.~Meerschaut, and J.~Rouxel, Inorg. Chem. 23 23
  (1984), p. 1221.

\bibitem{Gressier84b}
P.~Gressier, A.~Meerschaut, L.~Guemas, J.~Rouxel, and P.~Monceau, J. Solid
  State Chem. 51 (1984), p. 141.

\bibitem{Izumi84a}
M.~Izumi, T.~Iwazumi, T.~Seino, K.~Uchinokura, R.~Yoshizaki, and E.~Matsuura,
  Solid State Commun. 49 (1984), p. 423.

\bibitem{Hufner99}
S.~H{\"u}fner, R.~Claessen, F.~Reinert, T.~Straub, V.N. Strocov, and
  P.~Steiner, J. Electron Spectroscopy and Related Phenomena 100 (1999), p.
  191.

\bibitem{Voit00}
J.~Voit, L.~Perfetti, F.~Zwick, H.~Berger, G.~Margaritondo, G.~Gr{\"u}ner,
  H.~H{\"o}chot, and M.~Grioni, Science 290 (2000), p. 501.

\bibitem{Lee85}
\mbox{K.-B. Lee}, D.~Davidov, and A.J. Heeger, Solid State Commun. 54 (1985),
  p. 673.

\bibitem{Requardt98b}
H.~Requardt, J.E. Lorenzo, R.~Currat, P.~Monceau, B.~Hennion, H.~Berger, and
  F.~Levy, J. Phys.: Condens. Matter 10 (1998), p. 6505.

\bibitem{Favre-Nicolin01}
V.~Favre-Nicolin, S.~Bos, J.E. Lorenzo, \mbox{J.-L. Hodeau}, \mbox{J.-F.
  B{\'e}rar}, P.~Monceau, R.~Currat, F.~L{\'e}vy, and H.~Berger, Phys. Rev.
  Lett. 87 (2001), p. 015502.

\bibitem{Pouget85}
\mbox{J.-P. Pouget}, C.~Noguera, A.H. Moudden, and J.~Moret, J. Physique
  (Paris) 46 (1985), p. 1731.

\bibitem{Lorenzo88}
J.E. Lorenzo, R.~Currat, P.~Monceau, B.~Hennion, H.~Berger, and F.~Levy, J.
  Phys.: Condens. Matter 10 (1988), p. 5039.

\bibitem{Requardt96}
H.~Requardt, M.~Kalning, B.~Burandt, W.~Press, and R.~Currat, J. Phys.:
  Condens. Matter 8 (1996), p. 2327.

\bibitem{Gressier85}
P.~Gressier, L.~Guemas, and A.~Meerschaut, Mat. Res. Bull. 20 (1985), p. 539.

\bibitem{Izumi84b}
M.~Izumi, T.~Toshiaki, K.~Uchinokura, R.~Yoshizaki, and E.~Matsuura, Solid
  State Commun. 51 (1984), p. 191.

\bibitem{Staresinic06}
D.~Staresini{\v c}, P.~Lunkenheimer, J.~Hemberger, K.~Biljkakovic, and
  A.~Loidl, Phys. Rev. Lett. 96 (2006), p. 046402.

\bibitem{Meerschaut84}
A.~Meerschaut, P.~Gressier, L.~Guemas, and J.~Rouxel, J. Solid State Chemistry
  51 (1984), p. 307.

\bibitem{Roucau85}
{\rm C. Roucau and R. Ayrolles in {\it Lecture Notes in Physics ``Charge
  Density Waves in Solids"}} (1998).

\bibitem{Vucic96}
Z.~Vucic, J.~Gladic, C.~Haas, and \mbox{J.L. de Boer}, J. Phys. I (France) 6
  (1996), p. 265.

\bibitem{Monceau85}
\mbox{P. Monceau in ref. \cite{R16Monceau85}, Part II}  (1985), p. 139.

\bibitem{vanSmaalen86}
\mbox{S. van Smaalen}, K.D. Bronsema, and J.~Mahy, Acta Crystallogr. B 42
  (1986), p.~43.

\bibitem{Lorenzo92}
\mbox{J.E. Lorenzo-Diaz}, Thesis, University of Grenoble, unpublished,  1992.

\bibitem{Sekine87}
T.~Sekine, Y.~Kiuchi, E.~Matsuura, K.~Uchinokura, and R.~Yashizaki, Phys. Rev.
  B 36 (1987), p. 3153.

\bibitem{Nunez93}
M.~N{\'u}{\~n}ez-Regueiro, \mbox{J.-M. Mignot}, M.~Jaime, D.~Castello, and
  P.~Monceau, Synth. Metals 55-57 (1993), p. 2653.

\bibitem{Nunez92}
M.~N{\'u}{\~n}ez-Regueiro, \mbox{J.-M. Mignot}, and D.~Castello, Europhys.
  Lett. 18 (1992), p.~53.

\bibitem{Nunez11}
 The author acknowledges M.~N{\'u}{\~n}ez-Regueiro for having collected all his
  data \cite{Nunez92,Nunez93} in this single figure.

\bibitem{Ido90}
M.~Ido, Y.~Okayama, T.~Ijiri, and Y.~Okajima, J. Phys. Soc. Jpn 59 (1990), p.
  1341.

\bibitem{Yasuzuka05}
S.~Yasuzuka, K.~Murata, T.~Fujimoto, M.~Shimotori, and K.~Yamaya, J. Phys. Soc.
  Jpn 74 (2005), p. 1782.

\bibitem{Mori04}
N.~M{\^o}ri, H.~Takahashi, and N.~Takeshita, High Pressure Res. 24 (2004), p.
  225.

\bibitem{Comes79}
\mbox{R. Comes and G. Shirane in ref. \cite{R3Devresse79}} , p.~17.

\bibitem{Carneiro76}
K.~Carneiro, G.~Shirane, S.A. Werner, and S.~Kaiser, Phys Rev. B 13 (1976), p.
  4258.

\bibitem{Hennion92}
B.~Hennion, \mbox{J.-P. Pouget}, and M.~Sato, Phys. Rev. Lett. 68 (1992), p.
  2374.

\bibitem{Fujishita85}
H.~Fujishita, M.~Sato, S.~Sato, and S.~Hoshino, J. Phys. C: Solid State Phys.
  18 (1985), p. 1105.

\bibitem{Lorenzo98}
J.E. Lorenzo, R.~Currat, P.~Monceau, B.~Hennion, H.~Berger, and F.~L{\'e}vy, J.
  Phys.: Condens. Matter 10 (1998), p. 5039.

\bibitem{Monceau89}
P.~Monceau, L.~Bernard, R.~Currat, and F.~L{\'e}vy, Physica B 156-157 (1989),
  p.~20.

\bibitem{Saint-Paul88a}
\mbox{M. Saint-Paul}, P.~Monceau, and F.~L{\'e}vy, Phys. Rev. B 37 (1988), p.
  1024.

\bibitem{Saint-Paul88b}
\mbox{M. Saint-Paul}, P.~Monceau, and F.~L{\'e}vy, Solid State Commun. 67
  (1988), p. 581.

\bibitem{Lorenzo96}
J.E. Lorenzo, R.~Currat, A.J. Dianoux, P.~Monceau, and F.~L{\'e}vy, Phys. Rev.
  B 53 (1996), p. 8316.

\bibitem{Lorenzo93}
J.E. Lorenzo, R.~Currat, P.~Monceau, B.~Hennion, and F.~L{\'e}vy, Phys. Rev. B
  47 (1993), p. 10116.

\bibitem{Sekine85}
T.~Sekine, T.~Seino, M.~Izumi, and E.~Matsuura, Solid State Commun. 53 (1985),
  p. 767.

\bibitem{Degiorgi91a}
L.~Degiorgi, G.M. B.~Alavi, and G.~Gr{\"u}ner, Phys. Rev. B 44 (1991), p. 7808.

\bibitem{Degiorgi91b}
L.~Degiorgi and G.~Gr{\"u}ner, Phys. Rev. Lett. 44 (1991), p. 7820.

\bibitem{Monceau87a}
P.~Monceau, L.~Bernard, R.~Currat, F.~L{\'e}vy, and J.~Rouxel, Synth. Metals 19
  (1987), p. 819.

\bibitem{Requardt02a}
H.~Requardt, J.E. Lorenzo, P.~Monceau, R.~Currat, and M.~Krisch, Phys. Rev. B
  66 (2002), p. 214303.

\bibitem{Brill78}
J.W. Brill and N.P. Ong, Solid State Commun. 25 (1978), p. 1075.

\bibitem{Smontara96}
A.~Smontara, \mbox{J.-C. Lasjaunias}, and R.~Maynard, Phys. Rev. Lett. 77
  (1996), p. 5397.

\bibitem{Smontara98}
A.~Smontara, \mbox{J.-C. Lasjaunias}, R.~Maynard, H.~Berger, and F.~Levy, J.
  Low Temp. Phys. 111 (1998), p. 815.

\bibitem{Hogan69}
E.M. Hogan, R.A. Guyer, and H.A. Fairbank, Phys. Rev. 185 (1969), p. 356.

\bibitem{Berman76}
{\rm R. Berman, in} (1976).

\bibitem{Mezhov66}
\mbox{L.P. Mezhov-Deglin}, Sov. Phys. JETP 22 (1966), p. 47; 25 (1967) p. 568.

\bibitem{Saint-Paul96}
\mbox{M. Saint-Paul}, S.~Holtmeier, R.~Britel, P.~Monceau, R.~Currat, and
  F.~Levy, J. Phys.: Condens. Matter 8 (1996), p. 2021.

\bibitem{Graham66}
J.~Graham and A.D. Wadsley, Acta Crystallogr. 20 (1966), p.~93.

\bibitem{Ghedira85}
M.~Ghedira, J.~Chenavas, M.~Marezio, and J.~Marcus, J. Solid State Chem. 57
  (1985), p. 300.

\bibitem{Ando05}
H.~Ando, T.~Yokoya, K.~Ishizaka, S.~Tsuda, T.~Kiss, S.~Shin, T.~Eguchi,
  M.~Nohara, and H.~Takagi, J. Phys.: Condens. Matter 17 (2005), p. 4935.

\bibitem{Brun05}
C.~Brun, \mbox{J.-C. Girard}, Z.Z. Wang, J.~Marcus, J.~Dumas, and C.~Schlenker,
  Phys. Rev. B 72 (2005), p. 235119.

\bibitem{Schutte93}
W.J. Schutte and \mbox{J.L.D. Boer}, Acta Crystallogr. Sect. B: Struct. Sci. 49
  (1993), p. 579.

\bibitem{Machado06}
\mbox{E. Machado-Charry}, P.~Ordejon, E.~Canadell, C.~Brun, and Z.Z. Wang,
  Phys. Rev. B 74 (2006), p. 155123.

\bibitem{Whangbo86}
\mbox{M.-H. Whangbo} and L.F. Schneemeyer, Inorg. Chem. 25 (1986), p. 2424.

\bibitem{Mozos02}
\mbox{J.-L. Mozos}, P.~Ordejon, and E.~Canadell, Phys. Rev. B 65 (2002), p.
  233105.

\bibitem{Hufner95}
{\rm S. H{\"u}fner,} (1995).

\bibitem{Federov00}
A.V. Federov, S.A. Brazovskii, V.N. Muthikumar, P.D. Johson, m.~J.~Xue, K.E.
  Smith, W.H. McCaroll, M.~Greenblatt, and S.L. Hulbert, J. Phys.: Condens.
  Matter 12 (2000), p. L191.

\bibitem{Denoyer75}
F.~Denoyer, R.~Comes, A.F. Garito, and A.J. Heeger, Phys. Rev. Lett. 35 (1975),
  p. 445.

\bibitem{Kagoshima75}
S.~Kagoshima, H.~Anzai, K.~Kajimura, and T.~Ishiguro, J. Phys. Soc. Jpn 39
  (1975), p. 1143.

\bibitem{Jerome82}
D.~J{\'e}r{\^o}me and H.~Schulz, Adv. Phys. 31 (1982), p. 299.

\bibitem{Jerome04}
\mbox{For a recent review, D.J{\'e}r{\^o}me}, Chem. Rev. 104 (2004), p. 5565.

\bibitem{Wang03}
Z.Z. Wang, \mbox{J.-C. Girard}, C.~Pasquier, D.~J{\'e}r{\^o}me, and
  K.~Bechgaard, Phys. Rev. B 67 (2003), p. 121401.

\bibitem{Lacoe85}
R.C. Lacoe, H.J. Schulz, D.~J{\'e}r{\^o}me, K.~Bechgaard, and I.~Johannsen,
  Phys. Rev. Lett. 55 (1985), p. 2351.

\bibitem{Lacoe87}
R.C. Lacoe, J.R. Cooper, D.~J{\'e}r{\^o}me, F.~Creuzet, K.~Bechgaard, and
  I.~Johannsen, Phys. Rev. Lett. 58 (1987), p. 262.

\bibitem{ClaessenPRL02}
R.~Claessen, M.~Sing, U.~Schwingenschl{\"o}g, P.~Blaha, M.~Dressel, and C.S.
  Jacobsen, Phys. Rev. Lett. 88 (2002), p. 096402.

\bibitem{Claessen02}
R.~Claessen, U.~Schwingenschl{\"o}gl, M.~Sing, C.J. Jacobsen, and M.~Dressel,
  Physica B 312-313 (2002), p. 660.

\bibitem{Zwick98}
F.~Zwick, D.~J{\'e}r{\^o}me, G.~Margaritondo, M.~Onelllion, J.~Voit, and
  M.~Grioni, Phys. Rev. Lett. 81 (1998), p. 2974.

\bibitem{Almeida97}
{\rm M. Almeida and R. T. Henriques in {\it Organic conductor molecules and
  polymers}} (1997).

\bibitem{Alcacer80}
L.~Alc{\'a}cer, H.~Novais, F.~Pedroso, S.~Flandrois, C.~Coulon, D.~Chasseau,
  and J.~Gaultier, Solid State Commun. 35 (1980), p. 945.

\bibitem{Graf04a}
D.~Graf, E.S. Choi, J.S. Brooks, M.~Matos, R.T. Henriques, and M.~Almeida,
  Phys. Rev. Lett. 93 (2004), p. 076406.

\bibitem{Graf04b}
D.~Graf, J.S. Brooks, E.S. Choi, S.~Uji, J.C. Dias, M.~Almeida, and M.~Matos,
  Phys. Rev. B 69 (2004), p. 125113.

\bibitem{Gama93b}
V.~Gama, R.T. Henriques, M.~Almeida, and \mbox{J.-P. Pouget}, Synth. Metals
  55-57 (1993), p. 1677.

\bibitem{Lopes94}
E.B. Lopes, M.J. Matos, R.T. Henriques, M.~Almeida, and J.~Dumas, Eur. Phys.
  Letters 27 (1994), p. 241.

\bibitem{Gama93a}
V.~Gama, R.T. Henriques, G.~Bonfait, M.~Almeida, S.~Ravy, \mbox{J.-P. Pouget},
  and L.~Alcacer, Mol. Crys. Liq. Crys. 234 (1993), p. 171.

\bibitem{Henriques93}
R.T. Henriques, V.~Gama, G.~Bonfait, I.C. Santos, M.J. Matos, M.~Almeida, M.T.
  Duarte, and L.~Alcacer, Synth. Metals 55-57 (1993), p. 1846.

\bibitem{Bourbonnais91}
C.~Bourbonnais, R.T. Henriques, P.~Wzietek, D.~K{\"o}ngeter, J.~Voiron, and
  D.~J{\'e}r{\^o}me, Phys. Rev. B 44 (1991), p. 641.

\bibitem{Canadell04}
E.~Canadell, M.~Almeida, and J.~Brooks, Eur. Phys. J. B 42 (2004), p. 453.

\bibitem{Riess93}
W.~Riess and Br{\"u}tting, Physica Scripta T49 (1993), p. 721.

\bibitem{Brutting92}
W.~Br{\"u}tting, W.~Riess, and M.~Schwoerer, Ann. Physik 1 (1992), p. 409.

\bibitem{Ilakovac93}
V.~Ilakovac, S.~Ravy, \mbox{J.-M. Pouget}, W.~Riess, W.~Br{\"u}tting, and
  M.~Schwoerer, J. Phys. IV (France) C2 3 (1993), p. 137.

\bibitem{Riess91}
W.~Riess, W.~Schmid, J.~Gmeiner, and M.~Schwoerer, Synth. Metals 41-43 (1991),
  p. 2261.

\bibitem{Brun77}
G.~Brun, S.~Peytavin, B.~Liautard, E.T. Maurin, J.~Fabre, and L.~Giral, C.R.
  Acad. Sci. (Paris) 284 (1977), p. 211.

\bibitem{Galigne78}
\mbox{J.-L. Galigne}, B.~Liautard, S.~Peytavin, G.~Brun, \mbox{J.-M. Fabre},
  E.~Torreilles, and L.~Giral, Acta Cryst. B 34 (1978), p. 620.

\bibitem{Galigne79}
\mbox{J.-L. Galigne}, B.~Liautard, S.~Peytavin, G.~Brun, \mbox{J.-M. Fabre},
  E.~Torreilles, and L.~Giral, Acta Cryst. B 35 (1979), p. 2609.

\bibitem{Bechgaard81}
K.~Bechgaard, K.~Carneiro, M.~Olsen, F.B. Rasmussen, and C.S. Jacobsen, Phys.
  Rev. Lett. 46 (1981), p. 852.

\bibitem{Thorup81}
N.~Thorup, G.~Rindorf, H.~Soling, and K.~Bechgaard, Acta Cryst. B 37 (1981), p.
  1236.

\bibitem{Pouget96}
\mbox{J.-P. Pouget} and S.~Ravy, J. Phys. I (France) 6 (1996), p. 1501.

\bibitem{Giamarchi04}
T.~Giamarchi, Chem. Rev. 104 (2004), p. 5037.

\bibitem{Jerome91}
D.~J{\'e}r{\^o}me, Science 252 (1991), p. 1509.

\bibitem{Foury04}
P.~Foury-Leylekian, \mbox{D. Le Bolloc'h}, B.~Hennion, S.~Ravy, A.~Moradpour,
  and \mbox{J.-P. Pouget}, Phys. Rev. B 70 (2004), p. 180405(R).

\bibitem{Pouget06}
\mbox{J.-P. Pouget}, \mbox{P. Foury-Leylikian}, \mbox{D. Le Bolloc'h},
  B.~Hennion, S.~Ravy, C.~Coulon, V.~Cardoso, and A.~Moradpour, J. Low Temp.
  Phys. 142 (2006), p. 147.

\bibitem{Dressel07}
M.~Dressel, Naturwissenschaften 94 (2007), p. 527.

\bibitem{Mila95}
F.~Mila, Phys. Rev. B 52 (1995), p. 4788.

\bibitem{Castet96}
F.~Castet, A.~Fritsch, and L.~Ducasse, J. Phys. I (Paris) 6 (1996), pp. 583;
  Chem. Phys. 232 (1998) p. 37; A. Fritsch and L. Ducasse, J. Phys. I (Paris) 1
  (1991) p. 855.

\bibitem{Seo00}
H.~Seo, J. Phys. Soc. Jpn 69 (2000), p. 805.

\bibitem{Hiraki98}
K.~Hiraki and K.~Kanoda, Phys. Rev. Lett. 80 (1998), p. 4737.

\bibitem{Coulon85}
C.~Coulon, S.S.P. Parkin, and R.~Laversanne, Phys. Rev. B 31 (1985), p. 3583.

\bibitem{Javadi88}
H.H.S. Javadi, R.~Laversanne, and A.J. Epstein, Phys. Rev. B 37 (1988), p.
  4280.

\bibitem{Laversanne84}
R.~Laversanne, C.~Coulon, B.~Gallois, \mbox{J.-P. Pouget}, and R.~Moret, J.
  Physique Lett. 45 (1984), p. L393.

\bibitem{Pouget97}
\mbox{J.-P. Pouget} and S.~Ravy, Synth. Metals 85 (1997), p. 1523.

\bibitem{Nad98}
\mbox{F.Ya. Nad}, P.~Monceau, and \mbox{J.-M. Fabre}, Eur. Phys. J. B 3 (1998),
  p. 301.

\bibitem{Nad99}
\mbox{F.Ya. Nad}, P.~Monceau, and \mbox{J.-M. Fabre}, J. Phys. IV (France) 9
  (1999), pp. Pr10--361.

\bibitem{Nad06a}
\mbox{F.Ya. Nad} and P.~Monceau, J. Phys. Soc. Jpn 75 (2006), p. 051005.

\bibitem{Nad01a}
\mbox{F.Ya. Nad}, P.~Monceau, C.~Carcel, and \mbox{J.-M. Fabre}, J. Phys.
  Condensed Matter 13 (2001), p. L717.

\bibitem{Coulon82a}
C.~Coulon, P.~Delhaes, S.~Flandrois, E.~Bonjour, R.~Lagnier, and \mbox{J.-M.
  Fabre}, J. Phys. (Paris) 43 (1982), p. 1059.

\bibitem{Coulon82b}
C.~Coulon, A.~Maaroufi, J.~Amiell, E.~Dupart, S.~Flandrois, P.~Delhaes,
  R.~Moret, \mbox{J.-P. Pouget}, and \mbox{J.-P. Morand}, Phys. Rev. B 26
  (1982), p. 6322.

\bibitem{Nakamura07}
T.~Nakamura, K.~Furukawa, and T.~Hara, J. Phys. Soc. Jpn 76 (2007), p. 064715.

\bibitem{Fujiyama06}
S.~Fujiyama and T.~Nakamura, J. Phys. Soc. Jpn 75 (2006), p. 014705.

\bibitem{Yu04}
W.~Yu, F.~Zambarsky, B.~Alavi, A.~Baur, C.A. Merlic, and S.E. Brown, J. Phys.
  IV 114 (2004), p.~35.

\bibitem{Nakamura06}
T.~Nakamura, K.~Furukawa, and T.~Hara, J. Phys. Soc. Jpn 75 (2006), p. 013707.

\bibitem{Nogami02}
Y.~Nogami and T.~Nakamura, J. Phys. IV (Paris) 12 (2002), pp. Pr9--145.

\bibitem{Nogami05}
Y.~Nogami, T.~Ito, K.~Yamamoto, N.~Irie, S.~Horita, T.~Kambe, N.~Nagao,
  K.~Oshima, N.~Ikeda, and T.~Nakamura, J. Phys. IV (France) 131 (2005), p.~39.

\bibitem{Nad00}
\mbox{F.Ya. Nad}, P.~Monceau, C.~Carcel, and \mbox{J.-M. Fabre}, Phys. Rev. B
  62 (2000), p. 1753.

\bibitem{Nad00b}
\mbox{F.Ya. Nad}, P.~Monceau, C.~Carcel, and \mbox{J.-M. Fabre}, J. Phys.
  Condensed Matter 12 (2000), p. L435.

\bibitem{Nagasawa05}
M.~Nagasawa, \mbox{F.Ya. Nad}, P.~Monceau, and \mbox{J.-M. Fabre}, Solid State
  Commun. 136 (2005), p. 262.

\bibitem{Nad06b}
\mbox{F.Ya. Nad}, P.~Monceau, L.~Kaboub, and \mbox{J.-M. Fabre}, Europhys.
  Lett. 73 (2006), p. 567.

\bibitem{Jonscher83}
\mbox{A.K. Jonscher, in}, \emph{Dielectric Relaxation in Solids}, Chelsea
  Dielectric Press, London, 1983.

\bibitem{Lines77}
M.E. Lines and A.M. Glass, \emph{Principles and Application of Ferroelectrics
  and Related Materials}, Clarendon Press, Oxford, 1977.

\bibitem{Chow98}
D.S. Chow, P.~Wzietek, D.~Fogliatti, B.~Alavi, D.J. Tantillo, C.A. Merlic, and
  S.E. Brown, Phys. Rev. Lett. 81 (1998), p. 3984.

\bibitem{Zamborsky02}
F.~Zamborsky, W.~Yu, W.~Raas, S.E. Brown, B.~Alavi, C.A. Merlic, and A.~Baur,
  Phys. Rev. B 66 (2002), p. 081103(R).

\bibitem{Hirose10}
S.~Hirose, A.~Kawamoto, N.~Matsunaga, K.~Nomura, K.~Yamamoto, and K.~Yakushi,
  Phys. Rev. B 81 (2010), p. 205107.

\bibitem{Meneghetti84}
M.~Meneghetti, R.~Bozio, I.~Zanon, C.~Pecile, C.~Ricotta, and M.~Zanetti, J.
  Chem. Phys. 80 (1984), p. 6210.

\bibitem{Dumm05}
M.~Dumm, A.~Abaker, and M.~Dressel, J. Phys. IV (France) 131 (2005), p.~55.

\bibitem{Dumm06}
M.~Dumm, A.~Abaker, M.~Dressel, and L.K. Montgomery, J. Low Temp. Phys. 142
  (2006), p. 609.

\bibitem{Fujiyama05}
S.~Fujiyama and T.~Nakamura, J. Phys. IV (France) 131 (2005), p.~33.

\bibitem{Guionneau97}
P.~Guionneau, C.J. Kepert, G.~Bravic, D.~Chasseau, M.R. Truter, M.~Kurmoo, and
  P.~Day, Synth. Metals 86 (1997), p. 1973.

\bibitem{Monceau01}
P.~Monceau, \mbox{F.Ya. Nad}, and S.~Brazovskii, Phys. Rev. Lett. 86 (2001), p.
  4080.

\bibitem{Brazovskii02}
S.~Brazovskii, J. Phys. IV (France) 12 (2002), pp. Pr9--149.

\bibitem{Nad05}
\mbox{F.Ya. Nad}, P.~Monceau, T.~Nakamura, and K.~Furukawa, J. Phys.: Condens.
  Matter 17 (2005), p. L399.

\bibitem{Furukawa05}
K.~Furukawa, T.~Hara, and T.~Nakamura, J. Phys. Soc. Jpn 74 (2005), p. 3288.

\bibitem{Iwase09}
F.~Iwase, K.~Sugiura, K.~Furnkawa, and T.~Nakamura, J. Phys. Soc. Jpn 78
  (2009), p. 104717.

\bibitem{Horiuchi08}
S.~Horiuchi and Y.~Tokura, Nature Materials 7 (2008), p. 357.

\bibitem{Okamoto91}
H.~Okamoto, T.~Mitani, Y.~Tokura, S.~Koshihara, T.~Komatsu, Y.~Iwasa, T.~Koda,
  and G.~Saito, Phys. Rev. B 43 (1991), p. 8224.

\bibitem{Kishida09}
H.~Kishida, H.~Takamatsu, K.~Fujinuma, and H.~Okamoto, Phys. Rev. B 80 (2009),
  p. 205201.

\bibitem{Garcia05}
P.~Garcia, S.~Dahaoui, P.~Pertey, E.~Wenger, and C.~Lecomte, Phys. Rev. B 72
  (2005), p. 104115.

\bibitem{Girlando85}
A.~Girlando, C.~Pecile, and J.B. Torrance, Solid State Commun. 54 (1985), p.
  753.

\bibitem{Kagawa10a}
F.~Kagawa, S.~Horiuchi, H.~Matsui, R.~Kumai, Y.~Onase, T.~Hasegawa, and
  Y.~Tokura, Phys. Rev. Lett. 104 (2010), p. 227602.

\bibitem{Souza08}
\mbox{M. de Souza}, P.~Foury-Leylekian, A.~Moradpour, \mbox{J.-P. Pouget}, and
  M.~Lang, Phys. Rev. Lett. 101 (2008), p. 216403.

\bibitem{Foury10}
P.~Foury-Leylekian, S.~Petit, G.~Andr{\'e}, A.~Moradpour, and \mbox{J.-P.
  Pouget}, Physica B 405 (2010), p. S95.

\bibitem{Souza10}
\mbox{M. de Souza}, D.~Hofmann, P.~Foury-Leylekian, A.~Moradpour, \mbox{J.-P.
  Pouget}, and M.~Lang, Physica B 405 (2010), p. S92.

\bibitem{Kuwabara03}
M.~Kuwabara, H.~Seo, and M.~Ogata, J. Phys. Soc. Jpn 72 (2003), p. 225.

\bibitem{Coulon07}
C.~Coulon, G.~Lalet, \mbox{J.-P. Pouget}, P.~Foury-Leylekian, A.~Moradpour, and
  \mbox{J.-M. Fabre}, Phys. Rev. B 76 (2007), p. 085126.

\bibitem{Jerome80}
D.~J{\'e}r{\^o}me, A.~Mazaud, M.~Ribault, and K.~Bechgaard, J. Phys. Lett.
  (Paris) 41 (1980), p. L95.

\bibitem{Brazovskii85}
S.~Brazovskii and V.~Yakovenko, J. Physique Lett. 46 (1985), p. 111; JETP 62
  (1985) p. 1340.

\bibitem{Clay03}
R.T. Clay, S.~Mazumdar, and D.K. Campbell, Phys. Rev. B 67 (2003), p. 115121.

\bibitem{Henze03}
K.~Henz{\'e}, M.~Fourmigu{\'e}, P.~Batail, C.~Coulon, R.~Cl{\'e}rac,
  E.~Canadell, P.~Auban-Senzier, S.~Ravy, and D.~J{\'e}r{\^o}me, Adv. Mater 15
  (2003), p. 1251.

\bibitem{Zorina09}
L.~Zorina, S.~Simonov, C.~M{\'e}zi{\`e}re, E.~Canadell, S.~Suh, S.E. Brown,
  P.~Foury-Leylekian, P.~Fertey, \mbox{J.-P. Pouget}, and P.~Batail, J. Mat.
  Chem. 19 (2009), p. 6980.

\bibitem{Auban-Senzier09}
P.~Auban-Senzier, C.R. Pasquier, D.~J{\'e}r{\^o}me, S.~Suh, S.E. Brown,
  C.~M{\'e}zi{\`e}re, and P.~Batail, Phys. Rev. Lett. 102 (2009), p. 257001.

\bibitem{Nogami99}
Y.~Nogami, K.~Oshima, K.~Hiraki, and K.~Kanoda, J. Phys. IV (France) 9 (1999),
  pp. Pr10--357.

\bibitem{Kakiuchi07}
T.~Kakiuchi, Y.~Wakabayashi, H.~Sawa, T.~Itou, and K.~Kanoda, Phys. Rev. Lett.
  98 (2007), p. 066402.

\bibitem{Takahashi89}
T.~Takahashi, H.~Kawamura, T.~Ohyama, Y.~Maniwa, K.~Murata, and G.~Saito, J.
  Phys. Soc. Jpn 58 (1989), p. 703.

\bibitem{Takahashi86}
T.~Takahashi, Y.~Maniwa, H.~Kawamura, and G.~Saito, J. Phys. Soc. Jpn 55
  (1986), p. 1364; Physica B 143 (1986) p. 417.

\bibitem{Delrieu86}
\mbox{J.-M. Delrieu}, M.~Roger, and Z.~Toffano, J. Phys. (France) 47 (1986), p.
  839.

\bibitem{Le93}
\mbox{L.P. Le, A. Keren, G.M. Luke, B.J. Sternlieb, W.D. Wu, Y.J. Uemura, J.H.
  Brewer, T.M. Riseman}, R.K. Upasani, L.Y. Chiang, W.~Kang, P.M. Chaikin,
  T.~Csiba, and G.~Gr{\"u}ner, Phys. Rev. B 48 (1993), p. 7284.

\bibitem{Nakamura95}
T.~Nakamura, T.~Nobutoki, Y.~Kobayashi, T.~Takahashi, and G.~Saito, Synth.
  Metals 70 (1995), p. 1293.

\bibitem{Dumm00}
M.~Dumm, A.~Loidl, B.~Alavi, K.P. Starkey, L.K. Montgomery, and M.~Dressel,
  Phys. Rev. B 62 (2000), p. 6512.

\bibitem{Kagoshima99}
S.~Kagoshima, Y.~Saso, M.~Maeto, R.~Kondo, and T.~Hasegawa, Solid State Commun.
  110 (1999), p. 479.

\bibitem{Danneau03}
R.~Danneau, Doctorat, University of Grenoble, 2003 Annexe B, p. 107.

\bibitem{Currat03}
R.~Currat, R.~Danneau, and P.~Monceau, unpublished  (2003).

\bibitem{Wulfhekel07}
W.~Wulfhekel and J.~Kirschner, Annu. Rev. Mater. Res. 37 (2007), p.~69.

\bibitem{Wiesendanger09}
R.~Wiesendanger, Rev. Mod. Phys. 81 (2009), p. 1495.

\bibitem{Hanke05}
T.~H{\"a}nke, S.~Krause, \mbox{L. Berbil-Bautista}, M.~Bode, and
  R.~Wiesendanger, Phys. Rev. B 71 (2005), p. 184407.

\bibitem{Kawagoe05}
T.~Kawagoe, Y.~Iguchi, T.~Miyamachi, A.~Yamasaki, and S.~Suga, Phys. Rev. Lett.
  95 (2005), p. 207205.

\bibitem{Rodary11}
G.~Rodary, \mbox{J.-C. Girard}, L.~Largeau, C.~David, O.~Maugin, and Z.Z. Wang,
  Appl. Phys. Lett. 98 (2011), p. 082505.

\bibitem{Balicas94}
L.~Balicas, K.~Behnia, W.~Kang, E.~Canadell, \mbox{P. Auban-Senzier},
  D.~J{\'e}r{\^o}me, M.~Ribault, and \mbox{J.-M. Fabre}, J. Phys. I (France) 4
  (1994), p. 1539.

\bibitem{Jaccard01}
D.~Jaccard, H.~W{\"\i}lhem, D.~J{\'e}r{\^o}me, J.~Moser, C.~Carcel, and
  \mbox{J.-M. Fabre}, J. Phys.: Condens. Matter 13 (2001), p. L89.

\bibitem{Adachi00}
T.~Adachi, E.~Ojima, K.~Kato, and H.~Kobayashi, J. Am. Chem. Soc. 122 (2000),
  p. 3238.

\bibitem{Adachi01}
T.~Adachi, H.~Tanaka, H.~Kobayashi, and T.~Miyazaki, Rev. Sci. Instrum. 72
  (2001), p. 2358.

\bibitem{Itoi07}
M.~Itoi, M.~Kano, N.~Kurita, M.~Hedo, Y.~Uwatoko, and T.~Nakamura, J. Phys.
  Soc. Jpn 76 (2007), p. 053703.

\bibitem{Auban03}
\mbox{P. Auban-Senzier}, C.~Pasquier, D.~J{\'e}r{\^o}me, C.~Carcel, and
  \mbox{J.-M. Fabre}, Synth. Metals 133 (2003), p.~11.

\bibitem{Itoi08}
M.~Itoi, C.~Araki, M.~Hedo, Y.~Uwatoko, and T.~Nakamura, J. Phys. Soc. Jpn 77
  (2008), p. 023701.

\bibitem{Taniguchi03}
H.~Taniguchi, M.~Miyashita, K.~Uchiyama, K.~Satoh, N.~Mori, H.~Okamoto,
  Y.~Miyagawa, K.~Kanoda, M.~Hedo, and Y.~Uwatoko, J. Phys. Soc. Jpn 72 (2003),
  p. 468.

\bibitem{Salameh09}
B.~Salameh, \mbox{P. Auban-Senzier}, N.~Kang, C.R. Pasquier, and
  D.~J{\'e}r{\^o}me, Physica B 404 (2009), p. 476.

\bibitem{Vuletic02}
T.~Vuleti{\'c}, \mbox{P. Auban-Senzier}, C.~Pasquier, S.~Tomi{\'c},
  D.~J{\'e}r{\^o}me, M.~H{\'e}ritier, and K.~Bechgaard, Eur. Phys. J. B 25
  (2002), p. 319.

\bibitem{Kang10}
N.~Kang, B.~Salameh, \mbox{P. Auban-Senzier}, D.~J{\'e}r{\^o}me, C.R. Pasquier,
  and C.~Brazovskii, Phys. Rev. B 81 (2010), p. 100509.

\bibitem{Bjerkelund66}
E.~Bjerkelund, J.H. Fermor, and A.~Kjekhus, Acta Chem. Scand. 20 (1966), p.
  1836.

\bibitem{Bjerkelund64}
E.~Bjerkelund and A.~Kjekhus, Z. Anorg. Allg. Chem. B 228 (1964), p. 235.

\bibitem{Furuseth91}
S.~Furuseth and H.~Fjellvag, Acta Chem. Scand. 45 (1991), p. 694.

\bibitem{Gressier82}
P.~Gressier, L.~Guemas, and A.~Meerschaut, Acta Cryst. B 38 (1982), p. 2877.

\bibitem{Meerschaut77}
A.~Meerschaut, P.~Palvadeau, and J.~Rouxel, J. Solid State Chem. 20 (1977),
  p.~21.

\bibitem{Selte64}
K.~Selte and A.~Kjekhus, Acta Chem. Scand. 18 (1964), p. 690.

\bibitem{Boswell83}
F.W. Boswell, A.~Prodan, and J.K. Brandon, J. Phys. C 16 (1983), p. 1067.

\bibitem{Bohm99}
\mbox{H. B{\"o}hm in ref. \cite{R12Boswell99}} , p.~41.

\bibitem{Furuseth73}
S.~Furuseth, L.~Brattas, and A.~Kjekhus, Acta Chem. Scand. 27 (1973), p. 2367.

\bibitem{Sambongi77b}
T.~Sambongi, M.~Yamamoto, K.~Tsutsumi, Y.~Shiozaki, K.~Yamaya, and Y.~Abe, J.
  Phys. Soc. Jpn 42 (1977), p. 1421.

\bibitem{Wang83c}
Z.Z. Wang, H.~Salva, P.~Monceau, M.~Renard, C.~Roucau, R.~Ayrolles, F.~Levy,
  L.~Guemas, and A.~Meerschaut, J. Physique Lettres (Paris) 44 (1983), p. L311.

\bibitem{Roucau84}
C.~Roucau, R.~Ayrolles, P.~Gressier, and A.~Meerschaut, J. Phys. C 17 (1984),
  p. 2993.

\bibitem{Mahy83}
J.~Mahy, \mbox{J. van Landuyt}, and S.~Amelinckx, Phys. Stat. Solidi a 77
  (1983), p.~K1.

\bibitem{Dumas93}
J.~Dumas and C.~Schlenker, Int. J. Mod. Phys. B 7 (1993), p. 4045.

\bibitem{Monceau76}
P.~Monceau, N.P. Ong, A.M. Portis, A.~Meerschaut, and J.~Rouxel, Phys. Rev.
  Lett. 37 (1976), p. 602.

\bibitem{Fleming79}
R.M. Fleming and C.C. Grimes, Phys. Rev. Lett. 42 (1979), p. 1423.

\bibitem{Monceau80}
P.~Monceau, J.~Richard, and M.~Renard, Phys. Rev. Lett. 45 (1980), p.~43.

\bibitem{Ong77}
N.P. Ong and P.~Monceau, Phys. Rev. B 16 (1977), p. 3443.

\bibitem{Donovan94}
S.~Donovan, Y.~Kim, L.~Degiorgi, M.~Dressel, G.~Gr{\"u}ner, and
  W.~Wonnenberger, Phys. Rev. B 49 (1994), p. 3363.

\bibitem{Kim91}
T.W. Kim, S.~Donovan, G.~Gr{\"u}ner, and A.~Philipp, Phys. Rev. B 43 (1991), p.
  6315.

\bibitem{Richard82a}
J.~Richard, P.~Monceau, M.~Papoular, and M.~Renard, Phys. Rev. C 15 (1982), p.
  7157.

\bibitem{Bhatta87}
S.~Bhattacharya, J.P. Stokes, M.J. Higgins, and R.A. Klemm, Phys. Rev. Lett. 59
  (1987), p. 1849.

\bibitem{Monceau90}
\mbox{P. Monceau in Applications of Statistical and Field Theory Methods to
  Condensed Matter}, NATO ASI Series 218  (1990), p. 357.

\bibitem{Richard82}
J.~Richard, P.~Monceau, and M.~Renard, Phys. Rev. B 25 (1982), p. 931.

\bibitem{Shapiro72}
{\rm The steps in the ac-dc interference are called Shapiro steps in reference
  to the similar original experiment in vortex dynamics Shapiro }\mbox{S.
  Shapiro}, Phys. Rev. Lett. 11 (1963), p.~80.

\bibitem{Hall84}
R.P. Hall and A.~Zettl, Phys. Rev. B 30 (1984), p. 2279.

\bibitem{Thorne88}
R.E. Thorne, J.S. Hubacek, W.G. Lyons, J.W. Lyding, and J.R. Tucker, Phys. Rev.
  B 37 (1988), p. 10055.

\bibitem{Littlewood87b}
P.B. Littlewood, Phys. Rev. B 36 (1987), p. 3108.

\bibitem{Duggan85}
D.M. Duggan, T.W. Jing, N.P. Ong, and P.A. Lee, Phys. Rev. B 32 (1985), p.
  1397.

\bibitem{Brazovskii04}
S.~Brazovskii and T.~Nattermann, Adv. in Physics 53 (2004), p. 177.

\bibitem{Bardeen79}
J.~Bardeen, Phys. Rev. Lett. 42 (1979), p. 1498.

\bibitem{Bardeen79b}
\mbox{J. Bardeen in ref. \cite{R3Devresse79}} , p. 373.

\bibitem{Larkin79}
A.I. Larkin and \mbox{Yu. N. Ovchinnikov}, J. Low Temp. Phys. 34 (1979), p.
  409.

\bibitem{Lee79}
P.A. Lee and T.M. Rice, Phys. Rev. B 19 (1979), p. 3970.

\bibitem{Fukuyama78}
H.~Fukuyama and P.A. Lee, Phys. Rev. B 17 (1978), p. 535.

\bibitem{Lee78}
P.A. Lee and H.~Fukuyama, Phys. Rev. B 17 (1978), p. 542.

\bibitem{Efetov77}
K.B. Efetov and A.I. Larkin, Zh. Eksp. Teor. Fiz. 72 (1973), p. 7350.

\bibitem{Inui88}
M.~Inui, R.P. Hall, S.~Doniach, and A.~Zettl, Phys. Rev. B 38 (1988), p. 13047.

\bibitem{Gruner81}
G.~Gr{\"u}ner, A.~Zawadowski, and P.M. Chaikin, Phys. Rev. Lett. 46 (1981), p.
  511.

\bibitem{Wu84}
\mbox{Wei-Yu Wu}, L.~Mih{\'a}ly, G.~Mozurkewich, and G.~Gr{\"u}ner, Phys. Rev.
  Lett. 52 (1984), p. 2382.

\bibitem{Wu86}
\mbox{Wei-Yu Wu}, L.~Mih{\'a}ly, G.~Mozurkewich, and G.~Gr{\"u}ner, Phys. Rev.
  B 33 (1986), p. 2444.

\bibitem{Fisher83}
D.S. Fisher, Phys. Rev. Lett. 50 (1983), p. 1486.

\bibitem{Maki89a}
K.~Maki and A.~Virosztek, Phys. Rev. B 39 (1989), p. 9640.

\bibitem{Maki90b}
K.~Maki and A.~Virosztek, Phys. Rev. B 42 (1990), p. 655.

\bibitem{Maki90a}
K.~Maki and A.~Virosztek, Phys. Rev. B 41 (1990), p. 557.

\bibitem{Matsukawa88}
H.~Matsukawa, J. Phys. Soc. Jpn 57 (1988), p. 3463.

\bibitem{Abe85}
S.~Abe, J. Phys. Soc. Jpn 54 (1985), p. 22; 55 (1986) p . 1987.

\bibitem{Maki86}
K.~Maki and A.~Virosztek, Phys. Rev. B 33 (1986), p. 2852.

\bibitem{Maki89b}
K.~Maki and A.~Virosztek, Synth. Metals 29 (1989), p. F377.

\bibitem{Rice79}
T.M. Rice, P.A. Lee, and M.C. Cross, Phys. Rev. B 20 (1979), p. 1345.

\bibitem{Huang90}
X.~Huang and K.~Maki, Phys. Rev. B 42 (1990), p. 6498.

\bibitem{Fisher85}
D.S. Fisher, Phys. Rev. B 31 (1996), p. 1396.

\bibitem{Pietronero91}
L.~Pietronero and M.~Versteeg, Physica A 179 (1991), p.~1.

\bibitem{Parisi91}
G.~Parisi and L.~Pietronero, Physica A 179 (1991), p.~16.

\bibitem{Myers93}
C.R. Myers and J.P. Sethna, Phys. Rev. B 47 (1993), p. 11171.

\bibitem{Coppersmith90}
S.N. Coppersmith, Phys. Rev. Lett. 65 (1990), p. 1044.

\bibitem{Middleton93}
A.A. Middleton and D.S. Fisher, Phys. Rev. B 47 (1993), p. 3530.

\bibitem{Pietronero83}
L.~Pietronero and S.~Str{\"a}ssler, Phys. Rev. B 28 (1983), p. 5863.

\bibitem{Sibani90}
P.~Sibani and P.B. Littlewood, Phys. Rev. Lett. 64 (1990), p. 1305.

\bibitem{Narayan92}
O.~Narayan and D.S. Fisher, Phys. Rev. Lett. 68 (1992), p. 3615; Phys. Rev. B
  46 (1992) p. 11520.

\bibitem{Maki77}
K.~Maki, Phys. Rev. B 39 (1977), pp. 46, Phys. Rev. B 18 (1978) p. 1641.

\bibitem{Bardeen85}
J.~Bardeen, Phys. Rev. Lett. 55 (1985), p. 1010.

\bibitem{Bardeen89}
J.~Bardeen, Phys. Rev. B. 39 (1989), p. 3528.

\bibitem{Bardeen89b}
J.~Bardeen, Physica Scripta T 27 (1989), p. 136.

\bibitem{Bardeen90}
J.~Bardeen, Physics Today  (D{\'e}cembre 1990), p.~25.

\bibitem{Tucker79}
J.R. Tucker, I.E.E.E. Quantum Electron 15 (1979), p. 1234.

\bibitem{Barnes83}
S.E. Barnes and A.~Zawadowski, Phys. Rev. Lett. 51 (1983), p. 1003.

\bibitem{Tutto85}
I.~T{\"u}tto and A.~Zawadowski, Phys. Rev. B 32 (1985), p. 2449.

\bibitem{Tucker89}
J.R. Tucker, W.G. Lyons, and G.~Gammie, Phys. Rev. B 40 (1989), p. 5447.

\bibitem{Tucker93}
J.R. Tucker, Phys. Rev. B 47 (1993), p. 7614.

\bibitem{Takeda84}
S.~Takeda, J. Phys. Soc. Jpn 53 (1984), p. 2193.

\bibitem{Tua85a}
P.F. Tua and J.~Ruvalds, Phys. Rev. B 32 (1985), p. 4660.

\bibitem{Tua85b}
P.F. Tua and A.~Zawadowski, Phys. Rev. B 32 (1985), p. 2449.

\bibitem{Tutto88}
I.~T{\"u}tto and A.~Zawadowski, Phys. Rev. Lett. 60 (1988), p. 1442.

\bibitem{Tua84}
P.F. Tua, A.~Zawadowski, and J.~Ruvaldo, Phys. Rev. B 29 (1984), p. 6525.

\bibitem{Kang91}
W.~Kang, S.~Tomi{\'c}, and D.~J{\'e}r{\^o}me, Phys. Rev. B 43 (1991), p. 1264.

\bibitem{Tomic89}
S.~Tomi{\'c}, J.R. Cooper, D.~J{\'e}r{\^o}me, and K.~Beechgaard, Phys. Rev.
  Lett. 62 (1989), p. 462.

\bibitem{Sambongi89}
T.~Sambongi, K.~Nomura, T.~Shimizu, K.~Ichimura, N.~Kinoshita, M.~Tokumoto, and
  H.~Anzai, Solid State Commun. 72 (1989), p. 817.

\bibitem{Ido87}
M.~Ido, Y.~Okajima, H.~Wakimoto, and M.~Oda, J. Phys. Soc. Jpn 56 (1987), p.
  2503.

\bibitem{Mihaly88d}
G.~Mih{\'a}ly, P.~Beauch{\^e}ne, J.~Marcus, J.~Dumas, and C.~Schlenker, Phys.
  Rev. B 37 (1988), p. 1047.

\bibitem{Brill81}
J.W. Brill, N.P. Ong, J.C. Eckert, J.W. Savage, S.K. Khanna, and R.B. Somoano,
  Phys. Rev. B 23 (1981), p. 1517.

\bibitem{McCarten92}
\mbox{J. McCarten}, D.A. DiCarlo, M.P. Maher, T.L. Adelman, and R.E. Thorne,
  Phys. Rev. B 46 (1992), p. 4456.

\bibitem{Underweiser87}
M.~Underweiser, M.~Maki, B.~Alavi, and G.~Gr{\"u}ner, Solid State Commun. 64
  (1987), p. 181.

\bibitem{Schneemeyer84}
L.F. Schneemeyer, F.I. DiSalvo, S.E. Spengler, and J.V. Waszczak, Phys. Rev. B
  30 (1984), p. 4297.

\bibitem{Mutka84b}
H.~Mutka, S.~Bouffard, G.~Mih{\'a}ly, and L.~Mih{\'a}ly, J. Physique Lett. 45
  (1984), p. L113.

\bibitem{Mutka84a}
H.~Mutka, S.~Bouffard, J.~Dumas, and C.~Schlenker, J. Physique Lett. 45 (1984),
  p. 729.

\bibitem{Fuller81}
W.W. Fuller, G.~Gr{\"u}ner, P.M. Chaikin, and N.P. Ong, Phys. Rev. B 23 (1981),
  p. 6259.

\bibitem{Quelard76}
G.~Quelard and D.~Lesueur, Phys. Status Solidi a 36 (1976), p. 729.

\bibitem{Monceau81}
P.~Monceau, J.~Richard, and R.~Lagnier, J. Phys. C: Solid State Phys. 14
  (1981), p. 2995.

\bibitem{DiCarlo90}
\mbox{D. A. DiCarlo}, \mbox{J. McCarten}, T.L. Adelman, M.~Maher, and R.E.
  Thorne, Phys. Rev. B 42 (1990), p. 7643.

\bibitem{DiCarlo93a}
\mbox{D. A. DiCarlo}, \mbox{J. McCarten}, and R.F. Thorne, Phys. Rev. B 47
  (1993), p. 7618.

\bibitem{DiCarlo93b}
D.~DiCarlo, E.~Sweetland, M.~Sutton, J.D. Brock, and R.E. Thorne, Phys. Rev.
  Lett. 70 (1993), p. 845.

\bibitem{Tucker88}
J.R. Tucker, W.G. Lyons, and G.~Gammie, Phys. Rev. B 38 (1988), p. 1148.

\bibitem{Forro87}
L.~Forro, R.C. Lacoe, S.~Bouffard, and D.~J{\'e}r{\^o}me, Phys. Rev. B 35
  (1987), p. 5884.

\bibitem{Forro84}
L.~Forro, S.~Bouffard, and \mbox{J.-P. Pouget}, J. Phys. (Paris) Lett. 45
  (1984), p. L543.

\bibitem{Lopes97}
E.B. Lopes, M.J. Matos, R.T. Henriques, M.~Almeida, and J.~Dumas, Synth. Metals
  86 (1997), p. 2163.

\bibitem{Tsutsumi84}
K.~Tsutsumi, T.~Tamegai, S.~Kagoshima, H.~Tomozawa, and M.~Soto, J. Phys. Soc.
  Jpn 53 (1984), p. 3946.

\bibitem{Shimizu91}
T.~Shimizu, K.~Nomura, T.~Sambongi, H.~Anzai, N.~Kinoshita, and M.~Tokumoto,
  Solid State Commun. 78 (1991), p. 697.

\bibitem{Fleming86}
R.M. Fleming, R.J. Cava, L.F. Schneemeyer, E.A. Rietman, and R.G. Dunn, Phys.
  Rev. B 33 (1986), p. 5450.

\bibitem{Schlenker89b}
\mbox{C. Schlenker, J. Dumas, C. Escribe-Filippini, H. Guyot in ref.
  \cite{R9Schlenker89}} , p. 159.

\bibitem{Maeda85}
A.~Maeda, T.~Furuyama, and S.~Tanaka, Solid State Commun. 55 (1985), p. 951.

\bibitem{Saint-Paul09}
\mbox{M. Saint-Paul}, J.~Dumas, and J.~Marcus, J. Phys.: Condens. Matter 21
  (2009), p. 215603.

\bibitem{Kang90}
W.~Kang, S.~Tomi{\'c}, J.R. Cooper, and J{\'e}r{\^o}me, Phys. Rev. B 41 (1990),
  p. 4862.

\bibitem{Hoshikawa00}
A.~Hoshikawa, K.~Nomura, S.~Takasaki, J.~Yamada, S.~Nakatsuji, H.~Anzai,
  M.~Tokumoto, and N.~Kinoshita, J. Phys. Soc. Jpn 69 (2000), p. 1457.

\bibitem{Borodin86a}
D.V. Borodin, \mbox{S. V. Zaitsev-Zotov}, and \mbox{F.Ya. Nad}, JETP Lett. 43
  (1986), p. 625.

\bibitem{Yetman87}
P.J. Yetman and J.C. Gill, Solid State Commun. 62 (1987), p. 201.

\bibitem{Borodin86b}
D.V. Borodin, S.~Savitskaya, and \mbox{S. V. Zaitsev-Zotov}, Physica B 143
  (1986), p.~73.

\bibitem{McCarten89}
\mbox{J. McCarten}, M.~Maher, T.L. Adelman, and R.E. Thorne, Phys. Rev. Lett.
  63 (1989), p. 2841.

\bibitem{Tucker90}
J.R. Tucker, Phys. Rev. Lett. 65 (1990), p. 270.

\bibitem{Gill90a}
J.C. Gill, Phys. Rev. Lett. 65 (1990), p. 271.

\bibitem{Gill90b}
J.C. Gill, Europhys. Lett. 11 (1990), p. 175.

\bibitem{Lopes95}
E.B. Lopes, M.J. Matos, R.T. Henriques, and M.~Almeida, Phys. Rev. B 52 (1995),
  p. R2237.

\bibitem{Hundley89}
M.F. Hundley and A.~Zettl, Phys. Rev. B 39 (1989), p. 3026.

\bibitem{Sekine04}
T.~Sekine, N.~Satok, M.~Nakazawa, and T.~Nakamura, Phys. Rev B 70 (2004), p.
  214201.

\bibitem{Tomic88}
S.~Tomi{\'c}, F.~Fontaine, and D.~J{\'e}r{\^o}me, Phys. Rev. B 37 (1988), p.
  8468.

\bibitem{Kriza91b}
G.~Kriza, G.~Quirion, O.~Traetteberg, and D.~J{\'e}r{\^o}me, Phys. Rev. Lett.
  66 (1991), p. 1922.

\bibitem{Butaud90}
P.~Butaud, P.~S{\'e}gransan, A.~J{\'a}nossy, and C.~Berthier, J. Phys. (France)
  51 (1990), p.~59.

\bibitem{Brown85}
S.E. Brown and G.~Gr{\"u}ner, Phys. Rev. B 31 (1985), p. 8302.

\bibitem{Richard93}
J.~Richard, J.~Chen, and S.N. Artemenko, Solid State Commun. 85 (1993), p. 605.

\bibitem{Monceau83}
P.~Monceau, M.~Renard, J.~Richard, \mbox{M.-C. Saint-Lager}, H.~Salva, and Z.Z.
  Wang, Phys. Rev. B 28 (1983), p. 1646.

\bibitem{Artemenko84}
S.N. Artemenko, E.N. Dolgov, A.N. Kruglov, \mbox{Yu.I. Latyshev}, \mbox{Ya.S.
  Savitskaya}, and V.V. Frolov, Pisma Zh. ETF 39 (1984), p. 258.

\bibitem{Kriza91a}
G.~Kriza, G.~Quirion, O.~Traetteberg, and D.~J{\'e}r{\^o}me, Europhys. Lett. 16
  (1991), p. 585.

\bibitem{Nomura89}
K.~Nomura, T.~Shimizu, K.~Ichimura, T.~Sambongi, M.~Tokumoto, H.~Anzai, and
  N.~Kinoshita, Solid State Commun. 72 (1989), p. 1123.

\bibitem{Link88}
G.L. Link and G.~Mozurkewich, Solid State Commun. 65 (1988), p.~15.

\bibitem{Bhattacharya87}
S.~Bhattacharya, J.P. Stokes, M.J. Higgins, and R.A. Klemm, Phys. Rev. Lett. 59
  (1987), p. 1849.

\bibitem{Dumas95}
J.~Dumas, N.~Thirion, M.~Almeida, E.B. Lopes, M.J. Matos, and R.T. Henriques,
  J. Phys. I (France) 5 (1995), p. 539.

\bibitem{Preobrazhenskii99}
V.B. Preobrazhenskii, A.P. Grebenkin, \mbox{Yu.A. Danilov}, and \mbox{S.Yu.
  Shabanov}, Synth. Metals 103 (19999), p. 2608; 121 (2001) p. 1303.

\bibitem{Preobrazhenskii09}
V.B. Preobrazhenskii, A.P. Grebenkin, and \mbox{S.Yu. Shabanov}, Physica B 404
  (2009), p. 452.

\bibitem{Ross86}
J.H. Ross, Z.~Wang, and C.P. Slichter, Phys. Rev. Lett. 56 (1986), p. 663.

\bibitem{Ross90}
J.H. Ross, Z.~Wang, and C.P. Slichter, Phys. Rev. B 41 (1990), p. 2722.

\bibitem{Segransan86}
P.~S{\'e}gransan, A.~J{\'a}nossy, C.~Berthier, J.~Marcus, and P.~Butaud, Phys.
  Rev. Lett. 56 (1986), p. 1854.

\bibitem{Janossy87}
A.~J{\'a}nossy, C.~Berthier, P.~S{\'e}gransan, and P.~Butaud, Phys. Rev. Lett.
  59 (1987), p. 2348.

\bibitem{Clark94}
W.G. Clark, M.E. Hanson, W.H. Wang, and B.~Alavi, Phys. Rev. B 49 (1994), p.
  11895.

\bibitem{Wong93}
W.H. Wong, M.E. Hanson, B.~Alavi, W.G. Clark, and W.A. Hines, Phys. Rev. Lett.
  70 (1993), p. 1882.

\bibitem{Barthel93}
E.~Barthel, G.~Kriza, G.~Quirion, P.~Wzietck, D.~J{\'e}r{\^o}me, J.B.
  Christensen, M.~J{\o}rgensen, and K.~Bechgaard, Phys. Rev. Lett. 71 (1993),
  p. 2825.

\bibitem{Suh08}
S.~Suh, W.G. Clark, P.~Monceau, R.E. Thorne, and S.E. Brown, Phys. Rev. Lett.
  101 (2008), p. 136407.

\bibitem{Rouziere99}
S.~Rouzi{\`e}re, S.~Ravy, \mbox{J.-P. Pouget}, and R.F. Thorne, Phys. Rev. B 59
  (1999), p. 15121.

\bibitem{DeLand91}
S.M. DeLand, G.~Mozurkewich, and L.D. Chapman, Phys. Rev. Lett. 66 (1991), p.
  2026.

\bibitem{DiCarlo94}
D.~DiCarlo, R.E. Thorne, E.~Sweetland, M.~Sutton, and J.D. Brock, Phys. Rev. B
  50 (1994), p. 8288.

\bibitem{Girault88}
S.~Girault, A.H. Moudden, \mbox{J.-P. Pouget}, and \mbox{J.-M. Godard}, Phys.
  Rev. B 38 (1983), p. 7980.

\bibitem{Rouziere97}
S.~Rouzi{\`e}re, S.~Ravy, and \mbox{J.-P. Pouget}, Synthetic Metals 86 (1997),
  p. 2131.

\bibitem{Rouziere00}
S.~Rouzi{\`e}re, S.~Ravy, \mbox{J.-P. Pouget}, and S.~Brazovskii, Phys. Rev. B
  62 (2000), p. R16231.

\bibitem{Ravy06}
S.~Ravy, S.~Rouzi{\`e}re, \mbox{J.-P. Pouget}, S.~Brazovskii, J.~Marcus,
  \mbox{J.-F. B{\'e}rar}, and E.~Elkaim, Phys. Rev. B 74 (2006), p. 174102.

\bibitem{Brazovskii97}
S.~Brazovskii, \mbox{J.-P. Pouget}, S.~Ravy, and S.~Rouzi{\`e}re, Phys. Rev. B
  55 (1997), p. 3426.

\bibitem{Brill84}
J.W. Brill and W.~Roark, Phys. Rev. Lett. 53 (1984), p. 846.

\bibitem{Mozurkewich85}
G.~Mozurkewich, P.M. Chaikin, W.G. Clark, and G.~Gr{\"u}ner, Solid State
  Commun. 56 (1985), p. 421.

\bibitem{Xiang89}
\mbox{X.-D. Xiang} and J.W. Brill, Phys. Rev. B 39 (1989), p. 1290.

\bibitem{Xiang87}
\mbox{X.-D. Xiang} and J.W. Brill, Phys. Rev. B 36 (1987), p. 2969.

\bibitem{Brill86}
J.W. Brill, W.~Roark, and G.~Minton, Phys. Rev. B 33 (1986), p. 6831.

\bibitem{Tritt91}
T.M. Tritt, M.J. Skove, and A.C. Ehrlich, Phys. Rev. B 43 (1991), p. 9972.

\bibitem{Maclean92}
D.~Maclean, A.~Simpson, and M.H. Jericho, Phys. Rev. B 46 (1992), p. 12117.

\bibitem{Mozurkewich92}
\mbox{{\rm For a review: G. Mozurkewich in }``Perspectives in Physical
  Acoustics",} (1992).

\bibitem{Brill01}
\mbox{{\rm J.W. Brill in }``Handbook of Elastic Properties of Solids, Liquids
  and Gases"} (2001).

\bibitem{Mozurkewich90}
G.~Mozurkewich, Phys. Rev. B 42 (1990), p. 11183.

\bibitem{Maki87}
K.~Maki and A.~Virosztek, Phys. Rev. B 36 (1987), p. 6831; Synth. Met. 29
  (1989) p. F371.

\bibitem{Brown92}
S.E. Brown, B.~Alavi, G.~Gr{\"u}ner, and K.~Bartholomen, Phys. Rev. B 46
  (1992), p. 10483.

\bibitem{Hoen92}
S.~Hoen, B.~Burk, A.~Zettl, and M.~Inui, Phys. Rev. B 46 (1992), p. 1874.

\bibitem{Golovnya02}
A.V. Golovnya, \mbox{V.Ya. Pokrovskii}, and P.M. Shadrin, Phys. Rev. Lett. 88
  (2002), p. 246401.

\bibitem{Pokrovskii07}
\mbox{V.Ya. Pokrovskii}, S.G. Zybtsev, and I.G. Gorlova, Phys. Rev. Lett. 98
  (2007), p. 206404.

\bibitem{Day07}
C.~Day, Phys. Today 60(7) (2007), p.~24.

\bibitem{Zybtsev10}
S.G. Zybtsev, \mbox{V.Ya. Pokrovskii}, and S.V. Zaitsev-Zotov, Nature
  Communication 1 (2010), p. 85; Physica B 407 (2012) p. 1810.

\bibitem{Nichols09}
J.~Nichols, D.~Dominko, L.~Ladino, J.~Zhou, and J.W. Brill, Phys. Rev. B 79
  (2009), p. 241110.

\bibitem{Nichols10}
J.~Nichols, \mbox{C. Sandamali Weerasooriya}, and J.W. Brill, J. Phys.:
  Condens. Matter 22 (2010), p. 334224.

\bibitem{Anderson66}
P.W. Anderson, Rev. Mod. Phys. 38 (1966), p. 298.

\bibitem{Feinberg88}
D.~Feinberg and J.~Friedel, J. Physique 49 (1988), p. 485.

\bibitem{Feinberg89}
\mbox{D. Feinberg and J. Friedel, in \cite{R9Schlenker89}} , p. 407.

\bibitem{Dumas86}
J.~Dumas and D.~Feinberg, Europhys. Lett. 2 (1986), p. 555.

\bibitem{Bjelis89}
A.~Bjeli\u{s}, Physica Scripta T29 (1989), p.~62.

\bibitem{Gorkov84}
L.P. Gor'kov, JETP 59 (1984), p. 1057.

\bibitem{Langer67}
J.S. Langer and V.~Ambegaokar, Phys. Rev. 164 (1967), p. 498.

\bibitem{Ivlev78}
B.I. Ivlev and N.B. Kopnin, JETP Lett. 28 (1978), p. 592; J. Low Temp. Phys. 44
  (1980) p. 453.

\bibitem{Artemenko87}
S.N. Artemenko, A.F. Volkov, and A.N. Kruglov, Sov. Phys. JETP 64 (1987), p.
  906.

\bibitem{Batistic84}
I.~Batisti\'{c}, A.~Bjeli\u{s}, and L.P. Gor'kov, J. Physique 45 (1984), p.
  1049.

\bibitem{Jelcic91}
D.~Jelci{\'c} and A.~Bjeli\u{s}, Phys. Rev. B 43 (1991), p. 1735.

\bibitem{Ong84}
N.P. Ong, G.~Verma, and K.~Maki, Phys. Rev. Lett. 52 (1984), p. 663.

\bibitem{Ong85}
N.P. Ong and K.~Maki, Phys. Rev. B 32 (1985), p. 6582.

\bibitem{Gill01}
J.C. Gill, Phys. Rev. B 63 (2001), p. 125125.

\bibitem{Isakovic06}
A.F. Isakovic, P.G. Evans, J.~Kmetko, K.~Cicak, Z.~Cai, B.~Lai, and R.E.
  Thorne, Phys. Rev. Lett. 96 (2006), p. 046401.

\bibitem{Li99}
Y.~Li, S.G. Lemay, J.H. Price, K.~Cicak, K.~O'Neill, K.~Ringland, K.D.
  Finkelstein, J.D. Brock, and R.E. Thorne, Phys. Rev. Lett. 83 (1999), p.
  3514.

\bibitem{Neill04}
K.~O'Neill, K.~Cicak, and R.E. Thorne, Phys. Rev. Lett. 93 (2004), p. 066601.

\bibitem{LeBolloch05}
D.L. Bolloc'h, S.~Ravy, J.~Dumas, J.~Marcus, F.~Livet, C.~Detlefs, F.~Yakhou,
  and L.~Paolasini, Phys. Rev. Lett. 95 (2005), p. 116401.

\bibitem{Sutton91}
M.~Sutton, S.G.J. Mochrie, T.~Greytak, S.E. Nagler, L.E. Berman, G.A. Held, and
  G.B. Stephenson, Nature (London) 352 (1991), p. 608.

\bibitem{Ravy07}
S.~Ravy, D.L. Bolloc'h, R.~Currat, A.~Fluerasu, C.~Mocuta, and B.~Dkhil, Phys.
  Rev. Lett. 98 (2007), p. 105501.

\bibitem{Sutton02}
M.~Sutton, Y.~Li, J.D. Brock, and R.E. Thorne, J. Phys. IV (France) 12 (2002),
  pp. Pr9--3.

\bibitem{Jacques09}
\mbox{V.L.R. Jacques}, \mbox{D. Le Bolloc'h}, S.~Ravy, C.~Giles, F.~Livet, and
  S.B. Wilkins, Eur. Phys. J. B 70 (2009), p. 317.

\bibitem{LeBolloch08}
D.L. Bolloc'h, V.L.R. Jacques, N.~Kirova, J.~Dumas, S.~Ravy, J.~Marcus, and
  F.~Livet, Phys. Rev. Lett. 100 (2008), p. 096403.

\bibitem{Fawcett88}
E.~Fawcett, Rev. Modern Physics 60 (1988), p. 209.

\bibitem{Gill82}
J.C. Gill, Solid State Commun. 44 (1982), p. 1041.

\bibitem{Gorkov83}
L.P. Gor'kov, JETP Lett. 38 (1983), p.~87.

\bibitem{Gill96}
\mbox{J.C. Gill, in ref.~\cite{R10Schlenker96}} , p. 411.

\bibitem{Brazovskii91a}
S.N. Brazovskii and V.~Matveenko, J. Phys. (Paris) I 1 (1991), p. 269; 1173.

\bibitem{Brazovskii91b}
S.~Brazovskii and S.~Matveenko, JETP 72 (1966), p. 860; J. Phys. I (France) 2
  (1992) p. 409.

\bibitem{Saint-Lager88}
\mbox{M.-C. Saint-Lager}, thesis, University of Grenoble, unpublished  (1988).

\bibitem{Monceau86}
P.~Monceau, M.~Renard, J.~Richard, and \mbox{M.-C. Saint-Lager}, Physica B 143
  (1986), p.~64.

\bibitem{Mihaly83}
G.~Mih{\'a}ly, G.~Hutiray, and L.~Mih{\'a}ly, Phys. Rev. B 28 (1983), p. 4896.

\bibitem{Saint-Lager89}
\mbox{M.-C. Saint-Lager}, P.~Monceau, and M.~Renard, Europhys. Lett. 9 (1989),
  p. 585.

\bibitem{Verma84}
G.~Verma and N.P. Ong, Phys. Rev. B 30 (1984), p. 2928.

\bibitem{Zettl85}
A.~Zettl, K.~Kaiser, and G.~Gr{\"u}ner, Solid State Commun. 53 (1985), p. 649.

\bibitem{Mihaly84}
L.~Mih{\'a}ly and A.~J{\'a}nossy, Phys. Rev. B 30 (1984), p. 3530.

\bibitem{Lyding86}
J.W. Lyding, J.S. Hubacek, G.~Gammie, and R.E. Thorne, Phys. Rev. B 33 (1986),
  p. 4341.

\bibitem{Zhang86}
X.J. Zhang, N.P. Ong, and J.C. Eckert, Phys. Rev. Lett. 56 (1986), p. 1206.

\bibitem{Rideau01}
D.~Rideau, P.~Monceau, R.~Currat, H.~Requardt, \mbox{F.Ya. Nad}, J.E. Lorenzo,
  S.~Brazovskii, C.~Detlefs, and G.~Gr{\"u}bel, Europhys. Lett. 56 (2001), p.
  289; erratum.

\bibitem{Ramakrishna92}
S.~Ramakrishna, M.P. Maher, V.~Ambegaokar, and U.~Eckern, Phys. Rev. Lett. 68
  (1992), p. 2066.

\bibitem{Maher95}
M.P. Maher, T.L. Adelman, D.A. DiCarlo, J.P. McCarten, and R.E. Thorne, Phys.
  Rev. B 52 (1995), p. 13850.

\bibitem{Adelman96}
T.L. Adelman, \mbox{M.C. de Lind van Wijngaarden}, S.V. Zaitsev-Zotov,
  D.~DiCarlo, and R.E. Thorne, Phys. Rev. B 53 (1996), p. 1833.

\bibitem{Lemay98}
S.G. Lemay, \mbox{M.C. de Lind van Wijngaarden}, T.L. Adelman, and R.E. Thorne,
  Phys. Rev. B 57 (1998), p. 12781.

\bibitem{Brazovskii00}
S.~Brazovskii, N.~Kirova, H.~Requardt, \mbox{F.Ya. Nad}, P.~Monceau, R.~Currat,
  J.E. Lorenzo, G.~Gr{\"u}bel, and C.~Vettier, Phys. Rev. B 61 (2000), p.
  10640.

\bibitem{Requardt98a}
H.~Requardt, \mbox{F.Ya. Nad}, P.~Monceau, R.~Currat, J.E. Lorenzo,
  S.~Brazovskii, N.~Kirova, G.~Gr{\"u}bel, and C.~Vettier, Phys. Rev. Lett. 80
  (1998), p. 5631.

\bibitem{Rideau01b}
D.~Rideau, P.~Monceau, R.~Currat, H.~Requardt, \mbox{F.Ya. Nad}, J.E. Lorenzo,
  S.~Brazovskii, C.~Detlefs, and G.~Gr{\"u}bel, Nuclear Instruments and Methods
  in Physics Research A467-468 (2001), p. 1010.

\bibitem{Rideau01a}
D.~Rideau, Thesis, University of Grenoble, unpublished  (2001).

\bibitem{Requardt99}
H.~Requardt, \mbox{F.Ya. Nad}, P.~Monceau, R.~Currat, J.E. Lorenzo, D.~Rideau,
  D.~Smilgies, and G.~Gr{\"u}bel, J. Phys. IV (France) 9 (1999), pp. Pr10--133.

\bibitem{Requardt02}
H.~Requardt, D.~Rideau, R.~Danneau, A.~Ayari, \mbox{F.Ya. Nad}, J.E. Lorenzo,
  P.~Monceau, R.~Currat, \mbox{C. Detlefs, D. Smilgies}, and G.~Gr{\"u}bel, J.
  Phys. IV (France) 12 (2002), pp. Pr9--181.

\bibitem{Sweetland94}
E.~Sweetland, A.C. Finnefrock, W.J. Podulka, M.~Sutton, J.D. Brock, D.~DiCarlo,
  and R.E. Thorne, Phys. Rev. B 50 (1994), p. 8157.

\bibitem{Ringland99}
K.L. Ringland, A.C. Finnefrock, Y.~Li, J.D. Brock, S.G. Lemay, and R.E. Thorne,
  Phys. Rev. Lett. 82 (1999), p. 1923.

\bibitem{Skocpol74}
W.J. Skocpol, M.R. Beasley, and M.~Tinkham, J. Low Temp. Phys. 16 (1974), p.
  145.

\bibitem{Dolan77}
G.J. Dolan and L.D. Jackel, Phys. Rev. Lett. 39 (1977), p. 1628.

\bibitem{ESRF04}
P.~Monceau, \emph{Report of hs2030 experiment}, 2004. unpublished.

\bibitem{Itkis86}
M.E. Itkis, \mbox{F.Ya. Nad}, and V.~Pokrovskii, Sov. Phys. JETP 63 (1986), p.
  177.

\bibitem{Itkis95}
M.E. Itkis, B.M. Emerling, and J.W. Brill, Phy. Rev. B 52 (1995), p. R11545.

\bibitem{Bruisma80}
R.~Bruisma and S.E. Trullinger, Phys. Rev. B 22 (1980), p. 4543.

\bibitem{Orlov06}
A.P. Orlov, \mbox{Yu.I. Latyshev}, A.M. Smolovich, and P.~Monceau, JETP Lett.
  84 (2006), p.~89.

\bibitem{Kriza86}
G.~Kriza and G.~Mih{\'a}ly, Phys. Rev. Lett. 56 (1986), p. 2529.

\bibitem{Mihaly91}
G.~Mih{\'a}ly, Y.~Kim, and Gr{\"u}ner, Phys. Rev. Lett. 66 (1991), p. 2806.

\bibitem{Mihaly91a}
G.~Mih{\'a}ly, Y.~Kim, and Gr{\"u}ner, Phys. Rev. Lett. 67 (1991), p. 2713.

\bibitem{Turnhout75}
\mbox{J. van Turnhout}, \emph{Thermally Stimulated Discharge of Polymer
  Electrets}, Elsevier, Amsterdam, 1975.

\bibitem{Cava84a}
R.J. Cava, R.M. Fleming, E.A. Rietman, R.G. Dunn, and L.F. Schneemeyer, Phys.
  Rev. Lett. 53 (1984), p. 1677.

\bibitem{Staresinic99}
D.~Staresini{\v c}, K.~Biljakovi{\'c}, N.I. Baklanov, and S.V. Zaitsev-Zotov,
  Synth. Metals 103 (1999), p. 2610.

\bibitem{Fleming89}
\mbox{R.M. Fleming and R.J. Cava in ref. \cite{R9Schlenker89}} , p. 259.

\bibitem{Cava84b}
R.J. Cava, R.M. Fleming, P.B. Littlewood, E.A. Rietman, L.F. Schneemeyer, and
  R.G. Dunn, Phys. Rev. B 30 (1984), p. 3228.

\bibitem{Cava85}
R.J. Cava, R.M. Fleming, R.G. Dunn, and E.A. Rietman, Phys. Rev. B 31 (1985),
  p. 8325.

\bibitem{Cava86}
R.J. Cava, P.B. Littlewood, R.M. Fleming, R.G. Dunn, and E.A. Rietman, Phys.
  Rev. B 33 (1986), p. 2439.

\bibitem{Barisic87}
\mbox{S. Bari{\u s}i{\'c} in ref. \cite{R1Jerome77}} , p. 395.

\bibitem{Barisic89}
S.~Bari{\u s}i{\'c} and I.~Batistic, J. Physique France 50 (1989), p. 2717.

\bibitem{Baier90}
T.~Baier and W.~Wonnenberger, Z. Phys. B 79 (1990), p. 211.

\bibitem{Virosztek93}
A.~Virosztek and K.~Maki, Phys. Rev. B 48 (1993), p. 1368.

\bibitem{Wong87}
K.Y. Wong and S.~Takada, Phys. Rev. B 36 (1987), p. 5476.

\bibitem{Nakane85}
Y.~Nakane and S.~Takada, J. Phys. Soc. Jpn 54 (1985), p. 977.

\bibitem{Sneddon84}
L.~Sneddon, Phys. Rev. B 29 (1984), p. 719.

\bibitem{Takada85}
S.~Takada, K.Y.M. Wong, and T.~Holstein, Phys. Rev. B 32 (1985), p. 4639.

\bibitem{Biljakovic91a}
K.~Biljakovi{\'c}, \mbox{J.-C. Lasjaunias}, P.~Monceau, and F.~L{\'e}vy, Phys.
  Rev. Lett. 67 (1991), p. 1902.

\bibitem{Biljakovic93}
K.~Biljakovi{\'c}, in \textit{``Phase Transitions and Relaxation in Systems
  with Competing Energy Scales"}, NATO ASI Series C 415 (1993), p. 339.

\bibitem{Yang91}
J.~Yang and N.P. Ong, Phys. Rev. B 44 (1991), p. 7912.

\bibitem{Kriza91c}
G.~Kriza, Y.~Kim, A.~Beleznay, and G.~Mih{\'a}ly, Solid State Commun. 79
  (1991), p. 811.

\bibitem{Nad93a}
\mbox{F.Ya. Nad} and P.~Monceau, Solid State Commun. 87 (1993), p.~13.

\bibitem{Nad95a}
\mbox{F.Ya. Nad} and P.~Monceau, Phys. Rev. B 51 (1995), p. 2052.

\bibitem{Nad97}
\mbox{F.Ya. Nad} and P.~Monceau, JETP 84 (1997), p. 545.

\bibitem{Nad93b}
\mbox{F.Ya. Nad} and P.~Monceau, J. Physique IV, Colloque C2 3 (1993), p. 343.

\bibitem{Staresinic04}
D.~Staresini{\v c}, K.~Hasseini, W.~Briiting, K.~Biljakovi{\'c}, E.~Riedel, and
  \mbox{S. van Smaalen}, Phys. Rev. B 69 (2004), p. 113102.

\bibitem{Castaing91}
B.~Castaing and J.~Souletie, J. Phys. I (France) 1 (1991), p. 403.

\bibitem{Souletie94}
J.~Souletie, J. Appl. Phys. 75 (1994), p. 5513.

\bibitem{Ma81}
\mbox{S.K. Ma, in}, \emph{Modern Theory of Critical Phenomena}, Benjamin, New
  York, 1981.

\bibitem{Pytte87}
E.~Pytte and Y.~Imry, Phys. Rev. B 35 (1987), p. 1465.

\bibitem{Nad95b}
\mbox{F.Ya. Nad}, P.~Monceau, and K.~Bechgaard, Solid State Commun. 95 (1995),
  p. 655.

\bibitem{Littlewood88a}
P.B. Littlewood and R.~Rammal, Phys. Rev. B 38 (1988), p. 2665.

\bibitem{Erzan90}
A.~Erzan, E.~Veermans, R.~Heijungs, and L.~Pietronero, Phys. Rev. B 41 (1990),
  p. 11522.

\bibitem{Biljakovic98}
K.~Biljakovi{\'c}, D.~Staresini{\v c}, K.~Hosseinii, W.~Br{\"u}tting,
  H.~Berger, and F.~Levy, Physica B 244 (1998), p. 167.

\bibitem{Angell00}
C.A. Angell, K.L. Ngai, G.B. McKenna, P.F. McMillan, and S.W. Martin, J. Appl.
  Phys. 88 (2000), p. 3113.

\bibitem{Nad93c}
\mbox{F.Ya. Nad}, JETP Lett. 58 (1993), p. 111.

\bibitem{ZZ93}
\mbox{S. V. Zaitsev-Zotov}, Phys. Rev. Lett. 71 (1993), p. 605.

\bibitem{Hochli89}
U.T. H{\"o}chli and M.~Maglione, J. Phys.: Condens. Matter 1 (1989), p. 2241.

\bibitem{Math96}
C.~Math, W.~Br{\"u}tting, and W.~Riess, Europhys. Lett. 35 (1996), p. 221.

\bibitem{Staresinic02}
D.~Staresini{\v c}, K.~Biljakovi{\'c}, W.~Briiting, K.~Hasseini, P.~Monceau,
  H.~Berger, and F.~L{\'e}vy, Phys. Rev. B 65 (2002), p. 165109.

\bibitem{Larkin95}
A.I. Larkin and S.~Brazovskii, Solid State Commun. 93 (1995), pp. 275; Journal
  de Physique IV (France) 9 (1999) p. Pr10--77.

\bibitem{Larkin94}
A.I. Larkin, Sov. Phys. JETP 78 (1994), p. 971.

\bibitem{Volkov93}
A.F. Volkov, Phys. Letters A 182 (1993), p. 433.

\bibitem{Wonnenberger96}
W.~Wonnenberger, Solid State Commun. 97 (1996), p. 891.

\bibitem{Mihaly87}
G.~Mih{\'a}ly and P.~Beauch{\^e}ne, Solid State Commun. 63 (1987), p. 911.

\bibitem{Mihaly88c}
L.~Mih{\'a}ly, P.~Beauchene, T.~Chen, L.~Mih{\'a}ly, and G.~Gr{\"u}ner, Phys.
  Rev. B 37 (1988), p. 6536.

\bibitem{Mihaly88a}
G.~Mih{\'a}ly, P.~Beauch{\^e}ne, and J.~Marcus, Solid State Commun. 66 (1988),
  p. 149.

\bibitem{Tessema87}
G.X. Tessema and L.~Mih{\'a}ly, Phys. Rev. B 35 (1987), p. 7680.

\bibitem{Martin88}
S.~Martin, R.M. Fleming, and L.F. Schneemeyer, Phys. Rev. B 38 (1988), p. 5733.

\bibitem{Mihaly86}
L.~Mih{\'a}ly and G.X. Tessema, Phys. Rev. B 33 (1986), p. 5858.

\bibitem{Kim89}
Y.M. Kim, G.~Mih{\'a}ly, and G.~Gr{\"u}ner, Solid State Commun. 69 (1989), p.
  975.

\bibitem{Maeda90}
A.~Maeda, M.~Notomi, and K.~Uchinokura, Phys. Rev. B 42 (1990), p. 3290.

\bibitem{Ogawa05}
N.~Ogawa, K.~Miyano, and S.~Brazovskii, J. Phys. IV (France) 131 (2005), p.
  123; Phys. Rev. B 71 (2005) p. 075118.

\bibitem{Mihaly88b}
G.~Mih{\'a}ly, T.~Chen, T.W. Kim, and G.~Gr{\"u}ner, Phys. Rev. B 38 (1988), p.
  3602.

\bibitem{Chen88}
T.~Chen, L.~Mih{\'a}ly, and G.~Gr{\"u}ner, Phys. Rev. Lett. 60 (1988), p. 464.

\bibitem{Littlewood88b}
P.B. Littlewood, Solid State Commun. 65 (1988), p. 1347.

\bibitem{Littlewood89}
P.B. Littlewood, Synth. Metals 29 (1989), p. F531.

\bibitem{Zhilinskii83}
S.K. Zhilinskii, M.E. Itkis, I.U. Kalnova, \mbox{F.Ya. Nad}, and V.B.
  Preobrazhenskii, Sov. Phys.-JETP 58 (1983), p. 211.

\bibitem{Itkis91}
M.E. Itkis, \mbox{F.Ya. Nad}, and P.~Monceau, Synth. Met. 41-43 (1991), p.
  4037.

\bibitem{Itkis90}
M.E. Itkis, \mbox{F.Ya. Nad}, and P.~Monceau, J. Phys.: Condens. Matter 2
  (1990), p. 8327.

\bibitem{Nad92}
\mbox{F.Ya. Nad} and P.~Monceau, Phys. Rev. B 46 (1992), p. 7413.

\bibitem{Ivlev84}
B.I. Ivlev and N.B. Kopnin, Sov. Phys. Usp. 27 (1984), p. 206.

\bibitem{Borodin87}
D.V. Borodin, \mbox{S. V. Zaitsev-Zotov}, and \mbox{F.Ya. Nad}, Sov. Phys. JETP
  66 (1987), p. 793.

\bibitem{Gill86}
J.C. Gill, J. Phys. C 19 (1986), p. 6589.

\bibitem{ZZ97}
\mbox{S. V. Zaitsev-Zotov}, G.~Remenyi, and P.~Monceau, Phys. Rev. Lett. 78
  (1997), p. 1098.

\bibitem{Traetteberg92}
O.~Traetteberg, G.~Kriza, and G.~Mih{\'a}ly, Phys. Rev. B 45 (1992), p. 8795.

\bibitem{Caldeira81}
A.O. Caldeira and A.J. Leggett, Phys. Rev. Lett. 46 (1981), p. 211.

\bibitem{Larkin83}
A.I. Larkin and \mbox{Yu.N. Ovchinnikov}, JETP Lett. 37 (1983), p. 1085.

\bibitem{Larkin78}
A.I. Larkin and \mbox{Yu.N. Ovchinnikov}, Phys. Rev. B 17 (1978), p. 1586.

\bibitem{Miller12}
J.H. Miller, A.I. Wijesinghe, Z.~Tang, and A.M. Guloy, Phys. Rev. Lett. 108
  (2012), p. 036404.

\bibitem{Maki96}
K.~Maki, Ferroelectrics 176 (1996), p. 353.

\bibitem{Ovchinnikov96}
\mbox{Yu.N. Ovchinnikov}, K.~Biljakovi{\'c}, \mbox{J.-C. Lasjaunias}, and
  P.~Monceau, Europhys. Lett. 34 (1996), p. 645.

\bibitem{Hall88}
R.P. Hall, M.F. Hundley, and A.~Zettl, Phys. Rev. B 38 (1988), p. 13002\mbox{
  and references therein}.

\bibitem{Saint-Lager83}
\mbox{M.-C. Saint-Lager}, Thesis 3{\`e}me cycle, University of Grenoble,
  unpublished,  1983.

\bibitem{Li01}
Y.~Li, D.Y. Noh, J.H. Price, K.L. Ringland, J.D. Brack, S.G. Lemay, K.~Cicak,
  R.E. Thorne, and M.~Sutton, Phys. Rev. B 63 (2001), p. 041103.

\bibitem{Levy93}
J.~L{\'e}vy and M.S. Sherwin, Phys. Rev. B 48 (1993), p. 12223.

\bibitem{Hall86}
R.P. Hall, M.F. Hundley, and A.~Zettl, Phys. Rev. Lett. 56 (1986), p. 2399.

\bibitem{Adelman93}
T.L. Adelman, J.~McCarten, M.P. Maher, D.A. DiCarlo, and R.F. Thorne, Phys.
  Rev. B 47 (1993), p. 4033.

\bibitem{Lemay99}
S.G. Lemay, Y.L. R.~E.~Thorne, and J.D. Brock, Phys. Rev. Lett. 83 (1999), p.
  2793.

\bibitem{Thorne02a}
R.E. Thorne, K.~Cicak, K.O. Neill, and S.G. Lemay, J. Phys. IV (France) 12
  (2002), pp. Pr9--291.

\bibitem{Janssen07}
T.~Janssen, G.~Chapuis, and M.~\mbox{de~Boissieu,}, \emph{Aperiodic Crystals.
  From Modulated Phases to Quasicrystals}, University Press, Oxford, 2007.

\bibitem{Currat88}
R.~Currat and T.~Janssen, Solid State Physics - Advances in Research and
  Applications 41 (1988), p. 201.

\bibitem{deBoissieu08}
M.~\mbox{de~Boissieu}, Philosophical Magazine 88 (2008), p. 2295.

\bibitem{Aubry78}
\mbox{\rm S. Aubry, in} (1978).

\bibitem{Lefort96}
R.~Lefort, J.~Etrillard, B.~Toudi{\'c}, F.~Guillaume, T.~Breczewski, and
  P.~Bourges, Phys. Rev. Lett. 77 (1996), p. 4027.

\bibitem{Ollivier98}
J.~Ollivier, C.~Ecolivet, S.~Beaufils, F.~Guillaume, and T.~Beczewski,
  Europhys. Lett. 43 (1998), p. 546.

\bibitem{Currat02}
R.~Currat, E.~Kats, and I.~Luk'yanchuk, Eur. Phys. J. B 26 (2002), p. 339.

\bibitem{Poulet86}
H.~Poulet and \mbox{R.M. Pick in ref.~\cite{Blinc86}} Vol. 1 (1986), p. 315.

\bibitem{Currat89}
R.~Currat, Physica B 156-157 (1989), p.~1.

\bibitem{Currat00}
R.~Currat, Ferroelectrics 236 (2000), p.~11.

\bibitem{Bernard83}
L.~Bernard, R.~Currat, P.~Delamoye, C.M.E. Zeyen, S.~Hubert, and \mbox{R. de
  Kouchkovsky}, J. Phys. C: Solid State Phys. 16 (1983), p. 433.

\bibitem{Hubert81}
S.~Hubert, P.~Delamoye, S.~Lefrant, M.~Lepostollec, and M.~Hussonnais, J. Solid
  State Chem. 36 (1981), p.~36.

\bibitem{Hlinka02}
J.~Hlinka, J.~Petzelt, B.~Brezina, and R.~Currat, Phys. Rev. B 66 (2002), p.
  132302.

\bibitem{Denoyer80}
F.~Denoyer, A.H. Moudden, and M.~Lambert, Ferroelectrics 24 (1980), p.~43.

\bibitem{Denoyer86}
F.~Denoyer and \mbox{R. Currat in ref.~\cite{Blinc86}}  (1986), p. 129.

\bibitem{Cailleau80}
H.~Cailleau, F.~Moussa, C.M.E. Zeyen, and J.~Bouillot, Solid State Commun. 33
  (1980), p. 407.

\bibitem{Cailleau86}
\mbox{H. Cailleau in ref.~\cite{Blinc86}} tome 2 (1986), p.~71.

\bibitem{Moussa87}
F.~Moussa, P.~Launois, M.H. Lemee, and H.~Cailleau, Phys. Rev. B 36 (1987), p.
  8951.

\bibitem{Launois89}
P.~Launois, F.~Moussa, \mbox{M.-H. Lemee-Cailleau}, and H.~Cailleau, Phys. Rev.
  B 40 (1989), p. 5042.

\bibitem{Kurihara80}
Y.~Kurihara, J. Phys. Soc. Jpn 49 (1980), p. 852.

\bibitem{Artemenko89}
\mbox{S.N. Artemenko and A.F. Volkov in ref. \cite{R7Gorkov89}} , p. 365.

\bibitem{Ravy04}
S.~Ravy, H.~Requardt, D.L. Bolloc'h, \mbox{P. Foury-Leylikian}, \mbox{J.-P.
  Pouget}, R.~Currat, P.~Monceau, and M.~Krisch, Phys. Rev. B 69 (2004), p.
  115113.

\bibitem{Boriack78}
M.L. Boriack and A.W. Overhauser, Phys. Rev. B 18 (1978), p. 6454.

\bibitem{Kis99}
A.~Kis, D.~Pavici{\'c}, D.~Staresini{\v c}, K.~Biljakovi{\'c}, \mbox{J.-C.
  Lasjaunias}, and P.~Monceau, Fizika A (Zagreb) 8 (1999), p. 357.

\bibitem{Biljakovic86}
K.~Biljakovi{\'c}, \mbox{J.-C. Lasjaunias}, F.~Zougmore, P.~Monceau,
  F.~L{\'e}vy, L.~Bernard, and R.~Currat, Phys. Rev. Lett. 57 (1986), p. 1907.

\bibitem{Odin92}
J.~Odin, \mbox{J.-C. Lasjaunias}, A.~Berton, P.~Monceau, and K.~Biljakovi{\'c},
  Phys Rev. B 46 (1992), p. 1326.

\bibitem{Etrillard96}
J.~Etrillard, \mbox{J.-C. Lasjaunias}, K.~Biljakovi{\'c}, B.~Toudi{\'c}, and
  G.~Coddens, Phys. Rev. Lett. 76 (1996), p. 2334.

\bibitem{Yang99}
H.~Yang, \mbox{J.-C. Lasjaunias}, and P.~Monceau, J. Phys.: Condens. Matter 11
  (1999), p. 5083.

\bibitem{Lasjaunias02c}
\mbox{J.-C. Lasjaunias}, P.~Monceau, D.~Staresini{\v c}, and K.~Biljakovi{\'c},
  Eur. Phys. J. B 28 (2002), p. 187.

\bibitem{Odin01}
J.~Odin, \mbox{J.-C. Lasjaunias}, K.~Biljakovi{\'c}, K.~Hasselbach, and
  P.~Monceau, Eur. Phys. J. B 24 (2001), p. 315.

\bibitem{Dalhauser86}
K.~Dahlhauser, A.C. Anderson, and G.~Mozurkewich, Phys. Rev. B 34 (1986), p.
  4432.

\bibitem{Konate84}
K.~Konat{\'e}, Thesis 3e cycle, University Joseph Fourier, Grenoble,  1984,,
  unpublished.

\bibitem{Biljakovic11}
\mbox{K. Biljakovi{\'c} {\it et al.}}, \mbox{to be published}  (2012).

\bibitem{Etrillard00}
J.~Etrillard, \mbox{J.-C. Lasjaunias}, B.~Toudi{\'c}, F.~Guillaume, and
  T.~Breczewski, Europhys. Lett. 49 (2000), p. 610.

\bibitem{Requardt97}
H.~Requardt, R.~Currat, P.~Monceau, J.E. Lorenzo, A.J. Dianoux, \mbox{J.-C.
  Lasjaunias}, and J.~Marcus, J. Phys.: Condens. Matter 9 (1997), p. 8639.

\bibitem{Lorenzo02}
J.E. Lorenzo and H.~Requardt, Eur. Phys. J. B 28 (2002), p. 185.

\bibitem{Brown88}
S.E. Brown, J.O. Willis, B.~Allavi, and G.~Gr{\"u}ner, Phys. Rev. B 37 (1988),
  p. 6551.

\bibitem{Odin94}
J.~Odin, \mbox{J.-C. Lasjaunias}, K.~Biljakovi{\'c}, P.~Monceau, and
  K.~Bechgaard, Solid State Commun. 91 (1994), p. 523.

\bibitem{Lasjaunias94}
\mbox{J.-C. Lasjaunias}, K.~Biljakovi{\'c}, \mbox{F.Ya. Nad}, P.~Monceau, and
  K.~Bechgaard, Phys. Rev. Lett. 72 (1994), p. 1283.

\bibitem{Yang00}
H.~Yang, \mbox{J.-C. Lasjaunias}, and P.~Monceau, J. Phys.: Condens. Matter 12
  (2000), p. 7183.

\bibitem{Eldridge85}
J.E. Eldridge and G.S. Bates, Mol. Cryst. Liq. Cryst. 119 (1985), p. 183.

\bibitem{Ng83}
H.K. Ng, T.~Timusk, and K.~Bechgaard, J. Phys. Coll. 44 C3 (1983), p. 867.

\bibitem{Challener83}
W.A. Challener, P.L. Richards, and R.L. Greene, J. Phys. Coll. 44 C3 (1983), p.
  873.

\bibitem{Genensky57}
S.M. Genensky and G.F. Newell, J. Chem. Physics 26 (1957), p. 486.

\bibitem{Nicklow72}
R.~Nicklow, N.~Wakabayashi, and H.G. Smith, Phys. Rev. B 5 (1972), p. 4951.

\bibitem{Lasjaunias82}
\mbox{J.-C. Lasjaunias} and P.~Monceau, Solid State Commun. 41 (1982), p. 911.

\bibitem{Biljakovic91b}
K.~Biljakovi{\'c}, \mbox{J.-C. Lasjaunias}, and P.~Monceau, Phys. Rev. B 43
  (1991), p. 3117.

\bibitem{Lasjaunias99}
\mbox{J.-C. Lasjaunias}, K.~Biljakovi{\'c}, D.~Staresini{\v c}, P.~Monceau,
  S.~Takasaki, J.~Yamada, \mbox{S.-I. Nakatsuji}, and H.~Anzai, Eur. Phys. J. B
  7 (1999), p. 541.

\bibitem{Phillips96}
\mbox{For a review on Non Exponential Relaxation in Disorderd Systems, see J.C.
  Phillips}, Rep. Prog. Phys. 59 (1996), p. 1133.

\bibitem{Lasjaunias02b}
\mbox{J.-C. Lasjaunias}, P.~Monceau, D.~Staresini{\v c}, K.~Biljakovi{\'c},
  C.~Carcel, and \mbox{J.-M. Fabre}, J. Phys.: Condens. Matter 14 (2002), p.
  8583.

\bibitem{Kleiman87}
R.N. Kleiman, G.~Agnolet, and D.J. Bishop, Phys. Rev. Lett. 59 (1987), p. 2079.

\bibitem{Liu98}
L.~Xiao, P.D. Vu, R.O. Pohl, F.~Schiettekatte, and S.~Roorda, Phys. Rev. Lett.
  81 (1998), p. 3171.

\bibitem{Phillips81}
\emph{Amorphous Solids, Low Temperature Properties}, W.A. Phillips ed.,
  Springer Verlag, Berlin, 1981.

\bibitem{Hunklinger86}
{\rm S. Hunklinger and A. K. Raychandhari,} (1986).

\bibitem{Biljakovic89c}
K.~Biljakovi{\'c}, \mbox{J.-C. Lasjaunias}, P.~Monceau, and F.~L{\'e}vy,
  Europhys. Letters 8 (1989), p. 771.

\bibitem{Lasjaunias97}
\mbox{J.-C. Lasjaunias}, P.~Monceau, D.~Staresini{\v c}, K.~Biljakovi{\'c}, and
  \mbox{J.-M. Fabre}, J. Phys. I (France) 7 (1997), p. 1417.

\bibitem{Biljakovic03}
K.~Biljakovi{\'c}, M.~Miljak, D.~Staresini{\v c}, \mbox{J.-C. Lasjaunias},
  P.~Monceau, H.~Berger, and F.~L{\'e}vy, Europhys. Lett. 62 (2003), p. 554.

\bibitem{Biljakovic89b}
K.~Biljakovi{\'c}, \mbox{J.-C. Lasjaunias}, and P.~Monceau, Synth. Metals 29
  (1989), p. F289.

\bibitem{Nomura93}
K.~Nomura, N.~Keitoku, T.~Shimizu, T.~Sambongi, M.~Tokumoto, N.~Kinoshita, and
  H.~Anzai, J. Physique IV C2 (1993), p.~21.

\bibitem{Nomura95}
K.~Nomura, Y.~Hasokawa, N.~Matsunaga, M.~Nagasawa, T.~Sambongi, and H.~Anzai,
  Synth. Metals 70 (1995), p. 1295.

\bibitem{Kriza91d}
G.~Kriza, G.~Quirion, O.~Traetteberg, and D.~J{\'e}r{\^o}me, Europhys. Lett. 16
  (1991), p. 585.

\bibitem{Wonnenberger91}
W.~Wonnenberger, Solid State Commun. 80 (1991), p. 953.

\bibitem{Jones71}
G.O. Jones, \emph{\mbox{in }Glass\mbox{\rm , Chap. 4}}, Chapman and Hall eds,
  Berlin, 1971.

\bibitem{Coroneus93}
J.~Coroneus, B.~Alavi, and S.E. Brown, Phys. Rev. Lett. 70 (1993), p. 2332.

\bibitem{Kobayashi98}
N.~Kobayashi, M.~Ogata, and K.~Yonemitsu, J. Phys. Soc. Jpn 67 (1998), p. 1098.

\bibitem{Yoshioka01}
H.~Yoshioka, M.~Tsuchiizu, and Y.~Suzumura, J. Phys. Soc. Jpn 70 (2001), p.
  762.

\bibitem{Biljakovic89a}
K.~Biljakovi{\'c}, \mbox{J.-C. Lasjaunias}, P.~Monceau, and J.~L{\'e}vy, Phys.
  Rev. Lett. 62 (1989), p. 1512.

\bibitem{Lundgren83}
L.~Lundgren, P.~Svedlindk, P.~Nordblad, and O.~Beckman, Phys. Rev. Lett. 51
  (1983), p. 911.

\bibitem{Campbell90}
\emph{Relaxation in Complex Systems and Related Topics}, NATO Advanced Study
  Institutes, Ser. B, Vol. 222, I.A. Campbell and C. Giovanella eds, Plenum
  Press, New York, 1990.

\bibitem{Vincent72}
\mbox{E. Vincent, J. Hammann, and M. Ocio,}, \emph{Recent Progress in Random
  Magnets}, World Scientific, Singapore, 1972.

\bibitem{Monceau94}
P.~Monceau, \mbox{J.-C. Lasjaunias}, and K.~Biljakovi{\'c}, Physica B 194-196
  (1994), p. 403.

\bibitem{Lasjaunias96}
\mbox{J.-C. Lasjaunias}, K.~Biljakovi{\'c}, and P.~Monceau, Phys. Rev. B 53
  (1996), p. 7699.

\bibitem{Lasjaunias02a}
\mbox{J.-C. Lasjaunias}, \mbox{J.-P. Brison}, P.~Monceau, D.~Staresini{\v c},
  K.~Biljakovi{\'c}, C.~Carcel, and \mbox{J.-M. Fabre}, J. Phys.: Condens.
  Matter 14 (2002), p. 837.

\bibitem{Biljakovic09}
K.~Biljakovi{\'c}, \mbox{J.-C. Lasjaunias}, R.~Melin, P.~Monceau, G.~Remenyi,
  S.~Sahling, and D.~Staresini{\v c}, Synth. Metals 159 (2009), p. 2402.

\bibitem{Lasjaunias05a}
\mbox{J.-C. Lasjaunias}, R.~Melin, D.~Staresini{\v c}, K.~Biljakovi{\'c}, and
  J.~Souletie, Phy. Rev. Lett. 94 (2005), p. 245701.

\bibitem{Lasjaunias05b}
\mbox{J.-C. Lasjaunias}, K.~Biljakovi{\'c}, S.~Sahling, and P.~Monceau, J.
  Phys. I (France) 131 (2005), p. 193.

\bibitem{Lasjaunias05c}
\mbox{J.-C. Lasjaunias}, S.~Sahling, K.~Biljakovi{\'c}, P.~Monceau, and
  J.~Marcus, J. Magn. Magn. Mat. 290-291 (2005), p. 989.

\bibitem{Melin05}
R.~Melin, K.~Biljakovi{\'c}, and \mbox{J.-C. Lasjaunias}, Eur. Phys. J. B 43
  (2005), p. 489.

\bibitem{Melin02}
R.~Melin, K.~Biljakovi{\'c}, \mbox{J.-C. Lasjaunias}, and P.~Monceau, Eur.
  Phys. J. B 26 (2002), p. 417.

\bibitem{Bouchaud92}
\mbox{J.-P. Bouchaud}, J. Phys. I (France) 2 (1992), pp. 1705; \mbox{J.--P.
  Bouchaud}, E. Vincent, and J. Hamman, J. Phys. I (France) 4 (1995) p.~139; C.
  Monthus and \mbox{J.--P. Bouchaud}, J. Phys. A 29 (1996) p.~3847.

\bibitem{Fabrizio97}
M.~Fabrizio and R.~Melin, Phys. Rev. Lett. 78 (1997), p. 3382; Phys. Rev. B 56
  (1997) p. 5996.

\bibitem{Mozurkewich87}
G.~Mozurkewich and L.~Mih{\'a}ly, Phys. Rev. B 36 (1987), p. 6164.

\bibitem{Artemenko07}
S.N. Artemenko and T.~Nattermann, Phys. Rev. Lett. 99 (2007), p. 256401.

\bibitem{Artemenko05}
S.N. Artemenko and S.V. Remizov, Phys. Rev. B 72 (2005), p. 125118.

\bibitem{Artemenko09}
S.N. Artemenko, S.V. Remizov, D.S. Shapiro, and R.R. Vakhitov, Physica B 404
  (2009), p. 447.

\bibitem{Sahling07}
S.~Sahling, \mbox{J.-C. Lasjaunias}, R.~Melin, P.~Monceau, and G.~Remenyi, Eur.
  Phys. J. B 59 (2007), p.~9.

\bibitem{Melin06}
R.~Melin, \mbox{J.-C. Lasjaunias}, S.~Sahling, G.~Remenyi, and
  K.~Biljakovi{\'c}, Phys. Rev. Lett. 97 (2006), p. 227203.

\bibitem{Lasjaunias03}
\mbox{J.-C. Lasjaunias}, S.~Sahling, K.~Biljakovi{\'c}, and P.~Monceau, J. Low
  Temp. Phys. 130 (2003), p.~25.

\bibitem{Lasjaunias02d}
\mbox{J.-C. Lasjaunias}, S.~Sahling, K.~Biljakovi{\'c}, and P.~Monceau, J.
  Phys. IV (France) 12 (2002), pp. Pr9--27.

\bibitem{Sahling03}
S.~Sahling, \mbox{J.-C. Lasjaunias}, K.~Biljakovi{\'c}, and P.~Monceau, J. Low
  Temp. Phys. 133 (2003), p. 273.

\bibitem{Degiorgi95}
L.~Degiorgi, S.~Thieme, B.~Alavi, G.~Gr{\"u}ner, R.H. McKenzie, K.~Kim, and
  F.~L{\'e}vy, Phys. Rev. B 52 (1995), p. 5603.

\bibitem{Schwartz95}
\mbox{A. Schwartz, M. Dressel, B. Alavi, A. Blank, S. Dubois, G. Gr{\"u}ner,
  B.P. Gorshunov, A.A. Volkov}, G.V. Kozlov, S.~Thieme, L.~Degiorgi, and
  F.~L{\'e}vy, Phys. Rev. B 52 (1995), p. 5643.

\bibitem{Berner93}
D.~Berner, G.~Scheiber, A.~Gaymann, H.M. Geserich, P.~Monceau, and F.~L{\'e}vy,
  J. Phys. IV (France) C2 3 (1993), p. 255.

\bibitem{Creager91}
W.N. Creager, P.L. Richards, and A.~Zettl, Phys. Rev. B 44 (1991), p. 3505.

\bibitem{Rice76}
M.J. Rice, Phys. Rev. Lett. 37 (1976), p.~36.

\bibitem{Rice77}
M.J. Rice, L.~Pietronero, and P.~Br{\"u}esch, Solid State Commun. 21 (1977), p.
  757.

\bibitem{Rice78}
M.J. Rice, Solid State Commun. 25 (1978), p. 1083.

\bibitem{Rice75}
M.J. Rice, C.B. Duke, and N.O. Lipari, Solid State Commun. 17 (1975), p. 1089.

\bibitem{Kim93}
K.~Kim, R.H. McKenzie, and J.W. Wilkins, Phys. Rev. Lett. 71 (1993), p. 4015.

\bibitem{Sagar07}
D.M. Sagar, A.A. Tsvetkov, D.~Fausti, \mbox{S. van Smallen}, and \mbox{P.H.M.
  van Loosdrecht}, J. Phys.: Condens. Matter 19 (2007), p. 346208.

\bibitem{Shimatake06}
K.~Shimatake, Y.~Toda, and S.~Tanda, Phys. Rev. B 73 (2006; 75 (2007) p.
  115120), p. 153403.

\bibitem{Demsar99}
J.~Demsar, K.~Biljakovi{\'c}, and D.~Mihailovic, Phys. Rev. Lett. 83 (1999), p.
  800.

\bibitem{Tomeljak09}
A.~Tomeljak, H.~Sch{\"a}fer, D.~St{\"a}dter, M.~Beyer, K.~Biljakovi{\'c}, and
  J.~Demsar, Phys. Rev. Lett. 102 (2009), p. 066404.

\bibitem{Schafer10}
H.~Sch{\"a}fer, V.K. Kabanoo, M.~Beyer, K.~Biljakovi{\'c}, and J.~Demsar, Phys.
  Rev. Lett. 105 (2010), p. 066402.

\bibitem{Watanabe09}
S.~Watanabe, R.~Kondo, S.~Kagoshima, and R.~Shimano, Phys. Rev. B 80 (2009), p.
  220408.

\bibitem{Sagar08}
D.M. Sagar, D.~Fausti, S.~Yue, C.A. Kuntscher, \mbox{S. van Smaalen}, and
  \mbox{P.H.M. van Loosdretch}, New J. Phys. 10 (2008), p. 023043.

\bibitem{Yusupov10}
R.~Yusupov, T.~Mertelj, V.V. Kakanov, S.~Brazovskii, P.~Kusar, \mbox{J.-H.
  Chu}, I.R. Fisher, and D.~Mihailovic, Nature Phys. 6 (2010), p. 681.

\bibitem{Averitt10}
R.D. Averitt, Nature Phys. 6 (2010), p. 639.

\bibitem{Degiorgi04c}
\mbox{L. Degiorgi, in ref.~\cite{R11Baeriswyl04}}  (2004), p. 165.

\bibitem{Dressel96}
M.~Dressel, A.~Schwartz, G.~Gr{\"u}ner, and L.~Degiorgi, Phys. Rev. Lett. 77
  (1996), p. 398.

\bibitem{Vescoli00}
V.~Vescoli, F.~Zweck, W.~Henderson, L.~Degiorgi, M.~Grioni, G.~Gr{\"u}ner, and
  L.K. Montgomery, Eur. Phys J. B 13 (2000), p. 503.

\bibitem{Schwartz98}
A.~Schwartz, M.~Dressel, G.~Gr{\"u}ner, V.~Vescoli, L.~Degiorgi, and
  T.~Giamarchi, Phys. Rev. B 58 (1998), p. 1261.

\bibitem{Schonhammer02}
K.~Sch{\"o}nhammer, J. Phys.: Condens. Matter 14 (2002), p. 12783.

\bibitem{Giamarchi97}
T.~Giamarchi, Physica B 230-232 (1997), p. 975.

\bibitem{Degiorgi96}
L.~Degiorgi, M.~Dressel, S.~Schwartz, B.~Alavi, and G.~Gr{\"u}ner, Phys. Rev.
  Lett. 76 (1996), p. 398.

\bibitem{Vescoli99}
V.~Vescoli, L.~Degiorgi, M.~Dressel, A.~Schwartz, W.~Henderson, B.~Alavi,
  G.~Gr{\"u}ner, J.~Brinckmann, and A.~Virosztek, Phys. Rev. B 60 (1999), p.
  8019.

\bibitem{Grioni04}
\mbox{M. Grioni, in ref.~\cite{R11Baeriswyl04}}  (2004), p. 137.

\bibitem{Grioni09}
\mbox{For the more recent review, see M. Grioni, S. Pons, and E.
  Frantzeskakis}, J. Phys.: Condens. Matter 21 (2009), p. 023201.

\bibitem{Dardel91}
B.~Dardel, D.~Malterre, M.~Grioni, P.~Weibel, Y.~Baer, and F.~L{\'e}vy, Phys.
  Rev. Lett. 67 (1991), p. 3144.

\bibitem{Dardel93}
B.~Dardel, D.~Malterre, M.~Grioni, P.~Weibel, Y.~Baer, J.~Voit, and
  D.~J{\'e}r{\^o}me, Europhys. Letters 24 (1993), p. 687.

\bibitem{Hwu92}
Y.~Hwu, P.~Almeras, M.~Marsi, H.~Berger, F.~L{\'e}vy, M.~Grioni, D.~Malterre,
  and G.~Margaritondo, Phys. Rev. B 46 (1992), p. 13624.

\bibitem{Gweon01}
G.H. Gweon, J.D. Denlinger, J.W. Allen, R.~Claessen, C.G. Olson, H.~H{\"o}chst,
  J.~Marcus, C.~Schlenker, and L.F. Schneemeyer, J. Elect. Spect. and Rel.
  Phenomena 117-118 (2001), p. 481.

\bibitem{Zwick97}
F.~Zwick, S.~Brown, G.~Margaritondo, C.~Merli, M.~Onellion, J.~Voit, and
  M.~Grioni, Phys. Rev. Lett. 79 (1997), p. 3982.

\bibitem{Perfetti02b}
L.~Perfetti, S.~Mitrovic, and M.~Grioni, J. Elect. Spect. and Related Phenomena
  127 (2002), p.~77.

\bibitem{Castellani94}
C.~Castellani, \mbox{C. Di Castro}, and W.~Metzner, Phys. Rev. Lett. 72 (1994),
  p. 316.

\bibitem{Perfetti02a}
L.~Perfetti, S.~Mitrovic, G.~Margaritondo, M.~Grioni, L.~Forro, L.~Degiorgi,
  and H.~H{\"o}chst, Phys. Rev. B 66 (2002), p. 075107.

\bibitem{Perfetti01}
L.~Perfetti, H.~Berger, A.~Reginelli, L.~Degiorgi, H.~H{\"o}chst, J.~Voit,
  G.~Margaritondo, and M.~Grioni, Phys. Rev. Lett. 87 (2001), p. 216404.

\bibitem{Aubry89}
\mbox{S. Aubry and P. Quemerais, in ref.~\cite{R9Schlenker89}}  (1989), p. 295.

\bibitem{Aubry92}
S.~Aubry, G.~Abramovici, and \mbox{J.-L. Raimbault}, J. Stat. Phys. 67 (1992),
  p. 675.

\bibitem{Raimbault95}
\mbox{J.-L. Raimbault} and S.~Aubry, J. Phys.: Cond. Matter 7 (1995), p. 8287.

\bibitem{Aubry93}
S.~Aubry, J. Physique IV, Colloque C2 3 (1993), p. 349.

\bibitem{Brooks08}
J.S. Brooks, Rep. Prog. Phys. 71 (2008), p. 126501.

\bibitem{Gorkov84b}
L.P. Gor'kov and A.G. Lebed, J. Physique Lett. 45 (1984), p. 433.

\bibitem{Virosztek86}
A.~Virosztek, L.~Chen, and K.~Maki, Phys. Rev. B 34 (1986), p. 3371.

\bibitem{Heritier84}
M.~Heritier, G.~Montambaux, and P.~Lederer, J. Phys. Lett. 45 (1984), p. L943.

\bibitem{Lebed08}
\mbox{A. Lebed in ref. \cite{R14Lebed08}}  (2008), p. 127.

\bibitem{Chaikin96}
P.M. Chaikin, J. Physique I 6 (1996), p. 1875.

\bibitem{Montambaux91}
G.~Montambaux, Physica Scripta T35 (1991), p. 188.

\bibitem{Lebed04}
A.G. Lebed, N.N. Bamet, and M.J. Naughton, Phys. Rev. Lett. 93 (2004), p.
  157006.

\bibitem{Montambaux85}
G.~Montambaux, H.~H\'eritier, and P.~Lederer, Phys. Rev. Lett. 55 (1985), p.
  2078.

\bibitem{Naughton85}
M.J. Naughton, J.S. Brooks, L.Y. Chiang, R.V. Chamberlain, and P.M. Chaikin,
  Phys. Rev. Lett. 55 (1985), p. 969.

\bibitem{Pesty85}
F.~Pesty, P.~Garoche, and K.~Bechgaard, Phys. Rev. Lett. 55 (1985), p. 2495.

\bibitem{Danner96}
G.M. Danner, P.M. Chaikin, and S.T. Hannahs, Phys. Rev. B 53 (1996), p. 2727.

\bibitem{Chandrasekhar62}
B.S. Chandrasekhar, Appl. Phys. Lett. 1 (1962), p.~7.

\bibitem{Clogston62}
A.M. Clogston, Phys. Rev. Lett. 9 (1962), p. 266.

\bibitem{Dieterich73}
W.~Dieterich and P.~Fulde, Z. Physik A 265 (1973), p. 239.

\bibitem{Bonfait91}
G.~Bonfait, E.B. Lopes, M.J. Matos, R.T. Henriques, and M.~Almeida, Solid State
  Commun. 80 (1991), p. 391.

\bibitem{Matos96}
M.~Matos, G.~Bonfait, R.T. Henriques, and A.~Almeida, Phys. Rev. B 54 (1996),
  p. 15307.

\bibitem{McDonald04}
R.D. McDonald, N.~Harrison, L.~Balicas, K.H. Kim, J.~Singleton, and X.~Chi,
  Phys. Rev. Lett. 93 (2004), p. 076405.

\bibitem{Zanchi96}
D.~Zanchi, A.~Bjeli\u{s}, and G.~Montambaux, Phys. Rev. B 53 (1996), p. 1240.

\bibitem{Osada06}
T.~Osada and E.~Ohmichi, J. Phys. Soc. Jpn 75 (2006), p. 051006.

\bibitem{Bjelis99}
A.~Bjeli\u{s}, D.~Zanchi, and G.~Montambaux, J. Phys. IV (France) 9 (1999), pp.
  Pr10--203.

\bibitem{Fulde64}
P.~Fulde and A.~Ferrel, Phys. Rev. 135 (1964), p. A550.

\bibitem{Larkin64}
A.I. Larkin and \mbox{Yu.N. Ovchinnikov}, Sov. Phys. JETP 20 (1965), p. 762.

\bibitem{Grigoriev05}
P.D. Grigoriev and D.S. Lyubshin, Phys. Rev. B 72 (2005), p. 195106.

\bibitem{Brazovskii81}
S.A. Brazovskii, I.E. Dzlalovshinskii, and N.N. Kirova, Soviet Phys. JETP 54
  (1981), p. 1209.

\bibitem{Brazovskii84b}
S.~Brazovskii and N.~Kirova, Soviet Scientific Reviews\mbox{,} Sect. A 5
  (1984), p.~99.

\bibitem{Brazovskii84}
S.A. Brazovskii and S.F. Matveenko, Zh. Eksp. Teor. Fiz. 87 (1984), p. 1400.

\bibitem{Bjelis86}
A.~Bjeli\u{s} and S.~Bari{\u s}i{\'c}, J. Phys. C: Solid State Phys. 19 (1986),
  p. 5607.

\bibitem{Boucher96}
J.P. Boucher and L.P. Regnault, J. Phys. I (France) 6 (1996), p. 1939.

\bibitem{Palme96}
W.~Palme, G.~Ambert, \mbox{J.-P. Boucher}, G.~Dhalenne, and A.~Revcolevski, J.
  Appl. Phys. 79 (1996), p. 5384.

\bibitem{Meurdesoif99}
Y.~Meurdesoif and A.~Buzdin, Phys. Rev. B 59 (1999), p. 11165.

\bibitem{Fagot96}
Y.~Fagot-Revurat, M.~Horvatic, C.~Berthier, P.~S{\'e}gransan, G.~Dhalenne, and
  A.~Revcolevski, Phys. Rev. Lett. 77 (1996), p. 1861.

\bibitem{Horvatic99}
M.~Horvatic, Y.~Fagot-Revurat, C.~Berthier, G.~Dhalenne, and A.~Revcoleski,
  Phys. Rev. Lett. 83 (1999), p. 420.

\bibitem{Kiryukhin96}
V.~Kiryukhin, B.~Keimer, \mbox{J.-P. Hill}, and A.~Vigliante, Phys. Rev. Lett.
  76 (1996), p. 4608.

\bibitem{Coleman90}
R.V. Coleman, M.P. Everson, H.A. Lee, and A.~Johnson, Phys. Rev. B 41 (1990),
  p. 460.

\bibitem{Balseiro85}
C.A. Balseiro and L.M. Falicov, Phys. Rev. Lett. 55 (1985), p. 2236; Phys. Rev.
  B 34 (1986) p. 863.

\bibitem{Richard87}
J.~Richard, P.~Monceau, and M.~Renard, Phys. Rev. B 35 (1987), p. 4533.

\bibitem{Monceau78}
P.~Monceau and A.~Briggs, J. Phys. C: Solid State Phys. 11 (1978), p. L465.

\bibitem{Audouard93}
A.~Audouard, J.~Richard, S.~Dubois, \mbox{J.-P. Ulmet}, and S.~Askenazy, Synth.
  Metals 55-57 (1993), p. 2629.

\bibitem{Everson87}
M.P. Everson, A.~Johnson, \mbox{H-A Lu}, R.V. Coleman, and L.M. Falicov, Phys.
  Rev. B 36 (1987), p. 6953.

\bibitem{Sowa85}
E.C. Sowa and L.M. Falicov, Phys. Rev. B 32 (1985), p. 755.

\bibitem{Monceau88}
P.~Monceau and J.~Richard, Phys. Rev. B 37 (1988), p. 7982.

\bibitem{Kiryukhin98}
V.~Kiryukhin, D.~Casa, B.~Keimer, J.P. Hill, M.J. Higgins, and S.~Bhattacharya,
  Phys. Rev. B 57 (1998), p. 1332.

\bibitem{ESRF07}
ESRF (Grenoble) HE - 2420 report  (2007).

\bibitem{Monceau87b}
P.~Monceau, J.~Richard, and O.~Laborde, Synth. Metals 19 (1987), p. 801.

\bibitem{Lebed03}
A.G. Lebed, JETP Lett. 78 (2003), p. 138.

\bibitem{Andres03}
D.~Andres, M.V. Kartsovnik, P.D. Grigoriev, W.~Biberacher, and H.~M{\"u}ller,
  Phys. Rev. B 68 (2003), p. 201101.

\bibitem{Graf05}
D.~Graf, E.S. Choi, J.S. Brooks, J.C. Dias, R.T. Henriques, M.~Almeida,
  M.~Matos, and D.~Rickel, Synth. Metals 153 (2005), p. 361.

\bibitem{Lebed09}
A.G. Lebed, Phys. Rev. Lett. 103 (2009), p. 046401.

\bibitem{Lebed07}
A.G. Lebed, Phys. Rev. Lett. 99 (2007), p. 026402.

\bibitem{Harrison00}
N.~Harrison, L.~Balicas, J.S. Brooks, and M.~Tokumoto, Phys. Rev. B 62 (2000),
  p. 14212.

\bibitem{Kartsovnik09}
M.V. Kartsovnik, D.~Andres, W.~Biberacher, and H.~M{\"u}ller, Physica B 404
  (2009), p. 357.

\bibitem{Andres01}
D.~Andres, M.V. Kartsovnik, W.~Biberacher, H.~Weiss, E.~Balthes, H.~M{\"u}ller,
  and N.~Kushch, Phys. Rev. B 64 (2001), p. 161104.

\bibitem{Latyshev07}
\mbox{Yu.I. Latyshev}, P.~Monceau, A.P. Orlov, S.A. Brazovskii, and
  T.~Fournier, Supercond. Sci. Technol. 20 (2007), p. 587.

\bibitem{Orlov08}
A.P. Orlov, \mbox{Yu.I. Latyshev}, D.~Vignolles, and P.~Monceau, JETP Lett. 87
  (2008), p. 433.

\bibitem{Latyshev09}
\mbox{Yu.I. Latyshev}, A.P. Orlov, \mbox{A.Yu. Latyshev}, \mbox{A.-M.
  Smolovich}, P.~Monceau, and D.~Vignolles, Physica B 404 (2009), p. 399.

\bibitem{Latyshev11}
\mbox{Yu.I. Latyshev} and A.P. Orlov, JETP Letters 94 (2011), p. 481.

\bibitem{Guyot05}
H.~Guyot, J.~Dumas, J.~Marcus, C.~Schlenker, and D.~Vignolles, J. Phys. IV
  (France) 131 (2005), p. 261.

\bibitem{Balaska05}
H.~Balaska, J.~Dumas, H.~Guyot, P.~Mallet, J.~Marcus, C.~Schlenker, \mbox{J.-Y.
  Vuillen}, and D.~Vignolles, Solid State Sciences 7 (2005), p. 690.

\bibitem{Zant96}
\mbox{H.S. J. van der Zant}, O.C. Mantel, C.~Dekker, J.E. Mooij, and
  C.~Traeholt, Appl. Phys. Lett. 68 (1996), p. 3823.

\bibitem{Chrisey94}
\mbox{\rm For a review, see \textit{Pulse laser deposition of thin films}}
  (1994).

\bibitem{Mantel97}
O.C. Mantel, \mbox{H.S.J. van der Zant}, A.J. Steinfort, C.~Dekker,
  C.~Traeholt, and H.W. Zandberger, Phys. Rev. B 55 (1997), p. 4817.

\bibitem{Steinfort98}
A.J. Steinfort, \mbox{H.S.J. van der Zant}, A.B. Smits, O.C. Mantel, P.M.L.O.
  Scholte, and C.~Dekker, Phys. Rev. B 57 (1998), p. 12530.

\bibitem{Mantel99a}
O.C. Mantel, C.A.W. Bal, C.~Langezaal, C.~Dekker, and \mbox{H.S. J. van der
  Zant}, J. Applied Physics 86 (1999), p. 4440.

\bibitem{Mantel99b}
O.C. Mantel, C.A.W. Bal, C.~Langezaal, C.~Dekker, and \mbox{H.S.J. van der
  Zant}, Phys. Rev. B 60 (1999), p. 5287.

\bibitem{Dominko11}
\mbox{D. Dominko, D. Staresini{\v c}, K. Salamon}, \mbox{K. Biljakovi\'c, A.
  Tomeljak, H. Sch{\"a}fer, J. Demsar, G. Socol}, \mbox{C. Ristocu, I.N.
  Mihailescu, Z. Siketic, I. Bogdanovic Radovic, G. Pletikapic, V. Svetlicic,
  M. Delsic}, \mbox{H. Samic}, and \mbox{J. Marcus}, J. Appl. Phys. 110 (2011),
  p. 014907.

\bibitem{Demsar11}
J.~Demsar, 2011. private communication.

\bibitem{Mantel99c}
O.C. Mantel, F.~Chalin, C.~Dekker, \mbox{H.S. J. van der Zant}, \mbox{Yu.I.
  Latyshev}, B.~Pannetier, and P.~Monceau, Synthetic Metals 103 (1999), p.
  2612; Phys. Rev. Lett. 84 (2000) 538.

\bibitem{Ayari02}
A.~Ayari and P.~Monceau, Phys. Rev. B 66 (2002), p. 235119.

\bibitem{Sinchenko11}
A.A. Sinchenko, P.~Monceau, and T.~Crozes, JETP Lett. 93 (2011), p.~56.

\bibitem{Latyshev98}
\mbox{Yu.I. Latyshev}, B.~Pannetier, and P.~Monceau, Eur. Phys. J. B 3 (1998),
  p. 421.

\bibitem{Slot04}
E.~Slot, M.A. Holst, \mbox{H.S. J. van der Zant}, and S.V. Zaitsev-zotov, Phys.
  Rev. Lett. 93 (2004), p. 176602.

\bibitem{Inagaki05}
K.~Inagaki, T.~Toshima, S.~Tanda, and K.~Yamaya, Applied Phys. Lett. 86 (2005),
  p. 073101.

\bibitem{Latyshev05}
\mbox{Yu.I. Latyshev}, P.~Monceau, S.~Brazovskii, A.P. Orlov, and T.~Fournier,
  Phys. Rev. Lett. 95 (2005), p. 266402.

\bibitem{Latyshev06}
\mbox{Yu.I. Latyshev}, P.~Monceau, S.~Brazovskii, A.P. Orlov, and T.~Fournier,
  Phys. Rev. Lett. 96 (2006), p. 116402.

\bibitem{Kroto85}
H.W. Kroto, J.R. Health, S.C. Obrien, R.F. Curl, and R.E. Smalley, Nature 318
  (1985), p. 162.

\bibitem{Ijima91}
S.~Ijima, Nature 354 (1991), p.~56.

\bibitem{Tanda02}
S.~Tanda, T.~Tsuneta, Y.~Okajima, K.~Inagaki, K.~Yamaya, and N.~Hatakenaka,
  Nature 417 (2002), p. 397.

\bibitem{Tsuneta04}
T.~Tsuneta and S.~Tanda, J. Crystal Growth 264 (2004), p. 223.

\bibitem{Hayashi07}
M.~Hayashi, H.~Ebisawa, and K.~Kuboki, Phys. Rev. B 76 (2007), p. 014303.

\bibitem{Bogachek90}
E.N. Bogachek, I.V. Krive, I.O. Kulik, and A.S. Rozhavskii, Sov. Phys. JETP 70
  (1990), p. 336; Phys. Rev. B 42 (1990) p. 7614.

\bibitem{Latyshev97}
\mbox{Yu.I. Latyshev}, O.~Laborde, P.~Monceau, and S.~Klaum{\"u}nzer, Phys.
  Rev. Lett. 78 (1997), p. 919.

\bibitem{Latyshev99}
\mbox{Yu.I. Latyshev}, O.~Laborde, T.~Fournier, and P.~Monceau, Phys. Rev. B 60
  (1999), p. 14019.

\bibitem{Visscher98}
M.I. Visscher and B.~Rejaei, Europhys. Lett 43 (1998), p. 617.

\bibitem{Duhot07}
S.~Duhot and R.~Melin, Phys. Rev. B 76 (2007), p. 184503.

\bibitem{Altshuler81}
B.L. Al'tshuler, A.G. Aronov, and B.Z. Spivak, JETP Lett. 33 (1981), p.~94.

\bibitem{Tsubota09}
M.~Tsubota, K.~Inagaki, and S.~Tanda, Physica B 404 (2009), p. 416.

\bibitem{Tsubota12}
M.~Tsubota, K.~Inagaki, T.~Matsuura, and S.~Tanda, Eur. Phys. Lett. 97 (2012),
  p. 57011.

\bibitem{Nathanson92}
B.~Nathanson, \mbox{O. Entin-Wohlman}, and B.~M{\"u}hlschegel, Phys. Rev. B 45
  (1992), p. 3499.

\bibitem{Yi97}
J.~Yi, M.Y. Choi, K.~Park, and \mbox{E.-H. Lee}, Phys. Rev. Lett. 78 (1997), p.
  3523.

\bibitem{Matsuura09}
T.~Matsuura, K.~Inagaki, and S.~Tanda, Phys. Rev. B 79 (2009), p. 014304.

\bibitem{Zant01}
\mbox{H.S. J. van der Zant}, E.~Slot, S.V. Zaitsev-Zotov, and S.N. Artemenko,
  Phys. Rev. Lett. 87 (2001), p. 126401.

\bibitem{Artemenko03}
S.N. Artemenko, Phys. Rev. B 67 (2003), p. 125420.

\bibitem{Kasatkin84}
A.L. Kasatkin and E.A. Pashitskii, Sov. J. Low Temp. Phys. 10 (1984), p. 640;
  Sov. J. Low Temp. Phys. 27 (1985) p. 1448.

\bibitem{Andreev64}
A.F. Andreev, JETP 19 (1964), p. 1228; 22 (1966) p. 455.

\bibitem{Son87}
\mbox{P.C. van Son}, \mbox{H. van Kempen}, and P.~Wyder, Phys. Rev. Lett. 59
  (1987), p. 2226.

\bibitem{Latyshev02a}
\mbox{Yu.I. Latyshev} and A.A. Sinchenko, JETP Lett. 75 (11) (2002), p. 593.

\bibitem{Rejaei96}
B.~Rejaei and G.E.W. Bauer, Phys. Rev. B 54 (1996), p. 8487.

\bibitem{Duhot07b}
S.~Duhot and R.~Melin, Eur. Phys. J. B 55 (2007), p. 289.

\bibitem{Visscher97}
M.I. Visscher and B.~Rejaei, Phys. Rev. Lett. 79 (1997), p. 4461.

\bibitem{Visscher96}
M.I. Visscher and G.E.W. Bauer, Phys. Rev. B 54 (1996), p. 2798.

\bibitem{Artemenko84b}
S.N. Artemenko and A.F. Volkov, JETP 60 (1984), p. 395; JETP Lett. 37 (1983) p.
  368.

\bibitem{Artemenko97a}
S.N. Artemenko, JETP 84 (1997), p. 823.

\bibitem{Sinchenko96}
A.A. Sinchenko, \mbox{Yu.I. Latyshev}, S.G. Zybtsev, I.G. Gorlova, and
  P.~Monceau, JETP Lett. 64 (1996), p. 285.

\bibitem{Sinchenko98}
A.A. Sinchenko, \mbox{Yu.I. Latyshev}, S.G. Zybtsev, and I.G. Gorlova, JETP
  Lett. 67 (1998), p. 164; JETP 86 (1998) p. 1001.

\bibitem{Sinchenko03a}
A.A. Sinchenko and P.~Monceau, Phys. Rev. B 67 (2003), p. 125117.

\bibitem{Sharvin65}
Y.V. Sharvin, JETP 21 (1965), p. 655.

\bibitem{Sinchenko03}
A.A. Sinchenko and P.~Monceau, J. Phys.: Condens. Matter 15 (2003), p. 4153.

\bibitem{Blonder82}
G.E. Blonder, M.~Tinkham, and T.M. Klapwijk, Phys. Rev. B 25 (1982), p. 4515.

\bibitem{Sinchenko07}
A.A. Sinchenko and P.~Monceau, Phys. Rev. B 76 (2007), p. 115129.

\bibitem{Sinchenko99}
A.A. Sinchenko, \mbox{Yu.I. Latyshev}, S.G. Zybtsev, I.G. Gorlova, and
  P.~Monceau, Phys. Rev. B 60 (1999), p. 4624.

\bibitem{Fournel86}
A.~Fournel, \mbox{J.-P. Sorbier}, M.~Konczykowski, and P.~Monceau, Phys. Rev.
  Lett. 57 (1986), p. 2199.

\bibitem{Sorbier96}
\mbox{J.-P. Sorbier}, H.~Tortel, P.~Monceau, and F.~L{\'e}vy, Phys. Rev. Lett.
  76 (1996), p. 676.

\bibitem{Neill06}
K.O. O'Neill, E.~Slot, R.E. Thorne, and \mbox{H.S. J. van der Zant}, Phys. Rev.
  Lett. 96 (2006), p. 096402.

\bibitem{Escudero01}
R.~Escudero, A.~Briggs, and P.~Monceau, J. Phys.: Condens. Matter 13 (2001), p.
  6285.

\bibitem{Kulik75}
I.O. Kulik and A.N. Omel'yanchuk, JEPT Lett. 21 (1975), p. 96; Sov. J. Low
  Temp. Phys. 3 (1977) p. 459.

\bibitem{Ambegaokar63}
V.~Ambegaokar and A.~Baratoff, Phys. Rev. Lett. 10 (1963), p. 486; 11 (1963) p.
  104.

\bibitem{Slot02}
E.~Slot and \mbox{H.S. J. van der Zant}, J. Physique IV 12 (2002), pp.
  Pr9--103.

\bibitem{Latyshev03}
\mbox{Yu.I. Latyshev}, P.~Monceau, A.A. Sinchenko, L.N. Bulaevskii, S.A.
  Brazovskii, T.~Kawae, and T.~Yamashita, J. Phys. A: Math. Gen. 36 (2003), p.
  9323.

\bibitem{Kleiner94}
R.~Kleiner and P.~M{\"u}ller, Phys. Rev. B 49 (1994), p. 1327.

\bibitem{Dai92}
Z.~Dai, C.G. Slough, and R.V. Coleman, Phys. Rev. B 45 (1992), p. 9469.

\bibitem{Ekino94}
T.~Ekino and J.~Akimitsu, Physica B 194-196 (1994), pp. 1221; Jp. J. Applied
  Physics 26 (1987) suppl. 26--3 p. 625.

\bibitem{Schafer03}
J.~Sch{\"a}fer, M.~Sing, R.~Claessen, E.~Rotenberg, X.J. Zhou, R.~Thorne, and
  S.D. Kevan, Phys. Rev. Lett. 91 (2003), p. 066401.

\bibitem{Perruchi04}
A.~Perruchi, L.~Degiorgi, and R.E. Thorne, Phys. Rev. B 69 (2004), p. 195114.

\bibitem{Latyshev02b}
\mbox{Yu.I. latyshev}, A.A. Sinchenko, L.N. Bulaevskii, V.N. Pavlenko, and
  P.~Monceau, JETP Letters 75 (2) (2002), p.~93.

\bibitem{Brazovskii05}
S.~Brazovskii, \mbox{Yu.I. Latyshev}, S.I. Matveenko, and P.~Monceau, J. Phys.
  IV (France) 131 (2005), p.~77.

\bibitem{Brazovskii89}
\mbox{S. Brazovskii in Ref.~\cite{R7Gorkov89}} , p. 425.

\bibitem{Clem90}
J.~Clem and M.~Coffey, Phys. Rev.  (1990).

\bibitem{Coppersmith91}
S.N. Coppersmith and A.J. Millis, Phys. Rev. B 44 (1991), p. 7799.

\bibitem{Balents98}
L.~Balents, M.C. Marchetti, and L.~Radzihovsky, Phys. Rev. B 57 (1998), p.
  7705.

\bibitem{Giamarchi98}
T.~Giamarchi and \mbox{P. Le Doussal}, Phys. Rev. B 57 (1998), p. 11356.

\bibitem{Radzihovsky98}
L.~Radzihovsky and J.~Toner, Phys. Rev. Lett. 81 (1998), p. 3711.

\bibitem{Markovic00}
N.~Markovic, M.A.H. Dohmen, and \mbox{H.S. J. van der Zant}, Phys. Rev. Lett.
  84 (2000), p. 534.

\bibitem{Artemenko00}
S.N. Artemenko, S.V. Zaitsev-Zotov, V.E. Minakova, and P.~Monceau, Phys. Rev.
  Lett. 84 (2000), p. 5184.

\bibitem{Yue01}
S.~Yue, M.~Tian, and Y.~Zhang, Phys. Rev. B 64 (2001), p. 113102.

\bibitem{Ayari02a}
A.~Ayari, thesis, University of Grenoble, unpublished,  2002.

\bibitem{Adelman95}
T.L. Adelman, S.V. Zaitsev-Zotov, and R.E. Thorne, Phys. Rev. Lett. 74 (1995),
  p. 5264.

\bibitem{Slot05}
E.~Slot, M.A. Holot, and \mbox{H.S. J. van der Zant}, J. Phys. IV (France) 131
  (2005), p. 223.

\bibitem{Yamamoto09}
H.M. Yamamoto, M.~Hosoda, Y.~Kawasugi, K.~Tsukagoshi, and R.~Kato, Physica B
  404 (2009), p. 413.

\bibitem{Overhauser68}
A.W. Overhauser, Phys. Rev. 167 (1968), p. 691.

\bibitem{Amarasekara82}
C.D. Amarasekara and P.H. Keesom, Phys. Rev. B 26 (1982), p. 2720.

\bibitem{Giebultowicz86}
T.M. Giebultowicz, A.W. Overhauser, and S.A. Werner, Phys. Rev. Lett. 56
  (1986), p. 2228.

\bibitem{Pintschovius87}
L.~Pintschovius, O.~Blaschko, G.~Krexner, \mbox{M. de Posta}, and R.~Currat,
  Phys. Rev. B 35 (1987), p. 9330.

\bibitem{Blaschko88}
O.~Blaschko, \mbox{M. de Posta}, and L.~Pintchovius, Phys. Rev. B 37 (1988), p.
  4258.

\bibitem{Blaschko84}
O.~Blaschko and G.~Krexner, Phys. Rev. B 30 (1984), p. 1667.

\bibitem{Smith87}
H.G. Smith, Phys. Rev. Lett. 58 (1987), p. 1228.

\bibitem{Ernst86}
G.~Ernst, C.~Artner, O.~Blaschko, and G.~Krexner, Phys. Rev. B 33 (1986), p.
  6465.

\bibitem{Dressel03}
M.~Dressel, Naturwissenchaften 90 (2003), p. 337.

\bibitem{Deshpande10}
V.V. Deshpande, M.~Bockrath, L.I. Glazman, and A.~Yacoby, Nature 464 (2010), p.
  209.

\bibitem{Fiete07}
G.A. Fiete, Rev. Mod. Phys. 79 (2007), p. 801.

\bibitem{Moser98}
J.~Moser, M.~Gabay, P.~Auban-Senzier, D.~J{\'e}r{\^o}me, K.~Bechgaard, and
  \mbox{J.-M. Fabre}, Eur. Phys. J. B 1 (1998), p.~39.

\bibitem{Kim06}
B.J. Kim, H.~Kok, E.~Rotenberg, \mbox{S.-J. Olr}, H.~Eisaki, N.~Motoyama,
  S.~Uchida, T.~Tohyama, \mbox{S. Mackawa, Z.-X. Shen}, and C.~Kim, Nature
  Phys. 2 (2006), p. 397.

\bibitem{Wang09}
F.~Wang, J.V. Alvarez, J.W. Allen, \mbox{S.-K. Mo}, J.~He, R.~Jin, D.~Mandrus,
  and H.~H{\"o}chst, Phys. Rev. Lett. 103 (2009), p. 136401.

\bibitem{Bockrath99}
M.~Bockrath, D.H. Cobden, J.~Lu, A.G. Rinzler, R.E. Smalley, L.~Balents, and
  P.L. McEuen, Nature 397 (1999), p. 598.

\bibitem{Aleskin04}
A.N. Aleskin, H.J. Lee, Y.W. Park, and K.~Akagi, Phys. Rev. Lett. 93 (2004), p.
  19601.

\bibitem{Zaitsev-Zotov00}
\mbox{S.V. Zaitsev-Zotov}, Y.A. Kumzerov, Y.A. Firsov, and P.~Monceau, J.
  Phys.: Condens. Matter 12 (2000), p. L303.

\bibitem{Rodin10}
A.S. Rodin and M.M. Fogler, Phys. Rev. Lett. 105 (2010), p. 106801; Phys. Rev.
  B 80 (2009) p. 155435.

\bibitem{Ishii03}
\mbox{H. Ishii, H. Kataura, H. Shiozawa, H. Yoshioka, H. Otsubo, Y. Takayama,
  T. Miyahara}, \mbox{S. Suzuki , Y. Achiba, M. Nakatake, T. Narimura, M.
  Higashiguchi, K. Shimada, H. Namatame}, and M.~Taniguchi, Nature 426 (2003),
  p. 540.

\bibitem{Schulz93}
H.J. Schulz, Phys. Rev. Lett. 71 (1993), p. 1864.

\bibitem{Meyer09}
J.S. Meyer and K.A. Matseev, J. Phys.: Condens. Matter 21 (2009), p. 023203.

\bibitem{Meirav89}
U.~Meirav, M.A. Kastner, M.~Heiblum, and S.J. Wind, Phys. Rev. B 40 (1989), p.
  5871.

\bibitem{Scott-Thomas89}
\mbox{J.H.F. Scott-Thomas}, S.B. Field, M.A. Kastner, H.I. Smith, and D.A.
  Antoniadis, Phys. Rev. Lett. 62 (1989), p. 583.

\bibitem{Field90}
S.B. Field, M.A. Kastner, U.~Meirav, \mbox{J.H.F. Scott-Thomas}, D.A.
  Antoniadis, H.I. Smith, and S.J. Wind, Phys. Rev. B 42 (1990), p. 3523.

\bibitem{Matseev04}
K.A. Matseev, Phys. Rev. Lett. 92 (2004), p. 106801.

\bibitem{Glazman92}
L.I. Glazman, I.M. Ruzin, and B.I. Shklovskii, Phys. Rev. B 45 (1992), p. 8454.

\bibitem{Rahman07}
A.~Rahman and M.K. Sanyal, Phys. Rev. B 76 (2007), p. 045110.

\bibitem{Deshpande08}
V.V. Deshpande and M.~Bockrath, Nature Phys. 4 (2008), p. 314.

\bibitem{Nayak00}
C.~Nayak, Phys. Rev. B 62 (2000), p. 4880.

\bibitem{Dora04a}
B.~Dora, K.~Maki, A.~Vanyolos, and A.~Virosztek, Europhys. Lett. 67 (2004), p.
  1024.

\bibitem{Dora04b}
B.~Dora, K.~Maki, and A.~Virosztek, Modern Phys. Lett. B 18 (2004), p. 327.

\bibitem{Basletic02}
M.~Basletic, \mbox{B. Korin-Hamzi{\'c}}, and K.~Maki, Phys. Rev. B 65 (2002),
  p. 235117.

\bibitem{Basletic07}
M.~Basletic, \mbox{B. Korin-Hamzi{\'c}}, K.~Maki, and S.~Tomic, Phys. Rev. B 75
  (2007), p. 052409.

\bibitem{Ishioka10}
J.~Ishioka, Y.H. Liu, K.~Shimatake, T.~Kurosawa, K.~Ichimura, Y.~Toda, M.~Oda,
  and S.~Tanda, Phys. Rev. Lett. 105 (2010), p. 174401.

\bibitem{Ishioka11}
J.~Ishioka, T.~Fujii, K.~Katano, K.~Ichimura, T.~Kurosawa, M.~Oda, and
  S.~Tanda, Phys. Rev. B 84 (2011), p. 245125.

\bibitem{Guillamon11}
I.~Guillamon, H.~Suderow, J.G. Rodrigo, S.~Vieira, P.~Rodi{\`e}re, L.~Cario,
  \mbox{E. Navarro-Moratalla}, \mbox{C. Marti-Gastaldo}, and E.~Coronado, New
  J. of Physics 13 (2011), p. 103020.

\bibitem{Wezel12}
\mbox{J. van Wezel}, Europhys. Lett. 96 (2011), p. 67011; Physics B 407 (2012)
  p. 1779.

\bibitem{Tokura00}
 (2000).

\bibitem{Mori98}
H.~Mori, S.~Tanaka, and T.~Mori, Phys. Rev. B 57 (1998), p. 12023.

\bibitem{Bender84}
K.~Bender, I.~Henning, D.~Schweitzer, K.~Dietz, H.~Endres, and H.J. Keller,
  Mol. Cryst. Liq. Cryst. 108 (1984), p. 359.

\bibitem{Takahashi06}
T.~Takahashi, Y.~Nogami, and K.~Yakushi, J. Phys. Soc. Jpn 75 (2006), p.
  051008.

\bibitem{Miyagawa00}
K.~Miyagawa, A.~Kawamoto, and K.~Kanoda, Phys. Rev. B 62 (2000), p. 7679.

\bibitem{Chiba00}
R.~Chiba, H.~Yamamoto, K.~Hiraki, T.~Takahashi, and T.~Nokamura, J. Phys. Chem.
  Solids 62 (2001), p. 389.

\bibitem{Yamamoto02}
K.~Yamamoto, K.~Yakushi, K.~Miyagawa, K.~Kanoda, and A.~Kawamoto, Phys. Rev. B
  65 (2002), p. 085110.

\bibitem{Watanabe04}
Y.N. M.~Watanabe Y.~Noda and H.~Mori, J. Phys. Soc. Jpn 73 (2004), p. 116.

\bibitem{Takano01}
Y.~Takano, K.~Hiraki, H.M. Yamamoto, T.~Nakamura, and T.~Takahashi, J. Phys.
  Chem. Solids 62 (2001), p. 393.

\bibitem{Wojciechowski03}
R.~Wojciechowski, K.~Yamamoto, K.~Yakushi, M.~Inokuchi, and A.~Kawamoto, Phys.
  Rev. B 67 (2003), p. 224105.

\bibitem{Moroto04}
S.~Moroto, K.~Hiraki, Y.~Takano, Y.~Kubo, T.~Takahashi, H.M. Yamamoto, and
  T.~Nakamura, J. Phys. IV (France) 114 (2004), p. 339.

\bibitem{Yamamoto08}
K.~Yamamoto, S.~Iwai, S.~Boyko, A.~Kashiwazaki, F.~Hiramatsu, C.~Okabe,
  N.~Nishi, and K.~Yakushi, J. Phys. Soc. Jpn 77 (2008), p. 074709.

\bibitem{Yamamoto11}
K.~Yamamoto, private communication, August 2011 .

\bibitem{Yamamoto10}
K.~Yamamoto, A.A. Kowabka, and K.~Yakushi, Appl. Phys. Lett. 96 (2010), p.
  122901.

\bibitem{Rothaemel86}
B.~Rothaemel, L.~Forro, J.R. Cooper, J.S. Schilling, M.~Weger, P.~Bele,
  H.~Brunner, D.~Schweitzer, and H.J. Keller, Phys. Rev. B 34 (1986), p. 704.

\bibitem{Kobayashi07}
N.~Kobayashi, S.~Katayama, Y.~Suzumura, and H.~Fukuyama, J. Phys. Soc. Jpn 76
  (2007), p. 034711.

\bibitem{Novoselov05}
K.S. Novoselov, A.K. Geim, S.V. Morozov, D.~Jiang, M.I. Katsnelson, I.V.
  Grigorieva, S.V. Dubonas, and A.A. Firsov, Nature 438 (2005), p. 197.

\bibitem{Novoselov05b}
K.S. Novoselov, D.~Jiang, F.~Schedin, T.J. Booth, V.V. Khotkevich, S.V.
  Morozov, and A.K. Geim, Proc. Natl. Acad. Sci. U.S.A. 102 (2005), p. 10451.

\bibitem{Mori98Nature}
S.~Mori, C.H. Chen, and \mbox{S.-W. Cheong}, Nature 392 (1998), p. 473.

\bibitem{Dagotto01}
E.~Dagotto, T.~Hotta, and A.~Moreo, Phys. Reports 344 (2001), p.~1.

\bibitem{Dagotto05}
E.~Dagotto, Science 309 (2005), p. 257.

\bibitem{Vojta09}
M.~Vojta, Adv. in Physics 58 (2009), p. 699.

\bibitem{Tranquada12}
J.M. Tranquada, Physica B 407 (2012), p. 1771.

\bibitem{Kohsaka07}
Y.~Kohsaka, C.~Taylor, K.~Fujita, A.~Schmidt, C.~Lupien, T.~Hanaguri, M.~Azuma,
  and M.~Takano\mbox{, H. Eisaki, H. Takagi, S. Uchida and J.C. Davis}, Science
  315 (2007), p. 1380.

\bibitem{Hucker11}
M.~H{\"u}cker, \mbox{M.v. Zimmermann}, G.D. Gu, Z.J. Xu, J.S. Wen, G.~Xu, H.J.
  Kang, A.~Zheludev, and J.M. Tranquada, Phys. Rev. B 83 (2011), p. 104506.

\bibitem{Berg09}
E.~Berg, E.~Fradkin, and S.A. Kivelson, Phys. Rev. B 79 (2009), p. 064515.

\bibitem{Chakravarty01}
S.~Chakravarty, R.B. Laughlin, D.K. Morr, and C.~Nayak, Phys. Rev. B 63 (2001),
  p. 094503.

\bibitem{Varma06}
C.M. Varma, Phys. Rev. B 73 (2006), p. 155113.

\bibitem{Fradkin10}
E.~Fradkin, S.A. Kivelson, M.J. Lawler, J.P. Eisenstein, and A.P. MacKenzie,
  Annu. Rev. Condens. Matter Phys. 1 (2010), p. 153.

\bibitem{Hinkov08}
V.~Hinkov, D.~Haug, B.~Fauque, P.~Bourges, Y.~Sidis, A.~Ivanov, C.~Bernhard,
  C.T. Lin, and B.~Keimer, Science 319 (2008), p. 597.

\bibitem{Lawler10}
M.S. Lawler, K.~Fujita, J.~Lee, A.R. Schmidt, Y.~Kohsaka, C.K. Kim, H.~Eisaki,
  and S.~Uchida\mbox{, J.C. Davis, J.P. Sethna, and E.-A. Kim}, Nature 466
  (2010), p. 347.

\bibitem{Zhang12}
Y.~Zhang, C.~He, R.R. Ye, J.~Jiang, F.~Chen, M.~Xu, Q.Q. Ge, B.P. Xie, J.~Wei,
  and M.~Aeschlimann\mbox{, X.Y. Cui, M. Shi, J.P. Hu, and D.L. Feng}, Phys.
  Rev. B 85 (2012), p. 085121.

\bibitem{Chu10}
J.H. Chu, J.G. Analytis, \mbox{K. de Greve}, M.L. McMahon, Z.~Islam,
  Y.~Yamamoto, and I.R. Fisher, Science 329 (2010), p. 824.

\bibitem{Mesaros11}
A.~Mesaros, K.~Fujita, H.~Eisaki, S.~Uchida, J.C. Davis, S.~Sachdev, J.~Zaanen,
  M.J. Lawler, and \mbox{E.-A. Kim}, Science 333 (2011), p. 426.

\bibitem{Fauque06}
B.~Fauqu{\'e}, Y.~Sidis, V.~Hinkov, S.~Pailh{\`e}s, C.T. Lin, X.~Chaud, and
  P.~Bourges, Phys. Rev. Lett. 96 (2006), p. 197001.

\bibitem{Li08}
Y.~Li, V.~Baledent, N.~Bari{\u s}i{\'c}, Y.~Cho, B.~Fauqu{\'e}, Y.~Sidis,
  G.~Yu, X.~Zhao, P.~Bourges, and M.~Greven, Nature 455 (2008), p. 372.

\bibitem{Li10}
\mbox{Y. Li, V. Baledent, G. Yu, N. Bari{\u s}i{\'c}, K. Hradil, R. A. Mole, Y.
  Sidis, P. Steffens}, X.~Zhao, P.~Bourges, and M.~Greven, Nature 468 (2010),
  p. 283.

\bibitem{Blumberg02}
G.~Blumberg, P.~Littlewood, A.~Gozar, B.S. Dennis, N.~Motoyama, H.~Eisaki, and
  S.~Uchida, Science 297 (2002), p. 584.

\bibitem{Gorshunov02}
B.~Gorshunov, P.~Haas, T.~Room, M.~Dressel, T.~Vuleti{\'c}, \mbox{B.
  Korin-Hamzi{\'c}}, S.~Tomic, J.~Akimitsu, and T.~Nagata, Phys. Rev. B 66
  (2002), p. 060508.

\bibitem{Vuletic06}
T.~Vuleti{\'c}, \mbox{B. Korin-Hamzi{\'c}}, T.~Ivek, S.~Tomic, B.~Gorshunov,
  M.~Dressel, and J.~Akimitsu, Physics Reports 428 (2006), p. 169.

\bibitem{Iye85}
Y.~Iye and G.~Dresselhaus, Phys. Rev. Lett. 54 (1985), p. 1182.

\bibitem{Takahide10}
Y.~Takahide, M.~Kimata, K.~Hazama, T.~Terashima, S.~Uji, T.~Konoike, and H.M.
  Yamamoto, Phys. Rev. B 81 (2010), p. 235110.

\bibitem{Takahide06}
Y.~Takahide, T.~Konoike, K.~Enomoto, M.~Nishimura, T.~Terashima, S.~Uji, and
  H.M. Yamamoto, Phys. Rev. Lett. 96 (2006), p. 136602.

\bibitem{Nad08}
\mbox{F.Ya. Nad}, P.~Monceau, and H.M. Yamamoto, J. Phys.: Condens. Matter 20
  (2008), p. 485211.

\bibitem{Mori09}
T.~Mori, T.~Ozawa, Y.~Bando, T.~Kawamoto, S.~Niizeki, H.~Mori, and T.~Terasaki,
  Phys. Rev. B 79 (2009), p. 115108.

\bibitem{Mori07b}
H.~Mori, I.~Terasaki, and H.~Mori, J. Mat. Chem. 17 (2007), p.~1.

\bibitem{Chen97}
C.H. Chen and S.W. Cheong, J. Appl. Phys. 81 (1997), p. 4326.

\bibitem{Loudon05}
J.C. Loudon, S.~Cox, A.J. Williams, J.P. Attfield, P.B. Littlewood, P.A.
  Midgley, and N.D. Mathur, Phys. Rev. Lett. 94 (2005), p. 097202.

\bibitem{Cox08}
S.~Cox, J.~Singleton, R.D. McDonald, A.~Migliori, and P.B. Littlewood, Nature
  Materials 7 (2008), p. 25; 9 (2010) p. 689.

\bibitem{Wahl03}
A.~Wahl, S.~Mercone, A.~Pautrat, M.~Pollet, \mbox{Ch. Simon}, and
  D.~Sedmidubsky, Phys. Rev. B 68 (2003), p. 094429.

\bibitem{Barone09}
C.~Barone, A.~Galdi, N.~Lampis, L.~Maritato, \mbox{F. Miletto Granozio},
  S.~Pagano, P.~Perna, M.~Radovic, and \mbox{U. Scotti di Uccio}, Phys. Rev. B
  80 (2009), p. 115128.

\bibitem{Niizeki08}
S.~Niizeki, F.~Yoshikane, K.~Kohno, K.~Takahashi, H.~Mori, Y.~bando,
  T.~Kawamoto, and T.~Mori, J. Phys. Soc. Jpn 77 (2008), p. 073710.

\bibitem{Andrei97}
 (1997).

\bibitem{Grimes79}
C.C. Grimes and G.~Adams, Phys. Rev. Lett. 42 (1979), p. 795.

\bibitem{Jiang89}
H.W. Jiang and A.J. Dahm, Phys. Rev. Lett. 62 (1989), p. 1396.

\bibitem{Kristensen96}
A.~Kristensen, K.~Djerfi, P.~Fozooni, M.J. Lea, P.J. Richardson, \mbox{A.
  Santrich-Badal}, A.~Blackburn, and \mbox{R.W. van der Heijiden}, Phys. Rev.
  Lett. 77 (1996), p. 1350.

\bibitem{Glasson01}
P.~Glasson, V.~Dotsenko, P.~Fozooni, M.J. Lea, W.~Bailey, G.~Papageorgiou, S.E.
  Andersen, and A.~Kristensen, Phys. Rev. Lett. 87 (2001), p. 176802.

\bibitem{Shirahama95}
K.~Shirahama and K.~Kono, Phys. Rev. Lett. 74 (1995), p. 781.

\bibitem{Pudalov93}
V.M. Pudalov, \mbox{M. D'Iorio}, S.V. Kravchenko, and J.W. Campbell, Phys. Rev.
  Lett. 70 (1993), p. 1866.

\bibitem{Shayegan97}
{\rm M. Shayegan (chap. 9) and H. Fertig (chap. 5) in} (1997).

\bibitem{Goldman90}
V.J. Goldman, M.~Santos, M.~Shayegan, and J.E. Cunningham, Phys. Rev. Lett. 65
  (1990), p. 2189.

\bibitem{Jiang91}
H.W. Jiang, H.L. Stormer, D.C. Tsui, L.N. Pfeiffer, and K.W. West, Phys. Rev. B
  44 (1991), p. 8107.

\bibitem{Williams91}
F.I.B. Williams, P.A. Wright, R.G. Clark, E.Y. Andrei, G.~Deville, D.C. Glatti,
  O.~Probst, B.~Etienne, C.~Dorin, C.T. Foxon, and J.J. Harris, Phys. Rev.
  Lett. 66 (1991), p. 3285.

\bibitem{Ye02}
P.D. Ye, L.W. Engel, D.C. Tsui, R.M. Lewis, L.N. Pfeiffer, and K.~Weat, Phys.
  Rev. Lett. 89 (2002), p. 176802.

\bibitem{Li00}
C.C. Li, J.~Yoon, L.W. Engel, D.~Shahar, D.C. Tsui, and M.~Shayegan, Phys. Rev.
  B 61 (2000), p. 10905.

\bibitem{Chitra98}
R.~Chitra, T.~Giamarchi, and \mbox{P. Le Doussal}, Phys. Rev. Lett. 80 (1998),
  p. 3827; Phys. Rev. B 65 (2001) p. 035312.

\bibitem{Chitra05}
R.~Chitra and T.~Giamarchi, Eur. Phys. J. B 44 (2005), p. 455.

\bibitem{Chen03}
Y.~Chen, R.M. Lewis, L.W. Engel, D.C. Tsui, P.D. Ye, L.N. Pfeiffer, and K.W.
  West, Phys. Rev. Lett. 91 (2003), p. 016801.

\bibitem{Lilly99}
M.P. Lilly, K.B. Cooper, J.P. Eisenstein, L.N. Pfeiffer, and K.W. West, Phys.
  Rev. Lett. 82 (1999), p. 394; 83 (1999) p. 824.

\bibitem{Fogler96}
M.M. Fogler, A.A. Koulakov, and B.I. Shklovskii, Phys. Rev. B 54 (1996), p.
  1853.

\bibitem{Cooper03}
K.B. Cooper, J.P. Eisenstein, L.N. Pfeiffer, and K.W. West, Phys. Rev. Lett. 90
  (2003), p. 226803.

\bibitem{Sambandamurphy08}
G.~Sambandamurphy, R.M. Lewis, H.~Zhu, Y.P. Chen, L.W. Engel, D.C. Tsui, L.N.
  Pfeiffer, and K.W. West, Phys. Rev. 100 (2008), p. 256801.

\bibitem{Radisavljevic11}
B.~Radisavljevic, A.~Radinovic, J.~Brivio, and A.~Kis, Nature Nanotechn. 6
  (2011), p. 147.

\bibitem{Mak10}
K.F. Mak, C.~Lee, J.~Hone, J.~Shan, and T.F. Heinz, Phys. Rev. Lett. 105
  (2010), p. 136805.

\bibitem{DiMasi95}
\mbox{E. Di-Masi}, M.C. Aronson, J.F. Mansfield, B.~Foran, and S.~Lee, Phys.
  Rev. B 52 (1995), p. 14516.

\bibitem{Brouet08}
V.~Brouet, W.L. Yang, X.J. Zhou, Z.~Hussain, R.G. Moore, R.~He, D.H. Lu, Z.X.
  Shen, and J.~Laverock\mbox{, S.B. Dugdale, N. Ru, and I.R. Fisher}, Phys.
  Rev. B 77 (2008), p. 235104.

\bibitem{Ru08}
N.~Ru, C.L. Condon, G.Y. Margulis, K.Y. Shin, J.~Laverock, S.B. Dugdale, M.F.
  Toney, and I.R. Fisher, Phys. Rev. B 77 (2008), p. 035114.

\bibitem{Sinchenko12}
A.A. Sinchenko, P.~Lejay, and P.~Monceau, Phys. Rev. B 85 (2012), p. 241104(R).

\bibitem{Yao06}
H.~Yao, J.A. Robertson, \mbox{E.-A. Kim}, and S.A. Kivelson, Phys. Rev. B 74
  (2006), p. 245126.

\bibitem{DiSalvo80}
F.J. DiSalvo and R.M. Fleming, Solid State Commun. 35 (1980), p. 685.

\end{thebibliography}

\end{document}